\pdfoutput=1

\documentclass[10pt,encabcentrados,simple,unoymedio,emacs,index]{pcftesis}

\usepackage{colapso_cmds} \usepackage{colortbl}
\usepackage{ctable} \usepackage{multirow}
\usepackage{etex} 

 \ifpdf
\usepackage{pdflscape} \else \usepackage{lscape} \fi

\usepackage{srcltx}

\definechangesauthor[sudarsky]{sudarsky}{red}
\definechangesauthor[nano]{nano}{blue}

\autor{Adolfo J. De Unánue Tiscareño}
\numerocuenta{503011365} \planestudios{5009}
\titulo{Orígenes cuánticos de las asimetrías cosmológicas}
\tituloingles{The quantum origins of the cosmological
  asymmetry} \directorTesis{Daniel Eduardo Sudarsky}
\comitetutorala{Chryssommalis Chryssomalakos}
\comitetutoralb{Alejandro Corichi} \sinodald{Marcelo
  Salgado Rodríguez} \sinodale{Roberto Capovilla Chiariglione} \sinodalf{Alberto
  Guijosa Hidalgo} \sinodalg{Luis de la Peña Auerbachs} \sinodalh{Jerónimo
  Cortez Quezada} \sinodali{Manuel Torres} 
  \mesexamen{Enero}
  \anoexamen{2010}
\responsableVA{Dr. Isidro Ávila Martínez}
\tituloresponsableVA{Director General de Administración
  Escolar} \resumen{La versión estándar sobre el origen de
  la estructura cosmológica dada por el paradigma
  inflacionario, en el cual la estructura es resultado de
  las fluctuaciones cuánticas ocurridas durante la etapa
  inflacionaria, es poco satisfactoria: ¿Exactamente cómo el
  Universo transita de una época homogénea e isotrópica a
  una donde las incertidumbres cuánticas se convierten en
  fluctuaciones inhomogénas? El problema de la versión
  estándar es que las predicciones del modelo inflacionario
  respecto a este punto no se encuentran justificadas por
  ningún esquema de mecánica cuántica. Una propuesta de
  solución fue hecha en [A. Perez, H. Sahlmann and
  D. Sudarsky, Classical and Quantum Gravity \textbf{23},
  2317 (2006)]. En ese trabajo se propone un proceso similar
  al colapso de la función de onda mecanico-cuántica de los
  diversos modos del campo inflatónico. Esta idea fue
  inspirada por trabajos de R. Penrose en los cuales la
  gravedad cuántica tiene un papel en el rompimiento de la
  evolución unitaria estándar de la mecánica cuántica. En
  este trabajo de tesis proponemos y estudiamos un nuevo
  esquema de colapso que sigue las correlaciones indicadas
  en el funcional de Wigner del estado inicial. Se
  dedujeron, también, las fórmulas para múltiples colapsos y
  sus efectos en el espectro de potencias. Además, en esta
  tesis nos enfocamos en un estudio más detallado de los
  tres esquemas de colapso y de su posible traza en los
  datos observacionales.}

\abstract{The standard inflationary version of the origin of
  the cosmic structure as the result of the quantum
  fluctuations during the inflationary stage is less than
  fully satisfactory: how exactly does the Universe transit
  from a homogeneous and isotropic stage to one where the
  quantum uncertainties become actual inhomogeneous
  fluctuactions? The point is that the predictions of
  inflation in this regard cannot be fully justified in any
  known  scheme of quantum physics. A
  proposal was made in [A. Perez, H. Sahlmann and
  D. Sudarsky, Classical and Quantum Gravity \textbf{23},
  2317 (2006)] to solve this problem by invoking a process
  similar to the collapse of the quantum mechanical wave
  function of the various modes of the inflaton field. This
  in turn was inspired by ideas of R. Penrose about the
  roles that quantum gravity might play in bringing about
  such breakdown of standard unitary evolution of quantum
  mechanics. In this work we propose and study a new
  collapse scheme that follows the correlations indicated in
  the Wigner functional of the initial state. Furthermore,
  in this thesis work we will focus on a more detailed study
  of  three collapse schemes (robustness, multiple
  collapses) and on the traces they might leave on the
  observational data.}  

\dedicatoria{A mis padres por su confianza,\\ a mis hermanos
  por su apoyo, \\a mis amigos por su aliento.}

\agradecimientos{
Quisiera agradecer a todas las personas que
estuvieron conmigo todo este tiempo y no perdían la
confianza en que acabaría la tesis doctoral algún día. En
particular quiero agradecer a mis padres por educarme del
modo en el que soy y que eso me ayudó a terminar ¡Gracias
mamá!¡Gracias papá!  A mi hermano \textbf{Ernesto} que se \emph{chutó} toda
la parte dura del doctorado conmigo y me apoyó para
sobrevivir, sin él no hubiera sido posible nada de esto
(gracias mil, \emph{erni}). A
mi hermana \textbf{Daniela} que me ayudaba a mantener
la cordura. A mi otro hermano \textbf{Wicho} ¿Que hubiera
sido sin su apoyo?

 Quiero 
agradecer a mis amigos de mty (y de la infancia) \textbf{Juan Carlos}
(\emph{Pachón}), \textbf{Victor}
(\emph{Gigio}) y  a \textbf{Homero} (\emph{el
  ingeniero}), por darme consejos y 
palabras de aliento. A mi amiga del alma \textbf{Elena} que
como siempre me acompaña en todas mis locuras. A mis amigos
y compañeros de viaje \textbf{Alexander}, \textbf{Nacho} y
\textbf{Maria}\emph{neo}, con los que discutí, sufrí,
compartí miles de horas iluminadoras en varios aspectos de
mi vida. Quiero agradecer con todos los honores a mi asesor
y amigo \textbf{Daniel Sudarsky} que aconsejó y me enseñó a
ser mejor persona en varios aspectos, el me enseñó como ser
un excelente científico y ser humano y espero (\emph{sólo
  espero}) que se me haya quedado algo de lo mucho que me
enseñó. Muchísimas gracias por todo profesor (y gracias por
aguantarme, orientarme, soportarme, y un largo
etc. ). También quiero agradecer a mi
 asesor \textbf{Chryssommalis Chryssomalakos}, que con
su conocimiento y agudeza siempre me sorprendía, enseñaba y
me motivaba a ser mejor cada día. Quiero agradecer a los
sinodales por tomarse la molestia de leer este
\emph{mamotreto}\footnote{El profesor Capovilla
  acertadamente lo llamo \emph{War and Peace}, obviamente no por el esplendor
  literario que hay en la tesis.} lleno de errores
gramáticales y corregirlos 
(¡espero no haya más en los agradecimientos!). Gracias
también a \textbf{Yanalté} y \textbf{Manuel Torres} por su
paciencia y apoyo en todos los momentos de este \emph{laaaargo}
doctorado.

De verdad, sé que faltan personas, se que las que mencioné
no les he hecho justicia, pero, a todos ustedes, no saben el
inmenso bien que me ha hecho conocerlos. \textbf{¡Muchas Gracias!}

}
\archivobibTeX{bibliografia/tesis}

\ifpdf \hypersetup{%
  unicode=true, pdfauthor=Adolfo De Unanue, pdftitle
  =Origenes de las asimetrias cosmologicas, }%
\fi

\begin{document}
   \PCFfrontmatter

\begin{introduccion}




\section*{Inflación, el origen de la estructura y el
  Problema de la medición}

En los últimos años, se han producido avances espectaculares
en la cosmología física, resultantes del notable aumento de
la exactitud en las técnicas observacionales. Ejemplos de
estos éxitos son los estudios de las supernovas
\cite{Perlmutter96,Perlmutter97,Riess98,Riess04,WoodVasey07},
los estudios de estructura a gran escala
\cite{Tegmark97,Tegmark04,Tegmark04b,Gott05,Abazajian09,DGYork00}
y las observaciones extraordinariamente precisas de varios
estudios recientes del \acf{CMB} en particular los de
\acf{WMAP} \cite{wmap5}. En particular, las observaciones
llevadas a cabo por el WMAP han fortalecido la percepción
positiva de los escenarios inflacionarios entre los
cosmólogos.

Cabe señalar sin embargo que gran parte de la atención de la
investigación teórica en inflación se ha orientado hacia el
dilucidación de la forma exacta del modelo inflacionario (es
decir, el número de campos, la forma del potencial, y la
aparición de acoplamientos no-mínimos a la gravedad, por
mencionar algunos) y se ha prestado menos atención a las
cuestiones de principio, como pueden ser la determinación de
las condiciones iniciales, explicaciones de la baja entropía
del estado inicial, el mecanismo que ocasiona el tránsito de un
universo homogéneo e isotrópico a una etapa en que las
incertidumbres cuánticas se vuelven fluctuaciones
no-homogéneas, etc.\ . Por supuesto existen trabajos en los
cuales se abordan estas cuestiones
\cite{Polarski1996,Kiefer98b,Kiefer98a,Kiefer08,Halliwell89},
pero como se aclara en \cite{Sudarsky06a,Sudarsky06b,
  Sudarsky07,Sudarsky09} una explicación plenamente
satisfactoria parece requerir algo más allá de la
comprensión actual de las leyes de la física. Los autores
(\citeauthor*{Sudarsky06a}) propusieron la \textit{hipótesis
  del colapso} siguiendo unas ideas de Penrose
\cite{Penrose94,Penrose96,Penrose02,Penrose05}.

El punto neurálgico de la explicación tradicional es el
hecho de que las predicciones del paradigma inflacionario en
este sentido no pueden estar satisfactoriamente
justificadas en la mecánica cuántica estándar. La
Interpretación de Copenhague, por ejemplo, no es aplicable
en este caso, debido al hecho de que nosotros, los
observadores, somos parte del sistema, y para hacer
las cosas peores, de hecho, somos parte de el resultado del
proceso que queremos entender.

Aunque se tratarán con mayor extensión en el cuerpo principal
del trabajo de tesis (caps. \ref{cha:critica-al-origen} y
\ref{cha:analisis-esquemas}), veremos de manera sucinta
aquí las motivaciones de la \textit{hipótesis del colapso}
como una posible explicación para el origen de las semillas
de la estructura cósmica. Empezaremos revisando la
explicación estándar del paradigma inflacionario.

\begin{itemize}
\item El universo comienza homogéneo y isotrópico
  \footnote{Inflación podría funcionar aún si no comenzamos
    --estrictamente-- con esta condición, pero después de
    algunos \textit{e-foldings} el universo estará
    efectivamente homogéneo e isotrópico.}. El campo del
  inflatón es la especie dominante, y se encuentra en su
  estado de vacío cuántico, que también es \textit{homogéneo
    e isotrópico}. El campo del inflatón, se describe en
  términos de su valor esperado representado como un campo
  escalar que depende sólo del tiempo cósmico, pero no de
  las coordenadas espaciales, $\varphi = \varphi(t)$ y una
  fluctuación cuántica $\dphih$ que se encuentra en el
  estado de vacío adiabático, el cual también es homogéneo e
  isotrópico (algo que puede ser fácilmente verificado por
  la aplicación de los generadores de las rotaciones o las
  traslaciones a ese estado).

\item Las ``fluctuaciones'' cuánticas del inflatón actúan
  como perturbaciones \footnote{Encontramos esta redacción
    desafortunada porque podría inducir a pensar que algo
    está fluctuando en el sentido del movimiento browniano,
    mientras que el uso de palabras como ``incertidumbre
    cuántica'' evocan a la función de onda asociada al
    estado base del oscilador armónico, analogía más cercana
    de la situación.} del campo del inflatón y a través de
  las ecuaciones de Einstein como perturbaciones de la
  métrica.

\item Mientras la época inflacionaria continúa las
  longitudes de onda varios modos del campo inflatónico
  crecen hasta ser mayores que el radio de Hubble (En la
  literatura se conoce a esto como ``cruce del horizonte'' o
  ``salir del horizonte''\footnote{Otra redacción
    desafortunada, ya que identifica al \emph{horizonte de eventos} 
    con el \emph{radio de Hubble}. esto se explicará a más detalle en el capítulo 1.}), 
    las amplitudes cuánticas de los modos para modos mayores que
  el radio de Hubble  se "congelan" debido a la evolución y en este momento se considera a estos
  modos como fluctuaciones de un campo clásico. Más tarde
  las fluctuaciones ``reentran al radio de Hubble''
  transformándose en este momento en las semillas de la
  estructura cosmológica dejando su marca en el CMB.
\end{itemize}

El último paso es referido en la literatura como
\textit{transición cuántica-clásica}. Esta transición es
entendida de diferentes maneras por diferentes
investigadores, pero la principal línea de argumentación
está centrada en alguna forma de \textit{decoherencia}
\cite{Kiefer98b,Kiefer98b,Kiefer98a,Kiefer00a,Kiefer08,Halliwell89}. El
intento de resolver el problema de la medición en la
mecánica cuántica usando decoherencia (o algún derivado de
esta) ha sido criticada
en varios artículos \citep[por
ejemplo][]{Adler03,Ghirardi09,PearlePhil94}, pero como se
argumenta en
\cite{Sudarsky06a} el problema de la medición es más agudo
en el escenario cosmológico y por lo tanto es menos adecuado 
intentar resolverlo únicamente con decoherencia.

La postura adoptada en este trabajo de tesis (y concordante
con \cite{Sudarsky06a}) está basada en la idea de que toda la
física siempre es cuántica, y que el único papel que puede
tener una
descripción clásica es la de una aproximación en que las
incertidumbres del estado del sistema son insignificantemente
pequeñas y pueden ser ignoradas, permitiendo así que se
pueden tomar los valores de expectación 
cuánticos como una descripción razonable de los aspectos más
importantes de la situación estudiada. Sin embargo hay que
tener en cuenta que detrás de cualquier aproximación clásica
hay una descripción completamente cuántica y por lo tanto,
se debe rechazar cualquier descripción clásica en la que el
universo es heterogéneo y anisotrópico pero al nivel de
descripción cuántico se insiste en asociar al universo un
estado homogéneo e isotrópico en todo momento.

En este trabajo de tesis siguiendo el artículo de
\citeauthor*{Sudarsky06a} \cite{Sudarsky06a}, se introduce
una modificación al paradigma inflacionario para poder
generar las semillas de la estructura cósmica: \textit{la
  hipótesis del colapso auto-inducido}. Es decir, se
considera un régimen específico en el cual la función de
onda que describe a las fluctuaciones del inflatón en el
estado vacío sufre un colapso \textit{objetivo} que lo
sacará de este estado homogéneo e isotrópico a algún otro
estado con características diferentes. El mecanismo por el
cual se cree que este colapso ocurre está inspirado en ideas
de Roger Penrose
\cite{Penrose94,Penrose96,Penrose02,Penrose05} en
las cuales el colapso de 
la función de onda se supone causado de alguna manera por
aspectos gravitación cuántica.

La manera en que tratamos la transición de nuestro sistema
de un estado que es isotrópico y homogéneo a uno que no lo
sea, es asumir que a un determinado tiempo cósmico, ``algo''
induce el salto de un estado que describe un modo particular
del campo cuántico (de manera similar a la regla de la
mecánica cuántica que asocia un colapso de la función de
onda al realizar una medición del sistema, salvo que en la
situación cosmológica estudiada no hay un actor externo o
dispositivo de observación que realice la operación de
medición).

El principal objetivo de este trabajo de tesis es comparar
los resultados que surgen del análisis de diferentes
esquemas de colapso (los dos estudiados con anterioridad y
uno nuevo) y/o de diversas situaciones como colapsos
múltiples o tardíos, con las observaciones del CMB.

Este trabajo de tesis está organizado de la siguiente
manera: En el primer capítulo
(\nameref{cha:cosmologia-estandar}) se presenta una revisión
general del modelo cosmológico estándar, incluyendo
éxitos y sus problemas. Algunos de  los problemas del
modelo estándar cosmológico (los de ``naturalidad'') son resueltos 
al ser suplementado con una etapa
inflacionaria, pero como se verá, la importancia
fundamental del modelo inflacionario es la supuesta
explicación que ofrece del origen de las perturbaciones
primordiales. La aparición de las perturbaciones
primordiales y su evolución hasta el CMB será el tema de la
primera mitad del capítulo
``\nameref{cha:critica-al-origen}''; la segunda parte del
mismo capítulo \ref{cha:critica-al-origen} se centrará en la
crítica a la explicación estándar y se discutirá a detalle
la \textit{hipótesis de colapso} propuesta en
\cite{Sudarsky06a}. En el capítulo
\ref{cha:analisis-esquemas} titulado
\nameref{cha:analisis-esquemas} presentaremos los desarrollos
realizados en este trabajo de tesis: se estudiará un nuevo
esquema de colapso diferente a los propuestos en
\cite{Sudarsky06a} basado en el funcional de Wigner (sección
\ref{sec:colapso_wigner}), se comparará la robustez de las
predicciones hechas por cada esquema de colapso para un solo
colapso por modo del campo del inflatón durante la etapa
inflacionaria (subsección
\ref{sec:un-unico-colapso}). Además se estudiará
los casos de múltiples colapsos
por modo del campo suponiendo que el estado post-colapso es
un estado coherente ~\ref{sec:mult-colaps-coherente}. Se
finalizará el capítulo \ref{cha:analisis-esquemas}
reafirmando las diferencias conceptuales y predicciones de
la hipótesis del colapso con respecto a la explicación
estándar inflacionaria. Finalmente se
presentan las conclusiones del trabajo de tesis  en el capítulo \ref{cha:conclusiones}. 
Se incluyen
además apéndices que tratan sobre la teoría de
perturbaciones en Relatividad General (apéndice
\ref{cha:teor-de-pert}), comparaciones entre distintas
fórmulas que aparecen en la literatura inflacionaria
(apéndice \ref{cha:equiv-de-ecuac}),
y se recogen en los apéndices finales
algunas formulas, datos numéricos y significado de las
siglas usadas en este trabajo de tesis para comodidad del
lector. 

\section*{Convenciones seguidas en el trabajo de tesis}

\paragraph{Cantidades geométricas}
La notación usada para representar campos tensoriales será
la definida en \citep[][pags. 23-26]{Wald84}, conocida como
\textit{notación abstracta de índices}, e.g.\ el campo
tensorial (0,2), $\bm{T}$, se expresará con índices latinos
$T_{ab}$ cuando no este expresado en ningún sistema
coordenado y cuando lo queramos especificar en un sistema 
coordenado usaremos índices griegos $T_{\mu\nu}$.

Los índices griegos son índices espacio-temporales y pueden
tomar los valores de $\{1,2,3,4\}$, índices latinos en el
contexto de campos tensoriales referidos a un sistema
coordenado tales como $i, j, k$ representarán índices
espaciales.

La métrica del espacio-tiempo $g_{ab}$ es de cuatro
dimensiones  y tiene la signatura $(-,+,+,+)$. La métrica
$g_{ab}$ cumple con $g^{ac}g_{cb} = \delta\indices{^a_b}$. La métrica
también se utiliza para ``elevar'' y ``bajar''  índices
espacio-temporales, i.e.\ $T\indices{_b^a} \equiv g^{ac}
T\indices{_{bc}}$.

El tensor de \index{Riemann}Riemann $\Riemann$ sigue la
convención de signos 
dada por \cite[][pag. 38]{Wald84} y es definido de manera
libre de coordenadas mediante
\begin{equation}\tag{i.A}
  \label{eq:riemann_def}
  \left(\nabla_a\nabla_b - \nabla_b\nabla_a\right) t^c = -
  \Riemann\thinspace t^d
\end{equation}
en esta ecuación $\nabla_a$ es la derivada covariante de la
variedad $\mathcal{M}$ que es compatible con la métrica,
i.e.\ $\nabla_a g_{bc} = 0$. El tensor de Riemann cuando
está referido a un sistema  de coordenadas se expresa como
\citep[][pag. 48]{Wald84} 
\begin{equation*}\tag{i.B}
  R_{\mu\nu\rho}^{\quad\sigma}  = \partial_\nu
  \Gamma^\sigma_{\mu\rho} - \partial_\mu
  \Gamma^\sigma_{\nu\rho} + 
  \left(\Gamma^\alpha_{\mu\rho} \Gamma^\sigma_{\alpha\nu} -
    \Gamma^\alpha_{\nu\rho} \Gamma^\sigma_{\alpha\mu}\right)
\end{equation*}
donde los $\tensor{\Gamma}{^a_{bc}}$ son los llamados
símbolos de Christoffel y están definidos mediante
\begin{equation}\tag{i.C}
  \label{eq:simbolos_christoffel}
  \Gamma^a_{\;bc} := \frac{1}{2}g^{ad}(g_{b\,c,d} +
  g_{c\,d,b} - g_{b c, d}). 
\end{equation}

\paragraph{Transformaciones de Fourier} Nos apegaremos a la 
definición \textit{simétrica} de la Transformada de
Fourier, i.e.\ la transformada de Fourier $\hat{f}(\bvec{k})
\equiv  \mathcal{F}(f)(\bvec{k})$ de una función
$f(\bvec{x})$ es  
\begin{equation}\tag{i.D}
  \label{eq:transf_fourier}
  \hat{f}(\bvec{k}) = \frac{1}{(2\pi)^{3/2}} \int d^3x e^{-i
    \bvec{k}\cdot 
    \bvec{x}} f(\bvec{x}).
\end{equation}
La transformada de Fourier inversa es definida mediante
\begin{equation}\tag{i.E}
  \label{eq:transf_fourier_inversa}
  f(\bvec{x}) = \frac{1}{(2\pi)^{3/2}}\int d^3k e^{i
    \bvec{k}\cdot 
    \bvec{x}} \hat{f}(\bvec{k}).
\end{equation}
A lo largo de este trabajo de tesis realizaremos un abuso de
notación y usaremos el mismo símbolo para la transformada de
Fourier y su inversa (i.e.\ eliminaremos el ``gorro'',
$\hat{\{\}}$,
de las funciones transformadas), salvo cuando haya riesgo de
confusión. Esta decisión se tomó para no complicar la
notación a lo largo de la tesis. Si existiese algún riesgo
de confusión en el texto se volverá a reestablecer toda la
notación.

A veces, durante el texto, será conveniente expresar las
expansiones de Fourier en una sumatoria en lugar de una
expresión integral, esto corresponde a tomar las funciones
dentro de una caja cúbica con lados $L$ y volumen $V$. Para
recuperar las expresiones integrales será necesario hacer
las siguientes sustituciones
\begin{equation}\tag{i.F.a}
  \left(\frac{2\pi}{L}\right)^3 \sum_k \to \int d^3k,
\end{equation}
\begin{equation}\tag{i.F.b}
  \left(\frac{L}{2\pi}\right)^3 f_k \to
  \frac{1}{(2\pi)^{3/2}} f(k), 
\end{equation}
\begin{equation}\tag{i.F.c}
 \left(\frac{L}{2\pi}\right)^3 \delta_{kk'} \to
 \delta\left(\bvec{k} - \bvec{k}'\right).
\end{equation}

\paragraph{Unidades}

Como estamos interesados en esta tesis en los aspectos
cuánticos y gravitacionales, usaremos unidades apropiadas,
en las que la velocidad de la luz es $c=1$, pero la
constante de Planck $\hbar$ y la constante Gravitacional de
Newton $G$ conservan sus dimensiones. Por
ejemplo, el factor de expansión tiene unidades de $[a(\eta)]
= 1$ y $[\Hubble] = L^{-1}$.  De esta manera, las cantidades
que son expresables en unidades de longitud $[L]$, tiempo
$[T]$, se expresarán en potencias de longitud. En general,
para una cantidad con dimensiones $L^nT^mM^p$ en unidades
ordinarias, tendrá dimensión $L^{n+m}M^p$ en estas unidades,
y el factor de conversión entre ellas es $c^m$.

\end{introduccion}

   \PCFmainmatter

\chapter[Cosmología Estándar e
Inflación]{Cosmología  Estándar e Inflación}\label{cha:cosmologia-estandar}

\epigraph{Los inocentes hombres de mente ligera, que creen
  que la astronomía puede aprenderse mirando a las estrellas
  sin conocimiento matemático alguno, serán, en la siguiente vida,
  pájaros} {Platón, \textit{Timeo}}

\section{Introducción}
Cosmología es el estudio de la estructura a gran escala del
Universo, donde \textit{el Universo} es todo aquello que
existe en el sentido físico. Esta definición del
\textit{Universo} es poco precisa y es mejor afinarla un poco
diciendo que Cosmología \textit{principalmente} se encarga del
estudio de la estructura a gran escala en el \textit{Universo
  Observable}\footnote{ Hacemos énfasis en la palabra
  \textit{principalmente}, por que en teorías como Inflación el
  Universo Observable es una región pequeña comparada con el
  ``Universo'' que crece del parche que se infla y este
  ``parche'' a su vez pertenece a un ``Metaverso'' (o "Multiverso") mucho más
  grande, que se supone está inflando de manera caótica, pero esto
  se comprenderá más adelante.}, es decir, su campo de
estudio incluye la determinación de la naturaleza,
distribución, orígenes y su relación con la gran escala del
Universo, de las galaxias, \textit{clusters}\footnote{Aunque
  la tesis está escrita en español, mantendremos algunas
  palabras de uso común en inglés, aquí \textit{clusters} se
  puede traducir como ``agrupamientos'' o "cúmulos"} de galaxias,
objetos cuasi-estelares(QSO\footnote{Del inglés \textit{quasi-stellar
  objects}.}), etc.\ observados a través de diversas
longitudes de onda
electromagnéticas usando telescopios de todos los tipos y
que en un futuro se espera puedan ser utilizadas ondas
gravitacionales o neutrinos para el mismo fin.

La Cosmología se puede dividir a su vez, en \textit{Cosmología
  Observacional} cuya meta es determinar la geometría a gran
escala del Universo Observable y su distribución de
materia-energía a partir de las radiaciones emitidas por
objetos distantes; \textit{Cosmología física} que es el
estudio de las interacciones durante la época de expansión
muy temprana del \textit{Hot Big Bang}; y la \textit{Cosmología
  Astrofísica} que estudia el resultado del desarrollo
posterior de las grandes estructuras como galaxias y
clusters de las mismas. Dicho esto se puede decir que el
marco de desarrollo de este trabajo está contenido en el de
la cosmología física, la cual, de ahora en adelante la
llamaremos simplemente Cosmología.

\section{Cosmología Física}\label{sec:cosmologia-fisica}
El estudio de la Cosmología Física o simplemente Cosmología
parte de las siguientes suposiciónes:

\renewcommand{\labelenumi}{\alph{enumi}}
\begin{enumerate}\label{item:suposiciones}
\item La teoría de la Relatividad General de
  A. Einstein\footnote{Existen estudios en la actualidad
    donde la cosmología se basa en otras teorías que no son
    Relatividad General para describir la gravedad, un
    ejemplo son las teorías conocidas como $f(R)$ \cite{Sotiriou08}.}
  describe correctamente la evolución del espacio-tiempo a
  gran escala (es decir es la teoría correcta de
  gravitación) \citep{Wald84,HawkingEllis,dIverno,MTW}

\item Luego de ``promediar''\footnote{Es importante
    mencionar aquí que no existe un procedimiento bien
    definido para darle sentido riguroso a esta palabra en
    Relatividad General
    \citep{EllisG.F.05,Boersma98,Zalaletdinov97}.} a una
  escala lo suficientemente grande\footnote{Más adelante en
    el texto veremos a que nos referimos por
    ``suficientemente grande''.} existen supuestos
  \textbf{observadores fundamentales} que ven un
  \textit{universo isotrópico.}  Además si la isotropía es
  exacta, el universo será espacialmente homogéneo. Esta
  suposición es conocida como \textit{Principio Cosmológico}
  y fue introducida por A. Einstein en 1917.
\end{enumerate}

Con estas suposiciones es posible construir modelos
cosmológicos que se ajusten al universo observado. Estas
suposiciones fueron hechas inicialmente por A. Einstein y
W. de Sitter en 1917 y con ellas se dió inicio a la
cosmología moderna.  Idea que luego fue explotada por el
físico ruso A. Friedmann\footnote{En esta tesis se usará
  para este apellido la transliteración \textit{Friedmann}
  siguiendo la tradición occidental --que proviene de la
  traducción de sus artículos al alemán--. Su apellido es
  ruso y se escribe como \textit{Fridman}, hay autores que
  hacen un compromiso entre estas dos opciones y escriben
  \textit{Friedman}.}  y G. Lemaître en la
década de 1920 y puesta en sólidas bases geométricas por
H.P. Robertson y A.G. Walker. Un poco después, en el campo
observacional, E. Hubble (1929) estableció que las galaxias
mostraban un incremento sistemático del \textit{redshift} o
corrimiento al rojo con el aumento de la distancia hacia
nosotros y en la década de 1930, Eddington probó que los
modelos estáticos de Einstein eran inestables. Esto
estableció las bases de credibilidad de los universos
conocidos como universos de \textbf{\ac{FLRW}} \citep[ver
por
ejemplo][]{Dodelson,Mukhanov2005,Kolb1990,LythDavid93,Padmanabhan06,Padmanabhan1998,Padmanabhan96,Peebles93}.

De manera más formal se entiende que un \textbf{modelo
  cosmológico} \cite{Wainwright1997} representa el Universo
a cierta escala \citep{EllisG.F.05} y 
queda especificado por el par $\{(\mathcal{M}, \mathbf{g}),
\mathbf{u}\}$, es decir, especificando la geometría
espacio-temporal $(\mathcal{M}, \mathbf{g})$ y dando una
familia de observadores fundamentales, cuya congruencia de
líneas de mundo están descritas por el campo vectorial de
velocidad $\mathbf{u}$. Las variables $\{(\mathcal{M},
\mathbf{g}), \mathbf{u}\}$ determinan el comportamiento
cinemático del universo. En algunos modelos cosmológicos la
congruencia de observadores fundamentales se supone
expandiéndose en cierta época, pero existen modelos donde
esto no es el caso \cite{Wainwright1997}.  Para determinar
el comportamiento dinámico, es necesario especificar el otro
ingrediente importante de la Relatividad General, la
materia, representada --en alguna escala promediada-- por el
tensor de energía-momento $T_{ab}^{(A)}$ donde $A$ etiqueta
cada uno de los posibles tipos de materia presentes en el
espacio-tiempo. La interacción entre materia y
espacio-tiempo es gobernada mediante las \ac{ECE}:

\begin{equation}
  \label{eq:ece}
  G_{ab} \equiv R_{ab} - \frac{1}{2}\Ricci g_{ab} =  \kappa
  T_{ab}, \qquad T_{ab} =   \sum_{A} T_{ab}^{(A)}.
\end{equation}
\nomenclature{$G_{ab}$}{Tensor de Einstein}
\nomenclature{$R_{ab}$}{Tensor de Ricci}
\nomenclature{$\Ricci$}{Escalar de Ricci}
\nomenclature{$g_{ab}$}{Métrica del espacio-tiempo}
\nomenclature{$\kappa$}{Constante gravitacional de Einstein}
\nomenclature{$T_{ab}$}{Tensor de Energía-Momento}

En ellas $R_{ab} \equiv R\indices{_{acb}^c}$ es el tensor de
Ricci, $\Ricci \equiv R\indices{_{\ a}^a}$ es el escalar de
Ricci o escalar de curvatura, $\kappa \equiv 8\pi G$ es la
constante gravitacional de Einstein y $T_{ab}$ es el tensor
de energía-momento de todos los campos presentes en el
espacio-tiempo. El tensor de $\Riemann$ fue definido en la
introducción (\ref{eq:riemann_def}, página \pageref{eq:riemann_def}) .

El tensor de Ricci, $R_{ab}$, que se puede expresar en
términos de los símbolos de Christoffel
(\ref{eq:simbolos_christoffel}) mediante
\begin{equation}
  \label{eq:ricci_christoffel}
  R_{ab} = \Gamma^c_{ab,\,c} - \Gamma^c_{a\,c,b}+
  \Gamma^c_{d\,c}\Gamma^d_{ab} - \Gamma^c_{d\, b}\Gamma^d_{a\,c}.
\end{equation}

Las ECE implican la ecuación de conservación\footnote{Nótese
  que las ECE \textbf{no} garantizan la conservación de cada
  uno de los componentes de la materia, sólo el total.} a
través de las identidades de Bianchi
\citep{MTW,dIverno,Wald84}
\begin{equation}
  \label{eq:bianchi}
  \nabla_bG^{ab} = 0 \then  \nabla_bT^{ab} \equiv T^{ab}_{\quad;b} = 0.
\end{equation}

Como veremos posteriormente el modelo cosmológico actual
incluye un componente de \acfi{DE} \citep[ver][para otras
explicaciones posibles]{Marra07,Valkenburg09,Ellis08}, que
es representado en sus caracterizaciones más simples por una
\textbf{constante cosmológica} $\Lambda$, en ese caso las
ECE (\ref{eq:ece}) toman la forma
\begin{equation}
  \label{eq:ece_lambda}
  R_{ab} - \frac{1}{2}\Ricci g_{ab} =  \kappa T_{ab} - \Lambda g_{ab},
\end{equation}
\nomenclature{$\Lambda$}{Constante Cosmológica}

y esta ecuación (\ref{eq:ece_lambda}) es válida solamente si
$\Lambda$ es constante en tiempo y espacio, i.e.  satisface
$\nabla_a \Lambda = 0$. Como en este trabajo de tesis
consideramos --siguiendo la práctica moderna-- que $\Lambda$
es un tipo de materia exótico, se le incluirá dentro de
$T_{ab}$.

Es un procedimiento estándar descomponer al tensor de
energía-momento $T_{ab}$ respecto al campo vectorial
temporaloide $\mathbf{u}$ \citep{Ellis71}:
\begin{equation}
  \label{eq:tab_descomposicion}
  T_{ab} = \rho u_a u_b + 2 q_{(a}u_{b)} + p(g_{ab} + u_a
  u_b) + \pi_{ab},
\end{equation}
donde $\rho$ es la densidad de energía, $p$ es la presión
del fluido, $\pi_{ab}$ es la presión anisotrópica y por
último, $q_a$, aparece cuando la velocidad del fluido no
está alineada con el campo vectorial $u^a$. Estas variables
cumplen con las siguientes relaciones:
\begin{equation}
  q_au^a = 0, \quad \pi_{ab}u^b = 0, \quad, \pi_a^{\  a} =0,\quad
  \pi_{ab} = \pi_{ba}.
\end{equation}

\subsection{Caracterización de los modelos cosmológicos}
Es posible caracterizar al modelo cosmológico,
$(\mathcal{M}, \mathbf{g}, \mathbf{u})$, invariantemente de
acuerdo a  (a) sus propiedades cinemáticas, (b)  las
propiedades del tensor de Weyl de $(\mathcal{M},
\mathbf{g})$ y (c) sus simetrías.

\subsubsection{Propiedades
  cinemáticas}\label{sec:prop-cinem}
Las cantidades cinemáticas están asociadas con la
congruencia temporaloide $\mathbf{u}$  fueron introducidas
por Raychaudhuri a mediados de la década de 1950
\citep{Raychaudhuri55} y se construyen como a continuación
se muestra \citep{Wald84,Dadhich05}.

Un campo vectorial temporaloide $u^a$ determina un tensor de
proyección $h_{ab}$ de acuerdo a
\begin{equation}
  h_{ab} = g_{ab} + u_au_b  
\end{equation}
cuya acción es proyectar tensores al espacio tridimensional
ortogonal a $\mathbf{u}$. El tensor $h_{ab}$ posee las
siguientes características $h^a_{\ b} = g^{ab}h_{cb}$,
$h_a^{\ c}h_c^{\ b} = h_a^{\ b}$, $h_a^{\ b}u_b = 0$ y
$h_a^{\ a} = 3$.  Supongamos que $u^a$ es el vector tangente
a una congruencia temporaloide de geodésicas. El tensor
$h_{ab}$ permite descomponer la derivada covariante,
$\nabla_b u_a$, que es puramente espacial $\nabla_b u_a u^a
= \nabla_b u_a u^b = 0$ en \citep{Wald84}

\begin{equation}
  \label{eq:descomposicion_raychaudhuri}
  u_{a;b}= \frac{1}{3}\Theta
  h_{ab} + \sigma_{ab} + \omega_{ab},
\end{equation}

donde $\Theta = u^a_{\ ;a}$ es la expansión, $\sigma_{ab} =
u_{(a;b)} - \frac{1}{3} \Theta h_{ab}$ el corte
(\textit{shear}) y $\omega_{ab} = u_{[a;b]}$ el giro
(\textit{twist})\footnote{En la derivación original de
  Raychaudhuri, el parte de la congruencia de partículas con
  velocidad $u^a$ cayendo bajo su propia gravedad, luego, de
  mecánica de fluidos se sabe que sufrirá los efectos de (a)
  Contracción/Expansión, dada por la divergencia de $u^a$,
  $\Theta = u^a_{;a}$. (b) Corte, distorsión de la figura
  sin cambio en el volumen, dada por un tensor simétrico sin
  traza y ortogonal a $u^a$, definido por $\sigma_{ab} =
  u_{(a;b)} - \frac{1}{3}\Theta h_{ab} -
  \dot{u}_{(a}u_{b)}$. (c) Rotación/Vorticidad, rotación sin
  cambio en su forma, dado por un tensor antisimétrico
  ortogonal a $u^a$: $\omega_{ab} = u_{[a;b]} -
  \dot{u}_{[a}u_{b]}$ y (d) Aceleración debido a fuerzas no
  gravitacionales tales como gradientes de presión,
  definidos por $\dot{u}_a = u_{a;b}u^b$. Obteniendo la
  ecuación:
  \begin{equation}
    \label{eq:deduccion_raychaudhuri}
    u_{a;b} = \sigma_{ab} +\omega_{ab} + \frac{1}{3}\Theta
    h_{ab} + \dot{u}_a u_b.
  \end{equation}
  Como se podrá notar, Raychaudhuri no partió del supuesto
  de que la congruencia estaba formada por geodésicas, que
  es la suposición hecha en el texto principal.}. Es posible
encontrar una ecuación de evolución para la expansión
$\Theta$ \citep[ver pag. 218 de][]{Wald84} conocida como
\textit{la ecuación de Raychaudhuri}

\begin{equation}
  \label{eq:raychaudhuri}
  \dot\Theta = -\frac{1}{3}\Theta^2 - 2 (\sigma^2
  - \omega^2) - R_{cd}u^cu^d
\end{equation}

donde $\dot\Theta \equiv \Theta_{;a}u^a$, $\sigma^2 \equiv
1/2 \sigma_{ab}\sigma^{ab}$ y $\omega^2 \equiv 1/2
\omega_{ab}\omega^{ab}$.

Esta ecuación es de suma importancia por que es usada para
los lemas de enfocamiento --\citep[pag. 219-223]{Wald84}-- y
a través de estos a los teoremas de Penrose-Hawking de
singularidades \citep[ver \S 9.5 de][]{Wald84}.

Como se verá más adelante, en universos tipo
Friedmann-Lemaître, se definirá el parámetro de Hubble
mediante

\begin{equation}
  \label{eq:hubble_raychaudhuri}
  H \equiv \frac{1}{3}\Theta.
\end{equation}

\subsubsection{Tensor Conforme de Weyl}

La siguiente manera de caracterizar a los modelos
cosmológicos es usando el tensor de Weyl $C_{abcd}$ que se
obtiene restando las partes con traza del tensor de Riemann
$R_{abcd}$,

\begin{equation}
  \label{eq:weyl}
  C_{abcd} = R_{abcd} - \frac{1}{2}\left(g_{a[c}R_{d]b} -
    g_{b[c}R_{d]a} \right) - \frac{1}{3}\; \Ricci\;  g_{a[c}g_{d]b}.
\end{equation}
\nomenclature{$C_{abcd}$}{Tensor de Weyl}

El tensor de Weyl comparte todas las simetrías del tensor de
Riemann \citep{Wald84}. Usando las identidades de Bianchi y
las ECE, además de condiciones de frontera apropiadas,
$C_{abcd}$ es determinado de una manera no-local por la
materia en cualquier parte del universo, y representa así el
``campo gravitacional libre'' \citep{Ellis71}. Se le conoce
también como \textit{tensor conforme de Weyl} por que se
comporta de manera muy sencilla bajo transformaciones
conformes de la métrica\footnote{Estas transformaciones
  conformes quedan definidas \citep[apéndice D de
  ][]{Wald84} como sigue: \textit{Sea $\mathcal{M}$ una
    variedad con métrica $g_{ab}$. Si $\Omega$ es una
    función suave y estrictamente positiva, entonces la
    métrica $\tilde{g}_{ab} = \Omega^2 g_{ab}$ se dice que
    es derivada de $g_{ab}$ mediante una
    \textbf{transformación conforme.}} Es fácil probar que
  el tensor de Weyl, no cambia antes este tipo de
  transformaciones i.e.\ $\tilde{C}_{abcd} = C_{abcd}$.}.

Es útil separar a $C_{abcd}$ en sus partes ``eléctrica'', $E_{ab}$ y
``magnética'', $H_{ab}$, relativas a $\mathbf{u}$:
\begin{equation}
  \label{eq:weyl_magnetica_electrica}
  E_{ac} = C_{abcd}u^bu^d,\qquad H_{ac} = \ ^*C_{abcd}u^bu^d,
\end{equation}

donde $^*C_{abcd}$ es el dual del tensor de Weyl.

\begin{equation*}
  ^*C_{abcd} \equiv \frac{1}{2}\varepsilon\indices{_{ab}^{st}}C_{stcd}  
\end{equation*}

y $\varepsilon^{abcd}$ es el pseudo-tensor completamente
antisimétrico. Los tensores $E_{ab}$ y $H_{ab}$ son
simétricos y sin traza, además que si ambos son cero,
i.e. $E_{ab} = 0 = H_{ab} \then C_{abcd} = 0$. Se puede
demostrar que,

\begin{equation}
  \label{eq:weyl_relaciones}
  C_{abcd}C^{abcd} = 8\left(E_{ab}E^{ab} -
    H_{ab}H^{ab}\right),\quad  C_{abcd}\ ^*C^{abcd} = 16E_{ab}H^{ab}.
\end{equation}

En 1954 A.\ Z.\ Petrov \citep[reimpreso en][]{Petrov00} estudió
las simetrías algebraicas del tensor de Weyl, y desarrolló
un sistema para clasificarlas conocida ahora como
\emph{clasificación de Petrov}. Para lograrlo consideró al
tensor de Weyl, $C_{abcd}$, evaluado en algún evento del
espacio-tiempo, como actuando en el espacio de bivectores en
ese mismo evento, i.e.\ , $X^{ab} \to
\frac{1}{2}C^{ab}_{\;\;cd}X^{cd}$; de esta manera, es
natural considerar el problema de encontrar los eigenvalores
$\lambda$ y los eigenbivectores, $X^{ab}$ tales que

\begin{equation}
  \frac{1}{2}C^{ab}_{\;cd}X^{cd} = \lambda X^{ab}.
\end{equation}

Petrov encontró que sólo existen seis tipos de simetrías
algebraicas (relacionadas con la multiplicidad de los
eigenvalores) en las variedades Lorentzianas de cuatro
dimensiones. Así, es posible caracterizar a los modelos
cosmológicos usando los tipos de Petrov.
 
\subsubsection{Simetrías del espacio-tiempo}
Se define como isometría del espacio-tiempo $(\mathcal{M},
\mathbf{g})$, al mapeo $\phi: \mathcal{M} \to \mathcal{M}$
que deja $\mathbf{g}$ invariante, es decir, un difeomorfismo
$\phi$ tal que $(\phi^* g)_{ab} = g_{ab}$. Si el
difeomorfismo es un grupo uniparamétrico de isometrías,
$\phi_t^* g_{ab} = g_{ab}$, al campo vectorial $\xi^a$, que
 genera este grupo es conocido como \textit{campo
  vectorial de Killing}. Las órbitas de este grupo son las
curvas integrales de ${\xi^a}$. Para que este campo
vectorial transforme como una isometría, es necesario que la
derivada de Lie de la métrica con respecto a ${\xi^q}$ sea
cero,

\begin{equation}
  \label{eq:isometria}
  \Lie_\xi g_{ab} = 0.
\end{equation}

Ecuación que se conoce como \textit{ecuación de
  Killing}. Para ver las propiedades de estos vectores de
Killing consúltese \citep[][apéndice C]{Wald84}.

Para caracterizar modelos cosmológicos, según este criterio,
basta con especificar los vectores de Killing que posea.

\subsection{Modelos Cosmológicos de
  Friedman-Lemaitre-Robertson-Walker}
\label{sec:modelo-flrw}
Podemos describir el espacio-tiempo de nuestro universo a
cierta escala en la cual luce homogéneo e isotrópico
espacialmente mediante la métrica de \ac{RW}. En esta
geometría la materia se mueve sobre curvas geodésicas
irrotacionales y libres de corte ($\omega_{ab} = 0 =
\sigma_{ab}$) con una velocidad tangente $u^a$, que define
una variable temporal canónica $u_a = - t,_a$. Además, las
geometrías de RW tienen un tensor de Weyl igual a cero,
$C_{abcd} = 0$, i.e. todos los efectos no-locales como ondas
gravitacionales y fuerzas de marea están ausentes. El que
este tensor sea nulo nos indica que las geometrías RW son
conformalmente planas.  La homogeneidad y la isotropía se
pueden definir ahora de una manera precisa usando los campos
vectoriales de Killing y el concepto de isometrías. Un
espacio-tiempo es espacialmente homogéneo si existe una
familia uniparamétrica de hipersuperficies
$\Sigma_t$\label{def:sigma_t}
\nomenclature{$\Sigma_t$}{Hiper-superficies homogéneas e
  isotrópicas de las coordenadas de RW.}que folian el
espacio-tiempo tal que para cada $t$ y para cualquier par de
puntos $p$, $q$ $\in \Sigma_t$ existe una isometría de la
métrica $g_{ab}$ que lleva $p$ a $q$; por otra parte, se
dice que el espacio-tiempo es isotrópico, si en cada punto
existe una congruencia de curvas temporaloides u
observadores con tangentes $u^a$ que llenen el
espacio-tiempo, que satisfacen que, dado cualquier punto $p$
y dos vectores tangenetes unitarios $s_1^a$ y $s_2^a$ en $p$
ortogonales a $u^a$, existe una isometría de $g_{ab}$ que
deja a $p$ y $u^a$ en $p$ fijos, pero que rota $s_1^a$,
$s_2^a$ \cite{Wald84}.

El verificar que este modelo (u otros modelos) se ajuste al
Universo real es tarea de la Cosmología Observacional (cf.
\S \ref{sec:cosm-observ}).

\subsubsection{Cinemática: Geometría de Robertson-Walker}
La métrica de Robertson-Walker es la expresión más general
de la métrica para un espacio-tiempo tetradimensional que
está foliado por subespacios tridimensionales espaciales
(i.e.\ $t = $ constante) $\Sigma_t$, máximamente simétricos,
es decir, 
todas las propiedades geométricas del subespacio
tridimensional son iguales en todos las locaciones
espaciales y que esas propiedades geométricas no privilegian
dirección alguna en el espacio, en otras palabras
homogeneidad e isotropía. A la variable de foliación $t$ se
le llamará \textit{tiempo cosmológico}. La homogeneidad e
isotropía en el espacio tridimensional permite definir un
conjunto (infinito) de observadores preferenciales comóviles
(i.e.\ con $x^i = $ constante), que ven el Universo de
manera homogénea e isotrópica\footnote{Usando la ecuación
  geodésica, $u^a\nabla_a u^b = 0$, se puede ver que estos
  observadores son inerciales, ya que se mueven sobre curvas
  geodésicas.}. Entonces, usando las coordenadas de estos
observadores $(t, x^i)$ podemos escribir:

\begin{equation*}
  ds^2 = g_{\alpha\beta}\  dx^{\alpha}dx^{\beta} = g_{00}\ dt^2
  + \cancelto{0}{2\ g_{0i}\ dt dx^i} - \gamma_{ij} dx^idx^j,
\end{equation*}

donde el segundo término es cancelado por isotropía, ya que
si fuera $g_{i0} \neq 0$ se podría definir un vector $v_i =
g_{0i}$ que privilegiaría una dirección. Si usamos el tiempo
propio de los observadores preferenciales para etiquetar las
hipersuperficies, tendremos $g_{00} = 1$, $ds^2 = -dt^2 +
dl^2$. Al ser el espacio-tiempo homogéneo e isotrópico, la
curvatura del 3-espacio, $R\indices{^{(3)}_{ijk}^l}$ no
puede depender\footnote{Si dependiera de las derivadas de la
  métrica significaría que podríamos elegir un vector
  privilegiado en la hipersuperficie espacial rompiendo así
  la isotropía.} de las derivadas de la 3-métrica
$\gamma_{ij}$, entonces $R\indices{^{(3)}_{ijkl}} = q (\gamma_{ik}\gamma_{jl}
- \gamma_{il}\gamma_{kj})$ y su escalar de curvatura es
$\Ricci^{(3)} = 6q$, $q$ debe de ser una constante para cumplir
con la suposición de homogeneidad.  La métrica más general
que cumple con un Universo homogéneo e isotrópico en cada
instante del tiempo es entonces:

\begin{equation}
  \label{eq:metrica_rw}
  ds^2 = -dt^2 + S^2(t)\left[ \frac{dr^2}{1-K r^2} +
    r^2\left(d\theta^2 + \sin^2\theta d\phi^2\right)\right],
  \quad u^a = \delta^a_0,
\end{equation}

y está escrita en término de dos parámetros cosmológicos,
uno que describe la curvatura espacial del Universo ($K$) y
el segundo la expansión o contracción del Universo ($S(t)$),
conocido como \textit{factor de escala}.  Por último, el
vector tangente a las líneas de mundo $u^a = dx^a/dt$
representa la historia de los observadores fundamentales.

Podemos escalar la coordenada radial, $r$, de tal manera que
la constante de curvatura espacial $K$, solo tome los
valores $+1$, $-1$ y $0$, correspondiendo a geometrías
cerradas, abiertas y planas, respectivamente. Tomando este
escalamiento de $r$ es posible expresar la métrica como

\begin{equation}
  \label{eq:metrica_rw_2}
  ds^2 = -dt^2 + S^2(t)\left[d\chi^2 +
    f_k^2(\chi)\left(d\theta^2 + \sin^2\theta d\phi^2\right)\right]
\end{equation}

donde la función $f_k(\chi)$ es

\begin{equation}
  f_k(\chi) = \;
  \begin{cases}
    \; \sin\chi, & K=+1\\
    \; \chi,  & K=0\\
    \; \sinh\chi, & K=-1
  \end{cases}
\end{equation}

La tasa de expansión en cada tiempo $t$ está caracterizado
por el \textit{parámetro de Hubble} $H(t) = \dot S / S$,
donde $\dot{\{\}} \equiv d/dt\{\}$ representa derivadas
respecto a $t$, el tiempo cosmológico.
\nomenclature{$H(t)$}{Parámetro de Hubble\nomnorefeq}

Las dimensiones en la métrica están contenidas en el factor
de escala (que depende de $t$), ya que la $r$ y el
``ángulo'' $\chi$ son adimensionales. También es posible
tener un factor de escala adimensional mediante $a(t) =
S(t)/S_0$ donde $S_0 \equiv S(t_0)$, es $S$ en la época
actual, $t_0 =$ hoy; $a(t)$ determina las distancias
físicas entre dos puntos\footnote{En la literatura también se les llama
  distancias \textit{propias}, aunque este uso no siempre es
  consistente lo cual ha llevado a confusiones con el paso
  del tiempo \cite{DavisLineweaver}.} $l(t)$, a un tiempo
$t$, en términos de las distancias comóviles $l_0$ (estas
últimas no cambian en el tiempo, recuerde que los
observadores fundamentales son comóviles, i,e.\ sus
coordenadas $x^i$ están fijas)

\begin{equation}
  \label{eq:distancias}
  l(t) = l_0 a(t).
\end{equation}

Con esta nueva definición del factor de escala, el parámetro
de Hubble es ahora $H(t) \equiv \dot{a}/a$. En un universo
homogéneo e isotrópico, el parámetro de Hubble define las
única escala espacial significativa: el \textit{radio de
  Hubble} , $r_H \equiv cH^{-1}$.\nomenclature{$d_H$}{Radio
  de Hubble\nomnorefeq} El radio de Hubble representa la
distancia a la cual la velocidad de recesión de una galaxia
es igual a la velocidad de la luz \footnote{ El radio de
  Hubble está muy relacionado con el (cf. \S
  \ref{sec:cosm-observ} para una definición)
  \textit{horizonte de partículas} para ciertas épocas del
  universo, pero conceptualmente son muy diferentes, por lo
  que es muy importante no confundirlos; para una excelente
  discusión de esta y otras confusiones ver
  \citep{DavisLineweaver}.}. Para factores de escala que van
como $a(t) \propto t^p$, el radio de Hubble es $r_H \propto
\dfrac{t}{p}$, i.e.\ el radio de Hubble crece linealmente con
el tiempo.

Otra manera de expresar esta métrica se logra escalando la
coordenada temporal mediante 
\begin{equation}
\label{eq:conforme_cosmologico}
d\eta = \frac{dt}{a(t)},
\end{equation}
donde a $\eta$ se le conoce como \textit{tiempo conforme},
quedando,
\begin{equation}
  \label{eq:metrica_rw_3}
  ds^2 =  a^2(\eta)\left[ -d\eta^2 + d\chi^2 +
    f_k^2(\chi)\left(d\theta^2 + \sin^2\theta d\phi^2\right)\right].
\end{equation}

Por último, es posible reescribir la métrica en términos del
factor de escala, $a$ o del \textit{redshift}, $z = (a_0/a)
- 1$ \todo{checar redshift} como la coordenada temporal,
esta forma mostrará claramente que el único contenido
dinámico está en el parámetro de Hubble $H$: 

\begin{equation}
  \label{eq:metrica_rw_4}
  ds^2 = H^{-2}(a)\left(\frac{da}{a}\right)^2 - a^2dx^2 =
  \frac{1}{(1+z)^2} \left[H^{-2}(z) dz^2 - dx^2\right].
\end{equation}

\subsubsection{Dinámica: Universos de
  Friedmann-Lemaître}\label{sec:universo-fl} 
Una vez que se ha especificado la geometría del
espacio-tiempo usando la métrica de \acf{RW}, es útil
describir su dinámica mediante las ECE. Para hacerlo es
necesario conocer el contenido de materia-energía descrito
--a la escala determinada por la validez de la métrica RW--
mediante el tensor de energía-momento $T_{ab}$.

Debido a la isotropía local el tensor de energía momento,
$T_{ab}$, expresado como \eqref{eq:tab_descomposicion}, toma
necesariamente la forma de un fluido perfecto (i.e.\
$\pi_{ab} = 0 = q_a$) relativa a las líneas de mundo de los
observadores fundamentales, es decir el fluido es comóvil
con la expansión del universo,

\begin{equation}
  \label{eq:tab_fluido_perfecto}
  T_{ab} = (\rho + p )u_a u_b +  p g_{ab},
\end{equation}

la densidad de energía, $\rho$ y la presión, $p$ son los
eigenvalores temporaloides y espacialoides de $T_{ab}$. Para
encontrar las ecuaciones dinámicas de este modelo
cosmológico es necesario calcular $G_{ab}$ y $T_{ab}$ y
luego sustituirlos en las ECE\footnote{Para este modelo es
  posible obtener las ecuaciones de evolución sin tener que
  pasar por todo el tedioso procedimiento de calcular
  $g_{ab} \to \Gamma^a_{\;bc} \to R_{ab} \to \Ricci \to
  G_{ab}.$ Para lograrlo, primero, utilizando las
  identidades de Bianchi \eqref{eq:bianchi}, $\nabla_bT^{ab}
  = 0$, obtenemos,
  \begin{equation*}
    \dot\rho + 3\frac{\dot{a}(t)}{a(t)}(\rho + p) = 0.
  \end{equation*}

  Ahora, usando las propiedades cinéticas
  \ref{sec:prop-cinem} podremos escribir la ecuación de
  Raychaudhuri \eqref{eq:raychaudhuri} en la métrica de
  RW. Primero, sacando la traza de las ECE llegamos a
  \begin{equation}
    \Ricci = 4\Lambda -8\pi G T  
  \end{equation}
  donde $T$ es la traza de $T_{ab}$. Sustituyendo en la ECE,
  \begin{equation}
    R_{ab} = 8\pi G \left[T_{ab} - \frac{1}{2}
      \left(T+\frac{\Lambda}{4\pi G}\right) g_{ab}\right],
  \end{equation}
  en particular, para el fluido perfecto
  \eqref{eq:tab_fluido_perfecto}, tenemos $T = 3p -\rho$ y
  si multiplicamos la ecuación anterior por $u^au^b$,
  \begin{equation}
    R_{ab}u^au^b = 4\pi G (\rho + 3p) - \Lambda.
  \end{equation}

  Sustituyendo esto en la ecuación de Raychaudhuri; y
  notando que en las métricas RW, $\sigma_{ab} = \omega_{ab}
  = 0$, por lo tanto, la ecuación de Raychaudhuri ahora es,
  \begin{equation}
    \dot\Theta + \frac{1}{3}\Theta^2 = - 4\pi G (\rho + 3p) + \Lambda
  \end{equation}

  Sustituyendo \eqref{eq:hubble_raychaudhuri}, obtenemos que
  la ecuación de Raychaudhuri para métricas RW (cf.\ \ref{eq:ece_auxiliar}) es
  \begin{equation}
    \frac{\ddot{a}}{a} = - \frac{4\pi G}{3}(\rho + 3 p) +
    \frac{\Lambda}{3}\;, 
  \end{equation}

  Por último, calculando la primera integral de
estas dos últimas ecuaciones	
  obtenemos la ecuación controla la evolución del universo:
  la \textit{ecuación de Friedmann} (cf.\ \ref{eq:friedmann})
  \begin{equation}
    \left(\frac{\dot{a}}{a}\right)^2 =\frac{8\pi G}{3}\rho +
    \frac{\Lambda}{3} - \frac{K}{a^2}.
  \end{equation}
}.

Una de las muchas maneras de encontrar las expresiones buscadas es
empezar por el cálculo de los símbolos de Christoffel
(\ref{eq:simbolos_christoffel}), que para la métrica de RW
\eqref{eq:metrica_rw}, tienen la siguiente forma

\begin{alignat}{3}
  \label{eq:christoffel_fl}
  \Gamma^i_{j\,t} &= \delta^i_j \daa &\quad
  \Gamma^t_{r\,r} &= \delta_{ij}\frac{\dot{a}a}{1-Kr^2}   \nonumber  \\
  \Gamma^t_{\theta\,\theta} &= r^2\,\dot{a}{a} &\quad
  \Gamma^t_{\phi\,\phi} &= r^2\sin^2\theta\, \dot{a}{a} \nonumber\\
  \Gamma^r_{r\,r} &= \frac{kr}{1-Kr^2} &\quad
  \Gamma^r_{\theta\,\theta} &= -(1-Kr^2)r \\
  \Gamma^\theta_{\theta\, r} &= \frac{1}{r} =
  \Gamma^\phi_{\phi\,\phi} &\quad
  \Gamma^r_{\phi\,\phi} &= -(1-Kr^2)r \sin^2\theta  \nonumber\\
  \Gamma^\theta_{\phi\,\phi} &= -\sin\theta\cos\theta &\quad
  \Gamma^\phi_{\phi\,\theta} &=
  \frac{\cos\theta}{\sin\theta}. \nonumber
\end{alignat}


El siguiente paso es calcular el tensor de Ricci, $R_{ab}$,
al sustituir (\ref{eq:christoffel_fl}) en la fórmula
(\ref{eq:ricci_christoffel}) se llega a

\begin{equation}
  \label{eq:ricci_fl}
  R_t^{\;t} = 3\frac{\ddot a}{a}, \quad R_j^{\;i} =
  \delta_j^{\;i} \left[
    \frac{\ddot{a}}{a} + 2\left(\frac{\dot{a}}{a}\right)^2 +
    2\frac{K}{a^2} \right].
\end{equation}


Contrayendo el tensor de Ricci con la métrica, obtenemos el
escalar de Ricci, $\Ricci$,

\begin{equation}
  \label{eq:escalar_ricci}
  \Ricci = g^{ab}R_{ab}.
\end{equation}

Sustituyendo los valores encontrados en las secciones
anteriores,

\begin{equation}
  \label{eq:escalar_ricci_fl}
  \Ricci = 6\left[\frac{\ddot a}{a} + \left(\daa
    \right)^2 + \frac{K}{a^2} \right].
\end{equation}


Como último paso para completar el lado izquierdo de las
ECE, calculamos el tensor de Einstein $G_{ab}$, definido por

\begin{equation}
  \label{eq:einstein}
  G_{b}\,^a = R_{b}\,^a-\frac{1}{2}g_{b}\,^a\Ricci
\end{equation}

haciendo las sustituciones adecuadas, se obtiene

\begin{equation}
  \label{eq:einstein_fl}
  G_t^{\;t} = -3\left[ \left(\daa\right)^2 +
    \frac{K}{a^2}\right], \quad G_j^{\;i} =
  -\left[2\frac{\ddot a}{a} + \left(\daa\right)^2 + \frac{K}{a^2}\right]
\end{equation}


Por el lado de la materia, $T_{ab}$ tiene la forma
(\ref{eq:tab_fluido_perfecto}) en estas coordenadas,
\begin{subequations}
  \begin{equation}
    T_{t}^{\;t} = - \rho    ,
  \end{equation}
  \begin{equation}
    T_i^{\;j} =  \kronecker\, p
  \end{equation}
\end{subequations}

Entonces las ecuaciones de Einstein son

\begin{subequations}
  \label{eq:ece_fl}
  \begin{equation}
    \label{eq:friedmann}
    \left(\daa\right)^2 =  \frac{8 \pi \, G}{3}\rho  +
    \frac{\Lambda}{3} - 
    \frac{K}{a^2},
  \end{equation}
  \begin{equation}
    \label{eq:aceleracion}
    2\frac{\ddot a}{a} + \left(\daa\right)^2  = -\left(8\pi G  p
      + \frac{K}{a^2} -\Lambda \right).
  \end{equation}
\end{subequations}


Podemos observar que estas ecuaciones pueden ser unidas en
una sola si las restamos,

\begin{equation}
  \label{eq:ece_auxiliar}
  \frac{\ddot a}{a} = -\frac{4\pi G}{3}(\rho + 3p) +
  \frac{\Lambda}{3}. 
\end{equation}

El sistema de ecuaciones (\ref{eq:ece_fl}) está
subdeterminado, ya que en realidad es una sola ecuación
(\ref{eq:ece_auxiliar}) y posee tres incógnitas: $(a(t),
\rho, p)$. Para resolver este modelo cosmológico es
necesario encontrar dos ecuaciones más, la primera puede ser
obtenida utilizando la identidad de Bianchi
(\ref{eq:bianchi}) en el tensor de energía-momento,
$\nabla_bT^{ab} = 0$,

\begin{equation}
  \label{eq:conservacion_fl}
  \dot\rho + 3\frac{\dot{a}(t)}{a(t)}(\rho + p) = 0.
\end{equation}

Esta ecuación controla la densidad de la materia mientras el
universo se está expandiendo. La última ecuación necesaria
se obtiene suponiendo que la presión $p$ cumple con una
ecuación de estado, $p = p(\rho)$.

La existencia y unicidad de la solución $\{a(t), \rho(t)\}$
del conjunto de ecuaciones (\ref{eq:ece_fl}) queda
establecida si se da una ecuación de estado que describa el
contenido de materia (para uno o varios componentes) del
universo y un conjunto adecuado de condiciones
iniciales. Las condiciones iniciales se pueden dar a un
tiempo arbitrario, pero generalmente se dan en el tiempo
actual, $t_0$:

\begin{itemize}
\item La \textit{constante de Hubble}, i.e.\ el parámetro de
  Hubble al tiempo actual, $H_0 \equiv (\dot a/a)_0$;
\item Un \textit{parámetro de densidad} adimensional para
  cada una de los tipos de materia, $\Omega_{i0} \equiv
  \kappa \rho_{i0}/3H_0^2$;
\item Si $\Lambda \neq 0$, $\Omega_{\Lambda 0} \equiv
  \Lambda/3H_0^2$ o el \textit{parámetro de desaceleración}
  (adimensional) $q_0 :=$ $-(\ddot{a}/a)_0$ $H_0^{-2}$.
\end{itemize}

Los modelos cosmológicos que tienen una geometría
caracterizada por la geometría de Robertson-Walker y con una
evolución gobernada por las ecuaciones
\eqref{eq:conservacion_fl}, \eqref{eq:aceleracion} y
\eqref{eq:friedmann} son llamados \textit{universos de
  Friedmann-Lemaître} (FL).

\paragraph{\textsc{Addendum}: Ecuaciones en tiempo conforme}

En este trabajo de tesis estaremos usando el tiempo conforme
$\eta$, además del tiempo cosmológico $t$. La relación entre
estas dos coordenadas temporales está dada mediante
\begin{equation*}
  d\eta = \frac{dt}{a(t)}.
\end{equation*}
\nomenclature{$\eta$}{Tiempo conforme \nomnorefeq} 

En esta sub-sección expresaremos las ecuaciones más importantes
presentadas hasta ahora en estas nuevas coordenadas. Nótese
que para cualquier función $f(t)$ se cumple
\begin{eqnarray*}
  \dot{f}(t) &=&\frac{f'(\eta)}{a(\eta)},\\
  \ddot{f}(t)&=& \frac{f''(\eta)}{a^2(\eta)} -
  \mathcal{H}\frac{f'(\eta)}{a^2(\eta)},
\end{eqnarray*}
donde $\{\}' \equiv d/d\eta\{\}$ indica derivada respecto al
tiempo conforme. Estas relaciones nos permiten convertir
ecuaciones del tiempo cosmológico al conforme. En
particular, definimos $\HubbleComovil(\eta) \equiv
a'(\eta)/a(\eta) = a H$
\nomenclature{$\HubbleComovil$}{Parámetro de Hubble
  comóvil}, como el parámetro de Hubble comóvil.



La ecuación de Friedmann (\ref{eq:friedmann}), la de
aceleración (\ref{eq:aceleracion}) y la ecuación de
conservación (\ref{eq:conservacion_fl}) en la métrica RW
conforme \eqref{eq:metrica_rw_3}, son respectivamente:

\begin{subequations}
  \label{eq:friedmann_lemaitre_conforme}
  \begin{equation}
    \label{eq:friedmann_conforme}
    \HubbleComovil^2 = \left(\frac{a'}{a}\right)^2 =
    \frac{8\pi 
      G}{3}a^2\rho + \frac{a^2 \Lambda}{3}- K,
  \end{equation}
  \begin{equation}
    \label{eq:aceleracion_conforme}
    2 \HubbleComovil' + \HubbleComovil^2 =
    2 \frac{a''}{a} - \left(\aacomovil\right)^2 =
    -\left(8\pi G a^2 
      p -  a^2 \Lambda + K \right)
  \end{equation}
  \begin{equation}
    \label{eq:conservacion_conforme_fl}
    \rho' + 3 \HubbleComovil(\rho + p) = 0,
  \end{equation}
\end{subequations}

la ecuación auxiliar (\ref{eq:ece_auxiliar}) es

\begin{equation}
  \label{eq:ece_auxiliar_conforme}
  \HubbleComovil^{'} = -\frac{4\pi G a^2}{3}(\rho + 3p)+
  \frac{a^2 \Lambda}{3} .
\end{equation}

\subsubsection{Comportamiento de la materia en Universos de
  Friedmann-Lemaître}
La densidad total de materia $\Omega_{m0}$ en el presente,
está dada por la suma de las densidades de los diferentes
tipos de materia $\Omega_{i0}$, que suponemos existen,

\begin{equation}
  \label{eq:densidad_materia_total}
  \Omega_{m0} \equiv \Omega_{mat\thinspace 0} + 
  \Omega_{rad\thinspace 0} = \Omega_{B\,0} + \Omega_{DM\thinspace 0}  +
  \Omega_{\gamma\thinspace 0}  +
  \Omega_{\nu\thinspace 0},
\end{equation}
\nomenclature{$\Omega_{mat\thinspace 0}$}{Densidad de
  energía de la materia no relativista}
\nomenclature{$\Omega_{m0}$}{Densidad total de la materia
  hoy} \nomenclature{$\Omega_{B0}$}{Densidad de energía
  bariónica hoy} \nomenclature{$\Omega_{rad\thinspace
    0}$}{Densidad de energía de radiación hoy, i.e.\ materia
  ultra-relativista como fotones y neutrinos}
\nomenclature{$\Omega_{DM\thinspace 0}$}{Densidad de energía
  de la materia oscura hoy}
\nomenclature{$\Omega_{\nu\thinspace 0}$}{Densidad de
  energía de los neutrinos hoy}

donde $\Omega_{mat}$ incluye la materia no relativista (el
subíndice $B$ índica la materia bariónica y el subíndice
$DM$ representa a la materia oscura), y $\textit{rad}$
representa a la materia ultra-relativista o radiación (los
subíndices $\gamma$ y $\nu$ indican fotones y neutrinos
respectivamente).

El parámetro de densidad total $\Omega_0$ es la suma de la
densidad total de materia y la densidad de energía total de
la constante cosmológica

\begin{equation}
  \label{eq:densidad_total}
  \Omega_0 = \Omega_{m\,0} + \Omega_{DE\,0}.
\end{equation}

Usando estas variables, la ecuación de Friedmann
(\ref{eq:friedmann}) se expresa como

\begin{equation}
  \label{eq:friedmann_omega}
  \Omega_m + \Omega_{DE} -1 = \frac{K}{a^2H^2}.
\end{equation}

Otra forma muy útil de expresar la ecuación de Friedmann,
es en función de las densidades actuales,

\begin{equation}
  \label{eq:friedmann_omega_actual}
  H(a)^2 = H_0^2\left[\Omega_{DE\thinspace 0} + \Omega_{mat\thinspace 0}
    a^{-3} + \Omega_{rad\thinspace 0} a^{-4} - \left(\Omega_{0} -1\right)a^{-2}\right].
\end{equation}

Se supondrá regularmente que las ecuaciones de estado tienen
la forma

\begin{equation}\label{eq:estado_barotropica}
  p = (\gamma -1) \rho,    
\end{equation}

donde $\gamma$ es una constante. Desde un punto de vista
físico los casos más interesantes son $\gamma = 1$ (polvo),
$\gamma = \frac{4}{3}$ (radiación), $\gamma = 0$ (e.g.\
campo escalar dominado por el termino del potencial o una
constante cosmológica) y algunas veces $\gamma = 2$
(\textit{fluido duro}, e.g.\ ,\ un campo escalar dominado por el
término cinético \citep{Narlikar01, Wainwright1997}),
entonces el valor de $\gamma$ está en el rango dado por
\begin{equation*}
  0 \le \gamma \le 2.
\end{equation*}

Usando la ecuación de conservación
\eqref{eq:conservacion_fl} y la ecuación de estado
barotrópica (\ref{eq:estado_barotropica}), es posible
obtener la evolución de la densidad de energía
\begin{equation}
  \label{eq:relacion_densidad_factor_escala}
  \rho(t) = \frac{\rho_0}{a^{3\gamma}}.
\end{equation}




La materia no relativista (materia bariónica y materia
oscura) es considerada en estos modelos como ``polvo'', con
 una presión igual a cero, $p=0$,  por lo que su
densidad de energía varía como $\rho_m \propto a^{-3}$. La
radiación por su parte tiene la ecuación de
estado $p = \rho/3$, por lo que su densidad de energía
decae como\footnote{Este resultado también se puede obtener
  usando la ecuación geodésica para partículas sin masa}
$\rho \propto a^{-4}$. Estos resultados se pueden
interpretar de manera intuitiva de la siguiente manera: para
partículas no relativistas, su energía es igual a su masa en
reposo, la cual permanece constante en el tiempo. La
densidad de energía de muchas partículas no relativistas es
igual a su masa multiplicada por su densidad de número, el
cual varía inversamente proporcional al volumen ($a^3$),
i.e. $\rho_m \propto a^{-3}$; por su parte, los fotones
tienen una energía igual a $E_\gamma = k_B T$ y su longitud
de onda se expresa como $\lambda_\gamma = \hbar c /k_B
T$. En tiempos anteriores, la longitud de onda del fotón era
menor, ya que el factor de escala $a(t)$ era menor (ver
\ref{eq:distancias}), y debido a que la energía del fotón es
inversamente proporcional a su longitud de onda, obtenemos
que su energía debió de ser mayor en tiempos pasados por un
factor de $1/a(t)$ comparada con la energía actual. Usando
ambas relaciones obtenemos que la temperatura de un fotón en
función del tiempo es

\begin{equation}
  \label{eq:temperatura_foton}
  T(t) = \frac{T_0}{a(t)}
\end{equation}

Al considerar un plasma, su densidad de energía se puede
expresar como su densidad de número (que varía como
$a^{-3}$) multiplicada por su temperatura (cuya dependencia
temporal está dada por \ref{eq:temperatura_foton}). Este
último factor nos da un t\'ermino extra de $a^{-1}$ i.e.\
$\rho_{rad} \propto a^{-4}$. En la tabla
\ref{tab:comportamientos_factor_escala} se muestra un
resumen de lo tratado en esta sección (agregando el caso de constante cosmol\'ogica).

\begin{table}
  \begin{center}
    \begin{tabular}{c|c|c||c}
      \rowcolor[rgb]{0.8,0.8,0.8}  Tipo de Energía &  $a(t)$
      &  $a(\eta)$ &  $\rho$  \\  
      \hline 
      Radiación &  $t^{1/2}$ &  $\eta$ &  $a^{-4}$ \\ 
      Polvo &  $t^{2/3}$ &  $\eta^2$ &  $a^{-3}$  \\ 
      $\Lambda$ & $e^{Ht}$ & $-({H\eta})^{-1}$ & constante \\
      \hline 
    \end{tabular}
    \caption{Resumen sobre comportamientos del factor de
      escala en tiempo cósmico y tiempo conforme dependiendo
      de la época dominante (primera columna) y la evolución
      de la densidad de energía respecto al factor de
      escala}\label{tab:comportamientos_factor_escala}
  \end{center}
\end{table}

\paragraph{\textsc{Addendum}: Ecuaciones en tiempo conforme}
El comportamiento del factor de escala en el tiempo conforme
se puede calcular de manera análoga al caso del tiempo
cosmológico, resultando en $a(\eta) = C \eta^{2/(3\gamma
  -2)}$, siendo $C$ una constante de integración (ver la
tercera columna de la tabla
\ref{tab:comportamientos_factor_escala}).

\subsubsection{Singularidad inicial en Universos de
  Friedmann-Lemaître}\label{sec:singularidad-inicial}

De la ecuación (\ref{eq:ece_auxiliar}) es posible definir
una especie de \textit{masa gravitacional activa} de la
materia a $\mu_{grav} = \rho + 3p$ \citep{Ellis06}. Para
materia ordinaria, esta cantidad es positiva ya que
satisface la condición fuerte de energía
\citep[pags. 218-219]{Wald84}. La materia ordinaria entonces
desacelera la expansión del universo ($\ddot{a} < 0$). Una
constante cosmológica positiva causa una expansión acelerada
($\ddot{a} > 0$). La evolución final del universo dependerá
de cual de las dos tendencias es la dominante. Se puede
observar de (\ref{eq:ece_auxiliar}) que si $\Lambda \le 0$ y
$(\rho + 3p) > 0$ para todo tiempo $t$ y dado que en la
actualidad $a_0 > 0$ (por definición) y $H_0 > 0$ (ya que
observamos \textit{redshifts} y no \textit{blueshifts})
entonces de (\ref{eq:ece_auxiliar}) $\ddot{a} < 0$, por lo
tanto, existe un tiempo finito $t_* < t_0$ tal que $a(t) \to
0$ mientras que $t \to t_*$ (\ref{eq:ece_auxiliar}); y
$\lim_{t \to t_*^{+}} \rho = + \infty$
(\ref{eq:conservacion_fl}). Es decir el Universo ``inició''
en una singularidad \citep{Weinberg72, Ellis71, Wald84}.

La conclusión de que \emph{necesariamente} hubo un Big-Bang,
pudiera ser evitada con ayuda de una constante cosmológica
positiva ($\Lambda > 0$), pero debido a que sabemos que el
universo se ha expandido por lo menos en un radio de 11 (ya
que vemos objetos con un \textit{redshift}
de $10$). De la ecuación de Friedmann (\ref{eq:friedmann}) se
ve que $\Lambda$ tuvo que haber tenido una magnitud de por
lo menos $11^3$ veces la densidad de la materia actual para
dominar en esa época o tiempos anteriores, y evitar así la
singularidad, pero esto contradiría el valor dado por las
observaciones actuales \cite{Ellis06,EhlersRindler1989}.

Es de recalcar que, con esta conclusión, el universo alcanzó
temperaturas (o equivalentemente energías) del orden de la
temperatura de Planck ($T_{Planck} \equiv \sqrt{\frac{\hbar
    c^5}{Gk^2}} \approx 10^{32} \kelvin \sim
10^{19}\giga\electronvolt$) en las cuales la teoría General
de la Relatividad deja de ser válida, ya que esperamos que
los efectos cuánticos se vuelvan dominantes a esta escala.

\subsubsection{\label{sec:futuro-fl}Comportamiento Futuro de los Universos
  de Friedmann-Lemaître} 
Como vimos en la sección anterior, de acuerdo con la teoría
de Relatividad, los modelos cosmológicos de FLRW, con las
condiciones apropiadas inician en una singularidad. En esta
sección dividiremos el comportamiento posterior a la
singularidad inicial dependiendo de la materia dominante y
del valor de $\Lambda$. Primero, tomando en cuenta solamente
el valor de $\Lambda$ tenemos,

\begin{itemize}
\item $\mathbf{\Lambda = 0}.$ El Universo inicia en una
  singularidad, su futuro depende de la curvatura espacial
  $K$ o del parámetro de densidad $\Omega_0$. El Universo se
  expanderá por siempre si $K=0 \ssi \Omega =1$ o $K<0 \ssi
  \Omega < 1$, pero recolapsa en una singularidad
  (\textit{Big-Crunch}) si $K > 0 \ssi \Omega_0 > 1$. De
  este comportamiento, descubrimos que para universos FL con
  $\Lambda = 0$, $\Omega_0 = 1$ corresponde a la
  \textit{densidad crítica} que separa a los modelos que se
  expanden por siempre de los que recolapsan en el futuro:
  $\Omega_{critica} = 1 \ssi \kappa\rho_{critica} = 3H_0^2$
  (ver \ref{eq:friedmann_omega}), $\Omega_0 =
  \rho_0/\rho_{critica}$.
\item{$\mathbf{\Lambda < 0}. $} Todas las soluciones inician
  en singularidad y recolapsan.

\item{$\mathbf{\Lambda > 0}. $} Si
  $K = 0$ o $K=-1$ todas las soluciones inician en una
  singularidad y se expanden eternamente. Si $K = +1$
  existen modelos que inician en una singularidad y se
  expanden por siempre o colapsan en una singularidad
  futura, pero en esta situación existen también soluciones
  estáticas (e.g.\ Universo estático de Einstein), y soluciones
   en las cuales hay un ``rebote'': colapsan desde infinito, llegan a un radio mínimo y
  se expanden de nuevo (i.e.\ ,\ no tienen singularidad
  inicial).
\end{itemize}

\subsubsection{\label{sec:soluciones-fl}Soluciones Básicas
  de los Universos de Friedmann-Lemaître}

Estableceremos una clasificación de todas las soluciones de
los universos FL suponiendo que únicamente existe un
componente de materia y/o constante cosmológica. Esta
clasificación sigue la clasificación usada en
\cite{Rindler06}, para una clasificación más general
incluyendo universos que no son FL véase
\cite{Wainwright1997,Ellis99b}.

\begin{itemize}
\item \textsc{Modelos Estáticos} Los modelos estáticos
  fueron las primeras soluciones cosmológicas estudiadas y
  fueron propuestas por A. Einstein (1917). Es el único caso
  en el cual se deben de usar \textit{ambas} ecuaciones
  \eqref{eq:ece_fl}, en lugar de una sola de ellas. En ellas
  se pide que $\dot{a}(t) = 0$, lo cual implica que
  \begin{equation*}
    \frac{K}{a^2} = \Lambda = 4\pi G\rho .
  \end{equation*}
  Se pude ver inmediatamente (\ref{eq:conservacion_fl}) que
  $\rho$ es constante. Para tener un modelo realista es
  necesario que $\rho > 0$, por lo que $K=+1$. Los universos
  con estas características son conocidos bajo el nombre de
  \textit{Universos de Einstein}. Estos universos, como
  demostró Eddington poco después, son inestables ante
  pequeñas perturbaciones de $\rho$. Existen otras dos
  maneras de obtener un universo estático: (a) Sin constante
  cosmológica, i.e.\ ,\ $\Lambda = 0$, la única manera de obtener
  un universo estático es mediante presión negativa $p =
  -\rho/3$, de nuevo se tiene $K=+1$; y (b) $\Lambda = K =
  \rho = 0$, $\;a = $ cualquier constante, esto es (luego de
  una transformación que absorba la constante $a$) el
  espacio-tiempo de Minkowski, $\mathscr{M}^4$.

\item \textsc{Modelos vacíos} Son los universos en los
  cuales $\rho = \Lambda = 0$. Estos modelos no tienen mucho
  significado físico ya que son equivalentes a que la
  gravedad esté ``apagada'', i.e.\ $G = 0$. El
  espacio-tiempo de estos modelos sin materia es
  $\mathscr{M}^4$, sus diferencias se darán por la
  cinemática de las partículas de prueba en ellos (el
  \textit{substratum} en \cite[][pag. 358]{Rindler06}).  En
  el apartado anterior se mencionó uno de los dos modelos
  existentes: espacio-tiempo de Minkowski escalado. El otro
  es el modelo de Milne, descrito a continuación.

  \begin{itemize}
  \item{\textit{Modelo de Milne}}. Este modelo representa un
    universo plano, ($\Riemann = 0$ i.e.\ un espacio-tiempo
    de Minkowski, $\mathscr{M}^4$), visto por un conjunto
    uniforme de observadores en expansión\footnote{El por qué se están
      expandiendo y por que lo hacen desde un evento único
      no es especificado en el modelo.} en todas las
    direcciones y con todas las velocidades posibles desde un
    punto singular en $t=0$ 
    \citep[][pags. 360-363]{Rindler06}. Su expansión es
    lineal
    \begin{equation}
      \label{eq:factor_escala_milne}
      a(t) = Ct,
    \end{equation}
    y su edad es $\tau_0 = \frac{1}{H_0}$. El tiempo propio
    $\tau$ de los observadores comóviles, sirve para foliar
    el espacio-tiempo en hipersuperficies espaciales de
    curvatura negativa constante ($1/\tau^2$).

    Obviamente, en este modelo el ``Big-Bang'' es en
    realidad un evento puntual \textit{dentro} de o
    \textit{incluido} en el del espacio-tiempo. Este modelo puede
    ser una buena aproximación del futuro lejano del
    universo si $K <0$ y $\Lambda = 0$.
  \end{itemize}

\item \textsc{Modelos con materia} Son aquellos modelos en
  los cuales $\rho \neq 0$.

  \begin{itemize}
  \item \textit{Modelo con radiación}. En este modelo, un
    universo plano se supone lleno de radiación únicamente
    ($p = 1/3\rho, \Lambda = K = 0$). El factor de escala
    evoluciona de manera
    \begin{equation}
      \label{eq:factor_escala_radiacion}
      a(t) = C t^{1/2}.
    \end{equation}
    Con una edad en $a_0$ de $\tau_0 = \frac{1}{2H_0}$. Este
    es un buen modelo para describir lo que se conoce como
    etapa dominada por radiación.

  \item \textit{Modelos con ``polvo''}. El primer caso
    interesante es el conocido como \textit{Universo de
      Einstein-de Sitter} cuya composición es $p = \Lambda =
    K = 0 \then \Omega_0 = 1$. La condición $p=0$ indica que
    está formado por materia no relativista. El factor de
    escala,

    \begin{equation}
      \label{eq:factor_escala_materia}
      a(t) = C t^{2/3}.
    \end{equation}

    Su edad, cuando $H(t) = H_0$ es de $\tau_0 =
    \frac{2}{3H_0}.$ Es un buen modelo cuando la radiación
    ha dejado de ser dominante y antes de que la constante
    cosmológica domine. También es un buen modelo del
    comportamiento futuro del universo si los datos
    observacionales apuntaran a $K=\Lambda = 0$.

    Existen modelos con polvo con valores de $K \neq 0$ y
    $\Lambda = 0$. El caso con $K = -1$ fue durante mucho
    tiempo el considerado como el modelo correcto del
    universo actual. Su edad es \cite{Rindler06}
    \begin{equation*}
      \tau_0 = C_{mat} \left[\sqrt{\left(X + X^2\right)} -
        \sinh^{-1}\sqrt{X}\right], \quad X = \frac{a(t)}{C_{mat}}.
    \end{equation*}
    Para tiempos muy grandes $a(t) \propto t$ y se puede
    aproximar por un universo o modelo de Milne
    (cf. arriba). El caso con $K = +1$ es cíclico (en el
    sentido de que inicia en una singularidad y acaba en una
    singularidad futura) y si se define la masa total de un
    universo de este tipo mediante el producto de la
    densidad, $\rho$, y el volumen a un tiempo $t$,
    $2\pi^2a^3(t)$, tenemos $M := 2\pi^2a^3\rho = 3\pi
    C_{mat}/4G$. Esta característica impone una fuerte
    constricción ya que una masa finita $M$ implica $K=+1$,
    determina $C_{mat}$ y la historia entera de este
    espacio-tiempo. En este sentido es el modelo más
    \textit{Machiano} de la lista, ya que la masa determina
    completamente y de manera única al
    espacio-tiempo\cite{Rindler06}. Estos modelos poseen
    además de un horizonte de partículas, un horizonte de
    eventos (definido en la página \pageref{def:horizonte_eventos}).

  \item \textit{Modelos de de Sitter}. En estos modelos
    tienen $\Lambda > 0$ pero vacíos de cualquier otra
    materia. El espacio-tiempo de estos modelos es el
    espacio-tiempo de de Sitter $\bm{dS}^4$. Existen
    tres\footnote{Cuatro métricas diferentes si se toma en
      cuenta la métrica estática (que no es del tipo RW) que
      fue la descubierta por W. de Sitter
      \cite{Rindler06,Ellis06}.}  modelos o métricas
    diferentes \citep{HawkingEllis, Rindler06} dependiendo
    de como se elija la foliación del espacio-tiempo en
    hipersuperficies de curvatura constante positiva.  La
    diferencia entre estos modelos radica una vez más en el
    movimiento de las partículas de prueba.

    Para obtener los tres diferentes modelos debemos
    reescribir la ecuación (\ref{eq:friedmann}) con $\rho =
    0$ \cite{GronSigbjorn07},

    \begin{equation*}
      \dot{a}^2 - \omega^2 a^2 = -K
    \end{equation*}

    donde $\omega^2 = \frac{\Lambda}{3}$. Esta ecuación
    tiene para los distintos valores de $k$ las siguientes
    soluciones

    \begin{equation*}
      a(t)  = \, 
      \begin{cases}
        \; \cosh \omega t, & K=+1\\
        \; e^{\omega t},  & K=0\\
        \; \sinh\omega t, & K=-1
      \end{cases}
    \end{equation*}

    El caso con $\Lambda > 0, K = 0$, se conoce como
    \textit{Universo de de Sitter}. Como se acaba de mostrar
    este modelo tiene una expansión exponencial,

    \begin{equation}
      \label{eq:factor_escala_desitter}
      a(t) = C e^{Ht},
    \end{equation}

    donde $C$ y $H$ son constantes. El parámetro de Hubble
    es $H = \Lambda/3$.  Como la tasa de expansión es
    constante no tiene inicio (i.e.\ no tiene singularidad
    inicial para ningún $t$ finito)  su edad es
    infinita. Es un buen modelo del universo en el futuro si
    $\Lambda > 0$ (se puede entender observando la ecuación
    (\ref{eq:friedmann}) ya que para tiempo muy grandes el
    término que contiene $\Lambda$ dominará sobre los
    demás). También puede ser entendido como una solución
    con $\Lambda = 0$ y conteniendo materia que tiene una
    ecuación de estado $p = -\rho$, modelando así un periodo
    inflacionario muy temprano. También es el caso límite de
    todos los universos de FL con $\Lambda > 0$ que se
    expandan indefinidamente.

    El modelo con $K=-1$ inicia en una aparente
    singularidad, pero se puede demostrar que esta es una
    singularidad coordenada (cf. ver párrafo siguiente), es
    el análogo en $\bm{dS}^4$ del universo de Milne, pero
    aquí las partículas están acelerando debido a la
    constante cosmológica. El caso con $K=+1$ colapsa desde
    infinito a un tamaño mínimo distinto de cero y luego
    vuelve a expandirse.

    El espacio-tiempo de de Sitter es invariante ante
    transformaciones de Lorentz (esto se puede ver ya que es
    posible encajar la métrica este espacio-tiempo como un
    hiper-hiperboloide \cite{Rindler06,GronSigbjorn07} en el
    espacio-tiempo de Minkowski de 5 dimensiones,
    $\mathscr{M}^5$, las transformaciones de Lorentz dejan
    invariante al hiper-hiperboloide) siendo así un
    espacio-tiempo máximamente simétrico. A partir de esto
    se puede demostrar que las diferentes foliaciones del
    hiper-hiperboloide pueden ser transformadas una a la
    otra mediante combinaciones de rotaciones y
    transformaciones de Lorentz \cite{Rindler06}.
   
  \item \textit{Modelo de anti-de Sitter} El modelo con
    $\Lambda < 0$ es un espacio-tiempo de anti-de Sitter,
    $\bm{AdS}^4$. Este caso es análogo al universo de Milne
    en $\bm{AdS}^4$ pero los observadores son frenados por
    la atracción de $\Lambda$.
  \end{itemize}
\end{itemize}

Es importante notar que ninguno de estos modelos describe
por si solo al universo real. El universo real contiene una
mezcla de diversos tipos de materia y su combinación es la
que guiará la evolución del mismo, aunque existirán
intervalos de tiempo o épocas en las cuales su
comportamiento podrá ser descrito aproximadamente por alguno
de estos modelos.

\section{\label{sec:cosmologia-estandar}Cosmología Estándar}

Se entiende como modelo estándar de cosmología o cosmología
estándar a los modelos cosmológicos que (a) satisfacen las
dos suposiciones  mencionadas al inicio de \S
\ref{sec:cosmologia-fisica} (Relatividad General como teoría
correcta de la gravedad y Homogeneidad e Isotropía a cierta
escala), (b) tienen una estructura geométrica del tipo
Robertson-Walker, (c) poseen una dinámica dada por las ecuaciones
(\ref{eq:ece_fl},
\ref{eq:conservacion_fl}) con todas sus consecuencias
(subsecciones de \ref{sec:modelo-flrw}), incluyendo la
singularidad inicial
\cite{Peebles98,Peacock01,Maroto04,Scott05}. La cosmología
estándar está basada en el modelo estándar de partículas, el
cual ha sido probado como válido hasta energías de
$T \sim \unit{200} \giga\electronvolt$ mediante el
\acfi{LEP}\footnote{Se espera explorar energías más altas ($
  \unit{7} \tera\electronvolt$ por partícula o $\unit{574}
  \tera\electronvolt$ por núcleo) más en
  el \acfi{LHC} el cual se espera que esté en funcionamiento
  en Septiembre del 2009. Para obtener más información
  se puede visitar la página web localizada en
  \url{{http://lhc.web.cern.ch/lhc/}}.} 
\footnote{Consúltese su página web
  \url{http://delphiwww.cern.ch/offline/lepwgs.html} para
  más información} . 

Con estas suposiciones, la evolución de un universo 
en la cosmología estándar está caracterizado por tres épocas
importantes: (a) Una
época dominada por radiación ($\rho_{rad} \gg \rho_{m}$) que
ocurre a \textit{redshifts} mayores que $z_{eq} \approx
(\Omega_{DM}/\Omega_{rad}) \approx 10^4$. Para $z \ge
z_{eq}$, la densidad de energía es dominada por materia
relativística y el universo puede aproximarse mediante el
modelo con radiación (cf. pag.
\pageref{sec:soluciones-fl}). La segunda fase ocurre para $z
\ll z_{eq}$, en el cual es universo es dominado por materia,
que en los casos más sencillos se comporta como un modelo de
Einstein-de Sitter; (c) Observaciones
de supernovas \todo{Observaciones de supernovas} indican que
actualmente existe una dominación de constante cosmológica
sobre los otros tipos de materia.

En cualquier época, la temperatura va como $T \propto
a^{-1}$. Cuando la temperatura cae por abajo de $T \approx
\unit{10^3} \kelvin$, los átomos empiezan a formarse y los
fotones se desacoplan de la materia, es en este momento
($z_{dec} \approx 1100$) cuando se forma la \acfi{LSS},
emitiendo el 
\acfi{CBR}\footnote{También conocido como \ac{CMB} pero hay
  que notar que solamente en la actualidad las frecuencias
  de esos fotones se encuentran en el rango de las microondas.}, la cual
está formada por fotones que han  viajado sin dispersión 
desde entonces.  El CBR tiene un espectro de cuerpo negro
con una temperatura diferente de cero, este espectro de
cuerpo negro es una consecuencia de que los fotones y la
materia estuvieron en equilibrio térmico en épocas
anteriores al desacople (el equilibrio térmico se mantenía
ya que los fotones y electrones interactuaban fuertemente
mediante dispersiones de Thomson).

\subsection{\label{sec:historia-termica}Historia Térmica del
  universo en la Cosmología Estándar} 

La mayor parte de esta sección está basada en las siguiente
lista de referencias bibliográficas
\cite{Peacock98,Kolb1990,Peebles93,Bergstrom99,Dodelson}. Referencias
más puntuales se darán como citas a lo largo de la
sección. \todo{Hacer $k_B = 1$}

\paragraph{Preliminares.}  
Supongamos una caja cúbica de lado
$L$, con condiciones de 
frontera periódicas. Los campos cuánticos se expandirán en
ondas planas con números de onda $k$  dados por $k_i = n_i
(2\pi)/L$ con $i = \{x,y,z\}$.  Así, la densidad de estados
en el espacio de fase para una partícula con $g$ grados de
libertad internos (e.g.\ el espín) es

\begin{equation}
  \label{eq:densidad_estados}
  dN = \frac{g}{(2\pi\hbar)^3} V d^3k,
\end{equation}

Al número $g$ también se le conoce como \textit{factor de
  degeneración}. Esta cantidad es extensiva (ya que es
proporcional al volumen, $V$) y por lo tanto la densidad de
número $dn$ será independiente de $V$. De acuerdo con física
estadística el valor de expectación del número de ocupación
de un estado de energía $E$ o \textit{función de
  distribución} es

\begin{equation}
  \label{eq:funcion_distribucion}
  f(\bvec{p}) = \frac{1}{e^{(E-\mu)k_B T} \pm 1},
\end{equation}
\nomenclature{f(\bvec{p})}{Función de distribución}

donde $k_B$ es la constante de Boltzmann, y el signo
a usar depende del tipo de partículas ($+$
corresponde a fermiones y el $-$ a bosones); $\mu$ es el
potencial químico, definido por el cambio de energía
asociado con el cambio en el número de partículas, como se
puede apreciar de la relación termodinámica
\begin{equation}
  \label{eq:termo_fundamental}
  dE = TdS - PdV + \mu dN.
\end{equation}

Si las especies de partículas $i$
tienen la distribución 
mostrada arriba para algún $\mu_i$ y $T_i$, se dice que
están en \textit{equilibrio cinético}. Si todas las especies
están a la misma temperatura se dice que el sistema está en
\textit{equilibrio térmico}, este equilibrio requiere que
las interacciones entre los constituyentes del sistema
ocurran frecuentemente, si se cumple esto, podemos
describir al universo como evolucionando a través de una
secuencia de estados en equilibrio térmico y usar
 cantidades termodinámicas como la temperatura $T$,
presión $P$, densidad de entropía $s$ etc en cada tiempo
$t$. Si el sistema está en 
\textit{equilibrio químico}, los potenciales químicos de las
diferentes especies de partículas están relacionados de
acuerdo a las fórmulas de reacción, e.g. $i + j \leftrightarrow k +l$,
$\mu_i + \mu_j = \mu_k + \mu_l$, entonces, todos los
potenciales químicos pueden ser expresados en términos de
potenciales químicos de cantidades conservadas, e.g. el
potencial químico bariónico, $\mu_B$.

En el equilibrio térmico, $N$ se estabilizará alrededor de
su valor de equilibrio, por lo que esperamos que haya muy
pocos cambios en $N$, haciendo así que $\mu = 0$. Esto se
pude probar formalmente usando la definición de la
\textit{función de energía libre de Helmholtz}: $F = E -
TS$. Esta cantidad es minimizada en el estado de equilibrio
de un sistema a temperatura y volumen constante. Dado que
$dF = -SdT -PdV + \mu dN$, tenemos que $dF/dN = 0 \then \mu
= 0$.

La densidad de partículas en el espacio de fase es la
densidad de estados multiplicada por el número de ocupación,

\begin{equation}
  \label{eq:densidad_particulas_espacio_fase}
  \frac{g}{(2\pi \hbar)^3} f(\bvec{p}).
\end{equation}

La densidad de partículas en el espacio ordinario se obtiene
integrando sobre el espacio de momentos. 

\begin{equation}
  \label{eq:densidad_numero}
  n = \frac{g}{(2\pi\hbar)^3} \int_0^\infty f(\bvec{p}) d^3p.
\end{equation}

Otras cantidades de interés son la densidad de energía y la
presión
\begin{subequations}
  \begin{equation}
    \label{eq:densidad_energia_def}
    \rho = \frac{g}{(2\pi\hbar)^3} \int_0^\infty E(\bvec{p})
    f(\bvec{p}) d^3p,
  \end{equation}
  \begin{equation}
    \label{eq:presion_def}
    P = \frac{g}{(2\pi\hbar)^3} \int_0^\infty
    \frac{|\bvec{p}|}{3E}f(\bvec{p})d^3p .
  \end{equation}
\end{subequations}

Estas expresiones tienen dos límites muy importantes, el
primero ocurre cuando $T \gg m$, y es conocido como el
\textit{límite ultra-relativista}, en él podemos aproximar $E
= \sqrt{p^2 + m^2} \approx p$. En este límite también se
tiene que $T \gg |\mu|$, por lo que ambas variables pueden
ser despreciadas, obteniendo las siguientes fórmulas

\begin{subequations}
  \begin{equation}
    \label{eq:densidad_numero_ur}
    n = \frac{g}{(2\pi\hbar)^3} \int_0^\infty \frac{4\pi p^2
      dp}{e^{p/k_BT} \pm 1} = \begin{cases}\frac{3}{4\pi^2} \xi(3)
        gT^3, \quad \text{fermiones} \\
      \frac{1}{\pi^2}\xi(3)gT^3, \quad \text{bosones}\end{cases}.
  \end{equation}
  \begin{equation}
    \label{eq:densidad_energia_ur}
    \rho = \frac{g}{(2\pi\hbar)^3}\int_0^\infty \frac{4\pi
      p^3 dp}{e^{p/k_BT} \pm 1} = \begin{cases}
      \frac{7}{8}\frac{\pi^2}{30}gT^4,\quad \text{fermiones}
      \\
      \frac{\pi^2}{30}gT^4,\quad \text{bosones},
    \end{cases}
  \end{equation}
  \begin{equation}
    \label{eq:presion_ur}
    P = \frac{g}{(2\pi\hbar)^3} \int_0^\infty
    \frac{\frac{4}{3}\pi p^3 dp}{e^{p/k_BT} \pm 1} =
    \frac{1}{3}\rho ,
  \end{equation}
\end{subequations}

donde $\xi(3) \equiv \sum_{n=1}^\infty (1/n^3) = 1.20206$ es
la función zeta de Riemann y se usó el hecho de que las
funciones de distribución sólo dependen de $p \equiv |\bvec{p}|$, $d^3p
\to p^2dpd\Omega$. El otro límite es el conocido
como \textit{no relativista}, y está dado por (a) $T \ll m$,
i.e.\ las energías cinéticas típicas están muy por abajo de
la masa $m$, por lo que podemos aproximar a la energía por
$E = m + p^2/2m$; y (b) $T \ll m - \mu$, condición que lleva
a que el sistema esté \textit{diluido}, i.e.\ números de
ocupación $\ll 1$. Ambas condiciones permiten hacer la
aproximación siguiente

\begin{equation*}
  e^{(E-\mu)/k_BT} \pm 1 \approx e^{(E-\mu)/k_BT},
\end{equation*}

lo cual hace que las diferencias entre bosones y fermiones
desaparezcan (no sean importantes)  en este límite,

\begin{subequations}
  \begin{equation}
    \label{eq:densidad_numero_nr}
    n = g\left(\frac{mT}{2\pi\hbar}\right)^{3/2} e^{-(m-\mu)/T},
  \end{equation}
  \begin{equation}
    \label{eq:densidad_energía_nr}
    \rho = n \left( m + \frac{3}{2}T\right),
  \end{equation}
  \begin{equation}
    \label{eq:presion_nr}
    P = nT \ll \rho.
  \end{equation}
\end{subequations}

Las diferencias entre el caso ultra-relativista y el no
relativista (una caída exponencial) se debe principalmente a
que las partículas y las antipartículas se aniquilan unas a
las otras, a temperaturas altas estas reacciones ocurren
constantemente y son balanceadas por la producción de pares,
pero a bajas temperaturas la energía térmica de las
partículas en insuficiente para la producción por
pares. Para los casos intermedios ($T \sim m$) es 
necesario hacer las integrales numéricamente. 

Otra cantidad de suma importancia es la entropía $S(V,T)$,
la entropía es introducida como una ecuación central a la
termodinámica mediante el diferencial

\begin{equation}
  \label{eq:entropia_termo}
  dS(V,T) = \frac{1}{T} \left[ d(\rho(T)V) + P(T)dV\right], 
\end{equation}

Usando las ecuaciones 
(\ref{eq:densidad_energia_def}) y (\ref{eq:presion_def})
obtenemos, 

\begin{equation}
  \label{eq:presion_temperatura}
  \frac{d P(T)}{dT} = \frac{1}{T} (\rho(T) + P(T))
\end{equation}

e insertando esta relación en (\ref{eq:entropia_termo})
llegamos (salvo una constante de integración)  a 

\begin{equation}
  \label{eq:entropia}
  S(V,T) = \frac{V}{T}\left[ \rho(T) + P(T)\right].
\end{equation}

Es posible obtener una relación de conservación para la
entropía. Reescribiendo (\ref{eq:conservacion_fl}) como $a^3
dP/dT = d/dt [a^3(\rho +P)]$ ,  combinándola con
(\ref{eq:presion_temperatura}) e identificando a $V$ con
$a^3(t)$ llegamos a 

\begin{equation}
  \label{eq:conservacion_entropia}
  \frac{d}{dt}(S) = 0.
\end{equation}

Si definimos la densidad de entropía $s$ mediante $s = S/V$
la conservación se escribe como 

\begin{equation}
  \label{eq:conservacion_densidad_entropia}
  \frac{d}{dt} (a^3s) = 0.
\end{equation}

Para las especies relativistas tenemos

\begin{equation}
  \label{eq:densidad_entropia}
  s = \frac{\rho + P}{T} = \begin{cases} \frac{7\pi^2}{180}
    gT^3 \quad \text{fermiones} \\ \frac{2\pi^2}{45}gT^3
    \quad \text{bosones}.
  \end{cases}
\end{equation}

el caso no relativista no es importante y no se escribirá
aquí, ya que la entropía
de la radiación es abrumadoramente mayor que la entropía de
las especies no relativistas.

La entropía que se produce durante los diferentes procesos que
ocurren en el universo es insignificante comparada con la
entropía total del universo, por lo que se dice que 
el universo es \textit{adiabático} (cf. deducción arriba).

\paragraph{Condición de equilibrio térmico.} Si los
constituyentes del universo tienen una densidad de 
número $n$, velocidades típicas relativas $v$ e interactúan
mediante un proceso de dispersión que tiene una sección
efectiva de cruce\footnote{\textit{Cross-section} en
  inglés.} $\sigma$, la tasa de interacción por partícula
$\Gamma$ está dada por $\Gamma = \lambda_{fmp}^{-1} =
n\sigma v $, donde $\lambda_{fmp}$ es la distancia libre 
promedio\footnote{En ingleś \textit{free-mean path}}. La condición
que se dio arriba,
para que se mantenga el equilibrio térmico a través de las
interacciones es que su tasa de interacciones sea mucho más
grande que la tasa de expansión del universo:

\begin{equation}
  \label{eq:condicion_eq}
  \Gamma \gg H.
\end{equation}

La densidad de número decrece, regularmente,  con el tiempo a un mayor ritmo
que el parámetro de Hubble. Esto significa que en ciertas
épocas algunas especies abandonarán el equilibrio
térmico. Su densidad de número se ``congelará''\label{def:congelar} a cierto
valor y sólo cambiará al ser diluido por la expansión del
universo. Este ``congelamiento'' es uno de los actores más
importantes en la evolución del universo como veremos
adelante \cite{Bergstrom99}\footnote{Este argumento heurístico puede ser
  expuesto de una manera más rigurosa usando la ecuación de
  Boltzmann, la cual gobierna la abundancia de las especies
  de partículas en un universo que se expande, para esta
  demostración cf.\ \S 9.2 de \cite{Peacock98}.}.

\paragraph{Sopa primordial.} La sopa primordial estaba
compuesta por todas las diferentes especies de partículas
elementales. La masa de estas especies cubren un amplio
rango comprendido 
entre $m \sim \unit{175} \giga\electronvolt$ (quark top)
hasta el fotón $m = 0$. La expansión del universo está
gobernada por la densidad de energía total, $\rho(T) =
\sum_i \rho_i(T)$, donde la $i$ representa todas las
especies en la sopa primigenia. La densidad de energía de
las partículas ultra-relativistas (a las cuales llamaremos
genéricamente ``radiación'') es mayor que las de las
especies no-relativistas, por lo que, en el universo
temprano (i.e.\ dominado por radicación) es suficiente tomar
en cuenta a la radiación en el cálculo, así, la densidad de
energía es

\begin{equation}
  \label{eq:densidad_energia_universo_temprano}
  \rho(T) = \frac{\pi^2}{30}g_*(T) T^4,
\end{equation}

y la densidad de entropía,

\begin{equation}
  \label{eq:densidad_entropia}
  s(T) = \frac{2\pi}{45}g_*^s(T) T^3,
\end{equation}

donde $g_*(T) := \sum_{i =bosones}g_i(T_i/T)^4$\thinspace $+
\frac{7}{8}\sum_{j = fermiones} g_j(T_j/T)^4$ y $g_*^s :=
\sum_{i =bosones}g_i(T_i/T)^3$\thinspace + $ \frac{7}{8}$ $\sum_{j =
  fermiones}$ $g_j(T_j/T)^3$ . Las variables $T_{i,j}$ son temperaturas
específicas de cada una de las especies. Mientras todas las
especies tengan la misma temperatura y sean relativistas
(e.g. $P \sim (1/3)\rho$), $g_*(T) \simeq g_*^s(T)$. El
extraño factor de $\frac{7}{8}$ se arrastra desde la
ecuación (\ref{eq:densidad_energia_ur}). La presión es en
esta época $P(T) = \approx (1/3)\rho(T)$.

A temperaturas mayores que la masa del quark top $T > m_t
\approx \unit{175} \giga\electronvolt$ todas las partículas
son relativistas, sumando los grados de libertad internos de
las especies a esta época nos da $g_* = 106.75$. La
transición \acf{EW} ocurre aproximadamente a esta
temperatura ($T_{EW} \sim \unit{100}\giga\electronvolt$)
pero su discusión queda fuera de este trabajo de tesis. Poco
después de la transición EW, el quark \textit{top} está
aniquilándose, seguidos casi inmediatamente por el bosón de
Higgs y los bosones de norma $W^{\pm}, Z^0$. Para cuando la
temperatura llega a $T \sim
\unit{10}\giga\electronvolt$. tenemos $g_* = 86.25$. Los
siguientes en aniquilarse son los quarks $b$
(\textit{bottom}), $c$ (\textit{charm}) y luego el mesón
$\tau$, si el quark $s$ (\textit{strange}) tiene tiempo de
aniquilarse llegaremos a $g_* = 51.25$. A estas temperaturas
$T \sim \unit{150}\mega\electronvolt$ ocurre la
\textit{transición de confinamiento (o confinante) QCD}, en la cual, los quarks pierden su
\textit{libertad asintótica}. Las fuerzas \textit{fuertes}
se vuelve importante y mediante una transición de fase,
desaparece así el \textit{plasma de quarks-gluones} para
convertirse en un \textit{gas de hadrones}. Esto se debe a
que los quarks y gluones han formado sistemas de tres quarks
llamados \textit{bariones} y pares quarks-antiquark
conocidos como \textit{mesones}, Los bariones más ligeros,
el protón ($p^+$) y el neutrón ($n^0$) son conocidos con el
nombre común de \textit{nucleones}. A excepción de los
piones, todas estas partículas dejan de ser relativistas a
bajo de la temperatura de la transición de QCD. Las únicas
especies que quedan ultra-relativistas y en gran número son:
piones, muones, electrones ($e^-$), neutrinos ($\nu$) y
fotones ($\gamma$), por lo que al final de esta época $g_* =
17.25$.

\begin{table}
  \centering
  \begin{tabular}{c|c|c}
    \hline
    \rowcolor[rgb]{0.8,0.8,0.8}  Temperatura ($\giga\electronvolt$) & especies & $g_*$ \\
    \hline 
    $\sim 200$ & todas presentes & 106.75 \\
    $\sim 100$ & transición EW & sin efecto \\
    $ < 170$  & aniquilación del quark $t$  & 96.25 \\
    $ < 80 $ &  $W^{\pm}, Z^0, H^0$ & 86.25 \\
    $ < 4 $ & quark $b$ & 75.75 \\
    $ < 1 $ & quark $c$, $\tau^-$ & 61.75 \\
    $ \sim 150 \times 10^{-3}$ & transición QCD & 17.25 \;
    ($u,d,g \to \pi^{\pm,\thinspace  0}$) \\
    $< 100 \times 10^{-3}$ & $\pi^{\pm,\thinspace 0}, \mu^-$
    & 10.75 (sobreviven $e^{\pm}, \nu, \gamma$ )\\
    $< 500 \times 10^{-6}$ & aniquilación $e^{\pm}$ & 7.25
    \\
    \hline
  \end{tabular}
  \caption{Historia de $g_*(T)$}
  \label{tab:historia_g_estrella}
\end{table}

\paragraph{Desacople del neutrino.}  Los neutrinos sólo
son susceptibles a la fuerza débil. A temperaturas de $\unit{1}
\mega\electronvolt$ el neutrino se \textit{desacoplará},
i.e.\ no podrá mantenerse en equilibrio térmico con el resto
de las especies. Luego del
desacople los neutrinos se moverán libremente prácticamente
sin interacciones. 

La sección eficaz de las interacciones débiles se puede estimar
fácilmente \cite{Peacock98, Bergstrom99} y es proporcional a
$\alpha^2s/m_w^4$, donde $\alpha$ es la constante de
estructura fina ($\sim 1/137$), $s$ es la variable de
Maldestam $s= (p_1+p_2)$, y $m_w\sim
\unit{80}\giga\electronvolt$ es la masa del bosón 
$W^\pm$. Para neutrinos y leptones cargados relativistas tenemos $s \sim T^2$
ya que $s$ es proporcional al cuadrado de la energía que a
su vez es proporcional a $T$, además $|\bvec{v}| = c = 1$ y
$n \sim T^3$. Un
proceso típico que 
mantiene el equilibrio térmico en esta época es  $\nu_e + e^+ \to \nu_\mu
+ \mu^+$, cuya tasa de interacción es 
\begin{equation}
  \label{eq:tasa_ew}
  \Gamma_{ew} \sim \frac{\alpha^2T^5}{m_W^4}
\end{equation}

Durante el universo temprano la ecuación de Friedmann
(\ref{eq:friedmann}) es $H^2 = 2.76 g_* T^4/m_{planck}^2$ o 
\begin{equation}
  \label{eq:hubble_rad}
  H = 1.66 \sqrt{g_*} \frac{T^2}{m_{Planck}},
\end{equation}
la razón  entre las tasas es
\begin{equation}
  \label{eq:razon_neutrino}
  \frac{\Gamma}{H} \sim \frac{\alpha^2 m_{Planck}T^3}{m_W^4}.
\end{equation}

El desacople ocurrirá cuando esta razón sea menor que la
unidad, 

\begin{equation}
  \label{eq:desacople_neutrino}
  T_{\nu\thinspace d} \sim
  \left(\frac{m_W^4}{\alpha^2 m_{Planck}}\right)^{1/3} \sim 4 \mega\electronvolt.
\end{equation}

Una vez desacoplados los neutrinos, aunque ya no estarán en
equilibrio térmico con otras especies estarán en equilibrio
cinético \footnote{Esto puede mostrarse como sigue, el
  momentum de un neutrino libre (y de cualquier otra
  partícula) sufre un \textit{redshift} mientras el universo
se expande, $p(t_2) = (a_1/a_2)p(t_1)$. Al tiempo $t_1$ el
elemento en el espacio de fase $d^3p_1dV_1$ contiene
\begin{equation*}
  dN = \frac{g}{(2\pi\hbar)^3}f(\bvec{p_1}) d^3p_1 dV_1
\end{equation*}

Al tiempo dos, las mismas $dN$ partículas están en el
elemento del espacio de fase $d^3p_2 dV_2$, ahora la
función de distribución en $t_2$ está dada mediante
\begin{equation*}
  \frac{g}{(2\pi\hbar)^3} f(\bvec{p}_2) = \frac{dN}{d^3p_2dV_2},
\end{equation*}
La relación entre la función de distribución a dos
diferentes tiempos se puede obtener usando $d^3p_2 =
(a_1/a_2)d^3p_1$ y $dV_2 = (a_2/a_1)^3dV_1$, obteniendo
\begin{equation*}
  f(\bvec{p}_2) = \frac{1}{e^{(p_2-\mu_2)/k_BT_2} \pm 1},
\end{equation*}
donde $\mu_2 \equiv (a_1/a_2)\mu_1$ y $T_2 \equiv
(a_1/a_2)T_1$. Este análisis es válido para partículas
relativistas, para partículas no-relativistas existe un
resultado diferente.}. Las partículas mantendrán entonces la forma de una
distribución termal, pero con su temperatura y potencial
químico corridas al rojo por un factor $\propto
a^{-1}$. Esto parecería indicar que los neutrinos también seguirán
en equilibrio térmico, pero por esta época los electrones y
positrones empiezan a aniquilarse ($e^+e^- \to
\gamma\gamma$)   afectando así el valor de $g_*$ (ya que la
reacción inversa $\gamma\gamma \to e^+e^-$ ya no es
posible). Debido a esta aniquilación, debemos de empezar a
distinguir a $g_*(T)$ de $g_*^s(T)$. La manera más sencilla de
ver este cambio en la temperatura de las especies respecto a
la del neutrino es 
usando la conservación de la entropía  $g_*^s(T)a^3T^3 = $
constante. Antes de la aniquilación de $e^+e^-$ tenemos $g_*
= g_*^s = 2 + 3.5 + 5.25 = 10.75$, luego de la aniquilación
se tiene $g_*^s = 2 + 5.25 (T_\nu/T)^3$. Al reducirse el
número de grados de libertad relativistas la densidad de
energía y entropía se transfieren de los $e^+$ y $e^-$ a los
fotones, pero no a los neutrinos. Igualando el antes con el
después de la aniquilación $10.75 = 2(T/T_\nu)^3 + 5.25$, de
donde la temperatura de los neutrinos luego de la
aniquilación $e^+e^-$ será

\begin{equation}
  \label{eq:temperatura_neutrino}
  T_\nu = \left(\frac{4}{11}\right)^{1/3}\thinspace T= 0.714\thinspace
  T.
\end{equation}

Después de esta aniquilación, solo los fotones y los
neutrinos (suponiéndolos sin masa), permanecerán
relativistas.

\paragraph{Nucleosíntesis}\label{sec:nucleosintesis} Esta
etapa produce los elementos 
ligeros: $^2H (D), ^3He, ^4He$ y $^7Li$ 
al principio de la época dominada por radiación,
aproximadamente a temperaturas de $T_{ns}  \lesssim 
\unit{1}\mega\electronvolt$, correspondientes a una edad del
universo de aproximadamente $t \gtrsim \unit{1} \second$ 

La predicción de abundancia de dichos
elementos y su confirmación por las observaciones es uno de los 
grandes logros del modelo estándar cosmológico. Esta
predicción es impresionante ya que 
relaciones de las abundancias respecto a la abundancia  de
$H$ cubre varios órdenes de
magnitud: $^4$He/H $\sim 0.08$ hasta $^7$Li / H $\sim
10^{-10}$, así esta predicción y su confirmación ponen cotas
muy restrictivas a posibles desviaciones del modelo estándar
cosmológico \cite{Fields08}, ver figura
\ref{fig:nucleosintesis}.Se recomienda al lector revisar la
literatura \cite{Burles99,Fields08,GarciaBellido05}.  

\figura{imagenes/nucleosintesis}{width=10cm,height=14cm}{Constricciones
  sobre la densidad bariónica de la
  Nucleosíntesis. Son mostradas las predicciones de cuatro
  elementos (de arriba a abajo):  $^4$He (verde), $D$ (roja), $^3$He (azul) y
  $^7$Li (morada). Estas predicciones abarcan 10 ordenes de
  magnitud (ver texto). La banda sólida son las mediciones
  de $D$ primordial. Las cajas muestran observaciones
  sobre las abundancias de elementos, nótese que para el
  $^3$He sólo se conoce el límite superior. Figura tomada de
  \citeauthor{Burles99}, \citeyear{Burles99} 
  \cite{Burles99}.}{fig:nucleosintesis}

\paragraph{Recombinación\footnote{Este nombre, como muchas
    otros nombres en cosmología, no es el adecuado, ya que
    el plasma siempre ha estado ionizado hasta este tiempo y
    por lo tanto es la primera vez que se combinan los
    electrones con los protones para formar núcleos neutros,
    i.e.\ es una \textit{combinación} no una
    \textit{recombinación}. El nombre se importa de procesos
    estelares en los cuales el proceso de
    ionización-desionización ocurre varias veces.},
  Desacople y LSS}\label{def:recombinacion} No toda la
materia fue eliminada a través 
de la aniquilación de partículas/antipartículas, por algún
razón aún misteriosa existía un pequeño exceso de lo que
ahora llamamos materia sobre la antimateria. Esto nos indica
que $\mu_B$, el potencial químico bariónico, es diferente de
cero y positivo. Al ser los protones y neutrones los
bariones más ligeros, suponemos que la mayor parte de
$\mu_B$ se debe a los nucleones ($n_B \equiv n_n + n_p$). El
universo es eléctricamente neutro, por lo que el número de
electrones en el universo debe ser igual al número de los
protones($n_{e^-} = n_p$). Con estas simples observaciones
podemos continuar con los cálculos de la evolución del
universo.

Definamos un parámetro $\eta_{B\gamma}$, que es la razón
entre bariones y fotones hoy
\begin{equation}
  \label{eq:razon_barion_foton}
  \eta_{B\gamma} := \frac{n_B(t_0)}{n_\gamma(t_0)},
\end{equation}
\nomenclature{$\eta_{B\gamma}$}{Razón entre la densidad de
  número de los bariones y fotones hoy}
de las observaciones provenientes de nucleosíntesis, sabemos
que $\eta_{B\gamma} \approx 10^{-9}$. El número bariónico es
conservado por lo que $n_B V$ = constante, entonces $n_B
\propto a^{-3}$. Luego de la aniquilación $e^+e^-$,
$n_\gamma \propto a^{-3}$ también, entonces
\begin{equation}
  \label{eq:numero_barionico_nr}
  n_B(T) = \eta_{B\gamma}n_\gamma =
  \eta_{B\gamma}\frac{2\xi(3)}{\hbar^3\pi^2}T^3, \quad T \ll m_e.
\end{equation}

El universo contiene, en esta época ($T \sim
\unit{10}\kilo\electronvolt \to \unit{1}\electronvolt$), un
fondo relativista de neutrinos y fotones (``radiación'') y
electrones, protones y núcleos libres en forma no
relativista (``materia''). Aunque dominado por la densidad
de energía de la radiación, debido a sus diferentes
dependencias temporales ($\rho_{rad} \propto a^{-4}$ vs.\
$\rho_{mat} \propto a^{-3}$), la densidad de energía de la
radiación a partir del tiempo $t_{EQ}$ será menor que de la densidad de
energía de la materia\footnote{El \textit{tiempo de
    equilibrio} $t_{EQ}$ ocurre cuando la densidad de
  energía de la radiación y la densidad de energía de la
  materia son iguales, i.e. cuando $\rho_{rad} = \rho_{mat}$.} y el
universo  entrará en la \textit{época de dominación de la
  materia}. La evidencia observacional apunta a que 
hay más materia que los bariones, llamada \acf{DM} que no
hemos considerado en el análisis hasta este punto, pero la
misma evidencia observacional indica que la DM
interactúa débilmente, por lo que no afectará los cálculos
hasta aquí desarrollados (aunque si afectará la formación de
la estructura cf. capítulo \ref{cha:critica-al-origen}), su
efecto principal será adelantar el 
dominio de la materia en el universo, y será un ingrediente
fundamental en la formación de estructura.

La radiación y la materia permanecerán en equilibrio
termodinámico mientras haya muchos electrones libres. El
proceso que mantiene este equilibrio principalmente por la
dispersión de Thomson ($e^- + \gamma \to e^- + \gamma$) y la
fotoionización/recombinación del hidrógeno ($H + \gamma
\leftrightarrow p^+ + e^-$)

La temperatura seguirá disminuyendo y llegará a ser  lo suficientemente baja
para permitir que los electrones y nucleones formen núcleos
neutros --e.g.\ el proceso de fotoionización del hidrógeno
$H+\gamma \to p^+ + e^-$ -- ya no será posible
energéticamente). A este momento se le conoce como
\textit{recombinación}, provocando que la densidad de
electrones libres decaiga rápidamente y a la vez provocando
que \textit{el camino libre promedio} de los fotones se
vuelva mayor que el tamaño del horizonte.  Cuando esto
sucede el universo se vuelve transparente y se dice que los
fotones se han desacoplado. Estos fotones libres son los que
forman el CBR y tienen una temperatura actualmente (en $t_0$)
de $T(t_0) = T_0 = \unit{2.725}\kelvin$. Después del desacople la
temperatura de la materia caerá más rápido que la de los
fotones, pero la formación de estructura hará que la materia
se caliente a diferentes temperaturas en diferentes lugares.

La cosmología estándar, aunque predice el CBR, es incapaz de
predecir el valor de $T$ a $t = t_0$. Es un parámetro libre
de la teoría y se usará $\Omega_{rad} \propto T^4$ en lugar
de $T_0$ en este trabajo de tesis \citep{Narlikar01}.

Para simplificar la discusión sobre la recombinación,
supondremos que todos los núcleos son en realidad
protones. La densidad de número de los protones libres será
$n_p$, la de los electrones libres $n_e$ y la de los átomos
de hidrógeno por $n_H$. La reacción de recombinación es $p^+
+ e^- \to H + \gamma$, debido a que $\mu_\gamma = 0$, los
potenciales químicos están relacionados mediante $\mu_p +
\mu_e = \mu_H$, usando estos datos en
(\ref{eq:densidad_numero_nr}) obtenemos la relación
\begin{equation}
  \label{eq:densidad_hidrogeno}
  n_H =
  \frac{g_H}{g_pg_e}n_pn_e\left(\frac{m_em_pk_BT}{2\pi\hbar^2
      m_H}\right)^{-3/2} e^{B/k_B T}
\end{equation}
donde $B$ es la energía de enlace del hidrógeno, $B := m_p +
m_e - m_H = \unit{13.6}\electronvolt$, $g_e = g_p = 2$, $g_H
= 4$. Fuera del exponencial, podemos ignorar las diferencias
de masa entre el hidrógeno y el protón y hacer $m_p \simeq
m_H$, obtenemos la ecuación de Saha
\begin{equation}
  \label{eq:saha_1}
  \frac{n_H}{n_pn_e}  = \left(\frac{m_ek_BT}{2\pi\hbar^2
    }\right)^{-3/2} e^{B/k_B T},
\end{equation}

y definimos la ionización fraccionaria como

\begin{equation}
  \label{eq:ionizacion_fraccionaria}
  X := \frac{n_p}{n_p + n_H} = \frac{n_e}{n_e + n_H}
\end{equation}
de tal manera que en el momento cuando $X=1$ el sistema está
completamente ionizado, cuando $X=0$ es completamente
neutro. Usando esto en las ecuaciones (\ref{eq:saha_1}) y
(\ref{eq:numero_barionico_nr}) la ecuación de Saha
(\ref{eq:saha_1}) se puede reescribir como
\begin{equation}
  \label{eq:saha_2}
  \frac{1-X}{X^2} = n_p \left( \frac{m_e k_B
      T}{2\pi\hbar^2}\right)^{-3/2} e^{B/k_BT} =
  4  \xi(3) \sqrt{\frac{2}{\pi}} \left(\frac{k_B
      T}{m_e}\right)^{3/2} \eta_{B\gamma} e^{B/k_BT}.
\end{equation}

El factor $4\xi(3)\sqrt{2/\pi} \approx 3.84$. La ecuación de
Saha, es una ecuación cuadrática de la forma
$A(\eta_{B\gamma}, T) x^2 +x - 1 = 0$. Si definimos el
tiempo de recombinación $t_{rec}$ cuando $X = 1/2$,
tendremos $T_{rec} = \unit{0.323}\electronvolt =
\unit{3740}\kelvin$ y esto sucederá a un \textit{redshift}
$z_{rec} = 1370$. Durante este proceso el número de
electrones cae rápidamente, la tasa de interacción de los
fotones en función del \textit{redshift} con los electrones
es $\Gamma(z) = n_e(z) \sigma_{Thomson} =
X(z)(1+z)^3n_{B,\thinspace 0} \sigma_{Thomson}$, donde
$\sigma_{Thomson} = 8\pi\alpha^3/3m_e^2$. Usando datos
observacionales sabemos que $\Omega_{B,\thinspace 0} =
0.04$, entonces $\Gamma(z) = 4.4\times 10^{-21}\thinspace
\reciprocal\second X(z) (1+z)^3$. El universo se encuentra
en esta etapa dominado por materia, por lo que usando
(\ref{eq:friedmann_omega_actual})
\begin{equation}
  \label{eq:friedmann_omega_actual_mat}
  H(z) = H_o \sqrt{\Omega_{m,\thinspace 0} (1+z)^3}
\end{equation}
de nuevo usando las observaciones $\Omega_{m,\thinspace 0} =
0.3$, obteniendo $H(z) = 1.24 \times 10^{-18}\thinspace
\reciprocal\second (1+z)^{3/2}$. El tiempo de desacople
$t_d$ lo definiremos cuando $\Gamma = H$, obteniendo el
\textit{redshift} de $z_{d} = 1130$. Esta resultado arrastra
errores que se introdujeron en la simplificación que llevo a
la ecuación de Saha y además de estas se supone que la
fotoionización del hidrógeno está en equilibrio y esto
obviamente no ocurre cuando $\Gamma < H$. Realizando el
cálculo sin hacer aproximaciones tan groseras se obtiene
$z_d \approx 1100,\; T_d \approx \unit{3000}\kelvin$.

Por último, definimos como tiempo de LSS $t_{LSS}$, cuando
un fotón promedio choca con su último electrón. Durante $t
\to t+dt$ la probabilidad de un choque de un fotón con un
electrón es $dP = \Gamma(t)dt$, donde debido a la
recombinación $\Gamma(t)$ es
modificada por $X(t)$, $\Gamma(t) =
n(t)x(t)\sigma_{Thomson}$, si observamos un fotón al
tiempo $t_0$, el número de colisiones que experimenta desde
$t$ es la llamada \textit{profundidad
  óptica}\footnote{\textit{Optical depth} en inglés}

\begin{equation}
  \label{eq:optical_depth}
  \tau(t)  = \int_t^{t_0} \Gamma(t) dt
\end{equation}

Cuando $\tau = 1$ es el tiempo del LSS.  La tabla
\ref{tab:eventos_universo_mat} resume los eventos recién
comentados.

\begin{table}
  \centering
  \begin{tabular}{c|c|c}
    \hline
       \rowcolor[rgb]{0.8,0.8,0.8}  Evento & $T$ & $z$ \\
    \hline
    Transición EW & $\sim \unit{100}\giga\electronvolt$ & \, \\
    Transición de fase QCD & $\sim \unit{150} \mega\electronvolt$
    &\, \\
    Desacople de los $\nu$ & $\sim \unit{1} \mega\electronvolt$ &\,
    \\
    Aniquilación $e^-e^+$ & $< m_e \sim
    \unit{0.5}\mega\electronvolt$ &\, \\
    Nucleosíntesis & $\sim 50 - \unit{100} \kilo\electronvolt$ &\,
    \\
    Igualdad Rad-Mat & $\unit{9730}\kelvin$  & $3570$\\
    Recombinación & $\unit{3740}\kelvin$  & $1370$\\
    Desacople de los $\gamma$ & $\unit{3000}\kelvin$  & 1100 \\
    LSS  & $\unit{3000}\kelvin$ & 1100 \\
    \hline
  \end{tabular}
  \caption{Eventos más importantes del universo temprano.}
  \label{tab:eventos_universo_mat}
\end{table}

\paragraph{Época oscura} Es el periodo comprendido desde el
LSS hasta que las estrellas se formaron por
atracción gravitatoria y empezaron a brillar. El proceso de
formación de estrellas tiene un pico entre $z=2$, $z=1$. El
proceso de formación de estrellas y la física que ocurre
dentro de la época oscura, queda fuera del estudio de esta tesis.






\section{\label{sec:cosm-observ}Cosmología Observacional}

Los modelos cosmológicos, como todos los modelos estudiados
por la física, deben de ser confrontados por las
observaciones o experimentos. En el caso de la cosmología
deben de confrontarse con las observaciones
astronómicas. Las observaciones astronómicas pueden
clasificarse en dos tipos, aquellas que nos muestran que
está pasando muy lejos de nosotros y debido a la velocidad
finita de la luz, nos muestran lo que pasó hace mucho tiempo
(\textit{observaciones sobre el cono nulo}) y aquellas
observaciones de objetos cercanos (\textit{observaciones del
  tipo ``geológico''}), que cuando son relacionadas con
teorías acerca de los orígenes nos aportan información sobre
nuestra línea de mundo pasada hace mucho tiempo (por
ejemplo, la determinación local de las abundancias de
elementos, son relacionadas con los cálculos de
nucleosíntesis) \citep{Ellis06}.

Actualmente las observaciones son llevadas acabo usando 
 gran parte del espectro electromagnético, lentes
 gravitacionales, conteos
de fuentes, y por supuesto, las mediciones de alta precisión del
CBR.

Los estudios de
galaxias a nivel óptico, ultravioleta e infrarrojo han
detectado la existencia de galaxias (o proto-galaxias) a
distancias de $z \simeq 10$ (Hubble Space Telescope\footnote{Página Web:
  \url{http://hubble.nasa.gov/}}, W. Keck Observatory en
Hawaii \footnote{Página Web: 
  \url{http://www.keckobservatory.org/}}, European Southern 
Observatory en Chile\footnote{Página Web: 
  \url{http://www.eso.org/public/}}, etc.\ ), otras
observaciones como búsquedas con radio, rayos X y de rayos
gamma han identificado en las galaxias a QSOs y expulsores de
rayos gamma, GRB\footnote{Del inglés textit{Gamma
    Ray-Bursters}}, por último las técnicas que involucran
lentes gravitacionales han ayudado a detectar galaxias
gracias a las múltiples imágenes de sí misma 
encontradas en las placas debido a la lente
gravitacional entre nosotros y ellas). Las
observaciones del CBR y sus resultados serán discutidos 
en el capítulo \ref{cha:critica-al-origen}.

\subsubsection{\label{sec:distancias}Distancias en
  cosmología}

Debido a la importancia del campo electromagnético para
transmitir información hasta nosotros desde fuentes
distantes (e.g.\ los fotones que recibimos del CBR) ,
escribiremos algunas ecuaciones útiles sobre la 
transmisión de fotones sobre el fondo de RW. Los fotones
viajan sobre geodésicas nulas ($ds^2 = 0$), $x^a(\lambda)$
que tienen como vector nulo tangente $k^a = dx^a/d\lambda$,
siendo $\lambda$ el parámetro afín de la geodésica. Los
vectores tangentes de la geodésica  cumplen con $k^a_{\;
  ;b}\,k^b = 0$. Las simetrías de la 
geometría de RW nos permite concentrarnos en las geodésicas
radiales. La distancia comóvil $D_c$, recorrida por el fotón
desde su emisión al tiempo $t = t_{em}$ en $r = r_{em}$
hasta su detección en un tiempo $t_{obs}$,

\begin{equation}
  \label{eq:distancia_foton}
  D_c = \int_{t_{em}}^{t_{obs}}
  \frac{dt}{a(t)} = \int_0^{r_{em}} \frac{dr}{\sqrt{1-Kr^2}}
  = \int_{a(t_{em})}^{a(t_{obs})} \frac{da}{a\dot{a}} .
\end{equation}

Para el caso $K=0$ esta integral es simplemente el tiempo
conforme $\eta$.

Existen dos cantidades de interés observacional relacionadas
con $D_c$, la distancia luminosa, $d_L$ y la distancia
angular $d_A$. Para definirlas es necesario revisar los
conceptos de \textit{flujo} y \textit{luminosidad}. El flujo
es la medida cuantitativa de energía que pasa por unidad de
tiempo por unidad de superficie, y luminosidad es la
cantidad de energía recibida por unidad de tiempo. De las
definiciones se ve que estas cantidades están
relacionadas: el flujo $F$ de una fuente con luminosidad
$L$, a una distancia $d$ es $F \equiv L/4\pi d^2$. La
distancia luminosa $d_L$ es simplemente la adecuación $d \to
d_L$ de esta definición al caso de un universo
expandiéndose. Por otro lado, si un objeto tiene un tamaño
conocido $l$ subtendiendo un ángulo pequeño $\theta$, la
distancia angular a ese objeto está definida por $l=\theta
d_A$. Para el caso $K=0$ estas distancias son

\begin{equation}
  \label{eq:dl_da}
  d_L(z) = a_0 D_c (1+z), \quad d_A = a_0 D_c (1+z)^{-1}.
\end{equation}

Para una discusión más detallada sobre estas y otras
definiciones de distancias véase \citep{Hogg2000, Dodelson}.

El \textit{horizonte de partículas} de un evento $P$ está
relacionado con la integral (\ref{eq:distancia_foton}) y es
por definición la frontera entre las líneas de mundo que
pueden ser vistas en $P$ y aquellas que no lo son
\citep[][pags. 104-106]{Wald84},\citep[][pags. 376-382]{Rindler06}
\footnote{Existe otro tipo de horizonte: el
  \textit{horizonte de eventos} que es definido como la
  distancia de los objetos más lejanos que podremos observar
  en un futuro lejano. Este horizonte separa a aquellos
  objetos  que podremos observar en el futuro
  de aquellos que jamás podrán ser observados por hallarse
  fuera del cono de luz futuro del observador. Los modelos
  de universos que tienen como convergente a la integral
  \begin{equation*}
    d_{HE} = \int^t_{max} \frac{dt}{a}\;,
  \end{equation*}
  tienen un horizonte de eventos. con $t_{max}$ el tiempo de
  expansión futuro que puede ser infinito o finito y $t_0$
  la edad del universo en el momento de la observación.
}. Tomando en (\ref{eq:distancia_foton}) como tiempo de
emisión el Big-Bang ($t_{em} = 0$) y escogiendo nuestra
coordenada local cósmica como $r=0$, obtenemos la fórmula
para el horizonte de partículas\footnote{Debido a las
  simetrías de la métrica RW sólo la coordenada $r$ contiene
  información significativa, de ahí que solo se tomen en
  cuenta los fotones que viajan radialmente hacia nosotros
  (i.e.\ $\phi = \theta = 0$)\label{def:horizonte_eventos}}:

\begin{equation}
  \label{eq:horizonte_particulas}
  d^c_{HP} \equiv \int_{0}^{t_{obs}}
  \frac{dt}{a(t)} = \int_0^{r_{em}} \frac{dr}{\sqrt{1-kr^2}}.
\end{equation}

para $a(t) \sim t^p$\, con $0 < p <1$, el horizonte comóvil
de partículas será $d^c_{HP} = \frac{1}{1-p} t^{(1-p)} =
\frac{p}{1-p} \frac{1}{a} H^{-1}$, donde en la última
igualdad se utilizó el radio de Hubble resultante para este
factor de escala. La distancia física actual
(\ref{eq:distancias}) a la materia que compone el horizonte
es

\begin{equation}
  \label{eq:distancia_HP}
  d_{HP} = a_0 d^c_{HP}.
\end{equation}

Es importante recordar que estos horizontes aparecen como
resultado de dos factores: (a) la velocidad finita de la
luz, y (b) la edad finita del universo.

\subsection{Parámetros Observacionales de los Universos de
  Friedmann-Lemaître}\label{sec:param-observ}

Como se mostró en  \S \ref{sec:modelo-flrw}, la
geometría del universo de fondo está completamente
determinada si (a) la ecuación de estado de todos los
componentes de materia es especificada y (b) los valores de
$\Omega_0$ y $H_0$ son dados.

Las observaciones indican, que, para determinar $\Omega_0$,
deben de especificarse cinco parámetros relacionados con los
componentes del universo de fondo a energías debajo de los
\unit{200} \giga\electronvolt: $\Omega_B$, $\Omega_{rad}$,
$\Omega_\nu$, $\Omega_{DM}$ y $\Omega_\Lambda$ que describen
la contribución de los bariones, radiación, neutrinos,
materia oscura y constante cosmológica a $\Omega$
respectivamente \citep{Narlikar01}. Los primeros tipos
de materia estamos seguros que existen y los últimos dos son
sugeridos por las observaciones y no han sido refutados por
ninguna observación a la fecha. De todos ellos solamente
$\Omega_{rad}$ está bien restringido por las observaciones,
los valores de los demás contienen incertidumbres
considerables debido a los procesos de medición
utilizados. Con esto, se establece que el universo de fondo
en la cosmología estándar, es un modelo caracterizado por
estos 6 parámetros: $\{$ $H_0,\Omega_B,
\Omega_{rad},\Omega_\nu,\Omega_{DM},\Omega_\Lambda \}$.

Para que la teoría de la formación de estructura (cf.\ \S
\ref{sec:formacion_estructura}) de la cosmología estándar
pueda hacer predicciones es necesario proveerla con el
espectro de potencias de las inhomogeneidades\footnote{La palabra
  \textit{inhomogeneidades} no existe en español. La utilizo
  como una traducción directa de
  \textit{inhomogeneities}. La palabra correcta en español
  para no-homogéneo es \textit{heterogéneo}, pero esta
  palabra en español significa ``compuesto de diferentes
  partes'' lo cual no es la idea que quiero dar aquí. A lo
  largo de esta tesis por lo tanto usaremos la (incorrecta)
  palabra inhomogéneo.} iniciales, el
cual se supone no contiene una preferencia por ninguna
escala en particular, $P = Ak^n$, i.e.\ una ley de potencias
\cite{Peacock98,Longair00}. Lo cual agrega dos parámetros
más, ($A$, $n$) a los parámetros observacionales de la
cosmología.

Así, la cosmología estándar requiere la especificación
de ocho parámetros \footnote{En alguna literatura orientada
  a la parte observacional
  \cite{wmap5,wmap3yr-temperature,wmap3yr-implications,Freedman03}
  de la cosmología se llegan a incluir hasta 16 parámetros
  observacionales, a saber: $h$ (en lugar de $H_0$), el
  parámetro de desaceleración ($q_0$), la edad actual del
  universo ($t_0$), la ecuación de estado de la energía
  oscura ($w$), amplitud de la perturbación de densidad
  $\left(\sqrt{S}\right)$, amplitud de las ondas
  gravitacionales $\left(\sqrt{T}\right)$, fluctuaciones de
  la masa a $\unit{8}\mega\parsec$ ($\sigma_8$), índice
  tensorial ($n_T$) y el corrimiento del índice escalar
  ($dn/d\ln k$). Como se podrá apreciar muchos parámetros
  son redundantes, esta lista tiene dos objetivos, el
  primero es tener redundancia en los datos ya que se miden
  de diferentes maneras y así tener mayor certidumbre en sus
  valores, y segundo tratan de describir un universo más
  complejo que el tratado en esta sección (de ahí la
  inclusión de los modos tensoriales, el corrimiento del
  índice escalar y
  $\sigma_8$). \label{def:parametros_observacionales}}:
\{$H_0$, $\Omega_B$, 
$\Omega_{rad}$, $\Omega_\nu$, $\Omega_{DM}$, $\Omega_{DE}$,
$A$, $n$\}, los primeros seis relacionados con el universo de
fondo, perfectamente homogéneo e isotrópico y los últimos
dos especificando las inhomogeneidades iniciales del
universo perturbado.

Las observaciones han arrojado los siguientes resultados
respecto a los parámetros recién mencionados
\citep{Padmanabhan06} \todo{Agregar citas de este artículo}:

\begin{enumerate}
\item Nuestro universo tiene $0.98 \lesssim \Omega \lesssim
  1.08$. Este dato se puede obtener del espectro de
  anisotropías del CBR. Estas observaciones muestran que
  vivimos en un universo con densidad muy cercana a la
  crítica $\then K \approx 0$.

\item Observaciones del deuterio primordial producido en la
  nucleosíntesis (\pageref{sec:nucleosintesis})  junto con
  observaciones del CBR muestran que el 
  número total de bariones en el universo contribuye con
  $\Omega_B = (0.024 \pm 0.0012)h^{-2}$. Observaciones
  independientes fijan $h = 0.72 \pm 0.07$ y se concluye
  que $\Omega_B \simeq 0.04 - 0.06$. Estas observaciones
  incluyen todos los bariones sean luminosos o no, por lo
  que se puede deducir que la mayoría de la materia del
  universo es no-bariónica.
\item Los conteos de galaxias como función del
  \textit{redshift} muestran una homogeneidad espacial del
  universo y dan un estimado de la materia bariónica visible
  $\Omega_{Bv} \simeq 0.015 \ll \Omega_B$ con lo cual se
  concluye que la mayoría de los bariones del universo no
  son visibles encontrándose por ejemplo en estrellas
  muertas.

\item Conteos de fuentes de radio y de QSO's muestran una
  evolución del número/luminosidad de estos objetos,
  descartando así la primera versión del modelo cosmológico
  estático (\textit{steady state}) de Bondi, Gold y Hoyle
  \todo{Cita a steady state en Ellis06, Mover a soluciones
    de los universos de FL}.

\item Observaciones relacionadas con estructura a gran
  escala y su dinámica (curvas de rotación de galaxias,
  masas estimadas de los clusters, lentes gravitacionales)
  sugieren que el universo está poblado por un componente no
  luminoso de materia, \acfi{DM} formada con partículas
  masivas que interactuan débilmente y que se aglutina a
  escalas galácticas. Este componente contribuye con un
  $\Omega_{DM} \simeq 0.20-0.35$ de la energía
  total. ``Interactuar débilmente'' se interpreta en la
  actualidad a que sólo interacciona gravitacionalmente con
  otras partículas y por lo tanto se puede despreciar su
  presión (que proviene de colisiones entre partículas),
  teniendo así la simple ecuación de estado $p_{DM} \approx
  0$.

\item Usando está última observación junto con la primera,
  se concluye que debe de haber al menos un componente más
  de materia \todo{Citas a $f(R)$ en el pie de
    página}\footnote{ Podría darse el caso que Relatividad
    General no fuera la teoría correcta de la gravitación ,
    pero eso estaría en contra de las suposiciones de esta
    tesis y de los fundamentos del modelo estándar
    cosmológico.} que contribuya con un $70\%$ a la densidad
  de energía total. Las observaciones de
  supernovas\footnote{Esto pudo hacerse debido a que en las
    supernovas tipo Ia su luminosidad máxima está
    relacionada con el tiempo de decaimiento de la curva de
    luminosidad, volviéndose así ``velas estándar'' para
    galaxias a grandes distancias.} apuntan a que no se
  aglutina a ninguna escala y posee presión negativa.  Estas
  observaciones sugieren $\gamma -1 = p/\rho \lesssim -0.78$
  y contribuye con $\Omega_{DE} \simeq 0.60 - 0.75$. La
  elección más sencilla de esta \acfi{DE} es suponer una
  ecuación de estado $p_{DE} = -\rho_{DE}$, i.e.\ una
  constante cosmológica $\Lambda$, lo cual resulta al
  sustituir en (\ref{eq:conservacion_fl}) que $\rho_{DE}$ es
  constante durante la expansión del universo. Esta es la
  forma de materia dominante en la actualidad ya que las
  observaciones indican que la expansión está acelerando.

\item El universo también posee radiación que contribuye con
  una densidad de energía $\Omega_{rad} h^2 = 2.56 \times
  10^{-5}$. La mayor contribución a esta densidad proviene
  de los fotones del CBR.
 
\end{enumerate}

\section{Éxitos y problemas de la cosmología estándar}

\subsection{\label{sec:exitos-scc}Éxitos de la cosmología
  estándar}

Mencionaremos en esta sección la relación entre las
predicciones de la cosmología estándar y los valores de los
parámetros observacionales. El hecho de que existan
relaciones con un elevado grado de no trivialidad entre las
predicciones y los parámetros muestran la fuerza del modelo.

\begin{enumerate}
\item Mientras el universo se expande y se enfría ($T
  \propto a^{-1}$), diferentes interacciones físicas se
  ``congelan'' (definición en pag. \pageref{def:congelar}) en
  diferentes épocas, esto se debe 
  principalmente a que cuando el universo se enfría por la
  expansión la energía disponible localmente para las
  interacciones disminuye hasta que es insuficiente para
  continuar con las reacciones necesarias. La primera
  especie en desacoplarse es la de los neutrinos, al estar
  en equilibrio térmico comparte la misma temperatura que
  los fotones en esa época, pero al desacoplarse su
  temperatura no variará más mientras que la temperatura de
  los fotones aumentará por la aniquilación de pares de
  electrones y positrones hasta que la radiación se
  desacople también de la materia. Una de las predicciones
  del modelo estándar es que ambas temperaturas tendrán un
  radio de $(T_\nu/T_\gamma) = (4/11)^{1/3} \simeq
  0.71$. Esta predicción deberá ser probada en un futuro
  cercano \cite{Sarkar03}.

\item Cerca de la temperatura de unos pocos
  \mega\electronvolt, la nucleosíntesis ocurre
  \cite{Fields08}, produciendo los elementos ligeros a
  partir de los electrones y nucleones libres. Las
  abundancias de estos elementos dependen crucialmente de
  $\Omega_B$ y el número de especies de neutrinos. Además el
  $^3$He, D y $^7$Li dependen de maneras diferentes de estos
  parámetros: (a) la abundancia de He constriñe el número de
  especies de neutrinos a tres, resultado predicho por la
  cosmología antes de ser verificado en laboratorios
  \cite{Barrow89} usando el decaimiento de Z$^0$
  \cite{Schramm84} (para una revisión de este tema ver
  \cite{Dolgov02}).  Debe recalcarse que el modelo estándar 
  no fue ``diseñado'' para dar tres especies de
  neutrinos. (b) es posible escoger un rango de valores para
  $\Omega_B$ en la que las abundancias de He, D y Li pueda
  ser explicada, esta región de concordancia es un atributo
  no trivial de la cosmología estándar.

  \figura{imagenes/spectrum}{width=12cm,height=8cm}{ Espectro
    del fondo de radiación cósmica (CBR) correspondiente a
    la temperatura $T_0= \unit{2.725}\kelvin$ obtenido por
    el instrumento FIRAS del COBE
    \citeauthor{GarciaBellido05}, \citeyear{GarciaBellido05}
    \cite{GarciaBellido05}.}{fig:firas_cbr}

\item La cosmología estándar introduce una fase caliente
  dominada por la radiación en el universo temprano con una
  tasa entre los números de fotón\thinspace/\thinspace
  barión muy alta. Debido a la expansión 
  del universo los fotones se desacoplan de la materia
  cuando $T \approx$ \unit{10^3}\kelvin, formando así un
  fondo de radiación cósmica (CBR) con un espectro de Planck
  o de cuerpo negro (cf.\ \pageref{def:recombinacion}). Es importante
  mencionar que, aunque la teoría no puede decir cual es el
  valor actual de la temperatura del CBR \cite{Scott05},
  hace una predicción definitiva sobre su 
  existencia con un espectro Planckiano a una temperatura $T \neq 0$,
  ningún otro modelo cosmológico hizo esta predicción antes del
  descubrimiento del CBR en la década de 1960. La exactitud
  de esta predicción fue probada por el instrumento
  \acf{FIRAS} del satélite COBE \cite{COBE} (figura
  \ref{fig:firas_cbr}). Los modelos competidores de la
  cosmología estándar solo pudieron explicar el CBR
  retrospectivamente, es decir, mediante la agregación de
  procesos físicos ajenos a su teoría inicial.

  \figura{imagenes/SNIa-constraints}{width=12cm,height=13cm}{Intervalos
    de confianza dadas por las observaciones de supernovas
    de Tipo Ia (SN Tipo Ia) a los valores de $\Omega_{DM}$ y
    $\Omega_{DE}$. Los contornos sólidos son los datos del
    análisis del 2004, los contornos punteados son de la
    primeros resultados de los mismos autores de 1998.
    Figura tomada de \citeauthor{Riess04},
    \citeyear{Riess04} \cite{Riess04}.}{fig:omegadark}

\item El universo es dominado por $\Omega_B, \Omega_{DM}$ y
  $\Omega_{DE}$ en $z \ll 10^3$ y varias observaciones no
  relacionadas (Conteos de galaxias, Edad del Universo,
  Distribución estadística de lentes gravitaciones de QSO's)
  imponen restricciones a estos parámetros. La existencia de
  una zona consistente con las distintas observaciones en el
  espacio de parámetros es otro triunfo de la cosmología
  estándar (figura \ref{fig:omegadark}).

\item Hay dos predicciones falsificables que hace la
  cosmología estándar y que podrían estudiarse en un futuro:
  (a) la temperatura del fondo cósmico de neutrinos: $T_\nu
  =$ \unit{1.9}\kelvin \,  y (b) la predicción de que el
  universo se estuvo expandiendo durante todo el lapso de $0
  < z < 10^3$, lo cual resulta necesario para poder enfriar el
  CBR. Si una población sistemática de objetos distantes es
  encontrada con un espectro corrido al azul
  (\textit{blueshifted}) señalaría una etapa de contracción
  del universo y descartaría por completo a la cosmología
  estándar.

\item El paradigma de formación de estructura (cf.\ capítulo 
  \ref{cha:critica-al-origen}) también realiza predicciones
  que han sido confirmadas por observaciones. La primera y
  quizá la más importante es que debido a las fluctuaciones
  de la distribución de materia, pequeñas fluctuaciones
  deben de existir en el CBR. El espectro de potencias
  deberá de ser plano para poder reproducir la distribución
  de las estructuras cosmológicas que vemos hoy; y además,
  las anisotropías de la temperatura deben de ser de al
  menos el orden de $10^{-6}$ para que las estructuras
  tengan el tiempo de crecer y volverse no-lineales en $z =
  0$. Estas predicciones fueron comprobadas con la misión
  COBE usando el instrumento \acf{DMR} (figura
  \ref{fig:dmr-cbr}) a principios de 1990 \cite{COBE}. Otra
  predicción importante es que el espectro de potencias de
  las fluctuaciones primordiales debe de ser plano (es
  decir, todas las escalas tienen la misma potencia),
  propuesto --para explicar las distribuciones de
  galaxias y \textit{clusters} de galaxias observadas (otras
  consideraciones de consistencia fueron tomadas en cuenta e.g.\
  que no haya producción excesiva de agujeros negros
  \cite{Kolb1990}) -- de manera independiente por Zel'dovich
  \cite{Zeldovich72} y Harrison \cite{Harrison70}.

  \figura{imagenes/DMR}{width=12cm,height=15cm}{Espectro de
    potencias del CBR observado por el instrumento DMR de
    COBE en 1990. La figura superior muestra el monopolo
    ($T_0 = \unit{2.725}\kelvin$), la imagen central
    presenta el dipolo $\delta T_1 =
    \unit{3.372}\milli\kelvin$ y la figura inferior
    corresponde al cuadrupolo ($\delta T_2 =
    \unit{18}\micro\kelvin$ y multipolos superiores.  Figura
    tomada de \citeauthor{GarciaBellido05},
    \citeyear{GarciaBellido05}
    \cite{GarciaBellido05}.}{fig:dmr-cbr}

\item La forma final del espectro de potencias de las
  anisotropías de la temperatura del CBR, es debida a varios
  efectos que se discutirán en el capítulo siguiente (cf.\
  figura \ref{fig:wmap_5}, capítulo
  \ref{cha:critica-al-origen}) y depende de manera
  no-trivial de los parámetros observacionales de la
  cosmología estándar (e.g.\ la altura relativa del primer
  pico depende la densidad bariónica, $\Omega_B$ su posición
  depende de la geometría y por lo tanto de $\Omega_{DE}$ y
  $\Omega_{DM}$).
\end{enumerate}

\subsection{Problemas y debilidades de la cosmología
  estándar}\label{sec:problemas-scc}

La mayoría de los problemas \footnote{Ver los capítulos 15-17
de \cite{Peebles93} para un recuento histórico sobre como se identificaron
estos problemas en la década de 1970} que
presentaremos en esta sección son del tipo ``incomodidad con
ciertas condiciones iniciales'' \citep{Borner04}, que son en
realidad problemas de ajuste fino\footnote{En inglés:
  \textit{fine-tuning}.}. Llamarlos ``problemas'' puede
parecer rigorista, ya que cualquier solución de las
ecuaciones diferenciales de la cosmología estándar refleja
propiedades específicas de los datos iniciales: si
calculamos hacia atrás en el tiempo simplemente encontramos
las condiciones iniciales que fueron responsables por el
estado de las cosas como las vemos actualmente
\citep{Borner04}. Pero, aunque comparativamente hablando el
Modelo Estándar de la Cosmología
tiene muchos menos parámetros libres (7) que el Modelo
Estándar de partículas (23), al ser 
un modelo que intenta explicar los orígenes de todo lo
observable no es satisfactorio que dependa de condiciones
tan especiales para lograr tal cometido
\citep{Narlikar01}. Existe otra forma de ver estos
problemas: son problemas en el sentido de que existen
observaciones que no tienen una explicación natural dentro
del contexto de la teoría.

\begin{itemize}
\item\textit{Posible sobre-constricción de los parámetros
    observacionales.} Si tomamos en cuenta las múltiples
  observaciones independientes existentes, las restricciones
  en los parámetros cosmológicos es severa y se refleja en
  el hecho de que en el espacio de parámetros el área de
  concordancia es muy pequeña y para algunos críticos
  inexistente \citep{Narlikar01}.


\item\textit{Explicación física de $T_0$ en el CBR.} La
  predicción de la existencia del CBR es uno de los grandes
  éxitos del modelo cosmológico estándar, pero carece, a la
  fecha, de una interpretación sobre la relación entre la
  temperatura actual (i.e. $z = 0$) del CBR y otros procesos
  físicos.

\item\textit{Extrapolación de la física conocida en varios
    órdenes de magnitud.} En cosmología la extrapolación de
  las leyes físicas actuales para ser aplicadas en la época
  del universo muy temprano es algo rutinario, pero hay que
  notar la enormidad de la extrapolación: 17 órdenes de
  magnitud en temperatura si suponemos que la física
  conocida y comprobada en laboratorios (recordar que se ha
  probado hasta energías de $\approx \unit{10^2}
  \giga\electronvolt$) es válida hasta la temperatura de
  Planck \citep{Narlikar01}, por lo que un crítico podría
  preguntarse si aplicar una teoría especulativa a una época
  especulativa constituye física o algo más. Claro que la
  defensa que se establece es que los cálculos y su
  interpretación son más un ejercicio de consistencia que
  una descripción definitiva del universo muy temprano, pero
  lamentablemente esto no es siempre presentado así
  \citep{Padmanabhan96, Borner04, Ellis06}.

  Por otra parte, los fundamentos teóricos del modelo
  estándar pueden ser calificados de poseer una
  inconsistencia interna\footnote{La ``inconsistencia''
    del modelo estándar cosmológico se 
  puede calificar de transitoria ya que, cuando se tenga una
  teoría de la gravedad cuántica, la singularidad inicial
  podría ser removida. Pero esta defensa, para ser
  consistente debería de ir acompañada de la admisión de que
  cualquier física hecha a estas longitudes o densidades
  debería tomarse con extremo cuidado.} \citep{Narlikar99}, ya que las ECE
  de la cual son derivados predicen una singularidad inicial
  para ecuaciones de estado razonables (cf. \S
  \ref{sec:singularidad-inicial}). Esta singularidad marca
  el rompimiento de las suposiciones que nos llevan, entre
  otras, a la acción de los campos de materia, $S_{m}$
  involucrados y a la suposición de un espacio-tiempo
  continuo (En la cosmología, a diferencia de otras ramas de
  la física en las cuales esto hubiera sido una señal de que
  algo está muy mal, esto se ve como algo
  bueno que además es identificado con el ``mítico evento de
  la creación'' en palabras de \citeauthor*{Narlikar01}, \cite{Narlikar01}).
  
  \figura{imagenes/flatness_problem}{width=12cm,height=6.5cm}{El
    problema de la planitud mostrado en una gráfica de
    densidad relativa $\Omega$ contra el tiempo cosmológico
    $t$ (ningún eje está a escala). En esta imagen se
    muestra como la solución $\Omega = 1$ es repulsiva. Cada
    línea muestra un posible universo. Un universo similar
    al nuestro es la línea azul con $|\Omega -1| \sim 0$. El
    universo representado por la línea roja es uno que
    difirió mucho de las condiciones iniciales del universo
    representado por la línea azul. Figura tomada del
    Wikicommons
    \texttt{http:\//\//en.wikipedia.org\//wiki\//File:Flatness\_problem\_density\_graph.svg}}{fig:flatness-problem}

\item\textit{Problema de la
    Planitud.}\label{def:problema_planitud}  Sabemos que
  vivimos en 
  un universo en el cual $\Omega_0 \equiv \rho_0/\rho_{cri}
  \simeq 1$, i.e.\ un universo muy cercano a tener una
  geometría espacialmente plana. Podemos calcular que
  condiciones iniciales necesarias para que lleguemos a este
  estado. Tomemos la ecuación de Friedmann
  (\ref{eq:friedmann_omega}):

\begin{equation*}
  |\Omega - 1| = \frac{|k|}{a^2H^2} ;
\end{equation*}

durante la época dominada por radiación tenemos que $H^2
\propto a^{-4}$ entonces, $|\Omega - 1|_{rad} \sim a^2$; en
cambio en la época donde la materia es dominante $|\Omega
-1|_{mat} \sim a$, o sea que ambos disminuyen conforme
retrocedemos en el tiempo ($a \to 0$). Si actualmente esta
cantidad es casi cero, en el pasado debió de ser mas chica,
por ejemplo, la razón entre esta cantidad hoy y la evaluada
en tiempos del orden del tiempo de Planck, $t_{Planck}$, es
(estamos en el modelo estándar cosmológico, en épocas con $z
> 10^3$ la densidad de materia es dominada por radiación),

\begin{equation*}
  \frac{|\Omega -1|_{t=t_{Planck}}}{\Omega -1|_{t=T_0}}
  \simeq \left(\frac{a_{Planck}}{a_0}\right)^2 = \left(\frac{T_0}{T_{Planck}}\right)^2.
\end{equation*}

Sustituyendo el valor numérico de $T_{Planck} \sim$
\unit{10^{19}}\giga\electronvolt \; y $T_0 \sim $
\unit{10^{-13}} \giga\electronvolt se obtiene,

\begin{equation}
  \frac{|\Omega -1|_{t=t_{Planck}}}{\Omega -1|_{t=T_0}}
  \simeq 10^{-64}  ,
\end{equation}

Debido a que en la actualidad $\Omega \sim 1$, es necesario que en el tiempo
inicial las condiciones  hayan tenido precisión de
hasta $64$ cifras decimales. Aunque esta cifra es
impresionante, no es necesario retroceder hasta la época de
Planck para obtener este ajuste fino, por ejemplo, en la
época en la que sucede la nucleosíntesis , $T_N \approx
\unit{1}\mega\electronvolt$ se requería una precisión de 16
cifras decimales. Este problema es representado gráficamente
en la figura \ref{fig:flatness-problem}.

  \figura{imagenes/problema_horizonte}{width=10cm,height=7.5cm}{Problema
    del Horizonte. Puntos diametralmente opuestos del LSS
    (en la figura $\mathscr{P}$ y $\mathscr{Q}$) no
  han estado en contacto causal (i.e.\ sus conos de luz
  pasados no tienen puntos en común) y aún así tienen propiedades
similares.}{fig:problema_horizonte} 

\item\textit{Problema del Horizonte.} Como se vio antes, los
  modelos de FL tienen horizonte de partículas, es decir,
  hay en cualquier época regiones que no han estado en
  contacto causal. Con esto en mente el problema del
  horizonte (ilustrado en la figura
  \ref{fig:problema_horizonte}) se establece como sigue: Desde
  nuestro evento 
  observacional $\mathcal{R}$ en el espacio-tiempo que marca
  el ``aquí y el ahora'' en $t = t_0$, vemos en direcciones
  opuestas radiación proveniente del CBR, que fue emitida en
  los eventos $\mathscr{P}$ y $\mathscr{Q}$ en el LSS al
  tiempo de desacople $t = t_{des}$, $T_{des} \approx 3000
  K$. Los conos de luz pasados de estos eventos, son
  completamente disjuntos, ya que la singularidad inicial
  limita la extensión del espacio-tiempo hacia el
  pasado. Así, ninguna causa común, puede influenciar lo que
  pasa en $\mathscr{P}$ y $\mathscr{Q}$, por lo tanto,
  ningún mecanismo causal desde la creación del universo
  puede explicar las similares condiciones físicas de
  $\mathscr{P}$ y $\mathscr{Q}$ \cite{Ellis88}.

  Será de utilidad ver este problema analíticamente, el
  horizonte de partículas 
  (ecuación (\ref{eq:horizonte_particulas}),
  pag. \pageref{eq:horizonte_particulas}) del evento 
  $\mathcal{R}$
\begin{eqnref}{eq:horizonte_particulas}
  d_{HP}^c =\int_0^{t_0} \frac{dt}{a(t)},
\end{eqnref}
y la distancia física medida por un observador $\mathcal{O}$ en
$\mathcal{R}$ es
\begin{eqnref}{eq:distancia_HP}
  d_{HP} = a(t_0)d_{HP}^c.
\end{eqnref}

Definimos el horizonte visual $d_{HV}^c$ como la distancia
comóvil que mide $\mathcal{O}$ en el evento $\mathcal{R}$
hasta el LSS

\begin{equation}
  \label{eq:horizonte_visual}
  d_{HV}^c \equiv \int_{t_d}^{t_0} \frac{dt}{a(t)}, \quad
  d_{HV} = a(t_0) d_{HV}^c.
\end{equation}

Por otra parte el horizonte de partículas de $\mathscr{P}$
(o equivalentemente $\mathscr{Q}$), $\tilde{d}_{HP}$, medido
por un observador en $\mathscr{P}$ (o $\mathscr{Q}$)
corresponderá a la distancia medida en $t_0$ de

\begin{equation}
  \label{eq:hp_lss}
  \tilde{d}_{HP} = a(t_0) \int_0^{t_d} \frac{dt}{a(t)},
\end{equation}

mostrando que $d_{HP} = d_{HV} + \tilde{d}_{HP}$. Estas
funciones son $d_{HP}^c$ y $d_{HV}^c$ son funciones que
crecen con el tiempo de observación sin importar el
comportamiento de $a(t)$, ya que $a(t)$ es positiva; además
$d_{HP}$, $d_{HV}$ y $\tilde{d}_{HP}$ son funciones que
incrementan con el tiempo si $a(t)$ está creciendo también.

Así, si suponemos que el universo estuvo dominado por
radiación antes del tiempo de desacople $t_d$ y dominado por
materia después (i.e.\ ignorando los efectos de una época
intermedia) obtenemos, luego de un cálculo largo

\begin{subequations}
  \begin{equation}
    \label{eq:horizonte_lss_no_inf}
    \tilde{d}_{HP} = a_0 \left(\frac{1}{a_d H_d}\right) = a_0 \left[
      \frac{1}{a_0 H_0} 
      \left(\frac{a_d}{a_0}\right)^{1/2} \right]
  \end{equation}

  \begin{equation}
    \label{eq:horizonte_visual_no_inf}
    d_{HV} = a_0  \Bigg\{\frac{2}{a_dH_d}
    \left[\left(\frac{a_0}{a_d}\right)^{1/2} -1 \right]
    \Bigg\} = a_0 \Bigg\{
    \frac{2}{a_0 H_0} 
    \left(\frac{a_d}{a_0}\right)^{1/2} \left[\left(
        \frac{a_0}{a_d}\right)^{1/2} -1 \right]\Bigg\}
  \end{equation}
\end{subequations}

En donde $(a_o/a_d) \approx 10^3$, por lo que básicamente el
problema del horizontes es la siguiente desigualdad

\begin{equation}
  \label{eq:problema_horizonte}
  \tilde{d}_{HP} \ll d_{HV}
\end{equation}

Para otros enfoques de este problema se puede consultar
\citep{Brandenberer06,Shinji03,Kinney09,Brandenberger1989,Lazarides06}.

Será de utilidad más adelante en esta tesis, expresar este
problema en términos de ángulos del CMB. Usando coordenadas
comóviles podemos ver que la distancia al LSS (i.e.\ el
horizonte visual \ref{eq:horizonte_visual}) es

\begin{equation}
  \label{eq:horizonte_visual_com}
  d_{HV}^c \equiv \int_{t_d}^{t_0} \frac{dt}{a(t)} =
  \int_{\eta_d}^{\eta_0} d\eta = \eta_0 - \eta_d.
\end{equation}

Una escala física cualquiera, $\lambda$, es proyectada en el
LSS en una distancia angular (ignorando efectos de
curvatura)

\begin{equation}
  \label{eq:angulo_subtendido}
  \theta \approx \frac{\lambda}{\eta_0 - \eta_d}.
\end{equation}

Durante la época dominada por radiación, podemos estimar la
escala en la cual la materia interactúa usando la
``velocidad'' a la cual se mueven los fotones en el plasma,
conocida como \textit{velocidad del sonido} \footnote{La
  escala en la cual \textit{pudieron} haber interactuado
  está dada por los conos de luz pasados, pero esta escala
  no nos dice nada sobre \textit{si} interactuaron de una
  manera significativa. La escala de contacto causal
  significativo es menor que la escala de contacto causal
  posible (cf.\ \S 4 de \cite{Ellis88}). }, $c_s^2 = c
dp/d\mu$, que durante radiación es $c_s = c/\sqrt{3}$. El
``horizonte comóvil de sonido'' en el LSS es, $d_{HS}^c(t_d)
= c_s a(\eta_d) = c_s \eta_d$.

El ángulo subtendido por esta escala es

\begin{equation}
  \label{eq:angulo_hs_lss_1}
  \theta_{HS} \approx  c_s \frac{\eta_d}{\eta_0 - \eta_d}
  \approx c_s\frac{\eta_d}{\eta_0},
\end{equation}

donde en el último paso se usó el hecho de que $\eta_0 \gg
\eta_d$. El universo se puede considerar dominado por
materia desde el tiempo del LSS, revisando la tabla
\ref{tab:comportamientos_factor_escala} vemos que la
evolución del factor de escala es $a \simeq \eta^2 \simeq
T^{-1}$

\begin{equation}
  \label{eq:angulo_hs_lss_2}
  \theta_{HS} \approx c_s \left(\frac{T_0}{T_d}\right)^{1/2}
  \simeq 1^o
\end{equation}

El problema del horizonte es que, en el LSS, ángulos mayores
que $\theta_{HS}$ no estuvieron en contacto causal.

Debería de ser obvio que el problema del horizonte no es en
realidad un problema de los universos FL, ya que para estos
modelos la estructura del horizonte es una consecuencia
trivial la isotropía y homogeneidad supuestas. El problema
aparece cuando se busca una explicación para esta isotropía
\citep{Borner04}.

\item\textit{Origen de las fluctuaciones primordiales.} Este
  problema (el único con relevancia real según algunos
  autores y la ``única razón para tomarse en serio una época
  inflacionaria'' en palabras de \citeauthor{Padmanabhan06}
  en \citep{Padmanabhan06}) se puede describir sencillamente
  con el hecho de que no existe ningún mecanismo en la
  cosmología estándar para generar las fluctuaciones
  primordiales. Es justamente este problema el que se
  estudiará en este trabajo de tesis.

\end{itemize}

\section{Inflación}

Inflación es una propuesta teórica agregada al modelo
cosmológico estándar, que intenta solucionar los problemas de
Horizonte, Planitud y Orígenes de las fluctuaciones
primordiales\footnote{ Existían otros problemas relacionados
  a la existencia de reliquias provocadas por teorías de
  Gran Unificación -que se supone es un ingrediente del
  modelo inflacionario-, tales como la proliferación de
  monopolos magnéticos o paredes de dominio --\emph{domain
    walls}--, pero como bien apunta Penrose \citep{Penrose89}
  estos son problemas ``internos'' (i.e.\ auto-provocados) de las
  GUT y su solución es más bien un requerimiento de
  consistencia de la 
  teoría con la observación.} agregando una etapa previa
a la época dominada por radiación pero posterior a la
singularidad inicial \footnote{Aunque existen modelos
  inflacionarios en los cuales esta singularidad no existe,
  (eg.\ inflación caótica o inflación
  eterna
  \cite{Guth00,Aguirre07,Garriga07,Guth07,Huang07,Mersini-Houg07,Wang08}),
  pero siempre 
  es una época  agregada antes de la dominada por
  radiación.} en la cual 
el universo sufre una expansión exponencial por un periodo
pequeño de tiempo. La Inflación ocurre supuestamente a las
escalas de las teorías de Gran Unificación (GUT, por sus
siglas en inglés): $t_i \sim \,$
\unit{10^{-36}}\second\, a \,\unit{10^{-32}}\second.

La Inflación fue propuesta de manera independiente por Alexei
Starobinsky \citep{Starobinsky79} en
\citeyear{Starobinsky79} y Alan Guth \citep{Guth81} en
\citeyear{Guth81}, aunque Guth fue el primero en ensamblar
una imagen completa sobre este modelo. La idea original
(llamada ahora \textit{vieja inflación}) era que mientras el
universo se estaba enfriando quedó atrapado en un falso
vacío con una alta densidad de energía, muy parecida a una
constante cosmológica. Este falso vacío era metaestable, del
cual sólo se podría salir a través de un proceso de
nucleación de burbujas vía el efecto túnel cuántico
(\textit{tunneling}). Las burbujas de vacío verdadero se
formarían espontáneamente en el falso vacío y empezarían a
expandirse a la velocidad de la luz, con esta expansión se
resolvían los problemas de planitud y horizonte, y tanto Guth
como Starobinsky especulaban que esto podría de alguna
manera evitar la singularidad inicial.

El mismo Guth rápidamente reconoció que este modelo era
problemático ya que no recalentaba de la manera apropiada:
al nuclear, las burbujas no generaban radiación. La producción
de radiación sólo podía darse a través de colisiones entre
burbujas. El problema es que para que inflación pueda
resolver los problemas de condiciones iniciales del modelo
estándar debe de durar lo suficiente y si era este el caso,
las colisiones entre burbujas eran extraordinariamente
raras.

Este problema sería resuelto por los modelos conocidos como
\textit{nueva inflación} o de \textit{slow-roll}, propuestos
por Andrei Linde \citep{Linde82} e independientemente por
Andreas Albrecht y Paul Steinhardt \citep{Albrecht82}. Este
modelo, en lugar de efectuar un \textit{tunneling} desde un
falso vacío al vacío verdadero para lograr la expansión, esta es producida
por un campo escalar $\varphi$, que bajo condiciones
apropiadas ``rodaría 
lentamente''\footnote{\textit{Slow-roll} en inglés} 
pendiente abajo del potencial $V(\varphi)$; al ganar la
suficiente ``velocidad'' las condiciones necesarias para que
 exista la expansión acelerada dejarán
de ser válidas, iniciando la producción de partículas. Los
modelos inflacionarios de \textit{slow-roll}  serán los usados en
este trabajo de tesis.

\subsection{\label{sec:inflacion-problemas-estadar}
  Inflación y algunos problemas del modelo estándar
  cosmológico}

Antes de mostrar como se modela matemáticamente el campo
escalar en los modelos de \textit{slow-roll} de Inflación
estudiaremos cuál es el comportamiento que buscamos para
resolver, por lo menos los problemas de condiciones
iniciales (i.e.\ planitud y horizonte).

Como vimos en el problema de planitud, la cantidad $|\Omega
-1|$ crece con el tiempo, por lo que, si ahora tenemos
$\Omega \sim 1$ debió ser muy cercano a $1$ en épocas
pasadas. Este comportamiento se puede ver como sigue:
\begin{equation*}
  \dt{\,} |\Omega -1| =
  |K|\dt{\,}\left(\frac{1}{\dot{a}^2}\right) =
  -2\frac{|k|}{\dot{a}^3} \ddot{a}.
\end{equation*}

En épocas dominadas por radiación o materia, el universo se
está desacelerando (i.e.\ $\ddot{a} < 0$), entonces el
problema de la planitud se debe a que esta cantidad crece en
función del tiempo si el universo se está desacelerando:
\begin{equation}
  \dt{\,}|\Omega - 1| > 0 \ssi \ddot{a} < 0.
\end{equation}

Una de las maneras de solucionar el problema de la planitud,
es proponer una etapa cosmológica en la cual 
\begin{equation}
  \label{eq:condicion_planitud}
  \dt{\,}|\Omega - 1| < 0 \ssi \ddot{a} >  0.
\end{equation}

Por su parte, el problema del horizonte surge, en pocas
palabras, porque los conos de luz pasados de eventos
diametralmente opuestos en la esfera del LSS no contienen
eventos en común (i.e.\ eran disjuntos). 

El contacto causal queda determinado por el horizonte de
partículas (\ref{eq:horizonte_particulas}) el cual, para
materia ordinaria depende inversamente del parámetro de
Hubble: $d_{HP}^c \propto H^{-1}$. Se puede demostrar en
universos FL \cite{Rindler56}, que una vez que dos eventos
entren en contacto causal (i.e.\ , que el evento $\mathscr{Q}$
entre al cono de luz pasado de $\mathscr{P}$ ), no perderán
este contacto causal conforme pase el tiempo (i.e.\ ,
$\mathscr{Q}$ saldrá del cono de luz pasado de
$\mathscr{P}$). Esto se puede expresar en términos de
escalas físicas de la siguiente manera, una vez que una
escala (e.g.\ , la distancia entre dos fotones) entra al
horizonte de partículas durante una época dominada por
radiación o materia no relativista, el horizonte crecerá más
rápido que la escala en cuestión (que crece proporcional al
factor de escala, cf.\ (\ref{eq:distancias})) , por lo cual
siempre quedará adentro de este horizonte:

\begin{equation}
  \label{eq:razon_radio_hubble_escala}
  \frac{d}{dt} \left(\frac{\lambda}{H^{-1}}\right) \simeq
  \ddot{a} < 0.
\end{equation}

Si existiera una época dominada por un tipo de materia
(diferente a la radiación o la materia no relativista por
supuesto) en la cual, las longitudes físicas crecieran más
rápido que $H^{-1}$, existirían escalas que, dado el tiempo
suficiente, serían mayores que $H^{-1}$. Al acabar esta
época dominada por esta materia exótica, volveríamos  al
comportamiento (\ref{eq:razon_radio_hubble_escala}) en el
cual $H^{-1}$ crecerá más rápido que las escalas físicas  y
finalmente las ``engullirá''. Claramente, esta nueva
materia, en su periodo de dominación definirá  un horizonte
de eventos que no dependerá proporcionalmente de
$H^{-1}$, por lo que en realidad, a diferencia de lo que se
lee en la literatura estándar, no es que las escalas físicas
salgan del horizonte\footnote{En la literatura se hace
  comúnmente un
  abuso de lenguaje al usar la expresión ``salir /entrar del horizonte''
  refiriéndose a que las longitudes físicas son mayores o
  menores que el radio de Hubble $r_H = H^{-1}$. Este abuso
  de lenguaje se justifica de varias maneras, de las cuales
  podemos destacar las siguientes: (a) Para la
  época de radiación $d_{HP} \simeq r_H$; (b) En las
  ecuaciones perturbadas (cf. capítulo \ref{cha:critica-al-origen}) $H^{-1}$
  caracteriza el rango de las ``influencias causales'', ya
  que $kH^{-1}$ ``gobierna cuales términos de las ecuaciones
perturbadas son dominantes'' \cite{Bardeen80}; (c) ``El factor de escala
crece un \textit{e-folding} en un intervalo del orden
$H^{-1}$. Por esta razón, la microfísica solo puede operar
coherentemente en escalas físicas menores a
$H^{-1}$'' \cite{Bardeen83}  o en
palabras de \citeauthor{Brandenberger84} en
\cite{Brandenberger84}: ``el radio de Hubble, $H^{-1}$ es la
distancia máxima sobre la cual la microfísica puede actuar
coherentemente; \ldots para distancias mayores el tiempo que
le toma a la luz excede la tiempo de expansión
cosmológica característico''. Como se puede observar,
ninguna de estas justificaciones establece una relación
sólida entre el radio de Hubble y causalidad local. Para una
discusión extensa sobre este punto véase \cite{Ellis88}.} y por
supuesto nunca pierden contacto causal. 

Así, para obtener el comportamiento descrito en el párrafo
superior, necesitamos que 
\begin{equation}
  \label{eq:condicion_razon_radio_hubble_escala}
  \frac{d}{dt} \left(\frac{\lambda}{H^{-1}}\right) \simeq
  \ddot{a} > 0.
\end{equation}

Ambos problemas, el de planitud y el de horizonte, son
resultado, de la 
desaceleración del Universo. Así, si suponemos la existencia
de una época donde el factor de escala se esté acelerando
($\ddot{a} > 0$) y este periodo dura --por lo menos-- el
tiempo necesario\footnote{Expresado en \textit{e-foldings}
  el intervalo necesario es mayor a ($\sim 60$).  Un
  \textit{e-folding} es el periodo de tiempo en el cual una
  cantidad que está expandiéndose se incrementa por un
  factor $e$.  El número $\mathcal{N}$ de
  \textit{e-foldings} que pasaron entre dos tiempos $t_2 >
  t_1$ se define mediante la relación
  \begin{equation*}
    \mathcal{N} = \ln \left(\frac{a(t_2)}{a(t1)}\right).
  \end{equation*}
}, los problemas de planitud y de horizonte se
resolverán.

El problema de planitud desaparece ya que
podemos acercar $\Omega \sim 1$ con la precisión que se
desee y dejar  esa condición inicial a inicios de la época de
  radiación, por su parte, el problema del horizonte se
  resuelve, ya que este periodo 
inflacionario tiene como efecto desplazar hacia abajo
respecto a $t_d$ en el diagrama conforme 
a la singularidad inicial. Consecuentemente los eventos
$\mathscr{P}$ y $\mathscr{Q}$ diametralmente opuestos en el
LSS, ahora comparten un conjunto de eventos en el pasado que
pueden influenciar a ambos. Si el periodo inflacionario dura
mucho tiempo, el diagrama se extenderá más hacia abajo y la
mayoría de los eventos pasados de $\mathscr{P}$ y
$\mathscr{Q}$ será comunes. Aunque $d_{HV}$ sigue siendo el
mismo que antes (\ref{eq:horizonte_visual_no_inf}),
$\tilde{d}_{HP}$, el horizonte de partículas ahora es
(periodo inflacionario + periodo de radiación + periodo
entre el fin de radiación y el tiempo de desacople),

\begin{equation}
  \label{eq:horizonte_lss_inflacion}
  \tilde{d}_{HP} = a_0 \cdot
  \frac{1}{a_0H_0}\left(\frac{a_d}{a_0}\right)^{1/2} \Bigg\{1
    + 2 \frac{a_f}{a_d}\left[\left(\frac{a_f}{a_i}\right)
      -1\right] \Bigg\},
\end{equation}
donde $a_i \equiv a(t_i)$, $a_f \equiv a(t_f)$ son los
factores de escala al tiempo del inicio de inflación, $t_i$
y el fin de inflación $t_f$. Típicamente $(a_f/a_i) = 10^N$,
donde $N$ es el número de \textit{e-foldings} que dura la época
inflacionaria.  Por lo que ahora, tenemos 

\begin{equation}
  \label{eq:problema_horizonte_sol}
  \tilde{d}_{HP} \gg d_{HV}
\end{equation}

aunque, como se discute en \cite{Ellis88}, de esto no se
sigue que todos los puntos del LSS tengan datos iniciales
comunes al tiempo $t$, ya que para cualquier tiempo $t$
existen eventos en la hipersuperficie $t =$ constante, que
están en el pasado de $\mathscr{P}$ y no en el de
$\mathscr{Q}$ y viceversa. Lo que resuelve inflación es que
ahora tenemos la \textit{posibilidad} de explicar, debido a
la conexión causal de $\mathscr{P}$ y $\mathscr{Q}$ las
condiciones similares de ambos en el LSS.

  \figura{imagenes/solucion_horizonte}{width=10cm,height=7.5cm}{Solución
  propuesta por inflación al problema del horizonte. El
  efecto de la expansión acelerada es en cierta forma
  ``alejar'' la singularidad inicial del LSS. Esto permite
  que algunas partes de los conos pasados de luz de los
  puntos $\mathscr{P}$ y $\mathscr{Q}$ hayan estado en contacto
  causal.}{fig:solucion_horioznte}  

En base a la discusión anterior se definirá como época
inflacionaria a cualquier época en la cual tenemos aceleración positiva,
\begin{equation}
  \label{eq:condicion_inflacion_1}
  \ddot a > 0.
\end{equation}
De la ecuación
(\ref{eq:aceleracion}), obtenemos la condición para la
presión de la materia que generará inflación es
\begin{equation}
  \label{eq:condicion_inflacion_2}
  p < -\frac{1}{3}\rho.
\end{equation}

Respecto al origen de las anisotropías iniciales, Inflación
--teóricamente-- se apunta su mayor triunfo: las
anisotropías e inhomogeneidades primordiales son producto de
las fluctuaciones cuánticas del campo escalar, que al
evolucionar el universo en el periodo inflacionario son
``extendidas'' hasta poderlas 
considerar como perturbaciones clásicas que son observadas en la
estructura a gran escala del universo y en las anisotropías
de la temperatura del CBR. Esta explicación se tratará con
mayor detalle y se criticará en el capítulo siguiente.

\subsection{Modelado matemático: Campo
  escalar}\label{sec:model-matem-inflaton}
En el modelo inflacionario de \textit{slow-roll} la
condición (\ref{eq:condicion_inflacion_2}) se satisface
postulando la existencia de un campo escalar cuya ecuación
de estado es adecuada. La dinámica de un campo escalar acoplado
a la gravedad viene dada por la acción

\begin{equation}
  \label{eq:accion_campo_escalar}
  S_\varphi = \int d^4x\sqrt{-g} \left( -\frac{1}{2}
    g^{ab}\nabla_a \varphi \nabla_b \varphi -
    V(\varphi)\right). 
\end{equation}

El tensor de energía-momento del campo escalar se puede
calcular usando la definición de $T^{ab}$,

\begin{equation}
  \label{eq:def_tensor_energia_momento}
  T^{ab} = \frac{2}{\sqrt{-g}}
  \frac{\delta\left(\sqrt{-g}\mathcal{L}_{mat}\right)}{\delta g_{ab}}
\end{equation}

resultando en

\begin{equation}
  T_a\,^b =
  \nabla_a\varphi\nabla^b\varphi-\frac{1}{2}\delta_a\,^b(\nabla_c\varphi\nabla^c\varphi
  + 2V(\varphi))\label{eq:energia_momento_escalar}
\end{equation}

con componentes

\begin{equation}\label{eq:inflation_componentes}
  T_\eta\,^\eta = -\left(\frac{1}{2}\dot\varphi^2 +
    V(\varphi)\right), \quad 
  T_i\,^j = \left(\frac{1}{2}\dot\varphi^2 -
    V(\varphi)\right) \delta_i\,^j
\end{equation}


Comparando con el tensor de energía-momento de un fluido
perfecto podemos identificar a $T_\eta\,^\eta$ con la
energía, $\rho$, y a $T_i\,^j$ con la presión,
$p$. Recordando (\ref{eq:condicion_inflacion_2}) tenemos que
para que este campo genere inflación necesitamos que
$\dot\varphi^2 \ll V(\varphi)$.

La condición de que $V(\varphi) \gg \dot\varphi^2$ nos da
el requisito

\begin{equation}
  \label{eq:presion_densidad_inflaton}
  p_\varphi \simeq  -\rho_\varphi,
\end{equation}

i.e.\ la etapa inflacionaria es (o puede ser aproximada
mediante)  un espacio-tiempo de de Sitter.

La ecuación de movimiento del campo escalar se obtiene
variando la acción con respecto al campo escalar,

\begin{equation}
  \label{eq:kg_campo_Escalar}
  g^{ab}\nabla_a\nabla_b\varphi + \partial_\varphi V(\varphi) = 0.
\end{equation}

esta ecuación es del tipo de Klein-Gordon. En las
coordenadas de los observadores cosmológicos esta ecuación
toma la forma de

\begin{equation*}
  \ddot\varphi + 3H\dot\varphi + \partial_\varphi V(\varphi) = 0.
\end{equation*}

Podemos observar que el término $3H\dot\varphi$ actúa como
fricción, frenando así la evolución de $\dot\varphi$.

El conjunto de ecuaciones que controlan la evolución del
universo durante el régimen inflacionario son\todo{agregar
  la eq de aceleración}

\begin{subequations}\label{eq:inflacion_tiempo_cosmologico}
  \begin{equation}
    \label{eq:friedmann_escalar}
    H^2 = \frac{\kappa}{3} \left(\frac{1}{2}\dot\varphi^2 +
      V(\varphi)\right) , 
  \end{equation}
  \begin{equation}
    \label{eq:movimiento_escalar}
    \ddot\varphi + 3H\dot\varphi + \partial_\varphi V(\varphi) = 0, 
  \end{equation}
  \begin{equation}
    \label{eq:aux_escalar}
    \dot H = -4\pi G \dot\varphi^2
  \end{equation}
\end{subequations}

Este sistema de ecuaciones no siempre lleva a una expansión
acelerada del universo, para lograr esto  es
necesario que la energía potencial del campo escalar domine
sobre la energía cinética del mismo, específicamente,
implica que se desprecie el término cinético,
$\dot\varphi^2/2$, en (\ref{eq:friedmann_escalar}) y, debido
a que el potencial es plano, despreciar la aceleración,
$\ddot\varphi$, en (\ref{eq:movimiento_escalar}):

\begin{subequations}
  \begin{equation}
    \label{eq:friedmann_escalar_sra}
    H^2 \simeq \frac{\kappa}{3} V(\varphi) , 
  \end{equation}
  \begin{equation}
    \label{eq:movimiento_escalar_sra}
    3H\dot\varphi + \partial_\varphi V(\varphi) \simeq 0, 
  \end{equation}
\end{subequations}

Según el modelo de \textit{nueva inflación}, $\varphi$ está
inicialmente lejos del mínimo del potencial $V(\varphi)$. El
potencial entonces ``jala'' a $\varphi$ hacia su mínimo. Si
el potencial es lo suficientemente plano, el término de
fricción pronto hará a $\dot\varphi$ muy pequeño, logrando
que se cumpla $\dot\varphi^2 \ll V(\varphi)$, aún  cuando
originalmente esto no fuera así. Bajo estas condiciones el
espacio-tiempo es aproximadamente de Sitter
(cf.\  \pageref{sec:soluciones-fl}) , con $a(t) \sim
e^{Ht}$, $H \sim $ constante. 

Para expresar las condiciones de \textit{slow-roll} de
manera más precisa recurriremos a la definición de dos
parámetros que cuantifiquen que tan plano es el potencial
$\bm\epsilon$ y su curvatura $\bm{\eta}$ (no confundir con
$\eta$ el tiempo conforme). Usando la ecuación
(\ref{eq:movimiento_escalar_sra}) y derivándola
\begin{equation*}
  \ddot\varphi = H\dot\varphi\left[\bm\eta +
    \bm\epsilon\right] ,
\end{equation*}
donde
\begin{subequations}
  \label{eq:parametros_sra}
\begin{equation}
  \label{eq:parametros_sra_epsilon}
  \bm\epsilon(\varphi) = -\frac{\dot H}{H^2} =
  = 4 \pi G \frac{(\dot\varphi)^2}{H^2} =
  \frac{1}{4\pi G}\left(  
    \frac{\partial_\varphi V}{V}\right)^2, 
\end{equation}
\begin{equation}
  \label{eq:parametros_sra_eta}
\bm\eta(\varphi) = \frac{1}{8\pi G}
  \left(\frac{\partial^2_{\varphi} V}{V}\right) =
  \frac{1}{3}\frac{\partial^2_{\varphi} V}{H^2}, 
\end{equation}
\begin{equation}
  \label{eq:parametros_sra_delta}
  \bm\delta(\varphi) = \bm\eta - \bm\epsilon =
  -\frac{\ddot\varphi}{H\dot\varphi}\thinspace, 
\end{equation}
\end{subequations}

Las condiciones \textbf{necesarias} para que la aproximación
de \textit{slow-roll} sea válida y se tenga inflación son
\begin{equation}
  \label{eq:condiciones_inflacion_3}
  \bm\epsilon \ll 1,\quad |\bm\eta| \ll 1.
\end{equation}
Estas condiciones son válidas si $\dot\varphi^2 \ll V(\varphi)$ y
$|\ddot\varphi| \ll |3H\dot\varphi|$ respectivamente. Otra
forma de obtener este resultado es notando que 
\begin{equation*}
\frac{\ddot{a}}{a} = \dot{H} + H^2 = (1- \bm\epsilon)H^2,
\end{equation*}
por lo que inflación ($\ddot a > 0$) sólo se puede obtener
si $\bm\epsilon < 1$. De (\ref{eq:parametros_sra_epsilon})
se pude observar que 
$\bm\epsilon$ cuantifica cuanto cambia $H$ durante
inflación.

Con los parámetros de \textit{slow-roll} $\bm\epsilon$,
$\bm\eta$, $\bm\delta$ es posible escribir las
ecuaciones\footnote{Otras fórmulas útiles de escribir las
  ecuaciones exactas del sistema Einstein-Inflatón son 
  \begin{equation*}
    \dot{H} = -\frac{4\pi}{M_{Planck}^2} \dot\varphi^2,
    \quad \frac{dH}{d\varphi} = -\frac{4\pi}{M_{Planck}^2}
    \dot\varphi, \quad \left(\frac{dH}{d\varphi}\right)^2 = -
    \frac{12\pi}{M_{Planck}^2} H^2(\varphi) =
    -\frac{32\pi^2}{M_{Planck}^4}V(\varphi)
  \end{equation*}
  donde la última ecuación es la ecuación de Friedmann.
}\textit{exactas} (\ref{eq:friedmann_escalar}),
(\ref{eq:aux_escalar}): 
\begin{subequations}
  \begin{equation}
    \label{eq:friedmann_escalar_exacta_sra}
    H^2 = \frac{\kappa V}{3 -\bm\epsilon},
  \end{equation}
  \begin{equation}
    \label{eq:movimiento_escalar_exacta_sra}
    \dot\varphi = \frac{1}{(3-\bm\delta)H}\, \partial_\varphi V.
  \end{equation}
\end{subequations}

\subsection{Efectos de la inflación en las escalas físicas}

Durante inflación el factor de escala crece
exponencialmente, por lo que todas las escalas físicas
$\lambda = a(t) \lambda_c$, crecerán de la misma manera. El radio de Hubble
físico $d_H$ permanecerá constante durante esta misma
época. Entonces, se pueden tener situaciones en las que
una escala $\lambda' < d_H$ en la época inflacionaria, pero
al crecer exponencialmente se vuelva mayor que $d_H$ en el
transcurso de esta etapa. Al tiempo $t$ durante inflación
cuando $\lambda' = d_H$, se le denomina \textit{tiempo de
  salida} y lo denotaremos por $t_{salida}(\lambda'_c)$.

Al terminar inflación, durante la época dominada por
radiación $a, \lambda'  \propto t^{1/2}$, mientras que $d_H
\propto t$, por lo que (de mantenerse estas condiciones,
i.e.\ $a \propto t^n$  con $n < 1$ y $d_H \propto t$) en un
tiempo $t_{entrada}(\lambda'_c)$ volveremos a tener
$\lambda' = d_H$, a este tiempo se le llamará \textit{tiempo
de entrada}. Durante el análisis de formación de estructura,
capítulo \ref{cha:critica-al-origen}, será importante
relacionar la amplitud de la perturbación a la escala
$\lambda'$ al tiempo $t_{salida}(\lambda_c)$ con
la amplitud al tiempo $t_{entrada}(\lambda_c)$.

  \figura{imagenes/k_evolution}{width=10cm,height=8cm}{Evolución
    de las escalas y el radio de Hubble durante y después de
    inflación. En la imagen de arriba se muestra una escala
    física $\lambda$ y radio de Hubble. En la figura inferior la
    escala y el radio de Hubble comóviles.}{fig:k-evolucion}

\subsubsection{\textsc{Addendum}: Ecuaciones en tiempo
  conforme}

Las ecuaciones del campo inflatónico en tiempo conforme
serán entonces

\begin{subequations}\label{eq:einstein_inflaton_fondo}
  \begin{equation}
    \label{eq:friedmann_escalar_comovil}
    \HubbleComovil^2 =
    \frac{\kappa}{3}\left(\frac{1}{2a^2}(\varphi')^2 +
      V(\varphi)\right), 
  \end{equation}
  \begin{equation}
    \label{eq:acc_escalar_comovil}
    2\HubbleComovil' + \HubbleComovil^2 = -\kappa
    a^2\left(\frac{1}{2a^2} - V(\varphi)\right),
  \end{equation}
  \begin{equation}
    \label{eq:aux_escalar_comovil}
    \HubbleComovil^2 - \HubbleComovil' = 4\pi G (\varphi')^2,
  \end{equation}
  \begin{equation}
    \label{eq:movimiento_escalar_comovil}
    \varphi'' + 2\HubbleComovil\varphi' + a^2\partial_\varphi
    V(\varphi) = 0.
  \end{equation}
\end{subequations}

Siguiendo los pasos recién mostrados, es fácil obtener las
relaciones de \textit{slow-roll} \eqref{eq:parametros_sra}
para estas coordenadas. 
\begin{subequations}
  \label{eq:parametros_sra_comovil}
\begin{equation}
  \label{eq:parametros_sra_epsilon_comovil}
  \bm\epsilon(\varphi) = 1 -
  \frac{\HubbleComovil'}{\HubbleComovil^2} = 
  4\pi G\frac{(\varphi')^2}{\HubbleComovil^2},
\end{equation}
\begin{equation}
  \label{eq:parametros_sra_delta_comovil}
\bm\delta(\varphi) = 1 - \frac{\varphi''}{\HubbleComovil \varphi'}.
\end{equation}
\end{subequations}




\subsection{Críticas y problemas del modelo
  inflacionario}\label{sec:criticas-y-problemas-inflacion}

A pesar de que Inflación resuelve varios de los problemas
del modelos estándar cosmológico (como el de planitud y el
de horizonte), existen críticos a estas soluciones
propuestas por el modelo \citep{Penrose89,
  Borner04,Padmanabhan06,Sudarsky06a,Narlikar01,Brandenberger04}. Además,
inflación presenta nuevos problemas, que a continuación se enlistan:

\begin{itemize}
\item{\textit{Problema de la Fluctuación.}} Los modelos de
  inflación con campo escalar produce un espectro invariante
  de escala en las fluctuaciones cosmológicas y además
  predicen la amplitud del espectro. La dificultad es que sin
  introducir un problema de jerarquías\footnote{Un problema
    de jerarquías ocurre cuando los parámetros fundamentales
    (acoples o masas) de algún lagrangiano son diferentes
    que los parámetros medidos por el experimento. Estos
    problemas están relacionados con problemas de ajuste
    fino o con problemas de naturalidad del modelo.} en las
  escalas del modelo de física de partículas, la amplitud
  predicha es superior a la observada por varios órdenes de
  magnitud. Por ejemplo para un modelo inflacionario con
  $V(\varphi) = (1/4) \lambda \varphi^4$ se requiere que
  $\lambda \simeq 10^{-12}$ para recuperar la amplitud del
  espectro observada. Esta es una característica general de los
  modelos inflacionarios \citep{Brandenberger04} y dado que
  uno de los  principales objetivos de Inflación es eliminar
  los ajustes  finos en los parámetros cosmológicos, no es muy
  satisfactorio que ahora estos problemas 
  aparezcan en el sector de física de partículas.

\item{\textit{Problema Trans-Planckiano.}} En la mayoría de los
  modelos inflacionarios, el periodo de inflación dura mucho
  más que el número mínimo de \textit{e-foldings} requeridos
  para la solución de los problemas de horizonte y
  planitud. Esto tiene como consecuencia que los modos
  comóviles que corresponden 
  en la actualidad a las escalas de la estructura
  cosmológica fueran menores a la escala comóvil de la
  longitud de Planck al tiempo $t_i$ cuando inició la
  inflación \footnote{Es posible hacer un cálculo rápido que
  muestre este problema, supongamos que inflación cambia el
  factor de escala $a(t)$ en un factor $A\simeq 10^{30}$,
  entonces la escala física $\lambda(t)$ hoy $\lambda_0
  \equiv \lambda(t_0)$  al final de
  inflación $\lambda_e \equiv \lambda(t_e)$ tenía un tamaño
  $\lambda_e = \lambda_0\dfrac{a(t_e)}{a_0} =
  \lambda_0\dfrac{T_0}{T_e}$ $\simeq 10^{-28}$ suponiendo
  que inflación ocurre a escalas de gran unificación,
  entonces para una escala de $\lambda <
  \unit{3}\mega\parsec$ al inicio de inflación $t_i$,
  $\lambda_i \equiv \lambda(t_i) < L_{Planck}$.} 
. Entonces, ¿Cómo confiar en los cálculos que 
  hacemos para encontrar el espectro de potencias si estos
  cálculos dependen del comportamiento del espacio-tiempo y
  la materia a escalas de energía que no sabemos como
  describir? ¿Qué tan sensibles son las predicciones a esta física
  desconocida? Para una introducción ver
  \cite{Brandenberger03}, para revisiones más recientes e
  intentos de solución ver \cite{Martin00,Danielsson02,Shiu05}.

\item{\textit{Problema de la identidad del inflatón.}} Originalmente
  Guth supuso \cite{Guth81} que el inflatón podía ser
  identificado con el bosón de Higgs, pero la aparición de
  problemas de jerarquía mostraron que esta hipótesis no
  estaba fundamentada. En la actualidad, a pesar de los
  múltiples modelos del campo escalar, ninguno está fundado
  en física de partículas bien establecidas, por lo que la
  identidad de este campo escalar es una gran incógnita.

\item{\textit{Problema de la entropía.}}  La entropía del
  parche contenido en el radio de Hubble en la actualidad
  está dominada por la contribución de los fotones del
  CBR\footnote{La densidad 
  de entropía $s$ se obtiene de la ecuación
  (\ref{eq:densidad_entropia}) aplicada a los fotones ($g =
  2$) i.e.\ $s = \dfrac{4}{45}\pi^2T_{\gamma}^3$. La
entropía total 
  de una región está dada por $S=sV$. El volumen de la
  región contenida en un radio de Hubble puede aproximarse
  mediante $V_H = \dfrac{4}{3}\pi H^{-3}$.} y es del orden de
  $10^{88}$, la evolución del universo al ser un  sistema
  aislado, es adiabática, entonces ¿Por que es tan grande la
  entropía? \footnote{Esta pregunta da pie a varias otras
    preguntas como por ejemplo ¿Es válida la segunda ley
    durante toda la evolución del universo?¿Existen otras
    fuentes importantes a la entropía además de la del
    CBR?¿El campo gravitatorio es fuente de entropía?¿Cómo
    asociar entropía a un campo
    gravitatorio?
    \cite{Massimo08,Penrose89,Penrose05,Penrose94}}. Inflación
  resuelve este problema en el \textit{recalentamiento}, etapa en
  la cual la energía del inflatón es transferida a las
  partículas ordinarias generando una gran cantidad de
  entropía (consultar
  \cite{Mukhanov2005,Bergstrom99,Peacock98,Massimo08} para una 
  explicación más detallada).
 
  Algunos autores   \cite{Penrose89,Page83} opinan que
  inflaciónno resuelve  el problema si no que 
 de hecho lo agrava \citep[ver
 también][]{Penrose94,Penrose02,Penrose05}. 
  Esta argumentación se basa en el hecho de 
  tener condiciones iniciales de baja entropía es muy poco
  probable (entre más baja, más improbable), entonces, ya
  que el recalentamiento al 
  final de   inflación incrementa la entropía en varios ordenes de
  magnitud, hace necesario que 
  el estado del universo al principio de inflación fuera
  mucho más ordenado que en el modelo estándar cosmológico
  sin inflación.

\item{\textit{Problema del origen de las fluctuaciones cosmológicas.}}
  Este problema es presentado como resuelto por la
  inflación, pero como se verá en el capítulo siguiente esto
  no es del todo claro, ver también
  \cite{Sudarsky06a,Sudarsky06b,Proc2SISSA,Sudarsky07} 
\end{itemize}

\section{Modelo de Concordancia:
  $\Lambda$-CDM}\label{sec:modelo-de-concordancia}

Finalmente, en esta sección, expresaremos el modelo de
concordancia cosmológico y estableceremos las modificaciones
a las ecuaciones con las que estaremos trabajando.

Se le conoce como modelo de concordancia cosmológico al
modelo $\Lambda\thinspace$CDM, porque es el modelo que con las
características dadas abajo es el que mejor explica las
observaciones 
\cite{wmap3yr-temperature,wmap3yr-implications}. Este
paradigma cosmológico es, en pocas palabras, lo siguiente: \textit{una era de 
  gravedad cuántica de algún tipo , seguida por inflación;
  una época de big-bang caliente; desacople de materia y
  radiación; inestabilidad gravitacional que conlleva a la
  formación de cúmulos de galaxias que surgen de las
  ``semillas'' de perturbaciones de densidad de la LSS}
\citep{Ellis06}.

Observaciones recientes \citep{wmap3yr-implications}
concuerdan en que la composición del universo es
 $\Omega_{DE}\simeq 0.76$, $\Omega_{DM}
\simeq 0.20$, $\Omega_{B} \simeq 0.04$ y $\Omega_{rad}
\simeq 5\times 10^{-5}$. La constante de Hubble es $H_0 =
100\, {h}\, \kilo\meter \reciprocal\second \mega\parsec$ con
$h = 0.73 \pm 0.03$ \citep{Freedman01}. Además, la geometría
espacial es muy cercana a plana (y regularmente se supone
exactamente plana), y las perturbaciones iniciales son
Gaussianas,adiabáticas y con un espectro muy cercano a
invariante de escala (ver explicación del porqué de estas
características en el siguiente
capítulo) \citep{PDG08}.

\chapter[Crítica al orígen de la estructura propuesta por la
Inflación]{Análisis crítico a la formación de las
  fluctuaciones primordiales}
\label{cha:critica-al-origen}

\epigraph{In science, 'fact' can only mean 'confirmed to
  such a degree that it would be perverse to withhold
  provisional assent.' I suppose that apples might start to
  rise tomorrow, but the possibility does not merit equal
  time in physics classrooms.}{ Stephen Jay Gould (1941 -
  2002)}

\section{Introducción}

En el capítulo anterior se hizo una revisión con cierto
detalle de la física detrás de la cosmología moderna. Usando el
modelo cosmológico estándar y suplementándolo con el
paradigma inflacionario, se pudieron explicar fenómenos tales
como la expansión, la homogeneidad, la planitud, la
nucleosíntesis, el fondo de radiación cósmica, etc.

Basta observar alrededor nuestro para notar que la
aproximación hecha en el capítulo
\ref{cha:cosmologia-estandar} sobre la homogeneidad e
isotropía del universo no es válida a escalas menores que
$\unit{100} \mega\parsec$. Las existencia de estructuras
cosmológicas tales como \textit{clusters} de galaxias
observadas en la actualidad, se explican a través de la
teoría de formación de estructura. Esta teoría presupone que
en épocas muy tempranas (i.e.\ al principio o antes de la
época dominada por radiación) existían ciertas desviaciones
de la homogeneidad e isotropía que evolucionaron mediante
procesos físicos conocidos hasta convertirse en la
estructura que observamos actualmente. La época dominada por
radiación (y épocas anteriores) no puede ser observada
mediante ondas electromagnéticas debido a que el universo
era opaco (recuérdese que antes del desacople del fotón el
camino medio libre de esta partícula era despreciable) por
lo que la evidencia de la existencia de estas perturbaciones
primordiales debemos buscarla en la \acf{LSS} y en
particular en los fotones emitidos desde ella: el
\acf{CBR}. Estas desviaciones o fluctuaciones de la
homogeneidad e isotropía fueron observadas por primera vez
(fig. \ref{fig:dmr-cbr}) en el CBR por el \acf{DMR}
del satélite COBE a principios de la década de 1990
\cite{COBE}. Nótese que la teoría de formación de estructura
no dice nada sobre el origen de las desviaciones de
homogeneidad e isotropía primordiales, para ello necesitamos
considerar otra teoría que lo explique.

En este capítulo discutiremos el tratamiento de la formación
de la estructura. Debido a la extensión del tema, nos
limitaremos a darle una revisión relativamente rápida a los
resultados y desarrollos más importantes. Empezaremos,
primero, por mencionar los conceptos estadísticos que se
utilizan en las observaciones de la del fondo de radiación
cósmica y la física que se ve reflejada en su espectro de
potencias. Luego, repasaremos la teoría de perturbaciones en
Relatividad General aplicada a la cosmología. Usando la
teoría de perturbaciones a un orden lineal podremos obtener
las ecuaciones que rigen el crecimiento de las
perturbaciones primordiales en la densidad, velocidad y
entropía de la materia contenida por el universo. Por
último, presentaremos las distintas propuestas de solución a
la siguiente pregunta: en un universo homogéneo e
isotrópico, ¿Cómo surgen las perturbaciones primordiales?
Estudiaremos dos propuestas de solución. Ambas propuestas
estudiadas en este capítulo atribuyen el origen a la
evolución cuántica del campo inflatónico. La primera --la
estándar-- es parte del modelo inflacionario y atribuye la
aparición de las inhomogeneidades a una identificación --no
del todo justificada-- de cantidades cuánticas con
cantidades clásicas. En esta propuesta las ``fluctuaciones
cuánticas'' siguen únicamente la evolución (unitaria)
dictada por la mecánica cuántica, por lo que, además de la
dudosa identificación cuántica-clásica de algunas
cantidades, \textit{no} explica cómo se ``rompen'' las
simetrías de homogeneidad e isotropía del sistema
Einstein-inflatón ya que la 
evolución unitaria cuántica del sistema preserva dichas
simetrías. Estas críticas fueron presentadas \textit{in
  extensis} en \cite{Sudarsky06a} por
\citeauthor*{Sudarsky06a}, autores que además de presentar
los puntos oscuros de la explicación estándar, propusieron
una solución: el origen de las perturbaciones se logra
mediante la consideración en la dinámica del sistema
Einsten-inflatón del proceso de
reducción o colapso de la función de onda del inflatón. La
solución propuesta está basada en ideas de Roger Penrose
\cite{Penrose94,Penrose96,Penrose02,Penrose05} sobre la
interacción de la gravedad y mecánica cuántica, en
específico sobre el papel de la gravedad en la reducción de
la función de onda. La nueva física para generar las
semillas de la estructura esta contenida en la llamada
\textit{hipótesis del colapso}. Mostraremos cómo esta idea
puede resolver el problema de la formación de asimetrías del
universo. El desarrollo de esta idea es el tema principal de
este trabajo de tesis y se presentará en el capítulo
\ref{cha:analisis-esquemas}.

\section{\label{sec:fondo-de-radiacion}Anisotropía del Fondo
  de Radiación Cósmica}

Como se vio en el capítulo anterior (\S
\ref{sec:exitos-scc}), la simple existencia del CBR (o, como
es conocido comúnmente, \acf{CMB}\footnote{En la actualidad
  el CBR tiene su función de cuerpo negro tiene frecuencias
  en el régimen de las microondas, debido al
  \textit{redshift} ocasionado por la expansión del
  universo.} ) es una evidencia crucial para el modelo
cosmológico estándar, pero además de eso, el CBR se ha
convertido en el principal instrumento para estudiar el
universo \cite{COBE,wmap5}. De su observación se obtiene
información relativa a la curvatura de las hipersuperficies
espaciales $\Sigma_t$ (pag. \pageref{def:sigma_t}), a la
densidad de la materia bariónica $\rho_B$ y la densidad de
materia oscura $\rho_{DM}$, al parámetro de Hubble
normalizado ($h \equiv H_0/100$), la profundidad óptica
($\tau$) (pie de la página
\pageref{def:parametros_observacionales}) y a la
caracterización del espectro de potencia de las
fluctuaciones iniciales ($A$, $n$, $r$) (\textit{ibid.}). Hay
que notar que algunos de esos parámetros usando solamente el
estudio del CBR no son observacionalmente independientes.

La característica más importante del CBR es que presenta un
espectro de cuerpo negro a una temperatura de $T_0 =
\unit{2.725} \kelvin$. El espectro de la temperatura del CMB
a lo largo de cualquier dirección $\buvec{n}$ es el espectro
de un cuerpo negro con una temperatura $T(\buvec{n})$ con
fluctuaciones $\delta T (\buvec{n})$ en el rango de
$10^{-5}$ de la temperatura promedio\footnote{Esto es
  cierto luego de eliminar una perturbación dipolar que
  surge debido a nuestro movimiento peculiar a través del
  marco en reposo del CBR.} $T_0$.  Esta variación de la
temperatura fue descubierta por el satélite \acf{COBE}
\cite{COBE} en 1992 y luego estudiada con mayor precisión
por el \acf{WMAP}
\cite{wmap3yr-temperature,wmap3yr-implications,wmap5}.


\subsection{Descripción estadística del CMB}
Definimos como observable de las fluctuaciones del CMB a la
desviación \textit{relativa} de la temperatura o
\textit{contraste de la temperatura},
\begin{equation}
  \label{eq:contraste_temperatura}
  \Theta(\buvec{n}) \equiv \frac{\delta
    T(\buvec{n})}{T} = 
  \frac{1}{(2\pi)^{3/2}} \int d^3k \thinspace e^{i
    \bvec{k}\cdot\bvec{x}}\thinspace
  \Theta(\bvec{k}),  
\end{equation}
donde $\bvec{k}$ es el número de onda de la transformación
de Fourier, $\bvec{x} = R_D\buvec{n}$, la distancia
vectorial hasta la LSS y $\buvec{n}$ es un vector unitario
del momento del fotón. En esta expresión hemos ajustado el
sistema de coordenadas para que la posición del observador
sea $(\eta_{obs} = \eta_0$, $\bvec{x}_{obs} = \bvec{0})$. El
contraste de temperatura (\ref{eq:contraste_temperatura}) se
puede descomponer en armónicos esféricos
$Y_{lm}(\buvec{n})$,
\begin{equation}
  \label{eq:contraste_esfericos}
  \Theta(\buvec{n},\eta) = \sum_l\sum_m \alpha_{lm}
  Y_{lm}(\buvec{n}), 
\end{equation}
donde la sumatoria va de $l=1,2,\ldots, \infty$ y $m = -l,
\ldots, l$. Entonces se tienen $2l + 1$ valores de $m$ por
cada $l$.  Loas armónicos esféricos forman un conjunto
ortonormal de funciones sobre la esfera, por lo que podemos
calcular los coeficientes multipolares $a_{lm}$ mediante
\begin{equation}
  \label{eq:alm_esfericos_1}
  a_{lm} = \int d\Omega \thinspace Y_{lm}^*(\buvec{n})
  \Theta(\buvec{n}). 
\end{equation}
Podemos insertar en esta última ecuación, la expansión en el
espacio de Fourier de $\Theta(\buvec{n}, \eta)$ (segunda
igualdad de (\ref{eq:contraste_temperatura}),
\begin{equation}
  \label{eq:alm_esfericos_2}
  a_{lm} = \frac{1}{(2\pi)^{3/2}}\int d\Omega \int d^3k
  \thinspace Y_{lm}^*(\buvec{n}) 
  \Theta(\bvec{k})
  e^{i\bvec{k}\cdot\bvec{x}},
\end{equation}
usando el desarrollo de la onda plana en armónicos esféricos
\begin{equation}
  \label{eq:alm_esfericos_3}
  a_{lm} = \frac{4\pi}{(2\pi)^{3/2}}\int d\Omega \int d^3k
  \thinspace Y_{lm}^*(\buvec{n}) 
  \sum_{l'} \sum_{m'} i^{l'} j_{l'}(kR_D) Y_{l'm'}^*
  (\buvec{k}) Y_{l' m'}(\buvec{n})
  \Theta(\bvec{k}),
\end{equation}
donde $j_l(x)$ son las funciones esféricas de
Bessel,
\begin{equation}
  \label{eq:bessel_esfericos}
  j_l(x) = \sqrt{\frac{\pi}{2x}} \mathcal{J}_{l+1/2}(x),
\end{equation}
y $\mathcal{J}(x)$ son las funciones de Bessel de
primer tipo. Finalmente, empleando la ortogonalidad de los
armónicos esféricos llegamos a
\begin{equation}
  \label{eq:alm_esfericos}
  a_{lm} = \frac{4\pi}{(2\pi)^{3/2}} (i)^l\int d^3k
  Y_{lm}^*(\buvec{k}) j_l(kx) \Theta(\bvec{k}). 
\end{equation}


Modelaremos a $\Theta(\hat{\bvec{n}})$ como un campo
aleatorio\footnote{Es una generalización de un proceso
  estocástico (dado un espacio de probabilidad ($\Omega,
  \mathcal{F}, P$), un proceso estocástico (o proceso
  aleatorio) en el espacio de estados $X$ es una colección
  de variables aleatorias $X-$ valuadas indexadas por el
  conjunto $T$ (``tiempo''). Esto es, un proceso estocástico
  $F$ es la colección $\{F_t: t\in T\}$.) en el cual el
  parámetro $t$ no es un simple número real, sino que puede
  ser un campo vectorial multidimensional o inclusive una
  variedad.}, lo que significa que su valor en cada punto del
CMB es una variable aleatoria\footnote{De manera formal una
  variable aleatoria se puede definir como sigue: Sea
  $(\Omega, \mathcal{F}, \mathcal{P})$ un espacio de
  probabilidad (el triplete del espacio de probabilidad es
  $\Omega$ es el espacio de muestra, $\mathcal{F}$ es el
  $\sigma$-álgebra de subconjuntos de $\Omega$ y
  $\mathcal{P}$ la medida en $(\Omega, \mathcal{F})$) y $(Y,
  \Sigma)$ un espacio de medida.  Entonces la variable
  aleatoria $X$ es la \textit{función de medida} $X: \Omega
  \to Y$. Las pre-imágenes de los subconjuntos $Y$ ($\in
  \Sigma$) son eventos ($\in \mathcal{F}$) y tienen asignada
  una probabilidad $P$. Se puede ver que el nombre de
  \textit{variable} es incorrecto ya que en realidad es una
  función.}. Observaciones (e.g.\ \cite{wmap5}) indican que
las anisotropías se pueden modelar como un \textit{campo
  aleatorio gaussiano}, i.e.\ , las variables aleatorias del
campo aleatorio tienen una distribución de probabilidad
gaussiana.  Para campos aleatorios gaussianos \textit{toda
  la información} estadística está contenida en el primer
momento y en la función
de autocorrelación de dos puntos la cual da la covarianza
entre las variables aleatorias. Así la función de
correlación de dos puntos de la temperatura entre dos
direcciones denotadas por $\buvec{n}_1$ y $\buvec{n}_1$
puede ser escrita como sigue
\begin{equation}
  \label{eq:covarianza_cmb}
  C(\theta) =
  \overline{\Theta(\buvec{n}_1,\eta_0)\Theta(\buvec{n}_2,\eta_0)} 
\end{equation}
con $\cos\theta = \buvec{n}_1\cdot \buvec{n}_2$. La barra
horizontal indica promedio sobre un ensamble (teórico) de
universos\footnote{Para que la afirmación anterior tenga
  sentido es necesario considerar \textit{un ensamble de
    universos}, i.e.\ el campo $\delta$ en cada punto del
  espacio tendrá un valor diferente en cada uno de los
  universos miembros del ensamble, con una varianza
  $\expec{\delta^2}$ entre miembros del ensamble. La
  homogeneidad estadística del campo $\delta$ significa que
  la varianza es independiente de la posición. Obviamente el
  campo que estamos estudiando es un miembro de este
  ensamble i.e.\ una \textit{realización} del proceso
  estadístico.

  Si sólo tenemos acceso a un miembro (nuestro universo) de
  este ensamble de universos (suponiendo que el ensamble
  existe) ¿Cómo podemos medir la varianza
  $\expec{\delta^2}$? La mejor respuesta a este dilema es
  realizar mediciones en partes muy separadas del espacio,
  dado que los valores del campo $\delta$ estas diferentes
  partes del se supone que están causalmente
  desconectados. En otras palabras, estamos haciendo la
  identificación: promedio sobre el volumen $=$ promedio
  sobre el ensamble.}.

Nótese que la función de correlación
(\ref{eq:covarianza_cmb}) sólo depende de la separación de
las dos direcciones, i.e.\ no depende del ángulo
azimutal. Esto es un reflejo de que suponemos que la
estadística debería heredar las simetrías del espacio-tiempo
de fondo (i.e.\ , no perturbado) que en nuestro caso es la
métrica de \acf{RW} debe de ser
independiente de la posición espacial (homogeneidad) e
invariante ante rotaciones, i.e.\ ,  el promedio sobre el
ensamble de $\alpha_{lm}$ depende únicamente\footnote{De una
  manera intuitiva $l$ se refiere al tamaño angular de la
  anisotropía, mientras que $m$ está relacionado con la
  ``orientación''.} de $l$ y no de $m$,
\begin{equation}
  \label{eq:c_a_relacion}
  \overline{\alpha_{lm}\alpha_{l'm'}^*} = \delta_{ll'}\delta_{mm'}C_l,
\end{equation}
donde $C_l$ es conocido como el \textit{espectro de
  potencias angular} (teórico) de las anisotropías de la
temperatura. La relación (\ref{eq:c_a_relacion}) implica
\begin{equation}
  \label{eq:correlacion_legendre}
  C(\theta) = \frac{1}{4\pi}\sum_l (2l + 1) C_l
  \mathcal{P}_l(\buvec{n}_1\cdot\buvec{n}_2).
\end{equation}

Antes de proseguir es importante notar dos cosas en las
gráficas generadas (e.g.\ fig. \ref{fig:wmap_5}) del estudio
de las observaciones del CMB : (a) están presentadas en
términos de $C_l l (l+1) / 2\pi$, la explicación de esta
elección se debe a la siguiente relación aproximada
\[\sum_l \frac{2l +1}{4\pi} \thinspace C_l
\simeq \int \frac{l(l+1)}{2\pi}\thinspace C_l \thinspace
d\ln l\] i.e.\ $C_l l (l+1) / 2\pi$ es aproximadamente la
potencia por intervalo logarítmico de $l$; (b) el eje
vertical tiene unidades de $(\micro\kelvin)^2$, esto se debe
simplemente a que en lugar de estar trabajando con
$\Theta(\buvec{n})$ usan la fluctuación de temperatura
absoluta $\delta T = T_0 \thinspace \Theta$.

\paragraph{Discrepancia Cósmica.} La teoría predice los
valores de expectación $\overline{|a_{lm}|^2}$ de los
procesos estocásticos responsables de la anisotropía del
CMB, pero sólo podemos observar una realización de este
proceso aleatorio: el conjunto $\{a_{lm}\}$ de nuestro
CMB. Por lo que definimos el \textit{espectro de potencias
  angular observado} como el promedio
\begin{equation}
  \label{eq:espectro_angular_obs}
  \hat{C}_l = \frac{1}{2l +1} \sum_m |a_{lm}|^2
\end{equation}
de los valores de $a_{lm}$ observados. El valor de
expectación de $\hat{C}_l$ es igual a el espectro teórico
$C_l$, i.e.\ $\overline{\hat{C}_l} = C_l$, pero el valor de
una realización en general no es igual al valor
teórico. Podemos estimar el valor de expectación de la
diferencia cuadrática entre $\hat{C}_l$ y $C_l$:
\begin{equation}
  \label{eq:cosmic_variance}
  \frac{\sqrt{(\hat{C}_l - C_l)^2}}{C_l}  \simeq
  \sqrt{\frac{2}{2l+1}}.
\end{equation}

Podemos observar que esta cantidad es menor para $l$
grandes. Esto se debe a que la muestra estadística $(2l+1)$
de $a_{lm}$ crece conforme aumentamos $l$. Esta limitación,
que es grave para los multipolos bajos, recibe el nombre de
\textit{discrepancia cósmica}\footnote{En inglés es
  \textit{cosmic variance}}. Como es de esperarse, en la
práctica, nuestras observaciones distan del caso ideal y la
discrepancia es mayor que la dada por la relación
(\ref{eq:cosmic_variance}).

\subsection{Física de las anisotropías del
  CMB}\label{sec:fisica-cmb}

La física que da lugar a las inhomogeneidades y anisotropías
de la temperatura se puede entender fácilmente. Los
principales tres mecanismos para producir las anisotropías a
gran escala son: (a) \textit{diferencias en el potencial
  gravitatorio} entre un punto en particular en el espacio y
el observador. Esta diferencia de potencial resultará en que
la radiación propagándose entre el punto y el observador
tenga un corrimiento al rojo o al azul. Este mecanismo es
conocido como \acf{SW}.(b) \textit{Presión}: Los bariones
al caer en los pozos potenciales colisionan con los fotones
provocando oscilaciones de compresión y rarefacción. El
efecto de la compresión/rarefacción produce el cambio en la
temperatura de una manera análoga a los cambios de presión
adiabática de un gas ideal (c) \textit{velocidad}: Los
bariones se aceleran al caer en los pozos de gravitacional,
en el máximo de la compresión y/o rarefacción alcanzarán su
mínima velocidad, desplazándose su fase respecto a las
oscilaciones en $\pi/2$. Estas ondas desfasadas provocarán
un efecto Doppler en el campo de fotones. De los tres
mecanismos mencionados el más importante para este trabajo
de tesis es el efecto SW. La importancia del efecto SW se
debe a que contiene información de las perturbaciones
primordiales.

Los mecanismos que imprimen las anisotropías en la
temperatura del CMB, aunque bien entendidos, son difíciles de
calcular de manera analítica y por lo regular se usan
códigos numéricos, el más popular de ellos es el CMBFAST
\cite{CMBFAST}. Nótese que por su naturaleza los tres
mecanismos dominan en diferentes escalas: el efecto SW
domina a escalas mayores que el radio de Hubble y los otros
dos mecanismos dominan dentro del radio de Hubble, debido a
esto es conveniente dividir en tres rangos de escalas el
estudio del espectro \cite{Scott08,Anninos01}, justo como se
muestra en la figura \ref{fig:espectro_teorico}.
 
\figura{imagenes/espectro_teorico}
{width=12cm,height=8.5cm}{Espectro teórico de las anisotropías
  del CMB. Se muestran los diferentes procesos físicos que
  le dan su forma. También se muestra, con una línea
  punteada la contribución de las ondas
  gravitacionales. Esta imagen fue tomada de
  \citeauthor{Scott08}, \citeyear{Scott08}
  \cite{Scott08}.}{fig:espectro_teorico}

\paragraph{Efecto Sachs-Wolfe Integrado, $\left(l \lesssim
    10\right)$}
El \acfi{ISW}, se debe a variaciones temporales de las
perturbaciones de la métrica.

\paragraph{Efecto Sachs-Wolfe, $\left( 10 \lesssim l
    \lesssim100\right)$.} 
El horizonte de partículas en la época del LSS corresponde
aproximadamente\footnote{Esto se puede calcular de manera
  aproximada como sigue, usando
  (\ref{eq:horizonte_particulas}), para calcular el 
  horizonte de partículas en el tiempo de desacople $t_d$,
  obtenemos $d^c_{HP}(t_{d}) \simeq 0.168 h^{-1}
  \mega\parsec$, si calculamos su tamaño hoy, $\sim
  d_{HP}^c(t_d)(a_0/a_d) \simeq 184 h^{-1} \mega\parsec$,
  que comparado con la distancia que viajaron los fotones
  (i.e.\ la distancia al LSS) (\ref{eq:distancia_foton}),
  $D_c = a_0 \int_{t_d}^{t_0} (dt'/a(t')) \simeq \unit{6000}
  \mega\parsec$, el ángulo subtendido es $\theta \simeq
  184/6000 \sim 2\degree$, que a su vez está relacionado con
  el multipolo $l$ mediante $\theta \sim \pi/l$, i.e.\ $l
  \sim 100$. } a $l \simeq 100$, anisotropías mayores a esta
escala no han evolucionado significativamente y por lo
tanto, reflejan las ``fluctuaciones primordiales'' (las
cuales servirán como condiciones iniciales para las
ecuaciones de formación de la estructura
cf. \ref{sec:formacion_estructura}). El efecto del
\textit{redshift} gravitacional sobre la temperatura
(energía) de los fotones, es conocido como efecto de
Sachs-Wolfe, y su fórmula a está dada por
(\ref{eq:sachs-wolfe}) siendo $\psi$ la perturbación
gravitacional a primer orden.  Si el espectro de potencias
de las perturbaciones primordiales es invariante de escala
(plano), entonces $l(l+1)C_l \simeq$ constante para $l$'s
pequeños.

Esencialmente, la fórmula\footnote{Existen algunos artículos
  \citep[principalmente][]{White97} en los cuales se hacen
  aproximaciones heurísticas que en realidad no son
  aplicables a escalas superiores al radio de Hubble
  \citep[ver crítica en ][]{Hwang03}. Para expresiones
  invariantes de norma ver \cite{Hwang99}, si desea una
  demostración exclusivamente geométrica de este efecto ver
  \cite{MarceloCesar09,Merlin09}.}  del efecto de
Sachs-Wolfe \cite{Durrer01b,Hwang99,Hwang03} es simplemente
el \textit{redshift} que sufren los fotones en su trayecto
desde el LSS hasta el observador. La situación geométrica en
la cual vamos a realizar el cálculo es la siguiente. En la
esfera celeste que forma el LSS podemos descomponer la
temperatura en un punto $(\theta, \phi)$, o lo que es lo
mismo el punto señalado por el tres vector que parte de
nuestra posición $\buvec{n}$, en su parte promedio (o de
fondo) $\bar{T}(t)$ y la fluctuación en dicho punto $\delta
T(\bvec{x},t)$ : $T(\bvec{x}, t) = \bar{T}(t) + \delta
T(\bvec{x}, t)$, justo como hicimos en la sección
anterior. Luego del LSS los fotones no colisionan más con la
materia (i.e.\ , el camino libre medio de los fotones se
volvió más grande que el radio de Hubble) y ya no forman un
componente que ``genere gravitación'' de manera importante
i.e.\ ya no es la especie dominante en la época que sigue al
desacople. Los fotones, entonces se pueden considerar libres
y partículas de prueba, por lo que seguirán la geodésica
$x^a(\lambda)$ del espacio-tiempo real. El parámetro
$\lambda$ es el parámetro afín y el vector tangente a esta
geodésica es $k^a = dx^a/d\lambda$. Podemos aproximar la
geodésica del espacio-tiempo real como una perturbación de
primer orden de la geodésica del espacio-tiempo de FL. A
primer orden de perturbación, el vector tiene componentes
$k^0 = a^{-1}(\bar{\omega} + \delta\omega)$ y $k^i =
-\bar{\omega}a^{-1}(\bar{e}^i + \delta e^i)$, donde $\omega$
es la energía del fotón medida por un observador y $e^i$ es
el vector unitario espacial de propagación del fotón. Las
temperaturas del CMB en dos diferentes puntos a los que
llamaremos $\bm{E}$ (evento de emisión, determinado por la
intersección de la geodésica de emisión y el LSS) y $\bm{O}$
(evento de observación, aquí y ahora) a lo largo de una
geodésica nula en la dirección de observación son
\begin{equation}
  \label{eq:energia_emision_obs}
  \frac{T_{\bm{O}}}{T_{\bm{E}}} =
  \frac{(k^au_a)_{\bm{O}}}{(k^au_a)_{\bm{E}}} 
\end{equation}
donde $u^a$ en $\bm{O}$ y $\bm{E}$ es la cuatro-velocidad de
los observadores en el evento de observación y de emisión
respectivamente. Ahora descompondremos la temperatura
observada en una parte homogénea y su perturbación a primer
orden: $T(\bvec{x}_{\bm{O}}, t_{\bm{O}}; \buvec{e}_{\bm{O}})
= \bar{T}(\bvec{x}_{\bm{O}}, t_{\bm{O}}) + \delta
T(\bvec{x}_{\bm{O}}, t_{\bm{O}}; \buvec{e}_{\bm{O}})$

Con estas especificaciones el paso siguiente es escribir las
ecuaciones perturbadas de (i) la condición de vector nulo
$k^ak_a = 0$, (ii) la ecuación geodésica $k^a_{;b}k^b = 0$ y
la ecuación (\ref{eq:energia_emision_obs}).

Considerando únicamente las perturbaciones escalares ( cf.\
\cite{Unanue2009a} o ver apéndice \ref{cha:teor-de-pert}
para una explicación detallada de esta descomposición)

\begin{equation}
  \label{eq:sw_1}
  \frac{\delta T}{T}\Big|_{\bm{O}}  = \frac{\delta
    T}{T}\Big|_{\bm{E}}  - \frac{1}{k}
  v_{,\alpha}e^{\alpha}\Big|_{\bm{O}} + \frac{1}{k}
  v_{,\alpha}e^{\alpha}\Big|_{\bm{E}} - 
  \psi\Big|_{\bm{O}}+
  \psi\Big|_{\bm{E}} + 
  \int_{\bm{E}}^{\bm{O}} \left[\psi -
    \phi\right]'dy 
\end{equation}

donde $\frac{d}{dy} \equiv \dpar{}{\eta} -
\bar{e}^\alpha\dpar{}{x^\alpha}$, es decir la integral se
hace a lo largo de la geodésica nula. Podemos absorber el
término que no muestra dependencia angular i.e.\
$\psi\Big|_{\bm{O}}$ dentro de la definición de temperatura
de fondo y acomodar términos como sigue:

\begin{equation}
  \label{eq:sw_2}
  \frac{\delta T}{T}\Big|_{\bm{O}}  =  \left(\psi + \frac{\delta
      T}{T}\right)\Big|_{\bm{E}}  - \frac{1}{k}
  v_{,\alpha}e^{\alpha}\Big|_{\bm{O}} + \frac{1}{k}
  v_{,\alpha}e^{\alpha}\Big|_{\bm{E}} + 
  \int_{\bm{E}}^{\bm{O}} \left[\psi -
    \phi\right]'dy 
\end{equation}

Esta es la expresión general del efecto de Sachs-Wolfe. Al
primer término del lado derecho en la literatura se le llama
\textit{efecto de Sachs-Wolfe ordinario} y describe las
perturbaciones tanto gravitatorias como del plasma en la
LSS. El segundo y tercer término son \textit{efectos Doppler
  debidos al movimiento entre el observador y el plasma} y
el último término es el \textit{efecto integral de
  Sachs-Wolfe}. Es importante notar que esta descomposición
es heurística ya que no existe una manera genérica en
Relatividad General de distinguir los efectos de
\textit{redshift} debidos al efecto Doppler de los efectos
de \textit{redshift} gravitatorios. Para obtener la
expresión encontrada en la literatura se deben hacer las
suposiciones siguientes: (i) las perturbaciones primordiales
son adiabáticas, i.e.\ , $S= \delta_{DM} -
\frac{3}{4}\delta_{rad} = 0$; \footnote{ Las
  \textit{perturbaciones adiabáticas} se definen como
  aquellas que afectan a todas las especies de tal manera
  que las razones de los números de densidad permanecen sin
  perturbar: $\delta(n_X/n_Y) = 0$. En términos de las
  densidades de energía las perturbaciones adiabáticas están
  caracterizadas por las siguientes relaciones
  $(1/4)\delta_\gamma = (1/3)\delta_{DM} = (1/3)
  \delta_{B}$. Si la ecuación de estado es $p = p(\rho)$ las
  perturbaciones son necesariamente adiabáticas. Como las
  perturbaciones adiabáticas están asociadas a la
  perturbación de la curvatura vía las ECE, también se les
  conoce como \textit{perturbaciones de curvatura}. Por otro
  lado es posible perturbar --dado que hay múltiples
  componentes o especies-- a las densidades de energía de la
  materia sin perturbar la geometría, a este tipo de
  perturbación se le conoce como perturbación
  \textit{isocurvatura}.\label{def:adiabatico}} (ii) al
momento del desacople (\pageref{def:recombinacion}) los
bariones y los fotones forman un fluido ideal; (iii) el
universo es plano y no contiene energía oscura i.e.\
$K=\Lambda=0$; (iv) la ecuación de estado de la materia es
$p = 0$ al tiempo del desacople y posterior; y finalmente
(v) despreciar la solución decreciente (ver \S
\ref{sec:formacion_estructura}) de $\psi$
\cite{Hwang99,Hwang03} :

\begin{equation}
  \label{eq:sachs-wolfe}
  \frac{\delta T}{T} \Big|_{\bm{O}} = \frac{1}{3} \psi\Big|_{\bm{E}}.
\end{equation}

Es importante recalcar que el efecto Sachs-Wolfe
(\ref{eq:sw_2}) es el dominante a escalas mayores que el
radio de Hubble al momento del LSS, pero a escalas menores,
la física de plasmas domina y este análisis debe de ser
complementado con la ecuación de Boltzmann.



\paragraph{Picos acústicos, $\left(100 \lesssim l \lesssim
    1000\right)$.} La forma del espectro de potencias es
consecuencia de las oscilaciones provocadas por la gravedad
en el plasma antes de que los átomos se vuelvan neutrales
(para una explicación básica ver el artículo de divulgación
de \citeauthor{Hu04}, \citeyear{Hu04}).

Antes del tiempo de recombinación, las especies de elementos
se podían clasificar en dos grupos: un plasma formado por
materia bariónica (protones y electrones) y los fotones
fuertemente acoplados de tal manera que se pueden considerar
un fluido perfecto; y materia oscura la cual es la especie
dominante en cuanto a su densidad de energía y genera el
potencial gravitatorio en el cual está inmerso el plasma de
bariones y fotones. La materia oscura a través del potencial
gravitacional induce oscilaciones en el fluido
bariónico-fotónico, y por su parte, la presión de la
radiación actúa como fuerza restauradora. El papel de los
bariones es la de aportar una pequeña cantidad de inercia.

Al terminar la recombinación, la radiación (fotones) se
desacoplan de la materia y viajan libremente hacia
nosotros. Las fases de las oscilaciones se ``congelan''
(i.e.\ , no evolucionan más) y se proyectan en el CMB como una
serie de picos armónicos. El pico máximo es el modo que
alcanzó a recorrer un cuarto del periodo, alcanzando una
máxima compresión. Los picos pares corresponden a densidades
por abajo de la media y son regularmente menores que los
picos impares ya que en el ``rebote'' (restauración) deben
de sobreponerse también a la inercia de los bariones. Los
valles no alcanzan hasta cero debido al efecto Doppler.



\paragraph{Amortiguamiento, $\left(l \gtrsim 1000\right)$.}
La recombinación no es un proceso instantáneo. Su duración
se ve reflejada en el hecho que la LSS no es una superficie,
si no que posee un espesor. Las escalas menores que este
espesor se ven amortiguadas. Este efecto,conocido como
\textit{Efecto Silk}, virtualmente corta el espectro de
potencias en $l \sim 2000$. El efecto Silk se puede entender
como un efecto de difusión entre los bariones y fotones del
fluido. Otro efecto importante a estas escalas es el lente
gravitacional, pero este efecto se debe a estructuras
cercanas a nosotros (galaxias, clusters, etc.), por lo cual
se le clasifica como efecto no primordial o secundario.

\figura{imagenes/wmap5}{width=12cm,height=8cm}{Mapa (WMAP
  año 5)de fluctuaciones de la temperatura
  $\left(\dfrac{\delta T}{ T}\right)_{obs}$ luego de reducir
  el ruido y eliminar el dipolo. Los puntos amarillos y
  rojos son más calientes que el promedio, los azules más
  fríos que el promedio. Figura tomada de
  \citeauthor{wmap5}, \citeyear{wmap5}
  \cite{wmap5}.}{fig:wmap_5_2}

\subsubsection{Resultados del estudio del CMB.} A la fecha,
luego del experimento WMAP (ver figuras \ref{fig:wmap_5_2} y
\ref{fig:wmap_5}) y en espera del lanzamiento del satélite
PLANCK\footnote{Página web:
  \url{http://www.rssd.esa.int/index.php?project=Planck}},
el estudio del CMB ha dado los siguientes resultados
\cite{Scott08}:

\begin{itemize}
\item El universo pasó por una etapa de recombinación (cf.\
  \ref{sec:historia-termica}) en $z \simeq 1100$ y empezó a
  re-ionizarse en $z \simeq 10$

\item La geometría del universo es muy cercana a la plana,
$0.98 \lesssim \Omega \lesssim 1.08$

\item Materia oscura y Energía oscura son requeridas para
  explicar la geometría plana, la estructura del espectro de
  potencias del CMB (profundidad de los picos etc.) y la
  formación de estructura.

\item La inestabilidad gravitacional es suficiente para
  hacer crecer las fluctuaciones hasta las estructuras a
  gran escala observadas en la actualidad.

\item Las perturbaciones iniciales fueron
  \textit{adiabáticas} y con características de campos
  aleatorios gaussianos.
\end{itemize}

\figura{imagenes/wmap_5}{width=12cm,height=8cm}{Espectro de
  potencias de las anisotropías del CMB, tomadas por WMAP
  año 5. Figura tomada de \citeauthor{wmap5},
  \citeyear{wmap5} \cite{wmap5}.}{fig:wmap_5}

\section{Formación de
  estructura} \label{sec:formacion_estructura} Un universo
perfectamente homogéneo e isotrópico, permanecerá homogéneo
e isotrópico cuando es evolucionado con las \acf{ECE} ya que
su dinámica preserva estas simetrías. Para producir
desviaciones de la homogeneidad perfecta es necesario
suponer en el marco del modelo estándar de la cosmología la
existencia de pequeñas inhomogeneidades de la densidad de
energía en el universo temprano. El estudio de la
\textit{formación de estructura} se encarga de estudiar la
generación y la evolución de estas inhomogeneidades.  La
comprensión actual es que las
fluctuaciones primordiales se amplificaron debido a su
inestabilidad gravitacional y formaron la estructura a gran
escala que observamos en el presente. El problema de la
formación de estructura se puede dividir en dos partes: (a)
\textit{la generación de las inhomogeneidades primordiales},
y (b) el crecimiento debido a la evolución de esas
inhomogeneidades en la estructura que observamos.  

Esta segunda parte de la formación de estructura debe de
especificar la evolución o amplificación del contraste de
densidad o densidad relativa $\delta \equiv
\dfrac{\rho(\bvec{x}) - \bar{\rho}}{\bar{\rho}}$ desde una
amplitud $\delta \sim 10^{-5}$ en el LSS ($z\simeq 1100$)
hasta un contraste de densidad del orden de $\delta \sim
10^2$ para galaxias en $z << 1$, esto es un requerimiento
\textit{necesario} para una teoría consistente de formación
de estructura \citep*{Padmanabhan93}. Más adelante, en la \S
\ref{sec:pert-prim}, describiremos cómo pueden obtenerse
estas inhomogeneidades primordiales, pero por ahora nos
concentraremos únicamente en su evolución una vez que
aparecieron.

Mientras las perturbaciones permanezcan pequeñas es posible
usar teoría lineal de perturbaciones en Relatividad General,
desarrollada a detalle en el apéndice
\ref{cha:teor-de-pert}.  La teoría de perturbaciones en
Relatividad General presenta problemas de elección de norma
que dificultan la interpretación física de las
perturbaciones, es decir, es difícil saber si las
perturbaciones en la métrica o en el campo de densidad de
energía son físicos o aparecen por la elección de
coordenadas en las cuales se está calculando. Debido a este
problema es preferible expresar todas las ecuaciones de
manera invariante ante transformaciones de norma \todo{cita
  a definición del apéndice con pageref}, aunque existe la
posibilidad de elegir una norma en la cual se garantice la
interpretación física de todas las variables y no contenga
elementos espurios coordenados\footnote{Un ejemplo de una
  norma con problemas de grados de libertad no físicos de
  este tipo es la norma sincrónica (cf.\ pag.\ \pageref{pag:sincrona}) que
  lamentablemente fue la más usada al principio del
  desarrollo de la teoría de perturbaciones cosmológica. Por
  otro lado un ejemplo de elección de norma sin estos
  problemas es la norma de Poisson (cf.\ pag.\
  \pageref{pag:poisson}) }. En
esta sección no se discutirán los detalles y sutilezas del
método perturbativo en Relatividad y se remitirá al lector
interesado al apéndice \ref{cha:teor-de-pert} o al artículo
\cite{Unanue2009a}.

La idea básica de la teoría lineal de perturbaciones es
sencilla: perturbaremos la métrica $g_{ab} \to g_{ab} +
\delta g_{ab}$, y el tensor de energía-momento $T_{ab} \to
T_{ab} + \delta T_{ab}$. Sustituyendo estas perturbaciones
en las ECE, se obtienen ecuaciones perturbadas de la forma
$\mathrm{L}[g_{ab}] \delta g_{ab} = \delta T_{ab}$, donde
$\mathrm{L}$ es un operador diferencial de segundo orden
que depende de la métrica de fondo $g_{ab}$. En el caso
particular de los universos FL espacialmente planos ($K=0$)
es preferible descomponer las perturbaciones en modos de
Fourier,
\begin{equation}
  \label{eq:ece_perturbadas_fourier}
  \mathrm{L}(a(t), \bvec{k}) \delta g_{ab}(t, \bvec{k}) =
  \delta T_{ab}( t, \bvec{k}).
\end{equation}


Como se mostrará más adelante, el comportamiento de cada uno
de los modos depende del valor relativo de $\lambda(t)$
comparado con $d_H(t)$. Cuando $\lambda(t) = d_H(t)$, se
dice que \textit{el modo está entrando al radio de
  Hubble}. Las longitudes de onda pueden entrar al radio de
Hubble cuando el universo está dominado por radiación o por
materia. Definiendo a $\lambda_{eq}$ a la longitud que entra
al radio de Hubble cuando las densidades de materia y
radiación son iguales ($t = t_{eq}$), tenemos que si
$\lambda > \lambda_{eq}$ entonces el modo entra en la fase
dominada por materia, y en caso contrario ($\lambda <
\lambda_{eq}$) el modo entra en la época dominada por
radiación.

Para describir la formación de estructura en el universo, es
necesario encontrar cómo evolucionan las perturbaciones de
las tres especies de materia que conforman al
universo\footnote{A esta época la densidad de energía de la
  DE es despreciable comparada con la de las demás especies,
es por eso que es ignorada en lo que sigue.}:
bariones ($\rho_B$), materia oscura ($\rho_{DM}$) y
radiación ($\rho_R$). Para fines de cálculo, se supondrá que
la materia oscura está conformada por partículas de masa
$m_{DM}$, la materia oscura se desacopla del plasma a la
temperatura $T = T_{DM}^d$ y sus partículas se vuelven
no-relativistas después del tiempo $t^{nr}_{DM}$, $T(t^{nr}_{DM}) =
T_{DM}^{nr} \simeq m_{DM}$. Los modelos actuales de materia oscura
(cf. \cite{PDG08}) apuntan a que $T_{DM}^{nr} > T_{eq}$,
i.e.\ $t_{DM}^{nr} < t_{eq}$. Por su parte los bariones
permanecen acoplados con la materia hasta $t_{d}$ (ver
\ref{sec:historia-termica}). Es convencional en la
literatura introducir la velocidad de sonido de la especie
$X$ (bariones, DM, radiación) mediante $c_{s_X}^2 =
(\dot{p}_X/\dot{\rho}_X)$. 

Procederemos a elaborar formalmente lo dicho en los párrafos
anteriores sin llegar a las profundidades que se presentan
en el apéndice \ref{sec:teoria-perturbaciones}. El punto de
partida es la métrica perturbada de RW a primer orden,
\begin{equation}
  \label{eq:metrica_rw_perturbada}
  ds^2 = \left[ g_{\mu\nu} + \delta g_{\mu\nu}\right]
  dx^\mu dx^\nu = a^2(\eta)\left[ -d\eta^2 +
    \gamma_{ij}(\bvec{x})dx^idx^j + h_{\mu\nu}(\bvec{x},
    \eta)dx^\mu dx^\nu\right] 
\end{equation}

donde las perturbaciones de la métrica están dadas por
$h_{\mu\nu} = \delta g_{\mu\nu}/a^2$, $\gamma_{ij}$ es la
3-métrica de las hipersuperficies espaciales que folian el
espacio-tiempo de fondo $\Sigma_t$ (el espacio-tiempo sin
perturbar, i.e.\ el descrito por la métrica de RW) y
$\bvec{x}$ representa a las tres componentes espaciales:
$x^i$.

Las perturbaciones a primer orden $h_{\mu\nu}$ se pueden
separar de una manera $3+1$ como sigue:
\begin{equation}
  \label{eq:primer_orden}
  h_{00} = -2\phi, \quad h_{0i} = h_{i0} = w_i, \quad
  h_{ij} = 2 (\psi\gamma_{ij} + S_{ij}), \quad \text{con} \,
  \gamma^{ij} S_{ij} = 0.
\end{equation}
donde $\gamma^{ij}$ es la matriz inversa de $\gamma_{ij}$ y
$S_{ij}$ no tiene traza ya que fue puesta en $\psi$. En las
hipersuperficies espaciales, usamos a $\gamma_{ij}$ o
$\gamma^{ij}$ para bajar y subir índices espaciales. Además,
las derivadas espaciales se especificarán usando el operador
de derivada covariante 3-dimensional $\nabla_i$ definido
respecto a $\gamma_{ij}$. En este trabajo de tesis tratamos
con espacios-tiempos de RW que son espacialemente planos por
lo que en el sistema de coordenadas elegido en el capítulo
\ref{cha:cosmologia-estandar} $\gamma_{ij} = \delta_{ij}$ y
$\nabla_i = \partial/\partial x^i$.

Se puede demostrar (cf. apéndice
\ref{sec:teoria-perturbaciones} y \cite{StraumannN08}) que
los componentes de $h_{\mu\nu}$ se pueden descomponer en
este caso en partes que transformen ante rotaciones en las
hipersuperficies espaciales como escalares, vectoriales y
tensoriales. Además esta división en el caso cosmológico a
primer orden evolucionan de manera independiente. Este
procedimiento empieza con descomposición de cualquier vector
en sus partes \textit{longitudinales} y
\textit{transversales}:

\begin{equation}
  \label{eq:descomposicion_vectorial}
  w_i =  w_i^{\parallel}  + w_i^\perp = \partial_i w^\parallel +
  w_i^\perp, \quad \text{donde} \quad 
  \vec{\nabla} \times \bvec{w}^\parallel =
  \vec{\nabla}\cdot\bvec{w}^\perp = 0.
\end{equation}

Esta descomposición vectorial está basada en el teorema de
Helmholtz\footnote{El teorema de Helmholtz establece que
  cualquier campo vectorial suave en tres dimensiones se
  puede expresar como la suma de dos campos vectoriales: uno
  longitudinal (libre de rotación o \textit{irrotacional}) y
  otro transverso (libre de divergencia o
  \textit{solenoidal}). Como corolario se ve que el campo
  vectorial puede ser generado por un par de potenciales: un
  potencial escalar y uno vectorial.}  \cite{stewart08}. Los
nombres de \textit{longitudinal} y 
\textit{transversal} para la descomposición del vector
$\bvec{w}$ provienen de como es el vector $\bvec{w}$ es
paralelo o perpendicular respectivamente, respecto al
vector de onda $\bvec{k}$ de la transformación de
Fourier. Existe una descomposición similar  
para un tensor de dos índices:
\begin{equation}
  \label{eq:descomposicion_tensorial}
  S_{ij} = S_{ij}^\parallel + S_{ij}^\perp + S_{ij}^{TT},
\end{equation}
donde
\begin{equation}
  \label{eq:def_descomposicion_tensorial}
  S_{ij}^\parallel = \left(\delta_i\delta_j -
    \frac{1}{3}\nabla^2\right)\psi_S, \quad S_{ij} =
  \partial_i S_j^\perp + \partial_j S_i^\perp,
\end{equation}
y cumplen con $\delta^{ij}\partial_j S_i^\perp = 0$,
$\delta^{ik}\partial_k S_{ij}^{TT} = 0$ y
$\delta^{ij}S_{ij}^{TT} = 0$. Luego de estas
descomposiciones\footnote{ La nomenclatura expresada en
  términos de $h_{ab}$ es estándar en la presentación de
  teoría lineal de Relatividad General en libros de texto
  \cite{Wald84} y es la usada por \cite{NakamuraKo05}, pero
  lamentablemente no es estándar en la comunidad
  cosmológica. En \cite{Mukhanov90}, la signatura de la
  métrica es la usada por la comunidad de física de
  partículas y además los símbolos para identificar las
  perturbaciones están cambiados. En otras referencias
  \cite{RiottoAnto02,Langlois03,Martin04,Straumann05,StraumannN08}
  a las perturbaciones escalares las llaman ${A, B, C, D (o
    E)}$ que corresponderían (salvo signos) a $\{\phi,
  w^\parallel, \psi, \psi_s\}$, se le recomienda al lector
  ser cuidadoso en este enredo al consultar diversa
  bibliografía.}, podemos clasificar las variables de la
perturbación lineal como sigue: $S_{ij}^{TT} \equiv
h^{TT}_{ij}$ es la parte tensorial (o modo tensorial) de la
perturbación. La parte tensorial tiene dos grados de
libertad y físicamente representa a las ondas
gravitacionales. Los modos vectoriales son $w_i^\perp$ y
$S_{ij}^\perp$ poseen cada uno dos grados de libertad y
representan los efectos gravito-magnéticos. Finalmente los
modos escalares corresponden físicamente a los efectos
Newtonianos gravitacionales modificados
relativísticamente. Están representados por cuatro variables
$\psi$, $\phi$, $w^\parallel$ y $\phi_s$. En total tenemos
10 grados de libertad (2 tensoriales + 4 vectoriales + 4
escalares), de los cuales cuatro pueden ser eliminados
fijando una norma, i.e.\ fijando un sistema de coordenadas.

En este trabajo de tesis se elegirá la norma
\textit{Newtoniana Conforme}\footnote{La norma newtoniana
  conforme es la norma usada en los trabajos
  \cite{Dodelson,Langlois03,Mukhanov2005,Bertschinger01,RiottoAnto02,Martin04,Padmanabhan06},para
  comparaciones con otras normas ver
  \cite{Ma95,Mukhanov90}. Referencias usadas en esta tesis
  que no usan esta
  norma:\cite{Padmanabhan93,Padmanabhan96,Peebles93})}, en
la cual sólo se toman en cuenta las perturbaciones escalares
de la métrica y se hacen cero $\phi_s$ y $w^\parallel$. Esta
elección de norma es un caso especial de la norma de
\textit{Poisson} definida por $\delta^{ij}\partial_i w_j =
\delta^{ij}\partial_k S_{ij} = 0$ (cf.\
\ref{cha:teor-de-pert}).

En la norma Newtoniana Conforme, la métrica tiene la forma,
\begin{equation}
  \label{eq:newton-conforme}
  ds^2 = a^2(\eta) \left[-(1+2\phi)d\eta^2 +
    (1+2\psi)\delta_{ij}dx^i dx^j\right].
\end{equation}

Esta norma tiene como principales ventajas que las
expresiones obtenidas de las ECE se pueden interpretar
físicamente de manera relativamente directa y que sus
variables coinciden con las invariantes de norma (cf.\
apéndice \ref{cha:teor-de-pert} para ver más a fondo esta
cuestión).

Ahora sustituiremos la métrica (\ref{eq:newton-conforme}) en
las ECE (\ref{eq:einstein}) (haciendo $K=0$) y en
(\ref{eq:energia_momento_escalar}) e identificaremos los
órdenes cero y lineal. Así, los componentes del tensor de
Einstein a orden cero son
\begin{equation}
  G_{\eta}\,^\eta= -\frac{3}{a^2}\Hubble^2, \quad
  G_{i}\,^j = -\frac{1}{a^2}\delta_i^j\left(2\Hubble^{'} +
    \Hubble^2 \right), 
\end{equation}
mientras que las componentes del tensor de Einstein a primer
orden son
\begin{subequations}\label{eq:einstein_1er}
  \begin{equation}
    \delta G_\eta\,^\eta  =
    \frac{2}{a^2}\left\{(3\HubbleComovil\partial_\eta -
      \triangle)\psi + 
      3\HubbleComovil^2\phi\right\},
  \end{equation}
  \begin{equation}
    \delta G_i\,^\eta  =
    -\frac{2}{a^2}\partial_i\left(\partial_\eta\psi +
      \HubbleComovil\phi\right),
  \end{equation}
  \begin{equation}
    \delta G_\eta\,^i  =
    \frac{2}{a^2}\partial^i\left\{\partial_\eta\psi +
      \HubbleComovil\phi\right\},
  \end{equation}
  \begin{multline}
    \delta G_i\,^j =
    \frac{1}{a^2}\Bigg[\partial_i\partial^j\left(\psi-\phi\right)
    +\left\{(-\triangle + 2\partial_\eta^2 +
      4\HubbleComovil\partial_\eta )\psi +
      (2\HubbleComovil\partial_\eta +
      4\partial_\eta\HubbleComovil +
      2\HubbleComovil^2+\triangle)\phi \right
    \}\delta_i\,^j\Bigg].
  \end{multline}
\end{subequations}

Para modelar la materia se supondrá que la DM y la radiación
pueden ser tratados como un fluido perfecto. Por lo que 
se usará el tensor de energía-momento $T_{ab}$ dado por
la ecuación (\ref{eq:tab_fluido_perfecto}) 
\begin{eqnref}{eq:tab_fluido_perfecto}
  T_{ab} = (\rho + p )u_a u_b +  p g_{ab},
\end{eqnref}
expandiendo $T_a\,^b$ a primer orden obtenemos
\begin{subequations}
  \label{eq:perturbaciones_energia_momento}
  \begin{equation}
    T_0\,^\eta = -\rho(\bvec{x},\eta) = -[\overline{\rho}(\eta) +
    \delta\rho(\bvec{x}, \eta)], 
  \end{equation}
  \begin{equation}
    T_i\,^\eta = \left[\overline{\rho}(\eta) +
      \overline{p}(\eta)\right] 
    v_i(\bvec{x}, \eta) =  \left[\overline{\rho}(\eta) +
      \overline{p}(\eta)\right] 
    \nabla_i W, 
  \end{equation}
  \begin{equation}
    T_j\,^i = [\overline{p}(\eta) + \delta p(\bvec{x}, \eta)]
    \delta^i_j,
  \end{equation}
\end{subequations}
donde $\overline{\rho}$, $\overline{p}$ son la densidad de
energía y la presión en el espacio-tiempo de fondo con
métrica de RW, $v_i = \delta_{ij} dx^i/d\eta$ es la
3-velocidad del fluido (que se supone es
no-relativista). En este conjunto de ecuaciones
(\ref{eq:perturbaciones_energia_momento}) hemos hecho dos
aproximaciones, la primera
aproximación es que el campo de 3-velocidad de la materia es 
irrotacional $\bvec{v} = \bvec\nabla W$, i.e.\ $W$ es el
potencial de la velocidad; la segunda aproximación es que la
presión anisotrópica $\pi_j\,^i$ (la parte no-diagonal de
$T_a\,^b$) es despreciable comparada con la presión
$p$. Estas aproximaciones son válidas antes de la época de
formación de galaxias.
 
Igualando orden a orden las perturbaciones de la métrica
(\ref{eq:einstein_1er}) con las perturbaciones del tensor de
energía-momento~(\ref{eq:perturbaciones_energia_momento}) y
haciendo una pequeña manipulación algebraica, se obtienen  
las ecuación de Einstein en esta norma\footnote{Y bajo
  la transformación $\psi \to \Psi$, $\phi \to \Phi$ las ECE
  invariantes de norma. Se puede
  consultar el apéndice \ref{cha:teor-de-pert} para el cálculo
  invariante de norma, y haciendo la transformación inversa
  a la recién mencionada se obtienen estas ecuaciones.},
\begin{subequations}
  \label{eq:ece_fluido_primer_orden}
  \begin{equation}\label{eq:poisson_like}
    \nabla^2\psi = 4\pi G
    a^2 \left[ \delta\rho + 
      3\HubbleComovil (\overline{\rho}+\overline{p})
      W\right],
  \end{equation}
  \begin{equation}\label{eq:zeldovich_like}
    \partial_\eta \psi  + \HubbleComovil\psi = 4\pi G
    a^2(\overline{\rho}+\overline{p}) W,
  \end{equation}
  \begin{equation}\label{eq:evolucion_psi}
    \partial^2_\eta \psi + 3\HubbleComovil\partial_\eta\psi -
    \left(8\pi G a^2 \overline{p}\right) \psi = 4\pi G
    a^2\delta p ,
  \end{equation}
  \begin{equation}
    \label{eq:constriccion_anisotropica}
    \psi - \phi = 0.
  \end{equation}
\end{subequations}

Las ecuaciones de movimiento del fluido perfecto
\cite{Straumann05,Langlois03} se pueden obtener de la
ecuación de conservación (\ref{eq:conservacion_conforme_fl})
en la métrica (\ref{eq:newton-conforme}) son, la ecuación de
continuidad ($u^\mu\nabla_\nu T^\nu_\mu = 0$),
\begin{equation}
  \label{eq:continuidad_perturbada_grav}
  \delta\rho' +
  3\HubbleComovil\left(\delta\rho + \delta p\right) +
  (\bar\rho+\bar{p})\left(3\psi' + \nabla^2 W\right) = 0,
\end{equation}
y la ecuación de Euler (obtenida de la proyección espacial
de $\nabla_\nu T^\nu_\mu = 0$):
\begin{equation}
  \label{eq:euler_perturbada_grav}
  W' + \left(1-3c_s^2\right)\HubbleComovil W +
  \phi + \frac{\delta p}{\bar\rho + \bar{p}} = 0,
\end{equation}
donde $c_s^2 \equiv \dfrac{\bar{p}'}{\bar\rho'}$ es la
velocidad del sonido.

Hasta este momento, las ecuaciones
(\ref{eq:perturbaciones_energia_momento},
\ref{eq:ece_fluido_primer_orden},
\ref{eq:continuidad_perturbada_grav} y
\ref{eq:euler_perturbada_grav}) son las ecuaciones de dictan
la evolución de las perturbaciones y pueden aplicarse en
cualquier época cosmológica en las que sea valido aproximar a
la materia como un fluido.

\paragraph{Aproximación de dos fluidos.} Podemos especializar
estas ecuaciones al periodo que va desde el fin de inflación
hasta la recombinación. Supondremos que podemos describir
los componentes del universo por dos fluidos perfectos:
Materia Oscura (DM) y radiación\footnote{Radiación incluye a
  fotones, electrones y bariones, ya que todos ellos son
  partículas ultra-relativistas en esta época y están
  acopladas mediante dispersiones de Thomson}. La densidad
de energía en este periodo es $\delta\rho(\bvec{x}, \eta) =
\rho_{DM}\delta_{DM}(\bvec{x}, \eta) +
\rho_{rad}\delta_{rad}(\bvec{x}, \eta)$, en esta última
expresión hemos quitado las barras a las cantidades de fondo
($\overline{\rho} \to \rho$) y definimos los contrastes de
densidad $\delta \equiv \delta\rho/\rho$. La presión total
es $p(\bvec{x}, \eta) = \frac{1}{3}\rho_{rad}(a)(1+
\delta_{rad})$. Suponiendo que las ecuaciones de dinámica de
fluidos (\ref{eq:euler_perturbada_grav}),
(\ref{eq:continuidad_perturbada_grav}) son válidas para cada
especie, tenemos

\begin{subequations}
  \label{eq:mov_materia}
  \begin{equation}
    \label{eq:mov_dm}
    \delta_{DM}' + \nabla^2W_{DM} + 3\psi ' = 0, \quad W_{DM}'
    + \HubbleComovil W_{DM} + \psi = 0,
  \end{equation}
  \begin{equation}
    \label{eq:mov_rad}
    \frac{3}{4}\delta_{rad}' + \nabla^2W_{rad} + 3\psi' = 0,
    \quad W_{rad}' + \frac{1}{4}\delta_{rad} + \psi = 0.
  \end{equation}
\end{subequations}

Necesitamos elegir una ECE para completar el sistema de
ecuaciones ya que tenemos 5 incógnitas ($\delta_{rad},
\delta_{DM}, W_{rad}, W_{DM}$ y $\psi$) y cuatro ecuaciones
(\ref{eq:mov_materia}), por conveniencia, elegiremos a la
ecuación (\ref{eq:poisson_like}) para cerrar el sistema. El
sistema de ecuaciones se puede resolver de manera más fácil
transformándolas al espacio de Fourier\footnote{Esto se
  puede hacer en el universo que estamos estudiando ya que
  las hipersuperficies que folian el mismo, son planas. Si
  este no fuera el caso, esta transformación aún podría
  hacerse pero sólo tendría una validez local.}, quedando
\begin{subequations}
  \label{eq:mov_materia_fourier}
  \begin{equation}
    \label{eq:mov_dm_fourier}
    \delta_{DM}'  -k^2W_{DM} + 3\psi ' = 0, \quad W_{DM}'
    + \HubbleComovil W_{DM} + \psi = 0,
  \end{equation}
  \begin{equation}
    \label{eq:mov_rad_fourier}
    \frac{3}{4}\delta_{rad}' -k^2W_{rad} + 3\psi' = 0,
    \quad W_{rad}' + \frac{1}{4}\delta_{rad} + \psi = 0,
  \end{equation}
  \begin{equation}\label{eq:poisson_like_fourier}
    -k^2\psi = 4\pi G
    a^2 \left[ \delta\rho + 
      3\HubbleComovil (\rho+p) W\right].
  \end{equation}
\end{subequations}

Nótese que el vector de onda $\bvec{k}$ es comóvil, y por lo
tanto la longitud de onda física es $2\pi a /k$, con
$|\bvec{k}| = k$.

Resulta conveniente para los cálculos de formación de
estructura usar la variable $y = a/a_{eq}$ como variable
temporal en lugar del tiempo. La solución a la ecuación de
Friedmann (\ref{eq:friedmann}) para dos componentes
(radiación y DM) se puede expresar en términos de la nueva
variable $y$ de la siguiente manera \cite{Padmanabhan06}:
\begin{equation}
  \label{eq:def_y_x}
  y \equiv \frac{\rho_{DM}}{\rho_{rad}} = \frac{a}{a_{eq}}
  = \frac{\eta}{\eta_{eq}} +
  \left(\frac{\eta}{2\eta_e}\right)^2, \quad \eta_{eq} \equiv 
  \left(\frac{a_{eq}}{\Omega_{DM}}\right)^{1/2} H_0^{-1} =
  \frac{19}{\Omega_{DM}h^2} \mega\parsec
\end{equation}



La variable $\eta_{eq}$ define también al vector de onda
$k_{eq} = \eta_{eq}^{-1}$. Usando las nuevas variables,
manipulamos las ecuaciones (\ref{eq:zeldovich_like}),
(\ref{eq:poisson_like}), (\ref{eq:mov_materia_fourier}) y
(\ref{eq:poisson_like_fourier}) para eliminar las variables
de velocidad ($W_{DM}$, $W_{rad}$) y obtener tres ecuaciones
de evolución \cite{Padmanabhan06} para las tres incógnitas
restantes ($\psi$, $\delta_{DM}$, $\delta_{rad}$),
\begin{subequations}
  \label{eq:mov_x_y}
  \begin{equation}
    \label{eq:evolucion_psi_despues_inflacion}
    y\psi' + \psi +
    \frac{1}{3}\frac{k^2}{k^2_{eq}}\frac{y^2}{1+y}\psi =
    -\frac{1}{2}\frac{y}{1+y} \left(\delta_{DM} +
      \frac{1}{y}\delta_{rad}\right), 
  \end{equation}
  \begin{equation}
    (1+y)\delta_{DM}'' + \frac{2+3y}{2y}\delta_{DM}' =
    3(1+y)\psi'' + \frac{3(2+3y)}{2y} \psi' -
    \frac{k^2}{k_c^2}\psi,    
  \end{equation}
  \begin{equation}
    (1+y)\delta_{rad}'' + \frac{1}{2}\delta_{rad}'' +
    \frac{1}{3}\frac{k^2}{k_c^2}\delta_{rad} = 4(1+y)\psi ''
    + 2\psi' - \frac{4}{3}\frac{k^2}{k_c^2}\psi,
  \end{equation}
\end{subequations}
en estas ecuaciones $\{\}' \equiv d/dy$. Para resolver este
sistema de ecuaciones es necesario especificar as
condiciones iniciales, las cuales se suponen
dadas\footnote{El mecanismo generador de estas fluctuaciones
  primordiales se discutirá en \S
  \ref{sec:origen-tradicional} y \S
  \ref{sec:origen-colapso}.} al principio de la fase dominada
por radiación y para los modos que son más grandes que el
radio de Hubble, i.e.\ $y \ll 1$, $k\to 0$. En este límite
las ecuaciones (\ref{eq:mov_x_y}) son
\begin{equation}
  \label{eq:mov_k_y_iniciales}
  y\psi' + \psi \simeq -\frac{1}{2}\delta_{rad}; \quad
  \delta_{DM}'' + \frac{1}{y}\delta_{DM}' \simeq 3\psi'' +
  \frac{3}{y}\psi'; \quad \delta_{rad}'' +
  \frac{1}{2}\delta_{rad}' \simeq 4\psi''+ 2\psi'.
\end{equation}
\label{def:ci_estructura} El mecanismo que genera las
perturbaciones iniciales con un 
valor $\psi(y_i, l) = \psi_i(k)$. La primera ecuación
(\ref{eq:mov_k_y_iniciales}) da $\delta_{rad}(y_i,k) =
-2\psi_i$ cuando $y_i \to 0$. Si restamos las ecuaciones
para el contraste de densidad (\ref{eq:mov_materia_fourier})
vemos que en este límite ($k \to 0$) se conserva la
adiabaticidad\footnote{Cf.\ definición en el pie de la
  página \pageref{def:adiabatico}.},
$(\delta_{rad}-\delta_{DM})' = k^2((4/3)W_{rad}-W_{DM})$,
por lo que $\delta_{DM}(y_i,k) = (3/4)\delta_{rad} =
(3/2)\psi_i$. La ecuación
(\ref{eq:evolucion_psi_despues_inflacion}) determina $\psi$
dados ($\psi_i, \delta_{DM}, \delta_{rad}$). Finalmente,
usando las otras dos ecuaciones de (\ref{eq:mov_x_y})
tenemos $\delta_{DM}'(y_i,k) = 3\psi'$,
$\delta_{rad}'(y_i,k) = 4\psi'$ y obtenemos el conjunto total
de condiciones iniciales (ver tabla \ref{tab:condiciones}).

\ctable[pos=h, caption = {Condiciones iniciales para la
  época dominada por radiación. Se supone que el mecanismo
  generador de las perturbaciones iniciales asigna al
  potencial gravitacional $\psi$ el valor $\psi_i(y_i,k)$,
  cuando $y_i \to 0$.}, label =
tab:condiciones]{|c|c|c|c|c|}{}{ \hline
  $\psi$ & $\delta_{rad}$ & $\delta'_{rad}$ & $\delta_{DM}$ & $\delta'_{DM}$ \\
  \hline $\psi_i(k)$ & $-2\psi(y_i,k)$ & $3\psi'(y_i,k)$ &
  $\frac{3}{2}\psi(y_i,k)$ & $4\psi'(y_i, k)$ \\
  \hline }

El conjunto de ecuaciones
(\ref{eq:evolucion_psi_despues_inflacion}) puede ser
integrado fácilmente de manera numérica (cf.\
\cite{Padmanabhan06,Dodelson}), pero existen límites de
interés que son de utilidad para resolver al sistema de
ecuaciones de manera analítica. Estos límites y sus
soluciones se muestran en la tabla \ref{tab:soluciones}.

\ctable[ pos=htb, caption = {Soluciones analíticas en
  ciertos límites de interés del sistema de ecuaciones
  (\ref{eq:mov_x_y}). Las iniciales \textbf{MD} y
  \textbf{RD} significan ``época dominada por materia'' y
  ``época dominada por radiación'' respectivamente. Los
  límites se pueden entender como sigue: $y \equiv a/a_{eq}$
  funciona como una variable temporal, entonces $y \ll 1$
  representa tiempos muy tempranos (dentro de RD) y $y \gg
  1$ tiempos tardíos (dentro de MD). La variable $l$ está
  definida por $l^2 \equiv k^2/(3k_c^2)$, por lo que el
  límite $ly$ durante radiación marca el tiempo en el que
  entra la perturbación al radio de Hubble, $ly \gg 1$
  representa un tiempo posterior a la entrada de la
  perturbación al radio de Hubble. Finalmente $k\eta$,
  debido a que $d_H = (a'/a)^{-1} \propto \eta$ es
  simplemente el radio entre el radio de Hubble $d_H$ y el
  tamaño de la perturbación $\lambda$, $d_H/\lambda$,
  entonces el límite $\lambda \gg d_H$ se puede escribir
  como $k\eta \ll 1$ y $\lambda \ll d_H$ como $k\eta \gg
  1$.}, label = tab:soluciones ]{|c|c|c|p{6.8cm}|}{
  \tnote[$\circ$]{En este límite la adiabaticidad de la
    perturbación es respetada, entonces $\delta_{rad} \simeq
    (4/3)\delta_{DM}$.}  \tnote[$\star$]{Si la perturbación
    entra al radio de Hubble cuando se está en RD, no es
    posible despreciar los términos de presión.}
  \tnote[$\ast$]{Como $\delta_{DM}$ no es la especie
    dominante, el procedimiento estándar de resolución es
    resolver el sistema $\delta_{rad} - \psi$ y luego usar
    esas soluciones para determinar $\delta_{DM}$.}
  \tnote[$\dag$]{En MD se puede despreciar la dinámica de la
    radiación.}  }{ \hline
  \multicolumn{3}{|c|}{Aproximación} &
  \multirow{2}{*}{Soluciones} \\ \cline{0-2}
  Tamaño perturbación & \multicolumn{2}{c|}{Condiciones Extras}  & \,\\
  \hline \hline \multirow{2}{*}{$\lambda \gg
    d_H$\tmark[$\circ$]} & \multicolumn{2}{c|}{exacta} &
  $\psi = \psi_i \frac{1}{10 y^3}[16\sqrt{1+y} +
  9y^3+2y^2-8y-16]$ \qquad\qquad $\delta_{rad} \simeq 4\psi
  - 6\psi_i$ \\ \cline{2-4}
  &  \multicolumn{2}{c|}{$y \gg 1$} & $\psi \simeq
  \frac{9}{10} \psi_i$\\ 
  \hline \multirow{5}{*}{$\lambda \ll d_H$}
  &\multirow{3}{*}{\textbf{RD} \tmark[$\star$,$\ast$]} &
  exacta & $\psi = \psi_i \frac{3}{l^3y^3}[\sin(ly) -
  ly\cos(ly)]$ \\ \cline{3-4} &\,& $ly \gg 1$ & $\psi \simeq
  -3\psi_i(ly)^{-2}\cos(ly)$ , $\delta_{rad} \simeq
  6\psi_i\cos(ly)$ \\ \cline{3-4} &\,& $y \ll 1$ &
  $\delta_{DM} \simeq C_1 + C_2 \ln{y}$ \\ \cline{2-4}
  &\multicolumn{2}{c|}{\textbf{MD}\tmark[$\dag$]} & $\psi =
  \psi_\infty = \text{constante}$, \quad $\delta_{DM} =
  -2\psi_\infty - (2k^2/3k_c^2)\psi_\infty y \sim y \sim a$ \\
  \hline }

\subsection{Representaciones alternativas de las ecuaciones de
  evolución de las perturbaciones} 

La representación de las ECE y las ecuaciones dinámicas del
fluido (\ref{eq:mov_x_y}) no es única. Existen otras maneras
de expresar la evolución de las perturbaciones de materia
del universo \citep[por ejemplo
ver][]{Mukhanov90,Durrer01a,Durrer01b,Bertschinger01}. Cada
representación tiene sus ventajas y desventajas. En lo que
sigue mencionaremos tres opciones que serán de utilidad cuando
tratemos con el caso inflacionario y los mecanismos que dan
origen a las fluctuaciones primordiales.

\paragraph{Ecuación maestra para $\psi$.} En el caso más
general, la presión depende de la energía y de la entropía
por barión $S$, i.e.\ ,  $p = p(\rho, S)$. Cuando el fondo se
expande de manera reversible (i.e.\ los procesos de
aniquilación/creación de las especies de materia están
balanceados) $S$ es constante. Si el espacio tiempo tiene
una expansión adiabática, $p=p(\rho) \equiv w(\rho)\rho$. La
perturbación de presión es $\delta p = c_s^2 \delta\rho +
\tau \delta S$, donde $c_s^2 \equiv \left(\dfrac{\partial
    p}{\partial\rho}\right)_s$ es la velocidad isentrópica
de sonido y $\tau \equiv \left(\dfrac{\partial p}{\partial
    S}\right)_\rho$. Para un fondo que se expande
adibáticamente $c_s^2 \equiv \dfrac{p'}{\rho'}$.  Usando las
definiciones recién dadas y combinando las ecuaciones
(\ref{eq:poisson_like}), (\ref{eq:evolucion_psi}) obtenemos
la ecuación conocida como \textit{ecuación maestra}:

\begin{equation}
  \label{eq:maestra_hidro}
  \psi'' + 3\HubbleComovil(1+c_s^2)\psi' -c_s^2\nabla^2\psi
  + \left[2\HubbleComovil'+(1+3c_s^2)\HubbleComovil^2\right]\psi
  = \frac{\kappa}{2} a^2\tau \delta S.
\end{equation}

La ecuación maestra es la ecuación de movimiento para el
potencial métrico en el caso de un universo lleno de un
fluido perfecto. Es válida para cualquier época, excepto
para el caso muy especial de de Sitter exacto, ya que $\psi
= 0$. Esta ecuación se puede resolver fácilmente en dos
límites de interés:
\begin{itemize}
\item{\textit{Escalas mayores que el horizonte de sonido,
      ($kc_s^2 \ll \HubbleComovil$)}.}  Para el límite
  cuando $kc_s^2 \ll \HubbleComovil$, la ecuación maestra
  puede ser integrada de manera explícita:
  \begin{equation}
    \label{eq:maestra_hidro_1}
    \psi = \frac{\HubbleComovil}{a^2}\left[\bm\alpha\int d\eta
      \frac{a^2\left(\HubbleComovil' -
          \HubbleComovil^2\right)}{\HubbleComovil^2} +
      \bm\beta\right] ,
  \end{equation}
  donde $\bm\alpha$ y $\bm\beta$ son constantes de
  integración. En 
  este mismo límite si $a \propto t^p$, se obtiene
  \begin{equation}
    \label{eq:maestra_hidro_2}
    \psi = \frac{\alpha}{p+1} + \beta p t^{-p-1}.
  \end{equation}
  Esta solución muestra que existen dos modos uno creciente
  y uno decreciente en las perturbaciones del potencial
  gravitatorio. Podemos relacionar a $\psi$ en dos etapas
  cosmológicas diferentes con $a_1 \propto t^{p_1}$ y $a_2
  \propto t^{p_2}$ mediante
  \begin{equation}
    \label{eq:maestra_hidro_3}
    \psi_2 = \frac{p_2+1}{p_1+1}\psi_1.
  \end{equation}
\item{\textit{Una sola especie de materia.}} Para el caso en
  el que exista un solo tipo de materia\footnote{En este
    caso $\delta S = 0$ y $c_s^2 = w = $ constante. Por
    ejemplo, para radiación $w=c_s^2 = \dfrac{1}{3}$ y para
    materia no-relativista ``polvo'' $w=c_s^2 = 0$.} en el
  universo (o una especie de materia sea la dominante) con
  ecuación de estado $p=(\gamma -1)\rho$, la ecuación
  maestra   (\ref{eq:maestra_hidro})  se
  reduce a
  \begin{equation}
    \label{eq:maestra_hidro_uno}
    \psi'' + \frac{6\gamma}{(3\gamma - 2)}\frac{\psi'}{\eta}
    -(\gamma-1)\nabla^2\psi = 0,
  \end{equation}
  cuya  solución exacta en términos de
  funciones de Bessel.
  \begin{equation}
    \label{eq:maestra_hidro_dos}
    \psi(\eta) = \frac{C_1(\bvec{k})J_{\nu/2}(\sqrt{w}k\eta) +
      C_2(\bvec{k})Y_{\nu/2}(\sqrt{w}k\eta)}{\eta^{\nu/2}},
    \quad \nu = \frac{2+3\gamma}{3\gamma - 2}.
  \end{equation}
\end{itemize}

\paragraph{Ecuación del tipo oscilador armónico de masa
  variable.}  Es posible reescribir la ecuación maestra
(\ref{eq:maestra_hidro}) como una ecuación de oscilador
armónico con masa variable,
\begin{equation}
  \label{eq:oa_u}
  u'' - c_s^2\nabla^2 u -
  \left(\frac{\theta''}{\theta}\right) u = \mathscr{N},
\end{equation}
mediante el siguiente cambio de variables:
\begin{equation}
  \label{eq:u_hidro}
  \psi = 4\pi G (\rho +p)^{1/2}  u, \quad   \theta \equiv
  \dfrac{1}{a}\left(\dfrac{\rho}{\rho + 
      p}\right)^{1/2}, \quad \mathscr{N} \equiv a^2\sqrt{\rho +
    p}\sigma \delta S 
\end{equation}
Las soluciones a esta ecuación se pueden encontrar en sus
límites asintóticos como en la ecuación maestra y al igual
que esta, si sólo hay una especie, tiene una solución exacta
en términos de funciones de Bessel.

Como se verá en \S \ref{sec:evol-clasica} esta
ecuación nos permite estudiar la evolución de las
perturbaciones escalares de la métrica desde su generación
en inflación hasta su observación en el CMB  en la
actualidad\footnote{Aunque
  posiblemente exista un problema durante 
  el \textit{recalentamiento}. El potencial $\psi$ puede oscilar
  durante la etapa de \textit{recalentamiento} y se podría dar el
  caso que tuviera ceros que hicieran singular a la ecuación
  (\ref{eq:maestra_hidro}) (esto se ve claramente cuando
  especializamos la ecuación maestra para inflación cf.\
  ecuación (\ref{eq:u_inflaton})). En este trabajo de tesis
  se supondrá que la ecuación maestra no tiene estos puntos
  singulares.}.

\paragraph{Ecuación de conservación para modos que están
  fuera del radio de Hubble.}\label{def:xi} Podemos definir
la variable $\xi$ como
\begin{equation}
  \label{eq:integral_maestra_com}
  \xi \equiv \frac{2}{3\gamma}\frac{d}{d a} (a\psi) + \psi =
  \frac{2}{3} \frac{ \Hubble^{-1}\psi' + \psi}{\gamma} + \psi.
\end{equation}
Es fácil comprobar que $\xi$ es la primera integral de
(\ref{eq:maestra_hidro}) (i.e. $\xi' = 0$) para modos fuera
del radio de Hubble $k \ll \HubbleComovil$ si se cumple que
(a) la geometría espacial es plana $K=0$, (b) la
perturbación es adiabática ($\delta S = 0$) y (c) sólo se
toma en cuenta la solución creciente (i.e.\ donde
$c_s^2\nabla^2\psi$ es despreciable) \cite{Martin97}. Esta
variable fue introducida en por D. Lyth en 1985
\cite{Lyth85} y tiene una interpretación en la norma comóvil
\todo{Cita a la página de la norma comóvil}: es la
perturbación de la curvatura intrínseca \citep[cf.][para más
interpretaciones de esta variable.]{Liddle00}

También es posible relacionar a la variable $\xi$ con $u$
bajo las condiciones (a), (b) y (c) dadas arriba:
\begin{equation}
  \label{eq:xi_u}
  \xi = \theta^2\left(\frac{u}{\theta}\right)'.
\end{equation}

La importancia de la cantidad $\xi$ se debe a  de que
es una cantidad puramente geométrica, i.e.\ , el hecho de que
sea una integral de movimiento (bajo las condiciones
expresadas dos párrafos arriba) es válida para cualquier
tipo de materia y por lo tanto $\xi$ sirve para seguir el
comportamiento de las perturbaciones de la densidad de la
materia. 

Las variables $u$ y $\xi$ expresadas en el periodo
inflacionario serán de mucha ayuda para resolver el problema
del origen de las condiciones iniciales de las
perturbaciones de densidad de la energía en la explicación
estándar.

\section{Perturbaciones Primordiales}\label{sec:pert-prim}
El mecanismo \footnote{ Otro mecanismo propuesto para
  generar las perturbaciones primordiales es el de
  \textit{defectos topológicos} \citep[ver][ para
  exposisiones elementales]{Peacock98,Peebles93} y
  \citep[][para revisiones más actuales y/o
  completas]{Vachaspati93,Gangui03,Magueijo00,Brandenberger93}.
  Las predicciones de este mecanismo en cuanto a la
  formación de la estructura son distinguibles
  observacionalmente de la propuesta por inflación. Las
  perturbaciones generadas por los defectos topológicos son
  del tipo llamado \textit{isocurvatura}, mientras las
  generadas por el paradigma inflacionario son de tipo
  \textit{adiabático}. Uno de los \textit{test}
  observacionales más importantes para distinguirlos se basa
  en la posición del primer pico del espectro de potencias
  del CMB. El que este posicionado en el multipolo $l \simeq
  220$ indica que las perturbaciones primordiales son --por
  lo menos la componente dominante es-- adiabáticas.  Esta
  evidencia reciente apunta en contra de que el mecanismo de
  generación de perturbaciones sean los defectos topológicos
  \cite{Durrer00}} propuesto para generar las perturbaciones
primordiales se encuentra en el paradigma inflacionario. En
el contexto inflacionario la tarea de describir la creación
de estas fluctuaciones o inhomogeneidades se divide en dos
partes: (a) Las perturbaciones primordiales (que son
clásicas) surgen de las \textit{fluctuaciones cuánticas} del
inflatón $\varphi$ cuando el modo $k$ en cuestión ``sale'' del
radio de Hubble durante el periodo inflacionario. Esta
suposición es problemática a nivel conceptual y se analiza
dentro del paradigma inflacionario en \S
\ref{sec:origen-tradicional}. Debido a que la explicación
estándar no es satisfactoria se presenta la crítica y una 
hipótesis alternativa al origen de las perturbaciones en \S
\ref{sec:origen-colapso}. El desarrollo de esa hipótesis es
el tema central de esta tesis. (b) Suponiendo la existencia
de las perturbaciones \textbf{clásicas} del campo del
inflatón se encuentran sus ecuaciones dinámicas y se
evolucionan hasta convertirse en la perturbación inicial
gravitacional $\psi_i(k) \equiv \psi(y_i, k)$ de \S
\ref{sec:formacion_estructura},
pag. \pageref{def:ci_estructura}.  La evolución de las
perturbaciones primordiales clásicas se desarrolla de manera
casi idéntica a la ilustrada en \S
\ref{sec:formacion_estructura} y se expondrá a continuación.

Aunque ya se hizo hincapié en el capítulo
\ref{cha:cosmologia-estandar}, es importante recordarlo:
tenemos evidencia y por lo tanto confianza en que nuestra
imagen de la historia del 
universo es correcta para $T < \unit{1}\mega\electronvolt$,
pero tener esta confianza donde la evidencia o el conocimiento
teórico es más problemático y/o dudoso puede no estar
justificado. Por lo
tanto, \textit{suponer} que las perturbaciones primordiales
se originaron en la época inflacionaria, presupone de una
manera tácita que entendemos la evolución del universo entre
$H_{inf} \simeq 10^{-5}\thinspace M_{Planck} \simeq 10^{15}\thinspace
\giga\electronvolt$ y $H_{nucleosintesis} \simeq 10^{-31}\thinspace
\giga\electronvolt$ (esta última cifra es aproximadamente la
escala del parámetro de Hubble y es obtenido con la ecuación
(\ref{eq:hubble_rad})). Es importante tener en cuenta esta
advertencia siempre.

Las anisotropías del CMB 
en $l$ pequeños ($l < 100$) son reflejo de
las perturbaciones primordiales, y como vimos la  \S
\ref{sec:fondo-de-radiacion}, estas escalas están dominadas
por el efecto Sachs-Wolfe, el cual tiene una expresión
sencilla dada por la ecuación
(\ref{eq:sachs-wolfe}). Es importante deducir una expresión
que nos permita comparar las predicciones teóricas de los
modelos que dan cuenta del origen de las perturbaciones
primordiales y las anisotropías del CMB. Sustituyendo la ecuación
(\ref{eq:sachs-wolfe}) en (\ref{eq:alm_esfericos}) 
llegamos a
\begin{equation}
  \label{eq:alm_psi_materia}
  a_{lm} = \frac{4\pi}{(2\pi)^{3/2}} i^l \int d^3k\thinspace
  \frac{j_l(kR_D)}{3} Y_{lm}^*(\buvec{k})\psi(\bvec{k})
\end{equation}
donde hay que recordar que $\psi$ es la perturbación
gravitatoria \textit{al momento del desacople de los
  fotones}, momento que está dentro de la época cosmológica dominada
por la densidad de energía de la materia no-relativista. 
Usando (\ref{eq:c_a_relacion}) junto con
(\ref{eq:alm_psi_materia}) podemos calcular el espectro de
potencias angular $C_l$
\begin{equation}
  \label{eq:cl_psi_mat}
  \overline{a_{lm} a_{l' m'}^*} = \frac{(4\pi)^2}{(2\pi)^3}
  i^l (-i)^{l'} \int d^3k \int d^3k' \thinspace Y_{lm}^*(\buvec{k})
  Y_{l'm'} (\buvec{k}') \frac{j_l(kR_D)}{3}
  \frac{j_{l'}(kR_D)}{3} \thinspace \overline{\psi(\bvec{k}) \psi(\bvec{k}')}.
\end{equation}
Si $\psi(\bvec{k})$ son las componentes de Fourier de un
campo aleatorio gaussiano e isotrópico,
\begin{equation}
  \label{eq:power_spectrum}
  \overline{\psi(\bvec{k}) \psi(\bvec{k}')} =
  \frac{2\pi^2}{k^3} \bm{P}_\psi (k) \delta^{(3)} (\bvec{k}
  - \bvec{k}'), \quad  \bm{P}_\psi(k) \equiv
  \frac{k^3}{2\pi^2} \overline{\psi(\bvec{k})}\thinspace^2,
\end{equation}
donde se ha introducido el espectro de potencias de las
perturbaciones gravitatorias $\bm{P}_\psi$. Haciendo la
sustitución de (\ref{eq:power_spectrum}) en
(\ref{eq:cl_psi_mat})  y usando una vez más la ortogonalidad
de los armónicos esféricos tenemos
\begin{equation}
  \label{eq:cl_psi_power_spectrum}
  C_l = \frac{4\pi}{9} \int_0^\infty \frac{dk}{k}
  j_l^2(kR_D) \bm{P}_\psi(\bvec{k}). 
\end{equation}
Es importante notar que este espectro de potencias es el de
las perturbaciones gravitatorias durante la época del
desacople. Afortunadamente, como se verá en la siguiente
sección para $l$ pequeños la relación entre las
perturbaciones primordiales del inflatón y las
perturbaciones durante la materia tienen una expresión
sencilla si se utiliza la cantidad $\xi$
(pag. \pageref{def:xi}) la cual es constante fuera del radio
de Hubble (i.e.\ , para modos con una longitud de onda mayor
que el radio de Hubble).

Las observaciones llevadas a cabo por los experimentos  que
estudian el CMB (e.g.\ WMAP)  han supuesto que las
fluctuaciones a $l$ pequeños tienen una dependencia como ley
de potencias caracterizada por un único índice espectral $n$
\begin{equation}
  \label{eq:power_spectrum_power_law}
  \bm{P}_\psi =  \bm{A}
  \left(\frac{k}{k_p}\right)^{n-1} ,
\end{equation}
donde $k_p$ es una escala de pivote típica introducida
convencionalmente (para el WMAP se escogió $k_p =
\unit{0.05}\mega\parsec^{-1}$). Sustituyendo las funciones
esféricas de Bessel por (\ref{eq:bessel_esfericos}) e
introduciendo la variable $\bm{x} = kR_D$ se obtiene,
\begin{equation}
  \label{eq:cl_power_law}
  C_l = \frac{2\pi^2}{9} \bm{A}\cdot (k_pR_D)^{(1-n)}
  \int_0^\infty d\bm{x} \bm{x}^{n-3} \mathcal{J}^2_{l+1/2}(\bm{x}),
\end{equation}
la cual se puede resolver para $ -3 < n < 3$, obteniendo 
\begin{equation}
  \label{eq:cl_wmap}
  C_l = \frac{\pi^2}{36} \bm{A} \mathcal{Z}(n,l),
\end{equation}
donde,
\begin{equation}
  \label{eq:zl}
  \mathcal{Z}(n,l) =  (k_pR_D)^{(1-n)} 2^n
  \frac{\Gamma(3-n)\Gamma\left(l+ \frac{n}{2} -
      \frac{1}{2}\right)}{\Gamma^2
    \left(\frac{4-n}{2}\right)\Gamma\left(\frac{5}{2}
      + l  -\frac{n} {2}\right)}.
\end{equation}
Las observaciones indican que para $l$ pequeñas el espectro
de potencias angular 
es aproximadamente el espectro de Harrison-Zel'dovich, i.e.\ ,
$n \simeq 1$, en este caso 
\begin{equation}
  \label{eq:HZ_spectrum}
  \frac{l(l+1)}{2\pi} C_l = \frac{\bm{A}}{9},
\end{equation}
con $\bm{A} \sim 9\times 10^{-10}$ según las observaciones
del WMAP.

Aunque los cálculos en esta sección son válidos para $l <
100$, lo cual implica que estamos ignorando todos los
efectos de la física de plasmas (oscilaciones acústicas,
amortiguamiento, etc.\ , cf. \S \ref{sec:fisica-cmb}), en
este trabajo de tesis se considerará que este es el espectro
de potencias con el cual hay que comparar las predicciones
teóricas, ya que es el que refleja las perturbaciones
primordiales sin ``contaminación'' de otros efectos físicos.

\section{Evolución \textit{clásica} de las perturbaciones
  primordiales}\label{sec:evol-clasica}

Para escribir las ECE debemos complementar a las expresiones
a orden cero y orden lineal de las componentes del tensor de
Einstein (\ref{eq:einstein_1er}) con sus homólogos en el
tensor de energía-momento del campo escalar.

En la métrica (\ref{eq:newton-conforme}) las componentes del
tensor de energía-momento del inflatón a orden cero son
\begin{equation}
  T_\eta\,^\eta = -\left(\frac{1}{2a^2}(\varphi^{'})^2 +
    V(\varphi)\right), 
  \quad
  T_i\,^j = \left(\frac{1}{2a^2}(\varphi^{'})^2 -
    V(\varphi)\right) \delta_i\,^j,
\end{equation}
las ECE del sistema Einstein-Inflatón a orden cero son las
encontradas en el capítulo \ref{cha:cosmologia-estandar}
\begin{eqnref}{eq:friedmann_escalar_comovil}
  \Hubble^2 = \frac{8 \pi \, G}{3}a^2\rho = \frac{8 \pi \,
    G}{3}a^2\left(\frac{1}{2a^2}(\varphi^{'})^2 +
    V(\varphi)\right),
\end{eqnref}
\begin{eqnref}{eq:acc_escalar_comovil}
  2\Hubble^{'} + \Hubble^2 = -8\pi G a^2 p = -8 \pi G a^2
  \left(\frac{1}{2a^2}(\varphi^{'})^2 - V(\varphi)\right),
\end{eqnref}
además la suma de estas dos ecuaciones da
\begin{eqnref}{eq:aux_escalar_comovil}
  \Hubble^2 - \Hubble^{'} = 4\pi G a^2(\rho + p) = 4\pi G
  (\varphi^{'})^2,
\end{eqnref}
ecuación que será ser de utilidad más adelante.

También a orden cero obtenemos, usando la identidad de
Bianchi (\ref{eq:bianchi}) aplicada al campo escalar, la
ecuación de movimiento del inflatón,
\begin{eqnref}{eq:movimiento_escalar_comovil}
  \varphi'' + 2\HubbleComovil\varphi + a^2\partial_\varphi
  V(\varphi) = 0.
\end{eqnref}

El tensor de energía-momento del inflatón a primer orden es
\begin{subequations}\label{eq:tab_inflaton_1er}
  \begin{equation}
    \bm{\delta}\mathcal{T}_\eta\,^\eta =
    -\frac{1}{a^2}\left(\partial_\eta
      \delta\varphi\partial_\eta\varphi  
      - \phi(\partial_\eta\varphi)^2 +
      a^2\frac{dV}{d\varphi}\delta\varphi\right)
  \end{equation}
  \begin{equation}
    \bm{\delta}\mathcal{T}_i\,^\eta  =
    -\frac{1}{a^2}\partial_i\delta\varphi\partial_\eta\varphi
  \end{equation}
  \begin{equation}
    \bm{\delta}\mathcal{T}_\eta\,^i =
    \frac{1}{a^2}\partial_\eta\varphi\partial^i\delta\varphi
  \end{equation}
  \begin{equation}
    \bm{\delta}\mathcal{T}_i\,^j =
    \frac{1}{a^2}\delta_i\,^j
    \left( \partial_\eta\delta\varphi \partial_\eta\varphi    
      - \phi(\partial_\eta\varphi)^2 -
      a^2\frac{dV}{d\varphi}\delta\varphi\right).
  \end{equation}
\end{subequations}

Entonces, las ECE del sistema Einstein-Inflatón a primer
orden son
\begin{equation}
  (-3\HubbleComovil\partial_\eta + \triangle)\psi -
  3\HubbleComovil^2\phi = 4\pi\,G
  \left(\partial_\eta\delta\varphi\partial_\eta\varphi - 
    \phi(\partial_\eta\varphi)^2 +
    a^2\frac{dV}{d\varphi}\delta\varphi\right),
\end{equation}
\begin{equation}
  \partial_\eta\partial_i\psi +
  \HubbleComovil\partial_i\phi =
  4\pi\,G\partial_i\delta\varphi\partial_\eta\varphi
\end{equation}
\begin{multline}
  \partial_i\partial^j\left(\psi-\phi\right)
  +\left\{(-\triangle + 2\partial_\eta^2 +
    4\HubbleComovil\partial_\eta )\psi +
    (2\HubbleComovil\partial_\eta +
    4\partial_\eta\HubbleComovil +
    2\HubbleComovil^2+\triangle)\phi \right \}\delta_i\,^j =
  \\8\pi\,G \delta_i\,^j\left(\partial_\eta
    \delta\varphi\partial_\eta\varphi -
    \phi(\partial_\eta\varphi)^2 -
    a^2\frac{dV}{d\varphi}\delta\varphi\right).
\end{multline}

Resolviendo la última ecuación con las condiciones
apropiadas de frontera cuando $i \neq j$ se obtiene
\begin{equation}\label{eq:constriccion_fluido_perfecto}
  \psi = \phi,
\end{equation}
obsérvese que este resultado es independiente de que la
materia en el universo sea un campo escalar, es un
resultado que será verdadero mientras no haya anisotropías
en el fluido de la materia. Usando las
ecuaciones de Einstein de fondo
(\ref{eq:friedmann_escalar_comovil}),
(\ref{eq:acc_escalar_comovil}) y en especial
(\ref{eq:aux_escalar_comovil}) llegamos a la forma final de
las ecuaciones de Einstein para esta etapa cosmológica:

\begin{subequations}\label{eq:modo_escalar_inflaton}
  \begin{equation}\label{eq:modo_escalar_1_}
    (\triangle - 3\HubbleComovil\partial_\eta -
    \HubbleComovil' - 2\HubbleComovil^2)\psi = 4\pi
    G\left(\varphi'\delta\varphi'
      +
      a^2\frac{dV}{d\varphi}\delta\varphi\right) ,
  \end{equation}
  \begin{equation}\label{eq:modo_escalar_2_}
    \psi'+\HubbleComovil\psi =
    4\pi G \varphi'\delta\varphi ,
  \end{equation}
  \begin{equation} \label{eq:modo_escalar_3_}
    (\partial_\eta^2 + 3\HubbleComovil\partial_\eta +
    \HubbleComovil' + 2\HubbleComovil^2)\psi = 4\pi
    G\left(\varphi'\delta\varphi' -
      a^2\frac{dV}{d\varphi}\delta\varphi\right).
  \end{equation}
\end{subequations}

Es importante notar que sólo dos de estas ecuaciones son
independientes. La ecuación de evolución del campo escalar
se obtiene sustituyendo (\ref{eq:tab_inflaton_1er}) en las
identidades de Bianchi,
\begin{equation}
  \label{eq:mov_escalar_perturbada}
  \delta\varphi'' + 2\HubbleComovil \delta\varphi' -
  \nabla^2\delta\varphi + a^2\partial^2_{\varphi}V(\varphi)
  \delta\varphi - 4\psi'\varphi' + 2a^2\psi\partial_\varphi
  V(\varphi) = 0.
\end{equation}

Tomando las ecuaciones \eqref{eq:modo_escalar_1} y
\eqref{eq:modo_escalar_3} podemos eliminar el término
potencial del campo escalar
\begin{equation}\label{eq:Phi_1_y_varphi_1}
  (\partial_\eta^2 + \triangle)\psi =
  8\pi\,G\delta\varphi' \varphi',
\end{equation}
y usando \eqref{eq:modo_escalar_2} para eliminar
$\delta\varphi'$ de esta ecuación llegamos a
\begin{equation}\label{eq:maestra_inflaton}
  \left(\partial_\eta^2 + 2\left(\HubbleComovil -
      \frac{\varphi''}{\varphi'}\right)\partial_\eta -
    \triangle +2\left(\partial_\eta\HubbleComovil -
      \HubbleComovil\frac{\varphi''}{\varphi'}\right)\right)
  \psi = 0,
\end{equation}
ecuación que es conocida como la \textit{ecuación maestra}
para la perturbación de la métrica. Nótese que esta
ecuación se pudo haber obtenido sustituyendo los valores
apropiados de $\rho$ y $p$ del campo escalar, la velocidad
del sonido $c_s^2$ y $\tau\delta S$ para el inflatón
\footnote{Que se pueden calcular a partir de sus
  definiciones. Por ejemplo, $c_s^2 \equiv
  \left(\frac{p'}{\rho'}\right)$, entonces $c_s^2(\varphi) =
  \dfrac{1}{3}\left(1 +'\dfrac{2\varphi''}{\HubbleComovil
      \varphi'}\right)$.  De la misma manera para
  $\tau\delta S$, $\dfrac{\kappa}{2}a^2\tau\delta S =
  \left(1-c_s^2(\varphi)\right)\nabla^2\psi$.  }. Este hecho
es de suma importancia, pues nos permite basarnos en la
ecuación (\ref{eq:maestra_hidro}) para trazar la evolución
de las perturbaciones (clásicas) del potencial gravitatorio
en cualquier época.

Los dos términos de la ecuación (\ref{eq:maestra_inflaton})
se pueden interpretar directamente: el segundo término,
proporcional a $\psi'$ se comporta parecido a un término de
fricción y se le conoce como \textit{fricción de Hubble}; el
último término, proporcional a $\psi$ es el que genera la
inestabilidad gravitacional.

Podemos expresar esta ecuación en las variables $u$
(\ref{eq:u_hidro}) y $\xi$ (\ref{eq:integral_maestra_com})
definidas anteriormente en \S
\ref{sec:formacion_estructura}.

Cuando la especie dominante es el inflatón las variables
$u$, $\theta$ son
\begin{equation}
  \label{eq:u_inflaton}
  u \equiv \frac{\psi}{4\pi G \sqrt{\rho+p}} = \frac{a
  }{\varphi'}\psi =
  \frac{2}{3}\frac{a^2\theta}{\HubbleComovil}\sqrt\frac{3}{\kappa}\psi, \quad \theta \equiv
  \frac{1}{a}\left(\frac{\rho}{\rho + p}\right)^{1/2} = \sqrt\frac{3}{\kappa}\frac{\HubbleComovil}{a\varphi'},
\end{equation}
con esta definición se obtiene una ecuación como la
(\ref{eq:oa_u}), pero especializada a inflación
\begin{equation}
  \label{eq:oa_inflacion}
  u'' -\nabla^2 u - \left(\frac{\theta''}{\theta}\right)u = 0.
\end{equation}

No estamos interesados en este trabajo de tesis en el
solución exacta a esta ecuación, por lo que procederemos a
calcularla en el límite asintótico que nos interesa. Como
primer paso transformaremos esta última ecuación al espacio
de Fourier,
\begin{equation}
  \label{eq:oa_inflaton_fourier}
  u_k''+ k^2u_k - \left(
    \frac{\theta''}{\theta}\right)u_k = 0,
\end{equation}

Las soluciones asintóticas de (\ref{eq:oa_inflaton_fourier})
son las siguientes. Para perturbaciones que satisfacen $k
\gg (\theta''/\theta)$ se obtiene
\begin{equation}
  \label{eq:u_inflacion_sol_cortas}
  u \propto e^{\pm i k\eta}.
\end{equation}

Las perturbaciones $\psi$ y $\delta\varphi$ en este limite
presentan un comportamiento oscilatorio: $\psi \propto
\varphi' \times$ términos oscilatorios y $\delta\varphi
\propto \dfrac{k}{a} \times$ términos oscilatorios
\citep[ecuaciones 6.54 y 6.55 de
][pag. 243]{Mukhanov90}. Notemos que debido a que $\psi
\propto \varphi' \simeq -(V_{,\varphi} / 3\kappa V^{1/2})$
para perturbaciones de longitud de onda pequeña, el cambio
en $\psi$ mientras se satisfaga este límite es despreciable,
de hecho, para un potencial $V(\varphi) =
(1/2)m^2\varphi^2$, $\psi \propto m$ es constante. Por su
parte $\delta\varphi$ decrece como $a^{-1}$, por lo que
retrocediendo en el tiempo vemos que habrá un tiempo en el
cual, para un número de onda $k$ su amplitud será tan grande
que la teoría de perturbaciones no será válida.

Para longitudes de onda largas $k \ll \theta''/\theta$ se
puede despreciar el segundo término de la ecuación
(\ref{eq:oa_inflaton_fourier}) y obtener\footnote{El
  procedimiento de obtener la solución para $k^2 \ll
  \theta''/\theta$ es expandir la ecuación en potencias de
  $k^2$ \cite{Martin97}. Haciéndolo se obtiene la expresión
  para $u$ siguiente

  \begin{equation*}
    u = C_1 \theta \int \frac{1}{\theta^2} \left( 1-
      k^2\int^\eta \theta^2
      \int^{\bar{\eta}}\frac{1}{\theta^2}
      d\tilde{\eta}d\bar{\eta} + \mathscr{O}(k^4)\right)
    d\eta + C_2\theta\left(1- k^2\int^\eta \theta^2
      \int^{\bar{\eta}}\frac{1}{\theta^2}
      d\tilde{\eta}d\bar{\eta} + \mathscr{O}(k^4)\right),
  \end{equation*}

  tomando en cuenta sólo el orden significativo se reduce a
  la expresión (\ref{eq:u_largo}).  } la solución aproximada
\begin{equation}
  \label{eq:u_largo}
  u \simeq C_1\theta +
  C_2\theta\int^\eta\frac{d\eta}{\theta^2} .
\end{equation}
$C_1$ es la constante correspondiente al modo decreciente y
$C_2$ es la constante asociada con el modo creciente\footnote{Se
puede entender la elección de estos nombres analizando el
comportamiento de $\theta$ en una época cosmológica con una
especie dominante: el término proporcional a $C_1$  decrece
conforme el factor de escala $a$ crece, mientras que el
término proporcional a $C_2$ crece conforme a $a$.}.

Usando las definiciones (\ref{eq:u_inflaton}) y tomando en
cuenta solo la solución creciente tenemos que la
expresión para el potencial gravitatorio en este límite es
\begin{equation}
  \label{eq:psi_largo_a}
  \psi \simeq
  \frac{3}{2}C_2\frac{\HubbleComovil}{a^2}\int^\eta a^2
  \gamma d\bar\eta = C_2
  \frac{\HubbleComovil}{a^2} \int a^2
  \left(1-\frac{\HubbleComovil'}{\HubbleComovil^2}\right)
  d\eta.
\end{equation}

La expresión (\ref{eq:psi_largo_a}) coincide con
(\ref{eq:maestra_hidro_1}) si $\bm\alpha = (3/2)C_2$. Si en
una etapa cosmológica el factor de escala $a \propto
|\eta|^{1+\beta}$, y es dominado por una especie con
ecuación de estado $p = (\gamma -1)\rho$, $\gamma =
\dfrac{2}{3}\cdot\dfrac{2+\beta}{1+\beta}$. En esta época,
el modo ``creciente'' de la perturbación $\psi$ (afuera del
radio de Hubble) es en realidad constante
\begin{equation}
  \label{eq:psi_largo_etapa}
  \psi \simeq \frac{3}{2}C_2\frac{1+\beta}{3 + 2\beta}
  \gamma = \frac{3}{2}C_2 \frac{3\gamma}{2+3\gamma} .
\end{equation}

Entonces, podemos calcular el razón entre el valor de
$\psi_{inf}$ durante inflación ($\beta_{inf} \simeq 2$) en el
momento que salió del radio de Hubble y el  
valor de $\psi_{mat}$ en el momento cuando  entra al radio de
Hubble durante el periodo dominado por materia ($\beta_{mat}
= 1$),
\begin{equation}
  \label{eq:psi_inf_psi_mat}
  \frac{\psi_{mat}}{\psi_{inf}} \sim \frac{2}{5}
  \frac{1}{\gamma_{inf}}, 
\end{equation}
debido a que $\gamma_{inf} \simeq 0$, las perturbaciones escalares
gravitatorias son magnificadas por un enorme factor entre
estas dos épocas. 

Usando la relación entre $\xi$ y $\psi$
(\ref{eq:integral_maestra_com}) y la constancia respecto al
tiempo de $\psi$ para modos con longitudes de onda mayores
que el radio de Hubble (\ref{eq:psi_inf_psi_mat}) podemos
escribir
\begin{equation}
  \label{eq:psi_xi_2}
  \psi = \frac{3\gamma}{2+3\gamma} \xi,
\end{equation}
si además utilizamos (\ref{eq:modo_escalar_2}) tenemos que
para modos que salieron del radio de Hubble durante inflación
\begin{equation} 
  \label{eq:xi_inf_1}
  \xi_{inf} = \frac{4\pi G}{2\HubbleComovil}
  \left(\frac{2+3\gamma_{inf}}{3\gamma_{inf}} \right)
   \varphi' \dphi',
\end{equation}
la cual se puede escribir, luego de usar las ecuaciones de
movimiento de fondo (\ref{eq:einstein_inflaton_fondo}), como
\begin{equation}
  \label{eq:xi_inf_2}
  \xi_{inf}  = \frac{2 + 3\gamma_{inf}}{2} \cdot \HubbleComovil
  \left(\frac{\dphi}{\varphi'}\right).
\end{equation}
Con esta expresión y usando el hecho de $\xi$ es constante
en el tiempo (cf. pag. \pageref{def:xi}), podemos igualar
los valores de $\xi^k_{inf}$ cuando el modo con número de
onda $k$ sale del radio de Hubble
en $\eta^k_{salida}$ con el valor de $\xi^k_\gamma$ cuando
el modo entra al radio de Hubble en una época con una
especia dominante con ecuación de estado $p =
(\gamma-1)\rho$, i.e.\
$\xi^k_{inf}(\eta^k_{salida})$ $=
\xi^k_\gamma(\eta^k_{entrada})$ y obtendremos
\begin{equation}
  \label{eq:psi_xi_dos_epocas}
  \psi^k_\gamma = \frac{3\gamma}{2+3\gamma}\cdot
  \left[\HubbleComovil \frac{\dphi}{\varphi'} \right]_{\eta =
    \eta^k_{salida}},
\end{equation}
donde el super-índice $k$ indica que esta ecuación es válida
por modo $k$. En particular para el caso de perturbaciones
que entran al radio de Hubble en una época dominada por la
densidad de materia ($\gamma_{mat} = 1$)
\begin{equation}
  \label{eq:psi_xi_materia}
  \psi^k_{mat} = \frac{3}{5} \xi_{inf},
\end{equation}
por lo que sus espectros de potencia estarán relacionados
mediante 
\begin{equation}
  \label{eq:psi_xi_materia_power_spectrum}
  \bm{P}_{\psi_{mat}}(k) = \frac{9}{25}
  \bm{P}_{\xi_{inf}}(k) = \frac{9}{25}   \left[
    \frac{\HubbleComovil}{\varphi'} \right]^2_{\eta =
    \eta^k_{salida}} \overline{\dphi(k) \dphi(k')},
\end{equation}
donde $\overline{\dphi(k) \dphi(k')}$, es la cantidad a
determinar por el modelo de origen de fluctuaciones
primordiales (\S \ref{sec:origen-tradicional} y \S
\ref{sec:origen-colapso}).





\paragraph{Ecuaciones de movimiento en términos de $v$.}Aunque
$u$ es la variable tradicional para presentar la 
evolución de las fluctuaciones, es posible expresar las ECE
(\ref{eq:modo_escalar_inflaton}) y la ecuación de movimiento
de $\delta\phi$  (\ref{eq:mov_escalar_perturbada}) de una
manera más sencilla definiendo dos nuevas variables $v$ y
$z$:
\begin{equation}
  \label{eq:variables_mukhanov}
  v = a\delta\varphi + z\psi, \quad z =
  \frac{a}{\HubbleComovil} \varphi,
\end{equation}
$v$ es conocida como la variable de Mukhanov-Sasaki
\cite{Mukhanov88}. Nótese que la
variable $z$ está relacionada con la $\theta$
(\ref{eq:u_inflaton}) mediante $z = (1/\theta)$. Usando
estas variables el conjunto
de ecuaciones (\ref{eq:modo_escalar_inflaton}) y
(\ref{eq:mov_escalar_perturbada}) se puede escribir como
\begin{equation}\label{eq:vms_1}
  v'' - \nabla^2v - \frac{z''}{z} v = 0,
\end{equation}
\begin{equation}\label{eq:vms_2}
  \nabla^2\psi =
  \frac{\kappa}{2}\frac{\HubbleComovil}{a^2}\left(zv' - z'v\right),
\end{equation}
\begin{equation}\label{eq:vms_3}
  \left(\frac{a^2\psi}{\HubbleComovil}\right)' =
  \frac{\kappa^2}{2}zv.   
\end{equation}

Esta nueva manera de escribir las ECE del sistema
Einstein-Inflatón es la utilizada en la explicación
estándar del origen de las perturbaciones cosmológicas en el
paradigma inflacionario (cf.\ \S
\ref{sec:origen-tradicional}).

Como hicimos en el caso hidrodinámico
 (\S \ref{sec:formacion_estructura}), también podemos usar $\xi$
(\ref{eq:integral_maestra_com}) para describir la evolución
de las perturbaciones. En las variables $v$ y $z$ , $\xi$ se
escribe como $\xi = \dfrac{v}{z}$. Obviamente se obtendrán
los mismos resultados que con las otras variables. Para
reproducirlos se utilizará la constancia de $\xi$ respecto
al tiempo conforme, igualando
su valor cuando el modo alcanza  el radio de
Hubble durante inflación $\xi^k_{salida}$ y su valor cuando
reentra al radio de Hubble i.e.\ , $\xi^k_{salida} =
\xi^k_{entrada}$. En particular reproduciremos las relaciones
(\ref{eq:psi_inf_psi_mat}), (\ref{eq:psi_xi_dos_epocas}) y
(\ref{eq:psi_xi_materia_power_spectrum}). 

Conociendo las ecuaciones dinámicas de $u$ (ó $v$, ó $\xi$),
$\delta\varphi$ y $\psi$ podemos ver como las fluctuaciones
\textit{clásicas} evolucionan hasta convertirse en las
condiciones iniciales para las perturbaciones de DM,
radiación y bariones de \S \ref{sec:formacion_estructura}
que darán origen a la estructura que 
observamos actualmente. Ahora debemos resolver la pregunta
¿Cómo surgen estas fluctuaciones?. La solución propuesta en
el paradigma inflacionario es: \textit{las fluctuaciones
  gravitacionales son provocadas por ``fluctuaciones''
  cuánticas del campo del inflatón.}  Esta solución generará
nuevos problemas, que se tratarán en la siguiente sección.

\section{\label{sec:origen-tradicional}Orígen de las
  semillas cosmológicas: Enfoque tradicional}

El modelo inflacionario propone (y quizá este sea su mayor
mérito) una mecanismo para el origen de las fluctuaciones
primordiales: las perturbaciones primordiales del campo
$\varphi$ son generadas por las ``fluctuaciones''\footnote{
  Entrecomillamos la palabra ``fluctuaciones'' ya coloca, en
  el mismo estatus conceptual, a las perturbaciones o
  fluctuaciones clásicas y a las incertidumbres
  cuánticas. La palabra ``fluctuaciones'' pareciera invocar
  que ``algo''(¿El campo cuántico?) ``fluctúa'' en el
  sentido de algún proceso estocástico e.g.\ como el
  movimiento Browniano. Debido a esto consideramos
  desafortunado la generalización de este término en la
  comunidad cosmológica.} cuánticas dentro del radio de
Hubble --siendo así un mecanismo causal-- durante el periodo
inflacionario.

Podemos modelar matemáticamente (para ver el cálculo
detallado se invita al lector a consultar el apéndice
\ref{sec:cuant-mukhanov}) estas fluctuaciones cuánticas,
promoviendo a la perturbación $\delta\varphi$ a un campo
cuántico $\widehat\delta\varphi$. El proceso de cuantización
parte de la acción del sistema Einstein-inflatón
\cite{Mukhanov90,Mukhanov88}:
\begin{equation}
  \label{eq:accion_inflaton_einstein}
  S[\varphi, g_{ab}] = S_{grav} + S_{mat} = \int \left\{ R
    \sqrt{-g} \right\}d^4x + \int 
  \left\{\left[-\frac{1}{2}g^{ab}\nabla_a\varphi\nabla_b\varphi -
      V(\varphi)\right] \sqrt{-g}\right\}d^4x
\end{equation}

Como buscamos las ecuaciones de movimiento de la
perturbación a primer orden, necesitamos expandir la acción
a segundo orden en las variables métricas ($\psi$, $\phi$) y
en el campo ($\dphi$) alrededor de los campos de fondo. La
acción de la materia perturbada a segundo orden es
\begin{multline}\label{eq:accion_perturbada_2_c}
  \delta S^{(2)}_{mat} = \int \Big[\frac{a^2}{2}
  \Big(\delta\varphi'\,^2 - 4\delta\varphi'\phi\varphi' +
  4\phi^2\varphi'\,^2 -
  \delta\varphi_{,i}\delta\varphi_{,i} \\
  + 2\delta\varphi'\phi\varphi' -
  \frac{1}{2}\varphi'\,^2\phi^2 - 3\varphi'\,^2\psi -
  6\delta\varphi'\psi \varphi' - 3\varphi'\,^2\psi\phi +
  \frac{3}{2}\varphi'\,^2\psi^2\Big) \\
  - \frac{a^4}{2}\delta\varphi^2V_{\varphi\varphi} -
  a^4\delta\varphi\phi V + \frac{a^4}{2} \phi^2 V +
  3a^4\delta\varphi\psi V_\varphi + 3a^4 \phi\psi V -
  \frac{3a^4}{2}\psi^2 V \Big] d\,^4x
\end{multline}

Variando $\delta S^{(2)}_{mat}$
\eqref{eq:accion_perturbada_2_c} y $\delta S^{(2)}_{grav}$
(cf.\ ecuación 10.11 en \cite{Mukhanov90}, pag. 262)
respecto a $\psi$, $\phi$ y $\delta\varphi$, obtenemos
ecuaciones de movimiento de este sistema. Podemos obtener un
conjunto equivalente de ecuaciones si aplicamos las
constricciones (\ref{eq:constriccion_fluido_perfecto}) y
(\ref{eq:modo_escalar_2}) para eliminar dos de las tres
variables en la acción a segundo orden $\delta
S^{(2)}_{grav\, + \,mat}$ y usamos una vez más las
ecuaciones de la métrica de fondo, al final obtendremos la
acción para un sólo grado de libertad, la llamada variable
de Mukhanov-Sasaki \cite{Mukhanov88} $v$,
\begin{equation}\label{eq:vms_accion}
  \delta S^{(2)}_{grav\, + \,mat} = \frac{1}{2}\int d\,^4x\Big[v'\,^2 -
  (\triangle v)^2 + \frac{z''}{z}v^2\Big],
\end{equation}
con $z$ y $v$ definidas como antes
\begin{eqnref}{eq:variables_mukhanov}
  v \equiv a\left(\delta\varphi +
    \frac{\varphi'}{\Hubble}\psi\right), \quad z \equiv
  \frac{a\varphi'}{\HubbleComovil} .
\end{eqnref}

La ecuación de movimiento para $v$ se obtiene variando la
acción (\ref{eq:vms_accion}) respecto a $v$. Esta ecuación
tiene la forma de la ecuación de un oscilador armónico con
masa variable $m_v = z''/z$
\begin{equation}\label{eq:vms_evolucion}
  v'' - \triangle v - \frac{z''}{z} v = 0.
\end{equation}

La cuantización de la teoría definida por la acción
(\ref{eq:vms_accion}) prosigue de manera estándar, primero
calculamos el momento canónico conjugado de la teoría
\begin{equation}
  \label{eq:vms_pi}
  \piv (\eta, \bvec{x}) = \dpar{\mathcal{L}^{(v)}}{v'} = v'(\eta, \bvec{x}).
\end{equation}

El hamiltoniano se encuentra de manera inmediata
\begin{equation}
  \label{eq:vms_hamiltoniano}
  H = \int\left(v' \piv - \mathcal{L}^{(v)}\right)
  d^3x = \frac{1}{2}\int \left(\left(\piv\right)^2 +
    c_s^2\delta^{ij}v_{,i}v_{,j} - \frac{z''}{z} v^2\right) d^3x.
\end{equation}

El paso siguiente en el proceso de cuantización es promover
a las variable $v$ y $\piv$ a los operadores $\vms$ y
$\pivo$ e imponer que estos operadores cumplan con las
relaciones de conmutación estándar
\begin{equation}
  \label{eq:vms_conmutacion}
  \left[\vms(\eta, \bvec{x}), \vms(\eta,\bvec{x}')\right] =
  \left[\pivo(\eta, \bvec{x}), \pivo(\eta,\bvec{x}')\right] =
  0, \quad   \left[\vms(\eta, \bvec{x}),
    \pivo(\eta,\bvec{x}')\right] = i\hbar\delta(\bvec{x} -
  \bvec{x}'). 
\end{equation}

La ecuación de movimiento para $\vms$ es idéntica a
(\ref{eq:vms_evolucion}) cambiando $v \to \vms$ y se obtiene
de las ecuaciones de movimiento de Heisenberg
\begin{equation}
  \label{eq:vms_heisenberg}
  i\vms = [\vms, \hat{H}], \quad i\pivo = [\pivo, \hat{H}],
\end{equation}

donde $\hat{H}$ es el hamiltoniano $H$
(\ref{eq:vms_hamiltoniano}) escrito en términos de los
operadores $\vms$ y $\pivo$.

Descomponemos al operador $\vms$ en ondas planas
\begin{equation}
  \label{eq:vms_descomposicion_ondas}
  \vms = \int\frac{d^3k}{(2\pi)^{3/2}}\left[v_k(\eta) \ann_k
    \eikx + v_k^*(\eta)\cre_k\emikx \right]
\end{equation}

donde los operadores de creación ($\cre_k$) y aniquilación
($\ann_k$) satisfacen con las propiedades de conmutación
estándar
\begin{equation}
  \label{eq:vms_a_conmutacion}
  \left[\cre_k, \ann_{k'}\right] = \delta^{(3)}(\bvec{k} -
  \bvec{k}'), \quad \left[\cre_k,\cre_{k'}\right] =
  \left[\ann_k,\ann_{k'}\right] = 0,
\end{equation}
lo cual se puede verificar sustituyendo
(\ref{eq:vms_descomposicion_ondas}) en
(\ref{eq:vms_conmutacion}). Para que las relaciones de
conmutación (\ref{eq:vms_conmutacion}) y
(\ref{eq:vms_a_conmutacion}) sean consistentes entre sí, la
relación de normalización
\begin{equation}
  \label{eq:vms_normalizacion}
  v'_k(\eta)v_k^*(\eta) - v_k^*(\eta)v'_k(\eta) = i\hbar
\end{equation}
debe ser satisfecha. Para cada número de onda $k$, el modo
$v_k(\eta)$ cumple con la ecuación de evolución
\begin{equation}
  \label{eq:vms_modo_evolucion}
  v_k'' + \left(k^2 - \frac{z''}{z}\right) v_k = 0,
\end{equation}
la cual es la ecuación de un oscilador armónico con masa
variable, justo como se mencionó arriba.

Ahora debemos definir el estado de vacío como el estado que
es aniquilado por los operadores $\ann_k:$ $\ann_k\ket{0} =
0,  \, \forall k$. Como último paso de la cuantización de $\vms$ se deben
de fijar las constantes de normalización de los modos
$v_k(\eta)$.  Para esto elegiremos los modos $v_k(\eta)$ de
tal manera que para longitudes muy pequeñas (dentro del
radio de Hubble $\dfrac{k}{aH} \to \infty$) se aproximen a
las ondas planas de frecuencia positiva del espacio de
Minkowski, i.e.\ , elegimos en este límite el vacío de
Minkowski. A esta elección de vacío se le conoce como vacío
de \textit{Bunch-Davies} \cite{Birrel94,Mukhanov07}. Para el
caso de longitudes grandes $\left(\dfrac{k}{aH} \ll
  1\right)$ se puede despreciar el término que va como $k^2$
en (\ref{eq:vms_modo_evolucion}), entonces
\begin{equation}
  \label{eq:vms_asintotico}
  v_k(\eta) \propto \begin{cases}
    \dfrac{1}{\sqrt{2k}}e^{-ik\eta}, \quad \frac{k}{aH} \gg 1
    \\
    z, \quad \dfrac{k}{aH} \ll 1
  \end{cases}
\end{equation}

Como era de esperarse, para perturbaciones más grandes que
el radio de Hubble, $\xi = (v/z)$ es constante.

El siguiente paso es encontrar las soluciones de las
ecuaciones (\ref{eq:vms_modo_evolucion}), y la solución
queremos expresarla en función de variables que nos permitan
discriminar entre diferentes modelos
inflacionarios. Estos tendrán
diferentes predicciones y lo reflejarán en las variables de
\textit{slow-roll} (cf.\ \ref{eq:parametros_sra}), aunque en
este trabajo de tesis no estamos estudiando modelos
particulares de inflación, presentaremos los cálculos como
aparecen en la literatura. Primero encontraremos unas
relaciones de utilidad. La relación entre el tiempo
cosmológico $t$ y el tiempo conforme $\eta$ se escribió en
(\ref{eq:conforme_cosmologico}), si usamos las variables de
\textit{slow-roll} y suponemos que $\bm\epsilon$ es
constante, tenemos
\begin{equation}
  \label{eq:hubble_cosmologico_conforme}
  \eta = \int\frac{dt}{a(t)} = -\frac{1}{aH} +
  \bm\epsilon\int\frac{da}{a^2H}, \then aH = -
  \frac{1}{\eta(1-\bm\epsilon)}
\end{equation}
donde en la segunda expresión se usó el hecho que $\int dt/a
= \int da/(a^2H)$. Por otro lado, la \textit{masa variable}
en la ecuación (\ref{eq:vms_modo_evolucion}) se pude
escribir como
\begin{equation}
  \label{eq:diff2z_z}
  \frac{z''}{z} = \left(\frac{z'}{z}\right)^2 +
  \left(\frac{z'}{z}\right)'  = a^2\left[
    \left(\frac{\dot{z}}{z}\right)^2 + H \frac{\dot{z}}{z} +
    \dpar{}{t}\left(\frac{\dot{z}}{z}\right)\right].
\end{equation}
En la segunda igualdad se usó la definición de la variable 
$z = \dfrac{a\varphi'}{\HubbleComovil} =
\dfrac{a\dot{\varphi}}{H}$. Usando $\dfrac{\dot{z}}{z} = H +
\dfrac{\ddot{\varphi}}{\varphi} - \dfrac{\dot{H}}{H}$ y la
segunda expresión de 
(\ref{eq:hubble_cosmologico_conforme}) en la
última expresión se llega a
\begin{equation}
  \label{eq:diff2z_z_sra}
  \frac{z''}{z} = \frac{2 + 2\bm\epsilon - 3\bm\delta -
    \bm\epsilon\bm\delta + \bm\delta^2}{\eta^2 (1+\bm\epsilon)}.
\end{equation}

Es conveniente ahora expresar a $z''/z$ como
\begin{equation}
  \label{eq:diff2z_z_nu}
  \frac{z''}{z} = \frac{1}{\eta^2} \left(\nu^2 -
    \frac{1}{4}\right), \quad \text{con} \quad  \nu \equiv \frac{3
    + \bm\epsilon - 2\bm\delta}{2(1+\bm\epsilon)},
\end{equation} 
expresada de esta manera, podemos escribir la solución
\cite{Massimo08} de la
ecuación (\ref{eq:vms_modo_evolucion}) como
\begin{equation}
  \label{eq:solucion_v_k}
  v_k(\eta) = \frac{\mathcal{N}}{\sqrt{2k}} \sqrt{-k\eta}
  H_v^{(1)}(-k\eta),  \quad \mathcal{N} = \sqrt\frac{\pi}{2}
  e^{i\pi (2\nu +1)/4},
\end{equation}
donde $H_\nu^{(1)}(x) = J_\nu(x) + i Y_\nu(x)$ es la función
de Hankel de primer tipo o función de Bessel de tercer tipo
(cf.\ \citep[][capítulo 9]{Abramowitz65}).

Las expresiones para $\psi$ y $\delta\varphi$ se pueden
obtener sustituyendo la solución recién encontrada en
(\ref{eq:vms_3}), que como es de esperarse son las mismas
que las encontradas con la variable $u$ en
(\ref{eq:psi_largo_a}).

El primer problema al que nos enfrentamos ahora es ¿Cómo
obtenemos un campo $c-$\textit{número} $\delta\varphi$ de un
campo cuántico escalar $\widehat{\delta\varphi}$
\cite{Padmanabhan06} para poder utilizarlo en
(\ref{eq:psi_xi_materia_power_spectrum}) (o lo que es lo
mismo como extraemos $v$ de $\hat{v}$?) La respuesta
estándar es que la correlación de dos puntos del estado de
vacío del campo cuántico $\widehat{\delta\varphi}$ es el
campo $c-$\textit{número} buscado. Es decir la relación de
identificación siguiente es válida
\begin{equation}
  \label{eq:suposicion_estandar}
  \braket{\vms(\eta, \bvec{x})\vms(\eta,\bvec{y})}{0} \equiv
  \overline{v(\eta, \bvec{x})v(\eta,\bvec{y})}
\end{equation}

Este paso se justifica diciendo que al salir del radio de
Hubble, las amplitudes de las perturbaciones se ``congelan''
(cf.\ (\ref{eq:vms_asintotico})) y esto marca su transición
a lo clásico. Existen explicaciones más elaboradas, como
Decoherencia, pero básicamente
dicen lo mismo: \textit{es posible identificar las
  fluctuaciones cuánticas con fluctuaciones
  estadísticas}. Es importante notar que, aún fuera del
contexto cosmológico esta afirmación tiene una validez
cuestionable
\cite{PearlePhil94,Adler03,Zurek03,Schlosshauer04,
  Ghirardi09}.

Completemos aquí la receta de la obtención del espectro de
potencias cuántico usando para esto la cantidad
$\xi$. Nótese que debido a la cuantización de $\vms$, la
cuantización de $\xi = v/z$ ya fue realizada
automáticamente. El espectro de potencias está definido
mediante
\begin{equation}
  \label{eq:power_spectrum_def_xi}
  \frac{2\pi^2}{k^3}\bm{P}_{\xi}(k)\delta^{(3)}(\bvec{k} -
  \bvec{k}') \equiv \braket{\hat\xi_k\hat\xi^\dag_{k'}}{0} ,
  \quad \bm{P}_{\xi} = \frac{k^3}{2\pi^2z^2}|v_k(\eta)|^2
\end{equation}
En esta expresión utilizamos la suposición
(\ref{eq:suposicion_estandar}) para ligar las fluctuaciones
clásicas con las cuánticas. El módulo al cuadrado del modo
$v_k(\eta)$ es
\begin{equation}
  \label{eq:v_k_cuadrado}
  |v_k(\eta)|^2 =
  \frac{|\mathcal{N}|^2}{2k}(-k\eta)H_\nu^{(1)}(-k\eta)
  H_{\nu}^{(2)}(-k\eta)  \simeq \frac{\Gamma^2(\nu)}{4\pi
    k}2^{2\nu} (-k\eta)^{1-2\nu}
\end{equation}
donde hemos utilizado la aproximación de argumento
pequeño\footnote{Esta aproximación implica modos con
  longitud mayor al horizonte, ya que $k\eta \simeq
  \dfrac{k}{aH} = \dfrac{k}{\HubbleComovil}$.} de
las funciones de Hankel (cf.\ 9.1.9 de \cite{Abramowitz65},
pag. 360),
\begin{equation}
  \label{eq:hankel_small}
  H_\nu^{(1)} (-x) \simeq \frac{-i}{\pi}
  \Gamma(\nu)\left(\frac{-x}{2}\right)^{-\nu}, \quad |x| \ll 1.
\end{equation}

Entonces el espectro de potencias de $\xi$ es
\begin{equation}
  \label{eq:xi_v_power_spectrum_1}
  \bm{P}_\xi(k) = \frac{2^{2\nu - 3}}{\pi} \Gamma^2(\nu)
  (1-\bm\epsilon)^{1-2\nu}
  \left(\frac{k}{aH}\right)^{3-2\nu}
  \left(\frac{H^2}{\dot\varphi}\right)^2,
\end{equation}
expandiendo $v$ en el límite cuando $\bm\epsilon \ll 1$,
$\bm\delta \ll 1$, $\nu \simeq 3/2 + 2\bm\epsilon +
\bm\delta + \mathcal{O}(\bm\epsilon^2)$, por lo que a orden
más bajo $\nu \simeq 3/2$, además evaluando en el momento
cuando el modo sale del radio de Hubble, i.e.\ $k \simeq
aH$,
\begin{equation}
  \label{eq:xi_v_power_spectrum_2}
  \bm{P}_\xi(k) = \frac{1}{4\pi^2}
  \left(\frac{H^2}{\dot\varphi}\right)^2_{k\thinspace\simeq\thinspace
    aH}.   
\end{equation}

Finalmente debemos utilizar una vez más las variables de
\textit{slow-roll} (\ref{eq:parametros_sra}) y la ecuación
auxiliar (\ref{eq:aux_escalar}) para expresar $\dot\varphi$,
\begin{equation}
  \label{eq:xi_v_power_spectrum_3}
  \bm{P}_\xi(k) =
  \frac{\kappa^2}{24\pi^2}
  \left(\frac{V}
    {\bm\epsilon}\right)_{k\thinspace\simeq\thinspace aH}.    
\end{equation}



El espectro de potencias de $\psi$ se obtiene
usando  la ecuación  (\ref{eq:psi_xi_materia_power_spectrum}),
como era de esperarse, estas ecuaciones concuerdan con las
de la \S \ref{sec:evol-clasica}.


\section{\label{sec:origen-colapso}Orígen de las semillas
  cosmológicas: Hipótesis del colapso}

La explicación recién dada para el origen de las
perturbaciones cosmológicas depende de una paso ``confuso'':
no hay una justificación \textit{a priori} para la
identificación de las correlaciones de dos puntos cuánticas
con el espectro de potencias clásico (ecuación
(\ref{eq:suposicion_estandar}). Analizando más
detalladamente, nos enfrentamos a un problema
conceptualmente más profundo en la explicación estándar:
\textit{el análisis estándar empieza en un estado que es
  homogéneo e isotrópico (el estado de vacío) y termina con
  inhomogeneidades que están de acuerdo con las
  observaciones}. Este comportamiento no encuentra su
justificación en la mecánica cuántica estándar ya que la
dinámica del sistema preserva las simetrías iniciales. Estos
problemas ocurren también en la aplicación diaria en
experimentos en el laboratorio y son englobados en el \textit{problema de la
  medición de la mecánica cuántica}
\cite{againstmeasur,Bell}. Obviamente estos 
problemas han sido señalados por algunos miembros de la
comunidad cosmológica
\cite{Padmanabhan96,Padmanabhan06,Liddle00,Mukhanov2005} y
se han propuesto algunas soluciones basadas en las
planteado de solución del mencionado problema de la medición.  La
principal propuesta 
(si tomamos como indicador su popularidad en la
literatura científica) es la basada en  Decoherencia
\cite{Kiefer98a,Kiefer98b,Kiefer00a,Kiefer08,Feldman92},
aunque hay que mencionar que existen otras propuestas con
menos adherentes, pero aún significativas como las basadas
en \textit{Mundos Múltiples}. Estos intentos de solución
comparten un mismo problema: intentan justificar el primer
cuestionamiento (igualar correlaciones de dos puntos
cuánticas con espectros de potencias clásicos) dejando
intacto el segundo (la evolución de mecánica cuántica es
unitaria y en el caso inflacionario, esta evolución preserva
las simetrías del estado inicial).

Nos referiremos a ambos problemas como el
\textit{problema de la transición cuántica a clásica} de las
fluctuaciones cuánticas durante la inflación. Este problema en
el contexto de Fundamentos de Mecánica Cuántica es conocido
como \textit{problema de la medición} o  \textit{problema de la
  macro-objetificación}.

La raíz del problema de la macro-objetificación está en que
la mecánica cuántica tiene como rector dinámico a la
ecuación de Schrödinger, la cual establece una evolución
lineal, determinista y unitaria (a esta evolución se le
conoce como proceso \textbf{U} en \cite{Penrose02}),
evolución que preserva la superposición, y si no hay
fuentes de asimetrías externas, preserva también las
simetrías del sistema. Para recuperar el mundo como lo observamos, se
invoca al proceso \textbf{R} (\textit{ibid.}) el cual
``elige'' un eigenvalor de la función de onda $\ket{\Psi}
\to \ket{\psi_i}$, así, el proceso \textbf{R} es no-lineal
(rompe la superposición) y estocástico (genera o explica la
regla de probabilidad de Born). El problema de la medición se
puede reducir a cual es la física que explica al proceso
\textbf{R}, o si este proceso es siquiera necesario para una
explicación completa de la realidad \cite{Penrose94}.

En el escenario cosmológico, la solución al problema de la
macro-objetificación provista por la escuela de Copenhagen
\cite{Kiefer02,Grib99} , i.e.\ , \textit{colapso de la función
  de onda provocado por un observador al realizar la
  medición}, no es aplicable, ni siquiera en principio, ya
que hay varios elementos de la interpretación ortodoxa que
no están presentes: (a) ¿Quién realiza la medición?, (b)
¿Cuál es el conjunto de observables que se van a medir?
¿Quién decidió que justamente esos observables eran los que
se tenían que medir? y (c) ¿Cuándo es realizada la medición?
Claramente la interpretación ortodoxa es insostenible, a
menos que se defienda la posición de que \textit{la medición
  que desencadenó la reducción de la función de onda} la
realizó el COBE en 1992 y haya una especie de retro-causalidad,
que genere las perturbaciones clásicas primordiales que, al
evolucionar hacia adelante en el tiempo generarán cúmulos,
galaxias, estrellas, planetas y finalmente nosotros
\footnote{Aunque existe una interpretación de la mecánica
  cuántica con estas características, la interpretación
  transaccional de \citeauthor*{Cramer}, \cite{Cramer}. Esta
idea aplica la idea del teoría del absorbente
electromagnético de Wheeler-Feynman a la mecánica cuántica.}.

Con este problema en mente se puede apreciar que la
aseveración de que existe una predicción hecha por el modelo
inflacionario respecto a las anisotropías y las inhomogeneidades
primordiales de la densidad, no puede ser sostenida como
satisfactoria mientras el problema de la medición no sea
resuelto.

El problema de la medición en este contexto cosmológico
muestra en su desnudez las interpretaciones de la mecánica
cuántica tradicionales. Los enfoques tradicionales invocan
al observador, a la medición, al ambiente o al límite entre
el ``macro-mundo'' (esencialmente clásico) y el
``micro-mundo'' (completamente cuántico). Estas
explicaciones que invocan a algún actor o proceso sin
definición clara --o fuera-- de la teoría cuántica para
darle sentido a el mundo observado se topan con el problema
de marcar ``dónde'' hay que localizar la frontera entre los dos
tipos de mundos, ya que la teoría se queda en silencio en
este punto. Todas las soluciones de este tipo invocan a un
mecanismo parecido a una medición: la consciencia del
observador, el ambiente, el aparato macroscópico de medición
que perturba al sistema cuántico, etc.\ que de alguna manera
``escapa'' a la descripción mecánico-cuántica. Estas
soluciones al problema de la medición se ven confrontadas
con la siguiente cuestión: ¿La Mecánica Cuántica describe
todos los procesos físicos (i.e.\ ¿el universo es
esencialmente cuántico?) ? Si su respuesta es afirmativa, no
es posible resolver el problema de la macro-objetificación
como demuestran el argumento sencillo de
\citeauthor*{VonNeumann32} \cite{VonNeumann32} o la versión
no idealizada de \citeauthor*{Bassi2000} que se puede
encontrar en \cite{Bassi2000}. Ambos argumentos están
basados en la linealidad de la evolución dictada por la
ecuación de Schrödinger. Si su respuesta es negativa,
necesitan extender la física conocida. 

John S. Bell escribió \cite{BellJumps} que la única manera de resolver el
problema de la macro-objetificación es:

\begin{quote}
  \textit{La función de onda, descrita por la ecuación de
  Schrödinger, o no es todo, o no es
  correcta.}
\end{quote}

En el artículo \cite{Sudarsky06a}, \citeauthor*{Sudarsky06a}
proponen como solución al problema del origen de las
perturbaciones cosmológicas, la \textit{hipótesis del
  auto-colapso} siguiendo ciertas ideas de Roger Penrose
sobre el papel de gravedad para resolver el problema de la
macro-objetificación
\cite{Penrose96,Penrose02,Penrose05,Penrose94}. Esta
hipótesis supone que existe un mecanismo para producir la
evolución de tipo \textbf{R} y que este mecanismo involucra
\textit{nueva física},  i.e.\ , existe un mecanismo responsable
de la transición del estado de vacío homogéneo e isotrópico
a un nuevo estado que contiene las fluctuaciones que son
responsables de que existan desviaciones de la homogeneidad
e isotropía en el universo tipo Friedmann-Lemaître (cf.\ \S
\ref{sec:universo-fl}). A este mecanismo se le conoce como
\textit{colapso}. Entonces, podemos resumir los pasos esta
propuesta como sigue: (1) el colapso es especificado de una
manera puramente fenomenológica (para diferenciarlo de los
mecanismos de colapso, a estas descripciones fenomenológicas
se les denominará \textit{esquemas de colapso}), (2) el
universo será descrito por las ecuaciones de fondo hasta que
``la nueva física'' dispare el colapso (esta nueva física se
especula --siguiendo a Penrose-- que sea de origen
gravitatorio) y (3) es en ese momento cuando aparecen
perturbaciones métricas, debidas al acople originado por las
ecuaciones semi-clásicas\footnote{Nótese que estas
  ecuaciones son válidas salvo en el momento en el que
  ocurre el colapso.} de Einstein $G_{ab} = 8\pi G
\expec{T_{ab}}$.

Además del problema \textit{de la medición} o \textit{de la
  transición cuántica-clásica} que se podría clasificar como 
fundamental, existen otras partes dudosas en la
prescripción tradicional, también mencionados en
\cite{Sudarsky06a} relacionados con 
las cuestiones estadísticas de la prescripción estándar
respecto a valores de expectación e incertidumbres,
fluctuaciones y promedios, que son tratados como si no
existieran diferencias entre ellos.

\subsection{Formalismo de la hipótesis de
  colapso}\label{sec:modelado-matematico-colapso}  

\paragraph{Ecuaciones clásicas fundamentales.} A
continuación procederemos a establecer el formalismo
matemático de la hipótesis de colapso. Al igual que en el
tratamiento estándar, el punto de partida es la acción un
campo escalar acoplado mínimamente a la gravedad,
\begin{eqnref}{eq:accion_inflaton_einstein}
  S[\varphi, g_{ab}] = \int d^4x\sqrt{-g}\,\Big( \frac{1}{16\pi
    G} R[g_{ab}]- \\ \frac{1}{2}\nabla_a\varphi\nabla_b\varphi
  g^{ab} -V(\varphi)\Big).
\end{eqnref}
Recordemos que elegimos la norma Newtoniana Conforme
(\ref{eq:newton-conforme}) para realizar todas nuestras
descripciones físicas. Perturbando a segundo orden la
acción, llegaremos a las ecuaciones lineales
(\ref{eq:modo_escalar_inflaton}) y
(\ref{eq:mov_escalar_perturbada}).

Sustituyendo el valor de $\psi'$ (\ref{eq:modo_escalar_2_})
en (\ref{eq:modo_escalar_1_}) y acomodando términos obtenemos
una ecuación generalizada de Poisson para $\nabla^2\psi$

\begin{equation}
  \label{eq:classical_fundamental}
  \nabla^2\psi - \bm\mu \psi = 4\pi G\left(\bm\nu\delta\varphi +
    \varphi'\delta\varphi'\right) 
\end{equation}  

donde $\bm\mu \equiv \left(\HubbleComovil^2 -
  \HubbleComovil'\right)$ y $\bm\nu \equiv
\left(3\HubbleComovil \varphi' + a^2\partial_\varphi
  V\right) = \left(\HubbleComovil \varphi' -
  \varphi''\right)$. En esta última igualdad se usó la
ecuación de movimiento de fondo del campo escalar
\eqref{eq:movimiento_escalar_comovil}.

La ecuación (\ref{eq:classical_fundamental}) será nuestra
ecuación fundamental, ya que ella muestra lo que la
hipótesis de colapso considera: las fluctuaciones del campo escalar son
``fuente'' de las perturbaciones gravitatorias. Durante la
época inflacionaria $\bm\mu \simeq 0$ y $\bm\nu \simeq 0$. La
dinámica de la perturbación del inflatón está dada por la
ecuación (\ref{eq:mov_escalar_perturbada}), la cual se puede
escribir, luego de sustituir $\psi'$ usando
\eqref{eq:modo_escalar_2_}, como
\begin{equation}
  \label{eq:mov_escalar_perturbada_2}
  \delta\varphi'' + 2\HubbleComovil\delta\varphi' -
  \nabla^2\delta\varphi + a^2\partial_{\varphi\varphi}^2V
  \delta\varphi - 16\pi G\left(\varphi'\right)^2 -
  2\psi\HubbleComovil\varphi' = 0.
\end{equation}

En esta última ecuación los términos proporcionales al
potencial $\psi$ reflejan cómo la métrica responde
(\textit{back-reacts}) a las perturbaciones del
inflatón. Estos términos están suprimidos por un
factor de $G$ y por lo tanto los despreciaremos \todo{Cita a
  Mukhanov y sus discusión sobre el peligro de hacer
  esto}. Desde el punto de vista de la hipótesis del
colapso, esto no es tan grave, ya que la métrica de fondo no
es perturbada hasta que el estado del campo escalar
perturbado colapsa; sólo después de este evento $\psi$
aparecerá y podrá afectar la evolución de las perturbaciones
del inflatón (e.g.\ , es netamente un efecto de segundo
orden). Así, la ecuación fundamental para la evolución de
las perturbaciones del campo escalar es
\begin{equation}
  \label{eq:mov_escalar_perturbada_2}
  \delta\varphi'' + 2\HubbleComovil\delta\varphi' -
  \nabla^2\delta\varphi + a^2\partial_{\varphi\varphi}^2V
  \delta\varphi = 0.
\end{equation}

\paragraph{Aproximación de \textbf{slow-roll}.}

Durante la época inflacionaria las condiciones de
\textit{slow-roll}~(\ref{eq:condiciones_inflacion_3})  son
válidas, entonces en la ecuación
(\ref{eq:classical_fundamental}) $\bm\nu = \bm\mu = 0$ y en
la ecuación (\ref{eq:mov_escalar_perturbada_2})
$a^2\partial_{\varphi\varphi}V = 0$.

Así, la ecuación de Poisson (\ref{eq:classical_fundamental})
durante el \textit{slow-roll} se escribe como
\begin{equation}
  \label{eq:classical_fundamental_sra}
  \nabla^2\psi = 4\pi G \varphi' \delta\varphi',
\end{equation}
y la ecuación de movimiento de la perturbación del inflatón
(\ref{eq:mov_escalar_perturbada_2}),
\begin{equation}
  \label{eq:mov_escalar_perturbado_sra}
  \delta\varphi'' + 2\HubbleComovil\delta\varphi' -
  \nabla^2\delta\varphi = 0.
\end{equation}

Antes de cuantizar, es conveniente definir una nueva
variable $y$ como sigue $y= a\delta\varphi$. La ecuación
(\ref{eq:mov_escalar_perturbado_sra}) escrita con esta nueva
variable, en el espacio de Fourier es
\begin{equation}
  \label{eq:mov_y}
  y''- \left(\nabla^2  + \frac{a''}{a}\right)y = 0.
\end{equation}

Obviamente la cuantización $\y$ de $y$, implica la
cuantización $\dphih = \dfrac{1}{a}\y$ de $\dphi$.

\paragraph{Cuantizacióń del campo escalar $\y$} El siguiente
paso es cuantizar la parte fluctuante del inflatón. Como
dijimos antes es conveniente trabajar en el campo reescalado
$y=a\dphi$.

Haciendo esta sustitución en la acción del inflatón e
ignorando los términos del potencial ya que son
despreciables durante \textit{slow-roll}, tenemos
\footnote{Obsérvese que esta acción contiene un término de
  frontera. El término de frontera es $\partial_\eta
  \left(\HubbleComovil y^2\right)$. Usándolo la acción
  quedaría
  \begin{equation*}
    \delta S^{(2)}_y = \frac{1}{2}\int d\eta dx^3 \left[ 
      y'^2 -(\nabla y)^2 \frac{a''}{a}y^2\right].
  \end{equation*}
  Esto cambiará algunas definiciones correspondientes al
  momento conjugado, pero no afectará la física, como se
  puede comprobar fácilmente. En particular, lo que se gana
  de simplicidad en la expresión del momento conjugado, se
  pierde en la simplicidad de $g_k(\eta)$.  }

\begin{equation}
  \label{eq:accion_y}
  \delta S^{(2)}_y = \frac{1}{2}\int d\eta dx^3 \left[ 
    y'^2 -(\nabla y)^2 + +\HubbleComovil^2 y^2 
    - 2\HubbleComovil y y'
  \right].
\end{equation}

El momento conjugado del campo $y$ es

\begin{equation}
  \label{eq:y_pi}
  \pi^{(y)}\left(\eta, \bvec{x}\right) =
  \dpar{\mathcal{L}^{y}}{y'} = y' - \HubbleComovil y,
\end{equation}

el hamiltoniano de esta teoría es simplemente la integral
espacial de la densidad hamiltoniana, $\mathcal{H}$, $H
\equiv \int d^4x \mathcal{H} = \int d^3x \left[\pi^{(y)}y' -
  \mathcal{L}\right]$. Promovemos ahora a las variable $y$ y
$\pi^{(y)}$ a los operadores $\y$ y $\piy$. La cuantización
canónica consiste en imponer las siguientes relaciones de
conmutación

\begin{equation}
  \label{eq:y_conmutacion}
  \left[\y(\eta, \bvec{x}), \y(\eta,\bvec{x}')\right] =
  \left[\piy(\eta, \bvec{x}), \piy(\eta,\bvec{x}')\right] =
  0, \quad   \left[\y(\eta, \bvec{x}),
    \piy(\eta,\bvec{x}')\right] = i\delta(\bvec{x} - \bvec{x}').
\end{equation}

La ecuación de movimiento para $\y$ es idéntica a
(\ref{eq:y_evolucion}) cambiando $y \to \y$ y se obtiene de
las ecuaciones de movimiento de Heisenberg
\begin{equation}
  \label{eq:y_heisenberg}
  i\y = [\y, \hat{H}], \quad i\piy = [\piy, \hat{H}],
\end{equation}
donde $\hat{H}$ es el hamiltoniano $H$ escrito en términos
de los operadores $\y$ y $\piy$. Nótese que esto es
equivalente a imponer estas relaciones a $\dphi$ y su
momento conjugado $\pi = a^2\dphi$ \footnote{El lagrangiano
  durante el \textit{slow-roll} de la variable original
  $\dphi$ es $\mathcal{L} = \dfrac{a^2}{2}\left(\dphi^2 -
    (\nabla\dphi)^2\right)$, usando $\pi = \dfrac{\partial
    \mathcal{L}}{\partial \dphi'}$ se obtiene $\pi =
  a^2\dphi$.}.

Descomponemos ahora a los campos $y$ y $\pi^{(y)}$ en ondas
planas\footnote{Aunque es posible hacer todas las
  manipulaciones matemáticas usando formulaciones integrales
  o de suma, es usualmente más sencillo hacerlo con la
  suma. La simplicidad viene del hecho de que ya no aparecen
  deltas de Dirac en el espacio$-k$. Por ejemplo, si tenemos
  $f_k = \sum f_k \emikx$, la manera obvia de extraer $f_k$
  es vía $f_k = \dfrac{1}{L}\int f \eikx dx$, esto debido a
  que las condiciones de frontera armónicas hacen que todos
  los términos oscilatorios de la suma integran cero,
  dejando únicamente la integral de $0$ a $L$ de $f_k$.}
\begin{equation}\label{eq:y_ondas_planas}
  \y(\eta,x) = \frac{1}{L^3}\sum_k  \left\{
    \ann_ky_k(\eta) \eikx + \cre_k  y_k(\eta)^* \emikx \right\},
\end{equation}
donde $y_k(\eta)$ es solución de la ecuación
\begin{equation}\label{eq:y_evolucion}
  y_k''+\left( k^2 - \frac{a''}{a}\right)y_k = 0.
\end{equation}
y $\ann$ y $\cre$ son los operadores de creación y
aniquilación del campo. Por su parte, el momento conjugado
$\hat\pi(\eta,x)$
\begin{equation}\label{eq:pi_ondas_planas}
  \hat\pi(\eta,x) =  \frac{1}{L^3}\sum_k  \left\{
    \ann_kg_k(\eta) \eikx + \cre_k g_k(\eta)^* \emikx \right\},
\end{equation}
$g_k(\eta)$ en esta ecuación está relacionado con
$y_k(\eta)$ mediante
\begin{equation}\label{eq:g_y_relacion}
  g_k = y_k' - \Hubble y_k,
\end{equation}
En ambas expansiones $\hat{a}_k, \hat{a}_k^\dag$ son los
operadores de aniquilación y creación respectivamente. Estos
cumplen con las relaciones de conmutación estándar
\begin{equation}
\label{eq:ann_conn}
  [\hat a_k, \hat a_{k'}]  =     [\hat a^\dag_k,\hat
  a^\dag_{k'}] = 0, \quad 
  [\hat a_k,\hat a^\dag_{k'}]  =  \hbar L^3\delta_{kk'}.
\end{equation}

Las relaciones de conmutación (\ref{eq:y_conmutacion})
imponen una restricción a las soluciones $y_k,
g_k$:\footnote{Las unidades de los modos son:
  \begin{align*} [\hat{y}(\eta,x)]\, =
    \,\left(\frac{M}{L}\right)^{1/2}, & \qquad
    [\hat{\pi}(\eta,x)] \,=\, \frac{M^{1/2}}{L^{3/2}},
    \qquad
    [\hat{a}_k] \,=\, [\hat{a}_k] \,=\, M^{1/2}L^{2}, & \qquad \\
    [y_k(\eta)] \, =\, L^{1/2}, & \qquad [g_k(\eta)] \, = \,
    L^{-1/2}
  \end{align*}
}
\begin{equation}
  \label{eq:restriccion_normalizacion}
  y_kg_{k}^* - y_k^*g_{k} = i
\end{equation}

Para completar el proceso de cuantización necesitamos las
soluciones de la ecuación (\ref{eq:y_evolucion}). Durante la
etapa inflacionaria $\left(\dfrac{a''}{a} =
  \dfrac{2}{\eta^2}\right)$, quedando
\begin{equation}\label{eq:y_eq_evolucion_inf}
  y_k''+\left( k^2 - \frac{2}{\eta^2}\right)y_k = 0,
\end{equation}
escrita de esta manera, se puede encontrar la solución
exacta
\begin{equation}\label{eq:sol_general}
  y_k(\eta) = \alpha
  e^{-ik\eta}\left(1-\frac{i}{k\eta}\right) + \beta
  e^{ik\eta}\left(1+\frac{i}{k\eta}\right).
\end{equation}

La elección de una función específica $y_k(\eta)$
corresponde a una prescripción particular del vacío físico
$\ket{0}$ definido por
\begin{equation}
  a_k\ket{0} = 0,
\end{equation}
una elección diferente de $y_k(\eta)$ está asociada a una
descomposición diferente de modos de creación/aniquilación y
por lo tanto a un vacío diferente \todo{Cita a
  Mukhanov}. Obsérvese que este estado de vacío es
\textit{homogéneo} e \textit{isotrópico} en todas las
escalas, esto se puede ver aplicando  los
operadores de momento $\hat{\bm P}$ y momento angular
$\hat{\bm J }$ a $\ket{0}$.

Notemos que la longitud de onda asociada a un modo $k$
cualquiera, puede encontrarse siempre dentro del radio de
Hubble si se retrocede lo suficiente en el tiempo (i.e.\ , $k >
\Hubble$, o $k^{-1} < \Hubble ^{-1}$), ya que el radio
comóvil de Hubble, $\Hubble^{-1} = (aH)^{-1}$, disminuye
durante la etapa inflacionaria. Escogiendo un $|\eta|$ lo
suficientemente grande se obtiene $k|\eta| \gg 1$. Además
para una longitud de onda más pequeña que el radio de
Hubble, se pueden despreciar los efectos de curvatura y el
modo se comportara como si estuviera en un espacio-tiempo de
Minkowski\footnote{Esto se puede ver en la ecuación
  \eqref{eq:y_evolucion}, ya que la masa efectiva es
  despreciable cuando $k|\eta| \gg 1$.}.

La descripción física más natural es la de elegir la
solución que corresponda con la solución de frecuencias
positivas del vacío de Minkowski en el límite $k|\eta| \gg
1$.
\begin{equation}
  y_k^{Minkowski} \sim e^{-ik\eta},
\end{equation}
entonces $\beta = 0$ en \eqref{eq:sol_general}. Usando la
restricción  \eqref{eq:restriccion_normalizacion}
se obtiene $\alpha= \sqrt{\hbar/2k}$. Por lo que se
elige el estado de vacío que está relacionado a la solución
\begin{equation}
  \label{eq:vacio_BunchDavies}
  y_k(\eta) =
  \sqrt{\frac{1}{2k}}\left(1-\frac{i}{k\eta}\right)e^{-ik\eta}. 
\end{equation}

Como se mencionó anteriormente este estado de vacío es
conocido en la literatura como vacío 
de \emph{Bunch-Davies}.  Con esta elección de vacío
$g_k(\eta)$ es
\begin{equation}\label{eq:vacio_BunchDavies_momento}
  g_k(\eta) = -i\sqrt{\frac{ k}{2}} e^{-ik\eta}.
\end{equation}

Al definir los campos cuánticos y su vacío, la cuantización
es completada. Ahora debemos de especificar mediante la
\textit{hipótesis del colapso} como este vacío se convierte
en un estado no-homogéneo y anisotrópico.

\subsection{Hipótesis de Colapso}

El estado de vacío queda definido por la condición
$\ann_k\ket{0} = 0$ para todo valor de $k$. La
\textit{hipótesis del colapso}  o del 
\textit{auto-colapso} opera de una manera análoga a lo que
sería un proceso $\textbf{R}$.

En el artículo de \citeauthor*{Sudarsky06a} se hacen las
siguientes suposiciones sobre el colapso: (a) los modos son
``independientes''\footnote{El que los modos seas
  independientes significa que es posible (a) la descomposición del
operador de campo es una suma de ``operadores de modo'' que
conmutan, (b) laa descomposición ortogonal del espacio de
Hilbert de una partícula y (c) la descomposición de productos
tensoriales directos en el espacio de Fock.} y además  (b) se
requiere que el estado inicial del campo no esté enredado
(\textit{entangled}) respecto a esta descomposición, i.e.\ , 
el estado de vacío se puede escribir como el producto
directo de estados para los operadores de modo. De esta
manera, el colapso actúa a nivel del
modo $k$ y debido a las suposiciones tiene sentido hablar de
``un colapso individual por modo'', por lo que,
\textit{hipótesis del colapso} se debe de leer como
\textit{hipótesis del colapso por modo}. 

Bajo estas suposiciones escribiremos  el campo
mediante los ``operadores de modo'' \footnote{ Estos nuevos 
  operadores, $\hat y_k$ y $\hat\pi_k$ tienen unidades de
  $[\hat y_k ] \, = \, M^{1/2}L^{5/2}$ y $[\hat\pi_k] =
  M^{1/2}L^{3/2}$.}

\begin{subequations}
  \begin{equation}
    \hat{y}(\eta,x) = \frac{1}{L^3} \sum_k
    \hat{y}_k(\eta)e^{ikx}
  \end{equation}
  \begin{equation}
    \hat{\pi}(\eta,x) = \frac{1}{L^3} \sum_k
    \hat{\pi}_k(\eta)e^{ikx}
  \end{equation}
\end{subequations}
con
\begin{subequations}
  \begin{equation}
    \hat{y}_k = y_k(\eta)\hat a_k + y_k^*(\eta)\hat a_{-k}^\dag
  \end{equation}
  \begin{equation}
    \hat{\pi}_k = g_k(\eta)\hat a_k + g_k^*(\eta)\hat a_{-k}^\dag
  \end{equation}
\end{subequations}
usando estas definiciones y las relaciones de conmutación
(\ref{eq:y_conmutacion}) podemos deducir las de
$\hat{y}_k$ y $\hat\pi_k$: 
\begin{equation}
  \label{eq:rel_conmutacion_nueva}
  [\hat{y}_k, \hat\pi_{k'}] = i\hbar L^3 \delta_{kk'}.
\end{equation}

El siguiente paso, debido a que el colapso actúa como una
especie de ``medición'', debemos de  garantizar la
hermiticidad de los operadores usados. Obsérvese que los
``operadores de modo'', $\y_k$ y $\piy_k$, se pueden
descomponer en partes real e imaginaria: $\yRI$ y
$\pyRI$,

\begin{subequations}
  \begin{equation}
    \hat{y}_k(\eta) = \hat{y}_k^R(\eta) + i \hat{y}_k^I(\eta)
  \end{equation}
  \begin{equation}
    \hat{\pi}_k(\eta) = \hat{\pi}_k^R(\eta) + i \hat{\pi}_k^I(\eta)
  \end{equation}
\end{subequations}

donde los nuevos operadores se definen mediante

\begin{subequations}\label{eq:def_hermitiana}
  \begin{equation}
    \hat{y}_k^R(\eta) = \frac{1}{\sqrt{2}} \left(
      y_k(\eta)\hat{a}_k^R + y_k^*\hat{a}_k^{R\dag}\right),
\quad
    \hat{y}_k^I(\eta) = \frac{1}{\sqrt{2}} \left(
      y_k(\eta)\hat{a}_k^I + y_k^*\hat{a}_k^{I\dag}\right).
  \end{equation}
  \begin{equation}
    \hat{\pi}_k^R(\eta) = \frac{1}{\sqrt{2}} \left(
      g_k(\eta)\hat{a}_k^R + g_k^*\hat{a}_k^{R\dag}\right),
\quad
    \hat{\pi}_k^I(\eta) = \frac{1}{\sqrt{2}} \left(
      g_k(\eta)\hat{a}_k^I + g_k^*\hat{a}_k^{I\dag}\right).
  \end{equation}
\end{subequations}

Nótese que estos operadores son hermíticos. Los operadores
de creación y aniquilación son

\begin{equation}
  \hat{a}_k^R = \frac{1}{\sqrt{2}}\left(\hat{a}_k +
    \hat{a}_{-k}\right), \quad     \hat{a}_k^I = \frac{-i}{\sqrt{2}}\left(\hat{a}_k +
    \hat{a}_{-k}\right)
\end{equation}

Las relaciones de conmutación entre estos operadores de
aniquilación/creación son no estándares:
\begin{equation}
  [\hat{a}_k^R, \hat{a}_{k'}^R] =\hbar L^3\left( \delta_{k,k'} +
    \delta_{k,-k'}\right), \quad [\hat{a}_k^I, \hat{a}_{k'}^I] =
  \hbar L^3\left(\delta_{k,k'} - \delta_{k, -k'}\right),
\end{equation}
siendo las demás relaciones de conmutación cero. Esto indica
que los operadores que corresponden a $k$, $-k$ son
idénticos en el caso real e idénticos salvo un signo en el
caso imaginario. Será útil listar aquí los conmutadores de
$\hat{y}_k^{R,I}$ y $\hat\pi_k^{R,I}$
\begin{equation}
  [\hat{y}_k^R, \hat{\pi}_{k'}^R] =\frac{1}{2}i
  L^3\left( \delta_{k,k'} + 
    \delta_{k,-k'}\right), \quad [\hat{y}_k^I, \hat{\pi}_{k'}^I] =
  \frac{1}{2}i L^3\left(\delta_{k,k'} - \delta_{k, -k'}\right)
\end{equation}

los cuales obviamente no son estándar.

Introducimos ahora las siguientes definiciones, relacionadas
con los primeros momentos estadísticos, para un
estado arbitrario del campo $\ket{\Xi}$ \footnote{Es
  importante notar que dado que estamos en 
  la representación o imagen de Heisenberg, toda la
  dependencia temporal de los ``operadores de modo'' está en
  $y_k(\eta)$ y $g_k(\eta)$ y las cantidades  $d_k$, $c_k$ y
  $e_k$ son independientes del tiempo.}
\begin{subequations}
  \label{eq:def_d_c_e}
  \begin{equation}\label{eq:def_d}
    d_k^{(R,I)} \equiv \braket{\creRI_k}{\Xi} \equiv
    \left|d_k^{(R,I)}\right| e^{\imath\alpha_k},
  \end{equation}
  \begin{equation}\label{eq:def_c}
    c_k^{(R,I)} \equiv \braket{(\creRI_k)^2}{\Xi},
  \end{equation}
  \begin{equation}\label{eq:def_e}
    e_k^{(R,I)} \equiv
    \braket{\creRI_k\ann_k^{(R,I)}}{\Xi},
  \end{equation}
\end{subequations}
entonces, los valores de expectación del campo $\yRI$ se calcularían
usando estas cantidades como
\begin{align}
  \langle\hat{y}_k^{R,I}\rangle_\Xi &=
  \bra{\Xi}\hat{y}_k\ket{\Xi} \nonumber \\
  & = \frac{1}{\sqrt{2}}
  \left\{y_k(\eta)\bra{\Xi}\hat{a}_k^{R,I}\ket{\Xi} +
    y_k^*\bra{\Xi}\hat{a}_k^{R,I\,\dag}\ket{\Xi}\right\}
  \nonumber \\
  &= \sqrt{2} \Re \left(y_k(\eta)d_k^{R,I}\right)
\end{align}
y haciendo lo mismo para el momento conjugado $\pyRI$
obtenemos, que los valores de expectación para un estado
arbitrario $\ket{\Xi}$ son
\begin{subequations}\label{eq:valores_medios}
  \begin{equation}
    \langle\hat{y}_k^{R,I}\rangle_\Xi = \sqrt{2} \Re
    \left(y_k(\eta)d_k^{R,I}\right) 
  \end{equation}
  \begin{equation}
    \langle\hat{\pi}_k^{R,I}\rangle_\Xi = \sqrt{2} \Re
    \left(g_k(\eta)d_k^{R,I}\right)
  \end{equation}
\end{subequations}

La dispersión, $\Delta\hat{x}^2 \equiv
\langle\hat{x}^2\rangle_\Xi - \langle\hat{x}\rangle_\Xi^2$,
de los campos  para un estado cualquiera $\ket{\Xi}$,
\begin{subequations}\label{eq:dispersiones_generales}
  \begin{equation}\label{eq:dispersion_general_y}
    (\Delta y_k^{R,I})^2_\Xi = \Re \left(y_k^2
      c_k^{R,I}\right) + \frac{1}{2}|y_k|^2(\hbar
    L^3+2e_k^{R,I}) - 2\Re(y_k^{R,I}d_k^{R,I})^2 
  \end{equation}
  \begin{equation}\label{eq:dispersion_general_pi}
    (\Delta \pi_k^{R,I})^2_\Xi = \Re \left(g_k^2
      c_k^{R,I}\right) + \frac{1}{2}|g_k|^2(\hbar
    L^3+2e_k^{R,I}) - 2\Re(g_k^{R,I}d_k^{R,I})^2 
  \end{equation}
\end{subequations}

Estas expresiones demuestran que las cantidades
(\ref{eq:def_d_c_e}) especifican las principales cantidades
de interés que caracterizan al estado del campo. Obviamente
para el estado vacío $\ket{0}$ tenemos que el 
valor de expectación de los campos es
\begin{equation}\label{eq:expectacion_vacio}
  \langle\hat{y}_k^{R,I}\rangle_0 =
  \langle\hat{\pi}_k^{R,I}\rangle_0 = 0,
\end{equation}
y la dispersión, 
\begin{equation}\label{eq:dispecion_vacio}
  (\Delta y_k^{R,I})^2_0 = \frac{1}{2}|y_k|^2\hbar L^3, \qquad
  (\Delta \pi_k^{R,I})^2 = \frac{1}{2}|g_k|^2\hbar L^3.
\end{equation}

Así, podemos describir la operación del
\textit{auto-colapso} como sigue:  originalmente --i.e.\
antes del colapso -- \textit{no
  hay perturbaciones en la métrica}. Esto se puede observar
de la fórmula (\ref{eq:classical_fundamental}), que repetimos
aquí por comodidad del lector
\begin{eqnref}{eq:classical_fundamental}
  \nabla^2\psi - \bm\mu\psi  = 4\pi G (\bm\nu\dphi
  + \varphi'\dphi') = s\Gamma, 
\end{eqnref}
donde hemos introducido los símbolos $s \equiv 4\pi G
\varphi'$ y $\bm\Gamma \equiv \dphi' + \bm\nu$. Durante el
periodo inflacionario $\mu = 0$ y $\bm\Gamma = \dphi' = a^{-1}\piy$.
Ahora, procederemos a evaluar la perturbación en la métrica
usando una descripción semi-clásica de la gravedad
interactuando con campos cuánticos, justo como dictan las EFE
semi-clásicas: $G_{ab} = 8\pi G \expec{T_{ab}}$, de esta
manera, la ecuación (\ref{eq:classical_fundamental}) es
promovida a 
\begin{equation}
  \label{eq:semiclassical-fundamental}
  \nabla^2\psi - \bm\mu \psi = s \expec{\hat{\bm\Gamma}}_\Xi.
\end{equation}

Expresando esta ecuación en el espacio de Fourier y tomando
en cuenta la aproximación de \textit{slow-roll}, tenemos que
para un estado arbitrario, $\ket{\Xi}$,

\begin{equation}
  \label{eq:potencial_newtoniano_momentum_sra_inflacion}
  \psi_k =
  \frac{s}{k^2}\expec{\dphi_k'}_\Xi   
\end{equation}

Así, usando las ecuaciones
(\ref{eq:expectacion_vacio}) se puede comprobar la
afirmación de que antes de que ocurra el colapso \textit{no
  hay perturbaciones en la métrica}

\begin{equation}
  \label{eq:psi_antes_colapso}
  \psi_k = 0.
\end{equation}

Siguiendo con la descripción de la dinámica del colapso, a
un cierto tiempo $\eta_k^c$ la parte del estado que 
corresponde a el modo $k$ ``brinca'' a un nuevo estado
$\ket{\bm\Upsilon}$ que no es homogéneo e isotrópico, lo
cual debido a la ecuación
(\ref{eq:potencial_newtoniano_momentum_sra_inflacion}) es

\begin{equation}
  \label{eq:psi_despues_colapso}
  \psi_k = \frac{s}{k^2}\expec{\dphi'}_{\bm\Upsilon}.
\end{equation}

Lo que resta es especificar el colapso de los modos. ¿Cómo
ocurre el colapso?¿Cuáles son sus características? 
Para responder estas preguntas, es necesario especificar, o
una teoría completa como las \acf{DRM} entre las cuales se
incluye el \acf{GRW} \cite{Ghirardi1985,Ghirardi1986} o los \acf{CSL}
\cite{PearlePhil99} (cf.\ para una revisión completa
se puede consultar \cite{Bassi03,Bassi07}), o, como en el caso
de \cite{Sudarsky06a,Unanue2008} un modelo fenomenológico. A
las teorías completas las llamaremos en este trabajo de
tesis \textit{mecanismos de colapso}, a los segundos (los
modelos fenomenológicos) los denominaremos \textit{esquemas
  de colapso}.

\citeauthor*{Sudarsky06a} estudian dos esquemas de
colapso en \cite{Sudarsky06a}, llamados  \textit{esquema colapso
  independiente}  y \textit{esquema de colapso newtoniano},
ambos se describen a continuación.

\paragraph{Esquema de colapso independiente.} En este esquema
de colapso, tanto, el valor de expectación del campo $\y$ como
el valor de expectación del momento conjugado $\piy$ en el
estado post-colapso $\ket{\Upsilon}$ al tiempo $\tc$ están
distribuidos aleatoriamente en los respectivos rangos de las
incertidumbres del estado de vacío $\ket{0}$ y no se
encuentran correlacionados estadísticamente. Esto se
puede expresar de la siguiente manera
\begin{align}
  \label{eq:esquema_independiente}
  \ket{0} \to \ket{\Upsilon}:&\nonumber \\
  &\expec{\yRI(\eta_k^c)}_\Upsilon = 
  x_1^{(R,I)}\sqrt{\fluc{\yRI}_0}, \nonumber
  \\  &\expec{\pyRI(\tc)}_\Upsilon =
  x_2^{(R,I)}\sqrt{\fluc{\pyRI}_0}.
\end{align}
donde $x^{(R,I)}_{1,2}$ son variables aleatorias gaussianas
centradas en cero y con dispersión 1. El hecho de que el
colapso no correlacione los valores de expectación
post-colapso del campo y su momento conjugado se encuentra
en que los valores de las variables $x^{(R,I)}_1$ y
$x^{(R,I)}_2$  son eventos independientes. 

\paragraph{Esquema de colapso newtoniano.} Este esquema de
colapso está inspirado en que, a primer orden y durante el
\textit{slow-roll}  de la época inflacionaria, la ecuación
(\ref{eq:semiclassical-fundamental}) sólo depende de
$\pyRI$, i.e.\ , únicamente el momento canónico conjugado del
campo inflatónico aparece como fuente de la perturbación de
la métrica.
\begin{equation}
  \label{eq:esquema_newtoniano}
  \ket{0} \to \ket{\Upsilon}:\expec{\yRI(\eta_k^c)}_\Upsilon =
  0, \quad \expec{\pyRI(\tc)}_\Upsilon =
  x_2^{(R,I)}\sqrt{\left(\fluc{\pyRI}\right)_0^2} .
\end{equation}

\paragraph{Valores de expectación en el estado post-colapso
  de las variables originales } Ahora procederemos a
calcular el valor de expectación del momento conjugado
$\hat\pi$ del campo original $\dphih$, para ello será 
conveniente expresar a $y_k^{R,I}$ y  $g_k^{R,I}$ en 
forma polar compleja, entonces, $y_k^{R,I} =
|y_k(\eta)|\exp({\imath\beta_k^{R,I}})$ y $g_k^{R,I} =
|g_k(\eta)|\exp({\imath\gamma_k^{R,I}})$, donde
\begin{subequations}\label{eq:polar-form}
  \begin{equation}\label{eq:polar-form-y}
    |y_k(\eta)| =
    \sqrt{\frac{1}{2k}}\left(\frac{\sqrt{1+k^2\eta^2}}{k\eta}\right),
    \quad 
    \beta_k = -\left[k\eta +
      \arctan\left(\frac{1}{k\eta}\right)\right],
  \end{equation}
  \begin{equation}\label{eq:polar-form-pi}
    |g_k(\eta)| = \sqrt{\frac{k}{2}} , \quad \gamma_k =
    -\left(k\eta + \frac{\pi}{2}\right).
  \end{equation}
\end{subequations}

Así, podemos  expresar las ecuaciones
(\ref{eq:valores_medios}) de una manera más útil,

\begin{subequations}\label{eq:expectation_values}
\begin{equation}\label{eq:real_y}
  \expec{\yRI}=\sqrt{2}\Re(y_kd_k) =
  |y_k||d_k^{(R,I)}|\cos\left(\alpha_k^{(R,I)} 
    + \beta_k\right) ,
\end{equation}
\begin{equation}\label{eq:real_pi}
  \expec{\pyRI} = \sqrt{2}\Re(g_kd_k) =
  |g_k||d_k^{(R,I)}|\cos\left(\alpha_k^{(R,I)} 
    + \gamma_k\right) .
\end{equation}
\end{subequations}


Recordemos que en la ecuación
(\ref{eq:semiclassical-fundamental}) aparece $\expec{\dphi_k'}$, el
cual en función de los campos  $\pyRI$, es
\begin{equation}
  \label{eq:rel_dphi_pyRI}
  \expec{\dphih_k'}_\Upsilon  = \frac{1}{\sqrt{2}a}
  \left[\expec{\pyR_k(\eta)}_\Upsilon + i   
    \expec{\pyI_k(\eta)}_\Upsilon \right],
\end{equation}
tomando el valor de expectación en el estado post-colapso
$\ket{\Upsilon}$
\begin{equation}
  \expec{\dphih_k'}_\Upsilon = \frac{|g_k|}{a}
  \left[|d_k^R|\cos\left(\alpha^R_k + \gamma_k^c\right) + i
    |d_k^I|\cos\left(\alpha^I_k+\gamma_k^c\right)\right],
\end{equation}
sustituyendo el valor de $\gamma_k^c$
\begin{equation}
  \label{eq:expectacion_dphi_d_1}
  \expec{\dphih_k'}_\Upsilon=\frac{|g_k|}{a} \left[
    |d_k^R|\cos\left(\alpha_k^R - k\eta_c -
      \frac{\pi}{2}\right) + i|d_k^I|\cos\left(\alpha_k^I -
      k\eta_c - \frac{\pi}{2}\right)\right].
\end{equation}
Definiendo la variable que mide el tiempo transcurrido desde
el colapso, $\Delta_k = k (\eta - \eta_k^c)$.
\begin{equation}
  \label{eq:expectacion_dphi_d_2}
  \expec{\dphih_k'}_\Upsilon=\frac{|g_k|}{a} \left[
    |d_k^R|\cos\left(\alpha_k^R + \gamma_k +
      \Delta_k\right) 
    + i|d_k^I|\cos\left(\alpha_k^I + \gamma_k +
      \Delta_k\right)\right], 
\end{equation}
y usando una vez más las fórmulas trigonométricas de ángulos
múltiples
\begin{multline}
  \label{eq:expectacion_dphi_d_3}
  \expec{\dphi_k'}_\Upsilon=\frac{|g_k|}{a} \Big\{
  |d_k^R|\left[\cos\left(\alpha_k^R +
      \gamma_k\right)\cos\Delta_k -
    \sin(\alpha^{R}+\gamma_k)\sin\Delta_k\right] \\
  + i|d_k^I| \left[ \cos(\alpha_k^I + \gamma_k)
    \cos\Delta_k-\sin(\alpha^{I}+\gamma_k)
    \sin\Delta_k\right]\Big\}
\end{multline}

En esta última ecuación sólo faltan por especificar las expresiones
 de $d_k^{(R,I)}$ y $\alpha^{(R,I)}$, las cuales son calculadas
 a partir de la especificación del esquema de colapso. En el
artículo \cite{Sudarsky06a} se realiza un largo (y tortuoso)
proceso algebraico para poder obtener esta expresión para cada
esquema de colapso que no repetiremos aquí. En el capítulo
\ref{cha:analisis-esquemas} se presentan las ecuaciones
generales (\ref{eq:expectacion_y_general}) y
(\ref{eq:expectacion_pi_general}) para describir al valor de 
expectación de $\yRI$ y $\pyRI$ de manera independiente del
esquema y de la época cosmológica. Además, se presenta en 
\S \ref{sec:colapso_wigner} un procedimiento algebraico
parecido al llevado acabo en \cite{Sudarsky06a} para obtener
los valores de expectación.

\subsection{Comparación con las
  observaciones}\label{sec:comparacion-obs}

Una vez elegido el esquema de colapso, se procede a evaluar
la perturbación de la métrica usando las ecuaciones
semi-clásicas, que en este caso cosmológico a orden lineal
en las perturbaciones se reducen a
(\ref{eq:semiclassical-fundamental}).

En esa ecuación $\expec{\dphih'_k}_\Upsilon$ es el valor de
expectación del momento conjugado $\dphih_k' =
\piy_k/a(\eta)$ en el estado $\ket{\Upsilon}$ y caracteriza
la parte cuántica del campo inflatónico. Es importante
enfatizar que \textit{antes} de que el colapso ocurra
\textit{no hay} perturbaciones de la métrica (cf.\ ver
discusión arriba de la ecuación
(\ref{eq:psi_antes_colapso})) i.e.\ el lado derecho de la
ecuación es cero, entonces, sólo \textit{después} de que el
colapso ocurre las perturbaciones gravitacionales,
representadas por $\psi$ aparecen. El colapso de cada modo
representa el origen de la inhomogeneidad y la anisotropía
a la escala representada por el modo. Otro punto que es
importante recalcar, es que a todo tiempo, \textit{el
  universo será definido por un sólo estado
  $\ket{\Upsilon}$, y no por un ensamble de estados.} Los
aspectos estadísticos, entonces, surgirán debido a que no
medimos directamente y por separado cada uno de los modos
específicos representados por el vector de onda $\vec k$,
sino por la contribución de todos los modos en la
descomposición armónica esférica de las fluctuaciones de la
temperatura en la esfera celeste.

Ahora debemos de comprobar la hipótesis del colapso con las
observaciones. Las cantidades observadas (por ejemplo WMAP)
son las anisotropías del CMB, $\Theta(\buvec{n}) = \Delta T/
T (\buvec{n}) = \sum_l\sum_m \alpha_{lm}Y_{lm}(\bvec{n})$,
como vimos en \S \ref{sec:fisica-cmb}, el espectro de potencias
del CMB está relacionado con el espectro de potencias de
$\psi$ mediante (\ref{eq:cl_psi_power_spectrum}). Entonces,
debemos considerar la expresión para el potencial
gravitatorio $\psi$ dado por la ecuación
(\ref{eq:psi_despues_colapso}) en la LSS. En esta subsección
derivaremos de nuevo esa fórmula. El punto de partida es la
ecuación   \eqref{eq:alm_esfericos} 
\begin{equation}
  \label{eq:alm_hipotesis_colapso_1}
  a_{lm} = \frac{4\pi i^l}{(2\pi)^{3/2}} \int d^3k
  \thinspace \frac{\psi(\bvec{k})}{3} j_l(kR_D) Y_{lm}^*(\buvec{k}),
\end{equation}
en la cual hemos sustituido la relación de Sachs-Wolfe
(\ref{eq:sachs-wolfe}),  $\Theta = \dfrac{1}{3}\psi(\eta,
\bvec{x})$ y $\psi(\bvec{k})$ es el modo $k$ de la expansión
de $\psi(\eta, \bvec{x})$. Cada uno de los modos
$\psi(\bvec{k})$ está relacionado con la perturbación del
inflatón mediante \eqref{eq:psi_despues_colapso}, 
\begin{equation}
  \label{eq:fundamental}
 \psi(\eta,
\bvec{x})  =  \int d^3k \frac{s 
    \mathcal{T}(k)}{k^2} \expec {\dphih_k'}
  e^{i\bvec k \cdot \bvec x},
\end{equation}
donde hemos introducido el factor $\mathcal{T}(k)$ para
representar los efectos de los diversos mecanismos físicos
presentes entre el \textit{reheating} y el desacople. Estos
$\mathcal{T}(k)$ son conocidos como las \textit{funciones de
  transferencia}. Recuerde que la relación entre la
diferencial relativa de la temperatura $\Theta(\buvec{x})$ y
la perturbación del potencial gravitatorio dada por el
efecto Sachs-Wolfe (\ref{eq:sachs-wolfe}) está evaluada en
la emisión del fotón, esto es en la LSS, de ahí que
$\bvec{x} = R_D \buvec{x}$, siendo $R_D$ la magnitud de la
distancia entre el observador (e.g.\ WMAP)  y la LSS. Así,
la ecuación (\ref{eq:alm_hipotesis_colapso_1}) es
\begin{equation}
  \label{eq:alm_colapso}
 a_{lm} =  \frac{4\pi s  i^l}{3(2\pi)^{3/2}}
  \int d^3k \thinspace  j_l(k R_d)
  Y_{lm}^*(\buvec{k}) \mathcal{T}(k) 
  \frac{\expec{\dphih_k'}}{k^2}.
\end{equation}

Estamos interesados en el valor medio del cuadrado de esta
cantidad. Los aspectos estadísticos que permiten darle
sentido a la afirmación emergerán de la ecuación
(\ref{eq:fundamental}). Esta ecuación indica que la cantidad
de interés es el resultado de un infinito número de
osciladores armónicos cada uno contribuyendo con un número
complejo a la suma, lo que lleva, en efecto a una caminata
aleatoria bidimensional cuyo desplazamiento total medio
corresponde a la cantidad observacional. Como primer paso
tomaremos el cuadrado de (\ref{eq:alm_colapso})
\begin{equation}\label{eq:alpha_cuadrado}
 a_{lm}a_{l'm'}^* = \frac{(4\pi)^2 s^2 i^l (-i)^{l'}}{9
   (2\pi)^3} \int d^3k \int d^3k'
  \frac{\mathcal{T}(k)}{k^2}
  \frac{\mathcal{T}(k')}{k'^2} 
  \langle\delta\hat\varphi_k'\rangle
  \langle\delta\hat\varphi_{k'}'\rangle^* j_l(kR_d)j_l(k'
  R_d)Y_{lm}(\hat{k})Y_{lm}(\hat{k}'),
\end{equation}

Para obtener el valor ``más probable''\footnote{\textit{Most
    likely} en inglés} nos auxiliaremos de un ensamble
\textit{imaginario} de universos e identificaremos el valor
más probable con el valor medio del ensamble.

El valor medio del producto
$\expec{\delta\hat\varphi_k}
\expec{\delta\hat\varphi_{k'}^*}$, evaluado en los
estados post-colapso tiene una forma $
\overline{\expec{\dphih_k}\expec{\dphih_{k'}\thinspace^*}} =
\dfrac{\hbar L^3 k}{4 a^2} C(k) \delta_{\bvec{k}\bvec{k}'}
$, donde $C(k)$ es una función adimensional de $k$ el cual
codifica los aspectos físicos del esquema del
colapso. Entonces, la expresión para el valor ``más
probable'' (ML) de la cantidad de interés es:
\begin{equation}
  \label{eq:contacto_obs_0}
  |a_{lm}|^2_{ML}  = \frac{1}{9}\frac{s^2 \hbar }{2 \pi a^2}
  \int \frac{dk}{k} C(k) \mathcal{T}(k)^2
  j_l^2(|\vec{k}|R_D),
\end{equation}
expresión en la cual hemos usado las propiedades de
ortogonalidad de los armónicos esféricos. La ecuación
(\ref{eq:contacto_obs_0}) se puede escribir de una manera más
sencilla de integrar si hacemos el cambio de variable $x = k
R_d$:
\begin{equation}\label{eq:contacto_obs}
  |a_{lm}|^2_{ML}  = \frac{s^2\hbar}{2 \pi a^2}
  \int   \frac{C(x/R_D)}{x} \mathcal{T}(x/R_D)^2
  j_l^2(x)  dx.
\end{equation}

Esta es la expresión en la cual las predicciones del esquema
de colapso se pueden comparar con las
observaciones. Obsérvese en esta última ecuación que la
forma estándar del espectro i.e.\ , un espectro plano o de
Harrison-Zel'dovich, corresponde a reemplazar la función
$C(k)$ por una constante. Esto impondrá condiciones a los
diversos esquemas de colapso.

\paragraph{Esquemas de colapso y observaciones} Como se
mencionó arriba, toda la física de los esquemas de colapso
está codificada en la función $C(k)$, la cual es la parte
adimensional de $\overline{\expec{\dphih_k}
  \expec{\dphih_{k'}} }$. Las funciones $C(k)$ que se
obtienen de los esquemas de colapso son para el esquema
independiente (\ref{eq:esquema_independiente}) ,

\begin{equation}\label{eq:C_sudarsky_inflation}
  C_1(k) = 1 + \frac{2}{z_k^2}\sin^2\Delta_k + \frac{1}{z_k}\sin(2\Delta_k),
\end{equation}

y del esquema de colapso newtoniano
(\ref{eq:esquema_newtoniano}),

\begin{equation}\label{eq:C_2_inflation}
  C_2(k) = 1 + \sin^2\Delta_k \left(1 - \frac{1}{z_k^2}\right)
  - \frac{1}{z_k}\sin (2\Delta_k).
\end{equation}

Donde $z_k\equiv k\tc$ y $\Delta_k \equiv k(\eta - \tc)$. De
estas expresiones se puede observar que para recobrar el
caso $C =1$ es necesario que los tiempos de colapso vayan
como $\tc \propto k^{-1}$. Esta es una importante
restricción sobre cómo debe de ser el mecanismo de colapso.

En lo que resta de este trabajo de tesis se propondrá
(capítulo \ref{cha:analisis-esquemas}) un esquema adicional
de colapso basado en el funcional de Wigner y
desarrollaremos estudios para ver la robustez de los
esquemas de colapso ante desviaciones de la restricción $\tc
\propto k^{-1}$. Además estudiaremos los efectos físicos de
múltiples colapsos y colapsos tardíos.

\chapter[Estudio detallado de los esquemas de
colapso]{Estudio detallado de los esquemas de
  colapso.}\label{cha:analisis-esquemas} 

\epigraph{The abrupt change by measurement \ldots is the
  most interesting point of the theory}
{\textit{Erwin Schrödinger}}

Después del análisis y crítica de la propuesta inflacionaria
sobre el origen de la fluctuaciones cosmológicas, en este
capítulo analizaremos un nuevo \textit{esquema} de colapso
inspirado en las propiedades del funcional de Wigner,
haciendo luego un análisis comparativo de los tres esquemas
de colapso presentados hasta el momento. Se presentará
el análisis para el caso de varios colapsos y el
estudio de las implicaciones de un colapso tardío. Al final
se mostrarán algunas predicciones o características que
distinguen a la hipótesis del colapso sobre la explicación 
estándar inflacionaria.

\section{Ecuaciones
  generales}\label{sec:ecuaciones-generales}
A pesar de que en el capítulo
\nameref{cha:critica-al-origen} seguimos el procedimiento
de \citep{Sudarsky06a} para calcular el $C(k)$ de los dos
primeros esquemas de colapso, aquí desarrollaremos las
fórmulas generales, independientes del esquema de colapso y
de la época en la que el colapso ocurra.

Repetimos aquí las relaciones de los valores de expectación
del campo $\y_k \equiv a(\eta)\delta\hat\varphi$ y de su
momento conjugado $\piy_k$, así como de sus valores de
dispersión en un estado arbitrario. Primero recordamos la
definición de las siguientes cantidades asociadas a un
estado dado $\ket{\Omega}$;

\begin{eqnref}{eq:def_d}
  d_k^{(R,I)} \equiv \braket{\creRI_k}{\Omega} \equiv
  \left|d_k^{(R,I)}\right| e^{\imath\alpha_k},
\end{eqnref}
\begin{eqnref}{eq:def_c}
  c_k^{(R,I)} \equiv \braket{(\creRI_k)^2}{\Omega},
\end{eqnref}
\begin{eqnref}{eq:def_e}
  e_k^{(R,I)} \equiv
  \braket{\creRI_k\ann_k^{(R,I)}}{\Omega},
\end{eqnref}

de  esta manera  los valores de expectación están dados por,

\begin{eqnref}{eq:expectation_values}
  \expec{\yRI_k} = \sqrt 2 \, \Re \, \left(y_k
    d_k^{(R,I)}\right), \quad \expec{\pyRI_k} = \sqrt 2 \,
  \Re \, \left(g_k d_k^{(R,I)} \right),
\end{eqnref}

y sus dispersiones

\begin{eqnref}{eq:dispersion_general_y}
  \left(\Delta \hat y_k^{(R,I)}\right)^2_\Omega = \Re
  \left(y_k^2 c_k^{(R,I)}\right) + \frac{1}{2} \left| y_k
  \right|^2 \left(\hbar L^3 + 2 e_k^{(R,I)}\right) - 2
  \Re\left(y_kd_k^{(R,I)}\right)^2,
\end{eqnref}
\begin{eqnref}{eq:dispersion_general_pi}
  \left(\Delta \hat \pi_k^{(R,I)}\right)^2_\Omega = \Re
  \left(g_k^2 c_k^{(R,I)}\right) + \frac{1}{2} \left| g_k
  \right|^2 \left(\hbar L^3 + 2 e_k^{(R,I)}\right) - 2
  \Re\left(g_kd_k^{(R,I)}\right)^2.
\end{eqnref}

Recordemos también que el estado de vacío $\ket{0}$ está
caracterizado por que sus valores de expectación son cero y
sus incertidumbres son $\fluc{\yRI_k} = 1/2|y_k|^2(\hbar
L^3)$ y $\fluc{\pyRI_k} = 1/2|g_k|^2(\hbar L^3)$.

Por último, debemos expresar en forma polar compleja a las
variables del campo cuántico, $y_k^{R,I} =
|y_k(\eta)|\exp({\imath\beta_k^{R,I}})$ y $g_k^{R,I} =
|g_k(\eta)|\exp({\imath\gamma_k^{R,I}})$.

Para obtener las ecuaciones generales que describen la
evolución de los valores de expectación luego de un colapso,
procedemos como sigue.
Usando las ecuaciones (\ref{eq:real_y}, \ref{eq:real_pi}) pero evaluadas
en el tiempo de colapso $\eta_c$ tenemos

\begin{equation}
  \label{eq:d_en_terminos_colapso}
  \left|\dRI_k\right| =
  \frac{\expec{\pyRI_c}}{\sqrt{2}|g_c|\cos(\gamma_c +
    \alpha)}  =
  \frac{\expec{\yRI_c}}{\sqrt{2}|y_c|\cos(\beta_c + 
    \alpha)} ,
\end{equation}

donde $g_c \equiv g(\eta_c), y_c \equiv y(\eta_c)$ y los
ángulos están evaluados en el tiempo de colapso, $\gamma_c
\equiv \gamma(\eta_c) ,\,  \beta_c \equiv \beta(\eta_c)$
entonces, sustituyendo en (\ref{eq:expectation_values}) obtenemos,

\begin{equation}
  \label{eq:pi_1}
  \expec{\pyRI_k}(\eta) = G  \expec{\pyRI_{k,c}} \frac{\cos(\gamma(\eta)
    + \alpha)}{\cos(\gamma_c + \alpha)}.
\end{equation}

con $G$ como la razón entre las magnitudes de $\pyRI$, $G
\equiv \dfrac{|g(\eta)|}{|g_c|}$. Haciendo lo mismo para
$\expec{\yRI_k}$, tenemos

\begin{equation}
  \label{eq:y_1}
  \expec{\yRI_k}(\eta) = Y \expec{\yRI_{k,c}}  \frac{\cos\left(\beta(\eta) +
    \alpha\right)}{\cos(\beta_c + \alpha)},
\end{equation}

donde, $Y \equiv
\dfrac{|y(\eta)|}{|y_c|}$. Usando
(\ref{eq:d_en_terminos_colapso}) y despejando el ángulo
$\alpha$

\begin{equation}
  \label{eq:tg_alpha}
  \tan\alpha = \frac{\expec{\pi_c}|y_c|\cos\beta_c -
    \expec{y_c}|g_c|\cos\gamma_c}{\expec{\pi_c}|y_c|
    \sin\beta_c - \expec{y_c}|g_c|\sin\gamma_c}.
\end{equation}

Expandiendo la suma de ángulos en (\ref{eq:pi_1})

\begin{equation}
  \frac{\expec{\pyRI_k}(\eta)}{\expec{\pyRI_{k,c}}} = G\,
  \frac{\cos\gamma(\eta)  -\sin\gamma(\eta)\tan\alpha}{\cos\gamma_c - \sin\gamma_c\tan\alpha}.
\end{equation}

y sustituyendo (\ref{eq:tg_alpha}) en ella, llegamos a una
de las ecuaciones que nos interesa,

\begin{equation}
  \label{eq:expectacion_pi_general}
  \expec{\pyRI_k}(\eta) = G
  \frac{\expec{\pyRI_c}|y_c|\sin(\beta_c-\gamma) -
    \expec{\yRI_c}|g_c|\sin(\gamma_c-\gamma)}
  {|y_c|\sin(\beta_c-\gamma_c)}.
\end{equation}

Repitiendo los pasos anteriores para $\expec{\yRI_k}$,
 
\begin{equation}
  \label{eq:expectacion_y_general}
  \expec{\yRI_k}(\eta) = Y
  \frac{\expec{\pyRI_c}|y_c|\sin(\beta_c-\beta) +
    \expec{\yRI_c}
    |g_c|\sin(\beta-\gamma_c)}{|g_c|\sin(\beta_c
    -\gamma_c)}, 
\end{equation}

siendo válidas estas dos últimas ecuaciones en cualquier
época cosmológica posterior a un colapso\footnote{Y mientras
  no haya otro colapso (cf.\ sección
  \ref{sec:multiples-colapsos}) o interacción con agentes
  exteriores.} y para cualquier 
esquema de colapso. Las 
ecuaciones (\ref{eq:expectacion_pi_general}) y
(\ref{eq:expectacion_y_general}) muestran la evolución
temporal de las variables del campo luego de un colapso.

En particular, durante la época inflacionaria las
ecuaciones~(\ref{eq:expectacion_pi_general},
\ref{eq:expectacion_y_general}) se reducen a 
\begin{subequations}
\label{eq:expectacion_inflacion}
\begin{equation}
  \expec{\pyRI_k}(\eta) = \expec{\pyRI_c} \cos\Delta_k +
  \sin\Delta_k \left(\frac{\expec{\pyRI_c}}{z_c} - k
    \expec{\yRI_c}\right), 
\end{equation}
\begin{multline}
  \expec{\yRI_k}(\eta) = \cos\Delta_k
   \left[\frac{1}{z}\frac{\expec{\pyRI_c}}{k} +
    \left(\expec{\yRI_c} -
      \frac{\expec{\pyRI_c}}{kz_c}\right)\right] + \\ 
  \sin\Delta_k\left[\frac{1}{z}\left(\frac{\expec{\pyRI_c}}{kz_c}
      - \expec{\yRI_c} \right) + \frac{\expec{\pyRI_c}}{k}\right].
\end{multline}
\end{subequations}

Es posible expresar (\ref{eq:expectacion_pi_general}) sin
usar funciones trigonométricas, lo cual puede ser útil si
las expresiones se vuelven muy complicadas:

\begin{multline}
  \label{eq:expectacion_pi_sin_ang}
  \expec{\pyRI_k}(\eta) =
  \expec{\pyRI_c}\frac{\left(\Im(y_c)\Re(g(\eta)) - \Re(y_c)
      \Im(g(\eta)) \right)}{\left(\Im(y_c) \Re(g_c) -
      \Im(g_c) \Re(y_c) \right)}  \\
  - \expec{\yRI_c}\frac{\left(\Im(g_c) \Re(g(\eta)) -
      \Re(g_c) \Im (g(\eta))\right)} {\left(\Im(y_c)
      \Re(g_c) - \Im(g_c) \Re(y_c) \right)}.
\end{multline}

Sustituyendo esto en $\expec{\piy_k} = \expec{\pyR_k} + i
\expec{\pyI_k}$ llegamos a

\begin{equation}
  \label{eq:expectacion_pi_sin_ang_2}
  \expec{\piy_k}(\eta) = \frac{\left(\expec{\pyR_k} +
      i\expec{\pyI_k}\right) 
    \mathcal{A}(\eta,\eta_k)  - \left(\expec{\yR_k} +
      i\expec{\yI_k}\right)
    \mathcal{B}(\eta,\eta_k)}{\mathcal{C}(\eta_k^c)},
\end{equation}

donde las funciones del tiempo de colapso $\eta_k$,
$\mathcal{A}$, $\mathcal{B}$ y $\mathcal{C}$ son

\begin{align*}
  \mathcal{A}(\eta, \tc) &= \Im(y_c)\Re(g(\eta)) - \Re(y_c)
  \Im(g(\eta)) , \\
  \mathcal{B}(\eta,\tc) &= \Im(g_c) \Re(g(\eta)) -
  \Re(g_c) \Im (g(\eta)), \\
  \mathcal{C}(\tc) &= \Im(y_c) \Re(g_c) - \Im(g_c)
  \Re(y_c).
\end{align*}

Antes de proceder a calcular el valor medio del cuadrado del
valor de expectación del momento conjugado
(\ref{eq:expectacion_pi_sin_ang_2}), recordemos que todos
los esquemas de colapso 
propuestos en el capítulo \ref{cha:critica-al-origen} y el
propuesto en este capítulo en la \S
\ref{sec:colapso_wigner}, se pueden descomponer en una parte
aleatoria adimensional ($f,h$) y en una parte con unidades
($u,v$):
\begin{equation*}
  \expec{\pyRI}_k(\eta_k^c) = f^{R,I}_{k,c}u_{k,c}, \quad
  \expec{\yRI}_k(\eta_k^c) = h^{R,I}_{k,c}v_{k,c},
\end{equation*}
donde las partes aleatorias tienen como propiedad
(derivada de que $k$ y $-k$ no son independientes):
\begin{align*}
  \vme{f^{R}_{k,c}f^{R}_{k',c}} &=
  \vme{h^{R}_{k,c}h^{R}_{k',c}} = \delta_{k,k'} +
  \delta_{k,-k'}, \\
  \vme{f^{I}_{k,c}f^{I}_{k',c}} &=
  \vme{h^{I}_{k,c}h^{I}_{k',c}} = \delta_{k,k'} -
  \delta_{k,-k'}.
\end{align*}

Usando estas propiedades el valor medio del producto de dos
valores de expectación del momento conjugado es
\begin{equation}
  \label{eq:valor_medio_cuadrado_sin_angulos}
  \vme{\expec{\piy_k}\expec{\piy_{k'}}} =
  \frac{2}{|\mathcal{C}(\eta_k^c)|^2} \left(
    |\mathcal{A}_k|^2|u_{k,c}|^2 +
    |\mathcal{B}_k|^2|v_{k,c}|^2\right).
\end{equation}

De esta última ecuación se puede obtener (luego de sustituir
las funciones dependientes de la época cosmológica, i.e.\ ,  el
valor de $\dfrac{a''}{a}$ de la ecuación (\ref{eq:mov_y}) y las
prescripciones del esquema de colapso, e.g.\ , las ecuaciones
(\ref{eq:esquema_independiente}) ó  (\ref{eq:esquema_newtoniano}))
la función $C(k)$. 

\section{Evolución del campo escalar en diferentes épocas 
  cosmológicas}\label{sec:evol-diferentes-eras}

Procederemos ahora a especializar las fórmulas obtenidas
para diversas épocas cosmológicas, esto nos permitirá
estudiar secciones más adelante cómo se verán afectadas las
predicciones de los esquemas de colapso si el colapso
auto-inducido ocurre en otra época que no sea la etapa
inflacionaria o si el colapso ocurre en la época
inflacionaria pero nos interesa ver el sistema en otra
época. Con este fin dividiremos la historia del 
universo como sigue, desde $-\infty < \eta < \eta_{ei}$ es
la era inflacionaria; la era dominada por radiación ocurre en el
intervalo $\eta_{ei} \le \eta < \eta_{eq}$, y finalmente la
época dominada por la materia es desde $\eta_{eq}$ a la
fecha. Es importante notar que estamos ignorando --por
simplicidad-- la aparente
era actual de dominación de la energía oscura. \todo{Ver si se
  pueden reescribir en términos únicamente de cantidades
  actuales (Notas del problema del horizonte)}

La dinámica del campo escalar escalado $y_k(\eta)$ es
\begin{eqnref}{eq:y_evolucion}
  y_k'' + \left(k^2 - \frac{a''}{a}\right) y_k = 0,
\end{eqnref}
y usando la ecuación de Friedmann para el caso plano:
\begin{eqnref}{eq:friedmann_conforme}
  \left(\frac{a'}{a}\right)^2 = \frac{8 \pi G}{3} \rho a^2,
\end{eqnref}
junto con la ecuación de conservación
\begin{eqnref}{eq:conservacion_conforme_fl}
  \rho' + 3\frac{a'}{a}\left(\rho + p\right) = 0,
\end{eqnref}
se obtiene que $\rho a^4$ es constante durante la época
dominada por radiación y $\rho a^3$ lo es durante la época
dominada por materia. Entonces, el factor de escala está
descrito en las diferentes épocas como sigue: durante la
inflación
\begin{equation}
  a(\eta) = -\frac{1}{H_I\eta},
\end{equation}
donde $H_I$ es el parámetro de Hubble durante inflación (y
es casi constante) y $\eta_{ie}$ es el tiempo cuando inicia
la época inflacionaria. En la época dominada por radiación,
\begin{equation}
  \label{eq:scale_factor_radiation}
  a(\eta) = \sqrt{\frac{8\pi G}{3} \rho_{rad} a^4}
  \left(\eta - \eta_{ei}\right) + a_{ei},
\end{equation}
y en la época en la cual el fluido dominante es el polvo,
\begin{equation}
  \label{eq:scale_factor_matter}
  a(\eta) = \left[ \frac{1}{2} \sqrt{\frac{8\pi G}{3}
      \rho_{mat} a^3}\left(\eta - \eta_{eq}\right) +
    \sqrt{a_{eq}}\right]^2.
\end{equation}

Debe de ser claro de las expresiones del factor de escala
que estamos suponiendo que solamente un tipo de materia
domina durante las diversas épocas cosmológicas y que las
transiciones entre diferentes eras es continua pero que
ocurren en un tiempo despreciable comparado con la duración
de las épocas.

Durante la época inflacionaria la ecuación de la
perturbación del campo escalar toma la forma

\begin{equation}
  \label{eq:y_eq_evolucion_inf}
  y_k'' + \left(k^2 - \frac{2}{\eta^2}\right) y_k = 0,
\end{equation}

usando \eqref{eq:y_evolucion} y
\eqref{eq:scale_factor_radiation} obtenemos la dinámica del
campo en la era dominada por radiación

\begin{equation}
  \label{eq:y_eq_evolucion_radiacion}
  y_k'' + k^2 y_k = 0,
\end{equation}

y para la época de materia

\begin{equation}
  \label{eq:y_eq_evolucion_materia}
  y_k'' + \left\{ k^2 - \frac{2
      C_{mat}^2}{\left[C_{mat}\left(\eta - \eta_{eq}\right)
        + \sqrt{a_{eq}}\right]^2} \right\}y_k = 0,
\end{equation}

donde $C_{mat} \equiv \frac{1}{2}\sqrt{\frac{8\pi G}{3} \rho
  a^3}$. Podemos expresar la ecuación
\eqref{eq:y_eq_evolucion_materia}, de una manera similar a
\eqref{eq:y_eq_evolucion_inf}, usando el cambio de
variable $u = C_{mat}(\eta -\eta_{eq}) + \sqrt{a_{eq}}$,

\begin{equation}
  \label{field_evolution_matter_2}
  \frac{d^2 y}{du^2} +
  \left\{\kappa_k^2 -
    \frac{2}{u^2}\right\} y_k = 0
\end{equation}

con $\kappa_k \equiv \frac{k}{C_{mat}}$.

Todas estas ecuaciones tienen soluciones en sus respectivas
épocas, por lo que, la función completa para el modo $k$ del
campo escalar es

\begin{equation}
  \label{eq:y_evolucion_epocas}
  y_k(\eta) = \left\{ \begin{array}{lc}
      \alpha e^{-ik\eta}\left(1 - \frac{i}{k\eta}\right) +
      \beta   e^{ik\eta}\left(1 + \frac{i}{k\eta}\right)
      &\mbox{$-\infty<\eta < \eta_{ei}, $} \\
      \\
      -\frac{1}{2}\left(A e^{ik\eta} - B e^{ik\eta}\right)
      &\mbox{$\eta_{ei} 
        \le \eta < \eta_{eq}$} \\
      \\
      C  e^{-i\kappa_ku}\left(1 -
        \frac{i}{\kappa_k u}\right) +
      D   e^{i\kappa_ku}\left(1 +
        \frac{i}{\kappa_k u}\right) & \mbox{ $\eta_{eq} \le \eta$}
    \end{array} \right.
\end{equation}

Basado en puros argumentos físicos (ver capítulo 
\ref{cha:critica-al-origen}) se pueden determinar los
valores de $\alpha$ y $\beta$, el valor $\alpha =
1/\sqrt{2k}$ proviene de las restricciones que tienen que
satisfacer los modos del campo y su momento
conjugado (ecuación \ref{eq:restriccion_normalizacion}) y $\beta = 0$
del hecho de que queremos que el 
estado de vacío cuando $\eta \to -\infty$ sea el vacío de
Minkowski. Los coeficientes restantes pueden ser calculados
usando requisitos de continuidad, obteniendo,

\begin{equation}
  A =  -\frac{e^{-2ik\eta_{ei}}}{\sqrt{2k}k^2\eta_{ei}^2},
  \quad B =
  \sqrt{\frac{2}{k}}\left(1-\frac{i}{k\eta_{ei}}\right)-
  \frac{1}{\sqrt{2k}k^2\eta_{ei}^2},
\end{equation}

para la transición de las época inflación con la etapa dominada
por radiación. Para la transición de las etapas dominadas 
por radiación-materia,

\begin{gather}
  C=\frac{-ikB e^{-ik\eta_{eq}} - \displaystyle\frac{i
      C_{mat}^2}{2ka_{eq}\left(1+\frac{i}{\kappa
          \sqrt{a_{eq}}}\right)}
    \left(Ae^{ik\eta_{eq}}-Be^{-ik\eta_{eq}}\right)}
  {\displaystyle-2ike^{-i\kappa\sqrt{a_{eq}}}\left(1-  
      \frac{i}{\kappa\sqrt{a_{eq}}}\right) +
    \frac{ie^{-i\kappa \sqrt{a_{eq}}}C_{mat}^2}{k
      a_{eq}}\left(1 + \frac{\left(k\sqrt{a_{eq}} -
          iC_{mat}\right)^2}{k^2a_{eq} + C_{mat}^2}\right)}
  \\
  \nonumber \\
  D = \frac{\displaystyle-\frac{1}{2}Ae^{ik\eta_{eq}} +
    \frac{1}{2}Be^{-ik\eta_{eq}} - Ce^{-i\kappa u} \left(1 -
      \frac{i}{\kappa u}\right)}{\displaystyle e^{i\kappa
      u}\left(1+\frac{i}{\kappa u}\right)}
\end{gather}

Con esto completamos las ecuaciones generales que estaremos
usando a lo largo de este capítulo.

\section{Evolución con múltiples
  colapsos}\label{sec:multiples-colapsos}

Hasta ahora en esta investigación
\citep{Sudarsky06a,Unanue2008} se han investigado los
diferentes esquemas de colapso suponiendo que cada modo $k$
colapsa sólo una vez, esto ha facilitado los cálculos ya que
no ha habido necesidad de especificar la forma del estado
postcolapso, salvo su valor de expectación. En esta sección
desarrollaremos las fórmulas para dos o más colapsos sin
limitar la generalidad sobre el estado postcolapso. En la
sección \ref{sec:mult-colaps-coherente} se especializarán
las fórmulas desarrolladas aquí para el caso en que el
estado posterior al colapso sea un estado coherente, como el
estado inicial de vacío, esta suposición parece razonable ya
que los \textit{mecanismos de colapso}\footnote{No confundir
  \textit{mecanismos de colapso} con \textit{esquemas de
    colapso} , el primero es una descripción física de las
  ``entrañas'' del colapso -descripción fenomenológica o no-
  y el segundo es una ``receta'' de cómo podría
  pasar. Básicamente son dos niveles distintos de
  descripción.} preservan estas características del estado cuántico
resultante del colapso. 

Las ecuaciones para colapsos posteriores al primero se
pueden obtener como sigue. Primero, generalizaremos la
notación para poder tratar con los tres esquemas de colapso
a la vez. Las diversas ``recetas'' dadas por los esquemas de
colapso para el valor de expectación \textit{al tiempo de
  colapso} $\tc$ en el estado postcolapso $\ket{\xi}$,
pueden ser escritos de manera general para un colapso como

\begin{subequations}
  \label{eq:esquema_colapso_generalizado}
  \begin{equation}
    \expec{\yRI_k}^c \equiv
    \expec{\yRI_k\left(\eta_k^{c}\right)}_{\xi}   
    =  N_{y}^{R,I}\left(\tc\right) ,
  \end{equation}
  \begin{equation}
    \expec{\pyRI_k}^c \equiv
    \expec{\pyRI_k\left(\eta_k^{c}\right)}_{\xi} 
    =   N_{\pi}^{R,I}\left(\tc\right), 
  \end{equation}
\end{subequations}

donde $N_y$ y $N_\pi$ representan el valor aleatorio del
valor de expectación de $\yRI$ o $\pyRI$ que caracteriza al
estado postcolapso $\ket{\xi}$. Por ejemplo tomando el
esquema de colapso llamado ``independiente'' (ecuación
\ref{eq:esquema_independiente}) $N_y$ y $N_\pi$ están
definidos como

\begin{subequations}
  \label{eq:esquema_ejemplo}
  \begin{equation}
    \left\langle \hat
      y_k^{(R,I)}\left(\tc\right)\right\rangle_{\xi}
    =   \underbrace{(x_1')^{'(R,I)}
      \sqrt{\left(\Delta y_k^{(R,I)}\left(\tc\right)\right)_{0} ^2}}_{N_{y}},
  \end{equation}
  \begin{equation}
    \left\langle \hat
      \pi_k^{(R,I)}\left(\tc\right)\right\rangle_{\xi}
    =   \underbrace{(x_2')^{(R,I)}
      \sqrt{\left(\Delta
          \pi_k^{(R,I)}\left(\tc\right)\right)_{0}^2}}_{N_{\pi}}. 
  \end{equation}
\end{subequations}

Es importante enfatizar que el esquema de colapso gobierna a
qué  valor ``salta'' el valor de expectación en el nuevo
estado postcolapso. Luego del colapso, el valor de
expectación del modo $\yRI_k$ y $\pyRI_k$ evolucionarán
siguiendo las ecuaciones (\ref{eq:expectacion_pi_general}) y
(\ref{eq:expectacion_y_general}) respectivamente.

Existe la posibilidad que durante la evolución posterior al
primer colapso se diesen las condiciones para que ocurriese
un nuevo colapso. Las condiciones dependerán del
\emph{mecanismo de colapso} y no se especificarán aquí.  La
diferencia entre este nuevo colapso y 
el primero será que el valor de expectación del nuevo estado
no es cero. Esto se reflejará en las recetas dadas por los
esquemas para el colapso $\ket{\xi_1} \to \ket{\xi_2}$ como
sigue:

\begin{subequations}
  \label{eq:esquema_generalizado_2}
  \begin{equation}
    \expec{\yRI_k}_2^c \equiv  \left\langle \hat
      y_k^{(R,I)}\left(\eta_k^{c_2}\right)\right\rangle_{\xi_2}
    = N^{(R,I)}_{y_1}\left(\tcn{2}\right) +
    \expec{\yRI_k\left(\tcn{2}\right)}_{\xi_1}, 
  \end{equation}
  \begin{equation}
    \expec{\pyRI}_2^c \equiv \left\langle \hat
      \pi_k^{(R,I)}\left(\eta_k^{c_2}\right)\right\rangle_{\xi_2}
    =   N^{(R,I)}_{\pi_1}\left(\tcn{2}\right) +
    \expec{\pyRI\left(\tcn{2}\right)}_{\xi_1}.
  \end{equation}
\end{subequations}

Obsérvese que el lado izquierdo de la ecuación está en el
estado post-colapso $\ket{\xi_2}$, mientras que el lado
derecho está en el estado $\ket{\xi_1}$, i.e.\ , el nuevo
estado post-colapso depende del estado pre-colapso. Nótese
también que toda esta expresión está evaluada en $\tcn{2}$,
el tiempo del segundo colapso. El segundo término del lado
derecho es el valor de expectación posterior al primer
colapso \textit{evolucionado} desde $\tc$ hasta $\tcn{2}$
mediante las ecuaciones (\ref{eq:expectacion_y_general}) y
(\ref{eq:expectacion_pi_general}) respectivamente. Las
ecuaciones de evolución entonces reciben como
\textit{condición inicial} el valor del colapso al tiempo
$\tc$ y luego evolucionan sin sobresaltos hasta que ocurre
un nuevo colapso.

Siguiendo el mismo procedimiento descrito, podemos escribir
las ``recetas'' de colapso para el $n-$ésimo colapso
$\ket{\xi_{n-1}} \to \ket{\xi_n}$:

\begin{subequations}
  \label{eq:esquema_generalizado_n}
  \begin{equation}
    \expec{\yRI_k}_n^c
    = N^{(R,I)}_{y_{n-1}}\left(\tcn{n}\right) +
    \expec{\yRI_k\left(\tcn{n}\right)}_{\xi_{n-1}}, 
  \end{equation}
  \begin{equation}
    \expec{\pyRI}_n^c
    =   N^{(R,I)}_{\pi_{n-1}}\left(\tcn{n}\right) +
    \expec{\pyRI\left(\tcn{n}\right)}_{\xi_{n-1}}.
  \end{equation}
\end{subequations}

El segundo término del lado derecho de las ecuaciones
\eqref{eq:esquema_generalizado_n} son los valores de
expectación del $(n-1)$-ésimo colapso \textit{evaluado al
  $n-$ésimo tiempo de colapso } $\tcn{n}$. Es necesario
hacer hincapié en esto porque los valores de expectación del
estado $\ket{\xi_{n-1}}$ \textit{evolucionó} desde
$\tcn{n-1}$ hasta $\tcn{n}$ de acuerdo con las ecuaciones
(\ref{eq:expectacion_y_general}) y
(\ref{eq:expectacion_pi_general}) usando como
\textit{condición inicial el valor de expectación dado por
  el colapso} $n-2$, que a su vez evolucionó\footnote{Nótese
  que esto es exactamente la evolución de un sistema cuántico
  siguiendo las reglas estándar de mecánica cuántica, en el
  intervalo entre mediciones la función de onda evoluciona
  siguiendo la ecuación de Schrödinger y en los tiempos en
  los cuales ocurren las mediciones existen reducciones de
  la función de onda, que luego de la medición evolucionará
  siguiendo la ecuación de Schrödinger con la condición
  inicial del nuevo estado, etc.\ .} siguiendo, etc.\ . Esto
nos llevará a una relación recursiva de la ecuación de
movimiento para el valor de expectación de las variables del
campo luego de $n$ colapsos.

Las ecuaciones (\ref{eq:expectacion_y_general}) y
(\ref{eq:expectacion_pi_general}) son válidas en el periodo
entre dos colapsos consecutivos, e.g.\ ,  $n$ y $n-1$, lo que
significa que estas ecuaciones son válidas desde el tiempo
$\tcn{n-1}$ (su condición inicial) hasta $\tcn{n}$. Ahora
que ya tenemos la ``receta'' para el $n-$ésimo colapso,
podemos incluir \textit{todos} los colapsos, desde el
primero hasta el $n-$ésimo en las ecuaciones de evolución.
Las ecuaciones de evolución del valor de expectación del
campo y su momento conjugado \textit{después} del $n-$ésimo
colapso se pueden escribir como sigue
\begin{subequations}
  \label{eq:evolucion-ve-2}
  \begin{equation}
    \label{eq:evolucion-vepi-2}
    \left\langle \pi_k^{R,I} (\eta)\right\rangle =   \left\langle
      \pi_k^{R,I} \right\rangle_n^c \tilde{A}_n(\eta) +
    \tilde{B}_n(\eta) 
    \left\langle y_k^{R,I} 
    \right\rangle_n^c
  \end{equation}
  \begin{equation}
    \label{eq:evolucion-vey-2}
    \left\langle y_k^{R,I}(\eta) \right\rangle = \left\langle
      \pi_k^{R,I} \right\rangle_n^c \tilde{C}_n(\eta) +
    \tilde{D}_n(\eta) 
    \left\langle y_k^{R,I} 
    \right\rangle_n^c ,
  \end{equation}
\end{subequations}
donde las funciones $\tilde{A}_n, \tilde{B}_n, \tilde{C}_n$
y $\tilde{D}_n$ son las variables que marcan la evolución
temporal del sistema en el lapso que va desde $\tcn{n}$
hasta $\eta$. En particular, durante la época inflacionaria
(\ref{eq:expectacion_inflacion}) son
\begin{subequations}
  \label{eq:operadores_evolucion}
  \begin{equation}
    \tilde{A}_n(\eta) = \cos\Delta_n + \frac{\sin\Delta_n}{z_n} ,     
  \end{equation}
  \begin{equation}
    \tilde{B}_n(\eta) = - k\sin\Delta_n     
  \end{equation}
  \begin{equation}
    \tilde{C}_n(\eta) = \frac{\cos\Delta_n}{k}\left( \frac{1}{z} -
      \frac{1}{z_n}\right) +
    \frac{\sin\Delta_n}{k}\left(\frac{1}{z \cdot z_n} + 1
    \right),     
  \end{equation}
  \begin{equation}
    \tilde{D}_n(\eta) = \cos\Delta_n - \frac{\sin\Delta_n}{z}.    
  \end{equation}
\end{subequations}
donde $\Delta_n$ la variable del lapso transcurrido desde el
colapso $\tcn{n}$ hasta $\eta$ multiplicada por el vector de
onda $k$ i.e.\ $\Delta_n \equiv k(\eta - \tcn{n}) = z -z_n$,
$z \equiv k\eta$ y $z_n \equiv k\tcn{n}$. Para estados
intermedios, i.e.\ el estado $\ket{i}$ con $i \neq n$, hay
que notar que los $z = k\eta$ deben de sustituirse por
$z_{j=i+1}$ y las $\Delta_n = z - z_n$ por $\Delta_{ji} =
z_j-z_i$.

Para obtener la ecuación de evolución del valor de
expectación luego de $n$ colapsos sustituimos en las
ecuaciones de evolución (\ref{eq:evolucion-ve-2}), las
``recetas'' del colapso $n-$ésimo
(\ref{eq:esquema_generalizado_n}) en las cuales hay que
sustituir en el segundo término del lado derecho las
ecuaciones (\ref{eq:evolucion-ve-2}) pero ahora para el
colapso $n-1$ , etc.\ . Haciendo esto iterativamente
llegamos a la siguiente expresión para la evolución del
valor de expectación del momento conjugado $\pyRI_k$ luego
de la ocurrencia de $n$ colapsos\footnote{Hemos removido la
  tilde de las variables $\tilde{A}$, $\tilde{B}$,
  $\tilde{C}$ y $\tilde{D}$ para no cargar más la notación.}
\begin{align}
  \label{eq:multicollapse_mean_value_pi}
  \left\langle\hat\pi_k^{R,I}\left(\eta\right)\right\rangle
  &= N_{\pi_{n-1}}^{R,I} A_n  \nonumber \\
  &+ N_{\pi_{n-2}}^{R,I}\left(A_{n-1}A_n + C_{n-1}B_n\right)  \nonumber \\
  &+ N_{\pi_{n-3}}^{R,I}\left[A_{n-2}\left(A_{n-1}A_n +
      C_{n-1}B_n\right) + C_{n-2}\left(B_{n-1}A_n +
      D_{n-1}B_n\right)\right] \nonumber \\
  &+
  N_{\pi_{n-4}}^{R,I}\Big\{A_{n-3}\Big[A_{n-2}\left(A_{n-1}A_n
    + C_{n-1}B_n\right) + C_{n-2}\left(B_{n-1}A_n +
    D_{n-1}B_n\right)\Big]  \nonumber \\
  &\qquad + C_{n-3} \Big[B_{n-2}\left(A_{n-1}A_n +
    C_{n-1}B_n\right) + D_{n-2}\left(B_{n-1}A_n +
    D_{n-1}B_n\right)\Big]\Big\} \nonumber \\
  & + \ldots \nonumber \\
  &+ N_{y_{n-1}}^{R,I} B_n \nonumber \\
  &+ N_{y_{n-2}}^{R,I}\left(B_{n-1}A_n + D_{n-1}B_n\right) \nonumber \\
  &+ N_{y_{n-3}}^{R,I}\left[B_{n-2}\left(A_{n-1}A_n +
      C_{n-1}B_n\right) + D_{n-2}\left(B_{n-1}A_n +
      D_{n-1}B_n\right)\right] \nonumber \\
  &+
  N_{y_{n-4}}^{R,I}\Big\{B_{n-3}\Big[A_{n-2}\left(A_{n-1}A_n
    + C_{n-1}B_n\right) + C_{n-2}\left(B_{n-1}A_n +
    D_{n-1}B_n\right)\Big]  \nonumber \\
  &\qquad + D_{n-3} \Big[B_{n-2}\left(A_{n-1}A_n +
    C_{n-1}B_n\right) + D_{n-2}\left(B_{n-1}A_n +
    D_{n-1}B_n\right)\Big]\Big\} \nonumber \\
  & + \ldots \nonumber \\
\end{align}

con una expresión análoga para $\expec{\yRI_k(\eta_k)}$.
Debido a la forma particular de las ``variables de
evolución'' (\ref{eq:operadores_evolucion}) en la época
inflacionaria se puede demostrar que
\begin{equation}
  A_iA_j + C_iB_j =   \mathbf{A}_i \equiv \cos\Delta_i + \frac{\sin\Delta_i}{z_i}, 
\end{equation}
\begin{equation}
  B_iA_j + D_iB_j = \mathbf{B}_i \equiv -k \sin\Delta_i,
\end{equation}
para $j > i$. Nótese que las variables $A_n$ y
$B_n$ tienen la misma forma que $\mathbf{A}_n$ y
$\mathbf{B}_n$. Entonces, la ecuación
(\ref{eq:multicollapse_mean_value_pi}) se puede escribir
como
\begin{align}
  \label{eq:valor_expectacion_pi_multicolapso}
  \expec{\pyRI(\eta)} &= N_{\pi_{n-1}}^{R,I} \mathbf{A}_n +
  N_{y_{n-1}}^{R,I} \mathbf{B}_n  \nonumber \\
  &+ N_{\pi_{n-2}}^{R,I} \mathbf{A}_{n-1} +
  N_{y_{n-2}}^{R,I} \mathbf{B}_{n-1}  \nonumber \\
  &+ N_{\pi_{n-3}}^{R,I} \mathbf{A}_{n-2} +
  N_{y_{n-3}}^{R,I} \mathbf{B}_{n-2} +
  \ldots \nonumber \\
  &= \sum_{i=0}^n N_{\pi_{n-1-i}}^{R,I} \mathbf{A}_{n-i} +
  N_{y_{n-1-i}}^{R,I} \mathbf{B}_{n-i}.
\end{align}
La fórmula (\ref{eq:valor_expectacion_pi_multicolapso})
muestra la generalización de la ecuación
(\ref{eq:expectacion_pi_general}) para varios colapsos en la
época inflacionaria. La evolución para el valor de
expectación de $\pyRI$, es como la superposición de varias
evoluciones de un único colapso, donde en cada evolución
el colapso sucede a tiempos diferentes.

El momento $\dphih_k'$ está relacionado con $\piy_k$ mediante
$\dphih_k' = \piy_k/a$,
\begin{equation}
  \expec{ \dphih_k'}= \frac{1}{a(\eta)} \left[ \expec{\pyR_k}
    + i \expec{\pyI_k}\right].
\end{equation}

La cantidad que nos interesa es el valor medio del producto
de está última ecuación, ya que con ella podemos obtener las
predicciones de los esquemas de colapso.

Usando la misma descomposición que al final de la sección
\ref{sec:ecuaciones-generales}, $N_\pi^{R,I} = f_k^{R,I}
u_k$ y $N_y^{R,I} = h_k^{R,I} w_k$ donde $f,h$ son variables
aleatorias  adimensionales, y $u,w$ son la parte de la
receta de colapso que carga con las unidades (por ejemplo,
para el primer esquema de colapso en su primer colapso, $u_k
= \sqrt{\hbar L^3/2}|g_k(\eta)|$ y $w_k = \sqrt{\hbar
  L^3/2}|y_k(\eta)|$). Entonces, el valor medio del
ensamblen del cuadrado de $N_\pi^{R,I}$ are
$\expec{N_{\pi_k}^R N_{\pi_{k'}}^R\,^*} = (\delta_{k,k'} +
\delta_{k,-k'})u_ku_k^*$ y $\expec{N_{\pi_k}^I N_{\pi_{k'}}^I
  \,^*} = (\delta_{k,k'} - \delta_{k,-k'})u_ku_k^*$ y
expresiones similares para $N_y^{R,I}$.

Así, el valor medio del ensamble del cuadrado del momento
conjugado $\dphih_k'$ para $n$ colapsos es
\begin{equation}
  \label{eq:cuadrado_valor_medio_n_colapsos}
    \vme{\expec{\dphih_k\thinspace '}\expec{\dphih_{k'}\thinspace
        '\thinspace^*}} =  
 \frac{2}{a(\eta)^2} \sum_{i=0}^n \left[ |u_k|_{n-1-i}^2
 \mathbf{A}_{n-i,k}^2 + |w_k|_{n-1-i}^2 \mathbf{B}_{n_i,k}^2
 \right].
\end{equation}

En particular para dos colapsos
\begin{equation}
  \label{eq:valor_medio_dos_colapsos}
  \vme{\expec{\dphih_k\thinspace'}\expec{\dphih_{k'}\thinspace
      '\thinspace^*}} = 
  \frac{2}{a(\eta)^2} \Big\{\mathbf{A}_{2,k}^2|u_k|^2_1 +
  \mathbf{A}^2_{1,k}|u_k|_0^2 
  + \mathbf{B}_{2,k}^2|w_k|_1^2 +
  \mathbf{B}_{1,k}|w_k|_0^2\Big\},
\end{equation}
donde $|u_k|^2_n \equiv |u_k(\eta_k^{c_n})|^2$ y $|w_k|^2_n
\equiv |w_k(\eta_k^{c_n})|^2$ siendo $n$ el número de
colapsos ocurridos. Siguiendo la misma notación el valor
medio para tres colapsos es
\begin{equation}
  \label{eq:valor_medio_tres_colapsos}
  \vme{\expec{\dphih_k\thinspace
      '}\expec{\dphih_{k'}'\thinspace^*}} = 
  \frac{2}{a(\eta)^2} \Big\{ \mathbf{A}_3^2|u_k|_2^2 +
  \mathbf{A}_2^2|u_k|_1^2 + \mathbf{A}_1^2|u_k|_0^2 +
  \mathbf{B}_3^2|w_k|_1^2 + \mathbf{B}_2^2|w_k|_1^2 + 
  \mathbf{B}^2_1 |w_k|_0^2\Big\}.
\end{equation}

Si utilizamos los esquemas de colapso
(\ref{eq:esquema_independiente}) y  (\ref{eq:esquema_newtoniano})
en la ecuación (\ref{eq:cuadrado_valor_medio_n_colapsos}) para
el caso de un solo colapso obtenemos las ecuaciones
(\ref{eq:C_sudarsky_inflation}) y (\ref{eq:C_2_inflation})
respectivamente.

\section{Nuevo esquema de colapso: \textit{à la}
  Wigner\footnote{Toda está sección fue presentada en el
    artículo de A. De Un\'anue y D. Sudarsky publicado en
    \textit{Phys. Rev. D} 0801,
    2008.}}\label{sec:colapso_wigner}

Antes de indicar el esquema de colapso y sus consecuencias,
revisaremos algunos conceptos preliminares sobre el
funcional de Wigner, sus propiedades, su límite clásico y su
forma para campos cuánticos.

\subsection{Distribución de Wigner}

El concepto de \textit{espacio de fase} es muy útil en la
física clásica, el cual se puede entender como una
interpretación geométrica de las ecuaciones de Hamilton, al
considerar las variables canónicas generalizadas $(\irvec{q},\irvec{p})$
de un sistema de $n$ partículas como un solo punto
en una variedad $2n$-dimensional llamado espacio de fase
\cite{BlochWalecka00,JoseSaletan02}, así el problema de
mecánica clásica se ve reducido al estudio de la trayectoria
de un solo punto en el espacio de fase. Cuando el número de
partículas es muy grande (del orden de $n \simeq 10^{23}$)
es necesario utilizar métodos estadísticos, para esto, se
utiliza la \textit{función de densidad de probabilidad} o
\textit{función de distribución del espacio de fase}$f(\bvec{q},\bvec{p})$ la cual determina la probabilidad de
encontrar el conjunto de variables que describe el sistema
$(\bvec{q},\bvec{p})$ en una región infinitesimal de volumen
$d\lambda = \Pi_{i=1}^{n} dq_idp_i$, $dP =
f(\bvec{q},\bvec{p})d\lambda$. 

Sería deseable tener en
mecánica cuántica un ``espacio de fase cuántico''. Este
deseo presenta de inicio varias dificultades, por ejemplo,
debido al principio de incertidumbre de Heisenberg, no es
posible asignar a una partícula o sistema una posición $\bvec{q}$ y
un momento $\bvec{p}$ \textit{simultáneos}, dificultad que
se puede evadir  definiendo como soporte no puntos (como en
el caso clásico) si no áreas que dividan
el espacio de fase en áreas de orden $2\pi\hbar$
de tal manera que el producto de las incertidumbres  de $q$
y $p$ no exceda $1/2 \hbar$ \cite{Ballentine2000}. En el
artículo seminal \cite{Wigner1932}, E. P. Wigner 
propuso una función de distribución para describir un
sistema cuántico en el espacio de fase usando el operador de
densidad\footnote{En la referencia \cite{Hillery1984} la
  función de distribución de Wigner está definida con un
  factor de 2 extra:
  \begin{equation*}
  \Wigner(q,p) =
  \frac{1}{\pi\hbar}\int_{-\infty}^{\infty} d\,y \, \left\langle 
  q-y\Big|\hat{\rho}\Big|q+y\right\rangle 
  \exp{\left({\frac{2ipy}{\hbar}}\right)} 
  \end{equation*}

}\footnote{Es posible reescribir las reglas de
  mecánica cuántica usando el formalismo del operador de
  densidad \citep[sección 6.4 de ][]{dEspagnat99}. El operador de densidad
  $\hat\rho$ se define como sigue [\textit{ibid.}]:
  \textit{Sean $E_1, \ldots 
    E_\alpha \ldots E_\mu$ con  $\mu$ ensambles de sistemas
    físicos del mismo tipo, sea $N_\alpha$ el número de
    elementos de $E_\alpha$, sea $\tilde{E}$ el ensamble de
    todos los $N = N_1+ \ldots + N_\alpha +\ldots + N_\mu$
    elementos de los diversos $E_\alpha$. Supongamos además,
  que cada $E_\alpha$ puede ser descrito por un ket
  normalizado $\ket{\phi_\alpha}$, entonces el operador}
\begin{equation*}
  \hat\rho = \sum_{\alpha = 1}^\mu \ket{\phi_\alpha}
  \frac{N_\alpha}{N} \bra{\phi_\alpha}
\end{equation*}
\textit{es llamado ``el operador estadístico del ensamble
$\tilde{E}$'', también conocido como ``operador de
densidad'' o ``matriz de densidad''} El operador de densidad
tiene como propiedades
que (i) $\hat\rho$ es hermítico, (ii) $\hat\rho$ es positivo
definido, (iii) $Tr(\hat\rho) = 1$, (iv) $Tr(\hat\rho^2) \le
1$ donde la igualdad se cumple si $\hat\rho$ es un operador
de proyección. Si un ensamble $\tilde{E}$ se puede describir
con un vector $\ket{\Psi}$ entonces es llamado
\textit{estado puro} y puede ser descrito tanto por
$\ket{\Psi}$ como por $\hat\rho \equiv
\ket{\Psi}\bra{\Psi}$, el cual es un operador de
proyección. Si el ensamble $\tilde{E}$ contine elementos de
diversos sub-ensambles  $E_\alpha$ se le conoce como
\textit{mezcla} (mixture en inglés). 
} $\hat\rho$ \cite{Ballentine2000,Omnes94,dEspagnat99}

\begin{equation}\label{eq:wigner}
  \Wigner(q,p) =
  \frac{1}{2\pi\hbar}\int_{-\infty}^{\infty} dy \, \left\langle 
  q-\frac{1}{2}y\Big|\hat{\rho}\Big|q+\frac{1}{2}y\right\rangle 
  \exp{\left({\frac{ipy}{\hbar}}\right)} 
\end{equation}

donde $p$ es el momento conjugado de $y$. Si el estado es
puro, $\hat\rho \equiv 
|\Psi\rangle\langle\Psi|$ la función de distribución toma la
forma 

\begin{equation}\label{eq:wigner_fdo}
  \Wigner(q,p) =
  \frac{1}{2\pi\hbar}\int_{-\infty}^{\infty} dy\,
  \Psi\left(q-\frac{1}{2}y\right)\Psi^*\left(q+\frac{1}{2}y\right)
  \exp{\left({\frac{ipy}{\hbar}}\right)} 
\end{equation}

Estos resultados son para el caso unidimensional, la
generalización al caso multidimensional es directa: el
término $2\pi\hbar$ deberá ser reemplazado por
$(2\pi\hbar)^n$ donde $n$ es el número de variables de
$\Psi$, y el producto $py$ de la exponencial es cambiado por
el producto escalar de los vectores $n-$dimensionales $p$ y
$y$.

\paragraph{Propiedades de la función de Wigner.}
Se listan a continuación las propiedades de la distribución
de Wigner \cite{Hillery1984},

\renewcommand{\labelenumi}{\roman{enumi}.}

\begin{enumerate}
\item $\Wigner(q,p)$ con cumple
  \begin{subequations}
    \begin{equation}
      \int d\,p\Wigner(q,p) = |\Psi(q)|^2 =
      \bra{q}\hat\rho\ket{q} 
    \end{equation}
    \begin{equation}
      \int d\,q\Wigner(q,p) = \bra{p}\hat\rho\ket{p}
    \end{equation}
    \begin{equation}
      \int d\,p \int d\,q\Wigner(q,p) = Tr(\hat\rho) = 1
    \end{equation}
  \end{subequations}
\item $\mathcal{W}(q,p)$ es invariante ante
  transformaciones de Galileo, i.e.\ $\psi(q) \to \psi(q+a)$
  entonces $\Wigner(q,p) \to \Wigner(q+a, p)$ y si $\psi(q)
  \to e^{ip'q/\hbar}\psi(q)$ entonces $\Wigner(q,p) \to
  \Wigner(q, p-p')$.
\item $\mathcal{W}(q,p)$ debe de ser invariante ante
  inversiones temporales.
\item Si $\Wigner_\phi(q,p)$ y $\Wigner_\psi(q,p)$ corresponden a los
  estados $\phi(q)$ y $\psi(q)$, respectivamente, se tiene
  \begin{equation}
    \Bigg|\int\phi^*(q)\psi(q)\Bigg|^2 = (2\pi\hbar)\int
    d\,q \int d\,p \Wigner_\phi(q,p) \Wigner_\psi(q,p).
  \end{equation}
  Esta propiedad tiene dos consecuencias interesantes. Si
  $\phi(q) = \psi(q)$ entonces
  
    \begin{equation}
      \int d\,q  \int d\,p \, \left[\Wigner_\phi(q,p)\right]^2 =
      \frac{1}{2\pi\hbar},
    \end{equation}
  y si $\phi$ y $\psi$ son ortogonales  
    \begin{equation}
      \int d\,q \int d\,p \, \Wigner_\phi(q,p) \Wigner_\psi(q,p) = 0,
    \end{equation}
    lo cual implica que $\Wigner(q,p)$ no puede ser positiva
    en todos lados del espacio de fase. Esta conclusión es
    general \cite{Wigner1971}: si la función de distribución satisface la
    propiedad (i) entonces asumirá valores negativos para
    algunos $p$ y $q$ \footnote{La función de Husimi
      $\mathcal{H}(q,p)$  no
      satisface la propiedad (I) y es positiva en  todo el
      espacio de fase \citep[sección 15.3 de ][]{Ballentine2000}.}. 

\item La representación de Wigner de un operador cuántico
  $R_w(p,q)$ es
  \begin{equation} \label{eq:operadores_wigner} R_w(p,q) =
    \int
    \bbra{q-\frac{1}{2}y}\hat{R}\bket{q+\frac{1}{2}y}\,
    \exp{\left(\frac{ipy}{\hbar}\right)}. 
  \end{equation}
  Recordando que el promedio de la variable dinámica $R$ en
  el estado $\rho$ está dada por $\langle R \rangle =
  Tr(\hat{\rho}\hat{R})$
  \begin{equation}
    \langle R \rangle = Tr(\hat\rho \hat R) = \int\int
    \Wigner(q,p) R_w(q,p) dp \, dq.
  \end{equation}
\item Definiendo la transformada de Fourier de la función de
  onda
  \begin{equation}
    \phi(p) = \frac{1}{2\pi\hbar}\int d\, q \psi(q) \exp\left(\frac{-iqp}{\hbar}\right)    ,
  \end{equation}
  la ecuación (\ref{eq:wigner}) puede reescribirse de la
  manera
  \begin{equation}\label{eq:wigner_momento}
    \Wigner(q,p) = \frac{1}{2\pi\hbar}\int dk \,
    \bbraketd{\hat\rho}{p-\frac{1}{2}k}{p+\frac{1}{2}k}
    \exp{\left({\frac{-ipy}{\hbar}}\right)} 
  \end{equation}
\end{enumerate}

En \cite{Wigner1971} se muestra que las propiedades I-IV
determinan la función de distribución de manera única: la
dada por la ecuación (\ref{eq:wigner}).

\paragraph{Dinámica y límite clásico.}
La dinámica de la función de Wigner (toda esta sección
está basada en el capítulo 5 \cite{Ballentine2000} y la
sección 2.3 de \cite{Hillery1984}), puede
ser deducida de la ecuación de evolución del operador de
estado $\hat\rho$,

\begin{equation*}
  \frac{d\hat\rho}{dt} = \frac{i}{\hbar}[\hat\rho, \hat{H}] =
  \frac{i}{\hbar}(\hat\rho \hat{H} - \hat{H}\hat\rho), 
\end{equation*}

donde $\hat{H} = \hat{p}^2/2M + \hat{V}$ es el Hamiltoniano del sistema. Es
conveniente reescribir esta ecuación separando las
contribuciones de la energía cinética y la potencial.

\begin{equation*}
  \frac{d\hat\rho}{dt} \equiv \frac{\partial_K \hat\rho}{\partial t} +
  \frac{\partial_V  \hat\rho}{\partial t},
\end{equation*}

siendo

\begin{subequations}
  \begin{equation}\label{eq:wigner_cinetico}
    \frac{\partial_{K}\hat\rho}{\partial t} \equiv \frac{i}{2M\hbar}(\hat\rho \hat{p}^2 -\hat{p}^2\hat\rho),
  \end{equation}
  \begin{equation}\label{eq:wigner_potencial}
    \frac{\partial_{V}\rho}{\partial t} = \frac{i}{\hbar}(\hat\rho \hat{V}
    - \hat{V}\hat\rho),
  \end{equation}
\end{subequations}

Resulta conveniente evaluar \eqref{eq:wigner_cinetico} en el
espacio de momentos
\begin{align}
  \frac{\partial_K}{\partial t} \braketd{\hat\rho}{p}{p'}&=
  \frac{i}{2M\hbar}  \braketd{\hat\rho}{p}{p'}(p'^2-p^2) \\
  &= \frac{i}{2M\hbar}\braketd{\hat\rho}{p}{p'}(p'+p)(p'-p).
\end{align}

Usando \eqref{eq:wigner_momento} para transformar a la
representación de Wigner, obtenemos

\begin{equation}
  \frac{\partial_K}{\partial t}\Wigner(q,p,t) =
  \frac{i}{\hbar M} \int \bbraketd{\hat\rho}{p
    -\frac{1}{2}k}{p+\frac{1}{2}k} \,pk\,
  \exp\left(\frac{-iqk}{\hbar}\right) 
  d\,k,
\end{equation}

el factor de $k$ dentro de la integral puede ser reemplazada por el
operador $-\left(\dfrac{\hbar}{i}\right)\dfrac{\partial}{\partial q}$
obteniendo 

\begin{equation} \label{eq:wigner_dinamica_cinetico}
  \frac{\partial_K}{\partial t}\Wigner(q,p,t) =
  -\frac{p}{M}\dpar{}{q}\Wigner(q,p,t).
\end{equation}

Realizando pasos similares para $\dfrac{\partial_V \hat\rho}{\partial
t}$, pero en la representación de posiciones: 

\begin{equation*}
  \frac{\partial_V}{\partial t}\bra{x}\hat\rho\ket{x'} =
  \frac{i}{\hbar}\bra{x}\hat\rho\ket{x'}
  \left[V\left(x\right)-V\left(x'\right)\right],  
\end{equation*}
usando \eqref{eq:wigner}
\begin{equation*}
  \frac{\partial_V}{\partial t}\Wigner(q,p,t) =
  \frac{i}{\hbar(2\pi\hbar)}\int
  \bbraketd{\hat\rho}{q-\frac{1}{2}y}{p+\frac{1}{2}y}
  \left[V\left(q+\frac{1}{2}y\right)- 
    V\left(q-\frac{1}{2}y\right)\right]e^{ipy/\hbar} dy.  
\end{equation*}
Si $V(x)$ es analítica se puede expander en Series de Taylor
\begin{equation*}
V\left(q+\frac{1}{2}y\right)- 
    V\left(q-\frac{1}{2}y\right) =
  \sum_{n=impar}\frac{2}{n!}\,y^n\,\frac{d^nV(q)}{dq^n}
\end{equation*}
Como antes, podemos reemplazar $y^n$ dentro de
    la integral por
    $\left[\left(\dfrac{\hbar}{i}\right)\left(\dfrac{\partial}{\partial
          p}\right)\right]^n$, 
     fuera de la integral
\begin{equation}\label{eq:wigner_dinamica_potencial}
 \frac{\partial_V}{\partial t}
  \Wigner(q,p,t) =
  \sum_{n=impar}\frac{1}{n!}(-i\hbar)^{n-1}\frac{d^nV(q)}{dq^n}
  \frac{\partial^n}{\partial p^n} \Wigner(q,p,t) 
\end{equation}

Sumando las ecuaciones \eqref{eq:wigner_dinamica_cinetico} y
\eqref{eq:wigner_dinamica_potencial},

\begin{equation}
  \label{eq:wigner_temporal}
  \frac{\partial}{\partial t}\Wigner(q,p,t) =
  -\frac{p}{M}\frac{\partial}{\partial q}\Wigner(q,p,t) +
  \sum_{n=impar}\frac{1}{n!}(-i\hbar)^{n-1}
  \frac{d^nV(q)}{dq^n}\frac{\partial^n}{\partial   
    p^n}\Wigner(q,p,t),
\end{equation}

la suma de términos en potencias de $\hbar$ podría
sugerir que esta ecuación tiene un límite clásico fácil de
extraer. 

\begin{equation}
  \label{eq:wigner_temporal_liouville}
  \frac{\partial}{\partial t}\Wigner(q,p,t) = 
  -\frac{p}{M}\frac{\partial}{\partial q}\Wigner(q,p,t) +
  \frac{dV(q)}{dq}\frac{\partial}{\partial p}\Wigner(q,p,t)
  + \mathscr{O}(\hbar^2) 
\end{equation}

Despreciando la contribución de las correcciones
$\mathscr{O}(\hbar^2)$, esta es la ecuación de
Liouville. Pero esta apariencia clásica es engañosa, los
términos de orden $\hbar^n$ involucran la derivada $n-$ésima de
$\mathscr{W}(q,p,t)$ con respecto a $p$. Esto puede generar
factores de $1/\hbar$ y cancelar\footnote{El caso de dos ondas gaussianas
  separadas, es un ejemplo claro de esto:
  \begin{equation}
    \Psi(q) = \frac{N}{\sqrt{2}(2\pi a^2)^{1/4}}\left[
      e^{-(q-c)^2/4a^2} + e^{-(q+c)^2/4a^2} \right]
  \end{equation}
} así los factores explícitos
de $\hbar$. En estos casos las correcciones no desaparecen en el
límite $\hbar \to 0$, ver \citet{Ballentine2000} para una
discusión más extensa.

\paragraph{Distribución de Wigner para campos cuánticos.}

A cualquier estado de $N$ partículas $\ket{\Psi_N}$ en el
espacio de Fock le corresponde la función de onda de $N$
partículas \cite{Hillery1984} dada por

\begin{equation}\label{eq:funcion_onda}
  \Psi_N(r_1, r_2, \ldots, r_N) = \frac{1}{\sqrt{N!}}\bra{0}\hat\psi(r_1)\ldots\hat\psi(r_N) \ket{\Psi_N}
\end{equation}

donde $\hat\psi(r_i)$ es el operador de campo cuántico
\citep[ver][para una explicación más
detallada]{Peskin1995}. Nótese que la función de onda está
expresada en la base del espacio de Fock, en este caso
tenemos que la función de distribución para el estado
$\ket{\Psi_N}$, es

\begin{align}
  \label{eq:wigner_segunda_cuantizacion}
  \Wigner(r_1,\ldots r_N;p_1,\ldots p_N) &=
  \left(\frac{1}{2\pi\hbar}\right)^{3N}\int\cdots\int
  d^3\,y_1 \ldots d^3\,y_N \,
  \exp{\left[\frac{i}{\hbar}p\cdot
      y\right]}\times \\
  &\Psi_N^*(r_1+\frac{1}{2}y_1,\ldots
  r_N+\frac{1}{2}y_N)\Psi_N(r_1-\frac{1}{2}y_1,\ldots
  r_N-\frac{1}{2}y_N)
\end{align}

Como se puede observar, esta forma de la función de
distribución de Wigner es para estados con un número de
partículas definido. Lo que se necesita
es calcular la función de onda en la base de $\y$.

\subsection{Función de onda del campo escalar $\y$}
\label{sec:fdo-wigner}

Para obtener la distribución de Wigner del campo del
inflatón primero obtendremos la función de onda, recordemos
antes que los operadores hermitianos son $\yRI$ y $\pyRI$,
dados por \eqref{eq:def_hermitiana}.

Si usamos la relación \eqref{eq:restriccion_normalizacion}
es posible despejar $\annRI_k$.

\begin{equation}
  \label{eq:a_RI}
  \frac{i}{\sqrt{2}}\annRI_k =
  \pyRI_k(\eta)y_k^*(\eta)
  -\yRI_k(\eta)g_k^*(\eta) .
\end{equation}

Tomando el complejo conjugado de
\eqref{eq:vacio_BunchDavies} y
\eqref{eq:vacio_BunchDavies_momento} tenemos

\begin{equation*}
  \annRI_k  = \sqrt{2}\left\{ i
    \sqrt{\frac{1}{2k}}\left(1 + \frac{i}{k\eta}\right)
    e^{ik\eta}\pyRI_k - i^2 \sqrt{\frac{k}{2}} e^{ik \eta}\yRI_k\right\},
\end{equation*}

acomodando términos,

\begin{equation*}
  \annRI_k  = i \alpha \pyRI_k + \beta\yRI_k ,
\end{equation*}

donde

\begin{equation*}
  \alpha(\eta) = \sqrt{\frac{1}{k}}
  \left(1+\frac{i}{k\eta}\right)e^{ik\eta}, \qquad
  \beta(\eta) = 
  \sqrt{k}e^{ik\eta}.
\end{equation*}

Por definición, la aplicación del operador de aniquilación,
$\annRI_k$ al estado vacío, es cero

\begin{equation*}
  \hat{a}_k^{R,I}\ket{0} = 0,
\end{equation*}

entonces, usando esta propiedad y sustituyendo en ella a
\eqref{eq:a_RI}, obtenemos \cite{Schiff68}

\begin{equation*}
  \left\{ i \alpha\hat{\pi}_k^{R,I} + \beta \hat{y}_k^{R,I}(\eta)  \right\}
  \ket{0} = 0.
\end{equation*}

Es importante notar que al ser un operador hermitiano,
$\yRI_k$, tiene garantizada la existencia de ciertos
estados, $\ket{\bm{y}_k^{R,I}}$ que son sus eigenestados,
con eigenvalores\footnote{Las unidades de los eigenvalores
  son las mismas que las de $\y_k$, es decir $[y_k] \, = \,
  M^{1/2}L^{5/2}$. Es importante no confundir con las
  elecciones de vacío \eqref{eq:vacio_BunchDavies}.}
$\bm{y}_k^{R,I}$:

\begin{equation*}
  \yRI_k\ket{\bm{y}_k^{R,I}}
  = \bm{y}_k^{R,I}\ket{\bm{y}_k^{R,I}}.
\end{equation*}

Los eigenestados son ortonormales y completos. Proyectando
sobre estos estados

\begin{equation*}
  i\alpha\braketd{\pyRI_k}{\bm{y}_k^{R,I}}{0} + \beta
  \braketd{\yRI_k}{\bm{y}_k^{R,I}}{0} = 0 
\end{equation*}

Usando la relación de completitud de los eigenestados
$\ket{\bm{y}_k^{R,I}}$

\begin{equation*}
  \int\ket{\tilde{\bm{y}}_k^{R,I}}\bra{\tilde{\bm{y}}_k^{R,I}}\; d\tilde{\bm{y}}_k^{R,I} = 1
\end{equation*}

llegamos a

\begin{align}
  0 &=\int i \alpha
  \braketd{\pyRI_k}{\bm{y}_k^{R,I}}{\tilde{\bm{y}}_k^{R,I}}
  \bra{\tilde{\bm{y}}_k^{R,I}}0\,\rangle d\tilde{\bm{y}}_k'
  + \int \beta
  \braketd{\yRI_k}{\bm{y}_k^{R,I}}{\tilde{\bm{y}}_k^{R,I}}\bra{\tilde{\bm{y}}_k^{R,I}}0\,\rangle
  d\tilde{\bm{y}}_k' \\
  &= \int i \alpha
  \braketd{\pyRI_k}{\bm{y}_k^{R,I}}{\tilde{\bm{y}}_k^{R,I}}
  \bra{\tilde{\bm{y}}_k^{R,I}}0\,\rangle d\tilde{\bm{y}}_k'
  + \int \beta \bm{y}_k
  \bra{\bm{y}_k^{R,I}}\,\tilde{\bm{y}}_k^{R,I}\rangle\bra{\tilde{\bm{y}}_k^{R,I}}0\,\rangle
  d\tilde{\bm{y}}_k'
\end{align}

Como los eigenestados son ortonormales y normalizados
mediante

\begin{equation}
  \bra{\bm{y}_k^{R,I}}\tilde{\bm{y}}_k^{R,I}\rangle =
  \delta(\tilde{\bm{y}}_k^{R,I}-\bm{y}_k^{R,I}) 
\end{equation}

Además, de las relaciones de conmutación
\eqref{eq:rel_conmutacion_nueva}, se obtiene \footnote{ Los
  pasos de la deducción se muestran a continuación
  \begin{align}
    [\hat{y}_k^{R,I}, \hat\pi_k^{R,I}]  &= \frac{1}{2}i\hbar L^3 \\
    \bra{\tilde{\bm{y}}_k^{R,I}}\hat{y}_k^{R,I}\hat\pi_k^{R,I}
    - \hat\pi_k^{R,I}\hat{y}_k^{R,I}\ket{\bm{y}_k^{R,I}} &=
    \frac{1}{2}i\hbar L^3
    \bra{\tilde{\bm{y}}_k^{R,I}}\bm{y}_k^{R,I}\rangle =
    i\hbar L^3\delta\left(\tilde{\bm{y}}_k^{R,I} -
      \bm{y}_k^{R,I}\right) \\
    \left(\tilde{\bm{y}}_k^{R,I}-y_k^{R,I}\right)
    \bra{\tilde{\bm{y}}_k^{R,I}}\hat\pi_k^{R,I}\ket{y_k^{R,I}}
    &= \frac{1}{2}i\hbar
    L^3\delta\left(\tilde{\bm{y}}_k^{R,I} -
      \bm{y}_k^{R,I}\right) \\
    \bra{\tilde{\bm{y}}_k^{R,I}}\hat\pi_k^{R,I}\ket{y_k^{R,I}}
    &= \frac{1}{2}i\hbar L^3 \,
    \frac{\delta\left(\tilde{\bm{y}}_k^{R,I} -
        \bm{y}_k^{R,I}\right)}{\left(\tilde{\bm{y}}_k^{R,I}-\bm{y}_k^{R,I}\right)} \\
    \bra{\tilde{\bm{y}}_k^{R,I}}\hat\pi_k^{R,I}\ket{\bm{y}_k^{R,I}}
    &= -\frac{1}{2}i\hbar L^3 \delta '
    \left(\tilde{\bm{y}}_k^{R,I} - \bm{y}_k^{R,I} \right)
  \end{align}
  donde en el último renglón se usó una propiedad muy
  conocida de las deltas de Dirac. }

\begin{equation}
  \bra{\bm{y}_k^{R,I}}\hat{\pi}_k^{R,I}
  \ket{\tilde{\bm{y}}_k^{R,I}} 
  =  -\frac{1}{2}i\hbar L^3 s
  \delta'(\tilde{\bm{y}}_k^{R,I}-\bm{y}_k^{R,I}).
\end{equation}

Integrando

\begin{equation}
  \left(\frac{\alpha \hbar L^3}{2} \frac{d}{d\bm{y}_k^{R,I}}
    +    \beta \bm{y}_k^{R,I} 
  \right)\bra{\bm{y}_k^{R,I}}0\,\rangle = 0,
\end{equation}

dividiendo por $\alpha\hbar L^3/2$,

\begin{equation}
  \left(\frac{d}{d\bm{y}_k}  +
    \frac{2\beta}{ \hbar\alpha L^3}\bm{y}_k^{R,I}(\eta) \right)\bra{\bm{y}_k^{R,I}}0\,\rangle = 0.
\end{equation}

Esta ecuación tiene como solución\footnote{Notesé que $[y_k]
  = M^{1/2}L^{5/2}$, entonces $p_k$ en la última exponencial
  de la función de Wigner tiene unidades de $[p_k] =
  M^{1/2}L^{3/2}$.}

\begin{equation}\label{eq:funcion_de_onda}
  \bra{\bm{y}_k^{R,I}}0\,\rangle = A \exp\left({-\frac{\beta}{\alpha\hbar
        L^3}\left(\bm{y}_k^{R,I}\right)\,^2}\right)
\end{equation}

La constante $A$ se puede obtener de la condición de
normalización:

\begin{equation}
  \bra{0}\,\bm{y}_k^{R,I}\rangle\bra{\bm{y}_k^{R,I}}\,0\rangle \equiv
  \int_{-\infty}^\infty |\bra{0}\,\bm{y}_k^{R,I}\rangle|^2 d\bm{y}^{R,I} = 1
\end{equation}

Obteniendo

\begin{equation}
  A = \left( \frac{2\beta}{\alpha\pi\hbar L^3}\right)^{1/4}  =
  \left(\frac{2k}{\left(1+\displaystyle \frac{i}{k\eta}\right)\pi\hbar L^3}\right)^{1/4}.
\end{equation}

Si expresamos el valor de $\alpha$ y $\beta$ en
\eqref{eq:funcion_de_onda}

\begin{equation}
  \Psi^{R,I}\left(\bm{y}_k^{R,I}, \eta\right) \equiv \bra{\bm{y}_k^{R,I}}0\,\rangle =
  \left(\frac{2k}{\left(1+\displaystyle \frac{i}{k\eta}\right)\pi\hbar L^3}\right)^{1/4}
  \exp\left({-\frac{k}{\hbar L^3\left(1 + \displaystyle \frac{i}{k\eta}\right)}\left(\bm{y}_k^{R,I}\right)\,^2}\right)
\end{equation}

A partir de ahora dejaremos de usar el símbolo de $\bm{y}_k$
y por comodidad usaremos $y_k$, salvo en los casos que pueda
haber confusión.  Procedemos a escribir la función de onda
en una forma más cómoda para obtener su complejo conjugado,

\begin{equation}\label{eq:funcion_de_onda_final}
  \Psi^{R,I}\left(y_k^{R,I}, \eta\right)  = A
  \exp\left\{-\frac{k}{\hbar L^3}\cdot\frac{1-\displaystyle
      \frac{i}{k\eta}}{1+\displaystyle \frac{1}{k^2\eta^2}}
    \left(y_k^{R,I}\right)\,^2\right\},  
\end{equation}

su complejo conjugado es entonces

\begin{equation}
  \label{eq:funcion_de_onda_final_conjugada}
  \Psi^{R,I}\left(y_k^{R,I}, \eta\right)^*  = A^* 
  \exp\left\{-\frac{k}{\hbar L^3}\cdot\frac{1+\displaystyle
      \frac{i}{k\eta}}{1+\displaystyle \frac{1}{k^2\eta^2}}
    \left(y_k^{R,I}\right)\,^2\right\}. 
\end{equation}

\subsection{Funcional de Wigner para el modo $\y_k$ campo
  escalar   $\y$}

Evaluamos \eqref{eq:funcion_de_onda_final} y
\eqref{eq:funcion_de_onda_final_conjugada} en
$\left(y_k^{R,I}-\frac{1}{2}Y_k^{R,I}\right)$ y las
multiplicamos entre sí antes de sustituirlas en
\eqref{eq:wigner_fdo}\footnote{Nótese que dividimos el
  último exponencial de la función de Wigner
  \eqref{eq:wigner_fdo} por $L^3$}:

\begin{equation*}
  \Psi_k\left(y_k^{R,I}-\frac{1}{2}Y_k^{R,I}\right)\Psi_k^*\left(y_k^{R,I}+\frac{1}{2}Y_k^{R,I}\right)
  = AA^*\exp\left\{ -\frac{2k}{\hbar L^3}\left( y_k^2 +
      \frac{1}{4}Y_k^2 +
      \frac{i}{k\eta}y_kY_k\right)\right\},
\end{equation*}

obteniendo, luego de la sustitución,

\begin{multline*}
  \Wigner(y_k^{R,I},\pi_k^{R,I},\eta)= AA^* \exp\left\{
    -\frac{k}{\hbar L^3}\left(\frac{2}{1+\displaystyle
        \frac{1}{k^2\eta^2}}\right)\left(y_k^{R,I}\right)^2
  \right\}
  \times \\
  \int_{-\infty}^\infty
  \exp{\left\{-\frac{k\left(Y_k^{R,I}\right)^2}{2\hbar L^3
        \left(1+\displaystyle
          \frac{1}{k^2\eta^2}\right)}\right\}} \times
  \\
  \exp\left\{ -\frac{i}{ \hbar\eta L^3}\cdot
    \frac{2}{1+\displaystyle \frac{1}{k^2\eta^2}}\cdot
    y_k^{R,I}Y_k^{R,I} \right\}
  \exp{\left(\frac{i\pi_k^{R,I}Y_k^{R,I}}{\hbar
        L^3}\right)}dY_k^{R,I}
\end{multline*}

Usando la fórmula de Euler y definiendo una nueva variable
$\gamma$, en la penúltima exponencial,

\begin{equation*}
  \gamma(y_k^{R,I}, \eta) \equiv \frac{1}{\hbar\eta
    L^3}\cdot \frac{2}{1+\frac{1}{k^2\eta^2}}y_k^{R,I}
\end{equation*}

llegamos a

\begin{multline*}
  \Wigner(y_k^{R,I},\pi_k^{R,I},\eta)= AA^* \exp\left\{
    -\frac{2k}{\hbar L^3} \factorinvz
    \left(y_k^{R,I}\right)^2 \right\}
  \times \\
  \int_{-\infty}^\infty dY_k^{R,I}
  \exp{\left\{-\frac{k}{2\hbar L^3} \factorinvz
      \left(Y_k^{R,I}\right)^2 \right\}} \times
  \\
  \left[ \cos(\gamma Y_k^{R,I}) - i \sin(\gamma
    Y_k^{R,I})\right]
  \left[\cos\left(\frac{\pi_k^{R,I}Y_k^{R,I}}{\hbar
        L^3}\right) + i
    \sin\left(\frac{\pi_k^{R,I}Y_k^{R,I}}{\hbar
        L^3}\right)\right],
\end{multline*}

notando que la integral es sobre un intervalo simétrico, los
términos impares no contribuirán (serán cero) a la
integral. Utilizando la identidad de ángulos múltiples del
coseno:

\begin{multline*}
  \Wigner(y_k^{R,I},\pi_k^{R,I},\eta)= AA^* \exp\left\{
    -\frac{2k}{\hbar L^3} \factorinvz
    \left(y_k^{R,I}\right)^2 \right\}
  \times \\
  \int_{-\infty}^\infty dY_k^{R,I}
  \exp{\left\{-\frac{k}{2\hbar L^3} \factorinvz
      \left(Y_k^{R,I}\right)^2 \right\}}
  \cos\left[\left(\gamma - \frac{\pi_k^{R,I}}{\hbar
        L^3}\right)Y_k^{R,I}\right].
\end{multline*}

Para simplificar la notación introducimos

\begin{equation*}
  \xi = \frac{k}{2\hbar
    L^3\left(1+\frac{1}{k^2\eta^2}\right)} 
\end{equation*}

entonces el funcional de Wigner es hasta ahora,

\begin{multline*}
  \Wigner(y_k^{R,I},\pi_k^{R,I},\eta)= AA^* \exp\left(-4 \xi
    \left(y_k^{R,I}\right)^2\right)
  \times \\
  \int_{-\infty}^\infty dY_k^{R,I} \exp\left(-\xi
    \left(Y_k^{R,I}\right)^2\right) \cos\left[\left(\gamma -
      \frac{\pi_k^{R,I}}{\hbar L^3}\right)Y_k^{R,I}\right].
\end{multline*}

Es posible integrar ya que es simplemente la transformada de
Fourier de la función gaussiana, y usando

\begin{equation*}
  AA^* = \sqrt\frac{2k}{\hbar \pi L^3} \factorinvz^{1/4}  
\end{equation*}

llegamos al funcional de Wigner del estado vacío del campo
escalar escalado $\y_k$:

\begin{equation}
  \label{eq:wigner_inflaton}
  \Wigner(y_k^{R,I},\pi_k^{R,I},\eta)= 2 \factorz^{1/4}
  \exp\left( -4\,\xi \left(y_k^{R,I}\right)^2  \right)
  \exp\left(-\frac{\left(\gamma-\displaystyle
        \frac{\pi_k^{R,I}}{\hbar L^3}\right)^2}{4\xi}\right) 
\end{equation}

La función de Wigner calculada\footnote{La función de Wigner
  es adimensional. }, \eqref{eq:wigner_inflaton}, tiene una
forma parecida a la de una función gaussiana bidimensional
generalizada, como era de esperarse. Para hacer esto más
claro, acomodaremos términos\footnote{En este reacomodo se
  utilizó
  \begin{equation*}
    \gamma = \frac{4\xi}{k\eta}\left(y_k^{R,I}\right)^2.
  \end{equation*}
}

\begin{multline*}
  \Wigner(y_k^{R,I},\pi_k^{R,I},\eta) =
  2\left(1+\frac{1}{k^2\eta^2}\right)^{1/4}
  \exp\left(-\frac{2k}{\hbar
      L^3}\left(y_k^{R,I}\right)^2\right) \times \\
  \exp\left(\frac{2}{k\eta\hbar
      L^3}y_k^{R,I}\pi_k^{R,I}\right)
  \exp\left(-\frac{(1+k^2\eta^2)}{2\hbar L^3
      k^3\eta^2}\left(\pi_k^{R,I}\right)^2\right).
\end{multline*}

Comparando con

\begin{equation*}
  f(x,y) = A
  \exp\left(-\left(a(x-x_0)^2+b(x-x_0)(y-y_0)+c(y-y_0)^2
    \right)\right)  ,
\end{equation*}

vemos que la gaussiana está centrada en $(0,0)$, y sus
coeficientes son\footnote{Las unidades son: $[a] \,=\,
  (ML^5)^{-1}, [b]\, =\, (ML^4)^{-1} $ y $[c] \, = \,
  (ML^3)^{-1}$. Las unidades de $y_k^{R,I}$ y $p_k^{R,I}$
  son $[y_k^{R,I}]= M^{1/2}L^{5/2}$ y $[p_k^{R,I}] =
  M^{1/2}L^{3/2}$.}

\begin{equation*}
  a=\frac{2k}{\hbar L^3}, \quad
  b=-\frac{2}{k\eta\hbar L^3}, \quad c =
  \frac{(1+k^2\eta^2)}{2\hbar L^3 k^3\eta^2},
\end{equation*}

la aparición del término cruzado (es decir, $b \neq 0$) es
indicativo que los ejes principales de las secciones
transversales de la gaussiana están girados respecto a los
ejes $(y_k^{R,I}, \pi_k^{R,I})$. Estamos interesados en las
dispersiones de ambas variables ($\sigma_{y_k^{R,I}}$ y
$\sigma_{p_k^{R,I}}$).  Para obtenerlas rotaremos los ejes
para obtener una gaussiana bidimensional en su forma
normal\footnote{ Nótese que reescalamos  el eje  $\pi_k$ a $\Pi_k =
  \pi_k/k$ e hicimos una simple rotación (i.e.\  $y_k'^{R,I}
  = y_k^{R,I} \cos \Theta_k + \Pi_k^{R,I}\sin \Theta_k $,
  $\Pi_k'^{R,I} = \Pi_k^{R,I}\cos\Theta_k - y_k^{R,I} \sin
  \Theta_k$)} .

Es conocido que la ecuación de una cónica,

\begin{equation*}
  \mathcal{C} = a_{11}\left(y_k^{R,I}\right)^2 +
  a_{22}\left(\pi_k^{R,I}\right)^2 +
  2a_{12}y_k^{R,I}\pi_k^{R,I} + 
  2a_{01}y_k^{R,I} + 2a_{02}\pi_k^{R,I} + a_{00} = 0,
\end{equation*}

se puede expresar de una forma matricial

\begin{equation*}
  \mathcal{C} =
  \left(
    \begin{array}{c c c}
      1 & y_k^{R,I} & p_k^{R,I}
    \end{array}
  \right)
  \left(
    \begin{array}{ccc}
      a_{00} & a_{01} & a_{02} \\
      a_{01} & a_{11} & a_{12} \\
      a_{02} & a_{12} & a_{22} \\
    \end{array}
  \right)
  \left(
    \begin{array}{c}
      1 \\
      y_k^{R,I} \\
      p_k^{R,I}
    \end{array}
  \right),
\end{equation*}

A la matriz central, que llamaremos $\mathcal{A}$ se le
conoce como \emph{matriz de la cónica}. De aquí también se
puede definir la matriz $A_0$

\begin{equation*}
  A_0 =  \left(
    \begin{array}{cc}
      a_{11} & a_{12} \\
      a_{12} & a_{22}
    \end{array}
  \right),
\end{equation*}

Se puede demostrar que la matriz $A_0$ define el tipo de
cónica y los ejes principales de la misma. La ecuación
característica de $A_0$

\begin{equation}\label{eq:caracteristica}
  \det(\lambda \mathbf{I}- A_0) = \lambda^2-Tr(A_0)\lambda + det(A_0),
\end{equation}

En nuestro caso en particular

\begin{equation*}
  \mathcal{A} = \left(
    \begin{array}{ccc}
      a_{00} & 0 & 0 \\
      0 & a & b/2\\
      0 & b/2 & c\\
    \end{array}
  \right),  \qquad A_0 = \left(
    \begin{array}{cc}
      a & b/2 \\
      b/2 & c \\
    \end{array}
  \right),
\end{equation*}

donde $a, b$ y $c$ están dados arriba. La ecuación
característica (\ref{eq:caracteristica}) es

\begin{equation*}
  \lambda^2 - (a+c)\lambda + (ac - \frac{b^2}{4}) = 0.
\end{equation*}

Se puede observar que $ac$ y $b^2$ son dimensionalmente
compatibles, pero que $a$ y $c$ no lo son. Para arreglar
esto   multiplicáremos  y
dividiremos la variable $\pi_k^{R,I}$  por $k$. Ahora
definiremos una
nueva variable  $\Pi_k = 
\pi_k^{R,I}/k$, y los coeficientes $a, b$ y $c\,$ se
convierten en

\begin{equation*}
  a'=a=\frac{2k}{\hbar L^3}, \quad
  b'=-\frac{2}{\eta\hbar L^3}, \quad c' =
  \frac{(1+k^2\eta^2)}{2\hbar L^3 k\eta^2},
\end{equation*}

teniendo todos los coeficientes unidades de $(ML^5)^{-1}$.
La ecuación característica después del escalamiento es

\begin{equation}
  \lambda^2 - \frac{1}{2}\left(\frac{1+ 5 k^2\eta^2}{\hbar
      L^3 k \eta^2}\right)\lambda + \frac{k^2}{\hbar^2 L^6}
  = 0 ,
\end{equation}

con soluciones dadas por la fórmula estándar de las
ecuaciones cuadráticas\footnote{Las unidades son
  $[\lambda]\, = \, (ML^5)^{-1}$}:
 
\begin{equation}
  \label{eq:soluciones_rotacion}
  \lambda _{(\pm)} = \frac {1+5 k^2\eta^2\pm\sqrt
    {1+10 k^2\eta^2 + 9k^4\eta^4}}{4 \hbar L^3 k \eta^2}.
\end{equation}

El ángulo de rotación respecto al eje $y_k^{R,I}$, se puede
obtener mediante

\begin{equation*}
  \tan 2\Theta_k = \frac{b}{a-c},
\end{equation*}

resultando en

\begin{equation}
  \label{eq:angulo_rotacion}
  2  \Theta   =
  \arctan\left(\frac{4k\eta}{1-3k^2\eta^2}\right).
\end{equation}

La ecuación reducida está dada por

\begin{equation}
  \lambda_{(+)}y'\,_k^{R,I} + \lambda_{(-)} \Pi'\,_k^{R,I} +
  \frac{\det{\mathcal{A}}}{\det A_0} = 0,
\end{equation}

donde $\det \mathcal{A}/\det A_0 = a_{00}$. De aquí podemos
observar que la elipse en estas nuevas coordenadas es una
elipse vertical\footnote{Ya que el eigenvalor menor
  corresponde al eje mayor.  Esto se puede observar de la
  ecuación de la elipse vertical
  \begin{equation*}
    \frac{x^2}{b} + \frac{y^2}{a} = 1, \quad a > b
  \end{equation*}
}. La función de Wigner en estos nuevos ejes es

\begin{equation}
  \label{eq:wigner_rotada}
  \Wigner_{rot} =
  2\left(1+\frac{1}{k^2\eta^2}\right)^{1/4}\exp
  \left(-\lambda_{(+)}\left({y'}_k^{R,I}\right)^2\right)\,
  \exp\left(-\lambda_{(-)}\left({\Pi'}_k^{R,I}\right)^2\right),
\end{equation}

donde $\Wigner_{rot} \equiv \Wigner_{rot}(y'\,_k^{R,I},
\Pi'\,_k^{R,I}, \eta)$. Las dispersiones \footnote{La
  dispersión estadística de una variable $x$, tiene unidades
  de $[\sigma_x]\,=\, [x^2]$} pueden leerse directamente
siendo

\begin{equation}
  \sigma^2_{y'\,_k^{R,I}} = \frac{1}{\lambda_{(+)}},\qquad
  \sigma^2_{\Pi'\,_k^{R,I}} = \frac{1}{\lambda_{(-)}}.
\end{equation}

\subsection{Esquema de Colapso \textit{à la} Wigner}

En el capítulo ,
``\nameref{cha:critica-al-origen}''
(cap. \ref{cha:critica-al-origen})  , se propone que los valores
medios de $\yRI_k$ y $\pyRI_k$ en el estado posterior
al colapso adquieran valores aleatorios independientes,
estos esquemas de colapso obviamente estaban ignorando las
correlaciones entre las variables canónicas $\left(\yRI_k,
\pyRI_k\right)$ que impone la
mecánica cuántica a través del principio de incertidumbre de
Heisenberg. En esta sección 
se propondrá una relación 
más \emph{natural} que tomará en cuenta la correlación entre
las variables $\yRI$ y $\pyRI$. El nuevo esquema de colapso
esta basado en la recién calculada función de 
distribución de Wigner,  y propone un colapso donde los
valores medios de los estados post-colapso estén dados
por\footnote{Se agregó una $k$ en la segunda ecuación
  (\ref{eq:wigner_medio_pi}) por la definición que se hizo
  del momento conjugado $P_k$ para que tuviese las mismas
  unidades de $y_k$}

\begin{subequations}\label{eq:valores_medios_wigner}
  \begin{equation}\label{eq:wigner_medio_y}
    \langle\hat{y}_k^{R,I}\rangle_\Xi  = x^{(R,I)}_k \Lambda_k
    \cos\theta,
  \end{equation}
  \begin{equation}\label{eq:wigner_medio_pi}
    \langle\hat{\pi}_k^{R,I}\rangle_\Xi  = x^{(R,I)}_k
    \Lambda_k k  
    \sin\theta.
  \end{equation}
\end{subequations}

Es decir, colapsarán a un valor dado por $x^{(R,I)}$ - la
cual es una variable aleatoria gaussiana normalizada
centrada en cero - y con una dispersión $\Lambda_k \angle
\theta_k$, con $\Lambda_k$ dada por

\begin{equation}\label{eq:dispersion}
  \Lambda_k \equiv 2\, \sigma_{P'_k} = 2\,
  \sqrt{\frac{1}{\lambda_{(-)}}} 
  =  \frac{4\eta\sqrt{\hbar L^3 k}}{\sqrt{1+5k^2\eta^2
      -\sqrt{1+10k^2\eta^2+9k^4\eta^4}}},
\end{equation}

y $\Theta_k$ estará dada por (\ref{eq:angulo_rotacion}).La
dispersión es elegida basándose en el hecho de que la
distribución de Wigner del estado
vacío de un oscilador armónico es  una función gaussiana
bidimensional, entonces, 
$\Lambda_k$ está dada por el semi-eje mayor de la elipse
definida por la función gaussiana bidimensional. Esta elipse
corresponde a la frontera de la región en el ``espacio de
fase'' donde la función de Wigner tiene una magnitud de
$1/2$.

De esta manera, este esquema de colapso no ignora las
correlaciones entre las variables canónicas que estaban
presentes en el estado anterior del colapso (i.e.\ , el estado de
vacío y en todos los estados), ya que está información está
integrada en la función 
de distribución de Wigner. La elección de esta función para
describir estas correlaciones es justificada \citep[ver por
ejemplo][]{Ballentine2000, Wigner1971,Hillery1984}.

Igualando \eqref{eq:valores_medios} con
\eqref{eq:valores_medios_wigner} tenemos

\begin{subequations}
  \begin{equation}
    \sqrt{2}\Re\left(y_kd_k^{R,I}\right) = x^{(R,I)}_k
    \Lambda_k 
    \cos\theta,
  \end{equation}
  \begin{equation}
    \sqrt{2}\Re\left(g_kd_k^{R,I}\right) = x^{(R,I)}_k
    \Lambda_k k  
    \sin\theta,
  \end{equation}
\end{subequations}

Como vimos en el capítulo \ref{cha:critica-al-origen} es
deseable expresar en forma polar a  $d_k^{R,I} =
|d_k^{R,I}|e^{i\alpha_k}$, $\,y^{R,I}_k(\eta) = |y_k^{R,I}|e^{i\beta^{R,I}_k}$ y
$g_k^{R,I}(\eta) = |g_k^{R,I}|e^{i\gamma^{R,I}_k}$
en particular nos interesan estas ecuaciones al tiempo del
colapso: $\eta_c^k$, para indicar esto usaremos un
superíndice $c$ (de colapso) en los ángulos $\beta_k$ y
$\gamma_k$.

Recordando además que,

\begin{eqnref}{eq:real_y}
  \Re(y_kd_k) = |y_k||d_k^{(R,I)}|\cos\left(\alpha_k^{(R,I)}
    + \beta_k\right),
\end{eqnref}

\begin{eqnref}{eq:real_pi}
  \Re(g_kd_k) = |g_k||d_k^{(R,I)}|\cos\left(\alpha_k^{(R,I)}
    + \gamma_k\right),
\end{eqnref}

llegamos al par de ecuaciones

\begin{subequations}\label{eq:wigner_sistema_d}
  \begin{equation}
    |d_k^{R,I}|\cos(\alpha_k+\beta_k) = \frac{1}{\sqrt{2}|y_k|}x_k^{R,I} \Lambda_k
    \cos\theta,
  \end{equation}
  \begin{equation}
    |d_k^{R,I}|\cos(\alpha_k+\gamma_k) = \frac{1}{\sqrt{2}|g_k|}x_k^{R,I} \Lambda_k k
    \sin\theta.
  \end{equation}
\end{subequations}

Es posible reescribir $g_k(\eta)$ y $y_k(\eta)$ en forma
polar,

\begin{eqnref}{eq:polar-form-y}
  |y_k(\eta)| =
  \sqrt{\frac{1}{2k}}\left(\frac{\sqrt{1+k^2\eta^2}}{k\eta}\right),
  \quad \beta_k = -\left[k\eta +
    \arctan\left(\frac{1}{k\eta}\right)\right],
\end{eqnref}
\begin{eqnref}{eq:polar-form-pi}
  |g_k(\eta)| = \sqrt{\frac{k}{2}} , \quad \gamma_k =
  -\left(k\eta + \frac{\pi}{2}\right)
\end{eqnref}

Usando las identidades trigonométricas

\begin{equation}\label{eq:identidades_trigonometricas}
  \cos^2\left(\arctan x\right) = \frac{1}{1+x^2}, \qquad
  \sin^2\left(\arctan x\right) = \frac{x^2}{1+x^2}
\end{equation}

Las ecuaciones \eqref{eq:wigner_sistema_d} quedan ahora
\begin{subequations}
  \begin{equation}\label{eq:d_1}
    \frac{|d_k^{R,I}|}{\sqrt{1+k^2\eta^2}}\left\{\cos(\alpha_k^{R,I}-k\eta)k\eta
      + \sin(\alpha_k^{R,I}-k\eta) \right\} = \frac{1}{\sqrt{2}|y_k|}x_k^{R,I} \Lambda_k
    \cos\theta
  \end{equation}
  \begin{equation}\label{eq:d_2}
    |d_k^{R,I}|\sin\left(\alpha_k^{R,I}-k\eta\right) = \frac{1}{\sqrt{2}|g_k|}x_k^{R,I} \Lambda_k
    k \sin\theta
  \end{equation}
\end{subequations}

Sustituyendo \eqref{eq:d_2} en \eqref{eq:d_1}, obtenemos una
expresión para $|d_k^{R,I}|$:

\begin{equation}\label{eq:d_1_prima}
  |d^{R,I}_k|  =
  \frac{x_k^{R,I} \Lambda_k}{\sqrt{2}|y_k||g_k|} \cdot \frac{1}{
    \cos(\alpha_k^{R,I}-k\eta) k\eta} \left( \sqrt{1+k^2\eta^2}
    |g_k|\cos\theta - k|y_k|\sin\theta \right)
\end{equation}

insertando esta ecuación en \eqref{eq:d_2}

\begin{equation}
  (\alpha_k^{R,I} - k\eta)  =
  \arctan\left(\frac{k\eta\sin\theta}{k\eta\cos\theta - \sin\theta}\right),
\end{equation}

Usando de nuevo las identidades
\eqref{eq:identidades_trigonometricas}, podemos expresar
$\cos(\alpha-k\eta)$ y $\sin(\alpha-k\eta)$ de la forma
siguiente

\begin{subequations}
  \begin{equation}
    \cos(\alpha_k-k\eta) = \frac{k\eta\cos\theta -
      \sin\theta}{\sqrt{k^2\eta^2 -
        2k\eta\cos\theta\sin\theta + \sin^2\theta}},
  \end{equation}
  \begin{equation}
    \sin(\alpha_k-k\eta) = \frac{k\eta\sin\theta}
    {\sqrt{k^2\eta^2 - 
        2k\eta\cos\theta\sin\theta + \sin^2\theta}}.
  \end{equation}
\end{subequations}

Entonces podemos expresar \eqref{eq:d_1_prima} como

\begin{equation}\label{eq:d_1_prima_prima}
  |d_k^{R,I}| = \frac{x_k^{R,I}\sqrt{k}}{k\eta}\Lambda_k       \sqrt{k^2\eta^2 - 
    2k\eta\cos\theta\sin\theta + \sin^2\theta}
\end{equation}

El valor de expectación del momento conjugado de $\dphih_k$
en el estado post-colapso $\ket{\Omega}$ es

\begin{eqnref}{eq:expectacion_dphi_d_2}
  \expec{\dphih_k'}_\Omega=\frac{|g_k|}{a} \left[
    |d_k^R|\cos\left(\alpha_k^R + \gamma_k + \Delta_k\right)
    + i|d_k^I|\cos\left(\alpha_k^I + \gamma_k +
      \Delta_k\right)\right],
\end{eqnref}

donde será bueno recordar que $\gamma_k$ no está evaluada en
$\eta_c$ y $\Delta_k \equiv k(\eta -\eta^c_k)$ es el tiempo
transcurrido desde el colapso. Insertando 
\eqref{eq:d_1_prima_prima} en el valor de expectación del
estado post-colapso

\begin{multline}
  \expec{\dphih_k'}_\Omega= \frac{1}{\sqrt{2}a|y_k|}
  \frac{\Lambda_k}{k\eta}\left(x_k^R + i x_k^I\right) \times
  \\ \left[ k^2\eta|y_k|\cos\Delta_k \sin\theta +
    \sin\Delta_k\left( |g_k|\cos\theta\sqrt{1+k^2\eta^2} -
      k|y_k|\sin\theta\right)\right],
\end{multline}

usando una vez más las identidades trigonométricas
(\ref{eq:identidades_trigonometricas}) pero ahora aplicadas
al ángulo $\Theta_k$ (\ref{eq:angulo_rotacion}) y
sustituyendo \eqref{eq:dispersion} en esta expresión
llegamos a

\begin{multline}
  \label{eq:dphi_wigner}
  \langle\delta\hat\phi'_k\rangle_\Omega =
  \frac{2}{a(\eta_c)}\cdot\frac{k\eta_c\sqrt{\hbar L^3
      k}}{\left(1+10k^2\eta_c^2+9k^4\eta_c^4\right)^{1/4}}\cdot
  \frac{ x^R_k +i x_k^I
  }{\sqrt{1+5k^2\eta_c^2-\sqrt{1+10k^2\eta_c^2+9k^4\eta_c^4}}}  \\
  \Bigg\{\cos\Delta_k
  \sqrt{\sqrt{1+10k^2\eta_c^2+9k^4\eta_c^4} - 1
    +3k^2\eta_c^2 } \quad +    \\
  \sin\Delta_k
  \Bigg[\sqrt{\sqrt{1+10k^2\eta_c^2+9k^4\eta_c^4} + 1-
    3k^2\eta_c^2} - \\ \frac{1}{k\eta_c}
  \sqrt{\sqrt{1+10k^2\eta_c^2+9k^4\eta_c^4} - 1
    +3k^2\eta_c^2}\quad \Bigg] \Bigg\}. \\
\end{multline}

Ahora tomaremos el valor medio del ensamble
\textit{imaginario de universos} del cuadrado de
$\expec{\dphih_k'}_\Omega$ (cf.\ \S
\ref{sec:comparacion-obs} para una mayor explicación) ,
dejando afuera el factor de $\hbar L^3 k/4 a^2$,  y lo
denominaremos $C_{Wigner}(k)$ 
\begin{multline}\label{eq:C_wigner}
  C_{Wigner}(k) = \frac{32
    z_k^2}{\sqrt{1+10z_k^2+9z_k^4}}\cdot
  \frac{1}{1+5z_k^2-\sqrt{1+10z_k^2+9z_k^4}} \\
  \Bigg\{ \left[\sqrt{1+10z_k^2+9z_k^4} - 1
    +3z_k^2\right]\left(\cos\Delta_k
    -\frac{\sin\Delta_k}{z_k}\right)^2 + \\
  \sin^2\Delta_k \left[\sqrt{1+10z_k^2+9z_k^4} - 3z_k^2 - 7
  \right] + 8z_k\cos\Delta_k \sin\Delta_k \Bigg\},
\end{multline}
donde se ha reemplazado $k\eta_k^c(k)$ por $z_k$. Por lo
tanto la cantidad observacional \eqref{eq:contacto_obs} es
\begin{equation}
  |\alpha_{lm}|^2_{ML}  = \frac{s^2\hbar}{2 \pi a^2}
  \int   \frac{C_{Wigner}(x/R_D)}{x}\mathcal{T}(x/R_D)^2
  j_l^2(x)  dx.
\end{equation}

Con esta expresión podemos comparar las
predicciones de este esquema de colapso con las
observaciones, análisis que se hará en la sección
\ref{sec:comparacion-observaciones-esq}. Antes de hacerlo, es
preciso recordar que los resultados observacionales estándar
son obtenido si la función $C$ es una constante, y para
poder obtener una función constante, en este, o en los otros
esquemas estudiados hasta ahora, parace haber sólo una
opción: que $z_k$ sea independiente de $k$ indicando así,
que el tiempo de colapso para el modo $k$, $\eta_k^c$
dependa de inversamente de $k$, $\eta_k^c = z/k$, donde $z$
es constante e independiente de $k$.

\section{Comparación de los diferentes esquemas de colapso y
  las observaciones}\label{sec:comparacion-observaciones-esq}

Hemos estudiado tres diferentes esquemas de colapso,
teniendo todos ellos diferentes 
comportamientos  al tiempo de colapso $\eta_k^c$ sobre el
nuevo valor de expectación del estado posterior al colapso. En el
primer esquema propuesto en \cite{Sudarsky06a} y estudiado
en el capítulo \ref{cha:critica-al-origen}, ambos, el valor
de expectación del campo escalado $\yRI_k$ y el valor de
expectación del momento conjugado $\pyRI_k$ en el estado
post-colapso, $\ket{\Omega}$, están distribuidos
aleatoriamente en los respectivos rangos de las
incertidumbres del estado pre-colapso y no se encuentran
correlacionados. De una manera más precisa el valor al
tiempo $\eta_k^c$ está determinado por

\begin{equation}
  \label{eq:collapse_scheme_1}
  \left\langle \hat
    y_k^{(R,I)}\left(\eta_k^c\right)\right\rangle_\Omega =
  x_1^{(R,I)} 
  \sqrt{\left(\Delta y_k^{(R,I)}\right)_0^2}, \quad   \left\langle
    \hat \pi_k^{(R,I)} \left(\eta_k^c\right) \right\rangle_\Omega =   x_2^{(R,I)} 
  \sqrt{\left(\Delta \pi_k^{(R,I)}\right)_0^2} \, ,
\end{equation}

y su espectro de potencias \todo{Mirar lo de Gabriel}
resultante es
\begin{eqnref}{eq:C_sudarsky_inflation}
  C_1(k) = 1 + \frac{2}{z_k^2}\sin^2\Delta_k +
  \frac{1}{z_k}\sin(2\Delta_k).
\end{eqnref}

El segundo esquema de colapso, también propuesto en
\cite{Sudarsky06a}, tiene la forma

\begin{equation}
  \label{eq:collapse_scheme_2}
  \left\langle \hat
    y_k^{(R,I)}\left(\eta_k^c\right)\right\rangle_\Omega =
  0, \quad 
  \left\langle \hat
    \pi_k^{(R,I)}\left(\eta_k^c\right)\right\rangle_\Omega =
  x_2^{(R,I)}  
  \sqrt{\left(\Delta \pi_k^{(R,I)}\right)_0^2} \, ,
\end{equation}

está inspirado en el hecho, que en la ecuación
\eqref{eq:semiclassical-fundamental} sólo el momento
conjugado del campo inflatónico aparece como
fuente\footnote{Recuérdese que esta ecuación es sólo a
  primer orden, a órdenes posteriores ambos valores de
  expectación aparecen, ver \ref{cha:teor-de-pert}}, por lo
tanto, su valor de expectación cambia de cero a otro valor
dado por el rango de incertidumbre del momento conjugado en
el estado vacío, por otra parte, después del colapso, el
valor de expectación de $\y_k$ permanece sin cambio, i.e.\ ,
es cero. Su espectro de potencias es

\begin{eqnref}{eq:C_2_inflation}
  C_2(k) = 1 + \sin^2\Delta_k \left(1 - \frac{1}{z_k^2}\right)
  - \frac{1}{z_k}\sin (2\Delta_k).
\end{eqnref}

El último esquema estudiado (ver sección
\ref{sec:colapso_wigner}) y propuesto por vez primera en
\cite{Unanue2008} es, en el sentido siguiente, más natural,
ya que toma en cuenta las correlaciones entre los valores de
expectación del campo y de su momento conjugado en el estado
anterior al colapso, tal como es codificado por la
distribución de Wigner.

\begin{equation}
  \label{eq:collapse_scheme_wigner}
  \left\langle\hat{y}_k^{(R,I)}\left(\eta_k^c\right)\right\rangle_\Omega
  = x^{(R,I)}_k \Lambda_k 
  \cos\Theta_k\, ,
  \quad
  \left\langle\hat{\pi}_k^{(R,I)}\left(\eta_k^c\right)\right\rangle_\Omega
  = x^{(R,I)}_k \Lambda_k k 
  \sin\Theta_k,
\end{equation}

donde $\Lambda_k$ está dado por el semi-eje mayor de la
elipse caracterizada por la función bidimensional gaussiana
definida por la frontera en el ``espacio de fase'' donde la
función de Wigner tiene una magnitud mayor a $1/2$ y
$\Theta_k$ es el ángulo entre este eje y el eje de $y_k^{R,I
}$. Aquí repetimos su espectro de potencias (\ref{eq:C_wigner})
para ayudar a la comparación:

\begin{multline}\tag{\ref{eq:C_wigner}}
  C_{Wigner}(k) = \frac{32
    z_k^2}{\sqrt{1+10z_k^2+9z_k^4}}\cdot
  \frac{1}{1+5z_k^2-\sqrt{1+10z_k^2+9z_k^4}} \\
  \Bigg\{ \left[\sqrt{1+10z_k^2+9z_k^4} - 1
    +3z_k^2\right]\left(\cos\Delta_k
    -\frac{\sin\Delta_k}{z_k}\right)^2 + \\
  \sin^2\Delta_k \left[\sqrt{1+10z_k^2+9z_k^4} - 3z_k^2 - 7
  \right] + 8z_k\cos\Delta_k \sin\Delta_k \Bigg\}.
\end{multline}

Podemos ver que la expresión $C_{Wigner}$ (ecuación \ref{eq:C_wigner}) es más
complicada que la de $C_2$ (ecuación \ref{eq:C_2_inflation})
y que la de $C_1$ (ecuación \ref{eq:C_sudarsky_inflation}).

En todos los esquema $x_{1,2}^{(R,I)}$ son variables aleatorias
caracterizados por una distribución Gaussiana centrada en
cero y con dispersión uno.

A pesar del hecho de que la expresión para $C_{Wigner}$
aparece mucho más complicada que $C_2$, su dependencia en
$z_k$ es muy similar, excepto en la amplitud de las
oscilaciones.  (ver figuras \ref{fig:c_2} y
\ref{fig:c_wigner}).  Otro hecho interesante que puede ser
detectado en el comportamiento de los diferentes esquemas de
colapso se observa al considerar el límite $z_k \to \pm
\infty$. Así vemos que $C_1(k) \to 1$ recuperando el espectro invariante
de escala estándar, mientras que   $C_2(k)$ o
$C_{Wigner}(k)$ no tienen un límite bien definido (ver
figuras \ref{fig:c_2} y \ref{fig:c_wigner}).

\subsection{Un único colapso}\label{sec:un-unico-colapso}

Nuestro objetivo, en esta sección es comparar estas gráficas
con el espectro invariante de escala predicho por Harrison-
Zel'dovich y que es generado por varios modelos inflacionarios
(i.e.\ un valor constante para $2l(l+1)|\alpha_{lm}|^2$) y
no directamente con el espectro observado. En el análisis
siguiente nos concentraremos en el espectro primordial e ignoraremos los
efectos de la física que corresponde al recalentamiento y a
las oscilaciones acústicas (representadas mediante las
funciones de transferencia $\mathcal{T}(k)$, cf.\
\S \ref{sec:evol-clasica}). Un estudio considerando los efectos
tardíos con datos empíricos requeriría un análisis que está fuera del alcance
de esta tesis.

Recordemos que $C(k)$ encapsula todos los detalles del
esquema de colapso en el espectro de potencias
observacional.

  \figura{imagenes/c_1_}{width=11cm,height=8.5cm}{$C_1$
    del esquema de colapso 
  independiente, en el ambas variables de campo $\yRI_k$ y
  $\pyRI_k$ colapsan a un valor aleatorio de
  manera independiente de la 
  incertidumbre del estado de vacío.}{fig:c_1}   
  \figura{imagenes/c_2_}{width=11cm,height=8.5cm}{$C_{Wigner}$,
    esquema de colapso 
   de Wigner. Este esquema propone una correlación entre los
   valores posteriores al colapso determinada por la
   distribución de Wigner del estado de vacío.}{fig:c_2}  
  \figura{imagenes/c_wigner_}{width=11cm,height=8.5cm}{$C_{Wigner}$,
    esquema de colapso 
   de Wigner. Este esquema propone una correlación entre los
   valores posteriores al colapso determinada por la
   distribución de Wigner del estado de vacío.}{fig:c_wigner}


El espectro de potencias observado es recuperado tomando
$C(k) = 1$. Dado esto, podemos investigar en particular que
tan sensibles son las predicciones de los diferentes
esquemas, ante pequeños desvíos del caso en el cual $z_k$ es
independiente de $k$, que, como se argumentó arriba, nos
lleva a una concordancia precisa con el espectro de
potencias estándar. Para poder llevar a cabo este análisis,
debemos de evaluar las integrales \eqref{eq:contacto_obs}
para los esquemas de colapso caracterizados por las
funciones $C_1(k), C_2(k)$ y $C_{Wigner}(k)$. Será
conveniente definir la cantidad adimensional $\tilde{z}_x
\equiv x N(x)$, donde $x= k R_D$ y $N(x) \equiv
\eta_{k(x)}^c/R_D$. Se supondrá lo siguiente: (a) La
evolución de las cantidades de interés (e.g.\ $\expec{\dphih_k}$)
desde el colapso hasta el final de la inflación, es más
significativa que la evolución desde el fin de
inflación hasta nuestros días, entonces usaremos $\Delta_k =
-\tilde{z}_x$; (b) Para explorar la robustez del esquema
ante pequeñas desviaciones de ``la receta $z_k$ es
independiente de $k$'' considerando perturbaciones lineales
de esto, caracterizadas por $\tilde{z}_x$ as $\tilde{z}_x =
A + Bx$.  Nótese que en estas unidades $A$ y $B$ son
adimensionales.

\begin{figure*}
  \includegraphics[width=210mm,
  angle=90]{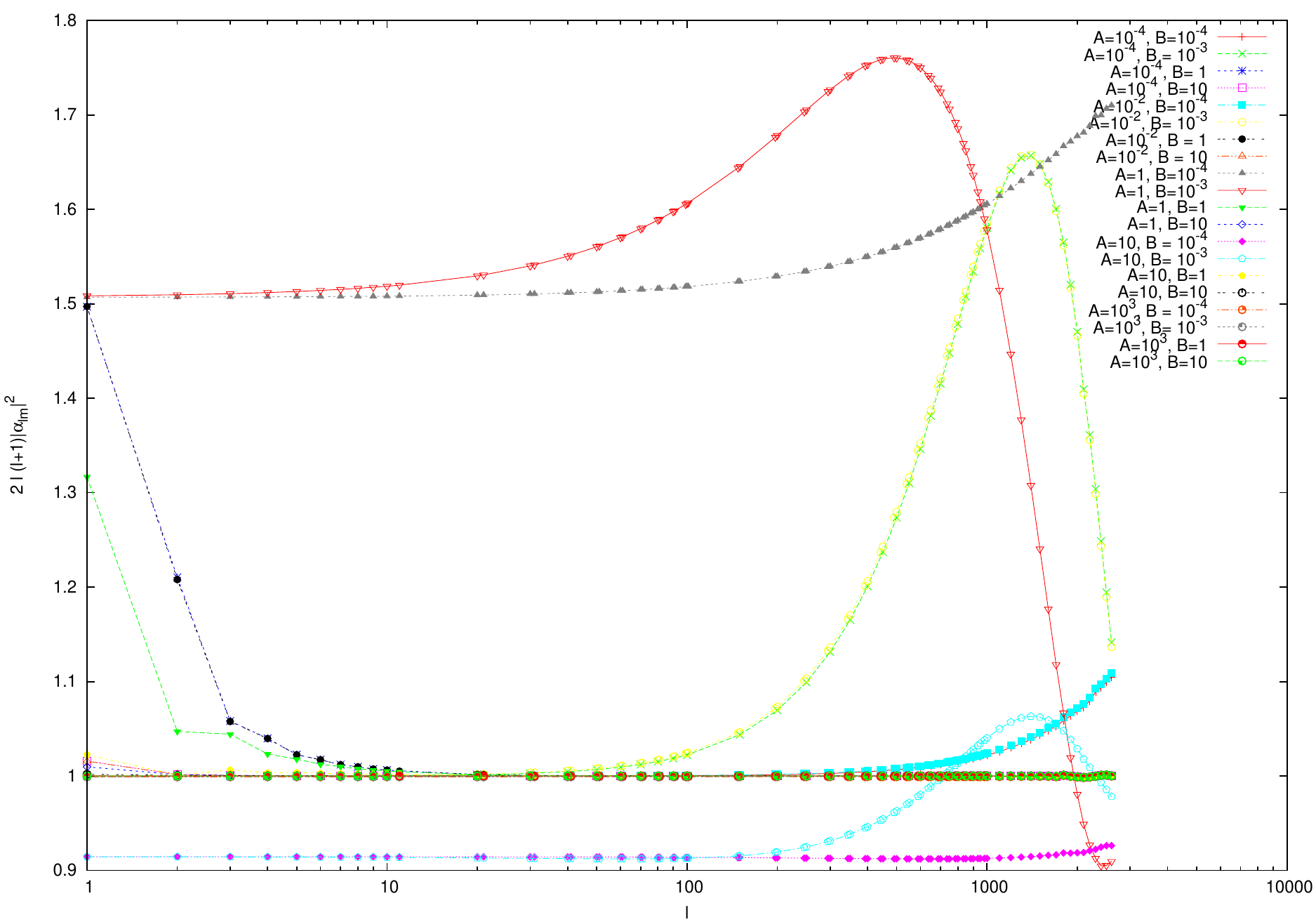}
  \caption{Gráfica semilogarítmica de
    $|\alpha_{lm}|^2(C_1(k))$ para diferentes valores de
    $(A,B)$ representando que tan robusto es el esquema de
    colapso cuando se desvía de $z_k$ constante. La abscisa
    es el multipolo $l$ hasta $l=2600$.}
  \label{fig:c1_log}
\end{figure*}

\begin{figure*}
  \includegraphics[width=210mm, angle=90]{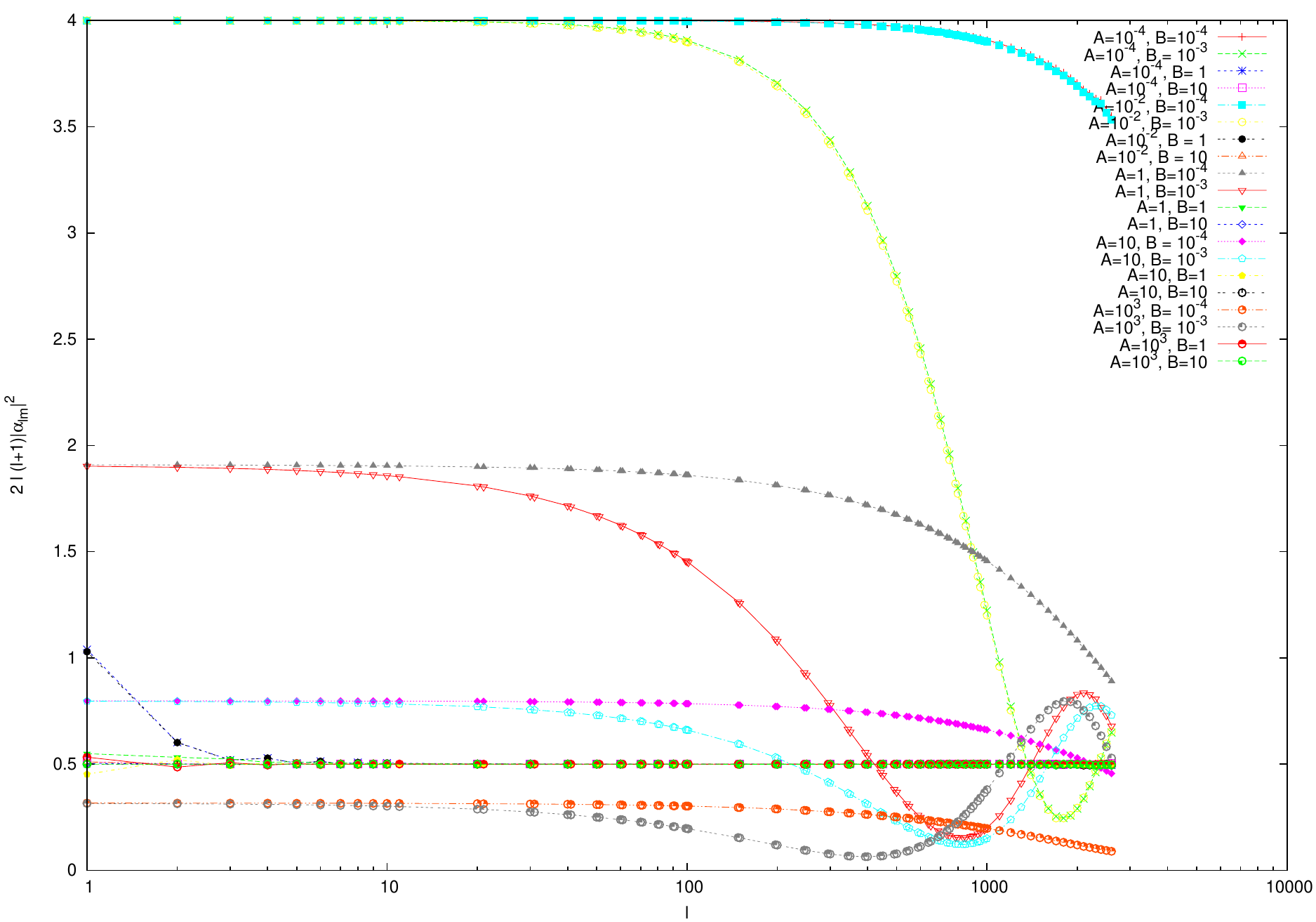}
  \caption{Gráfica semilogarítmica de
    $|\alpha_{lm}|^2(C_2(k))$ para diferentes valores de
    $(A,B)$ representando que tan robusto es el esquema de
    colapso cuando se desvía de $z_k$ constante. La abscisa
    es el multipolo $l$ hasta $l=2600$.}
  \label{fig:c2_log}
\end{figure*}

\begin{figure*}
  \includegraphics[width=210mm,
  angle=90]{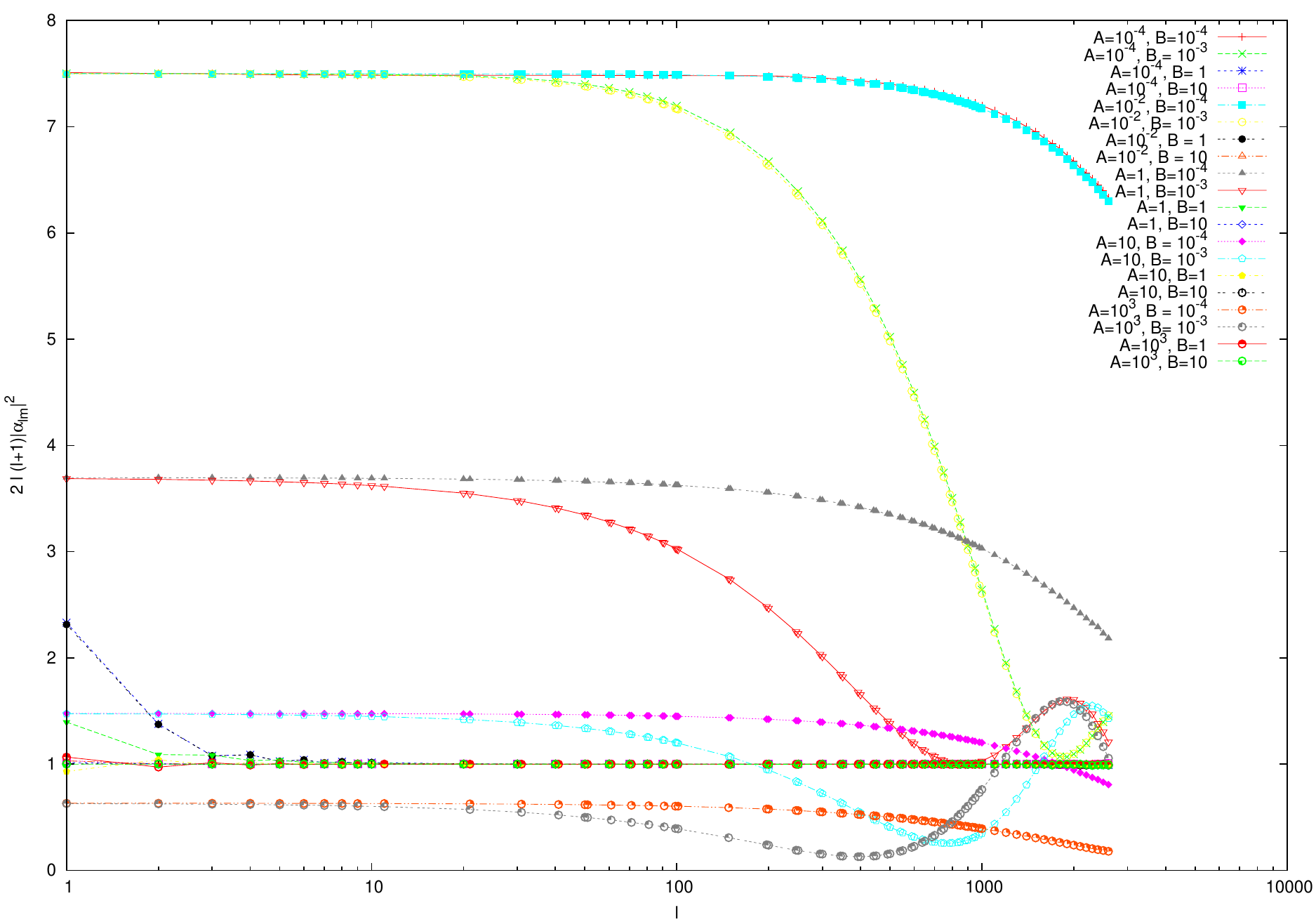}
  \caption{Gráfica semilogarítmica de  $|\alpha_{lm}|^2(C_{Wigner}(k))$ para
    diferentes valores de $(A,B)$, representando que tan
    robusto es el esquema de colapso cuando se desvía de  $z_k$
    constante. La abscisa es el multipolo $l$ hasta $l = 2600$.}
  \label{fig:c_wigner_log}
\end{figure*}

\begin{figure*}
  \centering \mbox { \subfigure[$C_{1}(k),\, A = 10^{-4}$]{
      \includegraphics[width=75mm,
      height=62mm]{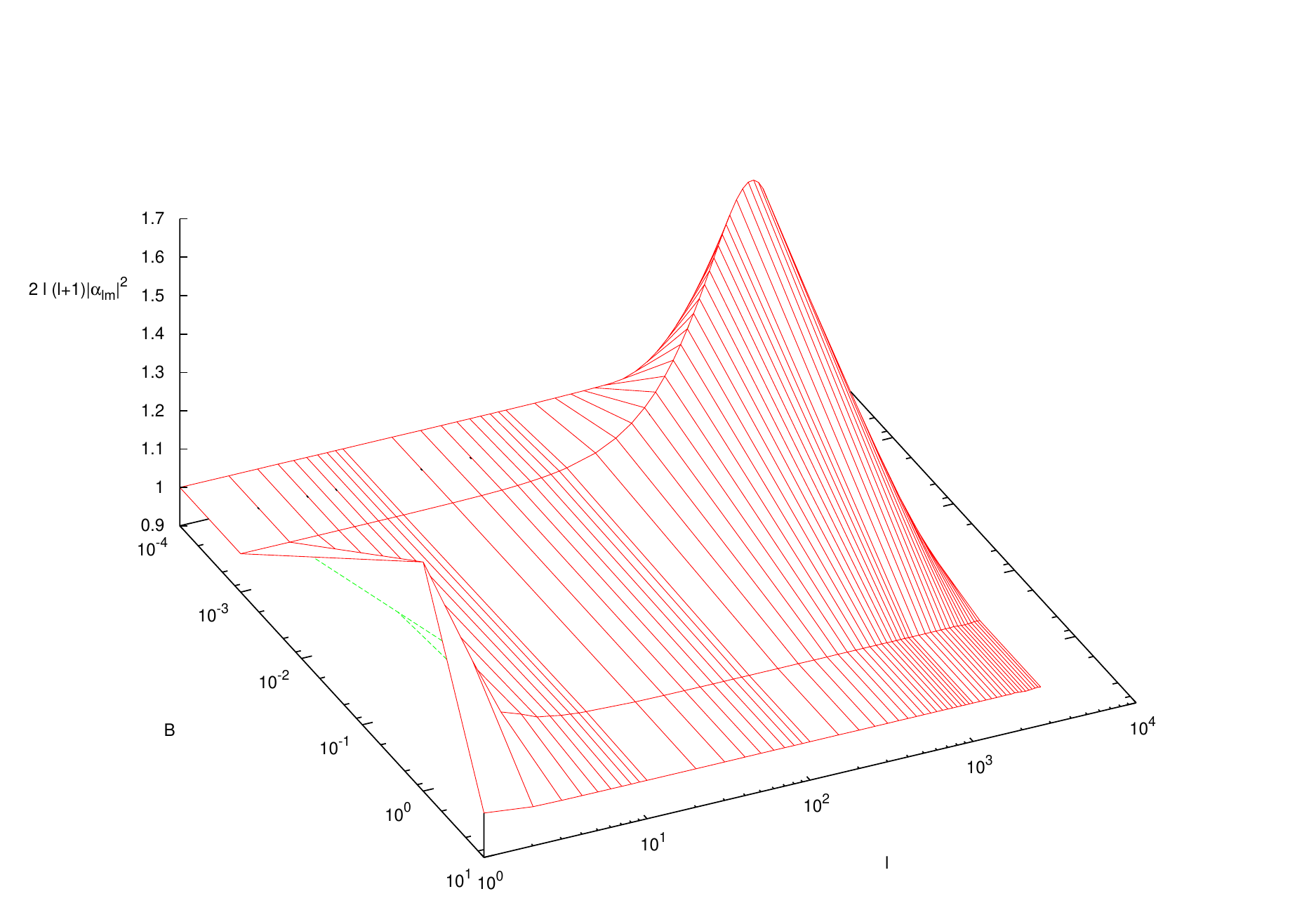}
      \label{fig:c1_a_10-4}} \subfigure[$C_{1}(k),\, A =
    10^{-3}$]{
      \includegraphics[width=75mm,
      height=62mm]{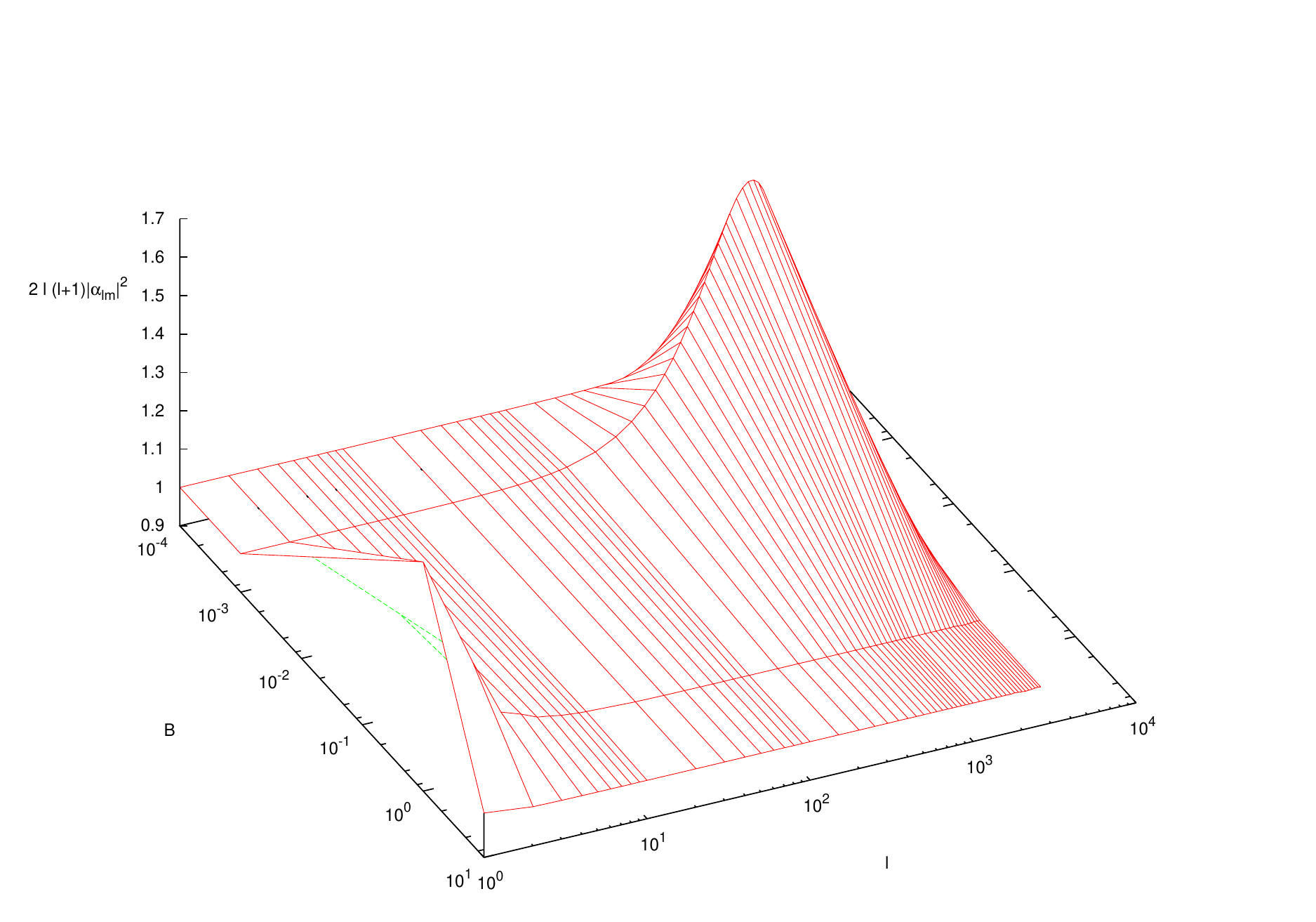}
      \label{fig:c1_a_10-2}} } \mbox {
    \subfigure[$C_{1}(k),\, A = 1$]{
      \includegraphics[width=75mm,
      height=62mm]{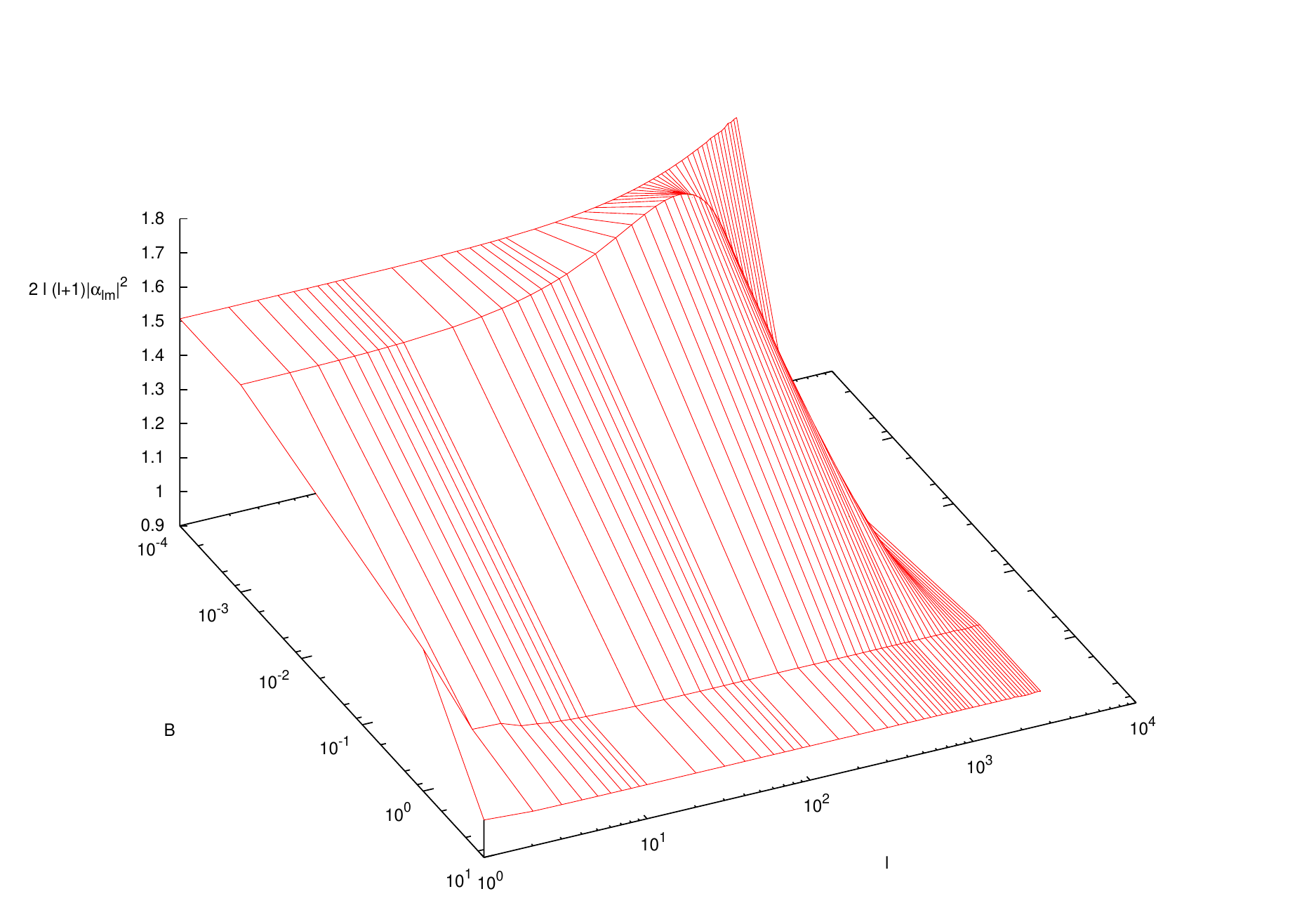}
      \label{fig:c1_a_1}}

    \subfigure[$C_{1}(k),\, A = 10$]{
      \includegraphics[width=75mm,
      height=62mm]{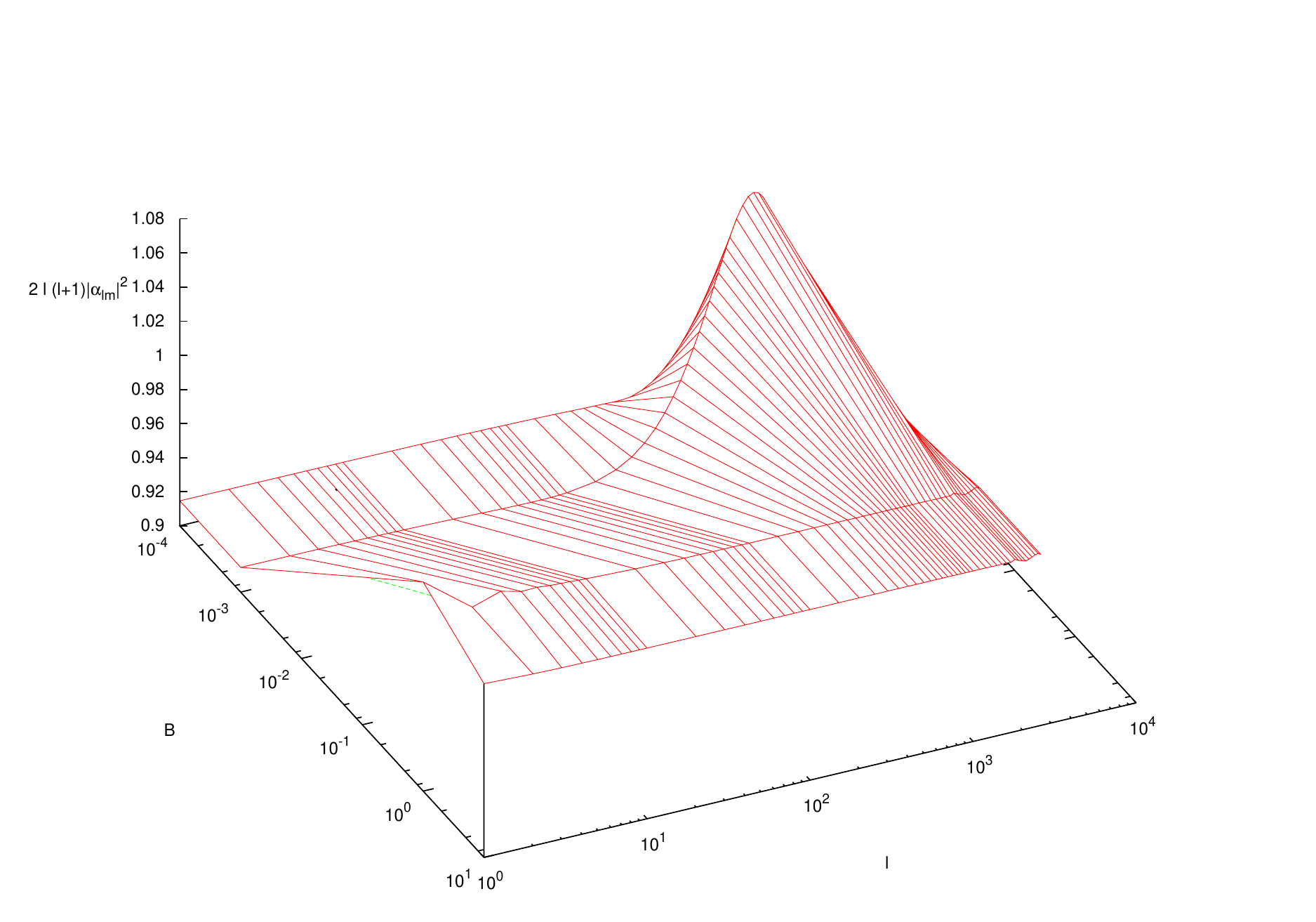}
      \label{fig:c1_a_10}
    } } \subfigure[$C_{1}(k),\, A = 1000$]{
    \includegraphics[width=75mm,
    height=62mm]{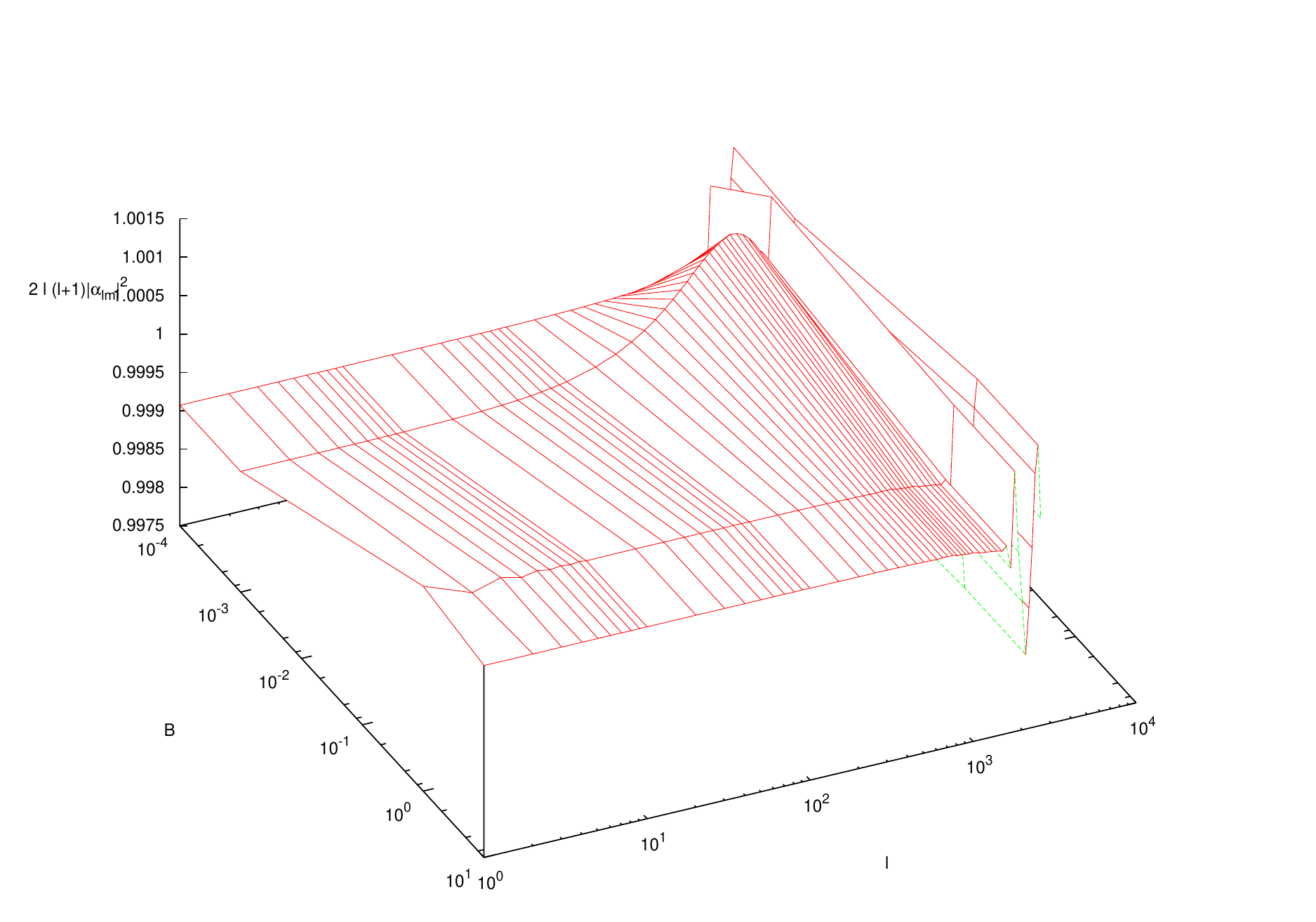}
    \label{fig:c1_a_1000}
  }

  \caption{Gráfica mostrando como la integral de 
    $|\alpha_{lm}|^2(C_{1})$ varia con respecto a cambios en
    $B$ ($10^{-4} - 10$), manteniendo $A$ fija. Ambos ejes $B$ y
    $l$ están en escala logarítmica. Ver el texto para una
    explicación más extensa.}  \label{fig:c1_3d}
\end{figure*}

\begin{figure*}

  \centering \mbox { \subfigure[$C_{2}(k),\, A = 10^{-4}$]{
      \includegraphics[width=75mm,
      height=62mm]{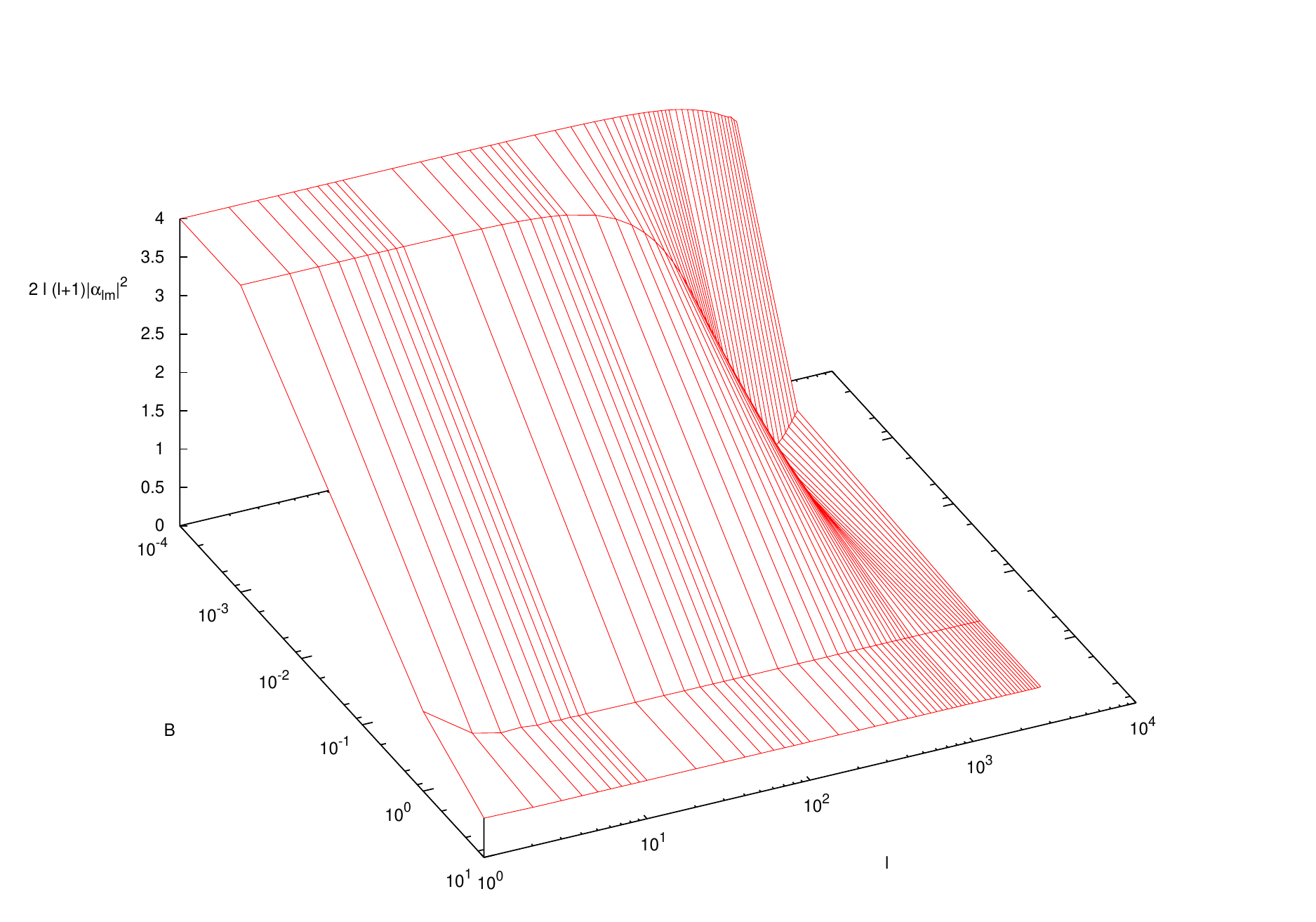}
      \label{fig:c2_a_10-4}} \subfigure[$C_{2}(k),\, A =
    10^{-3}$]{
      \includegraphics[width=75mm,
      height=62mm]{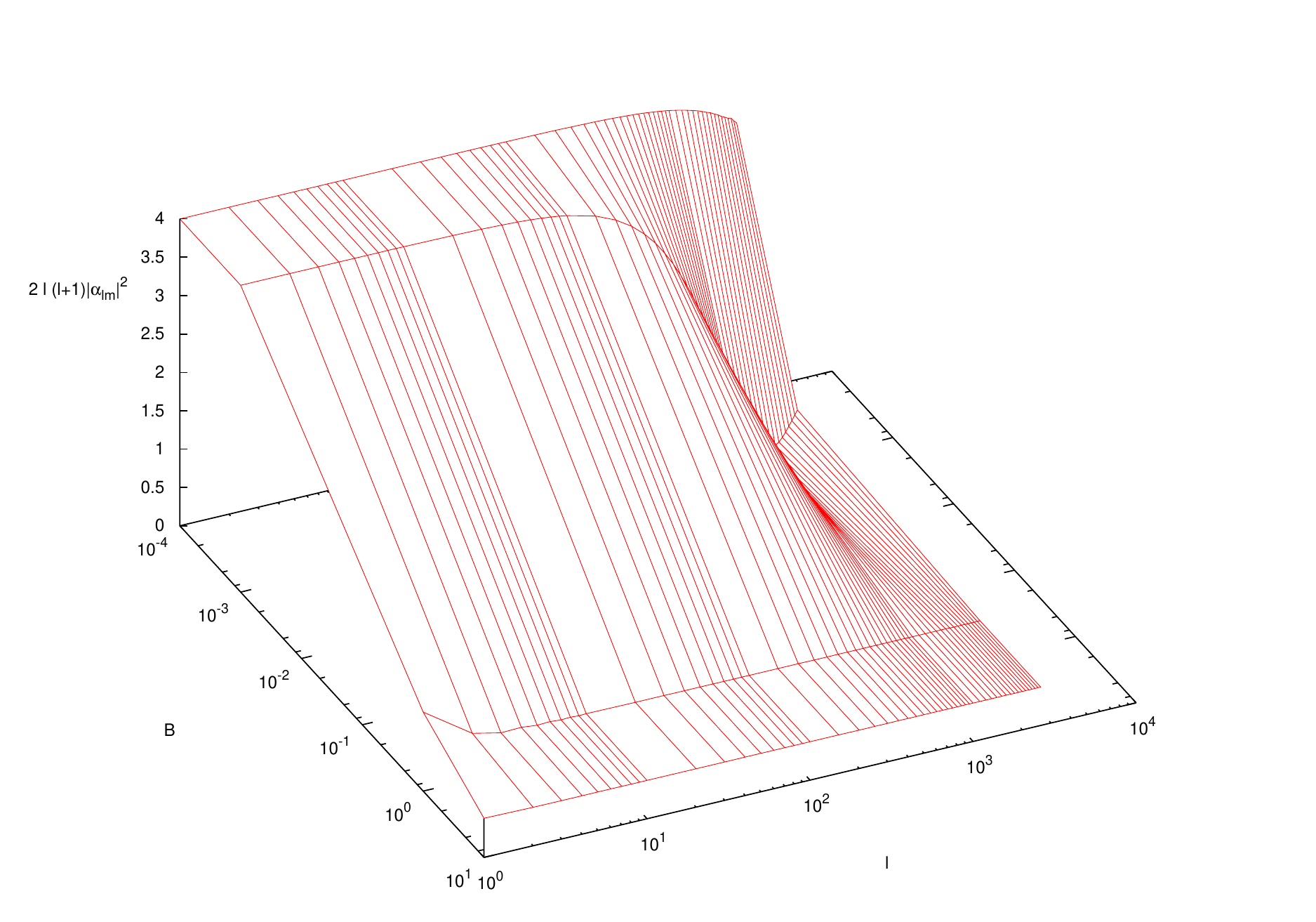}
      \label{fig:c2_a_10-2}} } \mbox {
    \subfigure[$C_{2}(k),\, A = 1$]{
      \includegraphics[width=75mm,
      height=62mm]{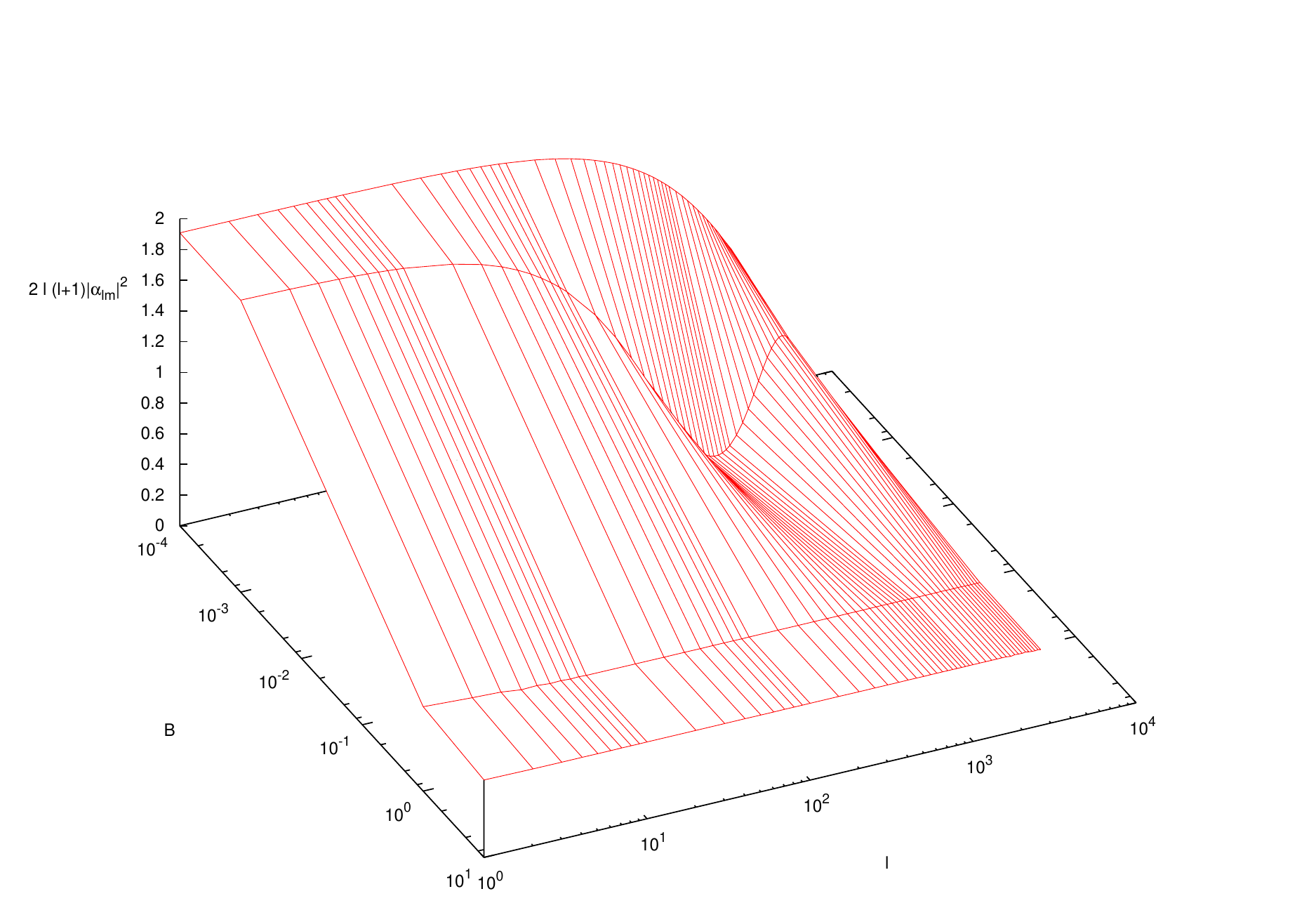}
      \label{fig:c2_a_1}}

    \subfigure[$C_{2}(k),\, A = 10$]{
      \includegraphics[width=75mm,
      height=62mm]{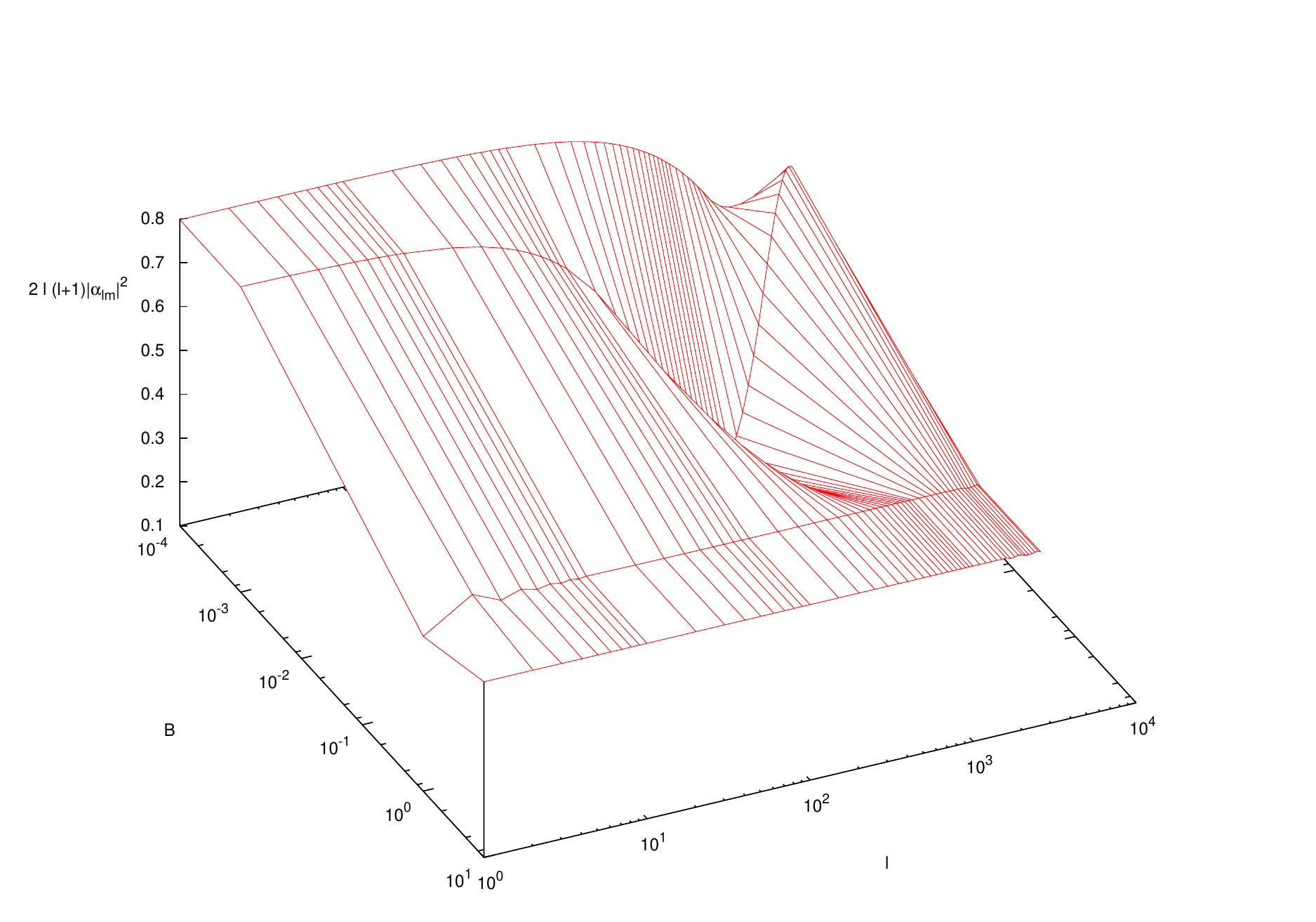}
      \label{fig:c2_a_10}
    } } \subfigure[$C_{2}(k),\, A = 1000$]{
    \includegraphics[width=75mm,
    height=62mm]{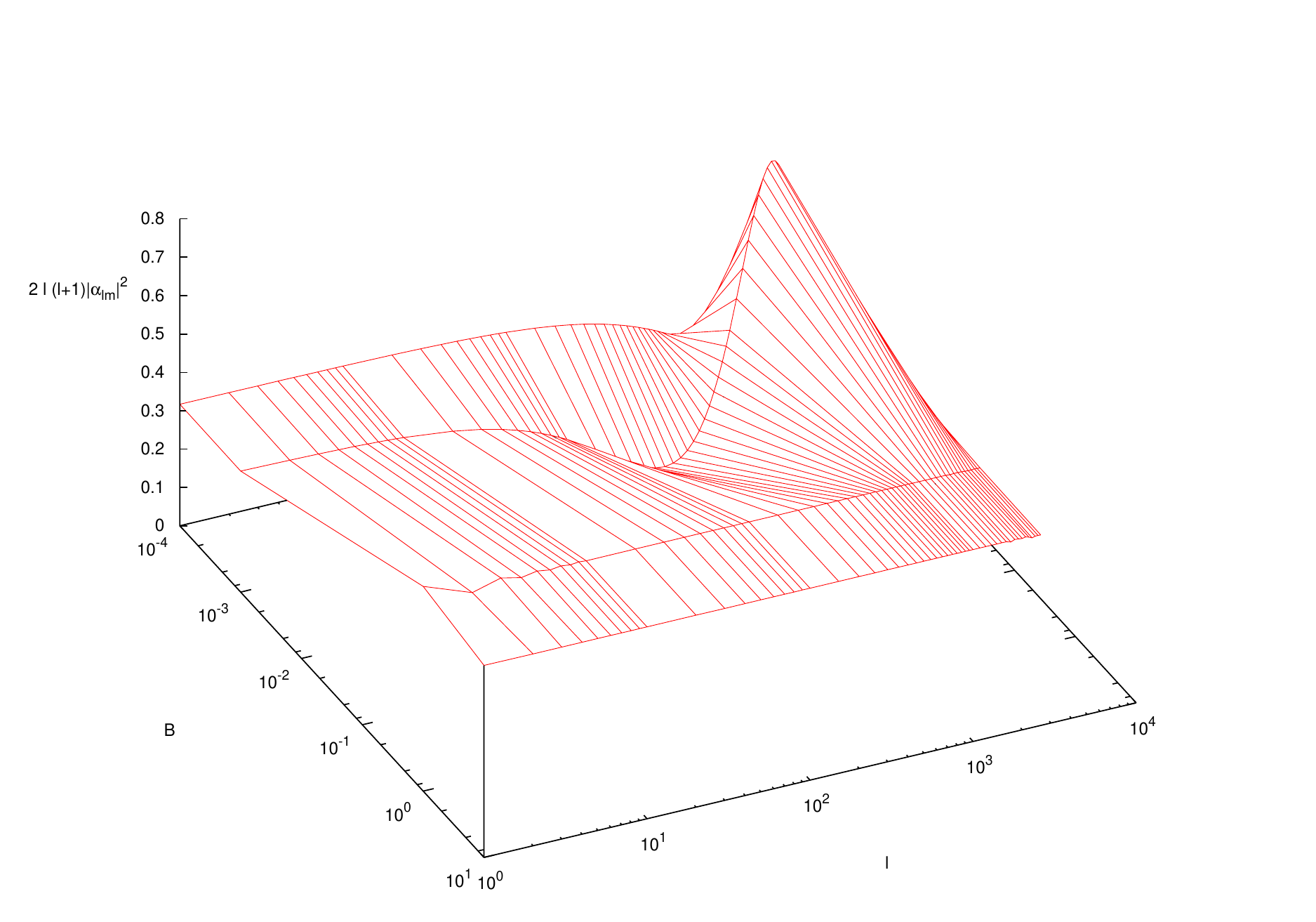}
    \label{fig:c2_a_1000}
  }

  \caption{Gráfica mostrando como la integral de 
    $|\alpha_{lm}|^2(C_{2})$ varia con respecto a cambios en
    $B$ ($10^{-4} - 10$), manteniendo $A$ fija. Ambos ejes $B$ y
    $l$ están en escala logarítmica. Ver el texto para una
    explicación más extensa.}  \label{fig:c2_3d}
\end{figure*}

\begin{figure*}

  \centering \mbox { \subfigure[$C_{Wigner}(k),\, A =
    10^{-4}$]{
      \includegraphics[width=75mm,
      height=62mm]{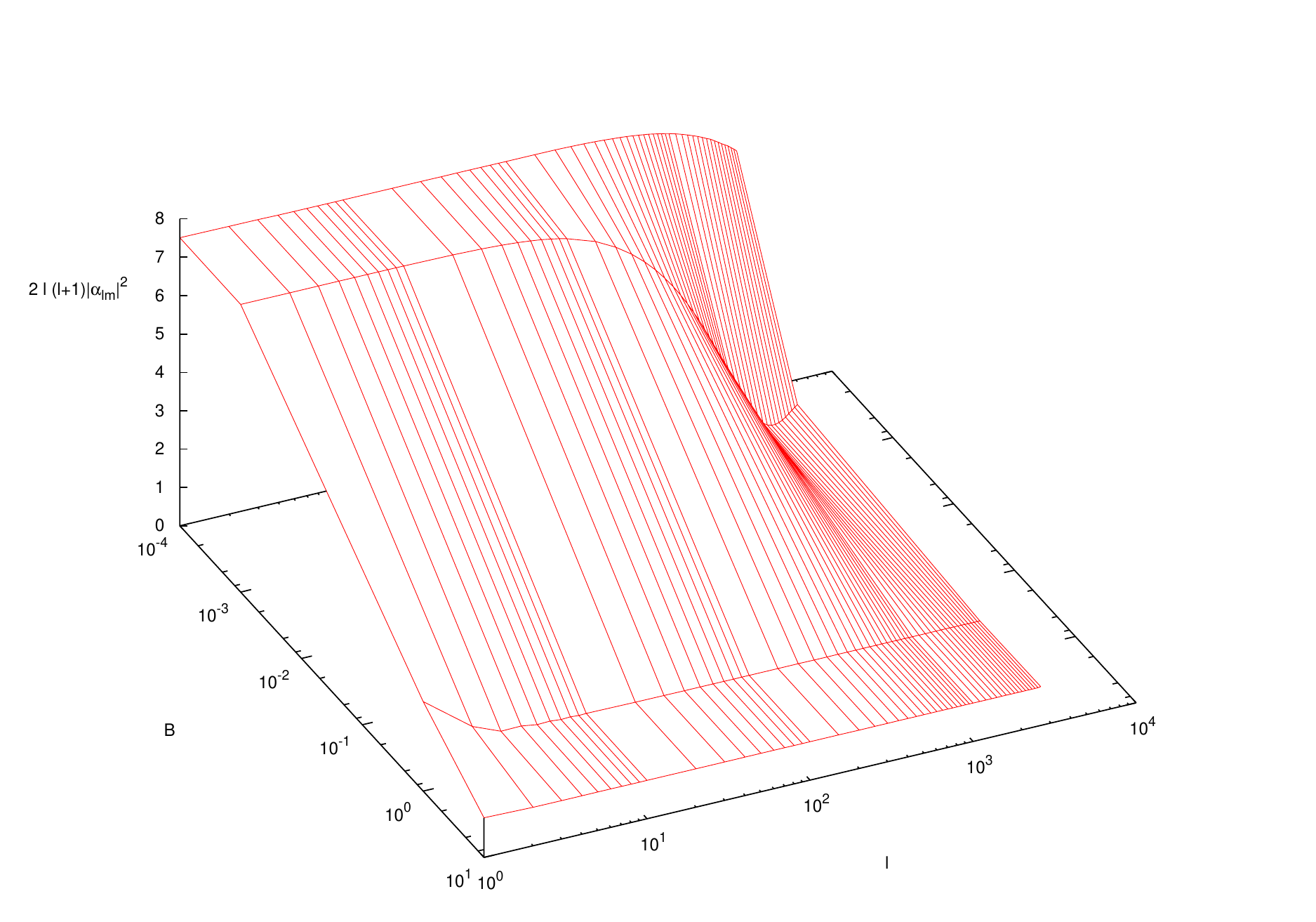}
      \label{fig:wigner_a_10-4}}
    \subfigure[$C_{Wigner}(k),\, A = 10^{-3}$]{
      \includegraphics[width=75mm,
      height=62mm]{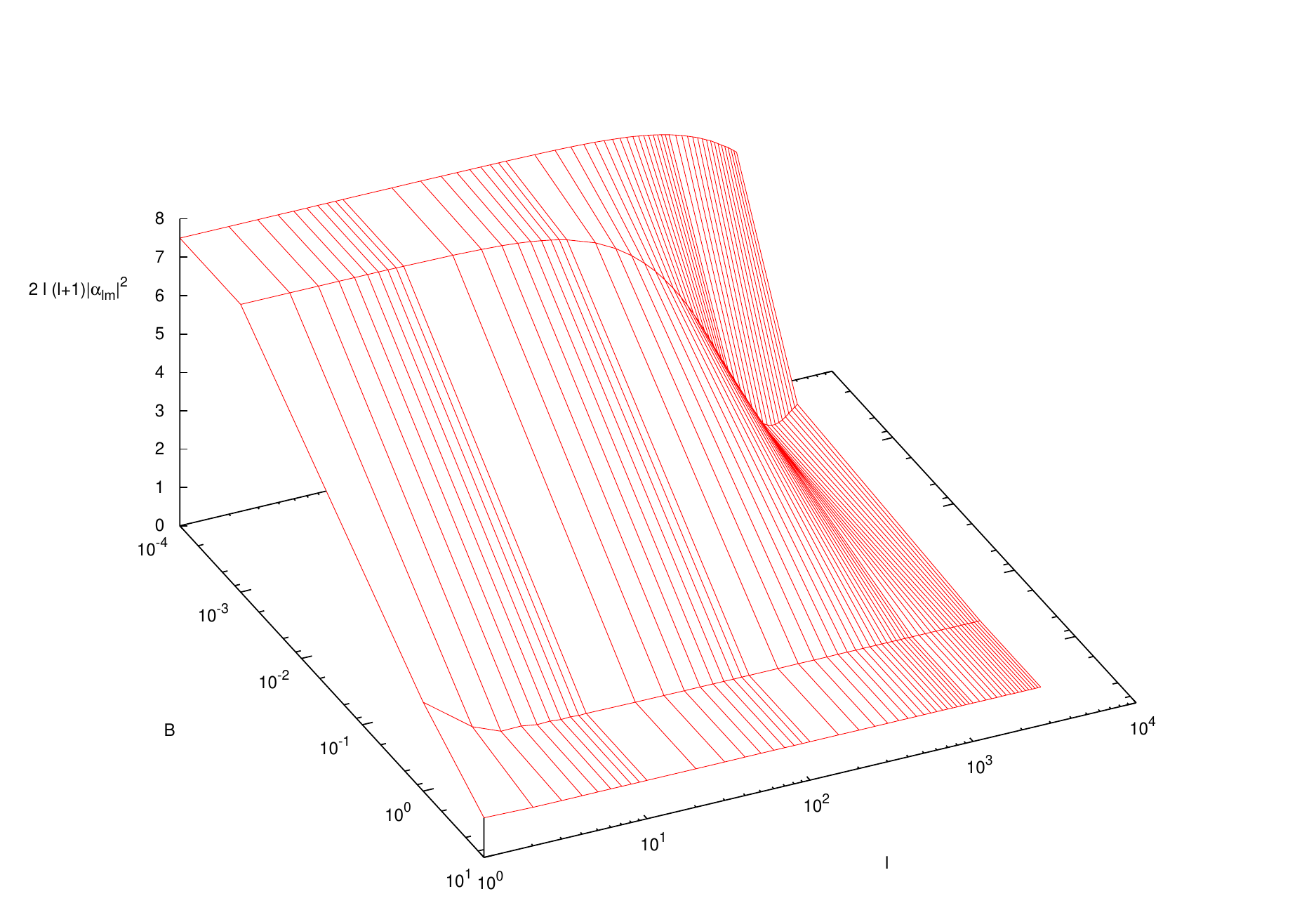}
      \label{fig:wigner_a_10-2}} } \mbox {
    \subfigure[$C_{Wigner}(k),\, A = 1$]{
      \includegraphics[width=75mm,
      height=62mm]{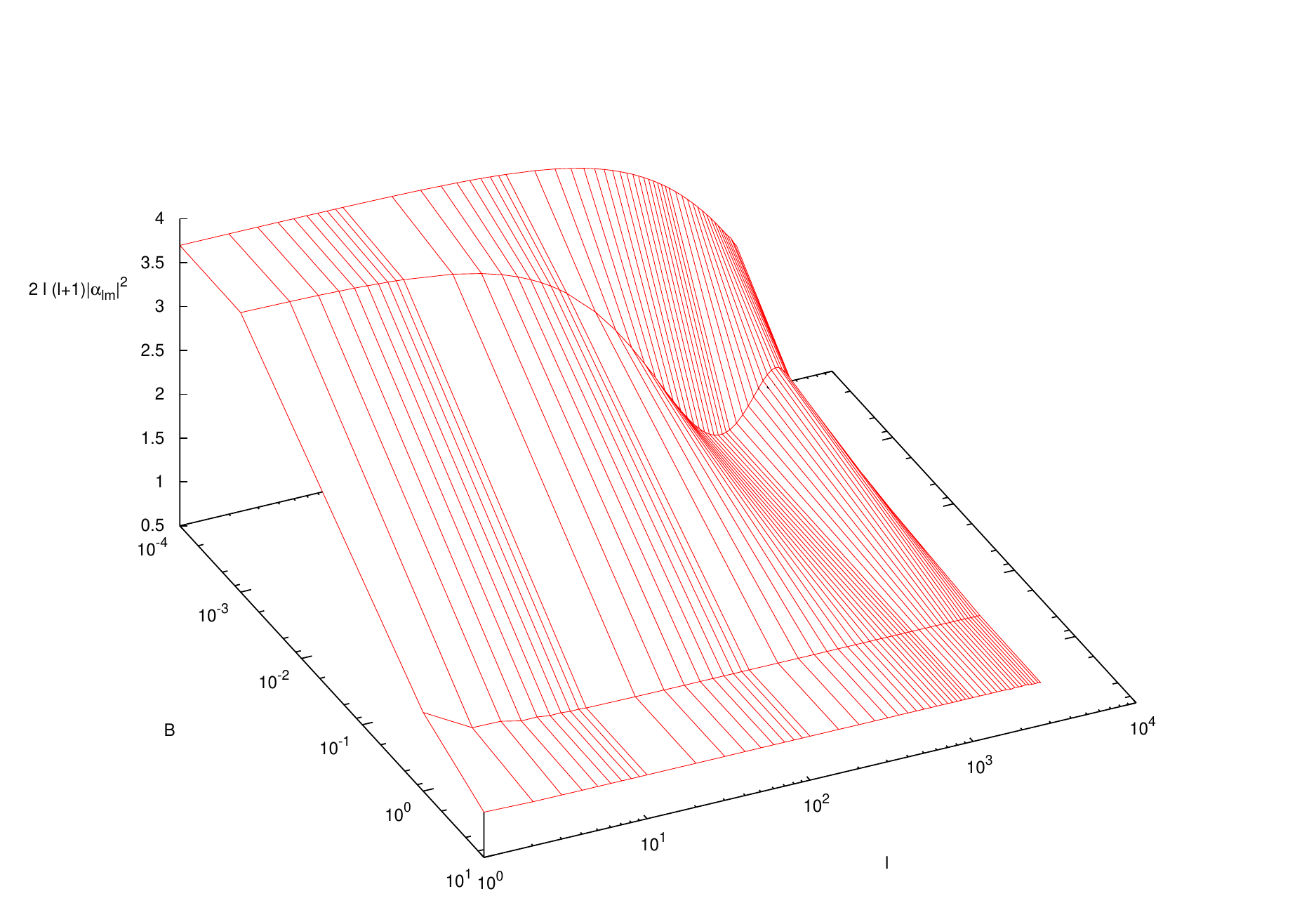}
      \label{fig:wigner_a_1}}

    \subfigure[$C_{Wigner}(k),\, A = 10$]{
      \includegraphics[width=75mm,
      height=62mm]{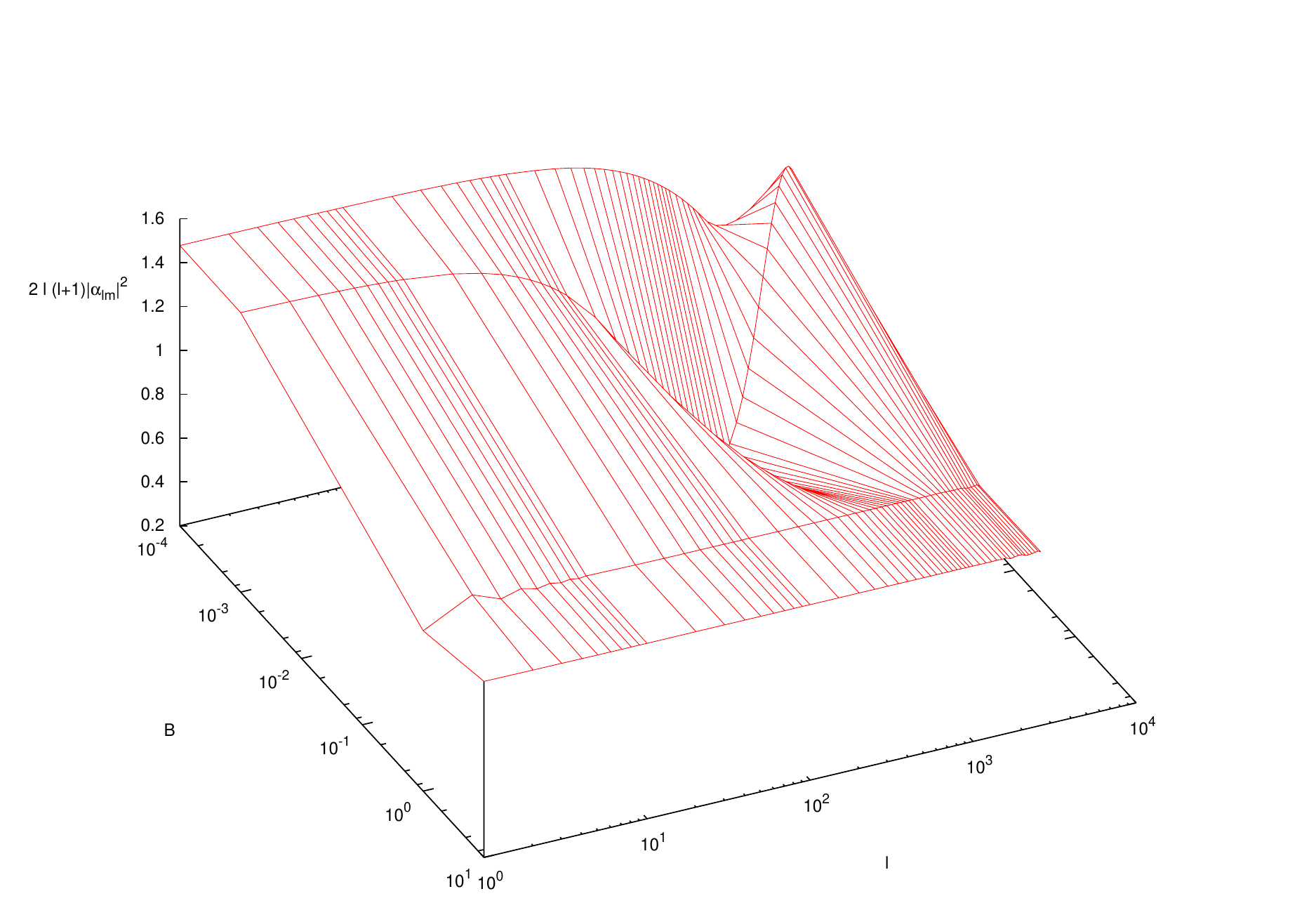}
      \label{fig:wigner_a_10}
    } } \subfigure[$C_{Wigner}(k),\, a = 1000$]{
    \includegraphics[width=75mm,
    height=62mm]{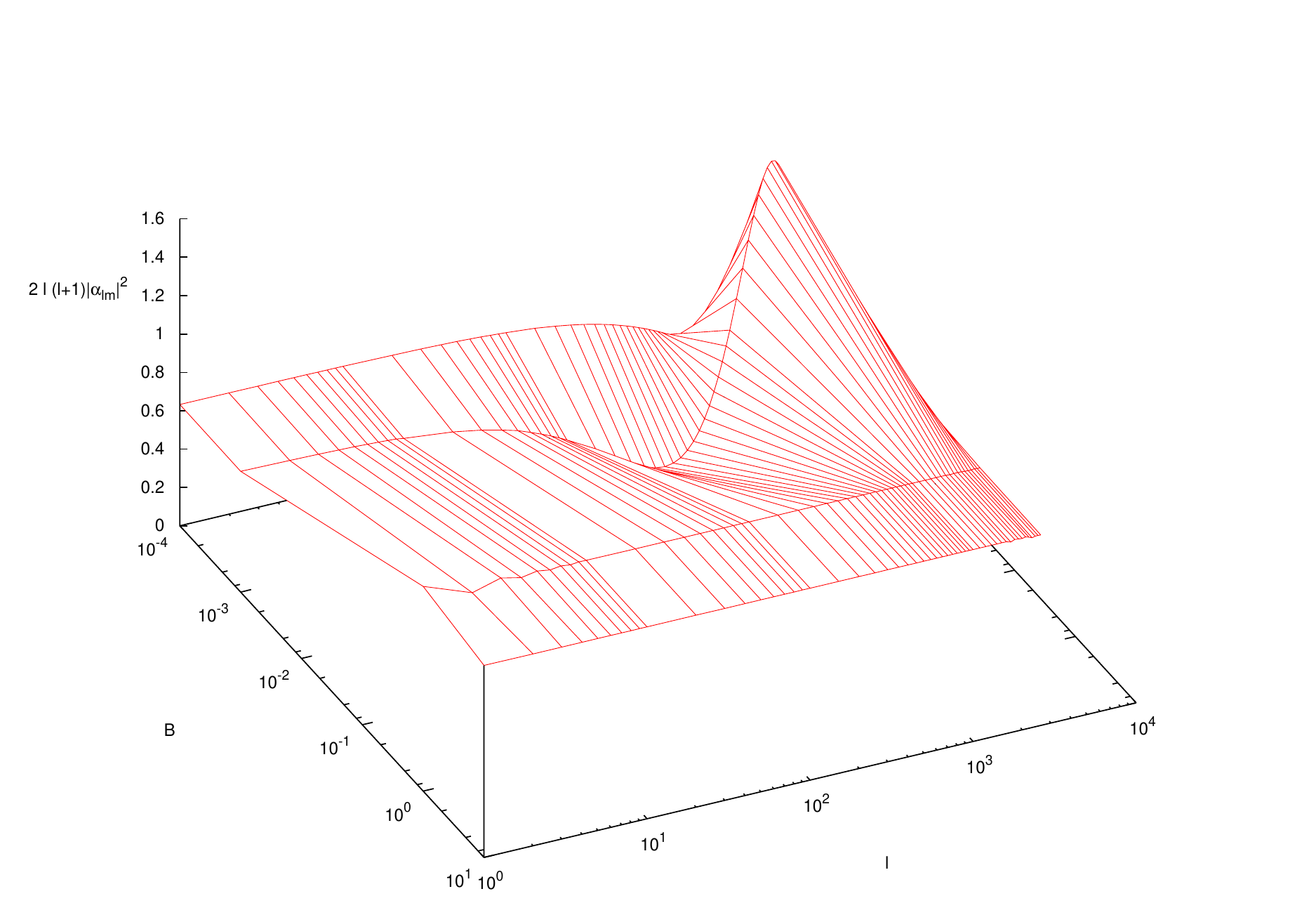}
    \label{fig:wigner_a_1000}
  }

  \caption{Gráfica mostrando como la integral de 
    $|\alpha_{lm}|^2(C_{Wigner})$ varia con respecto a cambios en
    $B$ ($10^{-4} - 10$), manteniendo $A$ fija. Ambos ejes $B$ y
    $l$ están en escala logarítmica. Ver el texto para una
    explicación más extensa.}  \label{fig:c_wigner_3d}
\end{figure*}

Las figuras \ref{fig:c1_log}, \ref{fig:c2_log} y
\ref{fig:c_wigner_log} reflejan la manera en la que el
espectro se comporta como función de $\, l\,$, debemos
recordar que la predicción estándar (ignorando física tardía
como las oscilaciones del plasma) es una línea
horizontal. Estas gráficas representan varios valores de $A$
y $B$ elegidos como muestra para cubrir un dominio
relativamente amplio. Las gráficas (\ref{fig:c1_3d},
\ref{fig:c2_3d} y \ref{fig:c_wigner_3d}) muestran la forma
del espectro para varias elecciones del valor $B$
manteniendo el valor de $A$ fijo.

Como se observó antes, el comportamiento de
$C_2$~(fig.\ \ref{fig:c2_log}) y $C_{Wigner}$ (fig.\
\ref{fig:c_wigner_log}) 
es cualitativamente similar, siendo su principal diferencia
las amplitudes de la oscilación del funcional.


De estos resultados podemos obtener algunas constricciones
razonables a los valores de $A$ y $B$ para los diferentes
esquemas de colapso. En el estudio numérico realizado se
exploraron diferentes 
valores de $A$ y $B$, $A = \{0.0001$, $0.01$, $1$, $10, 1000\}$ y
$B=\{0.0001, 0.001, 1, 10\}$. Para cuantificar la robustez
del esquema de colapso se utilizó simplemente una desviación
de la siguiente forma
\begin{equation}
  \label{eq:desviacion_observacion}
  \Delta_{l_{max}} =
  \frac{\sqrt{\frac{1}{l_{max}} \sum_{l=1}^{l=l_{max}}
      \left[ l(l+1) \cdot \frac{1}{(2 
          l+1)}\sum_m |\alpha_{lm}|_{colapso}^2 -S\right]^2}}{S},
\end{equation}
donde $S$ representa el espectro plano dado por $S\equiv
\frac{1}{lmax} \Sigma_{l=1}^{l=lmax} (l(l+1) \frac{1}{2l+1}
\Sigma_m |\alpha_{l,m} |^2)$.  Se considerará que una
desviación aceptable es aquella menor a 10\% , i.e.\
$\Delta_{l_{max}} < 0.1$, donde $l_{max} = 1500$. Haciendo
el análisis numérico, obtendremos, para los diversos
esquemas de colapso los rangos permitidos correspondientes
para $A$ y $B$.  Los resultados de este análisis son
presentados en la tabla \ref{tab:robustez}.

\ctable[ pos=hbp, caption = Valores de $A$ y $B$ para los
cuales se obtuvieron los mejores desempeños de la robustez
de los diferentes esquemas de colapso., label = tab:mejores
]{c||ccc}{ }{ \FL[0.09em] Esquema de colapso & A& B&
  $\Delta_{l_{max}}$ (\%) \ML[0.07em] $C_1$ & 10& 1&
  0.195523 \NN $C_2$ & 1000& 1& 0.208227 \NN $C_{Wigner}$ &
  0.0001& 10& 0.162458 \LL[0.05em] }

El mejor comportamiento de cada uno de los diferentes
esquemas de colapso se muestran en la tabla
\ref{tab:mejores}. Hay que mencionar que se obtuvieron
varios valores que cumplían con la condición
$\Delta_{l_{max}}< 0.1$: $C_1$ tuvo 15 (y se podría
considerar uno más que apenas sobrepaso el 10\% de
desviación), $C_2$ 12 y $C_{Wigner}$ 10 (aunque en este
último esquema de colapso hay tres valores más apenas por
encima del 10\%). Los peores valores de $A$ y $B$ son
mostrados en la tabla \ref{tab:peores}.

\ctable[ pos=hbp, caption = Valores de $A$ y $B$ para los
cuales se obtuvieron los peores desempeños de la robustez de
los diferentes esquemas de colapso., label = tab:peores
]{c||ccc}{ }{ \FL[0.09em] Esquema de colapso & A& B&
  $\Delta_{l_{max}}$ (\%) \ML[0.03em] $C_1$ & 0.0001 & 0.001
  & 28.3844 \NN $C_2$ & 1000 & 0.001 & 56.2369 \NN
  $C_{Wigner}$ & 1000 & 0.001 & 56.2646 \LL[0.07em] }


\begin{table}
  \centering
  \caption{\label{tab:robustez}Robustez del espectro de los
    diferentes esquemas de colapso cuando los parámetros
    $(A,B)$ son variados entre  $10^{-4} \leq A \leq 10^3$ y
    $10^{-4} \leq B \leq 10$. Los mejores valores están resaltados.}
  \begin{tabular}{|cc|ccc|}
    \multicolumn{2}{c|}{\,}&\multicolumn{3}{c|}{$\Delta_{lmax} \times 100$}\\
    \hline
    $A$&$B$&$C_1(k)$&$C_2(k)$&$C_{Wigner}(k)$\\
    \hline
    \hline
    0.0001 & 0.0001 & 6.63019 & 7.92849 & 10.0763  \\
    0.0001    & 0.001   &  28.3844   &  53.9872      & 47.3616  \\
    0.0001 &      1   &    0.288273 &     0.423473 &   0.506768\\
    0.0001 &  10 & 0.301883 & 0.249129 &  {\cellcolor[gray]{.8}}0.162458\\
    \hline
    0.01 & 0.0001   &  6.84475   &  8.12093 &    10.2874\\
    0.01 &  0.001   &  28.3706   &  54.2265 &     47.494\\
    0.01 &   1   & 0.282546   & 0.277929 &   0.359852\\
    0.01 &  10   & 0.301614   & 0.251313 &   0.165756\\
    \hline
    1 & 0.0001   &  10.1258   &  21.8266 &     18.445\\
    1 &  0.001   &  21.3117   &  50.6328 &    34.1731\\
    1 &   1   & 0.247444   & 0.312876 &   0.358535\\
    1 &  10   & 0.341509  & 0.443572 &   0.394309\\
    \hline
    10 & 0.0001   &  1.67782  &  18.4953 &    19.3128\\
    10 &  0.001   &  15.8869   &  46.1397 &    45.1946\\
    10 &   1   & {\cellcolor[gray]{.8}} 0.195523 & 0.917963 & 0.51842\\
    10 &  10   & 0.384265   & 0.445398   & 0.430548\\
    \hline
    1000 & 0.0001   &  0.44236   &  28.9085 &    28.9273\\
    1000 &  0.001   &  1.58567   &  56.2369  &    56.2646\\
    1000 &   1   & 0.394892   & {\cellcolor[gray]{.8}} 0.208227  &   0.197662\\
    1000 &  10   & 0.402706   & 0.434914  &   0.445794\\
    \hline
  \end{tabular}
\end{table}

\begin{figure}[htp]
  \centering
  \includegraphics[height=14cm,
  width=16cm]{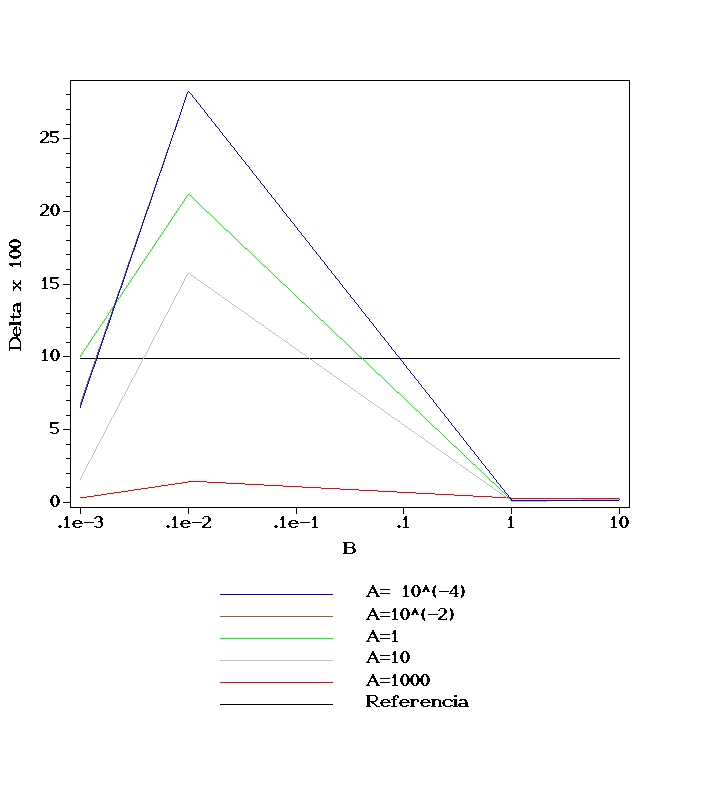}
  \caption{Desempeño del esquema de colapso 1. La figura
    muestra las desviaciones de $C(z_k)$, respecto al
    espectro plano, para diferentes valores de $A$ y $B$. El
    desempeño deseable está definido mediante
    $\Delta_{l_{max}}< 0.1$ que está representado en la
    figura con una línea recta horizontal con la leyenda
    ``Referencia''. Valores por abajo de la línea de
    referencia son desempeños
    aceptables.}\label{fig:performance_c1}
\end{figure}

\begin{figure}[htpb]
  \centering
  \includegraphics[height=14cm,
  width=16cm]{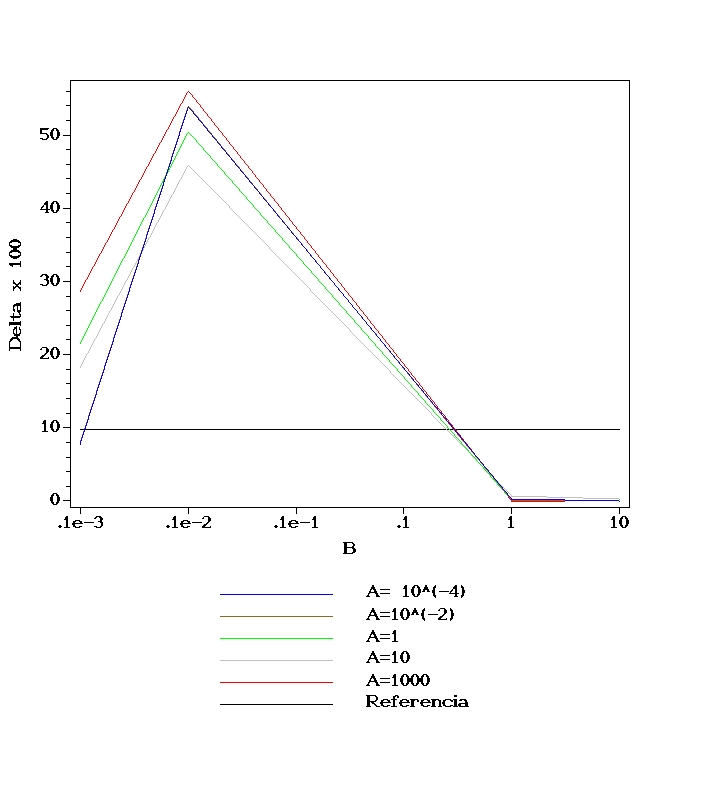}
  \caption{Desempeño del esquema de colapso 2. La figura
    muestra las desviaciones de $C(z_k)$, respecto al
    espectro plano, para diferentes valores de $A$ y $B$. El
    desempeño deseable está definido mediante
    $\Delta_{l_{max}}< 0.1$ que está representado en la
    figura con una línea recta horizontal con la leyenda
    ``Referencia''. Valores por abajo de la línea de
    referencia son desempeños
    aceptables.}\label{fig:performance_c2}
\end{figure}

\begin{figure}[tbp]
  \centering
  \includegraphics[height=14cm,
  width=16cm]{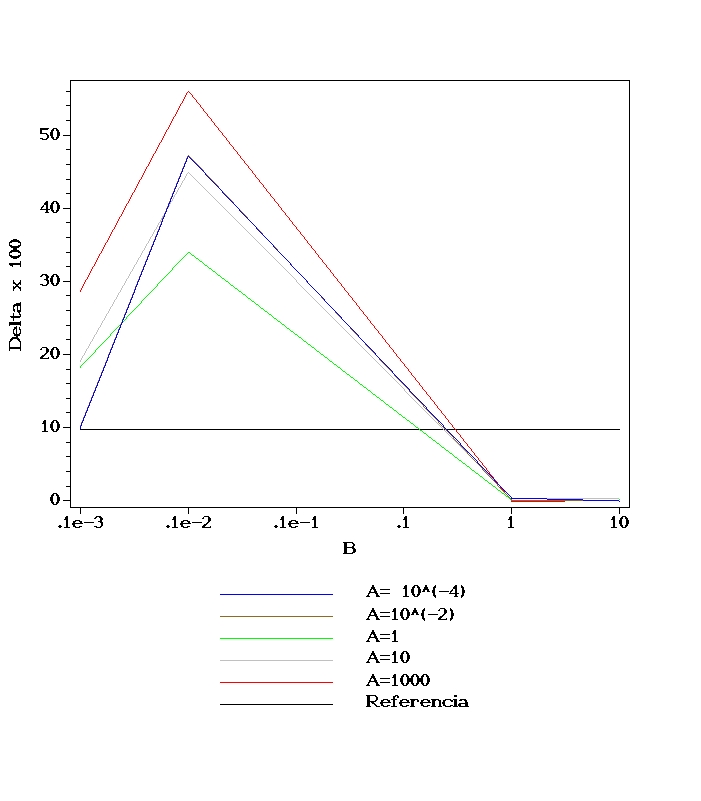}
  \caption{Desempeño del esquema de colapso ``de
    Wigner''. La figura muestra las desviaciones de
    $C(z_k)$, respecto al espectro plano, para diferentes
    valores de $A$ y $B$. El desempeño deseable está
    definido mediante $\Delta_{l_{max}}< 0.1$ que está
    representado en la figura con una línea recta horizontal
    con la leyenda ``Referencia''. Valores por abajo de la
    línea de referencia son desempeños
    aceptables.}\label{fig:performance_cwigner}
\end{figure}

Presentando la información de la tabla \ref{tab:robustez} en
gráficas por esquema de colapso se puede extraer más
información sobre los mismos. Las figuras
\ref{fig:performance_c1}, \ref{fig:performance_c2} y
\ref{fig:performance_cwigner} tienen escalado
logarítmicamente el eje $x$, el cual representa los
distintos valores de $B$. El eje vertical representa el
porcentaje de desviación (ecuación \ref{eq:desviacion_observacion}), i.e.\
$\Delta_{l_{max}} \times 100$.

\begin{figure*}

  \centering 
    \subfigure[$C_{newtoniano}(k)$]{
      \includegraphics[width=11.5cm,
      height=9.3cm]{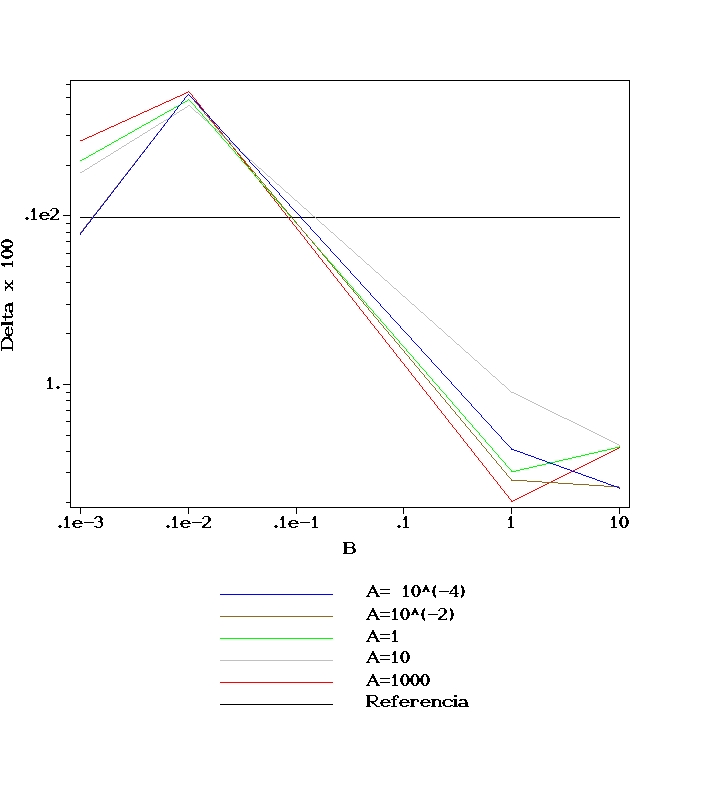}
      \label{fig:wigner_a_1}
    }

    \subfigure[$C_{Wigner}(k)$]{
      \includegraphics[width=11.5cm,
      height=9.3cm]{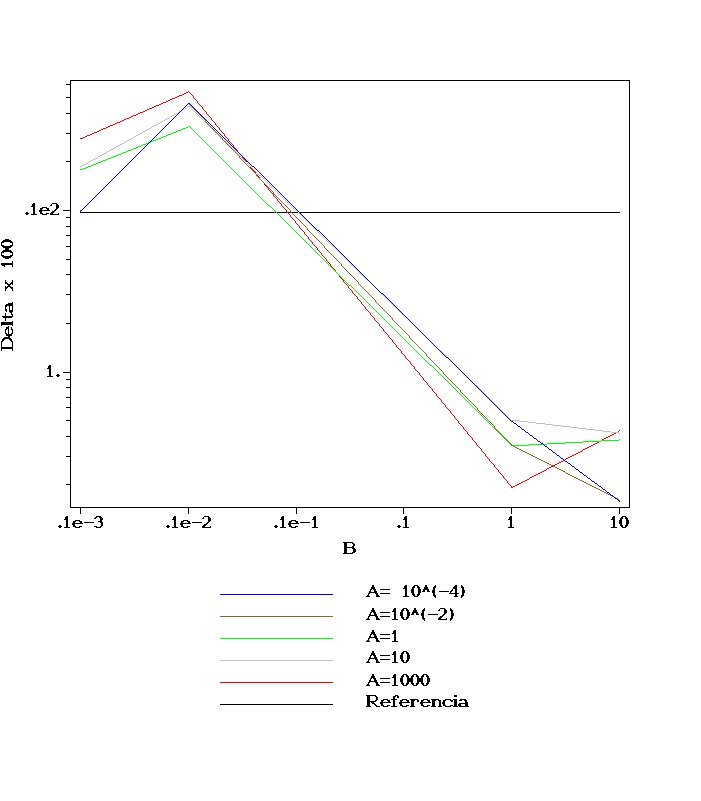}
      \label{fig:wigner_a_10}
    } 
 
  \caption{Desempeño de los esquemas de colapso newtoniano y ``de
    Wigner''. La figuras muestran las desviaciones de
    $C(z_k)$, respecto al espectro plano, para diferentes
    valores de $A$ y $B$.Ambos ejes están escalados
    logarítmicamente. El desempeño deseable está definido
    mediante $\Delta_{l_{max}}< 0.1$ que está representado
    en la figura con una línea recta horizontal con la
    leyenda ``Referencia''. Valores por abajo de la línea de
    referencia son desempeños
    aceptables.}\label{fig:loglog_performance}
\end{figure*}

Primero se presenta la gráfica que muestra el desempeño del
esquema de colapso uno (figura \ref{fig:performance_c1}). En
esta figura podemos notar que el peor comportamiento ocurre
en $B = 0.001$ indistintamente del valor tomado por $A$,
esta característica aparecerá en todos los esquemas de
colapso. También se advierte en esta gráfica la existencia
de un valor de $A$ ($A=1000$) el cual se ve poco afectada
por el cambio de $B$ en el rango estudiado ($B \in [0.0001,
10]$). Otra característica a observar es que el
comportamiento numérico de los valores de $A = 0.0001$ y $A
= 0.01$ son casi indistinguibles con la resolución mostrada
en la figura en rango de $B$ explorado (aunque muestra
ligeras desviaciones para valores de $B$ mayores, cf.\ ver
las gráficas \ref{fig:loglog_performance} ) este
comportamiento (justamente en los mismos valores de $A$)
estará presente en los esquemas de otros dos esquemas de
colapso, por último, en la gráfica de $C_1$ enseña que a
medida que $A$ toma valores mayores las restricciones sobre
$B$ se hacen más débiles.

La gráficas para los esquemas $C_2$ (figura
\ref{fig:performance_c2}) y $C_{Wigner}$ (figura
\ref{fig:performance_cwigner}) son muy parecidas entre sí
(como se comentó antes en este capítulo). Las diferencias
que se pueden mencionar en el desempeño del esquema de
Wigner respecto al del esquema de colapso dos son: (a) el
peor de los desempeños ocurre en $B=0.001$ en ambos esquemas
de colapso, pero el de $C_{Wigner}$ es mejor que el de $C_2$
por una diferencia del casi 10\%, en $A=\{0.0001$, $0.01\}$, de
casi el 15\% para $A=1$, pero para los valores superiores de
$A (10$ y $1000)$ esta brecha se cierra a 1\% en $A = 10$ y se
invierte (-1\%) en $A = 1000$, (b) para los valores de $B =
0.0001$ y cualquier valor de $A$ (exceptuando $A=1$) su
desempeño es ligeramente peor que el del esquema dos.

Al comparar las actuaciones de los esquemas de colapso dos y
de Wigner con el esquema de colapso uno se encuentran muchas
similitudes, pero algunas diferencias resaltan
inmediatamente: (a) el peor desempeño ante desviaciones de
la receta mencionada (que también se da en $B=0.001$, cf.\
tabla \ref{tab:peores}) es casi el doble comparado con el
obtenido con $C_1$ ( $\sim 55\%$ contra el $\sim 29\%$), (b)
la inexistencia de un valor de $A$ para el cual
independientemente del valor de $B$ -en el rango estudiado-
se obtenga un desempeño dentro de lo deseado
($\Delta_{l_{max}}<0.1$) y (c) la carencia de la relajación
del valor de $B$ necesario para cumplir con el desempeño
aceptado para valores mayores de $A$.

Para recuperar el rango de tiempos de colapso para los
diferentes valores de $A$ y $B$, para lograrlo, despejamos
$\eta_k^c$ de $N(x) = A/x + B$, obteniendo $|\eta_k^c(k)| =
A/k + R_D B$, $R_D$ es el radio \textit{comóvil} de la
LSS. Considerando geodésicas nulas radiales llegamos a
$R_{D} = \eta_0 - \eta_{d} $, donde $\eta_d$ es el tiempo
del desacople. El desacople ocurre bien entrada la época de
dominación de materia, por lo que podemos usar la expresión
de $R_D$ en términos del factor de escala de la época de
materia,

\begin{equation}
  R_D = \frac{2}{H_0}\left( 1 - \sqrt{a_d}\right),
\end{equation}

donde se normalizó el factor de escala de tal manera que en
la actualidad es $a_0 = 1$, entonces, $a_d \equiv a(\eta_d)
\simeq 10^{-3}$ y $H_0$ es la variable de Hubble hoy. Su
valor numérico es $\unit{5807.31} h^{-1} \mega \parsec$. Por
lo tanto,

\begin{equation}
  |\eta_k^c(k)| = \frac{A}{k} + \frac{2 B}{H_0}\left( 1 - \sqrt{a_d}\right).
\end{equation}

\begin{figure*}
  \centering \subfigure[$C_1$, with $A= 10, B = 1$.]{
    \label{fig:c1_tiempos_colapso}
    \includegraphics[width=100mm,
    height=60mm]{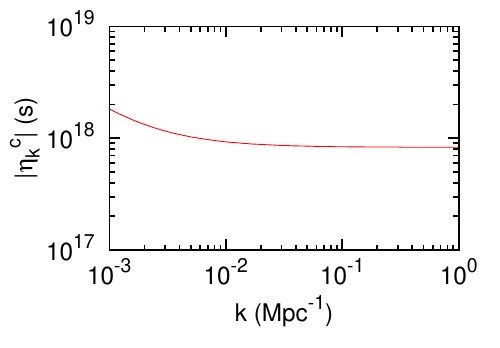}
  } \hspace{1cm} \subfigure[$C_2$, with $A=1000, B=1$.]{
    \label{fig:c2_tiempos_colapso}
    \includegraphics[width=100mm,
    height=60mm]{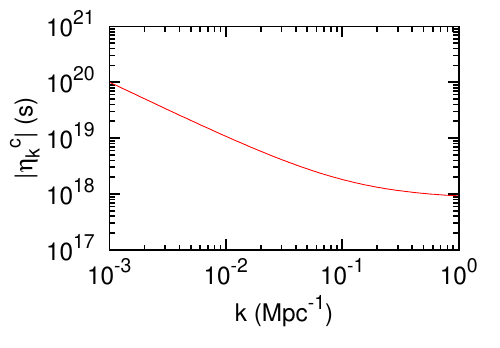}
  } \hspace{1cm} \subfigure[ $C_{Wigner}$, with $A=0.01,
  B=10$. Note how in this scheme almost all the modes must
  collapse at the same time.]{
    \label{fig:cwigner_tiempos_colapso}
    \includegraphics[width=100mm,
    height=60mm]{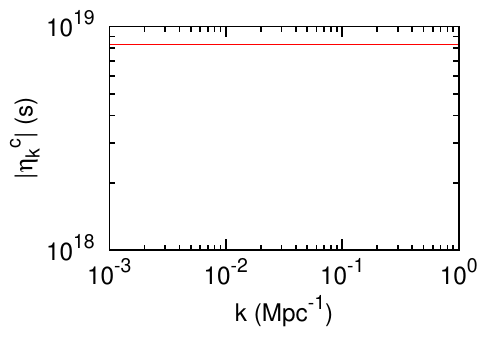}
  } \caption{Gráficas con escala logarítmica en ambos ejes,
    de los tiempos de colapso  $|\eta_k^c|$ (en segundos),
    para los tres esquemas tomando en cuenta solo los
    mejores valores de $(A,B)$ en el rango de  $10^{-3}$
    Mpc$^{-1}$ $< k < 1$ Mpc$^{-1}$. en estas gráficas: $h = 0.7$.}
  \label{fig:tiempos_colapso}
\end{figure*}

Entonces, podemos usar esta fórmula par calcular los tiempos
de colapso de los modos interesantes que observamos en el
CBR, en particular el rango entre $\unit{10^{-3}}
\mega\parsec ^{-1} \leq k \leq \unit{1}
\mega\parsec^{-1}$. Estos modos cubren el rango de
multipolos $l$ de interés: $1 \leq l \leq 2600$, donde
usamos la relación\footnote{La relación aproximada entere la
  escala angular $\theta$ y el multipolo $l$ es $\theta \sim
  \pi/l$. La escala angular comóvil $d_A$, entre nosotros y
  un objeto de tamaño físico lineal $L$, es $d_A =
  L/(a\theta)$. $L/a \sim 1/k$, $d_A = R_D$ si el objeto
  está en el LSS, y usando la primera expresión de este pie
  de página, obtenemos $l = k R_D $.} $\,l = k R_D$. Los
tiempos de colapso del modo $k$ indican el momento en que
las inhomogeneidades y anisotropías emergieron a la escala
correspondiente a $k$. Los tiempos de colapso se ven en la
figura (\ref{fig:tiempos_colapso}) para los mejores valores
de $(A,B)$ dados en las tablas \footnote{ El lector debe de
  recordar que nuestra parametrización del régimen
  infalcionario tiene el tiempo conforme desde números
  grandes negativos \textit{hacia} números pequeños
  negativos.} (\ref{tab:robustez}).

Podemos observar que  los esquemas de colapso newtoniano y
de Wigner, presentan una caída en el espectro de potencia
conforme $l$ se incrementa. Es importante notar que esta
caida es atribuida en la interpretación de los datos del CMB
\cite{wmap5}, a efectos de 
extinguimiento (\emph{Damping
  Effect}\footnote{Este efecto es básicamente un
  extinguimiento  para la densidad de fotones en la escala
  $k$ al tiempo de desacople por un factor de $e^{-k^2/k_D^2}$,
  donde  $k_D$  es una escala de difusión que depende en la
  física de las colisiones entre electrones y
  fotones. Acordemente, el espectro $C_l$ es extinguido como
  $e^{l^2/l_D^2}$ donde $l_D \sim k_D d_A(\eta_d) \sim
  1500$, para parámetros cosmológicos típicos.} en inglés), debidos a
que la superficie de última dispersión no es instantánea
\cite{Anninos01},  \citep[cap. 8]{Dodelson},  \citep[cap 15, 18]{Peacock98}.  Como se observa en la figuras (\ref{fig:c2_log},
\ref{fig:c_wigner_log}) para algunos valores de $(A,B)$
obtenemos una fuente adicional de ``extinguimiento'' 
debido a las fluctuaciones en el tiempo de colapso
alrededor de $\eta^c_k k =$ constante. Se espera que el
satélite PLANCK proverá con mayor información sobre el
espectro para mayores valores de $l$, lo cual proverá
constricciones en los parámetros $(A,B)$. Además, creemos
que es posible distinguir entre los dos efectos, debido  a
que en nuestro modelo para los parámetros en los cuales se
predice un extinguimiento adicional,  también predice  un
``rebote'' para valores aún mayores de $l$ (cf.\ figs.
\ref{fig:c1_log}, \ref{fig:c2_log}, \ref{fig:c_wigner_log}).

Dado un $k$ se puede comparar el valor del factor de escala en el
tiempo de colapso $a(\eta_k^c)$, con el factor de escala al
momento del ``cruce del horizonte''\footnote{En la
  explicación estándar marca ``la
transición de clásico a cuántico'' (capítulo
\ref{cha:critica-al-origen})}  $a_k^H$. El ``cruce del
horizonte'' ocurre cuando la longitud que corresponde al
modo $k$ tiene el mismo tamaño que el radio de Hubble,
$H_I^{-1}$ ( en coordenadas comóviles $k=aH_I$), entonces,
$a_k^H \equiv a(\eta_k^H) = \frac{k}{H_I} = \frac{3k}{8\pi G
  V}$. Entonces el radio entre el factor de escala en el
momento de cruzar el horizonte para el modo $k$ y factor de
escala evaluada en el tiempo de colapso para el mismo modo
es

\begin{equation}
  \frac{a^H_k}{a^c_k} = {k\eta_k^c(k)} = {A + B R_D k} = A+Bl.
\end{equation}

\begin{figure*}
  \centering \subfigure[$C_1$, with $A=10, B=1$.] {
    \label{fig:c1_e_foldings}
    \includegraphics[width=100mm,
    height=60mm]{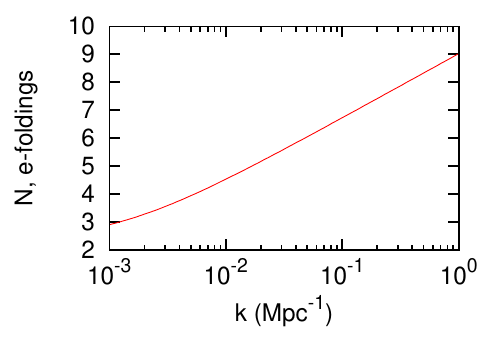}
  } \hspace{1cm} \subfigure[$C_2$, with $A=1000, B=1$.]{
    \label{fig:c2_e_foldings}
    \includegraphics[width=100mm,
    height=60mm]{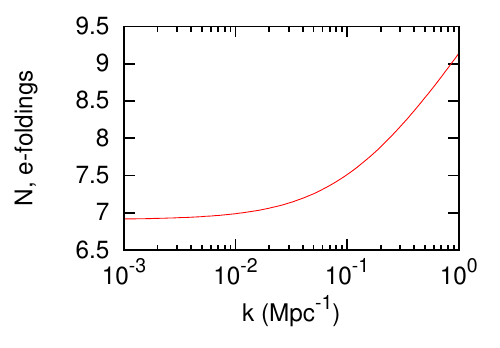}
  } \hspace{1cm} \subfigure[$C_{Wigner}$, with $A=0.01,
  B=10$.]{
    \label{fig:cwigner_e_foldings}
    \includegraphics[width=100mm,
    height=60mm]{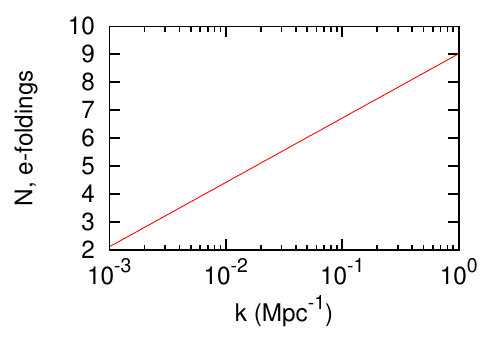}
  } \caption{Gráficas semi logarítmica del número de 
    e-foldings entre  $a^H_k$ y $a^c_k$ para los tres
    esquemas, tomando en cuenta solo los mejores valores
    $(A,B)$. El rango de interés es $10^{-3}$ Mpc$^{-1}$ $< k < 1$
    Mpc$^{-1}$. En estas gráficas: $h = 0.7$.}
  \label{fig:e_foldings}
\end{figure*}

Usando los mejores valores para los diferentes esquemas de
colapso, podemos graficar los \textit{e-folds} transcurridos
entre el colapso del modo y su cruce de horizonte. Como
podemos observar en la figura (\ref{fig:e_foldings}) esta
cantidad cambia --a lo mucho-- en un orden de magnitud en el
rango de $k$ para los valores de $A$ y $B$ que consideramos
más razonables, i.e.\
$a^H_k > a^c_k$, el tiempo de colapso $\eta_k^c \simeq
10^{-3}\eta_k^H$ en este rango.

\subsection{Múltiples Colapsos: caso coherente}\label{sec:mult-colaps-coherente}

En esta sección estudiaremos cómo afecta al espectro de
potencias predicho por cada esquema de colapso si ocurriesen
varios colapsos. Para simplificar el análisis supondremos
que el estado post-colapso es un estado coherente  y que
\textit{todos} los colapsos ocurren durante la época
inflacionaria. 

\paragraph{Estados Coherentes.}Un estado coherente es un
tipo específico de estado del oscilador armónico cuya
dinámica es muy parecida a la de un oscilador armónico
clásico. Los estados coherentes se definen como los
eigenestado del operador de aniquilación $\ann$, 
\begin{equation*}
  \ann \ket{\bm\xi} = \xi \ket{\bm\xi},
\end{equation*}
donde como $\ann$ no es hermítico $\xi$ es un número
complejo no necesariamente 
real y puede ser  representado por $\xi = |\xi|e^{i\chi}$, donde $|\xi|$ es la
amplitud y $\chi$ la fase del estado. Físicamente la fórmula
recién mostrada implica que un estado coherente no se ve
afectado por la detección o aniquilación de una
partícula. En los estados coherentes, el principio de
incertidumbre toma su mínimo valor, i.e.\ el producto de las
incertidumbres de $p$ y $q$ toman su mínimo valor $\Delta p
\Delta q = \dfrac{1}{2}\hbar$.

Exceptuando el estado de vacío, todos
los estados coherentes pueden ser obtenidos desplazando el
estado vacío $\ket{0}$ en el espacio de fase del modo $k$
usando el operador unitario $\hat D(\xi) = \exp(\xi \hat
a^\dagger - \xi^* \hat a)$:

\begin{equation*}
  \ket{\bm\xi}=\hat D(\xi)\ket{\bm{0}}.
\end{equation*}

\paragraph{Esquemas de colapso y estados coherentes} Usando
las simples propiedades de los estados coherentes podemos calcular los 
valores de $d_{k,\xi}^{(R,I)}, c_{k,\xi}^{(R,I)}$ y
$e_{k,\xi}^{(R,I)}$, (ecuaciones \ref{eq:def_d_c_e})) para
el caso cuando el estado postcolapso es un estado coherente $\ket{\bm\xi}$,

\begin{subequations}
  \begin{equation}
    d_{k,\bm\xi}^{(R,I)} = \xi_k^{(R,I)},
  \end{equation}
  \begin{equation}
    c_{k,\bm\xi}^{(R,I)} = \left(\xi_k^{(R,I)}\right)^2,
  \end{equation}
  \begin{equation}
    e_{k,\bm\xi}^{(R,I)} = \left|\xi_k^{(R,I)}\right|^2.
  \end{equation}
\end{subequations}

Como podemos observar, los cálculos que involucran los
valores de expectación o sus incertidumbres, son los mismos
para los estados coherentes, tan sólo haciendo la
sustitución $d_k^{(R,I)} \to \xi_k^{(R,I)}$, de hecho todas
las propiedades  de los estados coherentes están
determinadas por el valor $\xi_k$.  Entonces, los valores de
expectación para los estados coherentes son (ver la ecuación
\ref{eq:expectation_values})

\begin{subequations}
  \label{eq:coherente_expectation_values}
  \begin{equation}
    \langle \hat y_k^{(R,I)}\rangle_{\bm\xi} = \sqrt 2 \, \Re \, (y_k
    \xi_k^{(R,I)}) = \sqrt{2}\left|y_k^{(R,I)}\right|
    \left| \xi_k^{(R,I)}\right| \cos\left(\alpha_k^{(R,I)} +
      \beta_k\right),
  \end{equation}
  \begin{equation}
    \langle \hat \pi_k^{(R,I)} \rangle_{\bm\xi} = \sqrt 2 \,
    \Re \, (g_k \xi_k^{(R,I)} ) =   \sqrt{2}\left|g_k^{(R,I)}\right|
    \left| \xi_k^{(R,I)}\right| \cos\left(\alpha_k^{(R,I)} +
      \gamma_k\right),
  \end{equation}
\end{subequations}

y las dispersiones son (comparar con la ecuación
\ref{eq:dispersiones_generales})

\begin{subequations}\label{eq:dispersiones_coherentes}
  \begin{align}
    \left(\Delta \hat y_k^{(R,I)}\right)^2_{\bm\xi} &= \Re
    \left(y_k^2 \left(\xi_k^{(R,I)}\right)^2\right) +
    \frac{1}{2} \left| y_k \right|^2 \left(\hbar L^3 + 2
      \left|\xi_k^{(R,I)}\right|^2\right) - 2 \Re\left(y_k
      \xi_k^{(R,I)}\right)^2 \nonumber \\ &=
    \frac{1}{2}\left|y_k(\eta)\right|^2\hbar L^3 ,
  \end{align}
  \begin{align}
    \left(\Delta \hat \pi_k^{(R,I)}\right)^2_{\bm\xi} &= \Re
    \left(g_k^2 \left(\xi_k^{(R,I)}\right)^2\right) +
    \frac{1}{2} \left| g_k \right|^2 \left(\hbar L^3 + 2
      \left|\xi_k^{(R,I)}\right|^2\right) - 2 \Re\left(g_k
      \xi_k^{(R,I)}\right)^2 \nonumber \\ &=
    \frac{1}{2}\left|g_k(\eta)\right|^2\hbar L^3.
  \end{align}
\end{subequations}

Observando este último resultado podemos decir que las
incertidumbres de los estado coherentes son las mismas que
las del estado de vacío. Este resultado es independiente de
si se trata de un estado post-colapso o no. Otra característica
importante es que conforme el tiempo pasa, las
incertidumbres en la posición crecen, pero no así en el
momento conjugado (ver eq. \ref{eq:polar-form}).

\paragraph{Esquemas de colapso.} Para el esquema de colapso
independiente (\ref{eq:esquema_independiente}) en su
$n-$ésima colapso, tenemos que
\begin{equation*}
  N_{\pi_n}^{R,I} = x_{2,n}^{R,I} \sqrt{\frac{\hbar
      L^3}{2}} |g(\eta_k^{c_n})|  = \frac{1}{2} \sqrt{\hbar
    k L^3} , \quad 
  N_{y_n}^{R,I} = 
  x_{1,n}^{R,I} \sqrt{\frac{\hbar 
      L^3}{2}} |y(\eta_k^{c_n})| =  \frac{1}{2}
  \sqrt{\frac{\hbar L^3}{k}} \left(\frac{1+z_n^2}{z_n^2}\right).
\end{equation*}
Sustituyendo estas expresiones en el valor medio del
producto de valores de expectación del momento conjugado
para $n-$colapsos (\ref{eq:cuadrado_valor_medio_n_colapsos})
obtenemos, por ejemplo para dos colapsos
\begin{equation}
  \overline{\expec{\dphih_k'}\expec{\dphih_k'*}} = \frac{\hbar
    L^3}{2a(\eta)^2} \Bigg\{ \mathbf{A}_2^2 k +
  \mathbf{B}_2^k \frac{1}{k}
  \left(\frac{1+z_2^2}{z_2^2}\right) +  \mathbf{A}_1^1 k +
  \mathbf{B}_1^k \frac{1}{k}
  \left(\frac{1+z_1^2}{z_1^2}\right) \Bigg\}.
\end{equation}
Sustituyendo esta expresión en (\ref{eq:alm_colapso})
podemos observar que la cantidad $C(k)$ para dos colapsos en
este caso es la suma de dos funciones similares $C^1(k) +
C^2(k)$ del tipo (\ref{eq:C_sudarsky_inflation}), siendo la
diferencia entre $C^1(k)$ y $C^2(k)$ los tiempos de
colapso. Este comportamiento (superposición de dos funciones
ondulatorias) se repetirá para $n-$colapsos obteniendo $
C(k) = \sum_n C^n(k)$.

Para el esquema de colapso newtoniano, el resultado es
similar.  La ``receta'' para el $(n+1)$-ésimo queda definida
por
\begin{equation*}
  N_{\pi_n}^{R,I} = x_{2,n}^{R,I} \sqrt{\frac{\hbar
      L^3}{2}} |g(\eta_k^{c_n})|  = \frac{1}{2} \sqrt{\hbar
    k L^3} , \quad 
  N_{y_n}^{R,I} = 0,
\end{equation*}
usando la ecuación
(\ref{eq:cuadrado_valor_medio_n_colapsos}) se obtiene una
expresión muy sencilla
\begin{equation}
  \overline{\expec{\dphih_k'}\expec{\dphih_k'*}} = \frac{\hbar
    L^3 k}{2a(\eta)^2} \sum_n \mathbf{A}_n^2,
\end{equation}
la cual presentará el mismo comportamiento para $n$ colapsos
que el esquema independiente para $C(k)$, i.e.\ $C(k) = \sum_n
C^n(k)$.

El comportamiento  del espectro de potencias
$C(k)$ para $n$ colapsos con el esquema de Wigner
(cf.\ \S \ref{sec:colapso_wigner},  $N_n^{R,I}$ dados
por (\ref{eq:valores_medios_wigner}) ) presenta un comportamiento
similar al de los esquemas anteriores.

\cleardoublepage
\chapter{Conclusiones y trabajo
  futuro}\label{cha:conclusiones}




A lo largo de este trabajo de tesis se ha analizado el
surgimiento de las inhomogeneidades primordiales que dieron
origen a la estructura cosmológica que observamos hoy. La
explicación estándar es que las fluctuaciones cuánticas del
inflatón siguiendo la evolución del sistema
Einstein-Inflatón cruzan el radio de Hubble y en ese momento
el investigador iguala las incertidumbres cuánticas con
fluctuaciones estadísticas, siguiendo las ideas de
Decoherencia
\cite{Polarski1996,Kiefer98b,Kiefer98a,Kiefer08,Kiefer2007}. En
la literatura esto se conoce como \textit{la transición
  cuántico-clásica}. El lector reconocerá que este problema
está relacionado con el \textit{el problema de la medición
  de mecánica cuántica} también conocido como
\textit{problema de la macro-objetificación}.

A diferencia del enfoque basado en Decoherencia, en este
trabajo de tesis se estudio las diferentes implicaciones de
la \textit{hipótesis del colapso} propuesta por
\citeauthor*{Sudarsky06a} en \cite{Sudarsky06a}. El colapso
se supone inducido de alguna manera por la gravedad y
corresponde a un proceso \textbf{R} en la notación de
\citeauthor*{Penrose94}
\cite{Penrose94,Penrose96,Penrose02,Penrose05}. El colapso
imprime en el campo gravitatorio la inhomogeneidad que
veremos reflejada en el CMB, esto se logra ya que el colapso
rompe la simetría del estado vacío. Puesto de manera
diferente, en el análisis estándar se intenta justificar la
identificación de las correlaciones de dos puntos cuánticas
con correlaciones estadísticas clásicas, sin romper la
unitariedad de la evolución del sistema
Einstein-Inflatón. La evolución unitaria en este caso
preserva las simetrías originales del estado de vacío
(homogeneidad e isotropía), lo cual permite que uno se
pregunte entonces ¿Cómo se terminó en un estado inhomogéneo
y anisotrópico? Es aquí donde la hipótesis del colapso
muestra sus ventajas sobre el enfoque tradicional: no sólo
intenta explicar la aparición de una estadística clásica,
sino que además explica el orígen de la inhomogenidad (el 
estado post-colapso no tiene por qué  ser homogéneo e
isotrópico).

En este trabajo de tesis se ha considerado varios  esquemas
\textit{ad hoc} de colapso del campo
cuántico inflacionario como posibles explicaciones del
problema del origen de las semillas de la formación de
estructura cosmológica. Se estudiaron a mayor profundidad
los esquemas de colapso \textit{newtoniano} e
\textit{independiente} (cf. \S \ref{sec:origen-colapso} y
\cite{Sudarsky06a}) y además se propuso un nuevo esquema de
colapso basado en el funcional de Wigner (cf.\ \S
\ref{cha:analisis-esquemas}). El esquema de colapso
\textit{de Wigner} respeta la correlación que existe entra
variables conjugadas canónicas cuánticas y que fue ignorada
en los dos primeros esquemas de colapso, por lo que se puede
decir que tiene mayor motivación física que los anteriores.

Los colapsos, en todos los esquemas estudiados, se suponen
inducidos o provocados por un efecto gravitacional. El
rompimiento de la evolución unitaria que estos esquemas de
colapso conllevan, es congruente con la posición de que hay
que modificar a la mecánica cuántica estándar cuando se toma
en cuenta a la gravedad
\cite{Diosi1984,Diosi,DiosiLajos07,FrenkelAnd02,TumulkaRod06,PearlePhil95,PearlePhil07,Penrose94,Penrose96,Penrose02,Penrose05}. En
esta tesis no se discutieron con profundidad estos temas y solo
nos concentramos en los aspectos fenomenológicos del
problema.

Los diferentes esquemas de colapso fueron analizados y
desarrollados para que sus consecuencias observacionales
pudieran ser comparadas con los datos obtenidos de los
estudios del \acf{CMB}. Estas comparaciones ilustraron
varios puntos dignos de mencionar: Primero, dependiendo de
los detalles del esquema de colapso y de sus parámetros, su
efecto en el espectro de potencias resultante puede ser muy
diferente del espectro plano de potencias que se espera
resulte de inflación. Claro que esta diferencia o separación
del espectro de potencias plano se puede --y se obtiene-- en
los estudios de la comunidad inflacionaria con
modificaciones del \textit{slow-roll} o considerando
diversos potenciales, pero en este estudio las diferencias
surgen solo de considerar los detalles del esquema de
colapso cuántico, el cual, desde nuestro punto de vista es
necesario para comprender de una manera satisfactoria la
emergencia de la estructura desde las fluctuaciones
cuánticas.

El estudio de los esquemas de colapso al compararlo con las
observaciones (en particular con el espectro plano de
Harrison- Zel'dovich (HZ)\footnote{Como se comentó en el
  capítulo \ref{cha:critica-al-origen}, el espectro plano es
  observacionalmente visible para valores pequeños en el
  multipolo $l$ (ángulos grandes), para $l$ mayores aparecen
  efectos sobreimpuestos relacionados principalmente con
  física de plasmas.}) impuso condiciones sobre los tiempos
conformes de colapso para recuperar el espectro HZ:
\textit{los tiempos de colapso (en tiempo conforme) por modo
  deben de satisfacer} $\eta_k^c k = $ constante. Esta
restricción permitiría comprobar algunos \textit{mecanismos}
de colapso tal como se hizo preliminarmente en
\cite{Sudarsky06a}. La extensión de este estudio de
mecanismos de colapso se está haciendo actualmente por parte
del autor de esta tesis.

La naturaleza estocástica del colapso hace muy improbable
que se cumpla con exactitud la condición $\eta_k^c k = $
constante, por lo que hemos estudiado la robustez de los
esquemas de colapsos estudiados. Con este fin hemos
considerardo desviaciones lineales de comportamiento de
$\eta_k^c$ como función de $k$, i.e.\ , hemos explorado en los
tres esquemas de colapso los efectos de tener tiempos de
colapso dados por $\eta_k^c = A/k + B R_D$. Los resultados
de dicho estudio se presentaron en \S
\ref{sec:comparacion-observaciones-esq}, donde se mostró que
los diferentes esquemas de colapso se desvian de manera
diferente del espectro plano de potencias, por ejemplo el
esquema newtoniano y el de Wigner producen una caída natural
en el espectro para $l$ grande.

Se analizó también el efecto en el espectro de potencias
debido a múltiples colapsos ocurridos durante la época
inflacionaria. Para llevar a cabo este análisis se
encontraron las fórmulas para $n-$colapsos, de estas
expresiones se observó que el espectro de potencias para el
caso de $n-$colapsos es como la suma de $n$ expresiones de
espectros de potencias de un solo colapso ocurriendo a
diferentes tiempos de colapso.

Sin embargo, la conclusión más importante, ilustrada por
este trabajo de tesis, es que enfocándose en problemas que
se pueden pensar como filosóficos o de principios, se ha
encontrado la posibilidad de estudiar problemas que pueden
mostrar algunos aspectos de lo que posiblemente sea nueva
física. 

Actualmente se está extendiendo este análisis de diversas
maneras, entre las que podemos nombrar  colapsos múltiples
para el caso en que el estado 
post-colapso sea un estado \textit{apachurrado}
(\textit{squezee}) (a diferencia 
del trabajo presentado en esta tesis que fue para estados
post-colapso coherentes) \cite{Gabriel2009}, mecanismos de
colapso basados en las ideas de Penrose i.e.\ , provocados por
la gravedad \cite{Unanue2009b} y estudios detallados y
exactos sobre cómo 
se afecta el espectro de potencias predicho por los esquemas
de colapso luego de tomar en cuenta todos
los efectos de la física de plasmas \cite{Landau2009}.

\cleardoublepage

   \PCFapendices

\chapter{Teoría de Perturbaciones
  Cosmológicas}\label{cha:teor-de-pert} 

\section{Introducción}

La teoría de Relatividad General es conceptualmente sencilla
y particularmente elegante, pero presenta en la práctica
dificultades que tienen su raíz en que las ecuaciones de
campo de Einstein (ECE) forman un sistema no-lineal acoplado de 10
ecuaciones diferenciales parciales en cuatro dimensiones,
por lo tanto, la mayoría de los problemas que se estudian en
la teoría de Relatividad General son difíciles o imposibles
de solucionar de manera exacta. 

Una de las estrategias populares en la física 
para resolver problemas complicados es utilizar un análisis
perturbativo alrededor de una solución conocida. En la teoría 
de Relatividad General, sin embargo, el análisis 
perturbativo posee varias sutilezas que no comparten otras
áreas de la física. Estas 
sutilezas surgen en Relatividad General debido al principio
de covariancia general y al carácter dinámico del
espacio-tiempo: en Relatividad
General además de perturbar los campos de materia es
necesario perturbar la métrica del espacio-tiempo; pero,
debido a la covariancia intrínseca de la teoría, es
posible, perturbar la métrica de una
manera en la que se deforme el sistema de coordenadas sin
afectar la física subyacente al espacio-tiempo, esto tendría
como consecuencia 
la aparición de información ficticia 
(debida a la deformación del sistema coordenado y no al
fenómeno en sí) sobre el fenómeno físico estudiado. En este
apéndice trataremos de 
elucidar el papel de este principio y revisaremos tres
propuestas hechas en la literatura para establecer 
una teoría perturbativa de Relatividad General que permita
solucionar problemas de una manera confiable.

La teoría perturbativa de Relatividad General se aplica
actualmente en diferentes problemas como lo son Relatividad
Numérica (aparece por ejemplo en el planteamiento de las condiciones iniciales)
\cite{Alcubierre}, el estudio de ondas 
gravitacionales \cite{Wald84}, el análisis de la dinámica en
espacios-tiempos (cuerpos extendidos,  movimiento cerca de
agujeros negros \cite{Preston06,Martel05}, auto-fuerzas
\cite{Poisson04,Poisson03,Pfenning00}, 
problema de los tres cuerpos\cite{Burnell02,Imai07}, etc.),
el estudio del problema de 
los promedios \cite{EllisG.F.05,Boersma98,Zalaletdinov97} o  el análisis
de la formación de estructura en cosmología
\cite{Mukhanov90,Durrer01b,DURRER,Straumann05,Bertschinger01}.

En esta revisión se estudiarán las
diferencias que existen en el análisis perturbativo
de Relatividad General respecto al de  otras disciplinas de la
física y el origen de esta discrepancia (\S
\ref{sec:breve-repaso}) y se expondrá de una manera 
formal el problema y su tratamiento para el desarrollo de
una teoría de perturbaciones en Relatividad General (\S
\ref{sec:teoria-perturbaciones}). Finalmente se presentarán
los tres principales enfoques conocidos 
para hacer el análisis perturbativo (\S
\ref{sec:enfoques}), dándole énfasis a la teoría invariante
de norma expuesta principalmente por Kouji Nakamura,
\citep[por ejemplo ver][]{NakamuraKo07} (\S
\ref{sec:invariante-norma}). A lo largo del ensayo se estará
usando como ejemplo la métrica de Roberston-Walker de los
Universos de Friedmann-Lemaître con  un campo escalar
(presentada en la \S \ref{sec:flrw}). La elección de un
modelo cosmológico para el estudio de teoría de
perturbaciones se debe a dos motivos, el primero es una
conveniencia de cálculo,  ya que las simetrías de este
espacio-tiempo simplifican los cálculos que se mostrarán, 
segundo, la Cosmología de precisión \cite{wmap5,Dupac07}
es actualmente un tema candente en la física por lo cual
este trabajo podrá servir de apoyo o referencia al lector al revisar
la bibliografía cosmológica. 

En esta revisión se elegirán unidades en las que $c=1$, la
métrica tendrá la signatura $(-\;+\;\;+\;+)$ y se seguirá la
notación abstracta de índices y las definiciones del tensor
de Riemann dadas por Wald en \cite{Wald84}. Los índices
latinos ($i, j, k, $ etc). indican componentes espaciales
mientras que los griegos ($\mu, \nu,$ etc.)  representan las
cuatro componentes espacio-temporales.

\section{\label{sec:breve-repaso}Relatividad General y el
  principio de covariancia general}

La teoría de Relatividad General es conceptualmente 
sencilla y se puede resumir en pocas palabras citando
\citep{MTW} el famoso adagio ``El espacio le dice a la
materia como moverse y la materia le dice al espacio como
curvarse \footnote{En el idioma original: ``Space acts on
  matter, telling it how to move. In turn, matter reacts
  back on space, telling it how to
  curve''\citep[][pag. 5]{MTW}}'', o de una
 manera más formal: ``El espacio-tiempo es
una variedad $\mathscr{M}$, en la cual está definida una
métrica Lorentziana $g_{ab}$. La curvatura de $g_{ab}$ está
relacionada con la distribución de materia en el
espacio-tiempo mediante las ecuaciones de Einstein (ECE)''
\citep[][pag. 73]{Wald84}. La métrica $g_{ab}$ no sólo
representa las propiedades crono-geométricas del
espacio-tiempo, además define los potenciales del campo
gravitacional.

Las ecuaciones de Relatividad General, referidas en el
párrafo anterior como las ECE son
\begin{equation}
  \label{eq:ece}
  G_{ab} \equiv R_{ab} - \frac{1}{2}g_{ab}\Ricci = \kappa T_{ab}
\end{equation}
donde los símbolos representan lo que sigue: $G_{ab}$ es el
tensor de Einstein, $R_{ab}$ es el tensor de Ricci que está
relacionado con el tensor de Riemann, $\Riemann$ mediante una
contracción de índices, $R_{ac} \equiv R_{abc}^{\quad b}$
y $\Ricci$ es el escalar de curvatura o escalar de Ricci, que
se obtiene de la traza de $R_{ab}$, i.e.\ $\Ricci \equiv
R_a^{\;\; a}$. En el lado derecho de la ecuación tenemos a
la constante de Einstein $\kappa$ que está relacionada con
la constante gravitacional de Newton mediante $\kappa = 8\pi
G$ y para finalizar, $T_{ab}$, el tensor de energía-momento
que representa la distribución de materia en el
espacio-tiempo y cuya expresión exacta depende de la teoría
con la que se esté describiendo la materia. La información
geométrica de la teoría esta codificada en el lado izquierdo de
la ecuación \eqref{eq:ece} debido a sus relaciones con el
tensor de curvatura $\Riemann$.

Conceptualmente, Relatividad General, está basada en dos
principios importantes: (a) el \textit{principio de
  equivalencia}: el movimiento de los cuerpos\footnote{En
  realidad de las \emph{partículas de prueba}, definidas
  como aquellas partículas que sienten el efecto del campo
  gravitacional pero no lo afectan de manera alguna.} en un
campo gravitacional es independiente de su composición o su
masa; y (b) el \textit{principio de relatividad o
  covariancia general}: Las leyes de la naturaleza son
expresadas de manera natural mediante ecuaciones que sean
iguales en todos los sistemas de coordenadas, esto es, que
sean \textit{covariantes} con respecto a sustituciones
cualesquiera, i.e.\ ecuaciones tensoriales. Aparte de estos
principios, se podrían mencionar dos más de soporte o guía:
el \textit{principio de acople gravitacional mínimo} --una
versión de la navaja de Occam, i.e.\ una exigencia de
simplicidad-- y el \textit{principio de correspondencia}
--las ecuaciones de Relatividad General deben de coincidir
en el límite apropiado con Relatividad Especial y con la
Teoría Gravitacional Newtoniana-- \citep{dIverno, Wald84}.

El establecimiento de cual es el significado físico, la
importancia, el número\footnote{Por ejemplo, en
  \citep{dIverno} menciona el \textit{principio de Mach}
  --en realidad el principio de Mach está determinado por
  tres enunciados: (1) La distribución de materia determina
  la geometría, (2) Si no hay materia no hay geometría y (3)
  Un cuerpo en un universo vacío, no posee propiedades
  inerciales-- dentro de los principios fundamentales,
  aunque reconoce que quizá sólo sirva como principio guía
  para la formulación de Relatividad General.}  o
simplemente que es lo que dicen exactamente estos principios
\footnote{Como ejemplo, la expresión del \textit{principio
    de equivalencia} es diferente en
  \citep{MTW,Wald84,dIverno}; en
  \citep[][pags. 13-15]{Ciufolini95} se mencionan tres
  variantes distintas de este principio.} ha sido objeto de
debate a lo largo de décadas, debate cuyo inicio se puede
remontar al mismo Einstein y su lucha para establecer la
teoría tal como la conocemos actualmente \citep[ver][para
una clara y extensa discusión]{Norton93}.

El principio de relatividad o covariancia general es el que
más problemas causará cuando abordemos la cuestión de la
teoría perturbativa, por lo que, le dedicaremos un estudio
un poco más extenso para poder alcanzar a ver todas las
sutilezas que encierra.

Este principio, intuitivamente establece que todos los
observadores son equivalentes, o expresado de otra manera:
no existe un sistema coordenado privilegiado en la
naturaleza. Esto se puede entender de la siguiente manera:
los sistemas de coordenadas son simple 
etiquetas puestas a los puntos del espacio-tiempo y de
ninguna manera esta colocación de etiquetas afecta al
fenómeno natural. El principio de covariancia causó y causa
acaloradas discusiones en la literatura científica y en un
principio motivó que A. Einstein y D. Hilbert lo desecharan como
fundamento de la teoría debido al siguiente argumento
conocido como el \textit{argumento del
  agujero}\footnote{Originalmente, Einstein expresó este
  argumento usando sistemas de coordenadas, de la siguiente
  manera, Sea $G(x)$ el tensor métrico que satisface las ECE
  en el sistema de coordenadas $x$ y $G'(x')$ representa el
  mismo campo gravitacional en el sist. coordenado $x'$. Si
  suponemos covariancia general, entonces $G(x')$ (piénsese
  este cambio de coordenadas, únicamente en el sentido
  matemático, i.e.\ el cambio de $x \to x'$ no cambia el
  significado funcional de $G$, pero podríamos interpretarlo
  como un nuevo campo en $x'$). A partir de esto, se puede
  demostrar que no hay manera de especificar la métrica
  afuera y en la frontera de un ``agujero'' pude determinar
  el campo dentro del agujero. Invalidando así, la utilidad
  de las ECE. Aunque Einstein planteó este argumento, como
  un problema de frontera, Hilbert lo planteó, como un
  problema de valores iniciales\cite{Norton93},
  relacionándolo así con el \textit{Problema de Cauchy} de
  Relatividad General \cite{Wald84}.}  (para la versión
original de este argumento consúltese \cite{Norton93,
  Macdonald01}, para la versión moderna véase
\cite{Iftime05,Iftime06,sep-hole-arg}): Sea $\mathbf{g}$
--por el momento eliminaremos los subíndices para no
complicar la notación-- una solución de las ECE, entonces,
el \textit{pull-back}, $\phi^*(\mathbf{g})$, inducido por el
difeomorfismo $\phi$ en $\mathscr{M}$ también satisface las
ECE. La pregunta a responder es ¿Todas las métricas
$\phi^*(\mathbf{g})$ describen el mismo campo gravitacional?
Si suponemos el principio de covariancia general como
válido, la respuesta es si. Supóngase ahora que el
espacio-tiempo $(\mathscr{M}, \mathbf{g})$ contiene una
región abierta $H$ (el ``agujero'') vacía, es decir,
únicamente el campo $\mathbf{g}$ está presente y es solución
a las ECE en vacío. Debido a la covariancia general, aún
cuando encontremos una solución $\mathbf{g}$ fuera de $H$ y
especifiquemos las derivadas a un orden finito en la
frontera de $H$, para conectar la métrica fuera con la de
adentro de $H$ de una manera suave, la métrica dentro de $H$
no se puede determinar de manera única, no importa que tan
pequeño sea el ``agujero'', así, pareciera que las ECE no
poseen un problema de valores de frontera (o iniciales en la
versión del argumento del agujero de Hilbert
\cite{Norton93}) bien planteado si se sostiene como
verdadero el principio de covariancia general.

El argumento del agujero, básicamente nos enfrenta a la
decisión de declarar al espacio-tiempo como entidad ``real''
y con esto sacrificar la covariancia general, o, eliminar la
calidad de ente al espacio-tiempo y mantener la covariancia
general.

Tiempo después, Einstein resolvió esta aparente paradoja
quitándole ``realidad autónoma'' al espacio-tiempo si no
existe materia en él:

\begin{quote}
  En la base de la teoría de relatividad general \ldots
  espacio y ``lo que llena el espacio'' no tienen una
  existencia separada \ldots No hay tal cosa como el espacio
  vacío i.e.\ un espacio sin campo [gravitacional] \ldots
  [El] espacio-tiempo no reclama una existencia por sí
  mismo, si no únicamente como cualidad estructural del
  campo. \citep{Einstein52}
\end{quote}
esto se puede entender, si se recuerda que toda la
experiencia que obtenemos del espacio-tiempo está dada por
``coincidencias'' entre nuestros aparatos de medición y
otros campos, i.e.\ la ``realidad'' del espacio-tiempo está
en los \textit{eventos} no en los puntos de la
variedad. Otra forma de verlo, es que la variedad sin la
especificación de una métrica, con la cual se obtengan las
cantidades observables, no es un espacio-tiempo, ya que los
puntos de la variedad $\mathscr{M}$ no tienen propiedades
gravitacionales o crono-geométricas \textit{per se}. Lo que
conforma al espacio-tiempo es el par $(\mathscr{M},
g_{ab})$. Asumiendo esta postura, los \textit{pull-backs} de
la métrica no difieren físicamente de la métrica
original. Dos métricas $g_1$ y $g_2$ soluciones de las ECE
representan la misma solución si son \textit{isométricas}
entre sí, i.e.\ todas ellas forman una clase de
equivalencia. Este argumento puede extenderse con facilidad
al caso de que haya más campos dentro de $H$. Esto resuelve
el argumento del agujero, recuperando la covariancia general
y permite plantear el problema de condiciones iniciales en
Relatividad General de una manera satisfactoria.

A pesar de haberse resuelto el argumento del agujero, los
problemas no acabaron para el principio de covariancia, tan
pronto como Einstein publicó en 1916 la solución al
argumento del agujero y retomaba como principio fundacional
a la covariancia general, Kretschmann en 1917
\citep{Norton93}, utilizó el argumento de solución de
Einstein y lo volvió en su contra, estableciendo que el
principio de covariancia tenía significado físico vacío,
argumentando que cualquier teoría puede ser puesta de manera
covariante si es formulada correctamente\footnote{Einstein
  se defendió diciendo que el argumento de Kretschmann era
  falso, poniendo como ejemplo la teoría gravitacional de
  Newton ya que no podía escribirse de manera
  covariante. Poco tiempo después Cartan, escribió la teoría
  newtoniana en forma covariante \cite{MTW}. }. Esta
objeción, a la fecha, es tema de discusión académica, como
atestiguan artículos como \citep{Ellis95}.

Además de lo dicho, el principio de covariancia y su
interpretación está ligado con el significado de
``promediar'' en la teoría de Relatividad General
\cite{Ellis95} \citep[ver
también][]{Zalaletdinov97,Boersma98}, y como se muestra en
la siguiente sección es el causante de las complicaciones en
la teoría de perturbaciones en Relatividad General. Se
invita al lector interesado en estas sutilizas a consultar
las referencias hasta aquí expuestas.

\section{\label{sec:teoria-perturbaciones}Teoría de
  perturbaciónes en Relatividad General}

La teoría de perturbaciones de manera general --sin
restringirla a Relatividad General-- es el conjunto de
métodos matemáticos usados para encontrar la solución a un
problema que no puede ser resuelto de manera
exacta. Intuitivamente el análisis perturbativo, propone que
si se conoce la solución exacta de un problema, la solución
de un nuevo problema ligeramente diferente es una ``pequeña
modificación'' de la solución conocida.

Los métodos perturbativos sólo podrán ser aplicados si es
posible agregar una término ``pequeño'' a la solución exacta
del problema conocido, i.e.\ la nueva solución estará
expresada en términos de una serie de potencias en el
parámetro ``pequeño'', $\lambda$. El parámetro $\lambda$
cuantifica la desviación del nuevo problema respecto a el
problema que se sabe solucionar de manera exacta. Por
ejemplo si queremos obtener la cantidad $F$ a partir de la
solución exacta $F_0$ de un problema similar, $F$ estará
expresada a segundo orden en $\lambda$ como sigue:
\begin{equation*}
  F = F_0 + \lambda F_1 + \lambda^2 F_2 + \mathscr{O}(\lambda^3).
\end{equation*}

Estos métodos son usados con gran éxito en Mecánica Clásica,
Mecánica Cuántica, Electrodinámica, etc.\ Entonces, ¿De
dónde proviene la dificultad de usarla en Relatividad
General?

Los problemas con los que nos encontraremos en la teoría
perturbativa son varios. El primer problema proviene del
objeto dinámico de la teoría. En Relatividad General, el
espacio-tiempo no es un escenario estático, absoluto, sobre
el cual se pueda estar sin afectarlo, modificarlo; al
contrario, es un ente dinámico que actúa sobre y es actuado
por los participantes. Por lo tanto, es necesario perturbar,
además de los campos de materia (los participantes)
representados por $T_{ab}$ , al espacio-tiempo, representado
por $g_{ab}$ (el escenario) i.e.\ $g_{ab} = g_{ab}^0 +
\delta g_{ab}$; el segundo problema se observa al escribir
una cantidad física (i.e.\ representada por un tensor)
$\mathcal{Q}$ de manera perturbada (a primer orden):

\begin{equation}
  \label{eq:perturbacion_confusa_1}
  \mathcal{Q} = Q_0 + \lambda\, \delta\mathcal{Q}.
\end{equation}

Esta ecuación no es, en ningún sentido, precisa o correcta,
$\mathcal{Q}$ es un campo tensorial, el cual, para reflejar
el número que se va a medir u observar en un experimento
\citep[sec. 4.1 de][]{Wald84}, debe de estar evaluada en un
punto del espacio-tiempo (i.e.\ en un \textit{evento} del
espacio-tiempo), luego de esta corrección la ecuación es
\begin{equation}
  \label{eq:perturbacion_confusa_2}
  \mathcal{Q}(p) = Q_0(p) + \lambda\, \delta\mathcal{Q}(p).
\end{equation}

Aunque mejor, esta ecuación sigue siendo problemática, pues
el punto $p$ del lado izquierdo de la ecuación no es el
mismo punto $p$ del lado derecho, ya que, estos puntos se
encuentran en diferentes variedades $\mathcal{M}$ y
$\mathcal{M}_0$. Cuando se está realizando teoría de
perturbaciones en Relatividad General, se está trabajando
con dos variedades, una física, $\mathcal{M}$ (la solución
que intentaremos describir con perturbaciones), y otra de
fondo, $\mathcal{M}_0$, que es una variedad ficticia,
preparada por nosotros para desarrollar el análisis
perturbativo (la solución exacta que conocemos). Al
espacio-tiempo físico lo denotaremos de manera general por
$(\mathcal{M}, \overline{g}_{ab})$, y al espacio-tiempo de
fondo por $(\mathcal{M}_0, g_{ab})$. Además, denotaremos a
los campos tensoriales que estén sobre $\mathcal{M}$
mediante $\mathcal{Q}$ y los definidos sobre $\mathcal{M}_0$
por $\mathcal{Q}_0$. Así, la ecuación
\eqref{eq:perturbacion_confusa_2} debería de escribirse
como:
\begin{equation}
  \label{eq:perturbacion_confusa}
  \underbrace{\mathcal{Q}("p")}_{ \in (\mathcal{M},
    \overline{g}_{ab})} = \underbrace{\mathcal{Q}_0(p) + 
    \lambda\,\delta\mathcal{Q}(p)}_{\in (\mathcal{M}_0, g_{ab})} .
\end{equation}

La última ecuación nos muestra la aparición del problema
relacionado con el principio de covariancia: la ecuación
\eqref{eq:perturbacion_confusa} nos está dando la relación
entre variables definidas en dos variedades diferentes, y
para lograrlo \textbf{implícitamente} hemos identificado los
puntos en estas dos variedades. Es decir, asumimos que
$\exists \, \mathcal{M}_0 \to \mathcal{M}: p \in
\mathcal{M}_0 \mapsto "p" \in \mathcal{M}$. Esta
identificación no es única en la teoría de Relatividad
General debido a que está construida sobre el principio de
covariancia general (cf.\ \ref{sec:breve-repaso}).

Siendo conscientes de estos problemas procederemos a definir
conceptos que permitirán formalizar su planteamiento y
encontrar una solución. En teoría de perturbaciones a la
elección del mapa de identificación de puntos, i.e.\ $\chi :
\mathcal{M}_0 \mapsto \mathcal{M}$ se le nombra como
\textit{grado de libertad de norma} y esta libertad siempre
existe en teorías a las cuales se les imponga el principio
de covariancia general. Este grado de libertad no es físico,
si no espurio, i.e.\ no trae información relevante al
fenómeno estudiado. Por lo tanto, un enfoque correcto en la
teoría perturbativa en Relatividad General debe
\textit{eliminar} estos grados de libertad ficticios.

Es importante recalcar, que por la misma naturaleza del
análisis perturbativo (``desviaciones pequeñas de la
solución exacta conocida'', cf.\ arriba) los espacios-tiempo
de fondo y físico deben de ser lo ``suficientemente
iguales'' i.e.\ ``cuasi-isométricos''
\citep{PERTURBATIONS_ST_GR} y debido a los problemas con la
obtención de promedios de métricas en Relatividad General,
no existe una manera única de hacer esta declaración más
precisa\cite{PERTURBATIONS_ST_GR}.

Consideremos una variedad $\mathcal{N} = \mathcal{M} \times
\mathbb{R}$ con $\dim \mathcal{N} = \dim \mathcal{M} +
\dim\mathbb{R}$, que en el caso en particular de Relatividad
General $\dim \mathcal{N} = 5$. El parámetro infinitesimal
para la perturbación será denotado por $\lambda$. Entonces,
$\mathcal{M}_0 = \mathcal{N}|_{\lambda=0}$ y $\mathcal{M} =
\mathcal{M}_\lambda = \mathcal{N}|_{\mathbb{R}=\lambda}$,
con $\mathcal{M}_0, \mathcal{M} \subset
\mathcal{N}$. Además, denotemos los puntos en $\mathcal{N}$
por $r$, entonces, cada $ r \in \mathcal{N}$ será
identificado con $(p, \lambda)$, donde $p\in
\mathcal{M}_\lambda$ y cada punto de $\mathcal{M}_0$ se
identificará mediante $(p, 0)$. Esto implica, que hemos
construido un grupo uniparamétrico de difeomorfismos de
identificación de puntos entre las hipersuperficies
$\mathcal{M}_\lambda$ en $\mathcal{N}$, $\phi_\lambda:
\mathcal{N} \to \mathcal{N} \;|\; \mathcal{M}_0 \to
\mathcal{M}_\lambda$, con $p \in \mathcal{M}_0 \mapsto q =
\phi(p) \in \mathcal{M}$. Si asignamos un sistema de
coordenadas $x^\mu$ en $\mathcal{M}_0$, $\phi_\lambda(x)$
establecerá las coordenadas $x^\mu$ en $\mathcal{M}_\lambda$
y en $\mathcal{N}$, mediante $\phi: (p,0) \mapsto
(p,\lambda)$. El generador de este difeomorfismo $\phi$, es
$\left(\frac{\partial}{\partial\lambda}\right)^a$ el cual es
tangente a las órbitas de $\phi$.

De esta manera hemos foliado $\mathcal{N}$ mediante las
subvariedades $\mathcal{M}_\lambda$ para cada $\lambda$, que
son difeomórficas al espacio físico $\mathcal{M}$ y al
espacio de fondo $\mathcal{M}_0$. Es importante resaltar que
$\mathcal{N}$ tiene una estructura diferenciable ya que es
el producto tensorial directo de $\mathcal{M}$ y
$\mathbb{R}$ y por construcción las subvariedades
$\mathcal{M}_\lambda$ tienen una estructura
diferenciable. Es decir, por construcción, hemos requerido
que los puntos en diferentes subvariedades de $\mathcal{N}$
estén unidas por una curva contínua $\gamma \in
\mathcal{N}$. Podemos elegir entonces cartas con las
coordenadas $x^\mu (\mu=0,1,\ldots, m -1)$ , en cada
\textit{lámina} $\mathcal{M}_\lambda$ y que tienen a $x^m
\equiv \lambda$.  Las ECE se pueden escribir formalmente
como
\begin{equation}
  \label{eq:ece_formal}
  \mathcal{E}[\mathcal{Q}_\lambda] = 0
\end{equation}
en $\mathcal{M}_\lambda$ para $\mathcal{Q}_\lambda$ definido
en $\mathcal{M}_\lambda$. Es decir, cada
$\mathcal{M}_\lambda$ tiene su métrica $\mathbf{g}_\lambda$
y un conjunto de campos materiales $\mathbf{T}_\lambda$ que
satisfacen la ecuación anterior. Si el campo tensorial está
dado en cada $\mathcal{M}_\lambda$, $\mathcal{Q}_\lambda$ es
extendido a un campo tensorial en $\mathcal{N}$ mediante $
\mathcal{Q}(p, \lambda) \equiv \mathcal{Q}_\lambda(p)$ con
$p \in \mathcal{M}_\lambda$. Con esta extensión, la ecuación
\eqref{eq:ece_formal} se puede considerar una ecuación en
$\mathcal{N}$.

Con esta extensión tenemos que $\mathcal{E}[\mathcal{Q}] =
0$ es una ecuación en $\mathcal{N}$. Los campos tensoriales
en $\mathcal{N}$, como el anterior, son por construcción
\textit{tangentes} a cada $\mathcal{M}_\lambda$, i.e. su
componente normal a cada $\mathcal{M}_\lambda$ es cero.

La base del espacio tangente de $\mathcal{N}$, queda
establecida usando el generador del difeomorfismo
$\phi_\lambda$,
$\left(\frac{\partial}{\partial\lambda}\right)^a$ y su dual
$d\lambda$ que satisfacen
\begin{equation*}
  (d\lambda)_a\left(\frac{\partial}{\partial\lambda}\right)^a   = 1.
\end{equation*}

Por su construcción, la uno-forma $(d\lambda)_a$ y su dual
$(\partial/\partial\lambda)^a$ son normales a cualquier
tensor que es extendido del espacio tangente a cada
$\mathcal{M}_\lambda$. El conjunto de $(d\lambda)_a$ y
$(\partial/\partial\lambda)^a$ y la base del espacio
tangente a cada $\mathcal{M}_\lambda$ son considerados como
la base del espacio tangente de $\mathcal{N}$.

Entonces podemos empezar a considerar a las perturbaciones
del campo $\mathcal{Q}$ como las comparaciones de
$\mathcal{Q}$ en $\mathcal{M}_\lambda$ con $\mathcal{Q}_0$
en $\mathcal{M}_0$, por lo que es necesario identificar los
puntos de $\mathcal{M}$ con aquellos en $\mathcal{M}_0$. Con
este fin, se elegirá un nuevo difeomorfismo
$\varphi_\lambda$. Por simplicidad ese mapeo se puede elegir
como un mapeo exponencial\footnote{Esta restricción no
  afectará las ecuaciones importantes del formalismo
  invariante de norma, ver sección
  \ref{sec:invariante-norma} }. Este mapeo de identificación
de puntos, está representada por el mapeo $\varphi_\lambda:
\mathcal{N} \to \mathcal{N}$, tal que, $\varphi_\lambda:
\mathcal{M}_0 \to \mathcal{M}_\lambda$, y en general
$\phi(p) \neq \varphi(p)$, para $p \in \mathcal{M}_0$. El
\textit{grado de libertad de norma} $\varphi_\lambda$, es un
grupo uniparamétrico de difeomorfismos (e.g.\ un mapeo
exponencial) que satisface la propiedad
\begin{equation}
  \varphi_{\lambda_1+\lambda_2} =
  \varphi_{\lambda_1}\circ\varphi_{\lambda_2} =
  \varphi_{\lambda_2}\circ\varphi_{\lambda_1}.
\end{equation}
Este grupo uniparamétrico de difeomorfismos es generado por
el campo vectorial $^{\varphi}X^a_{\lambda}$. A este campo
vectorial se le conoce como \textit{generador} y es definido
por la acción del \textit{pull-back} para un campo tensorial
$\mathcal{Q}$ en $\mathcal{N}$:
\begin{equation}
  \pounds_X\mathcal{Q} \equiv \lim_{\lambda \to
    0}\frac{\varphi^*_\lambda\mathcal{Q} - \mathcal{Q}}{\lambda},
\end{equation}
y puede ser descompuesto usando la base de $\mathcal{N}$
como sigue
\begin{equation}
  ^\varphi X^a_\lambda =: \left( \frac{\partial}{\partial\lambda}\right)^a  +
  \theta^a, \quad \theta^a(d\lambda)_a = 0,\quad
  \pounds_{\frac{\partial}{\partial\lambda}}\theta^a=0.
\end{equation}
La segunda condición de esta relación se elige por
simplicidad, debido a que la elección de $\theta^a$ es
arbitraria salvo la condición de que sea tangente a
$\mathcal{M}_\lambda$. El campo vectorial $\theta^a$ está
definido en $\mathcal{M}_\lambda$ y exceptuando las
condiciones anteriores es arbitrario, i.e.\ la arbitrariedad
de la elección de norma está contenida en la componente
tangente del campo vectorial $X$, es decir en el campo
vectorial $\theta^a$.

El \textit{pull-back} $\varphi^*_\lambda\mathcal{Q}$ mapea
el campo tensorial $\mathcal{Q}$ en $\mathcal{M}_\lambda$ a
un tensor $\varphi^*_\lambda\mathcal{Q}$ en
$\mathcal{M}_0$. Entonces el \textit{pull-back} está
representado por la expansión de Taylor (ver apéndice
\ref{sec:series-de-taylor})\footnote{Nótese que
  $\delta^{(0)} \mathcal{Q} \equiv \mathcal{Q}_0$ y
  $\delta^{(1)} \mathcal{Q} = \delta\mathcal{Q}$} a segundo
orden mediante
  \begin{equation}\label{eq:taylor_1}
    \varphi^*_\lambda\mathcal{Q}(p)|_{\mathcal{M}_0} 
    = \sum_{k=0}^\infty \frac{\lambda^k}{k!}\, \Lie^k_X
    \mathcal{Q}|_{\mathcal{M}_0} =
    \mathcal{Q}(p)|_{\mathcal{M}_0} + \lambda\Lie_{X}Q|_{\mathcal{M}_0} 
    +\frac{1}{2}\lambda^2\pounds^2_X\mathcal{Q}|_{\mathcal{M}_0} +
    \mathscr{O}(\lambda^3).
  \end{equation}

Obsérvese que $\Lie_{X} \equiv \Lie_{^\varphi X_\lambda}$,
pero estamos simplificando la notación. Dado que $p \in
\mathcal{M}_0$, podemos considerar esta ecuación como
\begin{equation}\label{eq:expansion_segundo_orden}
  \varphi^*_\lambda\mathcal{Q}(p) = \mathcal{Q}_0(p) +
  \lambda\pounds_X \mathcal{Q}|_{\mathcal{M}_0}(p)  +
  \frac{1}{2}\lambda^2\pounds^2_X
  \mathcal{Q}|_{\mathcal{M}_0}(p) + O(\lambda^3)
\end{equation}
i.e.\ una ecuación en el espacio-tiempo de fondo
$\mathcal{M}_0$, con $\mathcal{Q}_0 =
\mathcal{Q}|_{\mathcal{M}_0}$ que es el valor de fondo de la
variable física $\mathcal{Q}$. Armados con esta definición,
la perturbacion $\Delta^\varphi\mathcal{Q}_\lambda$ del
campo tensorial $\mathcal{Q}$ bajo la elección de norma
$\varphi_\lambda$ es definida mediante:
\begin{equation}
  \label{eq:perturbacion}
  \Delta^\varphi\mathcal{Q}_\lambda \equiv
  \varphi^*_\lambda\mathcal{Q}|_{\mathcal{M}_0} -
  \mathcal{Q}_0.
\end{equation}
A diferencia de la ecuación \eqref{eq:perturbacion_confusa},
esta definición tiene todas sus variables definidas en
$\mathcal{M}_0$. Sustituyendo
\eqref{eq:expansion_segundo_orden} en esta última ecuación,
definimos las perturbaciones a primer y segundo orden del
campo tensorial $\mathcal{Q}$ bajo la elección de norma
$\varphi_\lambda$ mediante
\begin{equation}
  \label{eq:definicion_primeras_perturaciones}
  \delta\thinspace_\varphi\mathcal{Q} \equiv
  \pounds_X\mathcal{Q}|_{\mathcal{M}_0} , \qquad 
  \delta^2\thinspace_\varphi\mathcal{Q} \equiv
  \pounds^2_X\mathcal{Q}|_{\mathcal{M}_0}.
\end{equation}

Con las definiciones
\eqref{eq:definicion_primeras_perturaciones} podemos tratar
de establecer de una manera más exacta a lo que nos
referimos por ``desviaciones pequeñas'' de las métrica de
fondo. Serán ``desviaciones pequeñas'' $\delta g_{ab} =
(\varphi^*g_{ab}) - g_{ab}$, aquellas que $|\delta g_{ab}|
\ll 1$. Obviamente existen muchos difeomorfismos que no
cumplen con esto, por lo que, toda la discusión desarrollada
en este apéndice está limitada a los difeomorfismos que
satisfacen esta condición.  \cite{Carroll04}.

Giremos ahora nuestra atención ahora al problema de
transformaciones de norma. Sean dos campos
vectoriales $X, Y$ en $\mathcal{N}$. Sus curvas integrales
definen dos flujos $\varphi$ y $\psi$, respectivamente en
$\mathcal{N}$. Entonces $X$ y $Y$ son transversales a
cualquier $\mathcal{M}_\lambda$ y los puntos sobre la misma
curva integral son \textbf{considerados el mismo punto}
respecto a su norma. Entonces $\varphi$ y $\psi$ son mapeos
de identificación puntuales.  Si descomponemos los campos
vectoriales $X$ y $Y$ como lo hicimos anteriormente tenemos,

\begin{equation}
  ^\varphi X^a_\lambda = \left(\frac{\partial}{\partial\lambda}\right)^a +
  \theta^a \quad
  ^\psi Y^a_\lambda = \left(\frac{\partial}{\partial\lambda}\right)^a + \iota^a
\end{equation}

Cuando $\theta^a \neq \iota^a$ se dice que tenemos
\textit{dos elecciones distintas de norma}. Cada una de
estas elecciones de norma define un \textit{pull-back} de un
campo tensorial $\mathcal{Q}$ en $\mathcal{N}$ a otros dos
campos tensoriales, $\varphi^*_\lambda\mathcal{Q}$ y
$\psi^*_\lambda\mathcal{Q}$ para cualquier valor de
$\lambda$. Tomando en particular el espacio-tiempo de fondo
$\mathcal{M}_0$ vemos que tenemos tres campos tensoriales
asociados con $\mathcal{Q}$: $\mathcal{Q}_0$, el valor de
fondo de $\mathcal{Q}$, y los dos \textit{pull-back} de
$\mathcal{Q}$ desde $\mathcal{M}_\lambda$ por las dos
elecciones de norma diferentes. Estos dos últimos campos
tensoriales pueden ser expandidos en series de Taylor,

\begin{subequations}
  \begin{equation}
    \label{eq:expansion_norma_X}
    ^\varphi \mathcal{Q}_\lambda \equiv
    \varphi^*_\lambda\mathcal{Q}|_{\mathcal{M}_0} = 
    \sum^\infty_{k=0}\frac{\lambda^k}{k!}\;
    \left(\delta^{(k)}\thinspace_\varphi\mathcal{Q}\right) = \mathcal{Q}_0 
    + \Delta^\varphi\mathcal{Q}_\lambda,
  \end{equation}
  \begin{equation}
    \label{eq:expansion_norma_Y}
    ^\psi \mathcal{Q}_\lambda \equiv  \psi^*_\lambda\mathcal{Q}|_{\mathcal{M}_0} =
    \sum^\infty_{k=0}\frac{\lambda^k}{k!}\;
    \left(\delta^{(k)}\thinspace_\psi\mathcal{Q}\right) = \mathcal{Q}_0 
    + \Delta^\psi\mathcal{Q}_\lambda.
  \end{equation}
\end{subequations}
donde se ha usado las definición recién dada
\eqref{eq:perturbacion}. Como $\varphi_\lambda$ y
$\psi_\lambda$ son elecciones de norma que mapean el
espacio-tiempo $\mathcal{M}_0$ en $\mathcal{M}_\lambda$,
$^\varphi\mathcal{Q}_\lambda$ y $^\psi\mathcal{Q}_\lambda$
son representaciones en $\mathcal{M}_0$ del campo tensorial
$\mathcal{Q}$ en dos diferentes normas. Las cantidades
$\delta^k_\varphi\mathcal{Q}$ y $\delta^k_\psi\mathcal{Q}$
son las perturbaciones de orden $k$ en las normas $\varphi$
y $\psi$ respectivamente. Es muy importante notar que el
parámetro $\lambda$ usado para etiquetar las subvariedades
de $\mathcal{N}$ (recuérdese que proviene del difeomorfismo
original $\phi$) también sirve para realizar la expansión,
estableciendo así lo que significa ``perturbación a orden
$k$'', por lo que la división de $\Delta^\varphi
\mathcal{Q}_\lambda$ en perturbaciones de primer, segundo,
etc. orden no tiene un significado absoluto, ya que una
reparametrización de $\mathcal{N}$ (i.e.\ eligiendo un nuevo
mapeo primigenio) mezclará todos los ordenes. Lo único que
está definido de manera invariante, es la cantidad
$\Delta^\varphi\mathcal{Q}$.  Es importante notar que si el
campo tensorial $\mathcal{Q}$ es invariante de norma, su
representación en $\mathcal{M}_0$ no cambiará ante
transformaciones de norma, por definición.

Definimos que un campo tensorial $\mathcal{Q}$ en
$\mathcal{N}$ es \textit{totalmente invariante de norma}
(TIN) si $^\varphi\mathcal{Q}_\lambda
=\thinspace^\psi\mathcal{Q}_\lambda$ para cualquier par de
elecciones de norma $\varphi$ y $\psi$, implicando así que
$\delta^{(k)} \,^\varphi\mathcal{Q} = \delta^{(k)}
\thinspace^\psi\mathcal{Q}$ para todo $k$. Podemos relajar
este requerimiento y definir \textit{invariante de norma}
(IN) \textit{a orden $n$} si y sólo si para dos normas
cualesquiera $\varphi$ y $\psi$

\begin{align}
  \label{eq:invariantes_norma_orden_n}
  \delta^{(k)}\thinspace_\varphi\mathcal{Q} =
  \delta^{(k)}\thinspace_\psi\mathcal{Q} \qquad \forall k,
  \quad \mbox{con} \quad k < n.
\end{align}

Con esta definición es posible demostrar por inducción
\cite{BRUNI_1} que el campo tensorial $\mathcal{Q}$ es
invariante de norma a orden $n \geq 1$ si y sólo si
$\Lie_\xi\, \delta^{(k)} \mathcal{Q} = 0$ , para cualquier
campo vectorial $\xi^a$ definido en $\mathcal{M}_0$ y
$\forall k < n$. Esto generaliza el lema de Stewart-Walker
\citep[expuesto por vez primera en][]{PERTURBATIONS_ST_GR}:

\begin{definicion}[Lema de Stewart-Walker.]
  La perturbación a orden $n$ del campo tensorial
  $\mathcal{Q}$ es invariante de norma a orden $n$ si y solo
  si $\mathcal{Q}_0$ y todas las perturbaciones de orden
  menor a $n$, son, en cualquier norma:
  \begin{itemize}
  \item Cero
  \item Escalar constantes
  \item Combinaciones lineales de productos de deltas de
    Kronecker.
  \end{itemize}
\end{definicion}
La prueba de este lema es directa usando las definiciones
anteriores y se puede encontrar en
\cite{PERTURBATIONS_ST_GR}.

El cambio de elección de norma desde $\varphi_\lambda$ a
$\psi_\lambda$ está representado por el difeomorfismo
\begin{equation}
  \Phi_\lambda \equiv (\varphi_\lambda)^{-1} \circ \psi_\lambda
\end{equation}

El difeomorfismo $\Phi_\lambda$ es un mapeo
$\Phi_\lambda:\mathcal{M}_0 \to \mathcal{M}_0$ para cada
valor $\lambda \in \mathbb{R}$. Este difeomorfismo cambia el
punto de identificación, por lo tanto $\Phi_\lambda$ se
puede considerar como \textit{la transformación de norma}
$\Phi_\lambda: \varphi_\lambda \to \psi_\lambda$. Debemos
recalcar que $\Phi:\mathbb{R} \times \mathcal{M}_0 \to
\mathcal{M}_0$ \textit{no es} un grupo uniparamétrico de
difeomorfismos en $\mathcal{M}_0$ debido a que en general,
los generadores del difeomorfismo, $^\varphi X^a$ y $^\psi
Y^a$, no conmutan. Como consecuencia de esto, no se podrá expandir a
$\Phi_\lambda$ en una serie de Taylor usando la ecuación \eqref{eq:taylor_1}
debido a que esta sólo es válida para grupos uniparamétricos, sin
embargo, en el apéndice \ref{sec:dife-de-caballo} se muestra
que, a orden $n$ en $\lambda$ la familia uniparamétrica de
difeomorfismos $\Phi$ siempre puede ser aproximada por la
familia de diferomorfismos de caballo de rango $n$ (teorema
\ref{teo:generalizacion}), los cuales a su vez pueden ser
expresados en una serie de Taylor generalizada (lema
\ref{lemma:taylor_diff_caballo}).

La transformación de norma $\Phi_\lambda$ induce un
\textit{pull-back} desde la representación
$^\varphi\mathcal{Q}_\lambda$ del tensor perturbado en la
elección de norma $\varphi_\lambda$ a la representación
$^\psi\mathcal{Q}_\lambda$ en la elección de norma
$\psi_\lambda$. De esta manera estos tensores están
conectados por el mapeo lineal $\Phi^*_\lambda$ mediante
\begin{equation}
  \begin{split}
    ^\psi\mathcal{Q}_\lambda &=
    \psi^*_\lambda\mathcal{Q}|_{\mathcal{M}_0} \\ &= \left(
      \psi^*_\lambda\left(\varphi_\lambda\varphi^{-1}_\lambda
      \right)^*\mathcal{Q} \right)\Big|_{\mathcal{M}_0} \\
    &= \left(\varphi_\lambda^{-1}\psi_\lambda
    \right)^*\left(\varphi_\lambda^*\mathcal{Q}
    \right)\Big|_{\mathcal{M}_0} \\ &= \Phi^*_\lambda
    \,^\varphi\mathcal{Q}_\lambda
  \end{split}
\end{equation}

Usando el lema \ref{lemma:taylor_diff_caballo}, se expresa
la serie de Taylor de $\Phi^*_\lambda
\,^\varphi\mathcal{Q}_\lambda$, obteniendo
  \begin{equation}\label{eq:_expansion_cambio_norma}
    \Phi^*_\lambda \,^\varphi\mathcal{Q}_\lambda = \thinspace^\varphi \mathcal{Q} +
    \lambda\pounds_{\xi_{1}} \,^\varphi\mathcal{Q} +
    \frac{\lambda^2}{2}\left\{\pounds_{\xi_{2}}+\pounds^2_{\xi_{1}}\right
    \} \,^\varphi\mathcal{Q} +\frac{\lambda^3}{3!}\left\{\pounds^3_{\xi_1}
      +3\pounds_{\xi_1}\pounds_{\xi_2} + \pounds_{\xi_3}
    \right\} + \mathcal{O}(\lambda^4). 
  \end{equation}
donde $\xi_1$, $\xi_2$ y $\xi_3$ son los primeros tres
generadores de $\Phi_\lambda$. Si sustituimos
\eqref{eq:expansion_norma_X} y \eqref{eq:expansion_norma_Y}
en \eqref{eq:_expansion_cambio_norma} e igualando término a
término llegamos a que las relaciones entre el primer,
segundo y tercer orden de la perturbación del tensor
$\mathcal{Q}$ en dos normas diferentes es:
\begin{subequations}\label{eq:transformaciones_norma}
  \begin{equation}\label{eq:transformacion_1er}
    \delta^{(1)}_\psi\mathcal{Q} -
    \delta^{(1)}_\varphi\mathcal{Q} =
    \pounds_{\xi_1}\mathcal{Q}_0 ,
  \end{equation}
  \begin{equation} \label{eq:transformacion_2nd}
    \delta^{(2)}_\psi\mathcal{Q} -
    \delta^{(2)}_\varphi\mathcal{Q} = 2\pounds_{\xi_1}
    \delta^{(1)}_\varphi\mathcal{Q} +
    \left\{\pounds_{\xi_2}+\pounds^2_{\xi_1}\right\}\mathcal{Q}_0,
  \end{equation}
\end{subequations}
Estas relaciones son consistentes con la definición de
invariante de norma de orden $n$ dada anteriormente.

Para finalizar esta sección mostraremos la manera de obtener
los generadores del difeomorfismo $\Phi_\lambda$ en término
de las elecciones de norma $X$ y $Y$: sustituyendo
\eqref{eq:expansion_norma_X} y \eqref{eq:expansion_norma_Y},
en \eqref{eq:_expansion_cambio_norma},igualando términos en
orden $\lambda$, usando propiedades de la Derivada de Lie
(ver \ref{sec:derivadas-de-lie}) de un tensor y por último
considerando que $Y^m - X^m = 0$ (ambos tienen la misma
$m$-ésima coordenada: $\lambda$) obtenemos las igualdades
siguientes (ver \cite{BRUNI_1} para las demostraciones
detalladas de estas últimas dos afirmaciones):

\begin{subequations}\label{eq:relaciones_entre_generadores}
  \begin{equation}
    \xi_1^a = Y^a - X^a
  \end{equation}
  \begin{equation}
    \xi_2^a = [X,Y]^a
  \end{equation}
\end{subequations}

\section{\label{sec:flrw}Espacio-tiempo de ejemplo: el
  Universo FLRW} 

\subsection{Espacio-tiempo de fondo}

Para ejemplificar las diferentes aproximaciones al análisis
perturbativo en Relatividad General  usaremos el espacio-tiempo de
Friedmann-Lemaître-Roberston-Walker (FLRW).  Los Universos
FLRW describen la dinámica y la geometría de los universos con 
hipersuperficies espaciales máximamente simétricas. En particular
nos enfocaremos al caso espacialmente plano (i.e.\ $\Omega =
1$) de FLRW. Usaremos este espacio-tiempo por dos motivos,
el primero es que debido a sus 
características geométricas (simetrías y platitud espacial)
es fácil de manipular matemáticamente; y segundo, en la
actualidad, debido a los grandes avances observacionales en la
cosmología es posible extraer y comparar los cálculos 
hechos con la teoría de perturbaciones aplicada a estos
universos idealizados con los datos proporcionados por el
universo real, por ejemplo con los datos extraídos del
estudio del fondo de radiación cósmica \cite{wmap5}.

Este espacio-tiempo idealizado es homogéneo
e isotrópico espacialmente. Esto sólo se cumple -de una
manera aproximada y sin una clara justificación teórica- en
el universo real a una escala de aproximadamente $100$
Mpc. En este apéndice
supondremos que la geometría de las hiper superficies
espaciales es plana, por lo que su elemento de línea sin
perturbar es
\begin{equation}
  \label{eq:elemento-flrw}
  ds^2 = a^2(\eta)[-d\eta^2 + \delta_{ij}dx^idx^j],
\end{equation}
donde $a(\eta)$ es el factor de escala expresado en el
tiempo conforme $\eta$. La comilla, $\{\}'$, indica una
derivación respecto al tiempo conforme, i.e.\ $\{\}' \equiv
\frac{\partial}{\partial\eta}$. El tiempo conforme se
relaciona con el tiempo cosmológico $t$ (el tiempo medido por los
observadores fijos en las coordenadas comóviles espaciales
$x^i$), mediante $d\eta = \int \frac{dt}{a}$. Los símbolos de
Christoffel están definidos por \cite{Wald84}:
\begin{equation}
  \Gamma^a_{bc} \equiv \frac{1}{2}g^{ad}(g_{d\,b,c} + g_{c\,d,b} - g_{b c, d}).
\end{equation}

El inverso de la métrica $g^{ab}$ es usado para levantar
índices espacio-temporales, mientras que la 3-métrica
$\delta^{ij}$ ($\delta^{ij}\delta_{jk} = \delta^i_{\,k}$) es
usada para elevar los índices de los \textit{3-vectores} y
\textit{3-tensores}. Como el espacio es plano, la 3-derivada
covariante es $\nabla_i \equiv \partial_i$.

Para la métrica \eqref{eq:elemento-flrw} los símbolos de
Christoffel tienen una forma particularmente, aquellos
diferentes de cero son 
\begin{equation}
  \Gamma_{\eta\,\eta}^\eta = \boldsymbol{\mathscr{H}}, \quad
  \Gamma_{\eta\,i}^j = \delta_i^j\boldsymbol{\mathscr{H}}, \quad
  \Gamma_{i\,j}^\eta = \delta_{ij}\boldsymbol{\mathscr{H}},
\end{equation}
donde se ha introducido el parámetro de Hubble comóvil,
$\HubbleComovil \equiv \frac{a'(\eta)}{a(\eta)}$, que está
relacionado con el parámtero de Hubble\footnote{El nombre
  --claramente inapropiado-- de
  \textit{constante de Hubble} se reserva para el valor del
  parámetro de Hubble evaluado en este evento
  espacio-temporal (hoy, ahora), $H_0 \equiv H(t)$.}$H$, mediante
$\HubbleComovil = a H$. El tensor de Ricci se puede expresar
en términos de los símbolos de Christoffel:
\begin{equation}
  R_{ab} = \Gamma^c_{ab,\,c} - \Gamma^c_{a\,c,b}+
  \Gamma^c_{d\,c}\Gamma^d_{ab} - \Gamma^c_{d\, b}\Gamma^d_{a\,c},
\end{equation}
las componentes de este tensor en este
espacio son
\begin{equation}\label{eq:ricci_flrw}
  R_{\eta\eta} = 3\left(\Hubble - \frac{a^{''}}{a}\right),
  \quad
  R_{ij} = \left(\Hubble + \frac{a^{''}}{a}\right).
\end{equation}

El escalar de Ricci, $\Ricci$, por su parte, está definido
por la contracción total del tensor de Ricci con la métrica
$g_{\mu\nu}$, i.e.
\begin{equation}
  \Ricci = g^{\mu\nu}R_{\mu\nu},
\end{equation}
sustituyendo \eqref{eq:ricci_flrw} en esta última fórmula tenemos
\begin{equation}\label{eq:escalar_ricci_flrw}
  \Ricci = \frac{6}{a^2}\left(\Hubble^{'} - \Hubble^2 \right).
\end{equation}

El tensor de Einstein $G_{ab}$, está definido por la
ecuación

\begin{equation}
  G_{b}\,^a = R_{b}\,^a-\frac{1}{2}g_{b}\,^a\Ricci,
\end{equation}
usando \eqref{eq:ricci_flrw} y \eqref{eq:escalar_ricci_flrw}
obtenemos que sus componentes son
\begin{subequations}
  \begin{equation}
    G_{\eta}\,^\eta= -\frac{3}{a^2}\Hubble^2,
  \end{equation}
  \begin{equation}
    G_{i}\,^j = -\frac{1}{a^2}\delta_i^j\left(2\Hubble^{'} +
      \Hubble^2 \right) .
  \end{equation}
\end{subequations}

En este apéndice sólo estamos interesados en el caso
inflacionario, por lo que la materia en el universo estará
representada por un campo escalar, $\varphi =
\varphi(\eta)$, descrito por el tensor de energía-momento
\begin{equation}
  T_a\,^b =
  \nabla_a\varphi\nabla^b\varphi-\frac{1}{2}
  \delta_a\,^b(\nabla_c\varphi\nabla^c\varphi  
  + 2V(\varphi)),\label{eq:energia_momento_escalar}
\end{equation}
con componentes en las coordenadas definidas por
\eqref{eq:elemento-flrw}
\begin{subequations}
  \begin{equation}
    T_\eta\,^\eta = -\left(\frac{1}{2a^2}(\varphi^{'})^2 +
      V(\varphi)\right), 
  \end{equation}
  \begin{equation}
    T_i\,^j = \left(\frac{1}{2a^2}(\varphi^{'})^2 -
      V(\varphi)\right) \delta_i\,^j.
  \end{equation}
\end{subequations}

Comparando con el tensor de energía-momento de un fluido
perfecto podemos identificar a $T_\eta\,^\eta$ con la
energía, $\epsilon$, y a $T_i\,^j$ con la presión,
$p$. Entonces las ecuaciones de Einstein para el
espacio-tiempo de fondo $\mathcal{M}_0$ lleno con un campo
escalar, $\varphi$, es
\begin{subequations}\label{eq:einstein_fondo}
  \begin{equation}
    \Hubble^2 =  \frac{8 \pi \, G}{3}a^2\epsilon = \frac{8 \pi \, G}{3}a^2\left(\frac{1}{2a^2}(\varphi^{'})^2 + V(\varphi)\right),
  \end{equation}
  \begin{equation}
    2\Hubble^{'} + \Hubble^2 = -8\pi G a^2 p = -8 \pi G a^2 \left(\frac{1}{2a^2}(\varphi^{'})^2 - V(\varphi)\right),
  \end{equation}
\end{subequations}
en el trato perturbativo será útil la siguiente forma que se
obtiene sumando ambas ecuaciones
\begin{equation}\label{eq:einstein_auxiliar}
  \Hubble^2 - \Hubble^{'} = 4\pi G a^2(\epsilon + p) = 4\pi G (\varphi^{'})^2.
\end{equation}

\subsection{\label{sec:pert-prim-orden}Perturbaciones a
  primer orden en el espacio-tiempo FLRW}

Antes de introducir los esquemas desarrollados para lidiar
con los problemas de norma en la teoría de perturbaciones en
Relatividad General, perturbaremos de manera general a
primer orden la métrica. Las perturbaciones a segundo orden
se verán más adelante en el texto. Los componentes de la
perturbación lineal de la métrica \eqref{eq:elemento-flrw}
\begin{equation}
  \label{eq:h_ab_frw}
  \delta g_{ab} \equiv h_{ab} = \begin{pmatrix}
    h_{\eta\,\eta} & h_{\eta\,i} \\
    h_{i\,\eta} & h_{i\,j}
  \end{pmatrix}.
\end{equation}
El elemento de línea perturbado es entonces,
  \begin{equation}
    \label{eq:elemento-flrw-perturbado}
    ds^2 = a^2(\eta)\left[-(1+h_{\eta\eta})\;d\eta^2 + 2 h_{\eta\;i}
      d\eta\; dx^i + (\delta_{ij}+h_{ij})dx^idx^j\right].
  \end{equation}
Esta ecuación es completamente general: $g_{\mu\nu}$ tiene
10 componentes independientes y hemos introducido 10 campos
independientes que caracterizan la perturbación (1 de la
perturbación escalar + 3 de la perturbación vectorial + 6 de
la perturbación tensorial), pero téngase en cuenta que sólo
6 de estos campos representan grados de libertad físicos.

Debido a que la variedad $\left(\Sigma(\eta),
  a^2\delta_{i\,j}\right)$ es maximamente simétrica, es
posible encontrar una descomposición diferente a
${h_{\eta\,\eta}, h_{\eta\,i}, h_{ij}}$ de la perturbación a
primer orden que explote estas simetrías
\cite{StraumannN08}. Existen dos tipos de transformaciones
de coordenadas en teoría de perturbaciones en Relatividad
General: (a) \textit{Transformaciones de norma}, que son las
que han sido el tema central de este artículo en las cuales,
las coordenadas de $\mathcal{M}_0$ se mantienen fijas, y, al
elegir una correspondencia de puntos diferente entre
$\mathcal{M}_0$ y $\mathcal{M}$ las coordenadas de este
último cambian; y (b) \textit{Transformaciones en
  $\mathcal{M}_0$} que se diferencian de las anteriores, en
el hecho de que la norma se mantiene fija y hacemos cambios
de coordenadas en $\mathcal{M}_0$ que inducirá un cambio de
coordenadas en $\mathcal{M}$.  Estas últimas son las que nos
ayudarán a descomponer las perturbaciones de la
métrica. Como queremos preservar las simetrías de
$\mathcal{M}_0$, las transformaciones que podremos hacer en
esta variedad están restringidas a ser de dos tipos: (a)
Reparametrizaciones del Tiempo, (2) Transforamciones en
coordenadas espaciales. i.e.\ $x^{i'} = X^{i'}_k \, x^k$.
donde las únicas que preservan la simetría de
$\mathcal{M}_0$ son las rotaciones:
\begin{equation}
  X^{\mu'}_\nu =
  \begin{pmatrix}
    1 && 0 \\
    0 && \mathbf{R}^{i'}_k
  \end{pmatrix},
\end{equation}
La matriz $\mathbf{R}^{i'}_k$ es una matriz $\bm{O}(3)$ que
representa las rotaciones. Analizando como se transforman
las perturbaciones de la métrica antes estas
transformaciones observamos que $h_{\eta\,\eta}$ se
transforma como \textit{escalar}, $h_{\eta\,i}$ como
\textit{3-vector} y $h_{i\,j}$ como \textit{3-tensor}, donde
los términos \textit{escalar}, \textit{3-vector} y
\textit{3-tensor} se refieren a las propiedades de
transformación ante rotaciones en el espacio de
fondo\footnote{A veces en la literatura se denominan como
  variables de espín-0, espín-1 y espín-2,
  respectivamente.}. En particular \textit{escalar} en este
contexto NO significa que la perturbación es invariante
antes las transformaciones de norma, de hecho, \textit{las
  perturbaciones escalares no son invariantes de
  norma}. Estas denominaciones de \textit{escalar} y
\textit{3-vectorial} datan del artículo seminal de Lifshitz
\cite{Lifshitz46}.

Usando el teorema de Helmholtz podemos descomponer a la
perturbación vectorial en una parte potencial o libre de
rotación y en una parte libre de divergencia. Otras
denominaciones son \textit{longitudinal} para la libre de
rotacional y \textit{solenoidal} para indicar que la
divergencia de este término es cero.
\begin{equation}
  \label{eq:vector_descomposicion_1}
  h_{\eta\,i} = h_{i}^{(rot-free)} + h_{i}^{(div-free)} =
  h^{(SV)}_{,\,i} + h_{i}^{(V)}, 
\end{equation}
donde  $\delta^{ij}h_{i,\,j}^{(V)} = 0$; de manera similar, la
perturbación tensorial \footnote{Se agregó el factor de
  $a^2$ para simplificar cálculos más adelante.}:
\begin{equation}
  h_{ij} = a^2h^{(S)}\delta_{ij} + a^2h^{(T)}_{ij} ,
\end{equation}
con $ h^{(T)i}_i \equiv \delta^{ij}h^{(T)}_{ij} = 0$. A su vez,
$h_{ij}^{(T)}$ se puede descomponer en,
\begin{equation}
\label{eq:descomposicion_tensorial}
    h^{(T)}_{ij} = \left(\partial_i\partial_j -
      \frac{1}{3}\delta_{ij}\nabla^2\right)h^{(ST)} 
    -\frac{1}{2}\left(h^{(VT)}_{i,\,j}+h^{(VT)}_{j,\,i}\right)
    + h_{ij}^{(TT)},
\end{equation}
donde $\delta^{ij}h^{(VT)}_{i,\,j} = 0$,
$\delta^{ik}h^{TT}_{ij,\,k} = 0$, y $\delta^{ij}h^{TT}_{ij}
= 0$. En la literatura al primer término entre paréntesis
del lado derecho de la ecuación
\eqref{eq:descomposicion_tensorial} se le conoce como
``operador sin traza de doble gradiente''.

Tenemos entonces que la perturbación lineal de la métrica se
puede descomponer en tres conjuntos de variables, cada
conjunto definido por como se transforman sus
elementos\footnote{Es conveniente decir en este punto que la
  \textit{nomenclatura} de los símbolos con los cuales
  identificamos las perturbaciones dista de ser la estándar
  (salvo en el caso del $3-$tensor, $h^{TT}$), de hecho no
  hay acuerdo en la literatura sobre como nombrar a las
  variables. En este apéndice sigue las de
  \cite{NAKAMURA_4}}: los que transforman como escalar
$\left\{h_{\eta\,\eta}, h^{(S)},h^{(SV)}, h^{(ST)}\right\}$
que son aquellas que se pueden construir a partir de otros
escalares, sus derivadas y cantidades de la métrica de
fondo; aquellos que transforman como \textit{3-vectores},
$\,\left\{ h_{i}^{(V)},\, h_{i}^{(VT)}\right\}$, que tienen
como característica que su divergencia es cero y finalmente
un término que transforma como \textit{3-tensor} con
divergencia cero (o transverso) y con traza cero:
$\left\{h^{(TT)}\right\}$.

Esta descomposición tiene relaciones inversas que se
escriben a continuación:
\begin{subequations}\label{eq:inversion_perturbacion_1}
  \begin{equation}
    h^{(SV)} = \triangle^{-1} \delta^{ij}h_{\eta\,j,\,i}
  \end{equation}
  \begin{equation}
    h^{(V)}_i = h_{\eta\,i}- \left(\triangle^{-1}h_{\eta\,j\,,i}\right)^{\, , j}
  \end{equation}
  \begin{equation}
    h^{(S)} = \frac{1}{3a^2}h_i\,^i
  \end{equation}
  \begin{equation}
    h^{(T)}_{ij} = \frac{1}{a^2}\left( h_{ij}- \frac{1}{3}h_k\,^k\delta_{ij}\right)
  \end{equation}
  \begin{equation}
    h^{(ST)} = \frac{3}{2}\triangle^{-1}\triangle^{-1}h^{(T)}_{ij}\,^{,i\,,j}
  \end{equation}
  \begin{equation}
    h^{(VT)}_i= \triangle^{-1}h^{(T)}_{ik}\,^{,k} - \triangle^{-1}\left(
      \triangle^{-1}h^{(T)}_{kl}\,^{,\,k,\,l}\right)_{,\,i}
  \end{equation}
  \begin{equation}
      h^{(TT)}_{ij}  = h^{(T)}_{ij}  -
      \frac{3}{2}\Big(\partial_i\partial_j -
      \frac{1}{3}\delta_{ij}\triangle\Big)
      \triangle^{-1}\triangle^{-1}h^{(T)}_{kl}\,^{,k\,,l} 
      -2\partial_{(i}\triangle^{-1}\partial^kh^{(T)}_{j)k}
      +2\partial_{(i}\triangle^{-1}\partial_{j)}\triangle^{-1}
      h^{(T)}_{kl}\,^{,k\,,l}
  \end{equation}
\end{subequations}
donde $\triangle \equiv \nabla^2$, es el laplaciano de la
hipersuperficie espacial y estamos suponiendo que existe una
función de Green inversa del operador $\triangle$,
$\triangle^{-1}$, que garantice la correspondencia uno a uno
entre $\left\{h_{\eta\eta},\, h_{\eta\,i}, \,h_{ij}\right\}$
y $\left\{\left\{h_{\eta\,\eta},\, h^{(S)},\, h^{(SV)},\,
    h^{(ST)}\right\},\,\left\{ h_{i}^{(V)},\,
    h_{i}^{(VT)}\right\}, \left\{h^{(TT)}
  \right\}\right\}$. Estas funciones de Green existen si
especificamos el dominio de las perturbaciones (por ejemplo,
$L^2$ en el espacio $\Sigma(\eta)$) con las condiciones
apropiadas de frontera. Nótese que la suposición de la
existencia de estas funciones excluye al \textit{kernel} de
los operadores $\triangle$, por ejemplo, los campos
vectoriales de Killing, $v_i$\cite{NakamuraKo07} en
$\Sigma(\eta)$. El estudio de las perturbaciones que
pertenezcan al \textit{kernel} de $\triangle$ queda excluido
del alcance de esta revisión.

Una característica importante de esta descomposición, es que
las ecuaciones físicas construidas en esta métrica no
mezclarán, a primer orden, los elementos de un conjunto con
los de otro, de tal manera que pueden ser estudiados por
separado \footnote{Esta característica \textit{es cierta
    solo a primer orden} ya que a segundo orden todas las
  cantidades estarán acopladas, ver más adelante}. La
clasificación en escalares, vectores y tensores, es
importante ya que cada uno de ellos representan diferentes
fenómenos: la gravitación descrita por Newton es un fenómeno
escalar, pero los efectos relativistas se ven claramente (ya
que están ausentes en las ecuaciones gravitacionales de
Newton) en los comportamientos gravitomagnéticos (el
conjunto vectorial de las perturbaciones) y de ondas
gravitacionales (las pertubaciones tensoriales)
\cite{Bertschinger93t}. La métrica en esta nueva
descomposición se puede escribir como
\begin{equation*}
  ds^2 = g^{(0)}_{\mu\nu} + \delta g^{(S)}_{\mu\nu} + \delta
  g^{(V)}_{\mu\nu} + \delta g^{(T)}_{\mu\nu},
\end{equation*}
donde la perturbación lineal escalar ($\delta
g^{(S)}_{\mu\nu}$), vectorial ($ \delta g^{(V)}_{\mu\nu}$) y
tensorial ($ \delta g^{(T)}_{\mu\nu}$) son respectivamente:
\begin{subequations}
  \begin{equation}
        \delta g^{(S)}_{\mu\nu} = a^2(\eta) \left[ -h_{\eta\eta}
      d\eta^2 + 2 h^{(SV)}_{,i} d\eta dx^i + \left(h^{(S)}\delta_{ij} +
        h_{,i,j}^{(ST)}\right) dx^i dx^j \right],
  \end{equation}
\begin{equation}
  \delta g^{(V)}_{\mu\nu} = a^2(\eta) \left[ - 2 h_i^{(V)}
    d\eta dx^i + \left(h_{i,j}^{(VT)} + h_{j,i}^{(VT)}\right) dx^i dx^j\right],
\end{equation}
\begin{equation}
  \delta g^{(T)}_{\mu\nu} = h_{ij}^{(TT)} dx^i dx^j  .
\end{equation}
\end{subequations}

La regla de transformación de norma
\eqref{eq:transformacion_1er}, en el caso de FLRW es
\begin{equation}
  \label{eq:transformacion_h_ab}
  _{\varphi} h_{ab} - _{\psi} h_{ab} = \pounds_\xi g_{ab} = 2\nabla_{(a}\xi_{b)}.
\end{equation}

El generador $\xi^a$ de la transformación de norma es un
campo vectorial en el espacio-tiempo de fondo
$\mathcal{M}_0$. Este vector puede  descomponerse de una
manera 3+1:
\begin{equation}
  \xi_a = \xi_\eta\,(d\eta)_a + \xi_i(dx^i)_a,
\end{equation}

De esta manera la transformación de norma para las
perturbaciones originales, $\left\{h_{\eta\eta},\,
  h_{\eta\,i}, \,h_{ij}\right\}$ , es
\begin{subequations}\label{eq:descomposicion_3_1}
  \begin{equation}
    _\varphi h_{\eta\eta} - _\psi h_{\eta\eta} = 2(\partial_\eta - \Hubble)\xi_\eta
  \end{equation}
  \begin{equation}
    _\varphi h_{\eta\,i} - _\psi h_{\eta\,i} = \xi_{\eta,\,i} +
    (\partial_\eta - 2\Hubble)\xi_i
  \end{equation}
  \begin{equation}
    _\varphi h_{i\,j} - _\psi h_{i\,j} = 2 \xi_{(i,\,j)} -
    2\Hubble\delta_{ij}\xi_\eta 
  \end{equation}
\end{subequations}

Procedemos de la misma manera que con las perturbaciones a
primer orden y descomponemos a $\xi_i$ en sus partes
longitudinales y solenoidales usando el teorema de
Helmholtz:
\begin{equation}
  \xi_i = \xi^{(SV)}_{,\,i} + \xi^{(V)}_i \quad \text{con} \qquad \xi^i_{,\,i} = 0,
\end{equation}
entonces la descomposición \eqref{eq:descomposicion_3_1} ahora se escribe,
\begin{subequations}
  \begin{equation}
    _\varphi h_{\eta\eta} - _\psi h_{\eta\eta} = 2(\partial_\eta - \Hubble)\xi_\eta
  \end{equation}
  \begin{equation}
    _\varphi h_{\eta\,i} - _\psi h_{\eta\,i} = \left((\partial_\eta -
      2\Hubble)\xi^{(SV)} + \xi_\eta\right)_{,i}  +
      \left(\partial_\eta-2\Hubble\right)\xi_i^{(V)}
  \end{equation}
  \begin{equation}
    \begin{split}
    _\varphi h_{i\,j} - _\psi h_{i\,j} = 2\xi_{(i,\,j)}^{(V)} +
    2\left(\partial_i\partial_j -
      \frac{1}{3}\delta_{ij}\nabla^2\right)\xi^{(SV)} + 
    2\left(\frac{1}{3}\nabla^2\xi^{(SV)} -
      \Hubble\xi_\eta\right)\delta_{ij}
  \end{split}
\end{equation}
\end{subequations}
donde la última expresión se escribe de esa manera para
facilitar cálculos posteriores.

Utilizando las relaciones
\eqref{eq:inversion_perturbacion_1}, las reglas de
transformación de norma para la descomposición
$\Big\{\left\{h_{\eta\,\eta},\, h^{(S)},\, h^{(SV)},\,
  h^{(ST)}\right\},$\,$\left\{ h_{i}^{(V)},\,
  h_{i}^{(VT)}\right\}$, $\left\{h^{(TT)} \right\}\Big\}$
son para el conjunto escalar:

\begin{subequations}\label{eq:transformaciones_norma_escalar_1}
  \begin{equation}\label{eq:transformacion_escalar}
    _\varphi h_{\eta\eta} - _\psi h_{\eta\eta} = 2(\partial_\eta - \Hubble)\xi_\eta,
  \end{equation}
  \begin{equation}\label{eq:transformacion_escalar_vectorial}
    _\varphi h^{(SV)} - _\psi h^{(SV)} = \xi_\eta + (\partial_\eta-2\Hubble)\xi^{(SV)},
  \end{equation}
  \begin{equation}\label{eq:transformacion_hab_escalar}
    a^2\, _\varphi h^{(S)} - a^2\,_\psi h^{(S)} = -2\Hubble\xi_\eta +
    \frac{2}{3}\nabla^2 \xi^{(SV)},
  \end{equation}
  \begin{equation}\label{eq:transformacion_hab_escalar_tensorial}
    a^2\,_\varphi h^{(ST)} - a^2\,_\psi h^{(ST)} = 2\xi^{(SV)},
  \end{equation}
\end{subequations}
por su parte, para el conjunto de perturbaciones
vectoriales,

\begin{subequations}\label{eq:transformaciones_norma_vectorial_1}
  \begin{equation}
    \label{eq:transformacion_hab_vectorial}
    _\varphi h^{(V)}_i - _\psi h^{(V)}_i = (\partial_\eta - 2\Hubble)\xi^{(V)}_i,
  \end{equation}

  \begin{equation}\label{eq:transformacion_hab_vectorial_tensorial}
    a^2\,_\varphi h^{(VT)}_i - a^2\,_\psi h^{(VT)}_i =      \xi^{(V)}_i,
  \end{equation}
\end{subequations}
y para la parte tensorial,

\begin{equation} \label{eq:transformacion_hab_tensorial}
  a^2\,h^{(TT)}_{ij} - a^2\,h^{(TT)}_{ij} = 0.
\end{equation}

Sentadas las ecuaciones básicas de las perturbaciones de la
métrica de FLRW, podemos estudiar las diferentes
alternativas para tratar las perturbaciones en Relatividad
General, utilizándola como ejemplo.

\section{\label{sec:enfoques}Enfoques en el tratamiento
  perturbativo}

Otra forma de expresar el contenido del principio de
covariancia es diciendo que los observables físicos no
pueden depender de la elección de coordenadas, por lo tanto,
una característica deseable de una teoría perturbativa es
que especifique de una manera no ambigua tales observables,
es decir, los observables deben de ser invariantes de
norma. Básicamente existen dos maneras de extraer la
física (i.e.\ eliminar los grados de libertad espurios) en
el tratamiento perturbativo: (a) fijar los grados de
libertad de norma (opción conocida como \textit{eligir una
  norma}, o (b) \textit{extraer las partes invariantes de
  norma de las perturbaciones}. La opción (b) se puede
dividir a su vez en (b1) formalismo 1+3 covariante
invariante de norma y en (b2) formalismo invariante de
norma. Las opciones (a) y (b1) se mostrarán a grandes rasgos
en esta sección y la opción (b2) será el tema por el resto
de este reporte.

\subsection{Fijando la norma}\label{sec:fijando-la-norma}

La elección más sencilla para eliminar los problemas de
libertad de norma, es elegir o fijar una norma,
procedimiento conocido en inglés como \textit{gauge
  fixing}. Esto significa simplemente que se escogerá una
foliación particular espacial del espacio-tiempo. Esta ha
sido una de las opciones más populares de las décadas
pasadas y fué iniciada por Lifshitz en 1946
\cite{Lifshitz46}, cuando empezó a estudiar la estabilidad
(usando perturbaciones) en espacios-tiempos de FLRW.

Los principales problemas de esta aproximación son dos: (1)
el eliminar verdaderamente los grados de libertad espurios;
y (2) la interpretación física de los resultados obtenidos
en una norma en particular. Estos problemas fueron resueltos
de manera sistemática en cosmología por Bardeen
\cite{Bardeen80}. El procedimiento de Bardeen es el siguiente:
empezando en una norma, es posible definir cantidades
invariantes de norma mediante combinaciones algebraicas o
diferenciales de variables
dependientes de la norma. Esta ha sido la manera de resolver
las ambigüedades de la norma estándar en cosmología \cite{Mukhanov90}.

A continuación, mencionaremos algunas de las normas
favoritas elegidas por la comunidad cosmológica.

\subsubsection{Síncrona}\label{pag:sincrona}
La norma síncrona (\textit{synchronous gauge}) fue
introducida por Lifshitz en 1946 \cite{Lifshitz46}. La norma
síncrona está definida por  $h_{\eta\eta} = 0$, $h^{(V)} = 0$ y
$h^{(SV)} = 0$, i.e.\ dejamos sin perturbar a $g_{00}$ y a
$g_{0i}$. La métrica en la norma síncrona tiene la siguiente forma:
\begin{equation}
  ds^2 = a^2(\eta)\left[ -d\eta^2 + \left(\delta_{ij} +
      h_{ij}\right)dx^idx^j\right] .
\end{equation}
Nótese que para ser compatible con las métricas expresadas
en la literatura \cite{Lifshitz46, Peebles93} en esta norma
hemos regresado al uso de $h_{ij}$.

La norma síncrona es muy popular para los estudios
numéricos y es la norma usada en CMBFAST \cite{CMBFAST}.
Su principal problema es que no elimina completamente la
libertad de norma. Este problema, aunado a su popularidad ha
sido fuente de confusión en el pasado \cite{Bardeen80,Mukhanov90}.

Esta norma permite la existencia de un conjunto de 
observadores que caen libremente --i.e.\ se mueven sobre
geodésicas-- sin cambiar sus coordenadas
espaciales, llamados en la literatura ``observadores
fundamentales comóviles''\footnote{ El que los observadores comóviles
sigan geodésicas se puede comprobar usando la
ecuación geodésica \citep[la fórmula está en
][pags. 46-47]{Wald84} con las condiciones de esta norma,
llegando a que $u^i = 0$ es una geodésica}. Cada
observador está equipado con un reloj que mide el tiempo
conforme $\eta$ y permanece fijo en la
coordenada $x^i$. Los
observadores junto con sus relojes y etiquetas que
identifican su geodésica definen las
coordenadas en todo el espacio-tiempo. El grado de libertad
de norma residual (espurio)  se debe a que hay libertad para ajustar
las condiciones iniciales de los relojes y coordenadas
espaciales.  Otro problema de esta norma aparece cuando  las
perturbaciones son muy grandes, si es así, estas 
coordenadas pueden deformarse mucho e inclusive llegar a
intersectarse \cite{Bertschinger93t} causando singularidades
coordenadas. Las ecuaciones de Einstein y su relación con otra
norma, se pueden obtener de las relaciones derivadas en la
sección \ref{sec:teoria-perturbaciones} y se invita al
lector a verlas en \cite{Mukhanov90,
  Bertschinger93t, Peebles93}.

\subsubsection{Newtoniana Conforme o Longitudinal}\label{pag:poisson}

En la norma newtoniana solo son diferentes de cero las
perturbaciones escalares $h_{\eta\eta}$ y $h^{(S)}$. De esta
definición se puede ver que se eliminan todas las
perturbaciones de carácter vectorial y tensorial
\cite{Mukhanov90}, entonces la métrica en esta
norma es
\begin{equation}
  ds^2 = a^2(\eta) \left[-\left(1+h_{\eta\eta}\right)
    d\eta^2 + \left(1 + h^{(S)}\right)\delta_{ij}dx^i dx^j\right].
\end{equation}

Otra forma en la que esta norma es presentada en la
literatura esta norma es mediante $h_{\eta\:i} = h^{(T)} =
0$ \cite{Bertschinger93t}. 

Las condiciones que dan pie a esta norma sólo pueden ser
establecidas si el tensor de energía momento no contiene
partes vectoriales o tensoriales y no existen ondas
gravitacionales libres\footnote{Ignorar las perturbaciones
  vectoriales y tensoriales a segundo orden es inconsistente
  ya que, aún haciendo las perturbaciones vectoriales y
  tensoriales iniciales iguales a cero a primer orden, las
  perturbaciones a segundo orden servirán como fuente de
  estas perturbaciones lineales}. La norma newtoniana solo
tiene aplicabilidad a orden lineal perturbativo. Si se
ignorara esta limitante se estarían eliminando\footnote{Esta
  situación es análoga a hacer en electrodinámica
  $\vec{A}=0$.}  grados de libertad físicos y no grados de
libertad de norma \cite{Bertschinger93t}.

Esta norma se puede ver como una generalización cosmológica
relativista de las ecuaciones de gravitación
newtoniana. Permitiendo así una interpretación sencilla de
las perturbaciones: $h_{\eta\eta}$ es el análogo del
potencial gravitacional newtoniano $\Phi$.

Uno de sus atractivos es que las variables invariantes de
norma de primer orden \cite{Bardeen80, Mukhanov90,
  Bertschinger93t}, corresponden en esta norma a
$h_{\eta\eta}$ y $h^{(S)}$ lo cual permite dar una
interpretación física sencilla a las variables invariantes
de norma. Debido a esta propiedad en
\cite{Mukhanov90} se sugiere calcular las
ecuaciones en esta norma (debido a su sencillez relativa
respecto al cálculo invariante de norma) y luego hacer las
sustituciones de $h_{\eta\eta}$, $h^{(S)}$ por las variables
escalares invariantes de norma para tener las ecuaciones
invariantes de norma.

La norma newtoniana o longitudinal es un caso especial de
una norma más completa conocida como de \textit{norma de
  Poisson} \cite{Bertschinger93t}, la cual no tiene las
restricciones mencionadas arriba y que queda definida por
$h_{\eta\;i}^{\quad,i} = 0$ y $h_{ij}^{\quad,j} = 0$.

\subsubsection{Otros}

Existen otras normas populares en la literatura cosmológica,
como la (a) \textit{norma espacialmente plana}
$\left(h_{\eta\eta} = h^{(ST)} = h^{(VT)}_i=0\right)$ en
esta norma las hipersuperficies espaciales no son
perturbadas escalar o vectorialmente. En esta norma la
perturbación del campo escalar coincide con la variable de
Mukhanov-Sasaki \cite{Mukhanov88}; (b) la \textit{norma
  ortogonal comóvil} en las cual la 3-velocidad del fluido
se hace cero y (c) la \textit{norma de densidad uniforme}
que como nombre indica en sus hipersuperficies espaciales la
perturbación de la densidad es nula \cite{Malik08}.

\subsection{1+3 Covariante-Invariante de Norma}

Hasta el momento hemos considerado foliar el 
espacio-tiempo en hipersuperficies espaciales etiquetadas
por el tiempo $\eta$, esta división
se conoce como división 3+1 y recibe el nombre de
\textit{slicing} en inglés. La división 3+1 es la usada en la
formulación hamiltoniana de Relatividad General
\cite{Wald84}. Existen otras formas de hacer la separación
espacio-temporal,  en 1971  G.F.R. Ellis\cite{Ellis71} y
otros autores después de él, 
han desarrollado un formalismo basado en una divisón 1+3 de
las identidades de Bianchi y Ricci
\cite{GAUGE_INVARIANT_MEANING,Ellis99b,Enqvist06,Wainwright1997,Ellis89, 
  Ellis89b} llamado formalismo 1+3 Covariante-Invariante de Norma.

En el formalismo 1+3 Covariante-Invariante de Norma se ejecuta una
operación diferente al \textit{slicing}, llamada en inglés
\textit{threading} (literalmente ``hilado''). Los objetos
geométricos usados son una serie de líneas de mundo
temporaloides $x^\mu(\tau, \bm{q})$ donde $\tau$ es el
parámetro afín que mide el tiempo propio sobre la línea de
mundo y $\bm{q}$ es una etiqueta única para distinguir las
diferentes líneas de mundo. Los observadores en este
formalismo se mueven a lo largo de estas líneas \cite{Bertschinger93t}.
 
La herremienta principal en la construcción de este formalismo,
es el vector temporal ($u^\mu u_\mu = -1$) tangente a la
línea de mundo $u^\mu \equiv dx^\mu/d\tau$. Los tensores se
descomponen en una parte paralela al vector $u^\mu$ y otra
normal a este, usando el operador de proyección $P_{\mu\nu}
\equiv g_{\mu\nu} + u_\mu u_\nu$, que es la métrica espacial
de los observadores moviéndose con $u^\mu$. Las ecuaciones a
las que se les aplicará este procedimiento de descomposición
no son las ECE, si no ecuaciones definidas para el tensor de Riemann o el
tensor de Weyl y las ecuaciones de Raychaudhuri
\citep[ver][y referencias dadas arriba]{Ellis71, Ellis99b}.

Usando el Lemma de Stewart-Walker se escogen las variables
invariantes de norma, es decir se eligen las variables que
son cero de manera natural en el espacio-tiempo de fondo
FLRW (por ejemplo el tensor de Weyl --sus partes eléctrica
$E_{ab}$ y magnética $H_{ab}$ \cite{Wainwright1997} --, la
presión anisotrópica $\pi_{ab}$ del fluido, el corte de las
congruencias de geodésicas $\sigma_{ab}$, $\dot{u}^a$, etc
\citep[ver][para una lista exhaustiva]{Ellis99b}), por lo
tanto las ecuaciones de evolución 1+3 del universo
perturbado no poseen ambigüedades de norma. Además las
variables elegidas para caracterizar la evolución tienen una
interpretación física directa, a 
diferencia de las variables definidas en el  enfoque
Invariante de Norma (cf.\ \ref{sec:invariante-norma}). 

En este formalismo es fácil aplicar técnicas de sistemas
dinámicos a la cosmología \cite{Wainwright1997}.

\section{\label{sec:invariante-norma}Formalismo Invariante
  de Norma} 

En el formalismo Invariante de Norma --presentado en los
trabajos de J.M. Stewart y M. Walker 
\cite{PERTURBATIONS_ST_GR}, M. Bruni \textit{et al}.\ \cite{BRUNI_2} y luego desarrollado
extensamente en los trabajos de Kouji Nakamura
\cite{NAKAMURA_1,NAKAMURA_2,NakamuraKo07,NAKAMURA_4,
  Nakamura09a,Nakamura09b})-- se busca eliminar desde el 
inicio, cualquier posible elemento no físico de los grados
de libertad. Entre las ventajas de este formalismo, se pueden
mencionar que (a) no hace ninguna suposición acerca del
espacio-tiempo de fondo (a diferencia, por ejemplo,  del
método de elección de norma, cf.\ sección
\ref{sec:fijando-la-norma}); (b) no sólo se aplica a la teoría de
Relatividad General sino a toda teoría en la cual se exija
la validez del principio de covariancia general \citep[para
un ejemplo de esto ver][y las referencias ahí
dadas]{NAKAMURA_1}, y por último (c) si se cumple cierta
condición de separabilidad de la métrica perturbada a primer
orden propone un algoritmo para encontrar perturbaciones
invariantes de norma a cualquier orden superior al primero.

\subsection{\label{sec:algoritmo}Algoritmo}

El procedimiento para encontrar invariantes de norma a un
orden perturbativo arbitrario es el siguiente:

\begin{enumerate}

\item Se expande mediante la serie de
  Taylor~\eqref{eq:taylor_diff} a la métrica $\bar{g_{ab}}$
  de $\mathcal{M}_\lambda$ luego de aplicarle un
  \textit{pull-back} usando la elección de norma
  $\varphi_\lambda$ a $\mathcal{M}_0$, i.e.\

  \begin{equation}
    \label{eq:metrica_perturbada}
    \varphi^*_\lambda \overline{g_{ab}} = g_{ab}+\lambda \, _{\varphi}h_{ab} +
    \frac{\lambda^2}{2} \, _{\varphi}l_{ab} + \mathcal{O}^3(\lambda),
  \end{equation}

  donde $g_{ab}$ es la métrica de $\mathcal{M}_0$ y hemos
  definido a $\delta g_{ab} \equiv h_{ab}$ y a $\delta^2
  g_{ab} \equiv l_{ab}$ como las perturbaciones de la
  métrica a primer y segundo orden respectivamente. Las
  transformaciones de norma \eqref{eq:transformacion_1er} y
  \eqref{eq:transformacion_2nd} de la métrica a primer y
  segundo orden son

  \begin{subequations}\label{eq:transformacion_metrica}
    \begin{equation}
      \label{eq:transformacion_metrica_1er}
      \thinspace_\varphi h_{ab} - \thinspace_\psi h_{ab} =
      \Lie_{\xi_1} g_{ab},
    \end{equation}
    \begin{equation}
      \label{eq:transformacion_metrica_2do}
      \thinspace_\varphi l_{ab} - \thinspace_\psi l_{ab} =
      2\Lie_{\xi_1} h_{ab} + \left\{ \Lie_{\xi_2} + \Lie_{\xi_1}^2\right\}g_{ab}.
    \end{equation}
  \end{subequations}

\item Inspeccionando las reglas de transformación de norma
  \eqref{eq:transformacion_metrica} se procede a separar, en
  partes invariantes de norma y variantes de norma, a la
  primera perturbación de la métrica, $h_{ab}$.

  \begin{equation} \label{eq:gab_descomposcion_1} h_{ab} \equiv
    \,\mathcal{H}_{ab} + 2\nabla_{(a} \,X_{b)}.
  \end{equation}
  donde $\mathcal{H}_{ab}$ es la parte que es invariante de
  norma y $X^a$ es la parte variable. Es decir, ante
  transformaciones de norma $\Phi=\varphi^{-1}\circ\psi$,

  \begin{subequations}
    \begin{equation} \label{eq:transformacion_vin_1er}
      \,_\psi\mathcal{H}_{ab} - \,_\varphi\mathcal{H}_{ab} =
      0,
    \end{equation}
    \begin{equation}\label{eq:transformacion_vn_1er}
      _\psi \,X^a -\, _\varphi\,X^a = \xi_1^a.
    \end{equation}
  \end{subequations}

  Este procedimiento de separación se supone que se conoce y
  es la condición de entrada del algoritmo. Esta condición
  puede parecer muy restrictiva, ya que no existe una manera
  canónica de realizar esta descomposición.  En nuestro
  espacio-tiempo de ejemplo i.e.\ Universos FLRW, está
  descomposición se mostró en \ref{sec:pert-prim-orden}, a
  partir de esta primera descomposición se manipulan
  algebráicamente los elementos para llegar a
  \eqref{eq:gab_descomposcion_1} como se mostrará adelante.

  Al no existir una manera canónica de realizar la
  descomposición a primer orden
  \eqref{eq:gab_descomposcion_1} se pueden intentar algunas
  estrategias para lograrla, por ejemplo, si existen algunas
  simetrías de Killing en el espacio-tiempo de fondo, se
  puede intentar una descomposición armónica de
  las variables\footnote{Es importante mencionar que la expansión
  armónica depende fuertemente no solamente de las simetrías
  locales del espacio-tiempo de fondo, si no también en la
  topología global de la subvariedad en la cual los
  armónicos escalares, vectoriales y tensoriales son
  definidos \cite{NAKAMURA_1}.}.

\item Realice la misma descomposición en partes variantes e
  invariantes de norma para perturbaciones de orden superior
  de la métrica. El artículo \cite{NAKAMURA_1} mostró que
  esto se puede hacer siempre con algunas manipulaciones
  algebraicas que resultan de inspeccionar la reglas de
  transformación de norma a ordenes mayores que el segundo
  (que se vuelven más complicadas cada vez que aumentamos el
  orden perturbativo). A segundo orden, por ejemplo,
  definimos la variable basándonos en la forma de la
  ecuación \eqref{eq:transformacion_metrica_2do} para la
  elección de norma $\varphi$,
  \begin{equation}\label{eq:L_g_ab}
    _\varphi\hat{\mathcal{L}}_{ab} \equiv _\varphi l_{ab} -
    2\pounds_{_\varphi X}\, _\varphi h_{ab} + \pounds_{_\varphi X} g_{ab},
  \end{equation}
  y para la norma $\psi$,
  \begin{equation}
    _\psi\hat{\mathcal{L}}_{ab} \equiv _\psi l_{ab} -
    2\pounds_{_\psi X}\, _\psi h_{ab} + \pounds_{_\psi X} g_{ab}.
  \end{equation}

  Esta variable se transforma como mostraremos, primero
  restamos ambas variables,
  \begin{equation}
    \begin{split}
      _\psi\hat{\mathcal{L}}_{ab} -
      _\varphi\hat{\mathcal{L}}_{ab} &= _\psi l_{ab} - \,
      _\varphi l_{ab} \\ &- 2 \pounds_{_\psi X}\,_\psi
      h_{ab} + 2\pounds_{_\varphi X}\,\varphi h_{ab} \\ &+
      \pounds^2_{_\psi X} \, g_{ab} - \pounds^2_{_\varphi
        X}\, g_{ab},
    \end{split}
  \end{equation}
  luego, usando la linealidad de la derivada de Lie
  \ref{sec:derivadas-de-lie} y las reglas de transformación
  de norma a primer orden \eqref{eq:transformacion_metrica}
  y \eqref{eq:transformacion_vn_1er}
    \begin{equation}
      \begin{split}
        _\psi\hat{\mathcal{L}}_{ab} -
        _\varphi\hat{\mathcal{L}}_{ab} & =
        -2\pounds_{\xi_{1}}(\pounds_{\xi_{1}} g_{ab}) +
        \pounds_{\xi_2}g_{ab} + \pounds^2_{\xi_1}g_{ab} -
        2\pounds_{_\varphi X}\left(
          \pounds_{\xi_1}g_{ab}\right) +\pounds^2_{_\psi
          X}\, g_{ab} - \pounds^2_{_\varphi X}\,g_{ab} \\ &
        = \pounds_{\xi_2}\,g_{ab} +
        \Big(\pounds_{\xi_1}\pounds_{_\varphi X}
        - \pounds_{_\varphi X}\pounds_{\xi_1}\Big)g_{ab} \\
        & = \pounds_{\sigma_2}g_{ab}
      \end{split}
    \end{equation}
  donde $\sigma_2^a \equiv \xi^a_2 + \left[ \xi_1, \, _\varphi X
  \right]^a$.

  Entonces, la transformación de $\hat{\mathcal{L}}$ tiene
  la misma forma que la transformación de primer orden
  \eqref{eq:transformacion_metrica_1er}. Por lo tanto, como
  suponemos que existe un procedimiento para descomponer un
  campo tensorial de rango dos que se transforma como
  \eqref{eq:transformacion_metrica_1er} en
  \eqref{eq:gab_descomposcion_1}, podemos descomponer al
  tensor $\hat{\mathcal{L}}_{ab}$ en un campo tensorial
  $\mathcal{L}_{ab}$ y un campo vectorial $_\varphi Y^b$ de
  la siguiente manera:
  \begin{equation}\label{eq:L_descomposicion}
    _{\varphi}\hat{\mathcal{L}}_{ab} =: \mathcal{L}_{ab} +
    \nabla_{(a} \, _{\varphi}Y_{b)} 
  \end{equation}
  donde $\mathcal{L}_{ab}$ se transforma como la parte
  invariante de norma de $\hat{\mathcal{L}}_{ab}$
  \begin{equation}\label{eq:transformacion_vin_2do}
    _\psi\mathcal{L}_{ab} - _\varphi\mathcal{L}_{ab}  = 0,
  \end{equation}
  y el campo vectorial $Y^b$ como la parte variable de norma
  \begin{equation}\label{eq:transformacion_vn_2do}
    _{\psi} Y^a - _{\varphi} Y^a =   \xi^a_2 + \left[ \xi_1,
      \, _\varphi X \right]^a. 
  \end{equation}

  Usando este campo tensorial y la definición
  \eqref{eq:transformacion_metrica_2do}, la perturbación a
  segundo orden de la métrica es posible descomponerla en
  \begin{equation}\label{eq:gab_descomposicion_2}
    l_{ab}=\,\mathcal{L}_{ab} + 2 \pounds_{X}h_{ab} +
    \left(\pounds_{Y} - \pounds^2_{X}\right)g_{ab},
  \end{equation}
  donde $\mathcal{L}_{ab}$ es la parte invariante de norma y
  $Y^a$ es la parte variante de la perturbación a segundo
  orden.  El caso para ordenes superiores es completamente
  análogo al mostrado para el segundo orden, por ejemplo
  para tercer orden, el lector puede consultar
  \cite{NAKAMURA_1}.

\item Se pueden definir los invariantes de norma a orden $n
  < k$ de cualquier campo tensorial (excluyendo la métrica)
  usando las perturbaciones de la métrica. Las partes
  variantes de norma de la métrica ($X^a$, $Y^b$ a primer y
  segundo orden) no poseen nada de contenido físico, mas sin
  embargo, son importantes para definir las variables
  invariantes de norma de los campos tensoriales.  Las
  partes invariantes de norma $\mathbf{Q}$ de un campo
  tensorial arbitrario $\mathcal{Q}$ (sin incluir la
  métrica) están dadas por (a segundo orden)

  \begin{subequations}\label{eq:def_vin}
    \begin{equation}\label{eq:def_vin_1er}
      \,^{(1)}\mathbf{Q} \equiv \,^{(1)}\mathcal{Q} -
      \Lie_{X}\mathcal{Q}_0, 
    \end{equation}
    \begin{equation}\label{eq:def_vin_2do}
      \,^{(2)}\mathbf{Q} \equiv \,^{(2)}\mathcal{Q} -
      2\Lie_{X}\,^{(1)}\mathcal{Q}- \left\{
        \Lie_{Y}-\Lie^2_{X}\right\}\mathcal{Q}_0
    \end{equation}
  \end{subequations}
  
  Es fácil ver que estas variables son invariantes de norma
  si se usan las ecuaciones
  \eqref{eq:transformaciones_norma},
  \eqref{eq:transformacion_vn_1er} y
  \eqref{eq:transformacion_vn_2do}.

  En este momento podemos resolver la pregunta acerca de la
  generalidad de las ecuaciones \eqref{eq:def_vin}, esta
  pregunta es válida debido a que nuestras elecciones de
  norma eran mapeos exponenciales, si esto no fuese así, las
  ecuaciones afectadas serían
  \eqref{eq:expansion_segundo_orden} (la serie de Taylor del
  \textit{pull-back} de $\mathcal{Q}_\lambda$), las
  definiciones de las perturbaciones
  \eqref{eq:definicion_primeras_perturaciones} y las
  relaciones entre generadores de los mapeos y los
  generadores de la transformación de
  norma~\eqref{eq:relaciones_entre_generadores}, pero las
  ecuaciones más importantes \eqref{eq:def_vin} no se ven
  cambiadas, ya que son consecuencia directa de la expansión
  en serie de Taylor del difeomorfismo que define la
  transformaciones de norma, $\Phi_\lambda$
  \cite{NakamuraKo07}.

\end{enumerate}

En lo que resta de la sección se seguirá el esquema
siguiente, se presentan los cálculos y conceptos de manera
general en un apartado y luego se aplican en el apartado
siguiente al caso de FLRW. Entonces, el apartado siguiente
tratará sobre la descomposición de las partes invariantes de
norma de las perturbaciones de la métrica.

\subsection*{Perturbaciones de la métrica invariantes de
  norma en el Universo FLRW}

En esta sección procederemos a aplicar el algoritmo recién
dado (sección \ref{sec:algoritmo}) al espacio-tiempo de FLRW
presentado en la sección \ref{sec:flrw}.  Empezando por
descomponer la perturbación lineal de la métrica como en
\eqref{eq:gab_descomposcion_1}.

Analizando las transformaciones de norma
\eqref{eq:transformaciones_norma_escalar_1},
~\eqref{eq:transformaciones_norma_vectorial_1} y
~\eqref{eq:transformacion_hab_tensorial} podemos encontrar
las variables invariantes y variantes de norma a primer
orden en la perturbación de una manera sencilla. De la
ecuación \eqref{eq:transformacion_hab_tensorial} se ve que
la parte transversa y sin traza de la perturbación
tensorial, es invariante de norma, a la cual llamaremos
$\chi^{(1)}_{ij}$, donde el superíndice indica que es la
variable invariante de norma a primer orden,
\begin{equation}\label{eq:modo_tensorial_1}
  \chi^{(1)}_{ij} \equiv h^{(TT)}_{ij}, \quad \chi^{(1)}_{ij} = \chi^{(1)}_{ij},
  \quad \chi^{(1)\,i}\,_i = 0, \quad \chi^{(1)\, ,i}_{ij}= 0,
\end{equation}
podemos ver que $\chi^{(1)}_{ij}$ tiene dos componentes y es
llamado \textit{modo tensorial} en el contexto de las
perturbaciones cosmológicas. Usando ahora
\eqref{eq:transformacion_hab_vectorial} y
\eqref{eq:transformacion_hab_vectorial_tensorial} podemos
definir la variable
\begin{equation}\label{eq:modo_vectorial_1}
    a^2\nu^{(1)}_i \equiv h^{(V)}_i -
    (\partial_\eta-2\Hubble)\left(a^2h^{(VT)}_i\right)   =
    h^{(V)}_i - a^2\partial h^{(VT)}_i. 
\end{equation}
Esta variable también es invariante de norma y se le conoce
como \textit{modo vectorial} y es libre de divergencia,
i.e.\
\begin{equation}
  \nu^{(1)\, ,i }_i = 0,
\end{equation}
esta propiedad nos indica que el modo vectorial,
$\nu^{(1)}_i$, tiene dos variables independientes.

Por último, para obtener los \textit{modos escalares},
podemos definir la variable $\bar{X}_\eta$, a partir de las
ecuaciones \eqref{eq:transformacion_escalar_vectorial} y
\eqref{eq:transformacion_hab_escalar_tensorial} mediante

\begin{equation}\label{eq:bar_x}
    \bar{X}_\eta \equiv h^{(SV)} -
    \frac{1}{2}(\partial_\eta-2\Hubble)(a^2h^{(ST)})
    = h^{(SV)} - \frac{1}{2}a^2\partial_\eta h^{(ST)},
\end{equation}
esta variable transforma de la manera siguiente (i.e.\ es
variante de norma)
\begin{equation}
  _\varphi \bar{X}_\eta - _\psi \bar{X}_\eta = \xi_\eta ,
\end{equation}

Observando a \eqref{eq:transformacion_escalar} es fácil
definir al invariante de norma $\Phi$,

\begin{equation}\label{eq:modo_escalar_a_1}
  -2a^2\Phi^{(1)} \equiv h_{\eta\eta} - 2(\partial_\eta - \Hubble)\bar{X}_\eta.
\end{equation}

Además, de las transformaciones de norma
\eqref{eq:transformacion_hab_escalar},
\eqref{eq:transformacion_hab_escalar_tensorial} y de la
transformación de $\bar{X}_\eta$ definimos al invariante de
norma $\Psi$,
\begin{equation}\label{eq:modo_escalar_b_1}
  -2a^2\Psi^{(1)} \equiv
  a^2\left(h^{(S)}-\frac{1}{3}\nabla^2h^{(ST)}\right) +
  2\Hubble\bar{X}_\eta. 
\end{equation}

Tenemos entonces seis componentes invariantes de norma: 2
escalares $(\Phi^{(1)}, \Psi^{(1)})$, 2 vectoriales
$\nu^{(1)}_i$ y 2 tensoriales $\chi^{(1)}_{ij}$. Como la
parte perturbada de la métrica $h_{ab}$, tiene 10
componentes independientes las cuatro componentes restantes
son la parte variable de norma de la métrica. Para encontrar
sus expresiones, escribamos la descomposición 3+1 de la
métrica en términos de los invariantes de norma
$\left\{\Phi,\Psi, \nu_i, \chi_{ij}\right\}$:

\begin{subequations}
  \label{eq:sistema_1}
  \begin{equation}
    h_{\eta\eta} = -2a^2\Phi^{(1)} + 2(\partial_\eta-\Hubble)\bar{X}_\eta,
  \end{equation}
  \begin{equation}
    h_{\eta\,i} = a^2\nu^{(1)}_i + a^2\partial_\eta h^{(VT)}_i  + h^{(SV)}_{,i},
  \end{equation}
\end{subequations}

\begin{equation}\tag{72c}
    h_{ij} = -2a^2\Psi^{(1)}\delta_{ij} +
    a^2\chi^{(1)}_{ij} + a^2h^{(ST)}_{,i\,,j}  -
    2\Hubble\bar{X}_\eta\delta_{ij} +
    2a^2h^{(VT)}_{(j\,,i)},
\end{equation}

comparando estas ecuaciones con la descomposición en
términos de la separación \eqref{eq:gab_descomposcion_1}
\begin{subequations}
  \label{eq:sistema_2}
  \begin{equation}
    h_{\eta\eta} = \mathcal{H}_{\eta\eta} + 2(\partial_\eta-\Hubble)X_\eta
  \end{equation}
  \begin{equation}
    h_{\eta\,i} = \mathcal{H}_{\eta\,i}+ X_{\eta\,,i} + X_{i\,\eta} - 2\Hubble
    X_i
  \end{equation}
  \begin{equation}
    h_{ij} = \mathcal{H}_{ij}+ 2X_{(i\,,j)}-2\Hubble\delta_{ij}X_\eta
  \end{equation}
\end{subequations}
de aquí se sigue fácilmente que las partes invariantes de
norma de la perturbación a primer orden de la métrica son
  \begin{equation}
    \mathcal{H}_{\eta\eta} \equiv -2a^2\Phi^{(1)}, \quad
    \mathcal{H}_{\eta\,i}\equiv a^2\nu^{(1)}, \quad
    \mathcal{H}_{ij}\equiv -2a^2\Psi^{(1)}\partial_{ij}+a^2\chi^{(1)}_{ij}.
  \end{equation}

Igualando las partes variantes de norma de los conjuntos de
ecuaciones \eqref{eq:sistema_1} y \eqref{eq:sistema_2}
obtenemos las siguientes ecuaciones que hay que resolver
para obtener la parte variable de norma

\begin{subequations}\label{eq:eqs_x}
  \begin{equation}
    \label{eq:1}
    2(\partial_\eta-\Hubble)X_\eta =
    2(\partial_\eta-\Hubble)\bar{X}_\eta 
  \end{equation}
  \begin{equation}
    \label{eq:2}
    X_{i\,\eta} - 2\Hubble X_i    = a^2\partial_\eta h^{(VT)}_i  + h^{(SV)}_{,i}
  \end{equation}
  \begin{equation}
    \label{eq:3}
    \begin{split}
      2X_{(i\,,j)}-2\Hubble\delta_{ij}X_\eta =
      &a^2\chi^{(1)}_{ij} + a^2h^{(ST)}_{,i\,,j} \\ &-
      2\Hubble\bar{X}_\eta\delta_{ij} +
      2a^2h^{(VT)}_{(j\,,i)}
    \end{split}
  \end{equation}
\end{subequations}

De la ecuación \eqref{eq:1} obtenemos $X_\eta = \bar{X}_\eta
+ a \bar{C}_\eta$, donde $\bar{C}_\eta$ es una función
escalar que satisface $\partial_\eta \bar{C}_\eta =
0$. Usando esta solución, \eqref{eq:bar_x} en \eqref{eq:2}

\begin{equation}
  \label{eq:4}
  X_i = a^2\left(h^{(VT)}_i +
    \frac{1}{2}h^{(ST)}_{,i}\right) - \partial_i
  \bar{C}_\eta a^2 \int \frac{d\eta}{a} + a^2 \bar{C}_i,
\end{equation}
con $\partial\bar{C}_i = 0$. La última ecuación \eqref{eq:3}
da la siguiente relación de constricción

\begin{equation}
  \label{eq:5}
  a^2 \partial_{(i}\bar{C}_{j)} - \partial_i \partial_j
  \bar{C}_\eta a^2 \int\frac{d\eta}{a} - \HubbleComovil
  \gamma_{ij} a \bar{C}_\eta = 0
\end{equation}
definiendo $C_\eta \equiv a\bar{C}_\eta$ y $C_i
\equiv \partial_i \bar{C}_\eta a^2 \int \frac{d\eta}{a} +
a^2 \bar{C}_i$, podemos construir el vector $C_a \equiv
C_\eta(\eta) (d\eta)_a + C_i (dx^i)_a$, el cual es un vector
de Killing en el espacio-tiempo de fondo $\mathcal{M}_0$
como se puede verificar fácilmente. Por lo tanto la parte
variante de la perturbación de la métrica y su relación con
las perturbaciones $h^{(SV)}, h^{(ST)}, h^{(VT)}_i$ es única
salvo el grado de libertad proveniente del campo vectorial
de Killing. Además, como la parte variante contribuye a la
perturbación de la métrica mediante
\eqref{eq:gab_descomposcion_1}, el campo vectorial de
Killing $C_a$ no contribuye a las perturbaciones de la
métrica. Finalmente, para satisfacer la constricción
\eqref{eq:5}, tenemos $\bar{C}_\eta = 0$,
$\partial_{(i}\bar{C}_{j)} = 0$, con lo que $\bar{C}_j$ es
un vector de Killing en $\Sigma(\eta)$ y como se mencionó
antes estos vectores perteneces al \textit{kernel} de
$\triangle$ y quedan fuera del dominio de las
perturbaciones, entonces $\bar{C}_{j} = 0$. Claro que sería
posible incluir este \textit{kernel} en nuestras discusión
al extender el dominio de las perturbaciones, pero esto
queda fuera del alcance de este
artículo\cite{NAKAMURA_2,NakamuraKo07}. Es fácil comprobar que
la parte variante de la norma $X^a$ transforma mediante
\eqref{eq:transformacion_vn_1er}, como era de esperarse.

Las perturbaciones de la métrica a segundo orden se pueden
encontrar siguiendo los mismos argumentos y pasos
algebraicos que para el orden lineal, pero usando la
variable $\hat{\mathcal{L}}_{ab}$, definida por
\eqref{eq:L_g_ab}. Repitiendo todos los pasos con esta
variable (i.e. descomponiendola en su parte variable e
invariante de norma, descomponiendola de manera armónica,
etc) llegamos a esta expresión para la parte invariante de
norma de la perturbación a segundo orden de la métrica
\eqref{eq:gab_descomposicion_2}, $\mathcal{L}_{ab}$,
\begin{equation}\label{eq:L_ab}
  \begin{split}
    \mathcal{L}_{ab} &= -2a^2\Phi^{(2)}(d\eta)_a(d\eta)_b +
    2a^2\nu^{(2)}_i(d\eta)_{(a}(dx^i)_{b)}\\ &\quad +
    a^2\left(-2\Psi^{(2)}\delta_{ij} +
      \chi^{(2)}_{ij}\right)(dx^i)_a(dx^j)_b
  \end{split}
\end{equation}
donde $\nu^{(2)}$ y $\chi^{(2)}_{ij}$ satisfacen las
ecuaciones
\begin{equation}
  \nu^{(2)}_i\,^{,i} = \delta^{ij}\nu^{(2)}_{j\,,i} = 0  , \quad \chi^{(2)}_i\,^i
  = 0,\quad \chi^{(2)}_{ij}\,^{,i} = 0.
\end{equation}
Los invariantes de norma $\Phi^{(2)}, \Psi^{(2)}$ son los
\textit{modos escalares} de la perturbación a segundo orden,
y $\nu^{(2)}_i$ y $\chi^{(2)}_{ij}$ son los modos
\textit{vectoriales} y \textit{tensoriales} a segundo orden
respectivamente.

\subsection{Invariantes de norma de variables geométricas}

Luego de realizar los primeros tres pasos (encontrar las
partes invariantes de norma a primer y segundo orden del
\textit{pull-back} de la métrica del universo físico) del algoritmo
presentado en la sección \ref{sec:algoritmo}, el siguiente
paso (el cuatro) es encontrar las partes invariantes de
norma de las cantidad geométricas de interés.

\subsubsection{Tensor de curvatura}

La definición del tensor de curvatura $\bar{R}_{abc}\,^d$
\cite{Wald84} en el espacio-tiempo físico $(\mathcal{M},
\bar{g}_{ab})$ es
\begin{equation}\label{eq:Riemman}
  \left(\bar{\nabla}_a\bar{\nabla}_b -
    \bar{\nabla}_b\bar{\nabla}_a\right)\bar{\omega}_c =
  \bar\omega_d\bar{R}_{abc}\,^d 
\end{equation}
donde $\bar\nabla_a$ es la derivada covariante compatible
con $\bar{g}_{ab}$ y $\bar\omega_c$ es una uno-forma en el
espacio físico $\mathcal{M}$. Por otra parte, en el
espacio-tiempo de fondo $\mathcal{M}_0$, $R_{abc}\,^d$ esta
definido por
\begin{equation}
  \left(\nabla_a \nabla_b - \nabla_b\nabla_a\right)\omega_c
  = \omega_d R_{abc}\,^d 
\end{equation}
con $\nabla_a$ la derivada compatible con $g_{ab}$ y
$\omega_c$ una una-forma en $\mathcal{M}_0$. Introducimos el
operador $\varphi^*\bar\nabla_a(\varphi^{-1})^*$ en
$\mathcal{M}_0$, que podemos identificar con el
\textit{pull-back} de la derivada $\bar\nabla_a$ en
$\mathcal{M}$. Este operador depende de la elección de norma
$\varphi$.

La propiedad de que este operador derivada sea compatible
con la métrica está dada por
\begin{equation}
  \varphi^*\bar\nabla_a\left(\left(\varphi^{-1}\right)^*\varphi^*
    \bar{g}_{ab}\right)=0 
\end{equation}
donde $\varphi^*\bar{g}_{ab}$ es \textit{pull-back} de la
métrica en $\mathcal{M}$. Dado que el \textit{pull-back}
$\varphi^*\bar\nabla_a(\varphi^{-1})^*$ en $\mathcal{M}_o$
de la $\bar\nabla_a$ en $\mathcal{M}$ es lineal, satisface
la regla de Leibnitz, conmuta con la contracción y es libre
de torsión se puede considerar como operador derivada en
$\mathcal{M}_0$.

Aunque es obvio que los \textit{pull-backs} de las
cantidades geométricas del espacio-tiempo $\mathcal{M}$
dependen de la elección de norma, 
(e.g.\ de $\varphi$), a partir de este momento obviaremos a
$\varphi$ en las fórmulas, es decir, con el fin de no
complicar la notación
expresaremos a $\varphi^*\bar\nabla_a(\varphi^{-1})^*$
simplemente como $\bar\nabla_a$,  $\,^\varphi \bar{R}_{abc}\,^d$,
como $ \bar{R}_{abc}\,^d$ etc.

Como el operador $\bar\nabla_a$
se puede considerar como un operador derivada en
$\mathcal{M}_0$, existe entonces, un campo tensorial $C^a_{\;bc}$ en
$\mathcal{M}_0$ tal que,
\begin{equation}\label{eq:rel_nabla}
  \bar\nabla_a\omega_b = \nabla_a\omega_b - C^c_{\;ab}\omega_c,
\end{equation}
donde $\omega_c$ es una uno-forma arbitraria en
$\mathcal{M}_0$. Usando $\bar\nabla_a\bar{g}_{bc} = 0$ en
$\mathcal{M}$ el $C^a_{\;bc}$ en $\mathcal{M}_0$ está dado
por
\begin{equation}\label{eq:conexion_afin}
  C^c_{\;ab} =
  \frac{1}{2}\bar{g}^{cd}\left(\nabla_a\bar{g}_{db}+
    \nabla_b\overline{g}_{da}-\nabla_d\bar{g}_{ab}\right). 
\end{equation}

Debemos notar que toda la dependencia de la elección de
norma del operador $\bar\nabla_a$ en $\mathcal{M}_0$ está
incluida únicamente en $C^c_{\;ab}$.

Usando la definición del tensor de Riemann,
\eqref{eq:Riemman} y la ecuación \eqref{eq:rel_nabla}, por
un lado tenemos,
\begin{align}
  \bar{R}_{abc}\,^d = \bar\nabla_a\bar\nabla_b -
  \bar\nabla_b\bar\nabla_a &=
  \bar\nabla_a(\nabla_b-C^d_{bc})-\bar\nabla_b(\nabla_a-C^d_{bc})  \nonumber \\
  &=\nabla_a\nabla_b-C^e_{ab}\nabla_e-\bar\nabla_aC^d_{bc} -
  \nabla_b\nabla_a+C^e_{ba}\nabla_e-\bar\nabla_bC^d_{ac},
\end{align}
pero  aplicando la definición de $\bar\nabla$
\eqref{eq:rel_nabla} 
\begin{equation*}
\bar\nabla_aC^d_{bc}\omega_d =
  \nabla_aC^d_{bc}\omega_d - C^d_{ec}C^e_{ab}\omega_d -
  C^d_{be}C^e_{ac}\omega_d ,
\end{equation*}
obtenemos entonces,
\begin{equation}\label{eq:Riemman_dos_variedades}
  \bar{R}_{abc}\,^d = R_{abc}\,^d - 2\nabla_{[a}C^d_{b]c} +
  2C^d_{e[a}C^e_{b]c}.
\end{equation}

Así, hemos obtenido el tensor de curvatura en el
espacio-tiempo físico $\mathcal{M}$, en términos del tensor
de curvatura del espacio-tiempo de fondo
$\mathcal{M}_0$. 

Para obtener la expresión perturbada de la
curvatura $\bar{R}_{abc}\,^d$, es necesario calcular las
expansiones perturbativas de la inversa de la métrica,
$\bar{g}^{ab}$ y $C^c_{ab}$.

\paragraph{Perturbación de la inversa de la métrica}

Expandamos en serie de Taylor la inversa de la métrica
\begin{align*}
  \bar{g}^{ab} &=
  \sum^\infty_{k=0}\frac{\lambda}{k!}\, \delta^{(k)}\, _\varphi\bar{g}^{ab} 
 = g^{ab} + \lambda D^{ab} +
  \frac{\lambda^2}{2} E^{ab} + \mathscr{O}(\lambda^3),
\end{align*}
usando $\bar{g}^{ab}\bar{g}_{bc} = \delta^a_c$ tenemos
\begin{equation*}
  \begin{split}
  \delta^a_c &= \left(g^{ab} +
    \lambda C^{ab} + \frac{\lambda^2}{2} E^{ab}\right)
  \left(g_{bc} + \lambda h_{bc} +
    \frac{\lambda^2}{2} l_{ab}\right) \\
  &=\delta^a_c + \lambda \left(h^a_c +
    C^a_c\right) +
  \frac{\lambda^2}{2}\left(l^a_c+2h_{bc}C^{ab}+D^a_c\right)
\end{split}
\end{equation*}
de aquí se ve que $C^{ab} = -h^{ab}$. En el segundo orden 
\begin{align*}
  D^a_c &= -2h_{bc}\left(-h^{ab}\right) - l^a_c = 2h_{ec}h^{ae}-l^a_c
\end{align*}
obteniendo finalmente $D^{ab} = 2h^{ae}h_c^b - l^{ab}$.

Entonces, las perturbaciones a primer y segundo orden del
inverso de la métrica son

\begin{subequations}\label{eq:inversa_metrica_perturbada}
  \begin{equation} \label{eq:inversa_metrica_perturbada_1}
    \delta\bar{g}^{ab} \equiv -h^{ab}
  \end{equation}
  \begin{equation}
    \label{eq:inversa_metrica_perturbada_2}
    \delta^{(2)}\bar{g}^{ab} \equiv  2h^{ae}h_b^c - l^{ab}
  \end{equation}
\end{subequations}

\paragraph{Perturbaciónes de $C^a_{\;bc}$}

Sustituyendo en \eqref{eq:conexion_afin} las expansiones de
la métrica y la métrica inversa a segundo orden,
Se pueden obtener las expresiones perturbadas de$C^a_{bc}$ a primer y
segundo orden

\begin{subequations}\label{eq:afin_1}
  \begin{equation}
    \label{eq:afin_1_orden}
    \delta C^c_{\;ab} = \nabla_{(a}h^c_{b)} - \frac{1}{2}\nabla^ch_{ab},
  \end{equation}
  \begin{equation}
    \label{eq:afin_2_orden}
    \delta^{(2)}C^c_{\;ab} = \left[\nabla_{(a}l^{\,\;c}_{b)} -
      \frac{1}{2}\nabla^cl_{ab}\right] - 2 
    h^{cd}\left[\nabla_{(a}h_{b)d} - \nabla_dh_{ab}\right]
  \end{equation}
\end{subequations}

Definimos la variable $H_{ab}\,^c[A]$ donde $A = \{h, l\}$,

\begin{equation} \label{eq:H_ab_c_def}
  H_{ab}\,^c[A] \equiv \nabla_{(a}\,A_{b)}\,^c - \frac{1}{2}\nabla^c A_{ab},
\end{equation}
$H_{ab}\,^c[A]$ estará apareciendo repetidamente en los
cálculos que siguen con varios acomodos de índices, que
son definidos como
\begin{align*}
  H_{abc}[A] &\equiv g_{cd}H_{ab}\,^d[A], \\
  H_a\,^{bc}[A] &\equiv g^{bd}H_{ad}\,^c[A], \\
  H_a\,^b\,_c[A] &\equiv g_{cd}H_a\,^{bd}[A].
\end{align*}
escritas en estas nuevas variables las ecuaciones
\eqref{eq:afin_1} son
\begin{subequations}
  \begin{equation}
    \label{eq:afin_1_orden_2}
    \delta C^c_{\;ab} = H_{ab}\,^c[h],
  \end{equation}
  \begin{equation}
    \label{eq:afin_2_orden_2}
    \delta^{(2)}C^c_{\;ab} = H_{ab}\,^c[l] - 2 h^{cd}H_{abd}[h].
  \end{equation}
\end{subequations}

Ahora podemos expandir en serie de Taylor el tensor de
Riemann a segundo orden
\begin{equation}
  \bar{R}_{abc}\,^d = R_{abc}\,^d + \lambda \,
  ^{(1)}R_{abc}\,^d + \frac{\lambda^2}{2}\,^{(2)}R_{abc}\,^d .
\end{equation}
Comparando esta última ecuación con
\eqref{eq:Riemman_dos_variedades}, luego de sustituir
\eqref{eq:afin_1_orden_2} y \eqref{eq:afin_2_orden_2},
\begin{subequations}
  \begin{equation}
    \label{eq:riemann_1_orden}
    \delta\bar{R}_{abc}\,^d = - 2\nabla_{[a}H_{b]c}\,^d[h]
  \end{equation}
  \begin{equation}
    \label{eq:riemann_2_orden}
    \delta^{(2)}\bar{R}_{abc}\,^d = -2\nabla_{[a}H_{b]c}\,^d[l]
    + 4h^{de}\nabla_{[a}H_{b]ce}[h] +
    4H_{c[a}\,^e[h]H_{b]e}\,^d[h]
\end{equation}
\end{subequations}

Para descomponer el tensor de Riemann en sus partes 
invariante y variante de norma debemos escribir primero $H_{ab}^c[*]$
en términos de variables invariantes de norma, esto será
sencillo ya que hemos descompuesto a $h_{ab}$ como
\eqref{eq:gab_descomposcion_1}, 

\begin{align*}
  2H_{abc}[h] &= \left[\nabla_ah_{bc} + \nabla_bh_{ac} -
    \nabla_ch_{ab}\right],
\end{align*}
derivando $h_{ab}$,
\begin{align*}
  \nabla_ah_{bc} &=
  \nabla_a\mathcal{H}_{bc}+\nabla_a\nabla_b X_c +
  \nabla_a\nabla_c X_b \\
  \nabla_bh_{ac} &=
  \nabla_b\mathcal{H}_{ac}+\nabla_b\nabla_a X_c +
  \nabla_b\nabla_c X_c \\
  -\nabla_ch_{ab} &=
  -\nabla_c\mathcal{H}_{ab}-\nabla_c\nabla_a X_b -
  \nabla_c\nabla_b X_a 
\end{align*}
entonces, sumando estas expresiones,
\begin{equation*}
    2H_{abc}[h] = 2\nabla_{(a}\mathcal{H}_{b)c} -
    \nabla_c\mathcal{H}_{ab}+R_{acb}\,^dX_d   
  + R_{bca}\,^dX_x + (\nabla_a\nabla_b +
  \nabla_b\nabla_a)X_c + \nabla_b\nabla_aX_c -
  \nabla_b\nabla_aX_c 
\end{equation*}
en esta última ecuación hemos sumando y restando
    $\nabla_b\nabla_aX_c$. Ahora sumemos y restemos
    $R_{cab}\,^dX_d$
\begin{equation*}
    2H_{abc}[h] = 2\nabla_{(a}\mathcal{H}_{b)c} -
  \nabla_c\mathcal{H}_{ab} +R_{acb}\,^dX_d +
  R_{bca}\,^dX_x + R_{abc}\,^dX_d
  + 2\nabla_b\nabla_aX_c + R_{cab}\,^dX_d -
  R_{cab}\,^dX_d,
\end{equation*}
usando la primera identidad de Bianchi, $R_{a[bcd]} = 0$, 
\begin{equation*}
     2H_{abc}[h] =  2\nabla_{(a}\mathcal{H}_{b)c} - \nabla_c\mathcal{H}_{ab} +
  2\nabla_b\nabla_aX_c + R_{acb}\,^dX_d - R_{cab}\,^dX_d.
\end{equation*}
recordando las simetrías del tensor de Riemann
    $R_{acb}\,^d = -R_{cab}\,^d$, y usando por segunda vez
    la primera identidad de Bianchi, obtenemos finalmente
\begin{align*}  
  H_{abc}[h] &= \nabla_{(a}\mathcal{H}_{b)c} -
  \frac{1}{2}\nabla_c\mathcal{H}_{ab} +
  \nabla_a\nabla_bX_c + R_{bca}\,^dX_d  \\
  &= H_{abc}[\mathcal{H}]+ \nabla_a\nabla_bX_c +
  R_{bca}\,^dX_d
\end{align*}
Derivando esta última expresión obtenemos,
\begin{equation}
  \nabla_bH_{acd}[h] = \nabla_bH_{acd}[\mathcal{H}]
  +\nabla_b\nabla_a\nabla_cX_d 
  +X_e\nabla_bR_{cda}\,^e + R_{cda}\,^e\nabla_bX_e
\end{equation}
y sustituyéndola en \eqref{eq:riemann_1_orden}
\begin{align*}
  \delta\bar{R}_{abcd} &= -2\nabla_{[a}H_{b]cd}[\mathcal{H}] \\
  & \quad + \left(\nabla_b\nabla_a -
    \nabla_a\nabla_b\right)\nabla_cX_d +
  X_e\left(\nabla_bR_{cda}\,^e-\nabla_aR_{cdb}\,^e\right)  \\
  &\quad +  R_{cda}\,^e\nabla_bX_e-R_{cdb}\,^e\nabla_aX_e,  \\
  \intertext{usando
    $(\nabla_a\nabla_b-\nabla_b\nabla_a)Q_{cd} = R_{abc}\,^e
    Q_{ed} + R_{abd}\,^eQ_{ce}$ en esta última ecuación}
   \delta\bar{R}_{abcd} &=
   -2\nabla_{[a}H_{b]cd}[\mathcal{H}] +
   R_{bac}\,^e\nabla_eX_d \\ 
  &\quad + R_{bad}\,^e\nabla_cX_e +
  X_e\left(\nabla_bR_{cda}\,^e-\nabla_aR_{cdb}\,^e\right) +
  R_{cda}\,^e\nabla_bX_e-R_{cdb}\,^e\nabla_aX_e  \\
  &= -2\nabla_{[a}H_{b]cd}[\mathcal{H}] + R_{bac}\,^e\nabla_eX_d + R_{bad}\,^e\nabla_cX_e  \\
  &\quad+ R_{cda}\,^e\nabla_bX_e-R_{cdb}\,^e\nabla_aX_e +
  g^{fe}X_e\left(\nabla_bR_{cdaf}-\nabla_aR_{cdbf}\right),
  \intertext{una vez más aplicamos las simetrías del tensor
    de Riemann,}
    \delta\bar{R}_{abcd} &=
    -2\nabla_{[a}H_{b]cd}[\mathcal{H}] -
    R_{abc}\,^e\nabla_eX_d \\ 
  &\quad+ R_{abed}\nabla_cX^e +
  R_{aecd}\nabla_bX^e-R_{ebcd}\nabla_aX^e +
  X^e\left(\nabla_bR_{cdae}-\nabla_aR_{cdbe}\right) \\
  &= -2\nabla_{[a}H_{b]cd}[\mathcal{H}] +
  g_{hd}\left(R_{ebc}\,^h\nabla_aX^e +
    R_{aec}\,^h\nabla_bX^e + R_{abe}\,^h\nabla_cX^e -
    R_{abc}\,^e\nabla_eX^h\right)\\
  &\quad+ X^e(\nabla_bR_{aecd}+\nabla_aR_{ebcd}),
  \intertext{luego, empleando la segunda identidad de
    Bianchi, $\nabla_{[a}R_{bc]de} = 0$}
    \delta\bar{R}_{abcd} &=  -2\nabla_{[a}H_{b]cd}[\mathcal{H}] \\
  &\quad + g_{hd}\left(R_{ebc}\,^h\nabla_aX^e +
    R_{aec}\,^h\nabla_bX^e + R_{abe}\,^h\nabla_cX^e
    -  R_{abc}\,^e\nabla_eX^h \right) \\
  &\quad+ X^e(-\nabla_aR_{ebcd} -
  \nabla_eR_{bacd}+\nabla_aR_{ebcd}), \intertext{y utilizando
    la definición de la derivada de Lie \eqref{eq:Lie_def}
    sobre un campo tensorial de rango $(1,3)$}
    \delta\bar{R}_{abcd}&=-2\nabla_{[a}H_{b]cd}[\mathcal{H}] + + g_{hd}\left(
    \pounds_XR_{abc}\,^h-X^e\nabla_eR_{abc}\,^h\right) +
  X^e\nabla_eR_{abcd},
\end{align*}
llegamos finalmente a la perturbación lineal del tensor de
Riemann escrita en sus partes invariantes y variantes de norma,
\begin{equation}
  \label{eq:riemann_descomposicion_vin_1}
  \delta\bar{R}_{abc}\,^d =   -2\nabla_{[a}H_{b]c}\,^d[\mathcal{H}] + \pounds_XR_{abc}\,^d.
\end{equation}

Para la perturbación a segundo orden, usamos la variable
definida en \eqref{eq:L_g_ab}, y calculamos
$H_{ab}\,^c[\hat{\mathcal{L}}]$,

\begin{multline}
  H_{ab}\,^c[\hat{\mathcal{L}}] = H_{ab}\,^c[l] -
  \pounds_X\left(H_{ab}\,^c[h]+H_{ab}\,^c[\mathcal{H}]\right) + \\
  \left(H_{abd}[h]+H_{abd}[\mathcal{H}]\right)\pounds_Xg^{cd}
  -\left(h_c\,^d+\mathcal{H}_c\,^d\right)\left(\nabla_a\nabla_bX^d-R_{eab}\,^dX^e\right),
\end{multline}

despejando $H_{ab}\,^c[l]$ y sustituyendo en
\eqref{eq:riemann_2_orden}, llegamos a

\begin{multline}
  \delta^{(2)}\bar{R}_{abc}\,^d =
  -2\nabla_{[a}H_{b]c}\,^d[\hat{\mathcal{L}}] +
  4H_{[a}\,^{de}[\mathcal{H}]H_{b]ce}[\mathcal{H}] + \\
  2\mathcal{H}_e\,^d\left(\pounds_X R_{abc}\,^e -
    \delta\bar{R}_{abc}\,^e\right) + 2\pounds_X\left(
    \delta\bar{R}_{abc}\,^d-\frac{1}{2}\pounds_XR_{abc}\,^d\right),
\end{multline}
usando las fórmulas \eqref{eq:L_descomposicion} y
\eqref{eq:riemann_1_orden},
\begin{equation}
  \label{eq:riemann_descomposicion_vin_2}
  \delta^{(2)}\bar{R}_{abc}\,^d =
  -2\nabla_{[a}H_{b]c}\,^d[\mathcal{L}] +
  4H_{[a}\,^{de}[\mathcal{H}]H_{b]ce}[\mathcal{H}] + 
  4\mathcal{H}_e\,^d\nabla{[a}H_{b]c}\,^e[\mathcal{H}] +
  2\pounds_X ^{(1)}\bar{R}_{abc}\,^d + \left(\pounds_Y -
    \pounds^2_X\right) R_{abc}\,^d.
\end{equation}

Usando \eqref{eq:riemann_descomposicion_vin_1} y
\eqref{eq:riemann_descomposicion_vin_2} podemos definir los
tensores de curvatura invariantes de norma de la siguiente
manera

\begin{subequations}
  \begin{equation} \label{eq:riemann_vin_1}
    \bm{\delta}\mathbf{R}_{abc}\,^d \equiv
    \,\delta\bar{R}_{abc}\,^d - \pounds_X R_{abc}\,^d
  \end{equation}
  \begin{equation}
    \label{eq:riemann_vin_2}
    \bm{\delta}^{(2)}\mathbf{R}_{abc}\,^d \equiv
    \delta^{(2)}\bar{R}_{abc}\,^d - 2\pounds_X 
    \delta\bar{R}_{abc}\,^d - \left(\pounds_Y -
      \pounds^2_X\right)R_{abc}\,^d. 
  \end{equation}

\end{subequations}

\subsubsection{Perturbaciónes del tensor de Ricci}

Contrayendo los índices $b$ y $d$ en las ecuaciones
\eqref{eq:riemann_descomposicion_vin_1} y
\eqref{eq:riemann_descomposicion_vin_2}, podemos derivar las
fórmulas perturbativas de la curvatura de Ricci,
\begin{subequations}
  \begin{equation}
    \label{eq:ricci_1_orden}
    \delta\bar{R}_{ab} =
    -2\nabla_{[a}H_{c]b}\,^c\left[\mathcal{H}\right] +
    \pounds_XR_{ab} 
  \end{equation}
  \begin{equation}
    \label{eq:ricci_2_orden}
    \delta^{(2)}\bar{R}_{ab} =
    -2\nabla_{[a}H_{c]b}\,^c[\mathcal{L}] +
    4H_{[a}\,^{cd}[\mathcal{H}]H_{c]bd}[\mathcal{H}] +  
    4\mathcal{H}_d\,^c\nabla_{[a}H_{b]c}\,^d[\mathcal{H}] +
    2\pounds_X \delta\bar{R}_{ab} + \left(\pounds_Y -
      \pounds^2_X\right) R_{ab},
  \end{equation}
\end{subequations}
de estas ecuaciones definimos los invariantes de
norma del tensor de Ricci:
\begin{subequations}
  \begin{equation}
    \label{eq:ricci_1_vin}
    \bm{\delta}\mathbf{R}_{ab}=\delta\bar{R}_{ab} - \pounds_XR_{ab}
  \end{equation}
  \begin{equation}
    \label{eq:ricci_2_vin}
    \bm{\delta}^{(2)}\mathbf{R}_{ab}=\delta^{(2)}\bar{R}_{ab} -
    2\pounds_X \delta\bar{R}_{ab} - \left(\pounds_Y - \pounds^2_X\right)
    R_{ab}.
  \end{equation}
\end{subequations}

\subsubsection{Perturbaciones del escalar de Ricci}

El escalar de Ricci en el espacio-tiempo físico
$\mathcal{M}$ está dado por

\begin{equation}
  \label{eq:def_escalar_ricci}
  \bar\Ricci\equiv\bar{g}^{ab}\bar{R}_{ab}
\end{equation}
Expandiendo a primer orden ambas cantidades, usando
\eqref{eq:inversa_metrica_perturbada_1} y
\eqref{eq:ricci_1_orden}
\begin{equation}
  \bar{\Ricci} = \Ricci + \lambda^{(1)}\bar{\Ricci}= \left(g^{ab} - \lambda
    h^{ab}\right)\left(R_{ab}+\lambda\pounds_XR_{ab}-2\lambda\nabla_{[a}H_{c]b}\,^c[\mathcal{H}]\right)
\end{equation}

Reconociendo términos a primer orden
\begin{align}
  \delta\bar{\Ricci} &= g^{ab}\pounds_XR_{ab} -
  h^{ab}R_{ab} - 2\nabla_{[a}H_{c]}\,^{ac}[\mathcal{H}]
  \intertext{usando la regla de Leibnitz de la derivada de
    Lie,}   \delta\bar{\Ricci} &=\pounds_X\left(g^{ab}R_{ab}\right) -
  R_{ab}\pounds_X g^{ab} - h^{ab}R_{ab} -
  2\nabla_{[a}H_{c]}\,^{ac}[\mathcal{H}]
\end{align}
calculando la derivada de Lie de la métrica y
    usando la descomposición de $h^{ab}$
    \eqref{eq:gab_descomposcion_1}, llegamos a
\begin{equation}
  \label{eq:escalar_ricci_1}
  \delta\bar{\Ricci} =   \pounds_X \Ricci   -
  \mathcal{H}^{ab}R_{ab}   -
  2\nabla_{[a}H_{c]}\,^{ac}[\mathcal{H}] 
\end{equation}

Similarmente, usando \eqref{eq:inversa_metrica_perturbada_1},
\eqref{eq:inversa_metrica_perturbada_2},
\eqref{eq:ricci_1_orden} y \eqref{eq:ricci_2_orden} llegamos
a 
\begin{multline}
  \label{eq:escalar_ricci_2}
  \delta^{(2)}\bar{\Ricci} = -
  2\nabla_{[a}H_{c]}\,^{ac}[\mathcal{L}] +
  R^{ab}\left(2\mathcal{H}_{ca}\mathcal{H}_b\,^c-\mathcal{L}_{ab}\right) \\
  + 4 H_{[a}\,^{cd}[\mathcal{H}]H_{c]}\,^a\,_d[\mathcal{H}]
  +
  4\mathcal{H}_c\,^b\nabla_{[a}H_{b]}\,^{ac}[\mathcal{H}] \\
  + 4\mathcal{H}^{ab}\nabla_{[a}H_{d]}b\,^d[\mathcal{H}] 
  + 2 \pounds_X \delta\bar{\Ricci} + \left(\pounds_Y
    - \pounds^2_X\right)\Ricci,
\end{multline}
que es la perturbación a segundo orden del escalar de
Ricci. La definición de los escalares de Ricci invariantes
de norma es trivial:
\begin{subequations}
  \begin{equation}
    \label{eq:escalar_ricci_vin_1}
    \bm{\delta}\mathbf{R} \equiv \delta\bar{\Ricci}  -  \pounds_X \Ricci
  \end{equation}
  \begin{equation}
    \label{eq:escalar_ricci_vin_2}
    \bm{\delta}^{(2)}\mathbf{R} \equiv \delta^{(2)}\bar{\Ricci} - 2 \pounds_X
    \delta\bar{\Ricci}  - \left(\pounds_Y - \pounds^2_X\right)\Ricci.
  \end{equation}
\end{subequations}

\subsubsection{Perturbaciónes del tensor de Einstein}
La perturbación a primer orden el tensor de Einstein es
\begin{equation}
  \label{eq:einstein_1}
  \delta\bar{G}_{ab} =
  -2\nabla_{[a}H_{d]b}\,^d\left[\mathcal{H}\right] +
  g_{ab}\nabla_{[c}H_{d]}\,^{cd}\left[\mathcal{H}\right] 
  - \frac{1}{2}\mathcal{H}_{ab}\mathcal{R} +
  \frac{1}{2}g_{ab}R_{cd}\mathcal{H}^{cd} + \pounds_XG_{ab}
\end{equation}
y a segundo orden
\begin{multline}
  \label{eq:einstein_2}
  \delta^{(2)}\bar{G}_{ab} = -2\nabla_{[a}H_{c]b}\,^c
  \left[\mathcal{L}\right] +
  4H_{[a}^{cd}[\mathcal{H}]H_{c]bd}[\mathcal{H}]
  + 4\mathcal{L}_c\,^d\nabla_{[a}H_{d]b}\,^c[\mathcal{H}] \\
  -
  \frac{1}{2}g_{ab}\Bigg(-2\nabla_{[c}H_{d]}\,^{cd}[\mathcal{L}]
  +R_{de}\left(2\mathcal{H}_c\,^d\mathcal{H}^{ec}-\mathcal{L}^{de}\right) \\
  + 4H_{[c}\,^{de}[\mathcal{H}]H_{d]}\,^c\,_e[\mathcal{H}] +
  4\mathcal{H}_e\,^d\nabla_{[c}H_{d]}\,^{ce}[\mathcal{H}] +
  4\mathcal{H}^{ce}\nabla_{[c}H_{d]e}\,^d[\mathcal{H}]\Bigg) \\
  +2\mathcal{H}_{ab}\nabla_{[c}H_{d]}\,^{cd}[\mathcal{H}]+
  \mathcal{H}_{ab}R_{cd}\mathcal{H}^{cd}
  - \frac{1}{2}\mathcal{L}_{ab}\mathcal{R} \\
  + 2\pounds_X\delta\bar{G}_{ab} +
  \left(\pounds_Y-\pounds^2_X\right)G_{ab}.
\end{multline}

Es conveniente usar a  $\bar{G}_a\,^b \equiv   \bar{g}^{bc}\bar{G}_{ac}$, en
lugar de $\bar{G}_{ab}$. Entonces levantamos el índice en las ecuaciones
\eqref{eq:einstein_1} y \eqref{eq:einstein_2} con ayuda de
\eqref{eq:inversa_metrica_perturbada}:
\begin{subequations} 
\begin{equation}
  \delta\bar{G}_a\,^b = \bm{\delta}\mathcal{G}_a\,^b[\mathcal{H}]
  + \pounds_X G_a\,^b ,
\end{equation}
\begin{equation}
  \delta^{(2)}\bar{G}_a\,^b = \bm{\delta}\mathcal{G}_a\,^b[\mathcal{L}] +
  \bm{\delta}^{(2)}\mathcal{G}_a\,^b[\mathcal{H},\mathcal{H}] +
  2\pounds_X\delta\bar{G}_a\,^b +
  \left(\pounds_Y-\pounds^2_X\right)G_a\,^b ,
\end{equation}
\end{subequations}
donde se introdujeron las variables
\begin{align}
  \bm{\delta}\mathcal{G}[A]_a\,^b &\equiv ^{(1)}\Sigma_a\,^b[A] -
  \frac{1}{2}\delta_a^b
  \,^{(1)}\Sigma_c^c[A],     \label{eq:einstein_vin_1} \\
  \bm{\delta}^{(2)}\mathcal{G}[A]_a\,^b &\equiv ^{(2)}\Sigma_a\,^b[A,B] -
  \frac{1}{2}\delta_a^b
  \,^{(2)}\Sigma_c^c[A,B], \label{eq:einstein_vin_2}
\end{align}
\begin{equation}
  ^{(1)}\Sigma_a\,^b[A] \equiv -2\nabla_{[a}H_{d]}\,^{bd}[A] -
  A^{cb}R_{ac} +
  \frac{1}{2}\left(2\nabla_{[e}H_{d]}\,^{ed}[A]+
    R_{ed}A^{ed}\right),    
\end{equation}
\begin{multline}
^{(2)}\Sigma_a\,^b[A,B] \equiv 2R_{ad}B_c\,^{(b}A^{d)c} +
  2H_{[a}\,^{de}[A]H_{d]}\,^b\,_e[B] +
  2H_{[a}\,^{de}[B]H_{d]}\,^b\,_e[A] \\ +
  2A_e\,^d\nabla_{[a}H_{d]}\,^{be}[B] +
  2B_e\,^d\nabla_{[a}H_{d]}\,^{be}[A]  +
  2A_c\,^b\nabla_{[a}H_{d]}\,^{cd}[B] +
  2B_c\,^b\nabla_{[a}H_{d]}\,^{cd}[A].
\end{multline}
donde  $\bm{\delta}\mathcal{G}[A]_a\,^b$,
$\bm{\delta}^{(2)}\mathcal{G}[A]_a\,^b$ son las partes
invariantes de norma del tensor de Einstein a 
primer y segundo orden, respectivamente.

\subsubsection{Perturbaciónes de la divergencia de un tensor
  $(1,1)$}

Debido a la importancia de la identidad de Bianchi
$\nabla^b \bar{G}_b\,^a = 0$, en esta sección la
perturbación a primer y segundo orden de la divergencia de
un tensor $(1,1)$ se calculará  para un campo
tensorial arbitrario $\bar{T}_a\,^b$.  Luego de obtener las
expresiones generales se mostrará que las identidades de Bianchi se
cumplen orden a orden.

La divergencia de $\bar{T}_a\,^b$ se define como
\begin{equation}
  \bar{\nabla}_a\bar{T}_b\,^a = \nabla_a\bar{T}_b\,^a + C^a\,_{ce} \bar{T}_b\,^c
  - C^c\,_{ba}\bar{T}_c\,^a. 
\end{equation}
Expandiendo  en una serie de Taylor esta última ecuación,
\begin{equation}
  \bar{\nabla}_a\bar{T}_b\,^a = \sum_{k=0}^\infty \frac{\lambda^k}{k!}
  \quad \left[\delta^{(k)}\left(\bar{\nabla}_a\bar{T}_b\,^a\right)\right].
\end{equation}
A orden lineal encontramos
\begin{equation}
  \delta(\bar{\nabla}_a\bar{T}_b\,^a) =
  \nabla_a\bar{T}_b\,^a +
  (H_{ca}\,^a[\mathcal{H}] + \nabla_c\nabla_aX^a)T_b\,^c 
  -(H_{bac}[\mathcal{H}]+\nabla_b\nabla_aX_c +
  R_{abc}\,^eX_e)T^{ca},
\end{equation}
Expandimos ahora a el campo tensorial $\bar{T}_a\,^b$ usando
\eqref{eq:expansion_segundo_orden}. Siguiendo las
definiciones \eqref{eq:def_vin}, definimos las
perturbaciones invariantes de norma, $\mathcal{T}_b\,^a$,
como sigue
\begin{align}
  \bm{\delta}\mathcal{T}_b\,^a &\equiv \delta\bar{T}_b\,^a -
  \pounds_XT_b\,^a,  \label{eq:t_ab_vin_1}\\
  \bm{\delta}^{(2)}\mathcal{T}_b\,^a &\equiv
  \delta^{(2)}\bar{T}_b\,^a -
  2\pounds_X\,\delta\bar{T}_b\,^a -\left(\pounds_Y -
    \pounds^2_X\right)T_b\,^a \label{eq:t_ab_vin_2},
\end{align}
luego, usando \eqref{eq:t_ab_vin_1} y la fórmula del
apéndice \eqref{eq:lie_cov_rango_0_3}, obtenemos la división
de la parte invariante de norma de la divergencia de un
tensor $(1,1)$:
\begin{equation}
  \label{eq:t_ab_1}
  \delta(\bar{\nabla}_a\bar{T}_b\,^a) =
  \nabla_a\bm{\delta}\mathcal{T}_b\,^a +
  (H_{ca}\,^a[\mathcal{H}] + \nabla_c\nabla_aX^a)T_b\,^c 
  -\left(H_{ba}\,^c[\mathcal{H}]T_c\,^a +
    \pounds_X\nabla_aT_b\,^a\right).
\end{equation}

Podemos verificar la validez de la identidad de Bianchi del
tensor de Einstein a orden lineal perturbativo usando la
ecuación \eqref{eq:t_ab_1} con $\bar{T}_b\,^a \to
\bar{G}_b\,^a$:

\begin{equation}
  \label{eq:bianchi_1}
  \delta(\bar{\nabla}_a\bar{G}_b\,^a) =
  \nabla_a\bm{\delta}\mathcal{G}_b\,^a +
  (H_{ca}\,^a[\mathcal{H}] + \nabla_c\nabla_aX^a)G_b\,^c 
  -\left(H_{ba}\,^c[\mathcal{H}]G_c\,^a +
    \pounds_X\nabla_aG_b\,^a\right),
\end{equation}
necesitamos la derivada  de la parte invariante del tensor
de Einstein \eqref{eq:einstein_vin_1} 
\begin{equation}\label{eq:derivada_einstein_vin_1}
  \nabla_a\bm{\delta}\mathcal{G}_b\,^a   = -
  H_{ca}\,^a[\mathcal{H}]G_b\,^c+H_{ba}\,^c[\mathcal{H}]G_c\,^a 
\end{equation}
y sustituyéndola en \eqref{eq:bianchi_1} obtenemos la
divergencia del tensor de Einstein a primer orden
\begin{equation}
  \delta(\bar\nabla_a\bar{G}_b\,^a) = \pounds_X\nabla_aG_b\,^a,
\end{equation}
pero $\nabla_aG_b\,^a = 0$ en un espacio-tiempo arbitrario,
entonces a primer orden $\delta\bar\nabla_aG_b\,^a = 0$.

La perturbación a segundo orden de la divergencia del tensor
$\bar{T}_b\,^a$ se obtiene de manera similar que la
divergencia a orden lineal,
\begin{multline}
  \delta^{(2)}(\bar\nabla_a\bar{T}_b\,^a) =
  \nabla_a\bm{\delta}^{(2)}\mathcal{T}_b\,^a
  -(2H_{cad}[\mathcal{H}]\mathcal{H}^{da}-H_{ca}\,^a[\mathcal{L}])T_b\,^c
  +(2H_{bad}[\mathcal{H}]\mathcal{H}^{dc}-H_{ba}\,^c[\mathcal{L}])T_c\,^a\\
  -2H_{ba}\,^c[\mathcal{H}]\bm{\delta}\mathcal{T}_c\,^a +
  2H_{ca}\,^a[\mathcal{H}]\bm{\delta}\mathcal{T}_b\,^c
  +2\pounds_X(\bar\nabla_a\bar{T}_b\,^a) +
  (\pounds_Y-\pounds^2_X)(\nabla_aT_b\,^b).
\end{multline}

Con esta expresión comprobaremos la identidad de Bianchi a
segundo orden del tensor perturbado de Einstein
\begin{multline}
  \delta^{(2)}(\bar\nabla_a\bar{G}_b\,^a) =
  \nabla_a\left(\bm{\delta}\mathcal{G}_b\,^a[\mathcal{L}] +
    \bm{\delta}^{(2)}\mathcal{G}_b\,^a[\mathcal{H},\mathcal{H}]\right)
  -\left(2H_{cad}[\mathcal{H}]\mathcal{H}^{da}-
    H_{ca}\,^a[\mathcal{L}]\right)G_b\,^c
  + \\ \left(2H_{bad}[\mathcal{H}]\mathcal{H}^{dc} -
    H_{ba}\,^c[\mathcal{L}]\right)G_c\,^a 
  -2H_{ba}\,^c[\mathcal{H}]\bm{\delta}\mathcal{G}_c\,^a[\mathcal{H}]
  + \\
  2H_{ca}\,^a[\mathcal{H}]\bm{\delta}\mathcal{G}_b\,^c[\mathcal{H}]
  + 2\pounds_X(\bar\nabla_a\bar{G}_b\,^a) +
  (\pounds_Y-\pounds^2_X)(\nabla_aG_b\,^a).
\end{multline}
Usando la ecuación \eqref{eq:derivada_einstein_vin_1} con el
reemplazo $\mathcal{H} \to \mathcal{L}$ y calculando la
divergencia del invariante de norma a segundo orden
\begin{equation}
  \label{eq:derivada_einstein_vin_2}
  \nabla_a\bm{\delta}^{(2)}\mathcal{G}_b\,^a[\mathcal{H},\mathcal{H}]
  =
  -2H_{ca}\,^a[\mathcal{H}]\bm{\delta}\mathcal{G}_b\,^c[\mathcal{H}]
  + 2H_{ba}\,^e[\mathcal{H}]\bm{\delta}\mathcal{G}_e\,^a[\mathcal{H}] 
  -2H_{bad}[\mathcal{H}]\mathcal{H}^{dc}G_c\,^a
  -2H_{cad}[\mathcal{H}]\mathcal{H}^{ad}G_b\,^c,
\end{equation}
observamos que la identidad de Bianchi se cumple a segundo
orden también ya que $\bar\nabla_a\bar{G}_b\,^a = 0$ y
$\nabla_aG_b\,^a = 0$.

\subsubsection{Ecuaciones de Einstein perturbadas invariantes de norma}
Por último escribiremos las ecuaciones de Einstein
perturbadas $\delta^{(k)}G_a\,^b = 8\pi
G\,\delta^{(k)}T_a\,^b$. Usando las definiciones \eqref{eq:t_ab_vin_1} y
\eqref{eq:t_ab_vin_2} se puede mostrar que las ecuaciones de
Einstein se pueden expresar orden  por
orden .en términos de variables invariantes de
norma únicamente, 
\begin{subequations}
\begin{equation}
  \label{eq:eq_einstein_vin_1}
   \bm{\delta}\mathcal{G}_a\,^b[\mathcal{H}] = 8\pi G \,
    \bm{\delta}\mathcal{T}_a\,^b ,
\end{equation}
\begin{equation}
  \label{eq:eq_einstein_vin_2}
  \bm{\delta}\mathcal{G}_a\,^b[\mathcal{L}] +
   \bm{\delta}^{(2)}\mathcal{G}_a\,^b[\mathcal{H},\mathcal{H}] = 8\pi
  G\,  \bm{\delta}^{(2)}\mathcal{T}_a\,^b. 
\end{equation}
\end{subequations}

\subsection{Ecuaciones de Einstein invariantes de norma en
  el Universo FLRW} 

En esta sección aplicaremos las fórmulas recién deducidas al
espacio-tiempo de ejemplo \ref{sec:flrw}.

\paragraph{Tensor de energía-momento de un campo escalar.}
El tensor de energía momento de un campo escalar está dado
por la fórmula \eqref{eq:energia_momento_escalar}, para
obtener la ecuación perturbada debemos expandir al campo
escalar en series de Taylor
\begin{equation}\label{eq:varphi_taylor}
  \bar{\varphi} = \varphi + \lambda\delta\varphi +
  \frac{1}{2}\lambda^2\delta^{(2)}\varphi +
  \mathcal{O}(\lambda^3) ,
\end{equation}
donde $\varphi \equiv \varphi(\eta)$ es una función homogénea en
un universo homogéneo e isotrópico. El tensor de energía
momento también debe de ser descompuesto en la variadad de
fondo
\begin{equation}\label{eq:tab_escalar_taylor}
  \bar{T}_a\,^b = T_a\,^b + \lambda\delta T_a\,^b +
  \frac{1}{2}\lambda^2\delta^{(2)}T_a\,^b + \mathcal{O}(\lambda^3),
\end{equation}
donde $\delta T_a\,^b$ es lineal en $\delta\varphi$ y
$h_{ab}$ y $\delta^{(2)}T_a\,^b$ incluye las perturbaciones de
segundo orden $\delta^{(2)}\varphi$, $l_{ab}$ y los productos
cuadráticos de $\delta\varphi$ y $h_{ab}$. Sustituyendo
\eqref{eq:varphi_taylor} en \eqref{eq:tab_escalar_taylor} y
preservando sólo los términos hasta orden cuadrático, 
\begin{subequations}\label{eq:T_ab_escalar}
  \begin{equation}\label{eq:T_ab_1_escalar}
    \begin{split}
      \delta T_a\,^b&\equiv\nabla_a\varphi\nabla^b\delta\varphi
      -\nabla_a\varphi\,
      h^{bc}\nabla_c\varphi +  \nabla_a\delta\varphi\nabla^b\varphi  \\
      &\quad -\frac{1}{2}\delta_a\,^b\left(
        \nabla_c\varphi\nabla^c\delta\varphi -
        \nabla_c\varphi\, h^{dc}\nabla_d\varphi +
        \nabla_c\delta\varphi\nabla^c\varphi +
        2\delta\varphi\frac{\partial
          V}{\partial\varphi}\right),
    \end{split}
  \end{equation}
  \begin{equation}\label{eq:T_ab_2_escalar}
    \begin{split}
      \delta^{(2)}T_a\,^b&\equiv\nabla_a\varphi\nabla^b\delta^{(2)}\varphi -
      2\nabla_a\varphi\, h^{bc}\nabla_c\delta\varphi +
      \nabla_a\varphi \left(2h^{bd}h_d\,^c -
        l^{bc}\right)\nabla_c\varphi  \\
      &\quad +
      2\nabla_a\delta\varphi\nabla^b
      \delta\varphi-2\nabla_a\delta\varphi  
      h^{bc}\nabla_c\varphi
      + \nabla_a\delta^{(2)}\varphi g^{bc}\nabla_c\varphi \\
      &\quad
      -\frac{1}{2}\delta_a\,^b\Bigg(\nabla_c\varphi
      \nabla^c\delta^{(2)}\varphi 
      - 2\nabla_c\varphi\,h^{dc}\nabla_d\delta\varphi +
      \nabla_c\varphi\left(2h^{de}h_e\,^c -l^{dc}\right) \nabla_d\varphi \\
      &\quad +2\nabla_c\delta\varphi\nabla^c\delta\varphi -
      2\nabla_c\delta\varphi h^{dc}\nabla_d\varphi +
      \nabla_c\delta^{(2)}\varphi\nabla^c\varphi \\
      &\quad +2\delta^{(2)}\varphi\frac{\partial
        V}{\partial\varphi} +
      2(\delta\varphi)^2\frac{\partial^2V}{\partial\varphi^2}\Bigg).
    \end{split}
  \end{equation}
\end{subequations}

Siguiendo \eqref{eq:def_vin}, se puede
descomponer las perturbaciones del campo en su parte
invariante y variable de norma:
\begin{align}
  \delta\varphi&=:\bm\varphi^{(1)}+\pounds_X\varphi,\\
  \delta^{(2)}\varphi&=:\bm\varphi^{(2)}+2\pounds_X\delta\varphi +
  (\pounds_Y-\pounds_X^2)\varphi  ,
\end{align}
donde $\bm\varphi^{(1)}$ y $\bm\varphi^{(2)}$ son
las partes invariantes de norma a primer y segundo orden respectivamente.

Sustituyendo estas ecuaciones en las perturbaciones del
tensor de energía-momento \eqref{eq:T_ab_escalar},
y usando las descomposiciones de la 
métrica \eqref{eq:gab_descomposcion_1} y
\eqref{eq:gab_descomposicion_2}, llegamos a

\begin{subequations}
  \begin{equation}
    \begin{split}
      \bm{\delta}\mathcal{T}_a\,^b &=
      \nabla_a\varphi\nabla^b\bm\varphi^{(1)} -
      \nabla_a\varphi\mathcal{H}^{bc}\nabla_c\varphi +
      \nabla_a\bm\varphi^{(1)}\nabla^b\varphi + \\
      &\quad-\frac{1}{2}\delta_a\,^b
      \left(\nabla_c\varphi\nabla^c\bm\varphi^{(1)}  
        - \nabla_c\varphi\mathcal{H}^{dc}\nabla_d\varphi +
        \nabla_c\bm\varphi^{(1)}\nabla^c\varphi +
        2\bm\varphi^{(1)}\frac{\partial V}{\partial\varphi} \right),
    \end{split}
  \end{equation}
  \begin{multline}
    \bm{\delta}^{(2)}\mathcal{T}_a\,^b =
    \nabla_a\varphi\nabla^b\bm\varphi^{(2)} -
    2\nabla_a\varphi\mathcal{H}^{bc}\nabla_c\bm\varphi^{(1)} +
    2\nabla_a\varphi\mathcal{H}^{bd}\mathcal{H}_{dc}\nabla^c\varphi
    -
    \nabla_a\varphi g^{bd}\mathcal{L}_{dc}\nabla^c\varphi \\
    +2\nabla_a\bm\varphi^{(1)}\nabla^b\bm\varphi^{(1)} -
    2\nabla_a\bm\varphi^{(1)}\nabla^b\bm\varphi^{(1)} -
    2\nabla_a\bm\varphi^{(1)}\mathcal{H}^{bc}\nabla_c\varphi +
    \nabla_a\bm\varphi^{(2)}\nabla^b\varphi \\
    -\frac{1}{2}\Bigg(\nabla_c\varphi\nabla^c\bm\varphi^{(2)}-2\nabla_c\varphi\mathcal{H}^{dc}\nabla_d\bm\varphi^{(1)}
    +
    2\nabla^c\varphi\mathcal{H}^{de}\mathcal{H}_{ec}\nabla_d\varphi
    -
    \nabla^c\varphi \mathcal{L}_{dc}\nabla^d\varphi \\
    + 2\nabla_c\bm\varphi^{(1)}\nabla^c\bm\varphi^{(1)} -
    2\nabla_c\bm\varphi^{(1)}\mathcal{H}^{dc}\nabla_d\varphi +
    \nabla_c\bm\varphi^{(2)}\nabla^c\varphi 
    + 2\bm\varphi^{(2)}\frac{\partial V}{\partial\varphi} +
    2(\bm\varphi^{(1)})^2\frac{\partial^2V}{\partial\varphi^2}\Bigg).
  \end{multline}
\end{subequations}

Entonces, las componentes del tensor de energía-momento
invariantes de norma del campo escalar a primer orden son:
\begin{subequations}\label{eq:tab_vin_escalar_1er}
\begin{equation}
  \bm{\delta}\mathcal{T}_\eta\,^\eta =
  -\frac{1}{a^2}\left(\partial_\eta\bm\varphi^{(1)}\partial_\eta\varphi
    -
    \Phi^{(1)}(\partial_\eta\varphi)^2 +
    a^2\frac{dV}{d\varphi}\bm\varphi^{(1)}\right),
\end{equation}
\begin{equation}
\bm{\delta}\mathcal{T}_i\,^\eta  =
-\frac{1}{a^2}\partial_i\bm\varphi^{(1)}\partial_\eta\varphi,
\end{equation}
\begin{equation}
\bm{\delta}\mathcal{T}_\eta\,^i =
  \frac{1}{a^2}\partial_\eta\varphi\left(\partial^i\bm\varphi^{(1)}
    + (\partial_\eta\varphi)\nu^i\,^{(1)}\right),
\end{equation}
\begin{equation}
\bm{\delta}\mathcal{T}_i\,^j =
  \frac{1}{a^2}\delta_i\,^j\left(\partial_\eta\bm\varphi^{(1)}\partial_\eta\varphi
    - \Phi^{(1)}(\partial_\eta\varphi)^2 -
    a^2\frac{dV}{d\varphi}\bm\varphi^{(1)}\right),
\end{equation}
\end{subequations}
y a segundo orden,
\begin{subequations}\label{eq:tab_vin_escalar_2do}
\begin{multline}
    \bm{\delta}^{(2)}\mathcal{T}_\eta\,^\eta =
    -\frac{1}{a^2}\Bigg(\partial_\eta\varphi\partial_\eta\bm\varphi^{(2)}
    - 4\partial_\eta\varphi\Phi^{(1)}\partial_\eta\bm\varphi^{(1)}
    +
    4(\partial_\eta\varphi)^2\left(\Phi^{(1)}\right)^2 
    -\left(\partial_\eta\varphi\right)^2\Phi^{(2)} +\\
    (\partial_\eta\bm\varphi^{(1)})^2
    + \partial_i\bm\varphi^{(1)}\partial^i\bm\varphi^{(1)} +
    a^2\bm\varphi^{(2)}\frac{\partial V}{\partial\varphi} +
    a^2(\bm\varphi^{(1)})^2\,\frac{\partial^2
      V}{\partial\varphi^2}\Bigg),
  \end{multline}
  \begin{equation}
    \bm{\delta}^{(2)}\mathcal{T}_i\,^\eta
    =-\frac{1}{a^2}\Bigg(\partial_i\bm\varphi^{(2)}\partial_\eta\varphi +
    2\partial_i\bm\varphi^{(1)}\partial_\eta\bm\varphi^{(1)} - 4\partial_i\bm\varphi^{(1)}\Phi^{(1)}\partial_\eta\varphi\Bigg),
  \end{equation}
  \begin{equation}
    \bm{\delta}^{(2)}\mathcal{T}_\eta\,^i
    =\frac{1}{a^2}\Bigg(\partial_\eta\varphi\partial^i\bm\varphi^{(2)} +
    4\partial_\eta\Phi^{(1)}\partial^i\bm\varphi^{(1)} +
    \left(\partial_\eta\varphi\right)^2\nu^i\,^{(2)} + 2\partial_\eta\bm\varphi^{(1)}\partial^i\bm\varphi^{(1)}\Bigg),
  \end{equation}
  \begin{multline}
    \bm{\delta}^{(2)}\mathcal{T}_i\,^j =
    2\frac{1}{a^2}\partial_i\bm\varphi^{(1)}\partial^j\bm\varphi^{(1)} \\
    +
    \frac{1}{a^2}\delta_i\,^j\Bigg(\partial_\eta\varphi\partial_\eta\bm\varphi^{(2)}
    - 4\partial_\eta\varphi\Phi^{(1)}\partial_\eta\bm\varphi^{(1)}
    +
    4\left(\partial_\eta\varphi\right)^2\left(\Phi^{(1)}\right)^2
    -
    \left(\partial_\eta\varphi\right)^2\Phi^{(2)} \\
    +\left(\partial_\eta\bm\varphi^{(1)}\right)^2
    - \partial_k\bm\varphi^{(1)}\partial^k\bm\varphi^{(1)}
    -a^2\bm\varphi^{(2)}\frac{\partial V}{\partial\varphi} -
    a^2(\bm\varphi^{(1)})^2\frac{\partial^2
      V}{\partial\varphi^2}\Bigg).
  \end{multline}
\end{subequations}

\subsubsection{Ecuaciones de Einstein a primer orden}
Antes de poder escribir las componentes de la  ecuación
\eqref{eq:einstein_vin_1} a primer orden debemos
expresar las componentes de $H_{ab}\,^c$ dadas por la expresión
\eqref{eq:H_ab_c_def}:

\begin{align} \label{eq:H_ab_c_FLRW}
  H_{\eta\,\eta}\,^\eta[\mathcal{H}] &= \partial_\eta \Phi^{(1)} \\
  H_{i\,\eta}\,^\eta[\mathcal{H}] &= \partial_i\Phi^{(1)} + \Hubble\nu_i^{(1)} \\
  H_{i\,j}\,^\eta[\mathcal{H}]
  &=-\left(2\Hubble\left(\Psi^{(1)}+\Phi^{(1)}\right)+\partial_\eta\Psi^{(1)}\right)\delta_{ij}
  - \partial_{(i}\nu_{j)}^{(1)} +
  \frac{1}{2}\left(\partial_\eta +
    2\Hubble\right)\chi_{ij}^{(1)} \\
  H_{\eta\,\eta}\,^i[\mathcal{H}] &= \partial^i\Phi^{(1)} +
  \left(\partial_\eta +
    \Hubble\right)\nu^i\,^{(1)} \\
  H_{j\,\eta}\,^i[\mathcal{H}] &=
  -\partial_\eta\Psi^{(1)}\delta_j\,^i +
  \frac{1}{2}\left(\partial_j\nu^i\,^{(1)}
    - \partial^i\nu_j\,^{(1)}\right) +
  \frac{1}{2}\partial_\eta\chi_j\,^i\,^{(1)}\\
  H_{j\,k}\,^i[\mathcal{H}]&=\partial^i\Psi^{(1)}\delta_{kj}
  - 2\delta^i_{(k}\partial_{j)}\Psi^{(1)} -
  \Hubble\delta_{kj}\nu^i\,^{(1)}
  + \partial_{(j}\chi_{k)}\,^i\,^{(1)}
  -\frac{1}{2}\partial^i\chi_{kj}\,^{(1)}
\end{align}
Usando estas ecuaciones y la ecuación
\eqref{eq:einstein_vin_1},  las componentes del tensor de
Einstein invariante de norma a primer orden serán
\begin{subequations}\label{eq:Einstein_1_FLRW}
  \begin{equation}
    ^{(1)}\mathcal{G}_\eta\,^\eta[\mathcal{H}]  =
    -\frac{1}{a^2}\left\{(-6\Hubble\partial_\eta + 2\triangle)\Psi^{(1)} -
      6\Hubble^2\Phi^{(1)}\right\},
  \end{equation}
  \begin{equation}
    ^{(1)}\mathcal{G}_i\,^\eta[\mathcal{H}]  =
    -\frac{1}{a^2}\left(2\partial_\eta\partial_i\Psi^{(1)} +
      2\Hubble\partial_i\Phi^{(1)} -\frac{1}{2}\triangle\nu_i^{(1)}\right),
  \end{equation}
  \begin{equation}
    ^{(1)}\mathcal{G}_\eta\,^i[\mathcal{H}]  =
    \frac{1}{a^2}\left\{2\partial_\eta\partial^i\Psi^{(1)} +
      2\Hubble\partial^i\Phi^{(1)}+ \frac{1}{2}(-\triangle + 4\Hubble^2 -
      4\partial_\eta\Hubble)\nu^i\,^{(1)}\right\},
  \end{equation}
  \begin{multline}
    ^{(1)}\mathcal{G}_i\,^j[\mathcal{H}] =
    \frac{1}{a^2}\Bigg[\partial_i\partial^j
    \left(\Psi^{(1)}-\Phi^{(1)}\right)  \\ 
    +\Bigg\{(-\triangle + 2\partial_\eta^2 +
      4\Hubble\partial_\eta )\Psi^{(1)} + 
      (2\Hubble\partial_\eta + 4\partial_\eta\Hubble +
       2\Hubble^2+\triangle)\Phi^{(1)} \Bigg \}\delta_i\,^j \\
    -\frac{1}{2a^2}\partial_\eta\Bigg\{a^2\Bigg(\partial_i\nu^j\,^{(1)}
    + \partial^j\nu_i^{(1)}\Bigg)\Bigg\} 
    +\frac{1}{2}(\partial_\eta^2 + 2\Hubble\partial_\eta -
    \triangle)\chi_i\,^j\,^{(1)}\Bigg].
  \end{multline}
\end{subequations}

A continuación escribimos las componentes de las ecuaciones
de Einstein linearizadas 
\eqref{eq:eq_einstein_vin_1} de FLRW lleno con un campo
escalar (con perturbaciones lineales dadas por
\eqref{eq:tab_vin_escalar_1er}), la componente $\eta-\eta$
\begin{equation}
  (-3\Hubble\partial_\eta + \triangle)\Psi^{(1)} -
  3\Hubble^2\Phi^{(1)} = 4\pi\,G
  \left(\partial_\eta\bm\varphi^{(1)}\partial_\eta\varphi - 
    \Phi^{(1)}(\partial_\eta\varphi)^2 +
    a^2\frac{dV}{d\varphi}\bm\varphi^{(1)}\right) ,
\end{equation}
las componentes $i-\eta$ y $\eta-i$ son iguales en virtud de
\eqref{eq:einstein_auxiliar}
\begin{equation}
  \partial_\eta\partial_i\Psi^{(1)} +
  \Hubble\partial_i\Phi^{(1)} -\frac{1}{4}\triangle\nu_i^{(1)} =
  4\pi\,G\partial_i\bm\varphi^{(1)}\partial_\eta\varphi,
\end{equation}
y por último la componente espacial-espacial es
\begin{multline}
  \Bigg[\partial_i\partial^j\left(\Psi^{(1)}-\Phi^{(1)}\right) 
  +\left\{(-\triangle + 2\partial_\eta^2 +
    4\Hubble\partial_\eta )\Psi^{(1)} +
    (2\Hubble\partial_\eta + 4\partial_\eta\Hubble +
    2\Hubble^2+\triangle)\Phi^{(1)} \right \}\delta_i\,^j \\
  -\frac{1}{2a^2}\partial_\eta\Bigg\{a^2\Bigg(\partial_i\nu^j\,^{(1)}
  + \partial^j\nu_i^{(1)}\Bigg)\Bigg\} 
  +\frac{1}{2}(\partial_\eta^2 + 2\Hubble\partial_\eta -
  \triangle)\chi_i\,^j\,^{(1)}\Bigg] \\ = 8\pi\,G
  \delta_i\,^j\left(\partial_\eta\bm\varphi^{(1)}\partial_\eta\varphi
    - \Phi^{(1)}(\partial_\eta\varphi)^2 -
    a^2\frac{dV}{d\varphi}\bm\varphi^{(1)}\right).
\end{multline}
Podemos descomponer estas ecuaciones en sus modos escalares,
vectoriales y tensoriales, siguiendo los mismos pasos que en
la sección \ref{sec:pert-prim-orden}. Los modos escalares son
\begin{equation}
  (-3\Hubble\partial_\eta + \triangle)\Psi^{(1)} -
  3\Hubble^2\Phi^{(1)} = 4\pi\,G
  \left(\partial_\eta\bm\varphi^{(1)}\partial_\eta\varphi - 
    \Phi^{(1)}(\partial_\eta\varphi)^2 +
    a^2\frac{dV}{d\varphi}\bm\varphi^{(1)}\right),
\end{equation}
\begin{equation}
  \partial_\eta\partial_i\Psi^{(1)} +
  \Hubble\partial_i\Phi^{(1)} =
  4\pi\,G\partial_i\bm\varphi^{(1)}\partial_\eta\varphi,
\end{equation}
\begin{multline}
  (-\triangle + 2\partial_\eta^2 + 4\Hubble\partial_\eta
  )\Psi^{(1)} + (2\Hubble\partial_\eta +
  4\partial_\eta\Hubble + 2\Hubble^2+\triangle)\Phi^{(1)} \\
  = 8\pi\,G
  \left(\partial_\eta\bm\varphi^{(1)}\partial_\eta\varphi -
    \Phi^{(1)}(\partial_\eta\varphi)^2 -
    a^2\frac{dV}{d\varphi}\bm\varphi^{(1)}\right),
\end{multline}
\begin{equation}
  \partial_i\partial^j\left(\Psi^{(1)}-\Phi^{(1)}\right) = 0,
\end{equation}
de esta última ecuación vemos que
\begin{equation}
  \Psi^{(1)} = \Phi^{(1)}.
\end{equation}
Usando las ecuaciones de Einstein de fondo, integrando con
las apropiadas condiciones de frontera y usando
repetidamente \eqref{eq:einstein_fondo} y 
\eqref{eq:einstein_auxiliar} llegamos a
\begin{align}
  (\triangle - 3\Hubble\partial_\eta -\partial_\eta\Hubble -
  2\Hubble^2)\Phi^{(1)} &=
  4\pi\,G\left(\partial_\eta\bm\varphi^{(1)}\partial\varphi
    + a^2\frac{dV}{d\varphi}\bm\varphi^{(1)}\right), \label{eq:modo_escalar_1}\\
  \partial_\eta\Phi^{(1)}+\Hubble\Phi^{(1)} &=
  4\pi\,G\bm\varphi^{(1)}\partial_\eta\varphi, \label{eq:modo_escalar_2}\\
  (\partial_\eta^2 + 3\Hubble\partial_\eta
  + \partial_\eta\Hubble+2\Hubble^2)\Phi^{(1)} &=
  4\pi\,G\left(\partial_\eta\bm\varphi^{(1)}\partial_\eta\varphi -
    a^2\frac{dV}{d\varphi}\bm\varphi^{(1)}\right), \label{eq:modo_escalar_3}
\end{align}
es importante notar que sólo dos de estas ecuaciones son
independientes.  Para los modos vectoriales tenemos
\begin{align}\label{eq:modo_vectorial_2}
  \triangle\nu^i\,^{(1)} &= 0 \\
  \partial_\eta\Bigg\{a^2\Bigg(\partial_i\nu^j\,^{(1)}
  + \partial^j\nu_i^{(1)}\Bigg)\Bigg\} &= 0,
\end{align}
de donde podemos deducir que el campo escalar \textit{no
  produce} perturbaciones vectoriales si la condición
inicial no los incluye y finalmente, para los modos
tensoriales
\begin{equation}
  (\partial_\eta^2 + 2\Hubble\partial_\eta -
  \triangle)\chi_{ij}\,^{(1)}   = 0.
\end{equation}
Tomando las ecuaciones \eqref{eq:modo_escalar_1} y
\eqref{eq:modo_escalar_3} podemos eliminar el término
potencial del campo escalar
\begin{equation}\label{eq:Phi_1_y_varphi_1}
  (\partial_\eta^2 + \triangle)\Phi^{(1)} =
  8\pi\,G\partial_\eta\bm\varphi^{(1)}\partial_\eta\varphi,
\end{equation}
y si usamos \eqref{eq:modo_escalar_2} para eliminar
$\partial_\eta\bm\varphi^{(1)}$ de esta ecuación llegamos a
\begin{equation}
  \left(\partial_\eta^2 + 2\left(\Hubble -
      \frac{\partial_\eta^2\varphi}{\partial_\eta\varphi}\right)\partial_\eta -
    \triangle +2\left(\partial_\eta\Hubble -
      \Hubble\frac{\partial_\eta^2\varphi}{\partial_\eta\varphi}\right)\right)
  \Phi^{(1)} = 0,
\end{equation}
ecuación que es conocida como la \textit{ecuación maestra}
para la perturbación del modo escalar de las perturbaciones
cosmológicas en un universo lleno con un único campo
escalar.

\subsubsection{Ecuaciones de Einstein a segundo orden}
Para derivar las ecuaciones invariantes de norma de Einstein
a segundo orden, \eqref{eq:eq_einstein_vin_2}, es necesario
contar con $\mathcal{L}_{ab}$, \eqref{eq:L_ab}, y luego
calcular $H_{ab}\,^c[\mathcal{L}]$, \eqref{eq:H_ab_c_def},
$\bm{\delta}\mathcal{G}_a\,^b[\mathcal{L}]$,
\eqref{eq:einstein_vin_1},
$\bm{\delta}^{(2)}\mathcal{G}_a\,^b[\mathcal{H},\mathcal{H}]$,
\eqref{eq:eq_einstein_vin_2}, y por último
$\bm{\delta}^{(2)}\mathcal{T}_a\,^b$ ,
\eqref{eq:tab_vin_escalar_2do}. Todos estos cálculos
algebraicos son largos y tediosos, por eso, se creó una hoja
de cálculo de Maple que se puede descargar en la dirección
URL dada en el apéndice \ref{sec:codigo-maple}.

En el caso cosmológico regularmente se supone que las
condiciones iniciales de las perturbaciones vectoriales y
tensoriales a primer orden son cero, i.e.\
\begin{equation}
  \nu^i\,^{(1)} = 0, \qquad \chi_{ij}\,^{(1)} = 0,
\end{equation}
esto se puede justificar recordando (cf.\
\ref{eq:modo_vectorial_2}) que a primer orden no hay
generación de perturbaciones vectoriales, además, las
perturbaciones vectoriales están suprimidas por un factor de
$a^{-2}$ (segunda ecuación
\eqref{eq:modo_vectorial_2}). Para justificar el desprecio
del modo tensorial a primer orden usaremos el hecho que las
observaciones del CMB indican que la relación escalar-tensor
es mucho menor que la unidad. Esto simplificará mucho
nuestras expresiones. Si se quisieran las expresiones
completas (i.e. sin despreciar a primer orden los modos
vectoriales y tensoriales), la hoja de cálculo de Maple del
apéndice \ref{sec:codigo-maple}, se puede modificar
fácilmente para obtenerlas.

Las expresiones para $H_{ab}\,^c[\mathcal{L}]$ y
$\bm{\delta}\mathcal{G}_a\,^b[\mathcal{L}]$ se obtienen de
\eqref{eq:H_ab_c_FLRW} y de \eqref{eq:Einstein_1_FLRW} con
los siguientes reemplazos:
\begin{equation}
  \Phi^{(1)} \to \Phi^{(2)}, \quad \Psi^{(1)} \to \Psi^{(2)}, \quad
  \nu^i\,^{(1)} \to \nu^i \,^{(2)}, \quad \chi_{ij}\,^{(1)}
  \to \chi_{ij}\,^{(2)}, 
\end{equation}
entonces tenemos para
$\bm{\delta}\mathcal{G}_a\,^b[\mathcal{L}]$
\begin{subequations}\label{eq:Einstein_1L_FLRW}
  \begin{equation}
    \bm{\delta}\mathcal{G}_\eta\,^\eta[\mathcal{L}]  =
    -\frac{1}{a^2}\left\{(-6\Hubble\partial_\eta + 2\triangle)\Psi^{(2)} -
      6\Hubble^2\Phi^{(2)}\right\},
  \end{equation}
  \begin{equation}
    \bm{\delta}\mathcal{G}_i\,^\eta[\mathcal{L}]  =
    -\frac{1}{a^2}\left(2\partial_\eta\partial_i\Psi^{(2)} +
      2\Hubble\partial_i\Phi^{(2)} -\frac{1}{2}\triangle\nu_i^{(2)}\right),
  \end{equation}
  \begin{equation}
    \bm{\delta}\mathcal{G}_\eta\,^i[\mathcal{L}]  =
    \frac{1}{a^2}\left\{2\partial_\eta\partial^i\Psi^{(2)} +
      2\Hubble\partial^i\Phi^{(2)}+ \frac{1}{2}(-\triangle + 4\Hubble^2 -
      4\partial_\eta\Hubble)\nu^i\,^{(2)}\right\},
  \end{equation}
  \begin{multline}
    \bm{\delta}\mathcal{G}_i\,^j[\mathcal{L}] =
    \frac{1}{a^2}\Bigg[\partial_i\partial^j
    \left(\Psi^{(2)}-\Phi^{(2)}\right)\\ 
    +\left\{(-\triangle + 2\partial_\eta^2 +
      4\Hubble\partial_\eta )\Psi^{(2)} +
      (2\Hubble\partial_\eta + 4\partial_\eta\Hubble +
      2\Hubble^2+\triangle)\Phi^{(2)} \right \}\delta_i\,^j \\
    -\frac{1}{2a^2}\partial_\eta\Bigg\{a^2\Bigg(\partial_i\nu^j\,^{(2)}
    + \partial^j\nu_i^{(2)}\Bigg)\Bigg\}
    +\frac{1}{2}(\partial_\eta^2 + 2\Hubble\partial_\eta -
    \triangle)\chi_i\,^j\,^{(2)}\Bigg].
  \end{multline}
\end{subequations}

Las componentes del tensor invariante de norma
$\bm{\delta}^{(2)}\mathcal{G}_a\,^b[\mathcal{H},\mathcal{H}]$
son
\begin{subequations}
  \begin{equation}
    \bm{\delta}^{(2)}\mathcal{G}_\eta\,^\eta
    =-\frac{2}{a^2}\left\{3\partial_k\Phi^{(1)}\partial^k\Phi^{(1)} +
      3(\partial_\eta\Phi^{(1)})^2 + 8\Phi^{(1)}
      \triangle\Phi^{(1)} +
      12\Hubble^2(\Phi^{(1)})^2\right\},
  \end{equation}
  \begin{equation}
    \bm{\delta}^{(2)}\mathcal{G}_i\,^\eta =
    \frac{4}{a^2}\left(4\Hubble\Phi^{(1)}\partial_i\Phi^{(1)}
      - \partial_\eta\Phi^{(1)}\partial_i\Phi^{(1)}\right) ,
  \end{equation}
  \begin{equation}
    \bm{\delta}^{(2)}\mathcal{G}_\eta\,^i
    =\frac{4}{a^2}\left(\partial_\eta\Phi^{(1)}\partial^i\Phi^{(1)} +
      4\Phi^{(1)}\partial_\eta\partial^i\Phi^{(1)}\right),
  \end{equation}
  \begin{multline}
    \bm{\delta}^{(2)}\mathcal{G}_i\,^j = \frac{2}{a^2}\Bigg[
    2\partial_i\Phi^{(1)}\partial^j\Phi^{(1)} +
    4\Phi^{(1)}\partial_i\partial^j\Phi^{(1)} \\
    -\Bigg(3\partial_k\Phi^{(1)}\partial^k\Phi^{(1)} +
    4\Phi^{(1)}\triangle\Phi^{(1)} +
    \left(\partial_\eta\Phi^{(1)}\right)^2 + 8
    \Hubble\Phi^{(1)}\partial_\eta\Phi^{(1)} +
    4(2\partial_\eta\Hubble +
    \Hubble^2)(\Phi^{(1)})^2\Bigg)\delta_i\,^j\Bigg],
  \end{multline}
\end{subequations}
donde se usó $\Psi^{(1)} = \Phi^{(1)}$.  Con las ecuaciones
\eqref{eq:einstein_auxiliar} y \eqref{eq:modo_escalar_2} las
ecuaciones de Einstein $
\bm{\delta}\mathcal{G}_\eta\,^i[\mathcal{L}] +
\bm{\delta}^{(2)}\mathcal{G}_\eta\,^i[\mathcal{H},\mathcal{H}]
= 8\pi G\, \bm{\delta}^{(2)}\mathcal{T}_\eta\,^i$ y
$\bm{\delta}\mathcal{G}_i\,^\eta [\mathcal{L}] +
\bm{\delta}^{(2)}\mathcal{G}_i\,^\eta[\mathcal{H},\mathcal{H}]
= 8\pi G\, \bm{\delta}^{(2)}\mathcal{T}_i\,^\eta$ se reducen
a la ecuación
\begin{equation}\label{eq:Einstein2_A}
  2\partial_\eta\partial_i\Psi^{(2)} + 2\Hubble \partial_i\Phi^{(2)} -
  \frac{1}{2}\triangle\nu_i\,^{(2)} -
  8\pi\,G\partial_i\bm\varphi^{(2)}\partial_\eta\varphi = \Gamma_i,
\end{equation}
donde
\begin{equation}\label{eq:Einstein2_A_Gamma}
  \Gamma_i \equiv -4\partial_\eta\Phi^{(1)}\partial_i\Phi^{(1)} +
  8\Hubble\Phi^{(1)}\partial_i\Phi^{(1)} - 8
  \Phi^{(1)}\partial_\eta\partial_i\Phi^{(1)}+16\pi\,G\partial_i\bm\varphi^{(1)}\partial_\eta\bm\varphi^{(1)}.
\end{equation}
De la misma manera usando \eqref{eq:einstein_auxiliar} y
\eqref{eq:Phi_1_y_varphi_1} pero ahora con $
\bm{\delta}\mathcal{G}_\eta\,^\eta[\mathcal{L}] +
\bm{\delta}^{(2)}\mathcal{G}_\eta\,^\eta[\mathcal{H},\mathcal{H}]
= 8\pi G\, \bm{\delta}^{(2)}\mathcal{T}_\eta\,^\eta$,
obtenemos
\begin{equation}\label{eq:Einstein2_B}
  (-3\Hubble\partial_\eta + \triangle)\Psi^{(2)} + (-\partial_\eta\Hubble -
  2\Hubble^2)\Phi^{(2)}
  -4\pi\,G\left(\partial_\eta\varphi\partial_\eta
    \bm\varphi^{(2)}+a^2\bm\varphi^{(2)}\frac{\partial
      V}{\partial \varphi}\right) = \Gamma_0, 
\end{equation}
donde $\Gamma_0$ es
\begin{multline}\label{eq:Einstein2_B_Gamma}
  \Gamma_0 \equiv -2\Phi^{(1)} \partial^2_\eta\Phi^{(1)} -
  3\left(\partial_\eta\Phi^{(1)}\right)^2 -
  3\partial_k\Phi^{(1)}\partial^k\Phi^{(1)} -
  10\Phi^{(1)}\triangle\Phi^{(1)}
  -4(\partial_\eta\Hubble + 2\Hubble^2)\left(\Phi^{(1)}\right)^2 \\
  +4\pi\,G\left(\left(\partial_\eta\bm\varphi^{(1)}\right)^2
    +
    \partial_k\bm\varphi^{(1)}\partial^k\bm\varphi^{(1)} +
    a^2(\bm\varphi^{(1)})^2\frac{\partial^2
      V}{\partial\varphi^2}\right).
\end{multline}
Por último con \eqref{eq:einstein_auxiliar} y
\eqref{eq:Phi_1_y_varphi_1} pero ahora con $
\bm{\delta}\mathcal{G}_i\,^j[\mathcal{L}] +
\bm{\delta}^{(2)}\mathcal{G}_i\,^j[\mathcal{H},\mathcal{H}]
= 8\pi G\, \bm{\delta}^{(2)}\mathcal{T}_i\,^j$, llegamos a
\begin{multline}\label{eq:Einstein2_C}
  \partial_i\partial_j\left(\Psi^{(2)}-\Phi^{(2)}\right)
  +\Bigg\{(-\triangle + 2\partial_\eta^2 +
  4\Hubble\partial_\eta)\Psi^{(2)} +(2\Hubble\partial_\eta +
  2\partial_\eta\Hubble + 4\Hubble^2 +
  \triangle)\Phi^{(2)}\Bigg\}\delta_{ij} \\
  -\frac{1}{a^2}\partial_\eta\left(a^2\partial_{(i}\nu_{j)}\,^{(2)}\right)
  +\frac{1}{2}(\partial_\eta^2 +
  2\Hubble\partial_\eta-\triangle)\chi_{ij}\,^{(2)}
  -8\pi\,G\left(\partial_\eta\partial_\eta\bm\varphi^{(2)} -
    a^2\bm\varphi^{(2)}\frac{\partial
      V}{\partial\varphi}\right)\delta_{ij} = \Gamma_{ij},
\end{multline}
donde
\begin{multline}\label{eq:Einstein2_C_Gamma}
  \Gamma_{ij}\equiv
  -4\partial_i\Phi^{(1)}\partial_j\Phi^{(1)}
  -8\Phi^{(1)}\partial_i\partial_j\Phi^{(1)} +
  16\pi\,G\partial_i\bm\varphi^{(1)}\partial_j\bm\varphi^{(1)} \\
  +2\Bigg(8\Hubble\Phi^{(1)}\partial_\eta\Phi^{(1)} -
  2\Phi^{(1)}\partial_\eta^2\Phi^{(1)} +
  \left(\partial_\eta\Phi^{(1)}\right)^2 +
  3\partial_k\Phi^{(1)}\partial^k\Phi^{(1)} \\
  + 2\Phi^{(1)}\triangle\Phi^{(1)} + 4(\partial_\eta\Hubble
  + 2\Hubble^2)\left(\Phi^{(1)}\right)^2 +
  4\pi\,G\left((\partial_\eta\bm\varphi^{(1)})^2-\partial_k\bm\varphi^{(1)}\partial^k\bm\varphi^{(1)}-a^2(\bm\varphi^{(1)})^2\frac{\partial^2
      V}{\partial\phi^2}\right)\Bigg)\delta_{ij}.
\end{multline}
Tomando la divergencia de \eqref{eq:Einstein2_A} obtenemos
su parte escalar
\begin{equation}
  \label{eq:Einstein2_A_escalar}
  2\partial_\eta\Psi^{(2)}+2\Hubble\Phi^{(2)} -
  8\pi\,G\bm\varphi^{(2)}\partial_\eta\varphi =
  \triangle^{-1}\partial^k\Gamma_k\,  , 
\end{equation}
restándole esta ecuación a \eqref{eq:Einstein2_A} obtenemos
su parte vectorial
\begin{equation}
  \label{eq:Einstein2_A_vectorial}
  \nu_i\,^{(2)}=2\triangle^{-1}\left\{\partial_i\triangle^{-1}\partial^k\Gamma_k - \Gamma_i\right\}.
\end{equation}
Podemos dividir la ecuación \eqref{eq:Einstein2_C} en la
parte de traza (escalar), sin traza (escalar), vectorial y
tensorial como sigue. Tomando la traza de
\eqref{eq:Einstein2_C} llegamos a
\begin{multline}
  \label{eq:Einstein2_C_escalar_traza}
  \left(\partial_\eta^2+2\Hubble\partial_\eta -
    \frac{1}{3}\triangle\right)\Psi^{(2)} +
  \left(\Hubble\partial_\eta + \partial_\eta\Hubble +
    2\Hubble^2 + \frac{1}{3}\triangle\right)\Phi^{(2)} \\
  -4\pi\,G\left(\partial_\eta\varphi\partial_\eta\bm\varphi^{(2)}
    - a^2\bm\varphi^{(2)}\frac{\partial
      V}{\partial\varphi}\right) =\frac{1}{6}\Gamma_k\,^k,
\end{multline}
donde $\Gamma_k\,^k \equiv \delta^{ij}\Gamma_{ij}$. La parte
sin traza de \eqref{eq:Einstein2_C} es
\begin{equation}
  \label{eq:Einstein2_C_escalar_sintraza}
  \Psi^{(2)}-\Phi^{(2)} =
  \frac{3}{2}\triangle^{-1}\left\{\triangle^{-1}\partial^i\partial_j\Gamma_i\,^j -\frac{1}{3}\Gamma_k\,^k\right\},
\end{equation}
con $\Gamma_i\,^j\equiv\delta^{kj}\Gamma_{ik}$. La parte
vectorial de \eqref{eq:Einstein2_C}
\begin{equation}
  \label{eq:Einstein2_C_vectorial}
  \partial_\eta(a^2\nu_i^{(2)})=2a^2\triangle^{-1}\left\{\partial_i\triangle^{-1}\partial^k\partial_l\Gamma_k\,^l
    - \partial_k\Gamma_i\,^k\right\}.
\end{equation}
Nótese de las ecuaciones \eqref{eq:Einstein2_C_vectorial},
\eqref{eq:Einstein2_A_vectorial} que aún estableciendo como
las condiciones iniciales para las perturbaciones
vectoriales iguales a cero, estas se generarán debido a las perturbaciones
escalares a primer orden.

Finalmente, obtenemos la parte tensorial
\begin{multline}
  \label{eq:Einstein2_C_tensorial}
  (\partial_\eta^2+2\Hubble\partial_\eta-\triangle)\chi_{ij}\,^{(2)} = \\
  2\Gamma_{ij}-\frac{2}{3}\delta_{ij}\Gamma_k\,^k-3\left(\partial_i\partial_j-\frac{1}{3}\delta_{ij}\triangle\right)
  \triangle^{-1}
  \left(\triangle^{-1}\partial^k\partial_l\Gamma_k\,^l
    -\frac{1}{3}\Gamma_k\,^k\right) \\
  +4\left\{\partial_{(i}\triangle^{-1}\partial_{j)}\triangle^{-1}\partial^l\partial_k\Gamma_l\,^k
    - \partial_{(i}\triangle^{-1}\partial^k\Gamma_{j)k}\right\}.
\end{multline}

Es posible llegar a una \textit{ecuación maestra} para el
modo escalar de perturbación cosmológica de segundo orden
para un universo lleno con un único campo escalar. Para
lograrlo empezamos combinando \eqref{eq:Einstein2_B} con
\eqref{eq:Einstein2_C_escalar_traza}, obteniendo 
\begin{equation}\label{eq:Einstein2_B+C_1}
  \left(\partial_\eta^2-\Hubble\partial_\eta+
    \frac{2}{3}\triangle\right)\Psi^{(2)}  
  + \left(\Hubble\partial_\eta + \frac{1}{3}\triangle\right)\Phi^{(2)} 
  -8\pi\,G\partial_\eta\varphi\partial_\eta\bm\varphi^{(2)}=\Gamma_0+\frac{1}{6}\Gamma_k\,^k,
\end{equation}
y 
\begin{equation}\label{eq:Einstein2_B+C_2}
  \left(-\partial_\eta^2-5\Hubble\partial_\eta +
    \frac{4}{3}\triangle\right)\Psi^{(2)} 
  -\left(2\partial_\eta\Hubble + \Hubble\partial_\eta +
    4\Hubble^2 +
    \frac{1}{3}\triangle\right)\Phi^{(2)} 
  -8\pi\,Ga^2\bm\varphi^{(2)}\frac{\partial V}{\partial\varphi} =
  \Gamma_0 - \frac{1}{6}\Gamma_k\,^k.
\end{equation}

Manipulando \eqref{eq:Einstein2_A_escalar},
\eqref{eq:Einstein2_C_escalar_traza},
\eqref{eq:Einstein2_C_escalar_sintraza},
\eqref{eq:Einstein2_B+C_1} y \eqref{eq:Einstein2_B+C_2},
 se puede llegar a la
\textit{ecuación maestra} a segundo orden:
\begin{multline}\label{eq:master_2}
  \Bigg\{\partial_\eta^2+2\left(\Hubble -
    \frac{\partial_\eta^2\varphi}{\partial_\eta\varphi}\right)\partial_\eta
  - \triangle + 2\left(\partial_\eta\Hubble-
    \frac{\partial_\eta^2\varphi}{\partial_\eta\varphi}\Hubble\right)\Bigg\}\Phi^{(2)}
  \\
  =-\Gamma_0 - \frac{1}{2}\Gamma_k\,^k +
  \triangle^{-1}\partial^i\partial_j\Gamma_i\,^j +
  \left(\partial_\eta -
    \frac{\partial_\eta^2\varphi}{\partial_\eta\varphi}\right)\triangle^{-1}\partial^k\Gamma_k
  \\
  -\frac{3}{2}\left\{\partial_\eta^2-\left(2\frac{\partial_\eta^2\varphi}{\partial_\eta\varphi}-\Hubble\right)\partial_\eta\right\}
  \triangle^{-1}
  \left\{\triangle^{-1}\partial^i\partial_j\Gamma_i\,^j
    -\frac{1}{3}\Gamma_k\,^k\right\}.
\end{multline}
El sistema final de 10 ecuaciones para la perturbación de
segundo orden cosmológica para un Universo FLRW con campo
escalar está compuesto por \eqref{eq:Einstein2_A_escalar},
\eqref{eq:Einstein2_A_vectorial},
\eqref{eq:Einstein2_C_escalar_sintraza},~\eqref{eq:Einstein2_C_vectorial},
\eqref{eq:Einstein2_C_tensorial}, \eqref{eq:Einstein2_B+C_2}
y \eqref{eq:master_2}. Se puede observar en este conjunto de
ecuaciones que a pesar de que a primer orden las diferentes
ecuaciones para los modos escalares, vectoriales y
tensoriales no están acoplados, se acoplarán a segundo
orden; además, a pesar de hacer las perturbaciones
vectoriales y tensoriales a primer orden iguales a cero, el
acople entre diferentes generará estas perturbaciones.

\section{Conclusiones}

En este apéndice se discutieron las dificultades que
aparecen en el análisis perturbativo de Relatividad General.
En la discusión se identificó la raíz de este problema: el
principio de covariancia general. Debido al principio de
covariancia, es posible perturbar los ejes coordenados sin
perturbar las variables físicas, y esta ``perturbación
coordenada'' nublará los efectos físicos que se están
buscando. Por lo tanto, un buen método perturbativo en
Relatividad General elimina completamente estos grados de
libertad espurios.

Una vez identificado el problema se presentó un lenguaje
geométrico que permite plantear de una manera correcta el
análisis perturbativo. Se mostraron tres métodos diferentes
para eliminar los grados de libertad ficticios de la teoría
perturbativa de Relatividad General: (a) eligiendo o fijando
una norma, (b) el formalismo 1+3 covariante-invariante y (c)
el formalismo invariante de norma. De los tres enfoques
mostrados, se desarrolló en detalle el formalismo invariante
de norma \cite{NAKAMURA_1,NAKAMURA_4,Nakamura09a}. Se eligió
desarrollar este formalismo por ser el más general de los
tres ya que puede aplicar en espacios-tiempos generales y a
teorías covariantes sin restringirse a Relatividad General
si se cuenta con un procedimiento para extraer la parte
invariante de la perturbación de la métrica a primer
orden. Además, el formalismo invariante de norma cuenta con
un algoritmo para generar perturbaciones invariantes de
norma a cualquier orden superior al primero. Este formalismo
fue aplicado como ejemplo al espacio-tiempo de
Friedmann-Lemaître-Roberston-Walker (FLRW) con un campo
escalar.

Se incluye una hoja de cálculo en Maple para calcular las
ECE invariantes de norma a primer y segundo orden para la
métrica de FLRW usando el formalismo invariante de
norma. Debido a la generalidad del formalismo invariante de
norma, con pequeñas modificaciones se espera que pueda
generar ecuaciones invariantes en otras métricas, suponiendo claro,
que se tiene la descomposición de la perturbación de orden
lineal de la métrica de fondo.


\section{Fórmulas útiles}
\subsection{\label{sec:derivadas-de-lie}Derivada de Lie}
\begin{subequations}
  \begin{equation} \pounds_{t+u}W^a = \pounds_tw^a + \pounds_uw^a .
  \end{equation}
  \begin{equation} \pounds_{[v,w]}w^a = - \pounds^2_v w^a .
  \end{equation}
  \begin{equation} \pounds^2_{v+w}u^a = \left\{\pounds^2_v +
      2\pounds_v\pounds_w 
      + \pounds^2_w \right\}u^a.
  \end{equation}
  \begin{equation} \label{eq:Lie_def}
    \pounds_XT_{abc}\,^d =
    X^e\nabla_eT_{abc}\,^d+\nabla_aX^eT_{ebc}\,^d +\nabla_bX^eT_{aec}\,^d +
    \nabla_cX^eT_{abe}\,^d - \nabla_eX^dT_{abc}\,^e.
  \end{equation}
\end{subequations}
\subsection{Conmutación entre la derivada covariente y la
  derivada de Lie}
\begin{subequations}
  \begin{equation}
    \label{eq:lie_covariante_vector}
    \nabla_a\pounds_Xt_b = \pounds_X\nabla_at_b + X^cR_{acb}\,^dt_d + t_c\nabla_a\nabla_bX^c.
  \end{equation}
  \begin{equation}
    \label{eq:lie_cov_rango_0_2}
    \nabla_a\pounds_Xt_{bc} = \pounds_X\nabla_at_{bc} +
    X^dR_{adb}\,^et_{ec}+X^dR_{adc}\,^et_{be} 
    +t_{dc}\nabla_a\nabla_bX^d + t_{bd}\nabla_a\nabla_cX^d.
  \end{equation}
  \begin{equation}
    \label{eq:lie_cov_rango_1_1}
    \nabla_a\pounds_Xt_{b}\,^c = \pounds_X\nabla_at_{b}\,^c
    +
    X^dR_{adb}\,^et_{e}\,^c - X^dR_{ade}\,^ct_{b}\,^e 
    +t_{d}\,^c\nabla_a\nabla_bX^d -
    t_{b}\,^d\nabla_a\nabla_dX^c.
  \end{equation}
  \begin{multline}
    \label{eq:lie_cov_rango_0_3}
    \nabla_a\pounds_Xt_{bcd} = \pounds_X\nabla_at_{bcd} +
    X^eR_{aeb}\,^ft_{fcd} + X^eR_{aec}\,^ft_{bfd} +
    X^eR_{aed}\,^ft_{bcf}\\ + t_{ecd}\nabla_a\nabla_bX^e +
    t_{bed}\nabla_a\nabla_cX^e + t_{bce}\nabla_a\nabla_dX^e.
  \end{multline}
  \begin{multline}
    \label{eq:lie_cov_rango_1_2}
    \nabla_a\pounds_Xt_{bc}\,^d =
    \pounds_X\nabla_at_{bc}\,^d + X^eR_{aeb}\,^ft_{fc}\,^d +
    X^eR_{aec}\,^ft_{bf}\,^d -X^eR_{aef}\,^ft_{bc}\,^f\\ +
    t_{ec}\,^d\nabla_a\nabla_bX^e +
    t_{be}\,^d\nabla_a\nabla_cX^e +
    t_{bc}\,^e\nabla_a\nabla_eX^d.
  \end{multline}
\end{subequations}

\section{Series de Taylor de Campos
  Tensoriales}\label{sec:series-de-taylor}

  \begin{center}
    \textit{Esta apéndice y el siguiente están basados en el trabajo de
      Marco Bruni et al. \cite{BRUNI_1}.}
  \end{center}

Para funciones en $\mathbb{R}^m$, una expansión en series de
Taylor es esencialmente una manera conveniente de expresar
el valor de la función en un punto dado en términos de su
valor y el valor de todas sus derivadas en otro punto. Por
supuesto esta definición, tomada al pie de la letra,  es
imposible de llevar a cabo para un campo tensorial
$\mathcal{T}$ en una variedad $\mathcal{M}$, simplemente por
que $\mathcal{T}(p)$ y $\mathcal{T}(q)$ en $p, q \in
\mathcal{M}$, con $p \neq q$ pertenecen a diferentes
espacios y no pueden ser comparados directamente.

Una expansión de Taylor, por lo tanto, sólo puede ser
escrita si se es dado un mapeo entre dos campos tensoriales
en diferentes puntos de $\mathcal{M}$. El mapeo
$\phi_\lambda: \mathcal{M} \to \mathcal{M}$ es en
general una \textit{familia} uniparamétrica de
difeomorfismos generados por el campo vectorial $v^a$.  Un
caso especial se da cuando el mapeo es un \textit{grupo}
uniparamétrico de difeomorfismos. En este apéndice
consideremos la expansión de Taylor del \textit{pull-back}
de un tensor debido a un \textit{grupo} uniparamétrico de difeomorfismos,
como podría ser por ejemplo el mapeo exponencial. En el
siguiente apéndice se tratará el caso más general de una
familia uniparamétrica.





Sea $\mathcal{M}$ una variedad diferenciable y sea $\xi$ un
campo vectorial en $\mathcal{M}$, que genera un flujo $\phi:
\mathbb{R} \times \mathcal{M} \to \mathcal{M}$ donde
$\phi(0,p) = p$, $\forall p \in \mathcal{M}$. Para cualquier
$\lambda \in \mathbb{R}$ escribiremos $\phi_\lambda(p) \equiv
\phi(\lambda, p) \quad \forall p \in \mathcal{M}$. Sea
$\mathcal{T}$ un campo vectorial en $\mathcal{M}$. El mapeo
$\phi^*_\lambda$ define un nuevo campo $\phi^*_\lambda
\mathcal{T}$ llamado \textit{pull-back} de $\mathcal{T}$,
que es función de $\lambda$. Entonces en \cite{BRUNI_1} se
demuestra que

\begin{lemma}\label{lemma:taylor_diff}
  El campo $\phi_\lambda^*\mathcal{T}$ admite la expansión
  alrededor de $\lambda = 0$
  \begin{equation}\label{eq:taylor_diff}
    \phi^*_\lambda\mathcal{T} =
    \sum_{k=0}^{\infty}\frac{\lambda^k}{k!}\pounds_{\xi}^k\mathcal{T}.
  \end{equation}
\end{lemma}

\section{\label{sec:dife-de-caballo}Difeomorfismos de
  caballo}

Supóngase $\xi_{(1)}$ y $\xi_{(2)}$ en $\mathcal{M}$. Cada
uno de ellos generan los flujos $\phi^{(1)}$ y
$\phi^{(2)}$. Al combinarlos podemos formar la familia
uniparamétrica de difeomorfismos $\Psi : \mathbb{R} \times
\mathcal{M} \to \mathcal{M}$, cuya acción está dada mediante
\begin{equation}
  \label{eq:diff_caballo}
  \Psi_\lambda \equiv \phi^{(2)}_{\lambda^2/2} \circ \phi^{(1)}_\lambda
\end{equation}
Es decir, $\Psi_\lambda$ desplaza un punto en $\mathcal{M}$
un intervalo $\lambda$ a lo largo de las curva integral
$\xi_{(1)}$ y un intervalo $\lambda^2/2$ a lo largo de la
curva integral $\xi_{(2)}$. Por su similitud con el
movimiento de la pieza de ajedrez se le llama
\textit{difeomorfismo de caballo}.


Esta definición se puede generalizar a $n$ campos
vectoriales $\xi_{(1)}, ..., \xi_{(n)}$ en $\mathcal{M}$,
con flujos $\phi^{(1)},...,\phi^{(n)}$, de la siguiente
manera
\begin{equation}
  \label{eq:diff_caballo_generalizado}
  \Psi_\lambda \equiv \phi^{(n)}_{\lambda^n/n!} \circ ... \circ \phi^{(2)}_{\lambda^2/2}\circ\phi^{(1)}_\lambda.
\end{equation}
A los campos vectoriales $\xi_{(1)}, ..., \xi_{(n)}$ se les
denominan como los \textit{generadores} de $\Psi$.

Una cosa muy importante de notar es que $\Psi_\sigma \circ
\Psi_\lambda \neq \Psi_{\sigma+\lambda}$, ya que los
generadores no forman un grupo. Esto impide que podamos
aplicar el lema \ref{lemma:taylor_diff}, i.e.\ no podemos
expandir el \textit{pull-back} $\Psi^*_\lambda\mathcal{T}$
del tensor $\mathcal{M}$ definido $\mathcal{T}$. Pero como
veremos en el siguiente lema, el resultado
\ref{lemma:taylor_diff} puede ser generalizado.

\begin{lemma}\label{lemma:taylor_diff_caballo}
  El \textbf{pull-back} $\Psi^*_\lambda\mathcal{T}$ del
  campo tensorial $\mathcal{T}$ de la familia uniparamétrica
  de difeomorfismos de caballo $\Psi$ con generadores
  $\xi_{(1)}, ..., \xi_{(n)}$ se puede expandir alrededor de
  $\lambda = 0$ como
  \begin{equation}
    \label{eq:taylor_diff_caballo}
    \Psi^*_\lambda\mathcal{T}=\sum^\infty_{l_1=0}\sum^\infty_{l_2=0} \cdots \sum^\infty_{l_k=0} \cdots
    \frac{\lambda^{l_1+2l_2+\ldots+kl_k+\ldots}}{2^{l_2}\cdots(k!)^{l_k}\cdots
      l_1!l_2!\cdots l_k! \cdots}\pounds^{l_1}_{\xi_{(1)}}\pounds^{l_2}_{\xi_{(2)}}\cdots\pounds^{l_k}_{\xi_{(k)}}\cdots\mathcal{T}.
  \end{equation}
\end{lemma}

Los difeomorfismos de caballo parecen ser ejemplos muy
artificiales y de poca utilidad, pero como mostraremos en el
siguiente teorema, cualquier familia uniparamétrica de
difeomorfismos, puede siempre ser transformada o considerada
como una familia uniparamétrica de difeomorfismos de
caballo, de rango infinito en el caso general.

\begin{theorem}\label{teo:generalizacion}
  Sea $\Psi: \mathbb{R} \times \mathcal{M} \to \mathcal{M}$
  una familia de difeomorfismos. Entonces $\exists \quad
  \phi^{(1)}, \ldots \phi^{(k)}, \ldots$, grupos
  uniparamétricos de difeomorfismos en $\mathcal{M}$, tales
  que
  \begin{equation*}
    \Psi_\lambda = \cdots \circ \phi^{(k)}_{\lambda^2/k!} \circ \cdots \circ \phi^{(2)}_{\lambda^2/2}\circ\phi^{(1)}_\lambda.
  \end{equation*}
\end{theorem}

La demostración de este teorema se encuentra en
\citep*{BRUNI_1}.

\section{\label{sec:codigo-maple}Código}

El código que acompaña a este artículo ayuda a recuperar
todos las expresiones presentadas en el caso de FLRW. Se
proveen dos versiones ambas basadas en el lenguaje Maple,
una es directa, i.e.\ sin utilizar ninguna paquetería extra
y la segunda utilizando el paquete GRTensor II \cite{GRTensor}.  El código
en ambas versiones  se pueden encontrar en 
\url{http://nano.dialetheia.net/investigacion/codigo/perturbaciones}.

\chapter{Equivalencia de
  ecuaciones}\label{cha:equiv-de-ecuac}

Observando la literatura relacionada con perturbaciones
cosmológicas nos topamos con diferentes sistemas de
ecuaciones que están relacionados con la elección de una
norma. Estas ecuaciones pueden llegar a parecer muy
diferentes y es posible preguntarse si son equivalentes.

En este apéndice en particular compararemos las ecuaciones
(clásicas) a primer orden, obtenidas en la norma newtoniana
conforme o longitudinal y las ecuaciones invariantes de
norma.  Luego de mostrar su equivalencia compararemos su
cuantización con la establecida por Mukhanov.

\section{Introducción}

Un universo homogéneo e isotrópico (plano) es descrito por un
elemento de línea de Robertson-Walker
\begin{equation}
ds^2=a^2(\eta)\left(-d\eta^2+d\vec{x}^2\right)
\end{equation}
donde $a(\eta)$ es el factor de expansión, y $\eta$ es
el tiempo conforme, relacionado con el tiempo cósmico
mediante $d\eta = dt/a(\eta)$.

Este universo está lleno con un campo escalar,
$\varphi(\eta)$, este campo si cumple con que $p \simeq
-\rho$, i.e. que tenga presión negativa, 
provocará una etapa inflacionaria del universo.

Las ecuaciones a orden cero, conocidas como ecuaciones de Friedman-Lemaître, son:
\begin{subequations}
\begin{equation} \label{eq_friedman}
\Hubble^2 = -\frac{8\pi\,G}{3}a^2\rho =
-\frac{8\pi\,G}{3} a^2\left(\frac{1}{2a^2}\varphi'\,^2 + V(\varphi)\right)
\end{equation}
\begin{equation}\label{eq:aceleracion_flrw}
2\Hubble'+\Hubble^2 = -8\pi\,Ga^2p= -4\pi\,Ga^2\left(\frac{1}{2a^2}\varphi'\,^2 -
  V(\varphi)\right)
\end{equation}
\end{subequations}

De estas dos ecuaciones sólo una es no trivial,
\eqref{eq_friedman}. En las secciones siguientes será de
utilidad está expresión 
\begin{equation} \label{eq_cero_clave} 
\Hubble^2 - \Hubble ' = 4\pi\,G\varphi ' \,^2
\end{equation}

\subsection{Elemento de línea de FLRW perturbado}
La métrica de FLRW puede ser expandida a primer orden para
representar un universo perturbado ligeramente del estado
homogéneo e isotrópico:
\begin{equation}
  g_{ab} = \overline{g_{ab}} + h_{ab}
\end{equation}
Es una propiedad del universo FLRW que a primer orden se
pueda encontrar una descomposición tal que las
perturbaciones escalares, vectoriales y tensoriales se
desacoplen\footnote{Hay que recordar que estamos usando la
  terminología estándar y que perturbaciones escalares,
  vectoriales y tensoriales se refieren a sus propiedades de
  transformación ante rotaciones en el espacio de fondo.}
\ref{sec:pert-prim-orden}.

Aquí sólo consideraremos únicamente las perturbaciones
escalares. De esta manera la métrica perturbada escalarmente
es \footnote{En el apéndice \ref{cha:teor-de-pert} se
  acompaña con un factor extra de $a(\eta)^2$ a la variable
  perturbada $E$, que es necesaria para simplificar las
  expresiones vectoriales y tensoriales, que no son
  consideradas en este reporte, de ahí su omisión}:
\begin{equation}
  ds^2 = a(\eta)^2\Bigg\{-(1+A)d\eta^2 +
  2\partial_iBdx^id\eta \,+ 
  \Bigg[\left(1 + C -\frac{1}{3}\triangle
    E\right)\delta_{ij}+\partial_i\partial_j E
  \Bigg]dx^idx^j\Bigg\}.
\end{equation}
  
En este reporte utilizaremos sólo una norma: La Newtoniana
Conforme o longitudinal, que es caracterizada por la
siguiente elección:
\begin{equation}
  A = 2\phi, \qquad B = 0, \qquad C = -2\psi, \qquad E = 0, 
\end{equation}
entonces la métrica en la norma de newton conforme es
\begin{equation}
  ds^2 = a(\eta)^2\left\{-(1+2\phi)d\eta^2 + 
    (1 - 2\psi)\delta_{ij}dx^idx^j\right\}.
\end{equation}

Las variables escalares perturbadas invariantes de norma son
definidas mediante combinaciones lineales de las
perturbaciones $A, B, C, E$:
\begin{subequations}
  \begin{equation}
    \Phi  = \frac{1}{2}A - ( B - E')'+ \Hubble ( B - E')
  \end{equation}
  \begin{equation}
    \Psi = -\frac{1}{2}\left(C - \frac{1}{3}\triangle
      E\right) -   
    \Hubble(B- E')
  \end{equation}
  \begin{equation}
    \bm{\delta\varphi} = \delta\varphi + (B-E')\varphi'  .
  \end{equation}
\end{subequations}

\section{Ecuaciones en la Norma Newtoniana
  Conforme}\label{sec_nnc}
El sistema de ecuaciones de Einstein obtenidas en la norma
conforme de Newton, \citep[e.g]{Sudarsky06a}, son
\begin{subequations}
  \begin{equation}\label{ennc_1a}
    \triangle\psi - 3\Hubble\psi ' = 4\pi\,G\left(\varphi'\delta\varphi'+2a^2\phi
      V(\varphi) + a^2 \delta_\varphi
      V(\varphi)\delta\varphi\right) 
  \end{equation}
  \begin{equation}\label{ennc_2a}
    \partial_i\partial_\eta\psi + \Hubble\partial_i\phi = 4\pi\,G\varphi'\partial_i\delta\varphi
  \end{equation}
  \begin{multline}\label{ennc_3a}
    (\triangle-\partial_i\partial_j)(\phi-\psi)+2\psi''+2\left(2\Hubble'+\Hubble^2\right)(\psi+\phi)
    + 2\Hubble(\psi'+\phi') = \\
    8\pi\,G\left(-\phi\varphi'\,^2+\varphi'\delta\varphi'-\psi\left(\varphi'\,^2-a^2V(\varphi)\right) 
      - \frac{1}{2}a^2\partial_\varphi
      V(\varphi)\delta\varphi\right)
  \end{multline}
  \begin{equation}\label{ennc_4a}
    \partial_i\partial_j(\psi-\phi) = 0
  \end{equation}
\end{subequations}

De \eqref{ennc_4a} observamos que $\psi = \phi$. Usando este
resultado \eqref{ennc_2a} queda como
\begin{equation}\label{ennc_2b}
  \partial_i\left(\partial_\eta\psi + \Hubble\psi -
    4\pi\,G\varphi'\delta\varphi\right) = 0,
\end{equation}
de la cual se obtiene
\begin{equation}
  \partial_\eta\psi = - \Hubble\psi +
  4\pi\,G\varphi'\delta\varphi = 0.
\end{equation}

Sumando las ecuaciones \eqref{eq_friedman} y
\eqref{eq:aceleracion_flrw} obtenemos una expresión para el
potencial, $V(\varphi)$,
\begin{equation}\label{ennc_1-2_a}
  2a^2\,V(\varphi) = \frac{1}{4\pi\,G}(\Hubble'+2\Hubble^2).
\end{equation}

Usando las ecuaciones \eqref{ennc_2b} y \eqref{ennc_1-2_a}
la ecuación \eqref{ennc_1a} se convierte en
\begin{equation}
  \triangle\psi +\left(\Hubble^2 - \Hubble'\right)\psi =
  4\pi\, G\left[\varphi' \delta\varphi' + \left(3\Hubble\varphi' + 
      + a^2\partial_\varphi V(\varphi)\right) \delta\varphi\right]
\end{equation}

Definiendo $\mu\equiv \Hubble^2 - \Hubble'$ y
$u\equiv\left(3\Hubble\varphi' + + a^2\partial_\varphi
  V(\varphi)\right)$ llegamos a una ecuación tipo Poisson:
\begin{equation}\label{eq:poisson_nnc}
  \triangle\psi + \mu\psi = 4\pi\,G\left(u\delta\varphi +
    \varphi'\delta\varphi'\right).
\end{equation}

\section{Ecuaciones Invariantes de Norma}

El conjunto de ecuaciones invariantes de norma fué obtenido
en el apéndice \ref{cha:teor-de-pert} y se presenta a
continuación
\begin{subequations}
  \begin{equation}\label{ein_1}
    (-3\Hubble\partial_\eta + \triangle)\Psi -
    3\Hubble^2\Phi = 4\pi\,G \left(\partial_\eta\bm{\delta\varphi}\partial_\eta\varphi -
      \Phi(\partial_\eta\varphi)^2 + a^2\frac{dV}{d\varphi}\bm{\delta\varphi}\right)
  \end{equation}
  \begin{equation}\label{ein_2}
    \partial_\eta\partial_i\Psi +
    \Hubble\partial_i\Phi =
    4\pi\,G\partial_i\bm{\delta\varphi}\partial_\eta\varphi 
  \end{equation}
  \begin{multline}\label{ein_3}
    (-\triangle + 2\partial_\eta^2 + 4\Hubble\partial_\eta
    )\Psi \\+ (2\Hubble\partial_\eta + 4\partial_\eta\Hubble
    + 2\Hubble^2+\triangle)\Phi \\= 8\pi\,G
    \left(\partial_\eta\bm{\delta\varphi}\partial_\eta\varphi
      - \Phi(\partial_\eta\varphi)^2 -
      a^2\frac{dV}{d\varphi}\bm{\delta\varphi}\right)
  \end{multline}
  \begin{equation}\label{ein_4}
    \partial_i\partial^j\left(\Psi-\Phi\right) = 0
  \end{equation}
\end{subequations}
de esta última ecuación vemos que
\begin{equation}\label{eq:constriccion_dinamica_1}
  \Psi = \Phi  
\end{equation}

Reescribiendo las ecuaciones usando la igualdad anterior y
simplificando la notación $\partial_\eta\varphi \to
\varphi'$ y $V' \to V_\varphi$ obtenemos
\begin{subequations}
  \begin{equation}\label{ein_1b}
    -3\Hubble\Psi ' + \triangle\Psi -
    3\Hubble^2\Psi = 4\pi\,G \left(\bm{\delta\varphi} '
      \varphi ' - 
      \Psi\varphi '\,^2 + a^2V_\varphi\bm{\delta\varphi}\right)
  \end{equation}
  \begin{equation}\label{ein_2b}
    \partial_\eta\partial_i\Psi +
    \Hubble\partial_i\Psi =
    4\pi\,G\partial_i\bm{\delta\varphi}\varphi '
  \end{equation}
  \begin{equation}\label{ein_3b}
    \Psi '\,' + 3\Hubble\Psi ' +
    + 2\Hubble ' \Psi +
    \Hubble^2\Psi  = 4\pi\,G
    \left( \bm{\delta\varphi} ' \varphi ' -
      \Psi \varphi '\,^2 - a^2 V_\varphi
      \bm{\delta\varphi} \right).  
  \end{equation}
\end{subequations}

Integrando \eqref{ein_2b} respecto a $x_i$ y escogiendo como
cero la constante de integración.
\begin{equation}\label{eq:constriccion_dinamica_2}
  \Psi '  + \Hubble\Psi =
  4\pi\,G\bm{\delta\varphi}\varphi ',
\end{equation}
acomodando términos
\begin{subequations} \label{ein_base}
  \begin{equation}\label{ein_1c}
    -3\Hubble \left(\Psi ' +  \Hubble\Psi\right) +
    \triangle\Psi  
    = 4\pi\,G \left(\bm{\delta\varphi} ' \varphi ' -
      \Psi\varphi '\,^2 + a^2
      V_\varphi\bm{\delta\varphi}\right) 
  \end{equation}
  \begin{equation}\label{ein_2c}
    \partial_i\Psi ' = 4\pi\,G\partial_i\bm{\delta\varphi}\varphi ' - \Hubble\partial_i\Psi 
  \end{equation}
  \begin{equation}\label{ein_3c}
    \Psi '\,' + 3\Hubble\Psi ' +
    + 2\Hubble ' \Psi +
    \Hubble^2\Psi  = 4\pi\,G
    \left( \bm{\delta\varphi} ' \varphi ' -
      \Psi \varphi '\,^2 - a^2 V_\varphi \bm{\delta\varphi}
    \right)  
  \end{equation}
\end{subequations}

\subsection{Ecuación maestra }
En el apéndice \ref{cha:teor-de-pert} a partir del conjunto
de ecuaciones \eqref{ein_base} (o equivalentemente de las
ecuaciones de la \S \ref{sec_nnc}) se obtiene una sola
ecuación de segundo orden, conocida como \emph{ecuación
  maestra}. Aquí se reproduce el cálculo por razones de
completitud.

Sumemos \eqref{ein_1c} y \eqref{ein_3c}
\begin{equation}
  \Psi'' + 2\Hubble '\Psi  - 2\Hubble^2\Psi +
  \triangle\Psi = 8\pi\,G (\delta\varphi ' \varphi ' - \Psi\varphi'\,^2),
\end{equation}
sustituyendo \eqref{eq_cero_clave}
\begin{equation}\label{ein_3a+c}
  \Psi'' + \triangle\Psi = 8\pi\,G\delta\varphi ' \varphi '.
\end{equation}

Ahora manipulamos algebraicamente \eqref{ein_2c}, obteniendo
\begin{equation}\label{eq_perturbacion_evolucion}
  4\pi\,G\delta\varphi = \frac{\Psi ' +
    \Hubble\Psi}{\varphi '} ,
\end{equation}
obviamente este caso es válido, sólo si $\varphi ' \neq 0$.
Derivando respecto a $\eta$
\begin{equation}
  \delta\varphi ' =
  \frac{(\Psi''+\Hubble'\Psi+\Hubble\Psi')\varphi' -
    (\Psi'+\Hubble\Psi)\varphi''}{\varphi'\,^2},
\end{equation}
insertando esta ecuación en \eqref{ein_3a+c}
\begin{align*}
  \Psi'' + \triangle\Psi &=
  2\left\{\frac{(\Psi'' + \Hubble'\Psi +
      \Hubble\Psi')\varphi' -
      (\Psi'+\Hubble\Psi)\varphi''}{\varphi'}\right\}
  \nonumber \\
  \Psi''+\triangle\Psi &=
  2(\Psi''+\Hubble'\Psi+\Hubble\Psi') -
  2(\Psi'+\Hubble\Psi)\frac{\varphi''}{\varphi'}
  \nonumber
\end{align*}
reacomodando, llegamos
\begin{equation}\label{eq:maestra}
  \Psi'' + 2\left(\Hubble -
    \frac{\varphi''}{\varphi}\right)\Psi' -
  \triangle\Psi + 
  2\left(\Hubble'-\Hubble\frac{\varphi''}
    {\varphi}\right)\Psi = 0
\end{equation}

Nótese que para llegar a esta ecuación usamos el conjunto
completo de las ecuaciones \eqref{ein_base}. Una cosa
interesante es que si se define la variable\footnote{Estas
  son, salvo un factor de $a$ en $Y$ y $w$, las variables
  $u$ y $\theta$ de la \S \ref{sec:formacion_estructura}. }
$Y$,   
\begin{equation}\label{eq:variable_maestra}
  Y \equiv 2\frac{\Psi}{\varphi'}  
\end{equation}
La ecuación \eqref{eq:maestra} se puede reescribir como
\begin{equation}\label{eq:oa_grav}
  Y'' - \triangle Y - \frac{w''}{w} = 0  
\end{equation}
con $w \equiv \Hubble/\varphi'$. Aquí obtenemos por primera
vez en este reporte, una ecución del tipo oscilador armónico
con masa variable efectiva ($m^2_{variable} \equiv w''/w$).

\subsection{Ecuación de Poisson}
A diferencia de lo que se hizo en la subsección anterior,
en la cual se obtuvo la 
\emph{ecuación maestra}, a partir de las ecuaciones
(\ref{ein_base}), aquí 
se seguirá el método de \cite{Sudarsky06a}, i.e. obtener la
ecuación tipo Poisson.

Usando \eqref{eq_cero_clave} en \eqref{ein_1c}
\begin{equation}\label{ein_1d}
  \triangle\Psi - 3\Hubble\Psi ' - 2\Hubble^2\Psi  - \Hubble
  '\Psi = 4\pi\,G\left(\delta\varphi ' \varphi ' +
    a^2V_\varphi\delta\varphi\right).
\end{equation}

Sustiyendo en esta ecuación \eqref{ein_2c}
\begin{equation}\label{ein_1e}
  \triangle\Psi + \left(\Hubble^2 - \Hubble '\right)\Psi =
  4\pi\,G\left(3\Hubble\varphi ' +
    a^2V_\varphi\right)\delta\varphi + 4\pi\,G\delta\varphi '
  \varphi '.
\end{equation}

Reescribiendo \eqref{ein_1e}
\begin{equation}
  \triangle\Psi + \mu\Psi = 4\pi\,G (u\, \delta\varphi +
  \delta\varphi ' \varphi '),
\end{equation}
donde $\bm\nu$ se define como
\begin{equation}
  \bm\nu \equiv 3\Hubble\varphi ' +   a^2V_\varphi,
\end{equation}
y $\bm\mu$ está definida
\begin{equation}
  \bm\mu \equiv = \Hubble^2 - \Hubble'
\end{equation}

Esta ecuación tiene la misma forma que la obtenida en la
\S \ref{sec_nnc}. A diferencia del procedimiento para
derivar la ecuación 
maestra, aquí no se usó el conjunto completo de ecuaciones
\eqref{ein_base}: la ecuación \eqref{ein_3c} nunca es
utilizada.

\section{Cuantización de $y = a \dphi$}
En esta sección y la siguiente no mostraremos el proceso de
cuantización ya que es el método estándar, en su lugar
presentaremos de el punto de partida
tanto para  el proceso de cuantización  de $y = a\dphi$
\citep[ver][]{Sudarsky06a} como el la variable 
de Mukhanov \citep[ver][]{Mukhanov88}.

El lagrangiano $\mathcal{L}$ que describe el campo escalar
del inflatón es
\begin{equation}
  \mathcal{L} =
  -\frac{1}{2}g^{ab}\partial_a\varphi\partial_b\varphi -
  V(\varphi),
\end{equation}
para cuantizar utilizamos la acción $S$,
\begin{equation}
  S[\varphi,g_{ab}] = \int \sqrt{-g}\,\mathcal{L} \, d\,^4x  .
\end{equation}

Siguiendo \cite{Sudarsky06a} ignoraremos el
efecto de la perturbación de la métrica en este
reporte. Expandiendo la acción a primer orden en el campo
escalar, variando respecto a la perturbación,
$\delta\varphi$ y finalmente usando la ecuación dinámica del
campo inflatónico sin perturbar, obtenemos la ecuación de
movimiento de la perturbación \footnote{En el artículo
  \cite{Sudarsky06a} se desprecia el término
  del potencial ($a^2V_{\varphi\varphi}\delta\varphi$)
  también en base a la aproximación de \emph{slow-roll}.}
\begin{equation}\label{eq:inflaton_dinamica_perturbada}
  \delta\varphi'' + 2\Hubble\varphi'-\triangle\delta\varphi
  + a^2V_{\varphi\varphi}\delta\varphi = 0.
\end{equation}

De esta manera la dinámica del sistema perturbado queda
definida por las ecuaciones \eqref{eq:poisson_nnc} y
\eqref{eq:inflaton_dinamica_perturbada} mas las ecuaciones a
orden cero.

Si se define la variale $y \equiv a\delta\varphi$, la
ecuación \eqref{eq:inflaton_dinamica_perturbada}, se puede
reescribir como
\begin{equation}\label{eq:oa_sudarsky}
  y'' - \triangle y - \frac{a''}{a}y = 0  
\end{equation}
Esta ecuación una vez más tiene la forma de un oscilador
armónico con masa variable 
\begin{equation}
  m_{variable}^2 \equiv \frac{a''}{a}  
\end{equation}

\section{Cuantización de \emph{à la}
  Mukhanov}\label{sec:cuant-mukhanov} 
Empezamos con la acción
\begin{equation}
  S[\varphi, g_{ab}] = \int \left\{ R \sqrt{-g} \right\}d^4x
  + \int 
  \left\{\left[-\frac{1}{2}g^{ab}\partial_a
      \varphi\partial_b\varphi 
      -  V(\varphi)\right] \sqrt{-g}\right\}d^4x
\end{equation}
expandiendo a segundo orden en la métrica y en el campo
\begin{equation}
  S[\varphi + \delta\varphi, g_{ab} = \overline{g}_{ab} +
  h_{ab}] =
  S^{(0)}[\varphi,\overline{g}_{ab}]+S^{(1)}[\delta\varphi,h_{ab};
  \varphi,\overline{g}_{ab}]+S^{(2)}[\delta\varphi, 
  h_{ab};\varphi,\overline{g}_{ab}] 
\end{equation}
donde el primer término contiene sólo la parte homogénea,
$S^{(1)}$ está constituido por términos lineales en las
perturbaciones y $S^{(2)}$ está formado por terminos
cuadráticos en las perturbaciones
\cite{Mukhanov88,Mukhanov90}. 

Para escribir esta expresión se necesitarán las siguientes
expansiones:\footnote{Se usaron las conocidas expresiones:
  \begin{align*}
    (1+x)^{-1} &= 1  - x + x^2 \ldots & -1 \le x \le 1    \\
    (1+x)^{1/2} &= 1 + \frac{1}{2}x - \frac{1}{2 \times
      4}x^2 + \ldots & -1 \le x \le 1
  \end{align*}
}
\begin{subequations}
  \begin{equation}
    g_{00}= -a^2(1+2\Phi), \qquad  g_{ij} = a^2(1-2\Psi)
  \end{equation}

\begin{equation}
  g^{00} \simeq a^{-2}(-1+2\Phi-4\Phi^2), \qquad g^{ij}\simeq
  a^{-2}(\delta^{ij} + 2\Psi\delta^{ij} + 4\Psi^2\delta^{ij})
\end{equation}
\begin{equation}
  \sqrt{-g} \simeq a^4 \left(1+\Phi - \frac{1}{2}\Phi^2-3\Psi -
    3\Psi\Phi + \frac{3}{2}\Psi^2\right)
\end{equation}
\begin{multline}
  \mathcal{L} = -\frac{1}{2}a^{-2}\Big(-\varphi'\,^2 -
  2\delta\varphi'\varphi' - \delta\varphi'\,^2 +
  2\Phi\varphi'\,^2 + \\
  4\delta\varphi'\Phi\varphi'-4\Phi^2\varphi'\,^2+\delta\varphi_{,i}\delta\varphi_{,i}\Big)
  - V -\delta\varphi V_{\varphi} -
  \frac{1}{2}\delta\varphi^2 V_{\varphi\varphi}
\end{multline}
\end{subequations}

Entonces la parte de la acción que corresponde al campo
escalar es
\begin{multline}
  S_{mat} = \int d\,^4 x \Bigg[
  \frac{a^2}{2}\Big(\varphi'\,^2 + 2\delta\varphi'\varphi' +
  \delta\varphi'\,^2 -
  2\Phi\varphi'\,^2  \\
  -
  4\delta\varphi'\Phi\varphi'+4\Phi^2\varphi'\,^2-\delta\varphi_{,i}\delta\varphi_{,i}\Big)
  - a^4 V -a^4\delta\varphi V_{\varphi} -
  \frac{a^4}{2}\delta\varphi^2 V_{\varphi\varphi} \\
  +\frac{a^2}{2}\Big(\varphi'\,^2 \Phi +
  2\delta\varphi'\varphi'\Phi \Big)
  - a^4 V\Phi -a^4\delta\varphi \Phi V_{\varphi} \\
  -\frac{a^2}{4}\varphi'\,^2\Phi^2 + \frac{a^4}{2} V \Phi^2
  -\frac{3a^2}{2}\Big(\varphi'\,^2\Psi +
  2\delta\varphi'\varphi'\Psi \Big) + 3a^4V\Psi +
  3a^4\delta\varphi\Psi V_{\varphi} \\
  -\frac{3a^2}{2}\varphi'\,^2\Psi\Phi + 3a^4 V \Phi\Psi
  +\frac{3a^2}{4}\varphi'\,^2\Psi^2 - \frac{3a^4}{2} V\Psi^2
  \Bigg]
\end{multline}
En esta ecuación podemos identificar
\begin{equation}
S^{(0)}_{mat} = \int \left(
\frac{a^2}{2}\varphi'\,^2 - a^4V \right)d\,^4x
\end{equation}
Esta es la acción original (o la parte que describe el campo
escalar). Nada sorprendente.
\begin{multline}
  \delta S^{(1)}_{mat}   = \int \Big[\frac{a^2}{2}\left(
    \varphi'\,^2 - 
    2\delta\varphi'\varphi' - 2\Phi\varphi'\,^2
    +\varphi'\,^2\Phi - 3\varphi'\,^2\Psi 
\right) \\ + 3a^4V\Psi - a^4 V \Phi  -a^4\delta\varphi V_\varphi\Big] d\,^4x
\end{multline}
Integrando por partes $S^{(1)}$ y sustituyendo las
ecuaciones del universo FLRW
\eqref{eq_friedman},~\eqref{eq:aceleracion_flrw} y
\eqref{eq_cero_clave}, encontraremos que $S^{(1)} = 0$. Esto
no es sorprendente por que esa es la manera en la que
obtenemos las ecuaciones de Euler-Lagrange que al aplicar en
$\mathcal{L}^{(0)}$ nos da la ecuación dinámica del campo
homogéneo. 
\begin{multline}\label{eq:accion_perturbada_2}
  \delta S^{(2)}_{mat} = \int \Big[\frac{a^2}{2}
  \Big(\delta\varphi'\,^2 - 
    4\delta\varphi'\Phi\varphi' + 4\Phi^2\varphi'\,^2 -
    \delta\varphi_{,i}\delta\varphi_{,i} \\
   + 2\delta\varphi'\Phi\varphi' - 
    \frac{1}{2}\varphi'\,^2\Phi^2 - 3\varphi'\,^2\Psi -
    6\delta\varphi'\Psi \varphi' - 3\varphi'\,^2\Psi\Phi +
    \frac{3}{2}\varphi'\,^2\Psi^2\Big) \\
- \frac{a^4}{2}\delta\varphi^2V_{\varphi\varphi} -
a^4\delta\varphi\Phi V + \frac{a^4}{2} \Phi^2 V +
3a^4\delta\varphi\Psi V_\varphi + 3a^4 \Phi\Psi V  -
\frac{3a^4}{2}\Psi^2 V \Big] d\,^4x
\end{multline}
Este pedazo de la acción es el que nos interesa. Variando
\eqref{eq:accion_perturbada_2} juntamente con $\delta
S^{2}_{grav}$ respecto a $\Psi$, $\Phi$ y $\delta\varphi$,
obtenemos ecuaciones de movimiento que son equivalentes a
las obtenidas en secciones anteriores. Si utilizamos las
constricciones \eqref{eq:constriccion_dinamica_1} y
\eqref{eq:constriccion_dinamica_2} para eliminar dos de las
tres variables, y se aplican una vez más las ecuaciones de
FLRW\footnote{¡De verdad es una gran cantidad de trabajo
  algebraico!}, obtenemos una acción para un sólo grado de
libertad, la llamada variable de Mukhanov-Sasaki, $\nu$.

La variación de la acción ahora toma la forma
\begin{equation}
  \delta S^{(2)}_{grav\, + \,mat} = \frac{1}{2}\int d\,^4x\Big[\nu'\,^2 -
  (\triangle \nu)^2 + \frac{z''}{z}\nu^2\Big],
\end{equation}
con $z\equiv\varphi'/\Hubble$ y $\nu$ definida mediante
\begin{equation}\label{eq:mukhanov}
  \nu \equiv a\left(\delta\varphi   + \frac{\varphi'}{\Hubble}\Psi\right)
\end{equation}
La ecuación de movimiento para $\nu$ es 
\begin{equation}\label{eq:os_mukhanov}
  \nu'' - \triangle\nu - \frac{z''}{z}\nu = 0  
\end{equation}
Una vez más obtenemos la ecuación de un oscilador armónico
con masa variable, pero ahora esta masa es $m^2_{variable}
\equiv z''/z $.

\section{Conclusiones}

En este apéndice se mostró que el conjunto de ecuaciones
derivadas de las ecuaciones de Einstein para el caso FLRW
perturbado en la norma longitudinal conforme tiene la misma
forma que las obtenidas de manera independiente de la
elección de norma. Esta equivalencia estructural se mantiene ya que
$\Phi$, $\Psi$ invariantes se reducen a las newtonianas
cuando $B = E = 0$.  Esta equivalencia se mantiene tanto en
el formato \emph{ecuación tipo Poisson}, como en la llamada
\emph{ecuación maestra}. La diferencia de estas dos es que
en la última si se llega a una expresión que sólo incluye un
grado de libertad\footnote{Aquí hay que ser cuidadoso ya que
  aparecen en realidad dos ecuaciones: \eqref{eq:maestra} y
  \eqref{ein_3a+c}. En la literatura regularmente se hace la
aseveración existe  un sólo grado de libertad (en este caso
$\Psi$) y le llaman a la 
segunda ecuacuión \emph{ecuación de constricción}, en realidad esto
es otro nombre para indicar que ambas variables estan acopladas.}.

En el aspecto de cuantización se muestró el desarrollo para
llegar a la variable de Mukhanov, como único grado de
libertad, a diferencia, una vez más, del caso ecuación
Poisson/campo escalar. De hecho usando la definición de la
variable de Mukhanov, \eqref{eq:mukhanov}, las
constricciones \eqref{eq:constriccion_dinamica_1},
\eqref{eq:constriccion_dinamica_2} y
\eqref{eq:variable_maestra}, se demuestra
\cite{Deruelle92} que $Y$ y $\nu$ están
relacionados mediante
\begin{equation}
  \nu = Y' + \frac{z'}{z}Y  
\end{equation}
siendo estas variables equivalentes\footnote{La diferencia
  se notará en el momento de cuantizar, pero ese es tema del
  próximo reporte.}.

Pero para nuestros fines, es decir, estudiar el colapso de
la función de onda, causado por la gravedad, el sistema de
ecuaciones Poisson/escalar, es el más apropiado ya que
muestra la relación que existe entre estas variables
cuánticas y los grados de libertad de la métrica.

 \chapter{Línea Temporal}

A lo largo de este trabajo de tesis hicimos referencia a
varios momentos importantes de la evolución de nuestro
universo. Para facilitarle al lector la consulta de los
valores numéricos en varios tipos de unidades, hemos
resumido aquí esos datos.

Hemos dividido la historia de universo en tres etapas
principales: (a) época inflacionaria, (b) época dominada por
radiación y (c) época dominada por materia, e ignoramos la
época actual dominada por la energía oscura.

El factor de escala durante estas épocas tiene la ecuación
de evolución
\begin{equation}
  \label{eq:a_evolucion_epocas}
  a(\eta) = \left\{ \begin{array}{lc}
      -\frac{1}{H_I\eta},
      &\mbox{$-\infty<\eta < \eta_{ei}, $} \\
      \\
      C_{rad}
      \left(\eta - \eta_{ei}\right) + a_{ei},
      &\mbox{$\eta_{ei} 
        \le \eta < \eta_{eq}$} \\
      \\
      \left[ \frac{1}{2} C_{mat} \left(\eta - \eta_{eq}\right) +
        \sqrt{a_{eq}}\right]^2, 
      & \mbox{ $\eta_{eq} \le \eta$}
    \end{array} \right.
\end{equation}
donde las constantes (exceptuando $C_{mat}$ y $C_{rad}$) se
obtienen por continuidad de la función. Tomaremos como
normalización del factor de escala a
\begin{equation}
  \label{eq:normalizacion_factor_escala}
  a_0 \equiv a(\eta = \eta_0) = 1.
\end{equation}

Las constantes $C_{rad}$ y $C_{mat}$ de la ecuación
(\ref{eq:a_evolucion_epocas}) están dadas por
\begin{equation}
  \label{eq:constantes}
  C_{rad}^2 = \frac{8\pi G}{3} \left(\rho_{rad} a^4\right),
  \quad C_{mat}^2 = \frac{8\pi G}{3} \left(\rho_{mat} a^3\right),
\end{equation}
donde, las cantidades $(\rho_{rad} a^4)$ y $(\rho_{mat}
a^3)$ se conservan, como se puede constatar usando la
ecuación (\ref{eq:conservacion_conforme_fl}). En estas
constantes $\rho_{rad}$, $\rho_{mat}$ representan la
densidad de energía de la radiación y de la materia
respectivamente. Si usamos la conservación de $(\rho_{rad}
a^4)$, $(\rho_{mat} a^3)$ y el factor de normalización
(\ref{eq:normalizacion_factor_escala}) podemos expresar
(\ref{eq:constantes}) como sigue
\begin{equation}
  \label{eq:constantes_2}
  C_{rad}^2 = \frac{8\pi G}{3} \rho_{rad,0} = H_0^2 \Omega_{rad,0},
  \quad C_{mat}^2 \frac{8\pi G}{3}\rho_{mat,0} = H_0^2 \Omega_{mat,0},
\end{equation}
donde $H_0$ es la constante de Hubble y $\Omega{X,0} \equiv
\rho_{X,0} / \rho_c$ es la densidad de energía de la especie
en cuestión relativa a la densidad crítica $\rho_c$.

Con estas ecuaciones y relaciones podemos calcular el factor
de escala al tiempo $\eta_{eq}$ cuando la densidad de
energía de la radiación y de la materia son iguales, $a_{eq}
\equiv a(\eta = \eta_{eq})$
\begin{equation}
  \label{eq:a_eq}
  a_{eq} = \frac{\Omega_{rad,0}}{\Omega_{mat,0}}.
\end{equation}

Evaluando ahora la evolución del factor de escala durante
radiación (segunda ecuación de \ref{eq:a_evolucion_epocas})
en $\eta = \eta_{eq}$ llegamos a una expresión para
$\eta_{eq}$,
\begin{equation}
  \label{eq:eta_eq}
  \eta_{eq} = \frac{a_{eq} - a_{ei}}{C_{rad}} + \eta_{ei}.
\end{equation}

Propiamente, la expresión $(Ta) = $ constante, es válida
(tiene sentido) cuando (1) podemos hablar de temperatura en
el sistema, (2) existe radiación en la época en cuestión y
posee la mayoría de los grados de libertad del sistema. A
pesar de esto, en este apéndice supondremos la validez de
esa relación, aún en épocas en las cuales ni (1), ni (2) se
cumplen (e.g.\ durante la época inflacionaria) y habrá que
entenderlo como ``\textit{la temperatura hipotética que
  tendría la radiación a esa escala}''. Entonces, como $(Ta)
= $ constante, calculamos el factor de escala cuando
inflación estaba terminando para dar paso a al época
dominada por radiación,
\begin{equation}
  \label{eq:a_ei}
  a_{ei} = \frac{T_0}{T_{ei}}, \quad \eta_{ei} = -
  \frac{1}{a_{ei} H_{inf}},
\end{equation}
donde usamos la tercera expresión de
(\ref{eq:a_evolucion_epocas}). Para obtener un punto de
referencia se calculará el valor del factor de escala
durante la época de Planck,
\begin{equation}
  \label{eq:a_planck}
  a_{Planck} = \frac{T_0}{T_{Planck}}, \quad \eta_{Planck} = -
  \frac{1}{a_{Planck} H_{inf}}.
\end{equation}

Usando el requerimiento de que inflación dure al menos $80$
e-folds ($\sim 60$ es el mínimo necesario) podemos calcular
$a_{ii}$ y $\eta_{ii}$
\begin{equation}
  \label{eq:a_ii}
  \frac{a_{ei}}{a_{ii}} = e^{80}, \quad \eta_{ii} =
  -\frac{1}{a_{ii} H_{inf}}.
\end{equation}

Finalmente, durante la época inflacionaria $H_{inf}$ se
supone constante o muy cercano a serlo, su valor se puede
calcular de la ecuación de Friedmann,
\begin{equation}
  \label{eq:h_inf}
  H_{inf}^2 = \frac{8\pi G}{3} V(\varphi).
\end{equation}

Para obtener los valores numéricos mostrados en la tabla
\ref{tab:valores_numericos},
usamos los siguientes datos $\hbar = c = 1$, $\rho_c = 8.098
h^2 \times 10^{-14}\thinspace \electronvolt^4$, $T_0 \equiv T( \eta =
\eta_0) = 2.4 \times 10^{-13} \thinspace \giga\electronvolt$,
$\Omega_{mat,0} = 0.128 h^2 = 0.2275$, $\Omega_{rad,0} =
2.47\times 10^{-5} h^2 = 0.0000439$, $m_{Planck} = 1.221
\times 10^{19}\thinspace \giga\electronvolt$ y $G =
m_{planck}^2$. Supondremos que la escala energética del
inflatón al final del periodo inflacionario es $T_{ei} =
10^{15} \thinspace\giga\electronvolt$ y por lo tanto $V(\varphi)
\simeq (T_{ei})^4$. Los valores numéricos de las constantes
son $C_{rad} = 0.161 \times 10^{-5} \thinspace
\mega\parsec^{-1}$, $C_{mat} = \unit{0.00011}
\mega\parsec^{-1}$ y $H_{inf} = 0.36\times 10^{36} \thinspace
\second^{-1} = 0.2370 \times 10^{12} \thinspace
\giga\electronvolt = 0.36\times 10^{50} \thinspace \mega\parsec^{-1}$.

\begin{table}
  \begin{center}
    \begin{tabular}{|c|c|c|c|}
      \rowcolor[rgb]{0.8,0.8,0.8}  $\eta$ &  $\second$  
      &  $\mega\parsec$ &  $a(\eta)$\\  
      \hline 
      $\eta_{ii}$  &$-0.641\times
      10^{27}$ & $-0.64\times 10^{13}$ &$0.43 \times 10^{-62}$  \\
      $\eta_{Planck}$ & $-0.1157\times 10^{-3}$ &
      $-0.1157\times 10^{-17}$ &  $0.24 \times 10^{-31}$ \\
      $\eta_{ei}$ & $-0.1157\times 10^{-7}$ & $-0.1157 \times
      10^{-21}$ & $0.24\times 10^{-27}$  \\
      $ \eta = 0  $  & $0$ &  \, & $0.42 \times
      10^{-27}$ \\
      $\eta_{ns} $ & $15\times10^{9}$ & $0.00015$ & $0.24 \times
      10^{-9}$ \\
      $\eta_{eq}$ & $1.2 \times 10^{16}$ & $119.81$  &
      $0.00019296$ \\ 
      $\eta_{d}$  & $2.6 \times 10^{16}$ & $260.06$ & $0.0009$ \\
      $\eta_0$  &$8.6 \times 10^{17}$  & $8625.32$  & $1$ \\
      \hline 
    \end{tabular}
    \caption{Valores numéricos para distintos momentos
      importantes en la evolución del universo. La
      conversión se puede realizar usando $\unit{1} \second
      \simeq 10^{-14}\thinspace \mega\parsec$. Si requiere
      estas cantidades en $\giga\electronvolt$ use la
      relación de conversión $\unit{1} 
      \giga\electronvolt^{-1} = 6.6\times 10^{-25}\thinspace
      \second$. Para los valores de las constantes
      consúltese el texto
      principal.}\label{tab:valores_numericos}
  \end{center}
\end{table}

\chapter{Acrónimos}

\begin{acronym}[FLRW]
  \acro{CBR}{Fondo de Radiación Cósmica
    \acroextra{\textit{Cosmic Background Radiation}}}
  \acro{CMB}{Fondo Cósmico de Micro-ondas \acroextra{, del
      inglés \textit{Cosmic Microwave Background}}}
  \acro{COBE}{\textit{COsmic Background Explorer}}
  \acro{CSL}{Modelos de localización
    contínua \acroextra{del inglés \textit{Continuous
        Spontaneous Model}}}
  \acro{DE}{Energía oscura \acroextra{, del inglés
      \textit{dark energy}}}
  \acro{DM}{Materia oscura \acroextra{, del inglés
      \textit{dark matter}}}
  \acro{DMR}{Radiómetro de Microondas  Diferencial
    \acroextra{del inflés \textit{Differential Microwave
        Radiometers}}} 
  \acro{DRM}{Modelos de reducción dinámicos \acroextra{, del
      inglés \textit{Dynamical Reduction  Models}}}
  \acro{ECE}{Ecuaciones de Campo de Einstein}
  \acro{EW}{Electro-débil \acroextra{, del inglés
      \textit{electro-weak}}}
  \acro{FIRAS}{Espectrómetro Absoluto del Infrarojo Lejano
    \acroextra{, del inglés \textit{Far Infrared 
        Absolute Spectrophotometer}}} 
  \acro{FL}{Friedmann-Lemaître}
  \acro{FLRW}{Friedmann-Lemaître-Robertson-Walker}
  \acro{GRW}{Modelo de colapso de la función de onda de
    Ghirardi-Rimini-Weber} 
  \acro{ISW}{Efecto de Sachs-Wolfe Integrado \acroextra{, del
      inglés \textit{Integrated Sachs-Wolfe}}}
  \acro{LEP}{Gran colisionador de Electrones-Positrones
    \acroextra{del inglés \textit{Large Electron-Positron
        Collider}}}
  \acro{LHC}{Gran Colisionador de Hadrones \acroextra{, del
      inglés \textit{Large Hadron Collider}}}
  \acro{LSS}{Superficie de Última Dispersión \acroextra{, del
      inglés \textit{ Last Scattering Surface}}} 
  \acro{PMO}{Problema de la macro-objetificación}
  \acro{RW}{Robertson-Walker}
  \acro{SW}{Efecto de Sachs-Wolfe}
  \acro{WMAP}{\textit{Wilkinson Microwave Anisotropy Probe}}
\end{acronym}

   \PCFbackmatter
\end{document}